\pdfoutput=1 
\documentclass[11pt]{book}
\usepackage{a4wide}
\usepackage{epsfig,psfig,amsfonts,graphicx,rotating,multicol,color}
\usepackage{setspace,fancyheadings,chapterbib,longtable,amssymb}
\usepackage{array,colortbl,lscape}
\usepackage{astron,astbibref}

\def\arcsec{\ensuremath{\!\!^{\textrm{''}}\!}}
\def\HII{H{\sc ii}}

\def\uno{SDSS J145506.06+380816.6}  
\def\dos{SDSS J150909.03+454308.8}  
\def\tres{SDSS J152817.18+395650.4}  
\def\cuatro{SDSS J154054.31+565138.9}  
\def\cinco{SDSS J161623.53+470202.3}  
\def\seis{SDSS J165712.75+321141.4}  
\def\siete{SDSS J172906.56+565319.4}  

\def\unoc{SDSS J1455}  
\def\dosc{SDSS J1509}  
\def\tresc{SDSS J1528}  
\def\cuatroc{SDSS J1540}  
\def\cincoc{SDSS J1616}  
\def\seisc{SDSS J1657}  
\def\sietec{SDSS J1729}  

\def\whtuno{SDSS J002101.03+005248.1}   
\def\whtdos{SDSS J003218.60+150014.2}   
\def\whttres{SDSS J162410.11-002202.5}   

\def\whtunoc{SDSS J0021}   
\def\whtdosc{SDSS J0032}   
\def\whttresc{SDSS J1624}   

\def\farcs{\hbox{$.\!\!^{\prime\prime}$}}
\def\fsec{\hbox{$.\!\!^s$}}
\def\arcmin{\hbox{$^{\prime}$}}
\def\arcsec{\hbox{$^{\prime\prime}$}}

\graphicspath{{graphs/}{figures/}{chapter1/}{chapter2/}{chapter3/}{chapter4/}{chapter5/}{chapter6/}{chapter7/}{chapter9/}}

\begin{document}

                                   
\definecolor{dgreen}{rgb}{0,.5,.1} 
\definecolor{pink}{rgb}{.9,.2,.5}  
\definecolor{orange}{rgb}{.9,.4,0} 
                                   


\pagestyle{empty}

\pagestyle{empty}

\vspace*{.7cm}

\begin{center}

\begin{spacing}{2}
{\Huge \bf A comparative study of Star Formation processes in different
environments} 
\end{spacing}


\vspace*{4cm}

{\LARGE \bf
Guillermo F.\ H\"agele}
\medskip\medskip

\vspace*{1cm}

{\Large
Supervisor: \'Angeles I. D\'iaz Beltr\'an\\
Co-Supervisor: Elena Terlevich}

\vskip 1.7cm

\includegraphics[width=.21\textwidth,angle=0]{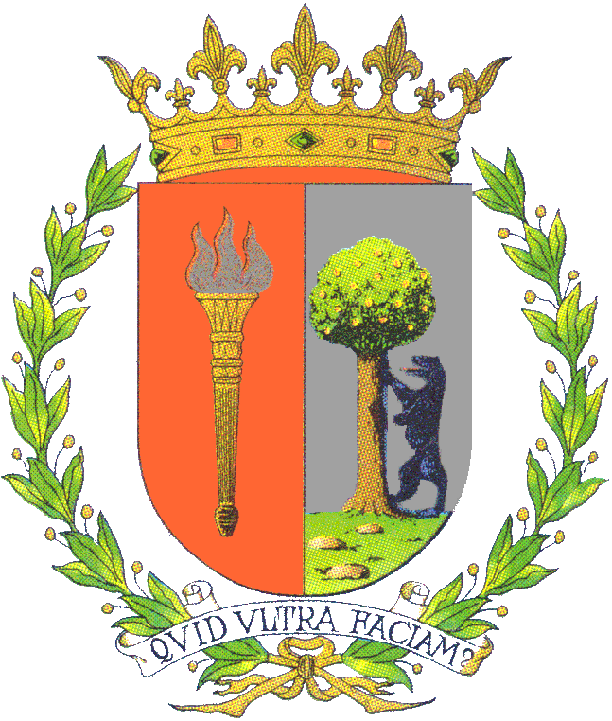}

{\large
Grupo de Astrof\'isica\\
Departamento de F\'isica Te\'orica\\
Facultad de Ciencias\\
Universidad Aut\'onoma de Madrid\\
\vskip 1.4cm
A thesis submitted for the admission to the degree of Doctor en Ciencias
F\'isicas \\
\medskip
2008}

\end{center}

\newpage

\hfill
\newpage

\pagestyle{empty}

\vspace*{20cm}




\begin{flushright}
 {\large to Moni and Vale, my two own bright stars}
\end{flushright}


\hfill
\newpage

\pagestyle{fancy}

\pagenumbering{roman}
\setcounter{page}{1}

\begin{spacing}{1.21}

\chapter*{Summary}
\label{summ}

Several aspects of the star formation processes can be studied from the
conditions of the ionized gas that surround the young massive stars formed in
the core of the giant molecular clouds in successive episodes. The emission
and absorption lines present in their spectra can give us clues about the
physical conditions of the gaseous media, such as metal abundances,
temperatures and ionization degree, as well as information about the ages,
masses and composition of the stellar populations and the properties of the
ionizing stellar clusters. Nowadays, with the advent of modern telescopes
and new instrumentation with greater capability to collect the information that
arrives to us from celestial bodies, we are able to use and develop innovative
techniques that allow us to study the star-forming regions in a way without
precedent in the history of Astrophysics.

The present thesis deals with the study of two very different environments
where star formation is taking place. One, associated with low metal irregular dwarf
galaxies, called \HII\ galaxies, has low density gas of relatively high
temperatures and shows evidence for at least one recent violent episode of star
formation. In contrast, the second environment, associated with circumnuclear
star-forming regions (CNSFRs) in ring patterns located in the central zones
($\sim$\,1\,kpc) of some early type barred spiral galaxies, has low
temperature and  relatively high density and metal rich gas. In Chapter \S
\ref{overv} we give a brief summary of the general properties of these two different king of objects 
and the studies made on them.

In Chapter \S \ref{HIIgal-obs} we propose a methodology to perform a
self-consistent analysis of the physical properties of the emitting gas of
\HII\ galaxies, adequate to the data that can be obtained with XXI century
technology. This methodology requires the production and calibration of
empirical relations between the different line temperatures that should
supersede currently used ones based on very simple, and poorly tested,
photo-ionization model sequences.  

As a first step to reach this goal, we have obtained simultaneous blue to far
red longslit spectra with the double-arm spectrograph of the William Herschel
Telescope (WHT) of three 
compact \HII\ galaxies selected from the  Sloan Digital Sky Survey (SDSS) Data
Release 2 (DR2) spectral catalog using the INAOE Virtual Observatory
superserver. These spectra cover the range from 3200 to 10500\,\AA, including
the Balmer jump, the [O{\sc ii}]\,$\lambda\lambda$\,3727,29\,\AA\ lines, the
[S{\sc iii}]\,$\lambda\lambda$\,9069,9532\,\AA\ doublet as well as various
weak auroral lines such as [O{\sc iii}]\,$\lambda$\,4363\,\AA\ and [S{\sc
iii}]\,$\lambda$\,6312\,\AA. In addition, we observed seven
luminous \HII\ galaxies using the 3.5\,m telescope at Calar Alto, obtaining
high signal-to-noise spectrophotometric observations. We also used a
double-arm spectrograph (TWIN) which provides 
spectra with a wide wavelength coverage, from 3400 to 10400\,\AA\ free of
second order effects, of exactly the same region of a given galaxy. 

The analysis of  these observations allowed the design of a methodology to obtain
accurate elemental abundances of oxygen, sulphur, nitrogen, neon, argon and
iron in the ionized gas. For all the objects we have measured at least four
line temperatures, T$_e$([O{\sc iii}]), T$_e$([S{\sc iii}]), T$_e$([O{\sc
    ii}]) and T$_e$([S{\sc ii}]), and an electron 
density, N$_e$([S{\sc ii}]), from the observed forbidden line ratios using a
five-level atom approximation. For our best objects, errors of  1\% in
T$_e$([O{\sc iii}]), 3\% in T$_e$([O{\sc ii}]) and 5\% in T$_e$([S{\sc iii}])
were achieved. For three objects also measured the Balmer
continuum temperature T(Bac). These
measurements and a careful and realistic treatment of the observational errors
yielded total oxygen abundances with accuracies between 5 and 9\%. These
accuracies are expected to improve as better calibrations based on more
precise measurements, both on electron temperatures and  densities, are
produced.

For the objects observed with the WHT we have compared the measurements obtained for our spectra with those performed on the spectra downloaded from the SDSS DR3 finding a satisfactory agreement. 

The ionization structure of the observed nebulae can be mapped by the derived
oxygen and sulphur ionic ratios, on the one side,  and the corresponding
observed emission line ratios, on the other --  the $\eta$ and $\eta$' plots
--. The combination of both is shown to provide a means to test
photo-ionization model sequences currently applied to derive elemental
abundances in \HII\ galaxies. 
The ionization structure found for the observed objects from the
O$^{+}$/O$^{2+}$ and S$^{+}$/S$^{2+}$ ratios points to high values of the 
ionizing radiation, as traced by the values of the ``softness parameter" 
$\eta$ which is less than one for all the objects. The use of line
temperatures derived from T([O{\sc iii}]) based on current photo-ionization
models yields for the two highest excitation objects much higher values of 
$\eta$ which would imply lower ionizing temperatures. This is however
inconsistent with the ionization structure as probed by the measured emission
line intensity ratios.

Finally, we have measured the T(Bac) for three of the \HII\ galaxies and
computed temperature fluctuations. Only for one of the objects, the
temperature  fluctuation is significant and could lead to higher oxygen abundances by about
0.20\,dex. 

In Chapter \S \ref{neon} we present a study of the strong optical collisional
emission lines of Ne and Ar in an heterogeneous sample of ionized gaseous
nebulae for which it is possible to derive directly the electron temperature
and hence the chemical abundances of neon and argon. We calculated, using a grid
of photo-ionization models, new ionization correction factors for these two
elements and we studied the behavior of the Ne/O and Ar/O abundance ratios with
metallicity. We find a constant value for Ne/O, while there seems to be some
evidence for the existence of negative radial gradients of  Ar/O over the
discs of some nearby spirals. We have studied the relation between the intensities of
the emission lines of [Ne{\sc iii}] at 3869\,\AA\ and [O{\sc iii}] at
4959\,\AA\ and 5007\,\AA. This relation can be used in empirical calibrations
and diagnostic ratios extending their applicability to bluer wavelengths and
therefore to samples of objects at higher redshifts. Finally, we propose a new
diagnostic tool using [O{\sc ii}], [Ne{\sc iii}] and H$\delta$ emission lines
to derive metallicities for galaxies at high z. 

In Chapter \S \ref{cnsfr-obs-kine} we present the measurements of
gas and stellar velocity dispersions in 17 circumnuclear
star-forming regions (CNSFRs) and the nuclei of three barred spiral galaxies:
NGC\,2903, NGC\,3310 and NGC\,3351 from high dispersion spectra. The stellar
dispersions have been obtained from the Ca{\sc ii} triplet (CaT) lines at
$\lambda\lambda$\,8494, 8542, 8662\,\AA, while the gas velocity dispersions
have been measured by Gaussian fits to the H$\beta$\,$\lambda$\,4861\,\AA\ and
to the [O{\sc iii}]\,$\lambda$\,5007\,\AA\ lines.  

The CNSFRs, with sizes of about 100 to 150\,pc in diameter, are seen to be
composed of several individual star clusters with sizes between 1.5 and
6.2\,pc on {\it Hubble Space Telescope} (HST) images. Using the stellar
velocity dispersions, we have 
derived dynamical masses for the entire star-forming complexes and for the
individual star clusters. Values of the stellar velocity dispersions are
between 31 and 73\,km\,s$^{-1}$. Dynamical masses for the whole CNSFRs are
between 4.9\,$\times$\,10$^6$ and 1.9\,$\times$\,10$^8$\,M$_\odot$ and
between 1.4\,$\times$\,10$^6$ and 1.1\,$\times$\,10$^7$\,M$_\odot$ for the
individual star clusters. 

We have found indications for the presence of two different kinematical
components in the ionized gas of the regions. The narrow component of the
two-component Gaussian fits 
seem to have a relatively constant value for all the studied CNSFRs, with 
estimated values close to 25\,km\,s$^{-1}$. This narrow component could be
identified with ionized gas in a rotating disc, while the stars and the
fraction of the gas (responsible for the broad component) related to the
star-forming regions would be mostly supported by dynamical pressure. 
To disentangle the origin of these two components it will be
necessary to map these regions with higher spectral and spatial resolution and
much better signal-to-noise ratio in particular for the O$^{2+}$ lines.

 The radial
velocity curves of the central zones of the studied galaxies seem to have
turnover points at the same positions as the star-forming ring, and the
velocity distribution is consistent with that expected  for this type of
galaxies.

In Chapter \S \ref{abundan} we present longslit observations in the optical
and near infrared of 12 circumnuclear \HII\ regions in the early type spiral
galaxies: NGC\,2903, NGC\,3351 and NGC\,3504 with the aim of deriving their
chemical abundances. Only for one of the regions, the [S{\sc
    iii}]\,$\lambda$\,6312\,\AA\ was detected providing, together with the
nebular [S{\sc iii}] lines at $\lambda\lambda$\,9069, 9532\,\AA, a value of
the electron temperature of T$_e$([S{\sc
    iii}])\,=\,8400$^{+4650}_{-1250}$\,K. A semi-empirical method for the
derivation of abundances in the high metallicity regime is presented.  

We obtain abundances which are comparable to those found in high metallicity 
disc \HII\ regions from direct measurements of electron temperatures and
consistent with solar values within the errors. The region with the highest
oxygen abundance is R3+R4 in NGC\,3504, 12+log(O/H)\,=\,8.85, about 1.6 solar
if the solar oxygen abundance is set at the value derived by Asplund et
al.\ (2005), 12+log(O/H)$_{\odot}$\,=\,8.66$\pm$0.05. Region R7 in NGC\,3351
has the lowest oxygen abundance of the sample, about 0.6 times solar. In all
the observed CNSFR the O/H abundance is dominated by the O$^+$/H$^+$
contribution, as is also the case for high metallicity disc \HII\ regions. For
our observed regions, however also the  S$^+$/S$^{2+}$ ratio is larger than
one, contrary to what is found in high metallicity disc \HII\ regions for
which, in general, the sulphur abundances are dominated by S$^{2+}$/H$^+$.  

The derived N/O ratios are in average larger than those found in high
metallicity disc \HII\ regions and they do not seem to follow the trend of N/O
vs.\ O/H which marks the secondary behavior of nitrogen. The
S/O ratios span a very narrow range between 0.6 and 0.8 of the solar value. 
 
As compared to high metallicity disc \HII\ regions, CNSFR show values of the
O$_{23}$ and the N2 parameters whose distributions are shifted to lower and
higher values respectively, hence, even though their derived oxygen and
sulphur abundances are similar, higher values would in principle be obtained
for the CNSFR  if pure empirical methods were used to estimate
abundances. CNSFR also show lower ionization parameters than their disc
counterparts, as derived from the [S{\sc ii}]/[S{\sc iii}] ratio. Their
ionization 
structure also seems to be different with CNSFR showing radiation field
properties more similar to \HII\ galaxies than to disc high metallicity \HII\
regions.

\bigskip

Large part of this thesis has been published in the Monthly Notices of the
Royal Astronomical Society (MNRAS) research journal and in conference 
proceedings:

\begin{itemize}

\item Almost all of Chapter \S \ref{HIIgal-obs} has been published in

{\bf The temperature and ionization structure of the emitting gas in HII
galaxies: Implications for the accuracy of abundance determinations.}\\
{\bf G. F. H\"agele}, E. P{\'e}rez-Montero, A.~I. D\'{\i}az,  
E. Terlevich and R. Terlevich. 2006, MNRAS, 372, 293.

{\bf Precision abundance analysis of bright HII galaxies.}\\ 
{\bf G. F. H\"agele}, A.~I.~D\'{\i}az, E. Terlevich, R. Terlevich,
E. P{\'e}rez-Montero and M.~V.~Cardaci. 2008, MNRAS, 383, 209.

Proceedings:

``The ionization structure of HII galaxies''.
{\bf G.F. H\"agele}, E. P\'erez-Montero, A.I. D\'{\i}az, E. Terlevich and
R. Terlevich.
{\it IV} Workshop Estallidos de Formaci\'on Estelar en Galaxias: Una
Aproximaci\'on Multifrecuencia, 2006 (CD-rom).

``On the accuracy in the derivation of elemental abundances in HII
galaxies''. {\bf G. F. H\"agele}, A. I. D\'{\i}az, E. P\'erez-Montero,
E. Terlevich and R. Terlevich.
``Galaxy Evolution Across the Hubble Time'' proc.\ of the IAU Symp.\ \#235
held during the IAU General Assembly in Prague, 2006, Cambridge University
Press. Fran\c coise Combes (Chief Editor) and Jan Palou\v s Eds.,
Pag.\ 103.

``On the accuracy in the derivation of elemental abundances of the
emitting gas in HII galaxies''.
{\bf G.F. H\"agele}, A.I. D\'{\i}az, E. P\'erez-Montero, E. Terlevich and
R. Terlevich.
``Highlights of Spanish Astrophysics IV'' Proceedings of the {\it VII}
Scientific Meeting of the Spanish Astronomical Society (SEA) held in
Barcelona, September 12-15, 2006, Springer. Eds.: F. Figueras, J.M. Girart,
M. Hernanz, C. Jordi. (6 pag., CD-rom).

``Effects of the temperature structure on the derivation of abundances
in HII galaxies''.
{\bf G.F. H\"agele}, E. P\'erez-Montero, A.I. D\'{\i}az, E. Terlevich and
R. Terlevich.
``From Stars to Galaxies: Building the pieces to build up the
Universe'', 2007, 374, 143, Astronomical Society of the Pacific Conference
Series. Antonella Vallenari, Rosaria Tantalo, Laura Portinari and Alessia
Moretti Eds. (2 pag.).

``Precision abundance analysis of bright HII galaxies''.
{\bf G.F. H\"agele}, A.I.~D\'{\i}az, E. Terlevich, R. Terlevich,
E. P{\'e}rez-Montero and M.V.~Cardaci.
``II Workshop ASTROCAM'', 2007, proceedings On-line.

\item Chapter \S \ref{neon} has been published in

{\bf Neon and Argon optical emission lines in ionized gaseous nebulae:
Implications and applications.}\\ 
E. P\'erez-Montero, {\bf G. F.~H\"agele}, T. Contini and
A.~I.~D\'iaz. 2007, MNRAS, 381, 125.

\item Part of Chapter \S \ref{cnsfr-obs-kine} has been published in

{\bf Kinematics of gas and stars in the circumnuclear starforming ring of
NGC\,3351} \\ 
{\bf G.~F.~H\"agele}, A.~I.~D\'iaz, M.~V.~Cardaci, E.
Terlevich and R. Terlevich. 2007, MNRAS, 378, 163.

{\bf Erratum: Kinematics of gas and stars in the circumnuclear starforming
  ring of NGC\,3351} \\ 
{\bf G.~F.~H\"agele}, A.~I.~D\'iaz, M.~V.~Cardaci, E.
Terlevich and R. Terlevich. 2008, MNRAS, 385, 543.

Proceedings:

``Velocity Dispersions in Circumnuclear Star Forming Regions''.
{\bf G.F. H\"agele}, A.I. D\'{\i}az, E. Terlevich and R. Terlevich.
The Many Scales in the Universe, JENAM 2004 Astrophysics Reviews, Springer,
J.C. del Toro Iniesta, E.J. Alfaro, J.G. Gorgas,
E. Salvador-Sol\'e y H. Butcher Eds.\ (CD-rom).

``Kinematics of metal-rich circumnuclear regions from future 8M class
observations''. E. Terlevich, {\bf G.F. H\"agele}, A.I. D\'{\i}az,
R. Terlevich and M.V. Cardaci. 
``First Light Science with the GTC'', 2006, Revista Mexicana de
Astronom\'ia y Astrof\'isica (Serie de Conferencias), 29,
Pag. 163. R. Guzm\'an, C. Packham, and J.M. Rodr\'iguez-Espinoza Eds.

``Kinematics of the Circumnuclear Region of NGC 3351''. A.I. D\'iaz, {\bf
  G.F. H\"agele}, M.V. Cardaci, E. Terlevich and R. Terlevich. 
``Galaxy Evolution Across the Hubble Time'' proc.\ of the IAU Symp.\ \#235
held during the IAU General Assembly in Prague, 2006, Cambridge University
Press. Fran\c coise Combes (Chief Editor) and Jan Palou\v s Eds., Pag.\ 308.

``Kinematics of gas and stars in the circumnuclear starforming ring of
NGC\,3351''. {\bf G.F. H\"agele}, A.I. D\'iaz, M.V. Cardaci, E. Terlevich and
R. Terlevich. 
``{\it V} Workshop Estallidos de Formaci\'on Estelar en Galaxias: Star
Formation and Metallicity'', 2007 (CD-rom).

``Kinematics of gas and stars in circumnuclear star-forming regions of
early type spirals''.
{\bf G.F. H\"agele}, A.I.~D\'{\i}az, M.V.~Cardaci, E. Terlevich and R.
Terlevich.
``II Workshop ASTROCAM'', 2007, proceedings On-line.

``Kinematics of gas and stars in circumnuclear star-forming regions of
early type spirals''.
{\bf G.F. H\"agele}, A.I.~D\'{\i}az, M.V.~Cardaci, E. Terlevich and R.
Terlevich.
``Young massive star clusters: Initial conditions and environments'', 2008,
Astrophysics \& Space Science, E. P\'erez, R. de Grijs and
R. Gonz\'alez-Delgado Eds (4 p\'ag.).

\item Chapter \S \ref{abundan} has been published in

{\bf The metal abundace of circumnuclear star forming regions in early type
spirals. Spectrophotometric observations.}\\ 
A.~I.~D\'iaz, E. Terlevich, M. Castellanos and {\bf G.~F.~H\"agele}. 2007,
MNRAS, 382, 251.

Proceedings:

``The metallicity of circumnuclear star forming regions''.
A.I. D\'iaz, E. Terlevich, M. Castellanos and {\bf G.F. H\"agele}.
``The Metal Rich Universe'', 2006, Cambridge University Press.
(4 pag.; astro-ph/0610787).

``The Metal Abundances  of Circumnuclear Star Forming Regions in Early
Type Spirals''. E. Terlevich, A.I. D\'iaz, {\bf G.F. H\"agele} and
M. Castellanos.
``Galaxy Evolution Across the Hubble Time'' proc.\ of the IAU Symp.\ \#235
held during the IAU General Assembly in Prague, 2006, Cambridge University
Press. Fran\c coise Combes (Chief Editor) and Jan Palou\v s Eds., Pag. 336.

``A spectrophotometric study of the physical parameters of circumnuclear
star forming regions''.
{\bf G.F. H\"agele}, A.I. D\'{\i}az, M.V. Cardaci, E. Terlevich, R. Terlevich
and M. Castellanos.
``Highlights of Spanish Astrophysics IV'' Proceedings of the {\it VII}
Scientific Meeting of the Spanish Astronomical Society (SEA) held in
Barcelona, September 12-15, 2006, Springer. Eds.: F. Figueras, J.M. Girart,
M. Hernanz, C. Jordi. (4 pag., CD-rom).

``Spectroscopy of Circumnuclear Star Forming Regions in Early Type
Spirals''. M.V. Cardaci, {\bf G.F. H\"agele}, A.I. D\'iaz, E. Terlevich,
R. Terlevich and M. Castellanos. 
``From Stars to Galaxies: Building the pieces to build up the
Universe'', 2007, Astronomical Society of the Pacific Conference
Series, 374, 137. Antonella Vallenari, Rosaria Tantalo, Laura Portinari and
Alessia Moretti Eds. (2 pag.).

``Physical Parameters in Circumnuclear Star Forming Regions''. {\bf
G.F. H\"agele}, M.V. Cardaci, A.I. D\'{\i}az, E. Terlevich, R. Terlevich and
M. Castellanos.
``Massive Stars: Fundamental Parameters and Circumstellar
Interactions'', 2007, Revista Mexicana de Astronom\'ia y Astrof\'isica
(Serie de Conferencias). P. Benaglia, G. Bosch and C.E. Cappa Eds. (1
pag.).

``Circumnuclear Regions of Star Formation in Early Type Galaxies''.
A.I. D\'iaz, E. Terlevich, {\bf G.F. H\"agele} and M. Castellanos.
``Pathways Through an Eclectic Universe'', 2007, Astronomical Society of
the Pacific Conference Series. Johan Knapen, Terry Mahoney and Alexandre
Vazdekis Eds.\ (4 pag.).

``Properties of the ionised gas of circumnuclear star-forming regions in
early type spirals''.
A.I.~D\'{\i}az, {\bf G.F. H\"agele}, E. Terlevich and R. Terlevich.
``Young massive star clusters: Initial conditions and environments'', 2008,
Astrophysics \& Space Science, E. P\'erez, R. de Grijs and
R. Gonz\'alez-Delgado Eds.\ (6 pag.).

\end{itemize}

\hfill

\chapter*{Resumen}
\label{ress}

Se pueden estudiar varios aspectos de los procesos de formaci\'on estelar a
partir de las condiciones del gas ionizado que rodea a las estrellas masivas y
j\'ovenes formadas en sucesivos episodios en los n\'ucleos de nubes moleculares gigantes. Las l\'ineas de emisi\'on y absorci\'on presentes en sus
espectros pueden darnos pistas acerca de las condiciones f\'isicas del
medio gaseoso, tales como la abundancia de metales, temperaturas y grado de
ionizaci\'on, as\'i como tambi\'en informaci\'on acerca de las edades, masas y
composici\'on qu\'\i mica de las poblaciones estelares y las propiedades de los c\'umulos
estelares ionizantes.
Hoy en d\'ia, con el advenimiento de los telescopios modernos y de nueva
instrumentaci\'on con mayor capacidad para recolectar la informaci\'on que nos
llega de los cuerpos celestes, estamos en condiciones de utilizar y
desarrollar t\'ecnicas innovadoras que nos permitan estudiar las regiones de
formaci\'on estelar de una manera sin precedente en la historia de la
Astrof\'isica. 

La presente tesis trata del estudio de dos entornos muy diferentes
donde est\'a teniendo lugar formaci\'on estelar a gran escala. Uno, asociado con galaxias
irregulares enenas de baja metalicidad, llamadas galaxias \HII, contiene gas de baja
densidad, temperatura relativamente alta y muestra al menos un episodio
de formaci\'on estelar violenta. Por el contrario, el segundo ambiente,
asociado con 
regiones circunnucleares de formaci\'on estelar (CNSFRs) en patrones 
anulares localizados en las zonas centrales ($\sim$\,1\,kpc) de algunas
galaxias espirales barradas de tipo temprano, contiene gas de temperatura baja y  densidad relativamente alta, rico en metales. En el Cap\'itulo \S
\ref{overv} se da un breve resumen de las propiedades generales y de los
estudios llevados a cabo en estas dos clases diferentes de objetos.

En el Cap\'itulo \S \ref{HIIgal-obs} se propone una metodolog\'ia para realizar
un an\'alisis autoconsistente de las propiedades f\'isicas del gas emisor en
galaxias \HII\ adecuado a los datos que se pueden obtener con la tecnolog\'ia
del siglo XXI. Esta metodolog\'ia requiere la producci\'on y la calibraci\'on
de relaciones emp\'iricas entre las diferentes temperaturas de l\'inea que
deber\'ia reemplazar los actualmente utilizados basados en secuencias de modelos de fotoionizaci\'on excesivamente simples y pobremente contrastados. 

Como primer paso para alcanzar esta meta hemos obtenido espectros de rendija
larga simult\'aneos desde el azul hasta el rojo lejano con el espectr\'ografo
de doble brazo del telescopio
William Herschel (WHT), de tres galaxias \HII\ compactas seleccionadas del 
cat\'alogo espectral Sloan Digital Sky Survey (SDSS) Data Release 2 (DR2)
utilizando el superservidor del observatorio virtual del INAOE. Estos
espectros cubren el rango desde 3200 hasta 10500\,\AA, incluyendo el salto de
Balmer, las l\'ineas de [O{\sc ii}]\,$\lambda\lambda$\,3727,29\,\AA, el
doblete [S{\sc iii}]\,$\lambda\lambda$\,9069,9532\,\AA\ como tambi\'en varias
l\'ineas aurorales d\'ebiles tales como [O{\sc iii}]\,$\lambda$\,4363\,\AA\ y
[S{\sc iii}]\,$\lambda$\,6312\,\AA. Adem\'as, observamos siete galaxias \HII\
luminosas usando el telescopio de 3.5\,m de Calar 
Alto, obteniendo observaciones espectrofotom\'etricas con alta relaci\'on
se\~nal a ruido. Nuevamente, utilizamos
un espectrografo de doble brazo (TWIN) que proporciona espectros con una cobertura amplia
en longitud de onda, desde 3400 a 10400\,\AA libre de efectos de segundo
orden, de exactamente la misma regi\'on de una galaxia dada.

El an\'alisis de estas observaciones nos ha permitido definir una metodolog\'ia para obtener abundancias elementales precisas de ox\'igeno, azufre, nitr\'ogeno,
ne\'on, arg\'on y hierro en el gas ionizado. Para todos los objetos se han
medido por lo menos cuatro temperaturas de l\'inea: T$_e$([O{\sc iii}]),
T$_e$([S{\sc iii}]), T$_e$([O{\sc ii}]) and T$_e$([S{\sc ii}]), y una densidad
electr\'onica, N$_e$([S{\sc ii}]), a partir de los cocientes de las l\'ineas
prohibidas observadas, utilizando una aproximaci\'on del \'atomo de cinco
niveles. Para nuestros mejores objetos, se han obtenido errores de 1\% en
T$_e$([O{\sc iii}]), 3\% en T$_e$([O{\sc ii}]) y 5\% en T$_e$([S{\sc iii}]). Para tres objetos se ha medido tambi\'enla temperatura del
continuo de Balmer, T(Bac). Con estas medidas y un tratamiento cuidadoso y
realista de los errores observacionales, se obtuvieron abundancias totales de
ox\'igeno con uan precisi\'on entre 5 y 9\%. Se espera que estas precisiones
mejoren cuando se produzcan mejores calibraciones basadas en medidas m\'as
precisas, tanto de temperaturas electr\'onicas como de densidades.

Para los objetos observados con el WHT, hemos comparado las medidas realizadas sobre nuestros espectros on las que hemos hecho sobre los descargados del SDSS DR3, encontando un acuerdo satisfactorio. 

La estructura de ionizaci\'on de una nebulosa puede trazarse mediante el cociente
de las abundancias i\'onica de ox\'igeno y azufre, por un lado, y el
correspondiente cociente de l\'ineas de emisi\'on observadas, por el otro --
los gr\'aficos de $\eta$ y $\eta$' --. Se muestra que la combinaci\'on de
ambos proporciona un modo de comprobar las secuencias de modelos de
foto-ionizaci\'on actualmente aplicadas para derivar abundancias elementales
en galaxias \HII. La estructura de ionizaci\'on encontrada para los objetos
observados, a partir de los cocientes O$^{+}$/O$^{2+}$ y S$^{+}$/S$^{2+}$,
apunta hacia altos valores de la temperatura de la radiaci\'on ionizante, como se ve a partir
de los valores del ``par\'ametro de suavidad" (``softness parameter") $\eta$
que es menor que uno para todos los objetos. El uso de temperaturas de
l\'inea derivadas a partir de T([O{\sc iii}]) basadas en modelos actuales de
foto-ionizaci\'on dan como resultado, para los dos objetos con mayor
excitaci\'on, valores mucho mayores de $\eta$ que podr\'ian implicar
temperaturas de ionizaci\'on menores. Sin embargo, como se comprueba a partir
de las intensidades medidas de las l\'ineas de emisi\'on. esto es inconsistente
con la estructura de ionizaci\'on.

Finalmente, hemos medido T(Bac) para tres galaxias \HII\ y calculado las
fluctuaciones de temperatura. Solamente para uno de los objetos 
la fluctuaci\'on de temperatura es significativa y puede dar como resultado
abundancias de ox\'igeno mayores en aproximadamente 0.20\,dex.

En el Cap\'itulo \S \ref{neon} presentamos un estudio de las l\'ineas 
de emisi\'on colisionales intensas en el \'optico de Ne y Ar en una muestra
heterog\'enea de nebulosas gaseosas ionizadas para las cuales es posible
derivar directamente la temperatura electr\'onica y por lo tanto las
abundancias qu\'imicas de ne\'on y arg\'on. Se han calculado nuevos factores de
correcci\'on de ionizaci\'on para estos dos elementos qu\'imicos utilizando un
conjunto de modelos de foto-ionizaci\'on y se ha estudiado el comportamiento de los
cocientes de abundancias Ne/O y Ar/O con la metalicidad. Mientras que se ha encontrado  un valor
constante para Ne/O, parece haber alguna evidencia de la existencia
de un gradiente radial negativo de Ar/O en los discos de algunas espirales
cercanas. Se ha estudiado tambi\'en la relaci\'on entre las intensidades de las l\'ineas de
emisi\'on de [Ne{\sc iii}] a 3869\,\AA\ y [O{\sc iii}] a
4959\,\AA\ y 5007\,\AA. Esta relaci\'on puede utilizarse en calibraciones
emp\'iricas y diagramas de diagn\'ostico, extendiendo su aplicaci\'on a
longitudes de onda m\'as azules y por lo tanto a muestras de objetos con
corrimientos hacia el rojo mayores. Finalmente, proponemos una nueva
herramienta de 
diagn\'ostico utilizando las l\'ineas de emisi\'on de [O{\sc ii}], [Ne{\sc
iii}] y H$\delta$ para derivar metalicidades para galaxias a alto z.

En el Cap\'itulo \S \ref{cnsfr-obs-kine} presentamos medidas de la
dispersi\'on de velocidades en 17 regiones de formaci\'on estelar
circunnucleares (CNSFRs) y los n\'ucleos de tres galaxias espirales barradas:
NGC\,2903, NGC\,3310 y NGC\,3351 a partir de espectros de alta
dispersi\'on. Las dispersiones estelares han sido 
obtenidas de las l\'ineas del
triplete de Ca{\sc ii} (CaT) a $\lambda\lambda$\,8494, 8542, 8662\,\AA,
mientras que las dispersiones de velocidades del gas han sido medidas mediante
ajustes gaussianos a las l\'ineas de H$\beta$\,$\lambda$\,4861\,\AA\ y
de [O{\sc iii}]\,$\lambda$\,5007\,\AA.

Las CNSFRs, con tama\~nos de alrededor de 100 a 150\,pc en diametro, parecen
estar compuestas por varios c\'umulos estelares individuales con tama\~nos
entre 1.5 y 6.2\,pc medidos sobre imagenes del Telescopio Espacial Hubble
(HST). Utilizando las dispersiones de velocidades estelares, hemos derivado
las masas din\'amicas para los complejos de formaci\'on estelar completos y
para los c\'umulos estelares individuales. Los valores de la dispersi\'on de
velocidades estelares est\'an entre 31 y 73\,km\,s$^{-1}$. Las masas
din\'amicas para las CNSFRs completas est\'an entre 4.9\,$\times$\,10$^6$ y
1.9\,$\times$\,10$^8$\,M$_\odot$ y entre 1.4\,$\times$\,10$^6$ y
1.1\,$\times$\,10$^7$\,M$_\odot$ para los c\'umulos estelares individuales.

Hemos encontrado indicaciones de la presencia de dos componentes cinem\'aticas
diferentes en el gas ionizado de las regiones. La componente estrecha proporcionada por el
ajuste de dos-componentes gausianas, parece tener un valor relativamente
constante para todas las CNSFRs estudiadas, con los valores estimados muy
pr\'oximos a  25\,km\,s$^{-1}$. Esta componente estrecha podr\'ia
identificarse con gas ionizado en un disco rotante, mientras que las estrellas
y la fracci\'on de gas (responsable de la componente ancha) relacionadas con
las regiones de formaci\'on estelar, estar\'ian mayormente soportadas por
presi\'on din\'amica. Para distinguir el origen de estas dos componentes habr\'a
que cartografiar estas regiones con alta resoluci\'on espectral y espacial con una relaci\'on 
se\~nal-ruido mucho mayor, particularmente para las l\'ineas de
O$^{2+}$. 

Las curvas de velocidad radial de las zonas centrales de
las galaxias estudiadas parecen tener m\'aximos y
m\'inimos relativos en la misma posici\'on que el anillo de formaci\'on estelar, y la
distribuci\'on de velocidades es consistente con la esperada para esta clase
de galaxias. 

En el Cap\'itulo \S \ref{abundan} se presentan observaciones de rendija larga
en el \'optico y en el infrarrojo cercano de 12 regiones \HII\ circunnucleares
en las galaxias espirales de tipo temprano: NGC\,2903, NGC\,3351 and NGC\,3504
con el prop\'osito de derivar sus abundancias qu\'imicas. S\'olo para una de 
las regiones, se detect\'o la l\'inea [S{\sc iii}]\,$\lambda$\,6312\,\AA\ 
proporcionando, junto con las l\'ineas nebulares de [S{\sc iii}] en
$\lambda\lambda$\,9069, 9532\,\AA, un valor de la temperatura electr\'onica de 
T$_e$([S{\sc iii}])\,=\,8400$^{+4650}_{-1250}$\,K. Se presenta un m\'etodo semi-emp\'irico
para la derivaci\'on de las abundancias en el r\'egimen de alta metalicidad.

Usando este m\'etodo, se han obtenido abundancias que son comparables con las encontradas en regiones
\HII\ de disco de alta metalicidad, con medidas directas de temperaturas
electr\'onicas y son consistentes con valores solares dentro de los errores. La
regi\'on con la abundancia de ox\'igeno m\'as alta es R3+R4 en NGC\,3504,
12+log(O/H)\,=\,8.85, alrededor de 1.6 veces solar si la abundancia de ox\'igeno
solar se fija en el valor derivado por Asplund et al.\ (2005),
12+log(O/H)$_{\odot}$\,=\,8.66$\pm$0.05. La regi\'on R7 en NGC\,3351 tiene la
abundancia de ox\'igeno m\'as baja de la muestra, alrededor de 0.6 veces la
solar. En todas las CNSFRs observadas la abundancia de O/H est\'a dominada por
la contribuci\'on de O$^+$/H$^+$, como es tambi\'en el caso para las regiones
\HII\ de disco de alta metalicidad. Sin embrago, para nuestras regiones observadas, tambi\'en el cociente S$^+$/S$^{2+}$ es mayor que uno, contrario a lo
que se encuentra en las regiones \HII\ de disco de alta metalicidad para las
cuales, en general, la abundancia de azufre est\'a dominada por
S$^{2+}$/H$^+$. 

El cociente de N/O derivado para las CNSFR es en promedio mayor que los encontrados en 
las regiones \HII\ de disco de alta metalicidad y no parecen seguir la
tendencia de N/O vs.\ O/H que marca el comportamiento secundario del
nitr\'ogeno. El cociente de S/O se comprende un rango muy
estrecho entre 0.6 y 0.8 veces el valor solar.

Comparadas con las regiones \HII\ de disco de alta metalicidad, las CNSFRs
muestran valores de los par\'ametros O$_{23}$ y N2 cuyas distribuciones
est\'an desplazadas hacia valores m\'as bajos y m\'as altos respectivamente;
por lo tanto, aunque sus abundancias derivadas de azufre y ox\'igeno son
similares, se btendr\'ian valores mayores para las CNSFRs si
se utilizaran m\'etodos puramente emp\'iricos para estimar las abundancias. Las
CNSFRs tambi\'en muestran par\'ametros de ionizaci\'on menores que los de sus
contrapartidas de disco, como se deriva a partir del cociente [S{\sc
    ii}]/[S{\sc iii}].  
Sus estructuras de ionizaci\'on tambi\'en parecen ser diferentes: las
CNSFRs muestran propiedades del campo de radiaci\'on m\'as similares a las
galaxias \HII que a las regiones \HII\ de disco de alta metalicidad.

\bigskip

Gran parte de esta tesis ha sido publicada en la revista de investigaci\'on
Monthly Notices of the Royal Astronomical Society (MNRAS) y en
proceedings de conferencias:

\begin{itemize}

\item Casi todo el Cap\'itulo \S \ref{HIIgal-obs} ha sido publicado en

{\bf The temperature and ionization structure of the emitting gas in HII
galaxies: Implications for the accuracy of abundance determinations.}\\
{\bf G. F. H\"agele}, E. P{\'e}rez-Montero, A.~I. D\'{\i}az,  
E. Terlevich and R. Terlevich. 2006, MNRAS, 372, 293.

{\bf Precision abundance analysis of bright HII galaxies.}\\ 
{\bf G. F. H\"agele}, A.~I.~D\'{\i}az, E. Terlevich, R. Terlevich,
E. P{\'e}rez-Montero and M.~V.~Cardaci. 2008, MNRAS, 383, 209.

Proceedings:

``The ionization structure of HII galaxies''.
{\bf G.F. H\"agele}, E. P\'erez-Montero, A.I. D\'{\i}az, E. Terlevich and
R. Terlevich.
{\it IV} Workshop Estallidos de Formaci\'on Estelar en Galaxias: Una
Aproximaci\'on Multifrecuencia, 2006 (CD-rom).

``On the accuracy in the derivation of elemental abundances in HII
galaxies''. {\bf G. F. H\"agele}, A. I. D\'{\i}az, E. P\'erez-Montero,
E. Terlevich and R. Terlevich.
``Galaxy Evolution Across the Hubble Time'' proc.\ of the IAU Symp.\ \#235
held during the IAU General Assembly in Prague, 2006, Cambridge University
Press. Fran\c coise Combes (Chief Editor) and Jan Palou\v s Eds.,
Pag.\ 103.

``On the accuracy in the derivation of elemental abundances of the
emitting gas in HII galaxies''.
{\bf G.F. H\"agele}, A.I. D\'{\i}az, E. P\'erez-Montero, E. Terlevich and
R. Terlevich.
``Highlights of Spanish Astrophysics IV'' Proceedings of the {\it VII}
Scientific Meeting of the Spanish Astronomical Society (SEA) held in
Barcelona, September 12-15, 2006, Springer. Eds.: F. Figueras, J.M. Girart,
M. Hernanz, C. Jordi. (6 pag., CD-rom).

``Effects of the temperature structure on the derivation of abundances
in HII galaxies''.
{\bf G.F. H\"agele}, E. P\'erez-Montero, A.I. D\'{\i}az, E. Terlevich and
R. Terlevich.
``From Stars to Galaxies: Building the pieces to build up the
Universe'', 2007, 374, 143, Astronomical Society of the Pacific Conference
Series. Antonella Vallenari, Rosaria Tantalo, Laura Portinari and Alessia
Moretti Eds. (2 pag.).

``Precision abundance analysis of bright HII galaxies''.
{\bf G.F. H\"agele}, A.I.~D\'{\i}az, E. Terlevich, R. Terlevich,
E. P{\'e}rez-Montero and M.V.~Cardaci.
``II Workshop ASTROCAM'', 2007, proceedings On-line.

\item El Cap\'itulo \S \ref{neon} ha sido publicado en

{\bf Neon and Argon optical emission lines in ionized gaseous nebulae:
Implications and applications.}\\ 
E. P\'erez-Montero, {\bf G. F.~H\"agele}, T. Contini and
A.~I.~D\'iaz. 2007, MNRAS, 381, 125.

\item Parte del Cap\'itulo \S \ref{cnsfr-obs-kine}  ha sido publicado en

{\bf Kinematics of gas and stars in the circumnuclear starforming ring of
NGC\,3351} \\ 
{\bf G.~F.~H\"agele}, A.~I.~D\'iaz, M.~V.~Cardaci, E.
Terlevich and R. Terlevich. 2007, MNRAS, 378, 163.

{\bf Erratum: Kinematics of gas and stars in the circumnuclear starforming
  ring of NGC\,3351} \\ 
{\bf G.~F.~H\"agele}, A.~I.~D\'iaz, M.~V.~Cardaci, E.
Terlevich and R. Terlevich. 2008, MNRAS, 385, 543.

Proceedings:

``Velocity Dispersions in Circumnuclear Star Forming Regions''.
{\bf G.F. H\"agele}, A.I. D\'{\i}az, E. Terlevich and R. Terlevich.
The Many Scales in the Universe, JENAM 2004 Astrophysics Reviews, Springer,
J.C. del Toro Iniesta, E.J. Alfaro, J.G. Gorgas,
E. Salvador-Sol\'e y H. Butcher Eds.\ (CD-rom).

``Kinematics of metal-rich circumnuclear regions from future 8M class
observations''. E. Terlevich, {\bf G.F. H\"agele}, A.I. D\'{\i}az,
R. Terlevich and M.V. Cardaci. 
``First Light Science with the GTC'', 2006, Revista Mexicana de
Astronom\'ia y Astrof\'isica (Serie de Conferencias), 29,
Pag. 163. R. Guzm\'an, C. Packham, and J.M. Rodr\'iguez-Espinoza Eds.

``Kinematics of the Circumnuclear Region of NGC 3351''. A.I. D\'iaz, {\bf
  G.F. H\"agele}, M.V. Cardaci, E. Terlevich and R. Terlevich. 
``Galaxy Evolution Across the Hubble Time'' proc.\ of the IAU Symp.\ \#235
held during the IAU General Assembly in Prague, 2006, Cambridge University
Press. Fran\c coise Combes (Chief Editor) and Jan Palou\v s Eds., Pag.\ 308.

``Kinematics of gas and stars in the circumnuclear starforming ring of
NGC\,3351''. {\bf G.F. H\"agele}, A.I. D\'iaz, M.V. Cardaci, E. Terlevich and
R. Terlevich. 
``{\it V} Workshop Estallidos de Formaci\'on Estelar en Galaxias: Star
Formation and Metallicity'', 2007 (CD-rom).

``Kinematics of gas and stars in circumnuclear star-forming regions of
early type spirals''.
{\bf G.F. H\"agele}, A.I.~D\'{\i}az, M.V.~Cardaci, E. Terlevich and R.
Terlevich.
``II Workshop ASTROCAM'', 2007, proceedings On-line.

``Kinematics of gas and stars in circumnuclear star-forming regions of
early type spirals''.
{\bf G.F. H\"agele}, A.I.~D\'{\i}az, M.V.~Cardaci, E. Terlevich and R.
Terlevich.
``Young massive star clusters: Initial conditions and environments'', 2008,
Astrophysics \& Space Science, E. P\'erez, R. de Grijs and
R. Gonz\'alez-Delgado Eds (4 p\'ag.).

\item El Cap\'itulo \S \ref{abundan} ha sido publicado en

{\bf The metal abundace of circumnuclear star forming regions in early type
spirals. Spectrophotometric observations.}\\ 
A.~I.~D\'iaz, E. Terlevich, M. Castellanos and {\bf G.~F.~H\"agele}. 2007,
MNRAS, 382, 251.

Proceedings:

``The metallicity of circumnuclear star forming regions''.
A.I. D\'iaz, E. Terlevich, M. Castellanos and {\bf G.F. H\"agele}.
``The Metal Rich Universe'', 2006, Cambridge University Press.
(4 pag.; astro-ph/0610787).

``The Metal Abundances  of Circumnuclear Star Forming Regions in Early
Type Spirals''. E. Terlevich, A.I. D\'iaz, {\bf G.F. H\"agele} and
M. Castellanos.
``Galaxy Evolution Across the Hubble Time'' proc.\ of the IAU Symp.\ \#235
held during the IAU General Assembly in Prague, 2006, Cambridge University
Press. Fran\c coise Combes (Chief Editor) and Jan Palou\v s Eds., Pag. 336.

``A spectrophotometric study of the physical parameters of circumnuclear
star forming regions''.
{\bf G.F. H\"agele}, A.I. D\'{\i}az, M.V. Cardaci, E. Terlevich, R. Terlevich
and M. Castellanos.
``Highlights of Spanish Astrophysics IV'' Proceedings of the {\it VII}
Scientific Meeting of the Spanish Astronomical Society (SEA) held in
Barcelona, September 12-15, 2006, Springer. Eds.: F. Figueras, J.M. Girart,
M. Hernanz, C. Jordi. (4 pag., CD-rom).

``Spectroscopy of Circumnuclear Star Forming Regions in Early Type
Spirals''. M.V. Cardaci, {\bf G.F. H\"agele}, A.I. D\'iaz, E. Terlevich,
R. Terlevich and M. Castellanos. 
``From Stars to Galaxies: Building the pieces to build up the
Universe'', 2007, Astronomical Society of the Pacific Conference
Series, 374, 137. Antonella Vallenari, Rosaria Tantalo, Laura Portinari and
Alessia Moretti Eds. (2 pag.).

``Physical Parameters in Circumnuclear Star Forming Regions''. {\bf
G.F. H\"agele}, M.V. Cardaci, A.I. D\'{\i}az, E. Terlevich, R. Terlevich and
M. Castellanos.
``Massive Stars: Fundamental Parameters and Circumstellar
Interactions'', 2007, Revista Mexicana de Astronom\'ia y Astrof\'isica
(Serie de Conferencias). P. Benaglia, G. Bosch and C.E. Cappa Eds. (1
pag.).

``Circumnuclear Regions of Star Formation in Early Type Galaxies''.
A.I. D\'iaz, E. Terlevich, {\bf G.F. H\"agele} and M. Castellanos.
``Pathways Through an Eclectic Universe'', 2007, Astronomical Society of
the Pacific Conference Series. Johan Knapen, Terry Mahoney and Alexandre
Vazdekis Eds.\ (4 pag.).

``Properties of the ionised gas of circumnuclear star-forming regions in
early type spirals''.
A.I.~D\'{\i}az, {\bf G.F. H\"agele}, E. Terlevich and R. Terlevich.
``Young massive star clusters: Initial conditions and environments'', 2008,
Astrophysics \& Space Science, E. P\'erez, R. de Grijs and
R. Gonz\'alez-Delgado Eds.\ (6 pag.).

\end{itemize}

\hfill

\end{spacing}

\tableofcontents
\listoffigures
\listoftables

\hfill
\newpage

\pagestyle{fancy}
\pagenumbering{arabic}
\setcounter{page}{1}


\begin{spacing}{1.21}

\chapter{Introduction}
\label{overv}

\section{Overview}

\label{general}

Gas content, masses, bar structure, and dynamical environment can strongly
influence the large-scale star formation rate (SFR) along the Hubble sequence
\cite{1998ARA&A..36..189K}. The variation of young stellar content and star
formation activity is one of the most conspicuous characteristic along this
sequence, and this variation in the young stellar population is part of the
basis of the morphological classification made by
\citetex{1926ApJ....64..321H}. The trend in SFRs and star formation histories
along the Hubble sequence was confirmed from evolutionary synthesis models of
galaxy colours by \citetex{1968ApJ...151..547T,1972A&A....20..383T} and
\citetex{1973ApJ...179..427S}. Later, the importance of the star formation
bursting mode in the evolution of low-mass galaxies and interacting systems
was studied by \citetex{1976PhDT.........6B,1977ApJ...217..928H} and
\citetex{1978ApJ...219...46L}.  

\begin{figure}
\centering
\includegraphics[width=1.\textwidth,angle=0]{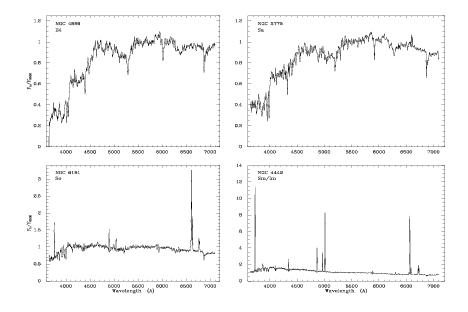}
\caption[Integrated spectra of different types of galaxies]{Integrated spectra
  of different types of galaxies from a 
  spectroscopic atlas by \citetex{1992ApJS...79..255K}:
  elliptical, two spirals and irregular. Figure 1 from
  \citetex{1998ARA&A..36..189K}. The fluxes are normalized to unity at
  5500\,\AA.} 
\label{spectragals}
\end{figure}

Due to their different average SFR, the integrated spectra of galaxies vary
considerably along the Hubble sequence. In Figure \ref{spectragals} we show a
comparison of the integrated spectra of one elliptical (E4), two spirals (Sa
and Sc) and one Magellanic irregular galaxy from
\citetex{1992ApJS...79..255K}. In these examples one can easily appreciate
that they are broadly different. From ellipticals to irregulars, as  SFR
increases, there is a rise in the blue continuum, a gradual change in the
dominant stellar composite spectrum in absorption (from K-giant to A-stars),
and a dramatical increase in the strengths of the nebular emission lines
\cite{1998ARA&A..36..189K}. 

The main trends in SFRs and star formation histories along the Hubble sequence
can be delineated from measurements of the integrated SFRs in hundreds of
nearby galaxies \cite{1998ARA&A..36..189K}. Large-scale star formation
processes occur in two very different environments, in the extended discs of
spiral and irregular galaxies and in compact, dense gas discs in the center of
galaxies. The total star formation in the local universe has a very important
contribution from these two mechanisms of stellar formation, which are traced
at different wavelengths and follow completely different patterns along the
Hubble sequence \cite{1998ARA&A..36..189K}.

\label{star formation}
\subsection{Gas ionization, metallicity and massive star formation}

The star formation processes depend strongly on the physical conditions of the
media in which they take place. Among these conditions, the most important are
the density of the molecular and gaseous material, its spatial distribution
and its metallicity, conditions that restrict the effectiveness and the star
formation rate, as well as its initial mass function. In particular, the
metallicity is a key parameter that controls many aspects in the formation and
evolution of stars and galaxies. 

There are many processes that control the metallicity in a galaxy or a gaseous
nebula, such as the galaxy formation and evolution, massive star formation,
stellar winds, chemical yields, outflows and inflows, etc
\cite{2000A&ARv..10....1K}. 

Due to the nature of the galaxies (except for the nearest systems) almost all
the information about their star formation properties is collected from
integrated light measurements of the ultraviolet continuum, far-infrared and
nebular emission recombination lines \cite{1998ARA&A..36..189K}. In
particular, the study of the gas ionized by the most massive stars can provide
information about their masses, temperatures and evolutionary state. This in
itself requires the use of photo-ionization, stellar atmosphere and stellar
evolution models. The subsequent link between the massive star properties and
those of the global star formation needs also the use of evolutionary
synthesis models. 

Photoionized gas shows a characteristic emission line spectrum and, in
principle, the spectra of  \HII\ galaxies, galactic \HII\ regions, giant
extragalactic \HII\ regions (GEHRs), circumnuclear and even nuclear regions,
can be analyzed in the same way. The emission line intensities are controlled
by: the energy distribution of the ionizing photons, the spatial configuration
of the ionized gas and its local properties, essentially its density and
metal content. In fact, this last parameter conforms the appearance of the
emission line spectra to such an extent that it imposes the methodology
required for its analysis. 

\begin{figure}
\centering
\hspace*{1.2cm}\includegraphics[width=.9\textwidth,angle=0]{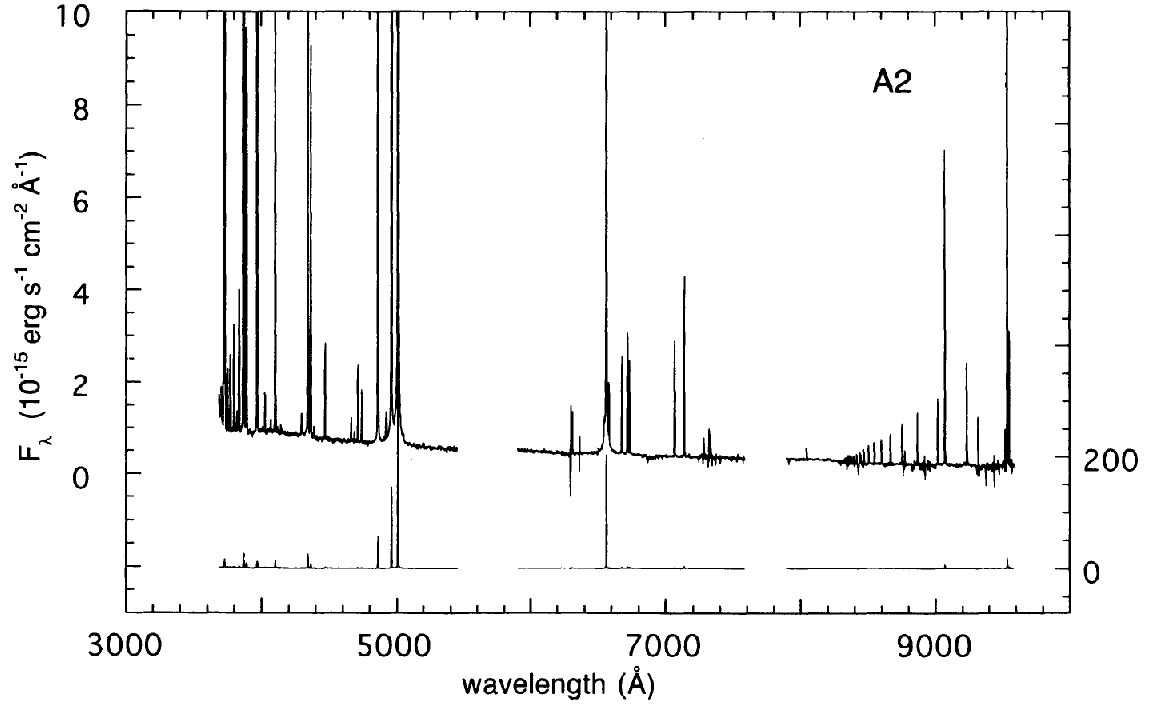}\vspace*{0.8cm}
\includegraphics[width=.9\textwidth,height=0.6\textwidth,angle=0]{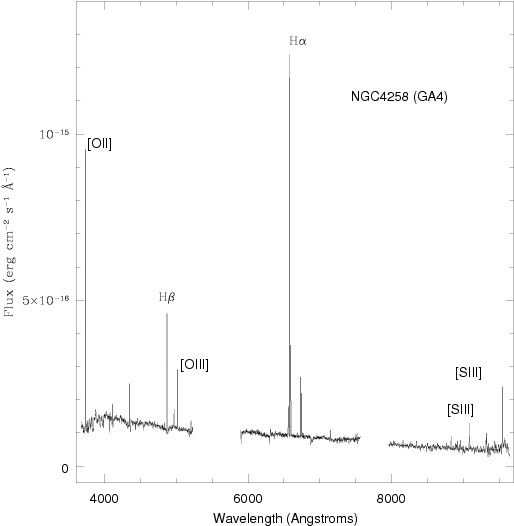}
\caption[Spectrum of knot A of NGC\,2363 and spectrum of region GA4 of
  NGC\,4258]{Upper panel: spectrum of knot A of NGC\,2363 from 
  \citetex{1994ApJ...437..239G}. Lower panel: spectrum of region GA4 in the
  spiral disc of NGC\,4258 from \citetex{2000MNRAS.318..462D}.}
\label{higZ-lowZ}
\end{figure}

Figure \ref{higZ-lowZ} shows typical spectra of two \HII\ regions of widely
different metallicities. The low metallicity one (upper panel), corresponding
to knot A of NGC\,2363 \cite{1994ApJ...437..239G}, shows very strong [O{\sc
    ii}], [O{\sc iii}] lines, the Balmer and Paschen series are clearly
visible and the weak auroral lines of [O{\sc ii}] and [S{\sc ii}] are
detectable and measurable. This makes possible the application of what is
called ``the direct method" for abundance determination (see for example
\citeplain{1992MNRAS.255..325P}). The high metallicity spectrum (lower panel),
corresponding to an \HII\ region (GA4) in the spiral disc of NGC\,4258
(D\'iaz et al., 2000), looks very different. The high oxygen content
provides a very efficient cooling of the region and therefore the emission
lines are,  in general, weaker. No intrinsically weak lines are detected and
the ``direct method" is not applicable.

\nocite{2000MNRAS.318..462D}

The application of these different methodologies to the optical spectroscopic
data through the years, which constitutes our main body of information about
global star formation in galaxies, has produced a strongly biased view. Due
both to selection effects and analysis requirements, most of the objects
observed and analyzed are those that show oxygen emission lines near to their
maximum intensity, which itself corresponds to a narrow range of abundances.  

One of the main aims of this work is to enlarge our view by first obtaining a
high quality set of data for low metallicity objects that allows to refine the
``direct method" for abundance determination increasing its accuracy, and
second obtaining a set of data on the highest abundance objects to allow the
definition of an appropriate scheme for abundance determinations for these
objects. Once this is made, the interpretation of the observed emission line
spectra in terms of the properties of star formation can be made. 

\HII\ galaxies and circumnuclear star-forming regions (CNSFRs) constitute two
obvious samples to serve our purposes since they represent two extreme classes
in metal content. \HII\ galaxies are among the lowest metallicity objects
known, whereas CNSFRs have high metallicities, between solar and twice solar,
as estimated from empirical calibrators (see for example
\citeplain{1991A&AS...91..285T} and \citeplain{1998ApJ...498..541K}). 

\begin{figure}
\centering
\hspace{-2cm}\includegraphics[width=1.1\textwidth,angle=0]{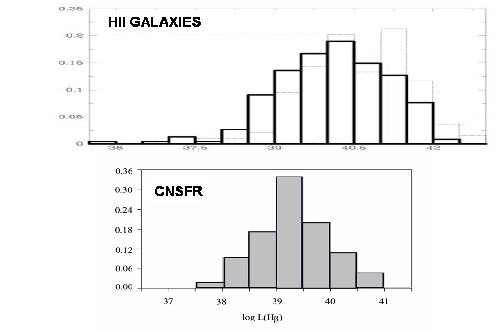}
\caption[H$\beta$ luminosity distribution for \HII\ galaxies and CNSFRs]{Upper
  panel: H$\beta$ luminosity distribution for \HII\ galaxies 
  from \citetex{2006MNRAS.365..454H}. Lower panel: H$\beta$ luminosity
  distribution for CNSFRs from \citetex{tesismar}.} 
\label{LHbeta}
\end{figure}

An interesting point in the comparison of the physical properties and the
ionizing stellar populations of these two kinds of objects is their H$\beta$
luminosity distribution. In Figure \ref{LHbeta} we show the histograms that
represent these distributions for \HII\ galaxies \cite{2006MNRAS.365..454H}
and CNSFRs \cite{tesismar} in the upper and lower panels respectively. The
\HII\ galaxies are further split into those with a measurable [O{\sc
iii}]\,$\lambda$\,4363\,\AA\ auroral emission line (solid line) and those
without (dashed line). We can see that the distributions differ and that the
study of the star formation properties of only the first of them would already
provide a biased view. 

In relation to the CNSFRs, the distribution of the \HII\ galaxies peaks at
greater values of the H$\beta$ luminosity and the width of the distribution in
order of magnitude is also larger. Yet, there is a rather significant overlap
which imply that, if the SFR is derived from the H$\alpha$ luminosities
\cite{1989ApJ...344..685K}, CNSFRs can form as many stars per year (the
equivalent in M$_\odot$\,yr$^{-1}$) as some \HII\ galaxies.

In what follows we describe some of the general properties of these objects
which are relevant to our study. 

\label{HII galaxies}

\subsection{\HII\ Galaxies: Star Formation in low metallicity environments} 

\citetex{1972ApJ...173...25S} reported the discovery of two extragalactic
objects with very low metal abundances: IZw18 and IIZw40, two very well studied
galaxies at present. They pointed out that these galaxies should be
either young, in the sense that most of their star formation has occurred in
recent times, or that the star formation in them should have occured in
intense bursts which 
are separated by long quiescent periods, since their derived abundances
are one order of magnitude lower than the solar value.

\begin{figure}
\centering
\vspace{2.2cm}
\includegraphics[width=1.\textwidth,angle=0]{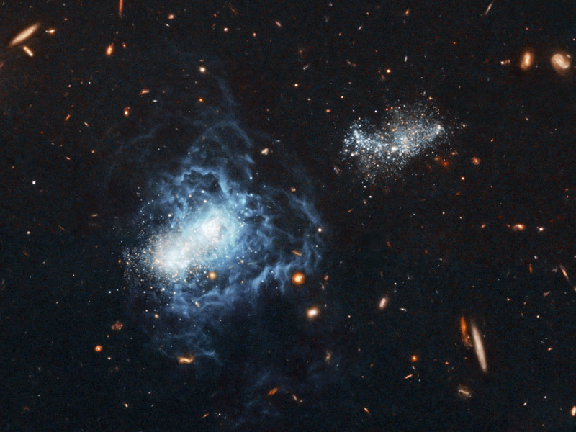}
\caption[False colour image of the low metallicity galaxy IZw18]{False colour
  image of the low metallicity galaxy IZw18 obtained with the HST
  [from  Astronomy Picture of the Day web page ({\it
  http://apod.nasa.gov/apod/ap071017.html}); credit: NASA, ESA, and A.\ Aloisi
  (ESA \& STScI)].} 
\label{izw18}
\end{figure}

In Figure \ref{izw18} we show a false colour image of IZw18
\cite{1966ApJ...143..192Z}. It was once thought to be one of the youngest
galaxies since its bright stars indicated 
an age of only 500 million years. The galaxy was also intriguing because it
resembled galaxies forming in the very early universe although it is a nearby
object (14.6\,$\pm$\,1.0\,Mpc; \citeplain{2000ApJ...529..786Mtot}) and is 
surrounded by galaxies that are significantly older. Relatively recent images
of IZw18 by the HST have helped 
resolve this mystery, by discovering a population of old 
faint stars intertwined with the bright star population. Therefore, IZw18
is now thought to be just as old as its neighbours, roughly 10$^{10}$ years
old, but with an intense episode of relatively new star formation
\cite{1999AJ....118..302A}. As was pointed out by Aloisi et al.,  the 
trigger for this recent episode of bright star formation is possibly the
changing 
gravitational influence of the smaller companion galaxy of IZw18, visible at
the upper right in the Figure.

IZw18 and IIZw40 are nowadays ascribed to the category of \HII\ galaxies.
\HII\ galaxies are a subclass of Blue Compact Dwarf galaxies (BCDs) which show
spectra with strong emission lines similar to those of GEHRs
(\citeplain{1970ApJ...162L.155S,1980ApJ...240...41F}), have the lowest metal
content of any star forming galaxy known, suggesting that they 
are among the youngest or less evolved galaxies
\cite{2007ApJ...654..226R,1972ApJ...173...25S}. 

\begin{figure}
\centering
\includegraphics[width=.9\textwidth,angle=0]{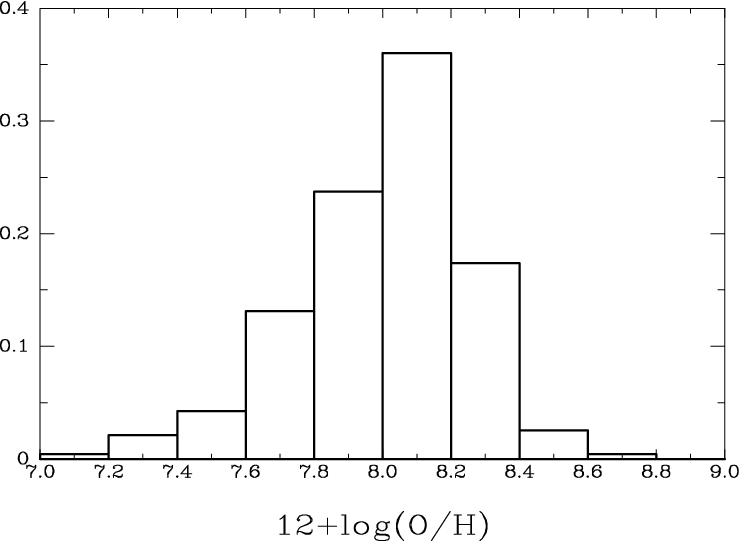}
\caption[Metallicity distribution of a sample \HII\ galaxies]{Metallicity
  distribution of a sample of \HII\ galaxies in the local 
  Universe with measurements of the [O{\sc iii}]\,$\lambda$\,4363\,\AA\
  auroral emission line. Figure 10 of \citetex{2006MNRAS.365..454H}.} 
\label{metdist}
\end{figure}

\begin{figure}
\centering
\includegraphics[width=1.3\textwidth,angle=-90]{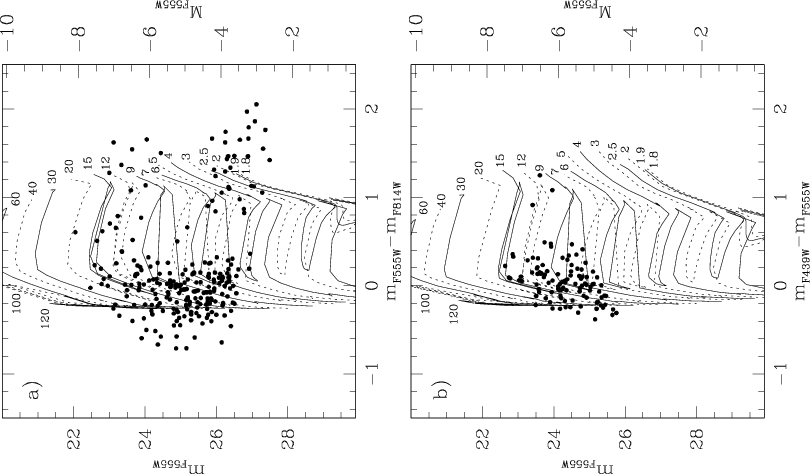}
\caption[Color-magnitude diagrams of IZw18 compared with Padova
  tracks]{Color-magnitude diagrams of IZw18 compared with Padova tracks with
  Z\,=\,0.0004 from \citetex{1999AJ....118..302A}, Figure 9 of that work. (a)
  V vs.\ V-I; (b) V vs.\ B-V. The stellar mass of the track is given in
  M$_\odot$.}
\label{colordiag}
\end{figure}

In general, BCDs are characterized by their compact aspect, very low
metallicities, gas richness and blue colors  
\cite{2000A&ARv..10....1K}. In Figure \ref{metdist} we show the oxygen
abundance 
distribution of a sample of \HII\ galaxies in the local Universe with
measurements of the [O{\sc iii}]\,$\lambda$\,4363\,\AA\ auroral emission line
from \citetex{2006MNRAS.365..454H}. Although, as mentioned above, these
properties  
have made of them
good candidates to host their first episodes of star formation, 
recently the detection in most of them of low surface-brightness elliptical
haloes, or the presence of stars belonging to older populations, have caused
this interpretation to be reconsidered.  Nowadays, only a few candidates remain
controversial. There are several works in the literature whose main aim is to
study the 
weight of these older stellar populations in BCDs and, hence, to find out the
actual age and evolutionary status of these objects. Among these studies, in
the sample of local objects, observations with enough spatial resolution to
provide photometry of the individual stars have allowed, by means of
colour-magnitude diagrams, to date some of the bursts (e.g.\ {\sc
VII}Zw403 by \citeplain{1998ApJ...493L..23S}; {\sc I}Zw18 by
\citeplain{1999AJ....118..302A}, see Figure \ref{colordiag}). In all these
works evidence for a stellar population older than 500 Myr 
has been found. Besides, old stellar low-surface brightness components have
been detected in  these local objects by studying the radial light
distribution in the optical (e.g.\ \citeplain{2003ApJS..147...29G}) and the
near infrared (e.g.\ \citeplain{2005A&A...429..115N}). 

After the findings that a
considerable number of the 
objects observed at intermediate and high redshifts seem to have properties
similar to the \HII\ galaxies we know in the Local  Universe, it has been
suggested that these objects might have been very common in the past and some
of them may have evolved to other kind of objects
\cite{1995ApJ...440L..49K}.

\label{CNSFRS}

\subsection{Circumnuclear Regions: Star Formation in high metallicity
  environments} 

The gas flows in disc of spiral galaxies can be strongly perturbed by the
presence of bars, although the total disc SFR does not appear to be
significantly affected by them \cite{1998ARA&A..36..189K}. These perturbations
of the gas flow trigger nuclear star formation in the bulges of some barred
spiral galaxies. There is only a modest effect of the spiral arm structure on
the global SFR of spiral galaxies
\cite{1986ApJ...311..554E,1986ApJ...311..548M,1990ApJ...349..497C,1992ApJ...385L..37K}.
Grand-design spiral galaxies (those that have strong two-armed spiral
patterns) show strong local enhancements of star formation in their
arms, without a corresponding excess in their total SFRs. This
suggests that the primary effect of the spiral density wave is to concentrate
star formation in the arms, but not to increase the global efficiency
\cite{1998ARA&A..36..189K}. 

\begin{figure}
\centering
\includegraphics[width=.9\textwidth,angle=0]{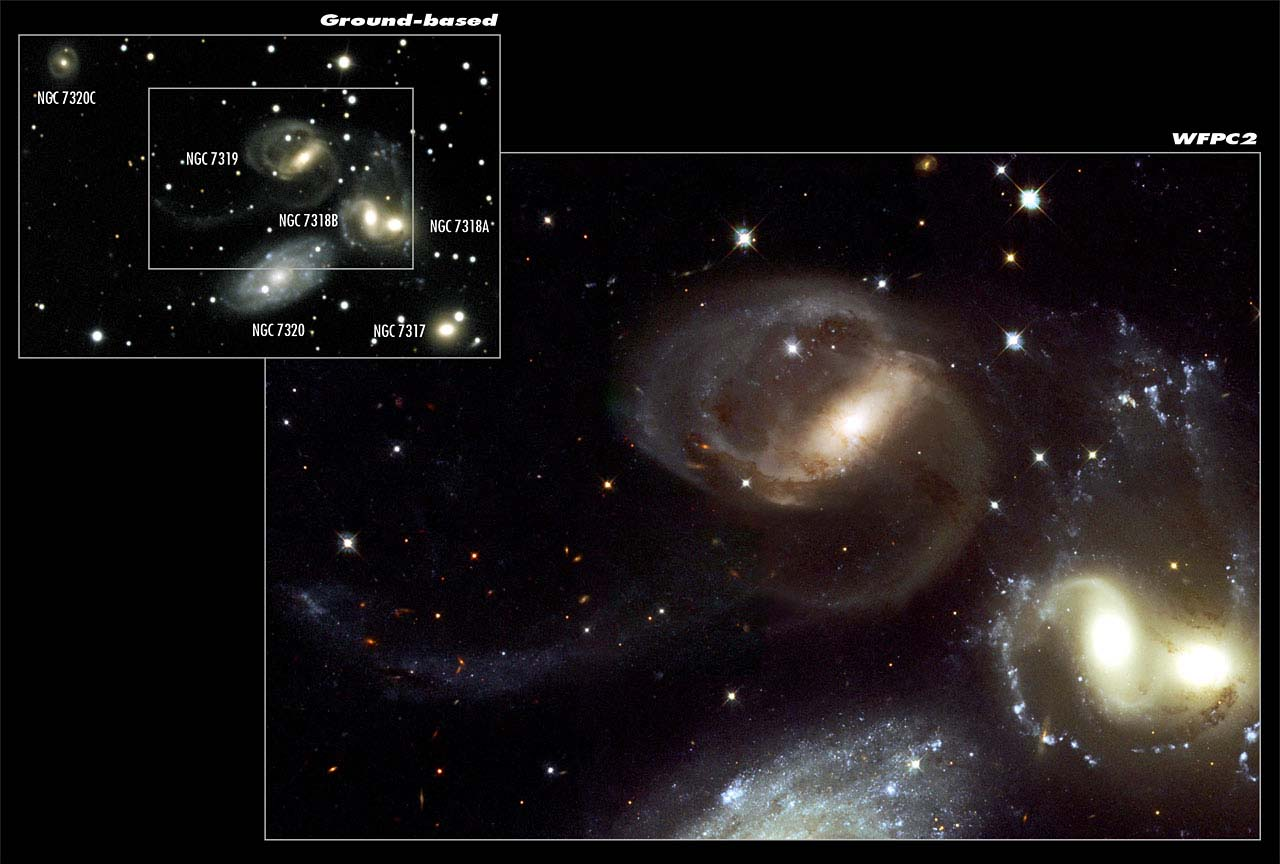}
\includegraphics[width=.9\textwidth,angle=0]{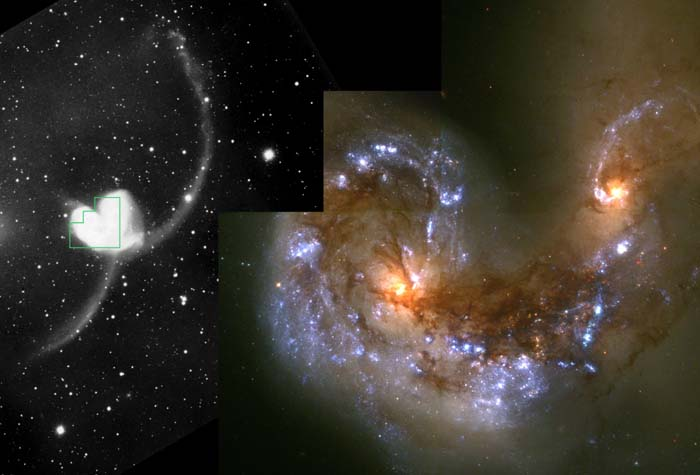}
\caption[False colour images of the Stephan's quintet and the Antennae
  galaxies]{Upper panel: false colour images of the Stephan's quintet 
  interacting system (ground base and WFPC2-HST; from the web page: {\it
  www.spacetelescope.org/news/html/heic0007.html}; credit: ESA \& NOAO). Lower
  panel: a ground-based view (left; black \& white) and a false
  colour image taken through the WFPC2-HST camera of the Antennae galaxies
  [from 
  Astronomy Picture of the Day web page ({\it
  http://apod.nasa.gov/apod/ap971022.html}); credit: NASA, ESA, and
  B. Whitmore (STScI), F. Schweizer (DTM), NASA].}
\label{interac}
\end{figure}

External environmental influences however, can
have much stronger effects on the SFR, among them, the most important by far,
are tidal interactions. There are several studies of these
interactions, in particular on the global H$\alpha$ and far-IR emission of
these interacting and merging systems, which show a strong excess of star
formation (see e.g.\
\citeplain{1987AJ.....93.1011K,1987ApJ...320...49B,1988ApJ...335...74B,1988ARA&A..26..343T,1991ApJ...374..407X,1995ApJ...450..547L,1998AJ....115..938B,2002ApJS..143...47D,2003ApJS..148..353T,2004A&A...425..813B,2005AJ....130.2117S,2006ApJ...642..158E,2006AJ....132..197W,2007AJ....134..527W}).
The young extragalactic star clusters belonging to these systems
have been the aim of different studies during the last decades
(e.g.\ \citeplain{1991MNRAS.253..245D,1992AJ....103..691H,1993AJ....106.1354W,1999AJ....118.1551W,1999AJ....118..752Z,2005A&A...443...41M,2005A&A...431..905B,2006A&A...448..881B,2007ApJ...664..284T,2008arXiv0805.2559M}).
The two most famous and emblematic examples of this phenomenon are the
Stephan's quintet (see \citeplain{2001AJ....122..163G}, and references
therein) and the pair called the Antennae galaxies (see
\citeplain{1995AJ....109..960W}, and references therein). In Figure
\ref{interac} we show two spectacular images of them. The enhancement of the
SFR is highly variable depending on the star formation conditions, the degree
of enhancement ranging from zero in gas-poor galaxies to around 
10-100 times in extreme cases \cite{1998ARA&A..36..189K}. Much larger
enhancements are often seen in the circumnuclear regions of strongly
interacting and merging systems
\cite{1998ARA&A..36..189K,1998ApJ...498..541K}.

\medskip

\noindent

\begin{figure}
\centering
\includegraphics[width=1.\textwidth,angle=0]{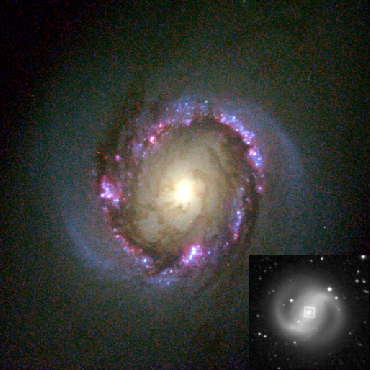}
\caption[False colour mosaic of the circumnuclear ring pattern of
  NGC\,4314]{False colour
  mosaic of the circumnuclear ring pattern of 
  star-forming regions in the dynamical center of NGC\,4314 taken with the
  WFPC2 camera on board the HST. The extension of the ring compared with the
  galaxy size is shown in the lower right panel, with the field represented in
  the enlargement marked with a box in a McDonald Observatory image [from
  Hubble Heritage Team web page ({\it http://heritage.stsci.edu/}); credit:
  NASA and ESA]).}
\label{ngc4314}
\end{figure}

Yet, the bulges of some nearby, non-interacting, spiral galaxies show intense
star-forming regions 
located in a  roughly annular pattern around their nuclei. Figure
\ref{ngc4314} shows the central part of NGC\,4314 in false colours with the
central zone marked in a complete image of the galaxy from McDonald
Observatory. 

In the middle of last century, \citetex{1958PASP...70..364M} 
classified a total of 608 galaxies from plates obtained mainly by
Edwin Hubble in the Mount Wilson-Palomar collection, using as the principal
classification 
criterion the degree of central concentration of light of each
galaxy. An apparent fairly common phenomenon in some types of galaxies was
pointed out by Morgan: their nuclear regions can consist of
an extremely brilliant, small nucleus superposed on a considerably fainter
background (NGC\,4051), or it may be made up of multiple ``hot-spots''
(NGC\,5248, 1808, 4321, and 3351). Almost a decade later,
\citetex{1965PASP...77..287S} suggested a relationship between the existence
of a bar and the presence of abnormal features in their nuclei for a survey of
35 bright southern galaxies. Inspecting the Hubble plate collection at
Pasadena, these authors extended the survey to the whole sky
\cite{1967PASP...79..152S}. They restricted their discussion to
galaxies brighter than 11.0 total photographic  magnitude
\cite{1963ApJS....8...31D} in order to include the southern objects previously
studied by them. The final sample consisted of 136 galaxies,  20 of which were
found to have peculiar nuclei, among them NGC\,2903, 3310 and 3351, the
galaxies studied in Chapter \S \ref{cnsfr-obs-kine}. They found that
$\sim$\,14\,\% of these galaxies presented peculiar nuclei. 

\begin{figure}
\centering
\includegraphics[width=.69\textwidth,angle=0]{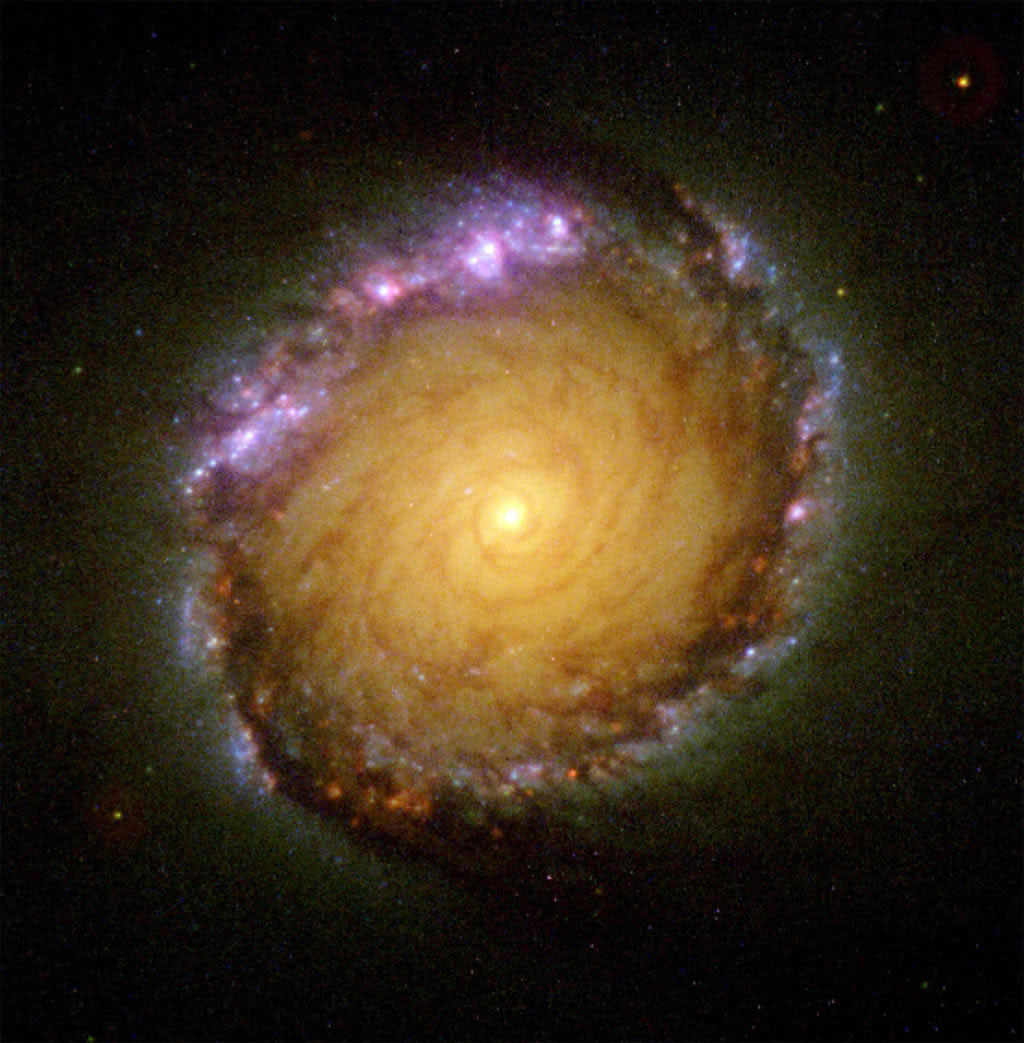}\\
\includegraphics[width=.69\textwidth,angle=0]{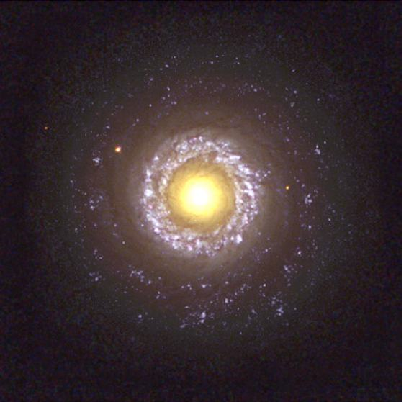}
\caption[Two examples of circular rings: NGC\,1512 and NGC\,7742 ]{Two
  examples of circular rings. Upper panel: NGC\,1512 (Figure 3 of 
  \citeplain{2001AJ....121.3048M}). Lower panel: NGC\,7742 [from Hubble
  Heritage Team web page ({\it http://heritage.stsci.edu/}); credit: NASA and
  ESA]).}  
\label{rings}
\end{figure}

\begin{figure}
\hspace{-1.cm}\includegraphics[width=.95\textwidth,angle=0]{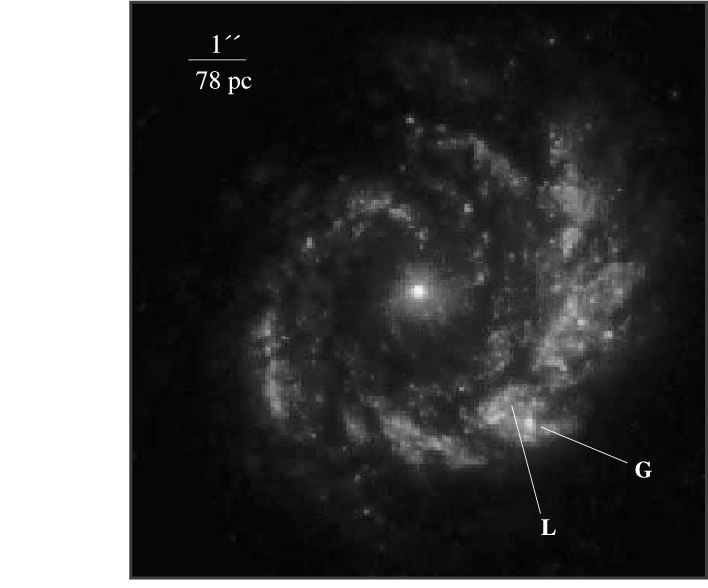}
\caption[STIS F25QTZ ultraviolet image of the central region of NGC\,4303]{STIS
  F25QTZ ultraviolet image of the NGC\,4303 nucleus and the surrounding
  star-forming structure (Figure 1 of \citeplain{2002ApJ...579..545C}). North
  is up, and east to the left. Cluster G and L detected in their STIS spectrum
  are marked.} 
\label{ngc4303}
\end{figure}

At optical wavelengths, these circumnuclear star-forming regions (CNSFRs) are
easily observable rings. In Figure \ref{rings} we show two false colour images
of the central zones of NGC\,1512 and NGC\,7742, upper and lower panels
respectively, which display very nice ring patterns. In the ultraviolet (UV),
massive stars dominate the observed circumnuclear emission even in the
presence of an active nucleus
\cite{1998ApJ...505..174G,2002ApJ...579..545C}. In Figure \ref{ngc4303} we
show the high-resolution (0.025\,\arcsec\,px$^{-1}$) deep STIS ultraviolet
image of the central zone of the spiral galaxy NGC\,4303 from
\citetex{2002ApJ...579..545C}. In this image we can observe in more 
detail the previously unresolved star forming knots (see the HST-WFPC2 data
from \citeplain{1997ApJ...484L..41C}), whose separation distances are less
than 0.2\,\arcsec\ ($\leq$15\,pc at the adopted distance for this galaxy). 
As was pointed out by Colina et al., the UV-bright nucleus remains a compact
source, showing properties very similar to the circumnuclear stellar clusters
G and L. 

CNSFRs are also present in galaxies with active nuclei. Cid Fernandes et al.\
\cite*{2001ApJ...558...81C}, for a representative sample 
of 35 Seyfert 2 galaxies,  find that about 40 per cent 
of them show unambiguous evidence of circumnuclear star formation within
300\,pc of the nucleus and that these star-forming regions contribute about 30
to 50 per cent to the H$\beta$ total emission of the central zone.  

 The distinctive nature (with respect
to the more extended star formation in discs) of the luminous nuclear
star-forming regions was fully revealed with the opening of the mid- and
far-IR spectral ranges (see for example
\citeplain{1972ApJ...176L..95R,1973ApJ...182L..89H,1978ApJ...220L..37R,1980ApJ...235..392T}).
In general, CNSFRs and giant \HII\ regions in the discs of galaxies are very
much alike, although  
the former look more compact and show higher peak surface brightness
\cite{1989AJ.....97.1022K} than the latter. 
Their large H$\alpha$ luminosities, typically higher than
10$^{39}$\,erg\,s$^{-1}$ (see Figure \ref{LHbeta}), point to relatively
massive star clusters as their 
ionization source, which minimizes the uncertainties due to small number
statistics when 
applying population synthesis techniques (see e.g.\
\citeplain{2002A&A...381...51C}). Added interest in the study of CNSFRs comes
from the fact that they are in general of high metal abundance
\cite{2006astro.ph..0787D}, therefore they 
provide clues for the understanding of star formation phenomena at large 
metallicities, and, being close to the nuclear regions, for the determination
of metallicity gradients in spiral galaxies.

In Table \ref{comparisonK} we list a brief comparative summary between the
characteristics of the more extended star forming discs of spiral galaxies and
the circumnuclear star-forming regions (Table 1 of
\citeplain{1998ARA&A..36..189K}). This shows that these two kinds of objects 
differ in many aspects. The CNSFRs are specially distinctive in terms of
the absolute range in SFRs, the much higher spatial concentration of gas and
stars, its burst-like nature and its systematic variation with galaxy type
\cite{1998ARA&A..36..189K}.

\begin{table}
\centering
\caption{Star Formation in Discs and Nuclei of Galaxies (Table 1
of Kennicutt, 1998a).}

\bigskip
\begin{tabular}{@{}lcc@{}}\hline\hline 


Property& Spiral Discs & Circumnuclear Regions \\

\hline 
Radius  &  1\,-\,30\,kpc & 0.2\,-\,2\,kpc \\
SFR &      0\,-\,20\,M$_\odot$~yr$^{-1}$ & 0\,-\,1000\,M$_\odot$~yr$^{-1}$ \\
Bolometric Luminosity   & 10$^6$\,-\,10$^{11}$\,L$_\odot$ & 10$^6$\,-\,10$^{13}$\,L$_\odot$ \\
Gas Mass   & 10$^8$\,-\,10$^{11}$\,M$_\odot$ & 10$^6$\,-\,10$^{11}$\,M$_\odot$ \\
Star Formation Timescale  & 1\,-\,50\,Gyr & 0.1\,-\,1\,Gyr \\
Gas Density  & 1\,-\,100\,M$_\odot$~pc$^{-2}$ & 10$^2$\,-\,10$^5$\,M$_\odot$~pc$^{-2}$ \\
Optical Depth (0.5\,$\mu$m) & 0\,-\,2 & 1\,-\,1000 \\
SFR Density  & 0\,-\,0.1\,M$_\odot$~yr$^{-1}$~kpc$^{-2}$ & 1\,-\,1000\,M$_\odot$~yr$^{-1}$~kpc$^{-2}$ \\
Dominant Mode & steady state & steady state $+$ burst \\[2pt]
\hline

Type Dependence? & strong & weak/none \\
Bar Dependence? & weak/none & strong \\
Spiral Structure Dependence?  & weak/none & weak/none \\
Interactions Dependence? & moderate & strong \\
Cluster Dependence? & moderate/weak & ? \\
Redshift Dependence? & strong & ? \\

\hline
\label{comparisonK}
\end{tabular}
\end{table}



\label{Objectives}

\section{Main objectives of this work}

Spectrophotometry of bright \HII\ galaxies in the Local Universe allows the
determination of abundances from methods that rely on the measurement of
emission line intensities and atomic physics (the  ``direct method" referred
to above).  
In the case of more distant or intrinsically fainter
galaxies, the low signal-to-noise obtained with current telescopes precludes
the application of this method and empirical ones based on the strongest
emission lines are required. The fundamental basis of these empirical methods
is reasonably well understood (see e.g. \citeplain{2005MNRAS.361.1063P}). The
accuracy of the results however depends on the goodness of their calibration
which in turn depends on a well sampled set of precisely derived abundances by
the ``direct method" so that interpolation procedures are reliable. Enlarging
the calibration range is also important since, at any rate, empirically 
obtained relations should never be used outside their calibration validity
range. 

The precise derivation of elemental abundances however is not a
straightforward matter. Firstly, accurate measurements of the emission lines
are needed. Secondly, a certain knowledge of the ionization structure of the
region is required in order to derive ionic abundances of the different
elements and in some cases photo-ionization models are needed to correct for
unseen ionization states. An accurate diagnostic requires the measurement of
faint auroral lines covering a wide spectral range and their accurate (better
than 5\%) ratios to Balmer recombination lines. These faint lines are usually
about 1\% of the  H$\beta$ intensity. The spectral range must include from
the UV [O{\sc ii}]\,$\lambda$\,3727\,\AA\ doublet, to the near IR [S{\sc iii}]
$\lambda\lambda$\,9069,9532\,\AA\ lines. This allows the derivation of the
different line temperatures: T$_e$([O{\sc ii}]), T$_e$([S{\sc ii}]),
T$_e$([O{\sc iii}]), T$_e$([S{\sc iii}]), T$_e$([N{\sc ii}]), needed in order
to study the temperature and ionization structure of each \HII\ galaxy
considered as a multizone ionized region. 

\begin{itemize}
\item[] {\bf One of the main objectives of this thesis has been to design a
  methodology to perform a self-consistent analysis of the emitting gas in
  \HII\ galaxies adequate to the data that can be obtained with the XXI
  century technology. This methodology requires the production and calibration
  of empirical relations between the different line temperatures that should
  replace the commonly used ones based on simplistic, and poorly contrasted,
  photo-ionization model sequences.} 
\end{itemize}


In many cases, CNSFRs show emission line spectra similar to those of disc
\HII\ regions. However, they show a higher continuum from background stellar
populations as expected from their circumnuclear location, often inside 500 pc
from the galaxy centre.
In early type spirals, CNSFRs are also expected to be
amongst the highest metallicity regions as corresponds to their position near
the galactic bulge. These facts taken together make the analysis of these
regions complicated since, in general, their low excitation makes any
temperature sensitive line too weak to be measured, particularly against a
strong underlying stellar continuum.  In fact, in most cases, the [O{\sc
    iii}]\,$\lambda$\,5007\,\AA\ line, which is typically  one hundred times
more intense than the auroral [O{\sc iii}]\,$\lambda$\,4363\,\AA\ one, can
barely be seen.

Accurate measures of elemental abundances of high metallicity regions are
crucial to obtain reliable calibrations of empirical abundance estimators,
widely used but poorly constrained, whose choice can severely bias results
obtained for quantities of the highest relevance for the study of galactic
evolution like the luminosity-metallicity (L-Z) relation for galaxies. CNSFRs
are also ideal cases to study the behavior of abundance estimators in the
high metallicity regime. 

\begin{itemize}
\item[] {\bf  A second objective has been to develop a semi-empirical method
  for the derivation of abundances in high metallicity \HII\ regions that can
  be applied to the CNSFRs of our study. Given the weakness of the oxygen
  emission lines in their spectra our study has been based mainly on the
  sulphur emission lines observed in the far red spectral region. }
\end{itemize}

Although CNSFRs are very luminous, not much is known about their 
kinematics or dynamics for both the ionized gas and the stars. In fact, the
most poorly known property of star forming clusters in galaxies is their
mass. 

There are different methods to estimate the mass of a stellar
cluster. Classically one assumes that the system is virialized and determines
the total mass inside a radius by applying the virial theorem to the observed 
velocity dispersion of the stars ($\sigma_{\ast}$). The stellar velocity
dispersion is however hard to measure in young stellar clusters (a few
million-years old) due to the 
shortage of prominent stellar absorption lines. The optical continuum between
3500 and 7000\,\AA\ shows very few lines since the light at  these wavelengths
is dominated by OB stars which have weak absorption lines at the same
wavelengths of the nebular emission lines (Balmer H and He{\sc i} lines).  A
better situation is encountered at longer wavelengths (far red). There, the
contamination due to nebular lines is much 
smaller and the stellar [Ca{\sc ii}]  absorption lines at
$\lambda\lambda$\,8498, 8542, 8662\,\AA\ (CaT), if present, can be used.  The
CaT lines in CNSFRs have previously been detected but at a spectral resolution
below that required to measure accurately their velocity dispersions
\cite{tesisdiego}.  


\begin{itemize}
\item[] {\bf A third objective of this work has been to measure the stellar
  velocity dispersions of selected CNSFRs from the CaT lines and derive their
  dynamical masses. The comparison of these masses with those inferred from
  the number of ionizing photons in the region will give an estimate of the
  contribution of the present star formation episode to the mass proceeding
  from previous stellar generations and will help to characterize the star
  formation history of these objects.}
\end{itemize}

A comparison between the velocity dispersions of gas and stars in CNSFRs is
also of the greatest importance for the interpretation of the motions of the
gas in the clusters and the influence of their gravitational fields. Also, the
investigation of the presence of gas infall or outflow in the central regions
of the galaxies where CNSFRs reside is important. 

\begin{itemize}
\item[] {\bf To perform this comparison has been another objective of this
  work. For this, we have used high dispersion spectroscopic observations and
  analysis techniques that involve the simultaneous fit of different velocity
  components.} 
\end{itemize}

\label{Structure}

\section{Structure of this thesis}

\noindent
This thesis is structured into two main parts. The first one is related to the
sample \HII\ galaxies, and the second to the CNSFRs. 

In Chapter \S \ref{HIIgal-obs} a methodology is proposed to perform a
self-consistent analysis of the physical properties of the emitting gas of
\HII\ galaxies, while Chapter \S \ref{neon} is focused on the study of the
strong optical 
collisional emission lines of Ne and Ar in ionized
gaseous nebulae for which new ionization correction factors for
these two elements are calculated. 

Chapters \S \ref{cnsfr-obs-kine} and \S \ref{abundan} deal with the study of
star formation in circumnuclear regions. The former is dedicated to the study
of the kinematical properties of gas and stars in a sample of CNSFRs, and the
derivation of their dynamical masses, the masses and some properties of their
ionizing stellar cluster, and the masses of the ionized gas, while the latter
is devoted to the derivation of  their chemical abundances by a 
semi-empirical method.

Finally, in the last Chapter of this thesis, \S \ref{genconc}, we present
the general conclusions of this work and we list some of the future projects
that stem from it.

\addcontentsline{toc}{section}{\numberline{}Bibliography}

\bibliographystyle{astron}
\bibliography{tesis}


\chapter{Star Formation in \HII\ Galaxies. Properties of the ionized gas}
\label{HIIgal-obs}

\section{Introduction}

\HII\ galaxies are low mass irregular galaxies with, at least, a recent episode
of  violent star formation (Melnick, Terlevich and Eggleton, 1985a; Melnick,
Terlevich and Moles 1985b)
\nocite{1985MNRAS.216..255M,1985RMxAA..11...91M}
concentrated in a few parsecs close to their cores. The ionizing fluxes
originated by these young massive stars dominate the light of this subclass of
Blue Compact Dwarf galaxies (BCDs) which show  emission line spectra very
similar to those of giant extragalactic  \HII\ regions (GEHRs; Sargent and
Searle, 1970; French 1980).
\nocite{1970ApJ...162L.155S,1980ApJ...240...41F}
Therefore, by applying the same 
measurement techniques as for \HII\ regions, we can derive the temperatures,
densities and chemical composition of  the interstellar gas in this type of
generally metal-deficient galaxies
\cite{1991A&AS...91..285T,2000A&ARv..10....1K,2006MNRAS.365..454H}. In some
cases, it is possible to detect in 
these objects, intermediate-to-old stellar populations which have a more
uniform spatial distribution than the bright and young stellar  populations
associated with the ionizing clusters \cite{1998ApJ...493L..23S}. This older
population produces a characteristic spectrum with absorption features which
mainly affect the hydrogen recombination emission lines
\cite{1988MNRAS.231...57D}, that is the Balmer and Paschen series in the
spectral range of our interest. In some cases, the underlying stellar
absorptions can severely affect the ratios of H{\sc i} line pairs and hence
the determination of the reddening constant [c(H$\beta$)]. They must therefore
be measured with special care (see discussion in \S \ref{results}).

A considerable number of the {blue} objects observed at intermediate
redshifts seem to have properties (mass, R$_e$, velocity width of  the emission
lines) similar to local \HII\ galaxies  
\cite{1994ApJ...427L...9K,1995ApJ...440L..49K,1996ApJ...460L...5G,1998ApJ...495L..13G}. 
In particular, those with $\sigma$\,$<$\,65\,km\,s$^{-1}$ follow the same
$\sigma - L_B\ {\rm and}\ L_{H\beta}$ relation as seen in \HII\ galaxies
\cite{2000MNRAS.311..629M,2003ApJ...598..858M,2003RMxAC..16..213T,2005MNRAS.356.1117S}.
Similar conclusion is drawn from recent studies on Lyman Break galaxies that
also suggest that strong narrow emission line galaxies  might have been 
very common in the past
(e.g.\ Pettini et al., 2000; Pettini et al.,2001; Ellison et al., 2001).
\nocite{2000ApJ...528...96P,2001ApJ...554..981P,2001A&A...379..393E}
To detect possible evolutionary effects like systematic 
differences in their chemical composition, accurate and reliable methods
for abundance determination are needed.

This is usually done by combining photo-ionization model results and observed
emission line intensity ratios.  There are  several major problems with this
approach that limit the confidence of  present results.
Among them: the effect of temperature structure in multiple-zone models
\cite{2003MNRAS.346..105P}; the presence of temperature fluctuations across the
nebula \cite{2003ApJ...584..735P};  collisional and density effects on ion
temperatures \cite{1999ApJ...527..110L,2003MNRAS.346..105P}; the
presence of neutral zones affecting the calculation of ionization correction
factors (ICFs; \nocite{2002ApJ...565..668P} Peimbert et al., 2002); the
ionization structure not adequately reproduced by current models
\cite{2003MNRAS.346..105P}; ionization vs.~matter bounded zones, affecting the
low ionization lines formed in the outer parts of the ionized regions
\cite{2002MNRAS.329..315C}. 
On the other hand, the understanding of the age and evolutionary state of \HII\
galaxies require the use of self-consistent models for the ionizing stars and
the ionized gas. However, model computed evolutionary sequences show important
differences with observations \cite{2003A&A...397...71S}, including: (a) He{\sc
  ii} is too strong in a substantial number of objects as compared to model
predictions; (b) [O{\sc iii}]/H$\beta$ vs.\ [O{\sc ii}]/H$\beta$ and [O{\sc
    iii}]/H$\beta$  vs.~[O{\sc i}]/H$\beta$ are not  well reproduced by
evolutionary model sequences in the sense that predicted collisionally excited
lines are too weak compared to observations; (c) there is a large spread in the
[N{\sc ii}]/[O{\sc ii}] values (more than an order of magnitude) for galaxies
with the same value of ([O{\sc ii}]+[O{\sc iii}])/H$\beta$ in the metallicity
range from 8 to 8.4 (see e.g.\ P\'erez-Montero and D\'iaz, 2005). 
\nocite{2005MNRAS.361.1063P}
\nocite{2002AJ....123..485S}

Substantial progress toward solving the problems listed above has to come from
the accurate measurement of weak emission lines  which will allow to derive 
[O{\sc ii}], [S{\sc ii}] and [S{\sc iii}] temperatures and densities allowing to
constrain the ionization structure as well as Balmer and Paschen discontinuities
which will provide crucial information about the actual values of temperature
fluctuations. It is possible that these fluctuations produce the observed
differences between the abundances relative to hydrogen derived from
recombination lines (RLs) and collisionally excited lines (CELs) when a constant
electron temperature is assumed
\cite{1967ApJ...150..825P,1969BOTT....5....3P,1971BOTT....6...29P}. These
discrepancies have been observed in a good sample of objects, such as galactic
\HII\ regions (e.g.\ Esteban et al., 2004; Garc\'ia-Rojas et al., 2005;
Garc\'ia-Rojas et al., 2006; Garc\'ia-Rojas et al., 2007, and references
therein), 
\nocite{2004MNRAS.355..229E,2005MNRAS.362..301G,2006MNRAS.368..253G,2007ApJ...670..457G}
\HII\ regions in the Magellanic Clouds (e.g.\ Peimbert et al., 2000; Peimbert,
2003; Tsamis et al., 2003, and references therein),
\nocite{2000ApJ...541..688P,2003ApJ...584..735P,2003MNRAS.338..687T}
extragalactic \HII\ regions and star-forming galaxies (e.g.\ Peimbert and
Peimbert, 2003; Peimbert et al., 2005; Guseva et al., 2006; Guseva et al.,
2007; Bresolin, 2007; Peimbert et al., 2007; Kewley and Ellison, 2008, and
references therein) 
\nocite{2003RMxAC..16..113P,2005ApJ...634.1056P,2006ApJ...644..890G,2007A&A...464..885G,2007ApJ...656..186B,2007RMxAC..29...72P,2008arXiv0801.1849K} 
and planetary nebulae (e.g.\ Rubin et al., 2002; Wesson et al., 2005; Liu et
al., 2006; Liu, 2006; Peimbert and Peimbert, 2006, and references therein).
\nocite{2002MNRAS.334..777R,2005MNRAS.362..424W,2006MNRAS.368.1959L,2006IAUS..234..219L,2006IAUS..234..227P}
Likewise, there are relatively recent theoretical works that study the possible
causes of these discrepancies in abundance determinations using
photo-ionization 
models of different complexity (e.g.\ Stasi\'nska, 2005; Jamet et al., 2005;
Tsamis and P\'equignot, 2005; Ercolano et al., 2007, and references therein).
\nocite{2005A&A...434..507S,2005A&A...444..723J,2005MNRAS.364..687T,2007MNRAS.379..945E}

Spectrophotometry of bright \HII\ galaxies in the Local Universe allows the
determination of abundances from methods that rely on the measurement of
emission line intensities and atomic physics. This is referred to as the
"direct" method. In the case of more distant or intrinsically fainter
galaxies, the low signal-to-noise obtained with current telescopes precludes
the application of this method and empirical ones based on the strongest
emission lines are required. The fundamental basis of these empirical methods
is reasonably well understood (see e.g.\ P\'erez-Montero and D\'iaz, 2005).
\nocite{2005MNRAS.361.1063P}
The accuracy of the results however depends on the goodness of their calibration
which in turn depends on a well sampled set of precisely derived abundances by
the "direct" method so that interpolation procedures are reliable. Enlarging
the calibration range is also important since, at any rate, empirically 
obtained relations should never be used outside their calibration validity
range. 

The precise derivation of elemental abundances however is not a
straightforward matter. Firstly, accurate measurements of the emission lines
are needed. Secondly, a certain knowledge of the ionization structure of the
region is required in order to derive ionic abundances of the different
elements and in some cases photo-ionization models are needed to correct for
unseen ionization states. An accurate diagnostic requires the measurement of
faint auroral lines covering a wide spectral range and their accurate (better
than 5\%) ratios to Balmer recombination lines. These faint lines are usually
about 1\% of the  H$\beta$ intensity. The spectral range must include from
the UV [O{\sc ii}]\,$\lambda\lambda$\,3727,29\,\AA\ doublet, to the near IR
[S{\sc iii}] 
$\lambda\lambda$\,9069,9532\,\AA\ lines. This allows the derivation of the
different line temperatures: T$_e$([O{\sc ii}]), T$_e$([S{\sc ii}]),
T$_e$([O{\sc iii}]), T$_e$([S{\sc iii}]), T$_e$([N{\sc ii}]), needed in order
to study the temperature and ionization structure of each \HII\ galaxy
considered as a multizone ionized region.

Unfortunately most of the available starburst and \HII\ 
galaxy spectra have only a restricted wavelength  range (usually from about 3600
to 7000 \AA), consequence of observations with single arm spectrographs, and
do not have the adequate signal-to-noise ratio (S/N) to accurately measure the 
intensities of the weak diagnostic emission lines. Even the Sloan Digital Sky
Survey (SDSS) spectra (Stoughton et al.\ 2002) do not cover simultaneously the
[O{\sc ii}]\,$\lambda\lambda$\,3727,29 and the [S{\sc iii}]\,$\lambda$\,9069\,\AA\
lines, they only represent an average inside a 3\,arcsec fibre and reach the
required signal-to-noise ratio only for the brightest objects.
We have therefore undertaken a project with the aim of obtaining a database of
top quality line ratios for a sample that includes the best objects for the
task. The data is collected using exclusively two arm  spectrographs in order to
guarantee both high quality spectrophotometry in the whole spectral range from
3500 to 10500\,\AA\, approximately, and good spectral and spatial
resolution. In this way we 
are able to vastly improve constraints on the photo-ionization
models including the  mapping of the ionization structure and the measurement
of temperature fluctuations about which very little is known.
\nocite{2002AJ....123..485S}

It is important to realize that the combination of accurate spectrophotometry
and wide spectral coverage cannot be achieved  using single arm spectrographs
where, in order to  reach the necessary spectral resolution, the wavelength
range must be split into several independent observations. In those cases, the
quality of the spectrophotometry is at best doubtful mainly because the
different spectral ranges are not observed simultaneously. This problem
applies to both objects and calibrators. Furthermore one can never be sure of
observing exactly the same region of the nebula in each spectral range. To
avoid all these problems the use of double arm spectrographs is required. 

In this chapter we present simultaneous blue and red observations of three
\HII\ galaxies obtained with the ISIS double-beam spectrograph mounted on the
4.2m William Herschel Telescope (WHT) and seven galaxies observed with the
double arm TWIN spectrograph at the 3.5m 
telescope of the Calar Alto Observatory at the Complejo Astron\'omico Hispano
Alem\'an (CAHA). All of these objects were selected from
the SDSS, and  are of a sufficient quality as to allow
the detection and measurement of several temperature sensitive lines and add
to the still scarce base of precisely derived abundances. Moreover, in the case
of the WHT observations we are able to measure the Balmer jump, and then we
can estimate the temperature fluctuations.
Details regarding the selection of the objects as well as the
observations and data reduction are given in \S \ref{observations}. Section \S
\ref{results} presents the results including 
line measuring techniques. The methodology for the derivation of gaseous
physical conditions and elemental abundances is presented in Sections \S
\ref{physiHIIgal} and \S \ref{chem-abund-der}, respectively. Section \S
\ref{discu} is devoted to the discussion of our results and finally, our
conclusions are summarized in Section \S \ref{summ2}.

\section{Observations and data reduction}
\label{observations}

\subsection{Object selection}

SDSS constitutes a very valuable base for statistical studies of the
properties of galaxies. At this moment, the Sixth Data
Release\footnote{http://www.sdss.org/dr6/} (DR6), the last one up to now,
represents the completion of the SDSS-I project
\cite{2008ApJS..175..297A}. The DR6 contains five-band photometric data 
for  about 2.87$\times$10$^8$  objects selected over 9583\,$deg^2$ and more
than 1.27 million spectra of galaxies, quasars and stars selected from
7425\,$deg^2$. The spectroscopic data have a resolution (R) of 1800-2100
covering a spectral range from 3800 to 9200\,\AA, with a single 3\,arcsec
diameter aperture. The SDSS data were reduced and flux-calibrated using
automatic pipelines.
However, when we selected our WHT objects on July 2004 and our CAHA objects on
June 2006, the DR2\footnote{http://www.sdss.org/dr2/}
\cite{2004AJ....128..502A} and the 
DR4\footnote{http://www.sdss.org/dr4/} \cite{2006ApJS..162...38A},
respectively, were just 
available. All the SDSS data releases, contain the same type of objects
observed in the 
same five photometric bands and using the same spectroscopic configuration. 
All the objects belonging to one data release are also included in the next
ones, but some objects could have been re-calibrated or re-observed somehow.

Using the implementation of the SDSS database in the INAOE Virtual Observatory
superserver\footnote{http://ov.inaoep.mx/}, we selected the brightest nearby
narrow emission line galaxies with very strong lines and large equivalent
widths of the H$\alpha$ line from the whole SDSS data release available when we
plan each observing run. Specifically our selection criteria were: 
\begin{itemize}
\item[-] H$\alpha$ flux, F(H$\alpha$)\,$>$\,4\,$\times$\,10$^{-14}$\,erg\,cm$^{-2}$\,s$^{-1}$
\item[-] H$\alpha$ equivalent width, EW(H$\alpha$)\, $>$\,50\,\AA
\item[-] H$\alpha$ width, 2.8\,$<$\,FWHM(H$\alpha$)\,$<$\,7\,\AA
\item[-] redshift, {\it z}, 10$^{-3}$\,$<\,z\,<$\,0.2
\end{itemize}
These preliminary lists were then processed using BPT
\nocite{1981PASP...93....5B}(Baldwin, Phillips and Terlevich, 1981) diagnostic
diagrams to remove AGN-like objects. 

The final list obtained on July 2004 consisted of about 200 bonafide bright
\HII\ galaxies. They show spectral properties indicating a wide range of
gaseous abundances and ages of the underlying stellar populations
\nocite{jesustesis} (L\'opez, 2005). From this list, the final set was
selected by further 
restricting the sample to the largest H$\alpha$ flux and highest
signal-to-noise ratio objects. 

From the 10 brightest of the final set we selected three \HII\ galaxies to be
observed in our one single night observing run. This final selection was 
made based on the relative positions of the sources in the sky allowing to
optimize the observing time. Subsequently we have used the
explore tool\footnote{http://cas.sdss.org/astro/en/tools/explore/} implemented
in the DR3\footnote{http://www.sdss.org/dr3/} \cite{2005AJ....129.1755A},
which was  available at the time of 
analysis, to extract again the three object SDSS spectra  for comparison
purposes. 

On June 2006, the obtained list contains about 10500 \HII\
like objects  \cite{jesustesis}. Again, they  show spectral properties
indicating a wide range of gaseous abundances and ages of the underlying
stellar populations. 
The objects with the highest (H$\alpha$) fluxes and equivalent widths
observable from the Calar Alto Observatory at the epoch of observation were
selected and for seven of them the corresponding data were secured. 

\begin{figure}
\centering
\includegraphics[width=.49\textwidth]{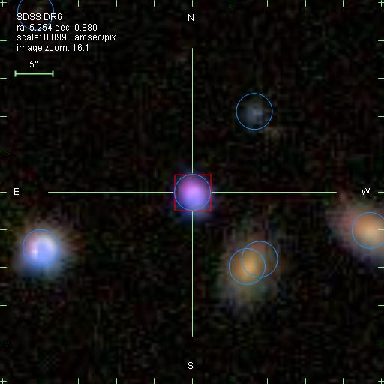}
\includegraphics[width=.49\textwidth]{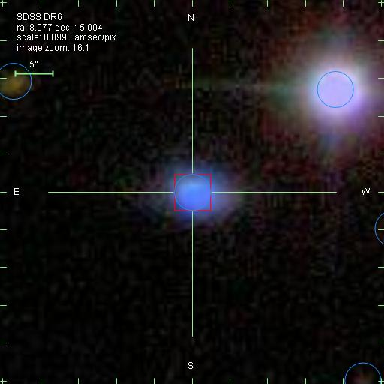}\\
\vspace{0.1cm}
\includegraphics[width=.49\textwidth]{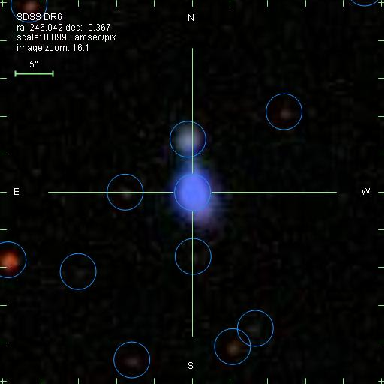}\\
\caption[False colour images of the WHT objects]{False colour images of
  the WHT objects, \whtunoc\ (upper left), \whtdosc\ (upper right) and
  \whttresc\ (lower). These images were obtained using the SDSS explore
  tools. Circles and squares represent the photometric and spectroscopic SDSS
  targets, respectively.}
\label{images-wht}
\end{figure}

For both observing runs, by an independent visual inspection of each object we
have selected the final sample. In Figures \ref{images-wht} and
\ref{images-caha} we show the false colour images of the observed objects used
to select them.
The journal of observations is given in Table \ref{jour}  and some general
characteristics of the objects from the SDSS web page are listed in Table
\ref{obj}. Column 3 of Table \ref{jour} gives the short name by which we will
refer to the observed  \HII\ galaxies in what follows.

%
%

\begin{table*}
\centering
{\scriptsize
\caption[Journal of observations of \HII\ galaxies]{Journal of observations.}
\label{jour}
\begin{tabular} {@{}c c c c c c}
\hline
Object  ID  &  spSpec SDSS   & hereafter ID &  Date &  Exposure  & Seeing\\
   &   &   &   &   (sec)   &   (\arcsec)   \\
\hline
 \multicolumn{6}{c}{WHT objects} \\[2pt]
\whtuno & spSpec-51900-0390-445  & \whtunoc & 2004 July 18  &  1\,$\times$\,1200\,+\,2\,$\times$\,2400 & 0.5-0.8\\
\whtdos & spSpec-51817-0418-302  & \whtdosc & 2004 July 18  &  1\,$\times$\,1200\,+\,2\,$\times$\,2400 & 0.5-0.8\\
\whttres& spSpec-52000-0364-187  & \whttresc& 2004 July 18  &  1\,$\times$\,1200\,+\,3\,$\times$\,1800 & 0.5-0.8\\[2pt]
 \multicolumn{6}{c}{CAHA objects} \\[2pt]
\uno    & spSpec-52790-1351-474  & \unoc    & 2006 June 25  &  5\,$\times$\,1800  & 0.9-1.2 \\
\dos    & spSpec-52721-1050-274  & \dosc    & 2006 June 23  &  4\,$\times$\,1800  & 0.8-1.1 \\
\tres   & spSpec-52765-1293-580  & \tresc   & 2006 June 22  &  4\,$\times$\,1800  & 0.8-1.2 \\
\cuatro & spSpec-52072-0617-464  & \cuatroc & 2006 June 24  &  6\,$\times$\,1800  & 1.0-1.4 \\
\cinco  & spSpec-52377-0624-361  & \cincoc  & 2006 June 23  &  5\,$\times$\,1800  & 0.8-1.1 \\
\seis   & spSpec-52791-1176-591  & \seisc   & 2006 June 25  &  5\,$\times$\,1800  & 0.9-1.2 \\
\siete  & spSpec-51818-0358-472  & \sietec  & 2006 June 22  &  5\,$\times$\,1800  & 0.8-1.1 \\[2pt]
\hline
\end{tabular}
}
\end{table*}

%
%

\begin{table*}
\centering
{\footnotesize
\caption[Right ascension, declination, redshift and SDSS photometric
  magnitudes]{Right ascension, declination, redshift and SDSS photometric
  magnitudes obtained using the SDSS explore tools$^a$.}
\label{obj}
\begin{tabular} {l c c c c r r c c}
\hline
 \multicolumn{1}{c}{Object  ID}   & \multicolumn{1}{c}{RA} & \multicolumn{1}{c}{Dec} &  redshift  & \multicolumn{1}{c}{u} & \multicolumn{1}{c}{g} & \multicolumn{1}{c}{r} & \multicolumn{1}{c}{i}   &    \multicolumn{1}{c}{z} \\
\hline
 \multicolumn{9}{c}{WHT objects} \\[2pt]
\whtunoc &  00$^h$\,21$^m$\,01\fsec00 & 00$^{\circ}$\,52\arcmin\,48\farcs10 &  0.098 & 17.56 & 17.35 & 17.51 & 16.98 & 17.45 \\
\whtdosc &  00$^h$\,32$^m$\,18\fsec60 & 15$^{\circ}$\,00\arcmin\,14\farcs20 &  0.018 & 17.04 & 16.49 & 16.53 & 16.74 & 16.65 \\
\whttresc&  16$^h$\,24$^m$\,10\fsec10 &-00$^{\circ}$\,22\arcmin\,02\farcs50 &  0.031 & 17.07 & 16.46 & 16.91 & 16.80 & 16.74 \\
 \multicolumn{9}{c}{CAHA objects} \\[2pt]
\unoc    &  14$^h$\,55$^m$\,06\fsec06 & 38$^{\circ}$\,08\arcmin\,16\farcs67 &  0.028 & 18.25 & 17.57 & 17.98 & 18.23 & 18.18 \\
\dosc    &  15$^h$\,09$^m$\,09\fsec03 & 45$^{\circ}$\,43\arcmin\,08\farcs88 &  0.048 & 18.57 & 17.72 & 18.19 & 17.87 & 17.94 \\
\tresc   &  15$^h$\,28$^m$\,17\fsec18 & 39$^{\circ}$\,56\arcmin\,50\farcs43 &  0.064 & 18.54 & 17.88 & 18.17 & 17.52 & 17.99 \\
\cuatroc &  15$^h$\,40$^m$\,54\fsec31 & 56$^{\circ}$\,51\arcmin\,38\farcs98 &  0.011 & 19.11 & 18.91 & 18.97 & 19.53 & 19.46 \\
\cincoc  &  16$^h$\,16$^m$\,23\fsec53 & 47$^{\circ}$\,02\arcmin\,02\farcs36 &  0.002 & 16.84 & 16.45 & 16.77 & 17.35 & 17.43 \\
\seisc   &  16$^h$\,57$^m$\,12\fsec75 & 32$^{\circ}$\,11\arcmin\,41\farcs42 &  0.038 & 17.63 & 17.03 & 17.27 & 17.15 & 17.15 \\
\sietec  &  17$^h$\,29$^m$\,06\fsec56 & 56$^{\circ}$\,53\arcmin\,19\farcs40 &  0.016 & 18.05 & 17.26 & 17.21 & 17.38 & 17.24 \\
\hline
\multicolumn{9}{l}{$^a$http://cas.sdss.org/astro/en/tools/explore/obj.asp}
\end{tabular}
}
\end{table*}


\begin{figure}
\centering
\includegraphics[width=.49\textwidth]{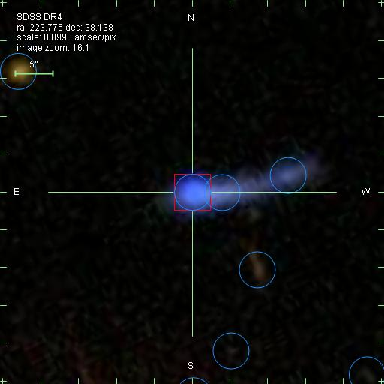}
\includegraphics[width=.49\textwidth]{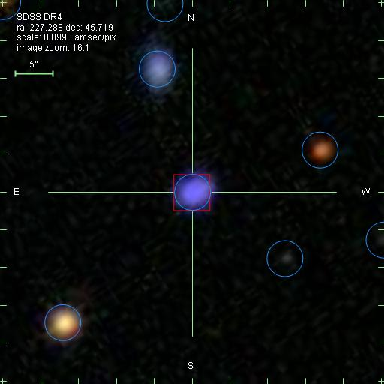}\\
\vspace{0.1cm}
\includegraphics[width=.49\textwidth]{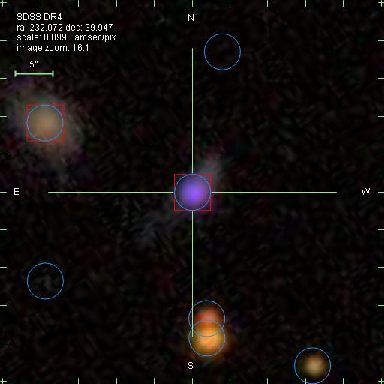}
\includegraphics[width=.49\textwidth]{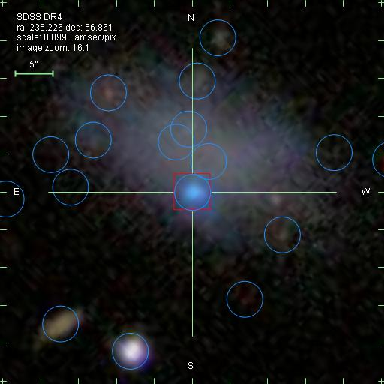}\\
\caption[False colour images of the CAHA objects]{False colour images of
  the CAHA objects, \unoc\ (upper left), \dosc\ (upper right), \tresc\ (lower
  left) and \cuatroc\ (lower right). These images were obtained using the SDSS
  explore tools. Circles and squares represent the photometric and
  spectroscopic SDSS targets, respectively.}
\label{images-caha}
\end{figure}

\setcounter{figure}{1}

\begin{figure}
\centering
\includegraphics[width=.49\textwidth]{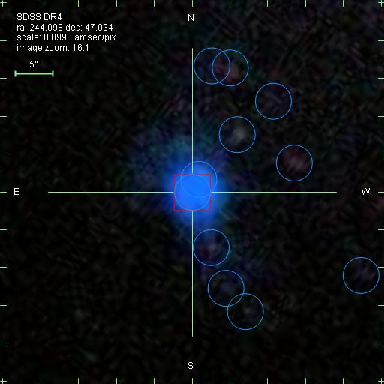}
\includegraphics[width=.49\textwidth]{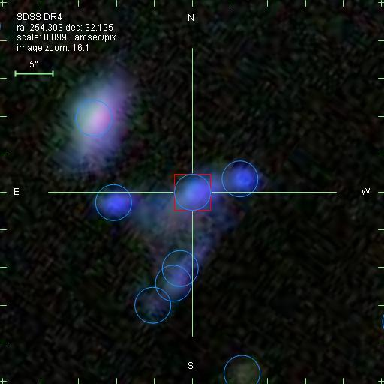}\\
\vspace{0.1cm}
\includegraphics[width=.49\textwidth]{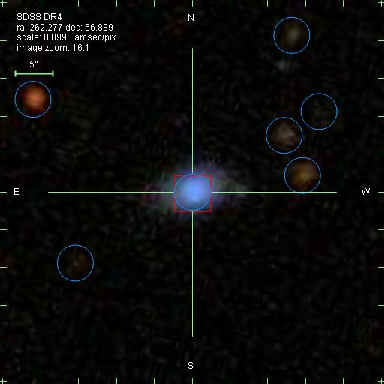}\\
\caption[($cont$) False colour images of the CAHA objects]{({\it cont})
  False colour images of 
  the CAHA objects, \cincoc\ (upper left), \seisc\ (upper right) and
  \sietec\ (lower). These images were obtained using the SDSS explore
  tools. Circles and squares represent the photometric and spectroscopic SDSS
  targets, respectively.}
\label{images-caha}
\end{figure}


\subsection{WHT observations}


\begin{table}
\centering
\caption[Instrumental configurations for the observations.]{Instrumental
  configurations for the observations.} 
\label{config}
\begin{tabular} {@{}l c c c c}
\hline
\hline
 & Spectral range  &       Disp.           & R$_{\mathrm{FWHM}}^a$   & Spatial res.         \\
 &      (\AA)      & (\AA\,px$^{-1}$)        &           & (\arcsec\,px$^{-1}$)      \\
\hline
 \multicolumn{5}{c}{WHT - ISIS} \\[2pt]
blue & 3200-5700   &       0.86              &  1780     &   0.2                \\
red  & 5500-10550  &       1.64              &  1670     &   0.2                \\
 \multicolumn{5}{c}{CAHA - TWIN} \\[2pt]
blue & 3400-5700   &       1.09              &  1420     &   0.56                \\
red  & 5800-10400  &       2.42              &  1160     &   0.56                \\

\hline
\multicolumn{5}{l}{$^a$R$_{\rm{FWHM}}$\,=\,$\lambda$/$\Delta\lambda_{\rm{FWHM}}$}
\end{tabular}
\end{table}

The blue and red spectra were obtained simultaneously using the Intermediate
dispersion Spectrograph and Imaging System (ISIS) double beam
spectrograph mounted on the 4.2m William Herschel Telescope (WHT) of the Isaac
Newton Group (ING) at the Roque de los Muchachos Observatory, on the Spanish
island of La Palma. They were acquired on July the 18th 2004 during one
single night observing run and under photometric conditions. EEV12 and Marconi2
detectors were attached to the blue and red arms of the spectrograph,
respectively. The R300B grating was used in the blue covering the wavelength
range 3200-5700\,\AA\ (centered at $\lambda_c$\,=\,4450\,\AA), giving a spectral
dispersion of 0.86\,\AA\,pixel$^{-1}$. On the red arm, the R158R grating was
mounted  providing a spectral range from 5500 to 10550\,\AA\
($\lambda_c$\,=\,8025\,\AA) and a spectral dispersion of
1.64\,\AA\,pixel$^{-1}$. In order to reduce the readout noise of our images we
have taken the observations with the `SLOW' CCD speed. The pixel size for this
set-up configuration is 0.2 arcsec for both spectral ranges.  The slit width
was $\sim$0.5 arcsec, which, combined with the spectral dispersions, yielded
spectral resolutions of about 2.5 and 4.8\,\AA\ FWHM in the blue and red arms 
respectively. All observations were made at paralactic angle to avoid effects of
differential refraction in the UV. The instrumental configuration, summarized in
Table \ref{config},was planned in order to cover the whole spectrum from 3200 to
10550\,\AA\  providing at the same time a moderate spectral resolution.   This
guarantees the simultaneous measurement  of the Balmer discontinuity and the
nebular lines of [O{\sc ii}]\,$\lambda\lambda$\,3727,29 and [S{\sc
    iii}]\,$\lambda\lambda$\,9069,9532\,\AA\  at both ends of the spectrum, in
the very same region of the galaxy. A good signal-to-noise ratio was also
required to allow the detection and measurement of weak lines such as  [O{\sc
    iii}]\,$\lambda$\,4363, [S{\sc ii}]\,$\lambda\lambda$\,4068, 6717 and 6731,
and [S{\sc iii}]\,$\lambda$\,6312. The signal-to-noise ratios attained for
each final WHT spectrum are given in the top part of Table \ref{snr}.  

Several bias and sky flat field frames were taken at the beginning and at the end
of the night in both arms. In addition, two lamp flat fields and one calibration
lamp exposure were performed at each telescope position. The calibration lamp
used was CuNe+CuAr. 

\subsection{CAHA observations}

Blue and red spectra were obtained simultaneously using the  double beam
Cassegrain Twin Spectrograph (TWIN) mounted on the 3.5m telescope of the Calar
Alto Observatory at the Complejo Astron\'omico Hispano Alem\'an (CAHA),
Spain. They were acquired in June 2006, during a four night observing run and
under excellent seeing and photometric conditions. Site\#22b and Site\#20b,
2000\,$\times$\,800\,px 15\,$\mu$m, detectors were attached to the blue and
red arms of the spectrograph, respectively. The T12 grating was used in the
blue covering the wavelength range 3400-5700\,\AA\ (centered at
$\lambda_c$\,=\,4550\,\AA), giving a spectral dispersion of
1.09\,\AA\,pixel$^{-1}$ (R\,$\simeq$\,4170). On the red arm, the T11 grating
was mounted  providing a spectral range from 5800 to 10400\,\AA\
($\lambda_c$\,=\,8100\,\AA) and a spectral dispersion of
2.42\,\AA\,pixel$^{-1}$ (R\,$\simeq$\,3350).  
The pixel size for this set-up configuration is 0.56\,arcsec for both spectral
ranges.  The slit width was $\sim$1.2\,arcsec, which, combined with the
spectral dispersions, yielded spectral resolutions of about 3.2 and 7.0\,\AA\
FWHM in the blue and the red respectively. Again, all observations were made at
paralactic angle to avoid effects of differential refraction in the UV. The
instrumental configuration, summarized in the second part of Table
\ref{config}, covers the whole 
spectrum from 3400 to 10400\,\AA\ (with a gap between 5700 and 5800\,\AA)
providing at the same time a moderate spectral resolution. As in the case of
the WHT data, this guarantees the 
simultaneous measurement  of the nebular lines from [O{\sc
    ii}]\,$\lambda\lambda$\,3727,29 to [S{\sc
    iii}]\,$\lambda\lambda$\,9069,9532\,\AA\  at both ends of the spectrum, in
the very same region of the galaxy, and a good signal-to-noise ratio was also
required to allow the detection and measurement of the weak lines. The
signal-to-noise ratios of the CAHA spectrum are given in the last part Table
\ref{snr}. 
Unfortunately, although the spectral range allowed the observation of the
Balmer discontinuity, we do not have enough signal-to-noise ratio to
measure this jump with acceptable accuracy.

Again, several bias and sky flat field frames were taken at the beginning and
end of each night, and two lamp flat fields and one calibration lamp 
exposure were performed at each telescope position. In these cases the
calibration lamp used was HeAr. 

%
%

\begin{table}
\centering
{\footnotesize
\caption[Signal-to-noise ratio attained for each final spectrum]{Signal-to-noise ratio attained for each final spectrum.}
\label{snr}
\begin{tabular} {l cccccccc}
\hline
Object  ID & 5100-5200 & 6000-6100 & $\lambda$\,4068 & $\lambda$\,4363 & $\lambda$\,5755 & $\lambda$\,6312 & $\lambda$\,7319 & $\lambda$\,7330 \\
           &(continuum)&(continuum)& [S{\sc ii}]     & [O{\sc iii}]    & [N{\sc ii}]     & [S{\sc iii}]    & [O{\sc ii}]     & [O{\sc ii}]     \\
\hline
 \multicolumn{9}{c}{WHT - ISIS} \\[2pt]
 \whtunoc &  25 & 45 & 27 & 250 & 51  & 122 & 225 & 151 \\
 \whtdosc &  30 & 35 & 46 & 170 & --- & 181 & 175 & 134 \\
 \whttresc&  40 & 65 & 50 & 281 & 20  & 216 & 315 & 260 \\
 \multicolumn{9}{c}{CAHA - TWIN} \\[2pt]
 \unoc    &  20 & 50 & 27 & 180 & --- & 157 & 160 & 126 \\
 \dosc    &  15 & 40 & 25 & 40  & --- & 95  & 99  & 82  \\
 \tresc   &  10 & 33 & 18 & 60  & --- & 125 & 91  & 73  \\
 \cuatroc &  10 & 35 & 15 & 24  & --- & 92  & 78  & 62  \\
 \cincoc  &  15 & 40 & 12 & 87  & --- & 61  & 82  & 53  \\
 \seisc   &  15 & 35 & 21 & 64  & --- & 107 & 67  & 47  \\
 \sietec  &  15 & 35 & 12 & 77  & 29  & 105 & 147 & 122 \\
\hline
\end{tabular}
}
\end{table}

\subsection{Data reduction}
\label{data}

All the images were processed and analyzed with
IRAF\footnote{IRAF: the Image Reduction and Analysis Facility is distributed by
  the National Optical Astronomy Observatories, which is operated by the
  Association of Universities for Research in Astronomy, Inc. (AURA) under
  cooperative agreement with the National Science Foundation (NSF).} routines in 
the usual manner. The procedure includes
the removal of cosmic rays, bias subtraction, division by a normalized flat
field and wavelength calibration. Typical wavelength fits were performed using 30-35
lines in the blue and 20-25 lines in the red and polynomials of second to
third order. These fits have been done at 117 different locations along the slit
in each arm for the WHT-ISIS observations (beam size of 10 pixels) and at 80
different locations for the CAHA-Twin observations (beam size of 10 pixels) obtaining
rms residuals between $\sim$0.1 and $\sim$0.2\,pix for the WHT data and
between $\sim$0.1 and $\sim$0.3\,pix for the CAHA data.  

In the last step, the spectra were corrected for atmospheric extinction and
flux calibrated. 
For the blue WHT spectra, four standard star observations were used, allowing a good
spectrophotometric calibration with an estimated accuracy of about 5\%. 
Unfortunately, only one standard star could be used for the calibration of
the red WHT spectra. Nevertheless, after flux calibration, in the overlapping 
region of the spectra taken with each arm, the agreement  in the average
continuum level was good.
For the CAHA data, four standard star observations were performed each night
at the same time for both arms,
allowing a good spectrophotometric calibration with an estimated accuracy of
about 3\%, estimated from the differences between the different standard star
flux calibration curves.

\section{Results}
\label{results}

\begin{figure}
\includegraphics[width=.48\textwidth,height=.31\textwidth,angle=0]{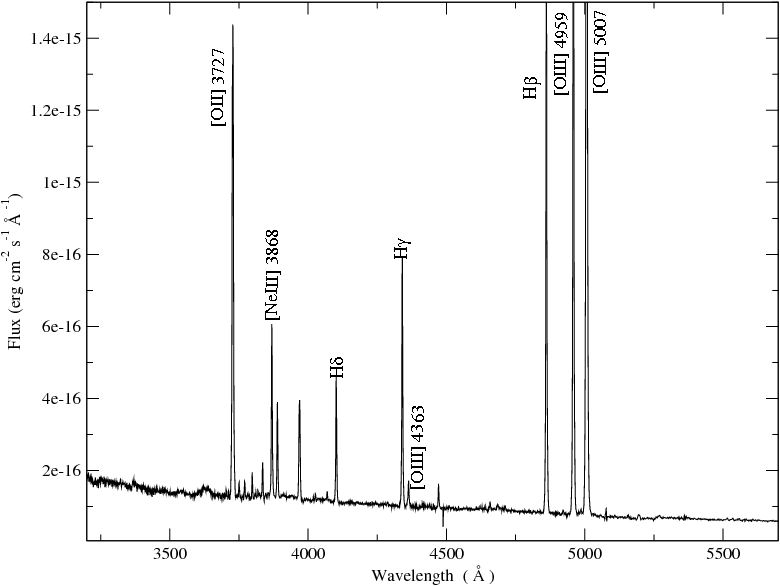}\hspace{0.5cm}
\includegraphics[width=.48\textwidth,height=.31\textwidth,angle=0]{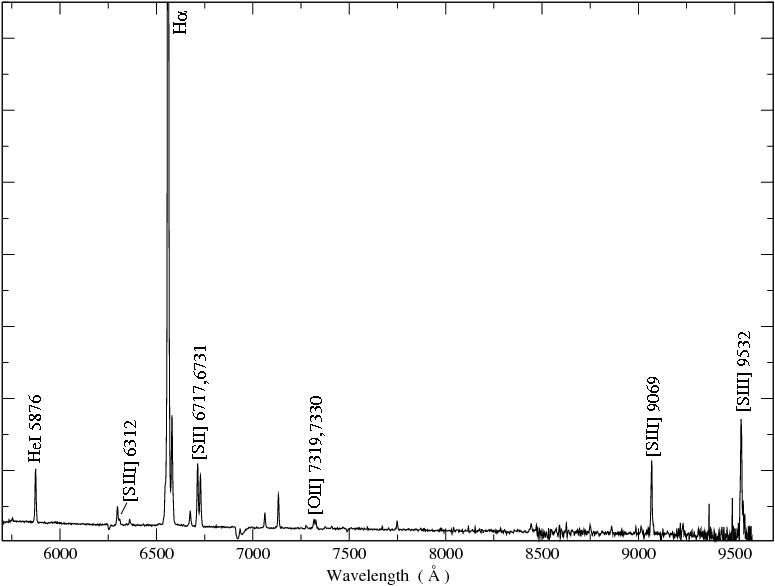}\\
\vspace*{0.2cm}
\includegraphics[width=.48\textwidth,height=.31\textwidth,angle=0]{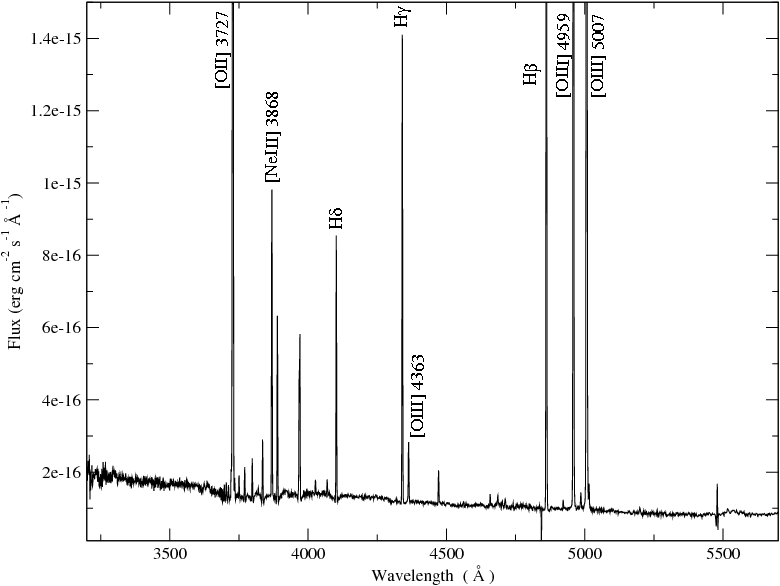}\hspace{0.5cm}
\includegraphics[width=.48\textwidth,height=.31\textwidth,angle=0]{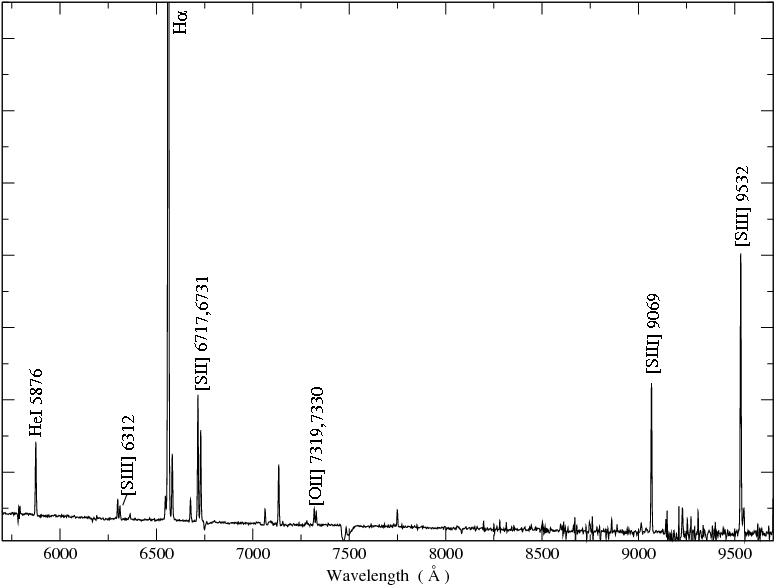}\\
\vspace{0.2cm}
\includegraphics[width=.48\textwidth,height=.31\textwidth,angle=0]{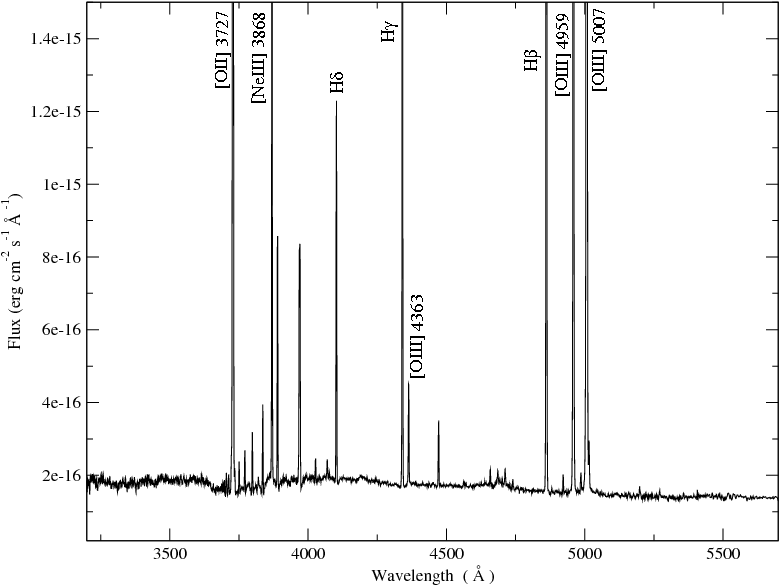}\hspace{0.5cm}
\includegraphics[width=.48\textwidth,height=.31\textwidth,angle=0]{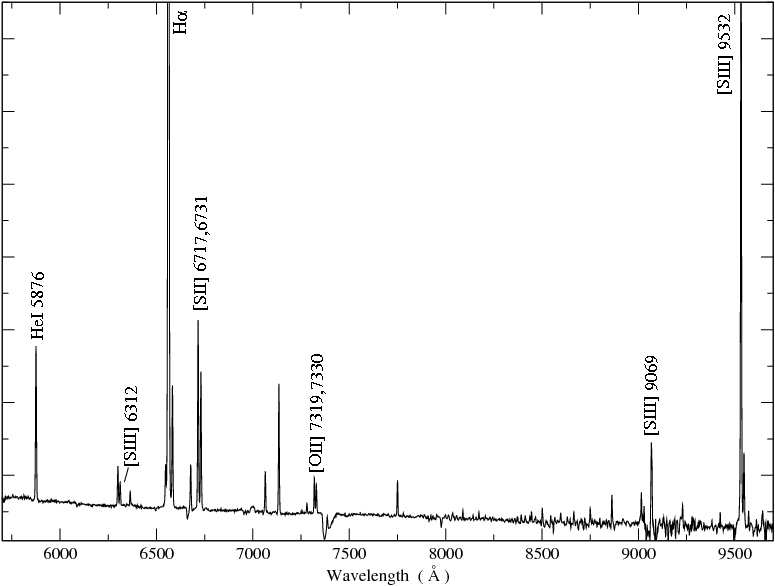}
\caption[Blue and red WHT spectra]{Blue and red WHT spectra of \whtunoc, \whtdosc\ and \whttresc\
  in the rest frame. The flux scales are the same in both spectral ranges.}
\label{whtspect}
\end{figure}

\begin{figure}
\includegraphics[width=.48\textwidth,height=.31\textwidth,angle=0]{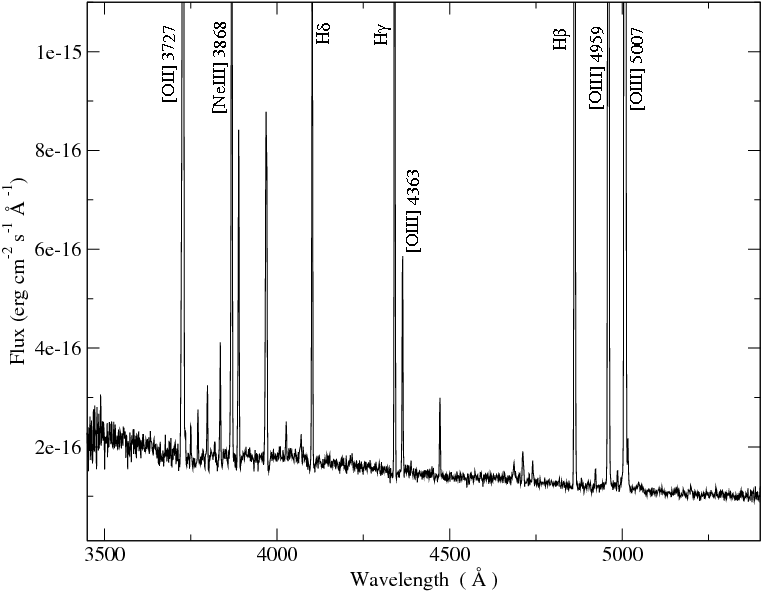}\hspace{0.5cm}
\includegraphics[width=.48\textwidth,height=.31\textwidth,angle=0]{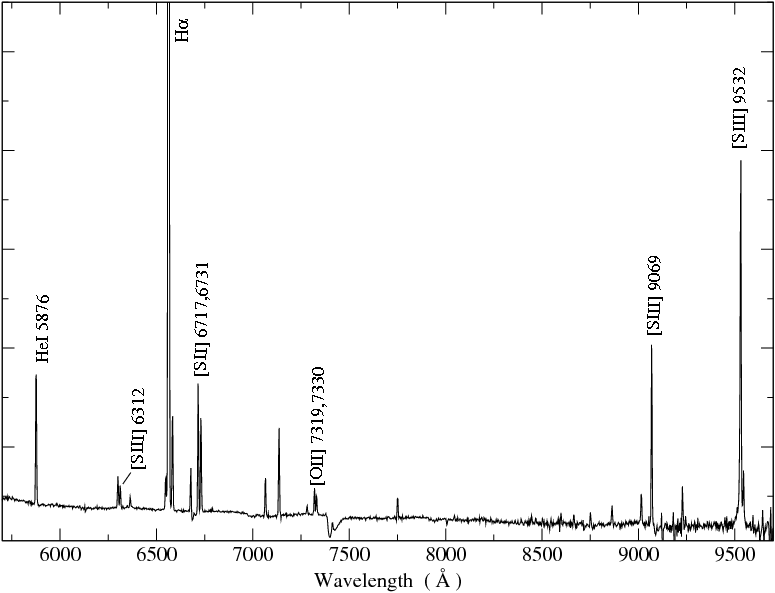}\\
\vspace{0.2cm}
\includegraphics[width=.48\textwidth,height=.31\textwidth,angle=0]{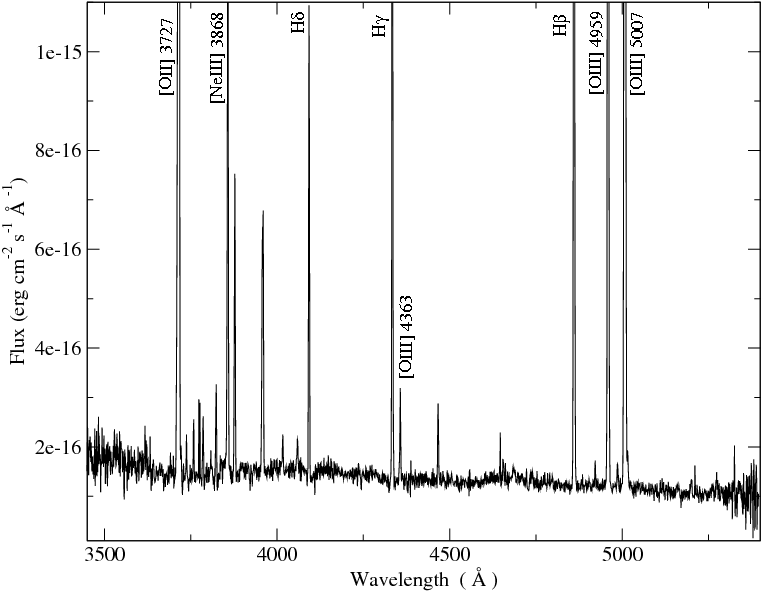}\hspace{0.5cm}
\includegraphics[width=.48\textwidth,height=.31\textwidth,angle=0]{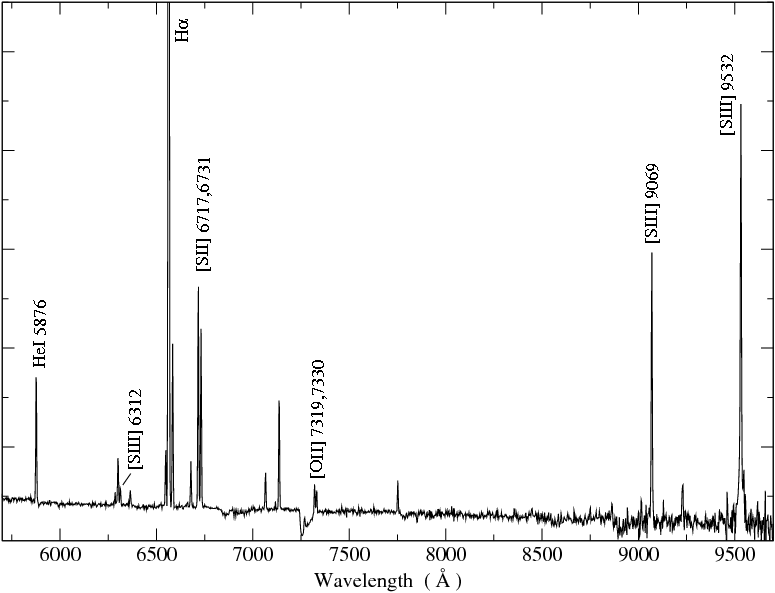}\\
\vspace{0.2cm}
\includegraphics[width=.48\textwidth,height=.31\textwidth,angle=0]{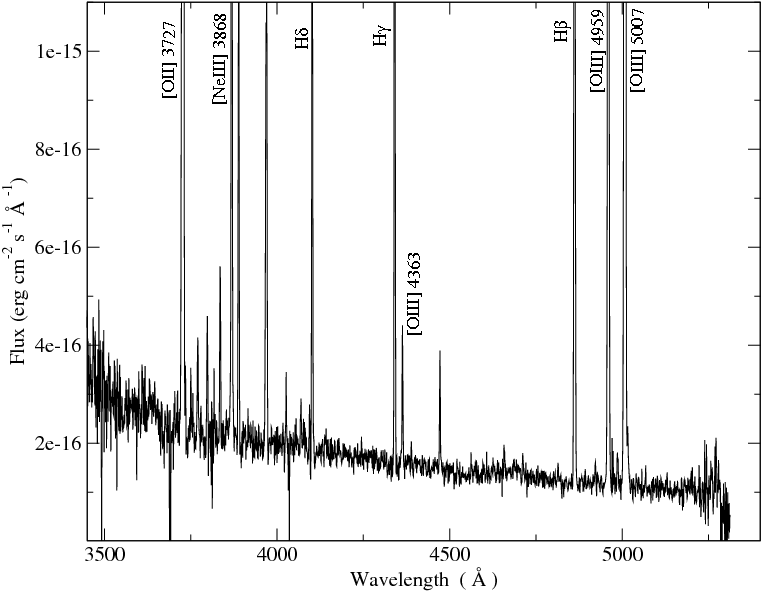}\hspace{0.5cm}
\includegraphics[width=.48\textwidth,height=.31\textwidth,angle=0]{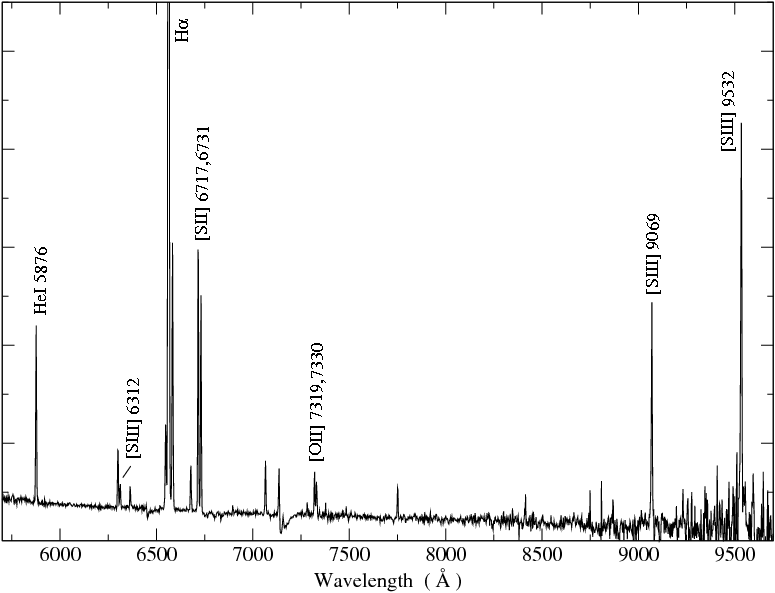}\\
\vspace{0.2cm}
\includegraphics[width=.48\textwidth,height=.31\textwidth,angle=0]{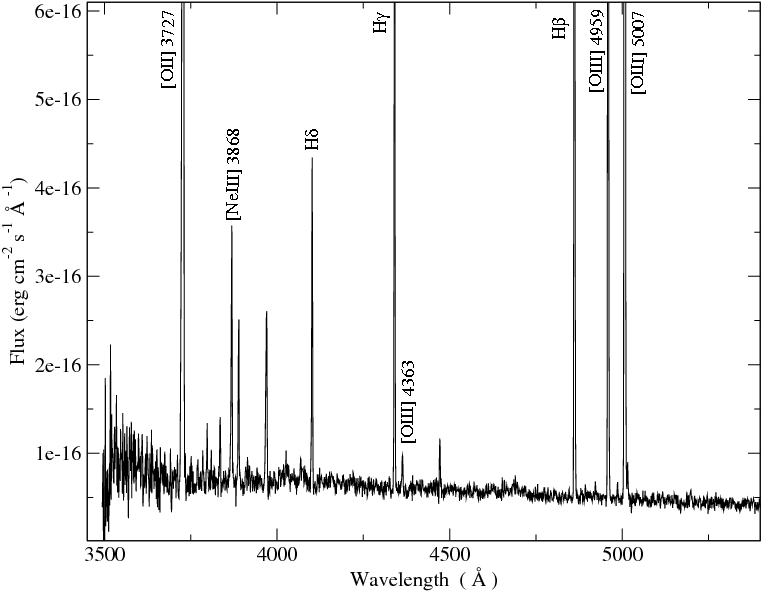}\hspace{0.5cm}
\includegraphics[width=.48\textwidth,height=.31\textwidth,angle=0]{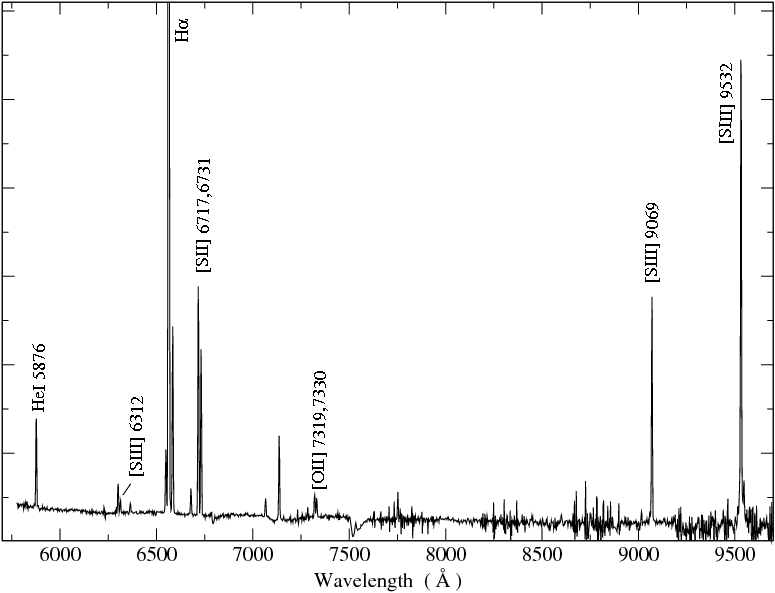}
\caption[Blue and red CAHA spectra]{ Blue and red CAHA spectra of \unoc,
\dosc, \tresc\ and \cuatroc\ in the rest frame. The flux scales are the same
in both spectral ranges.} 
\label{cahaspect}
\end{figure}

The spectra of the observed \HII\ galaxies with some of the relevant
identified emission lines are shown in Figures \ref{whtspect} and
\ref{cahaspect}, WHT and CAHA data, respectively. The spectrum of each
observed galaxy is split into two panels, with the blue part on the left-hand
side and the red part on the right-hand side.

\setcounter{figure}{3}
\begin{figure}
\includegraphics[width=.48\textwidth,height=.31\textwidth,angle=0]{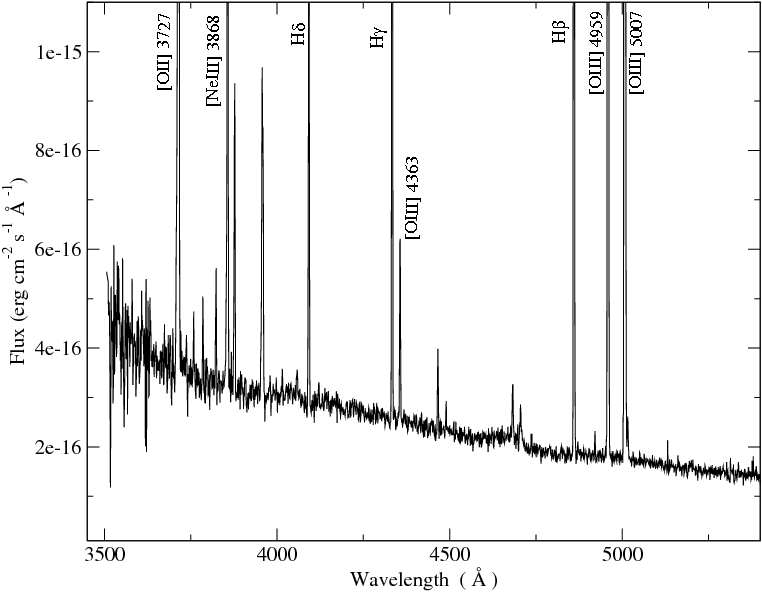}\hspace{0.5cm}
\includegraphics[width=.48\textwidth,height=.31\textwidth,angle=0]{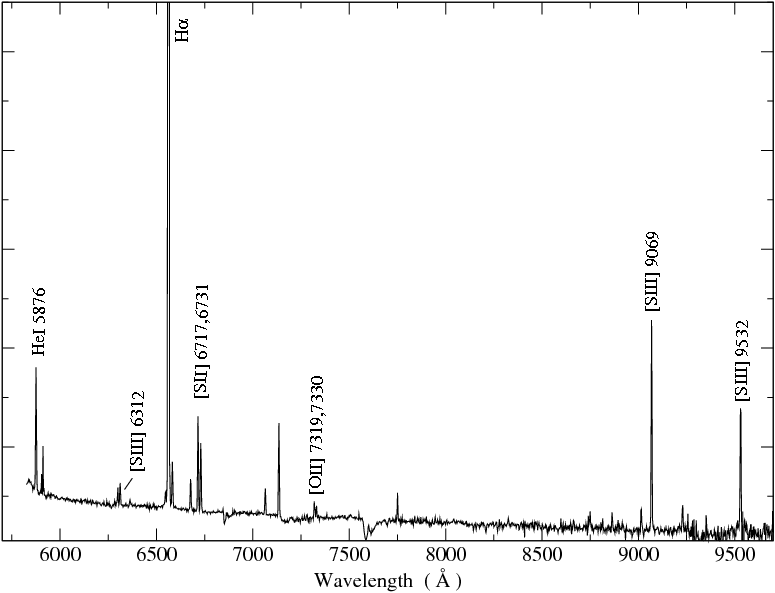}\\
\vspace{0.2cm}
\includegraphics[width=.48\textwidth,height=.31\textwidth,angle=0]{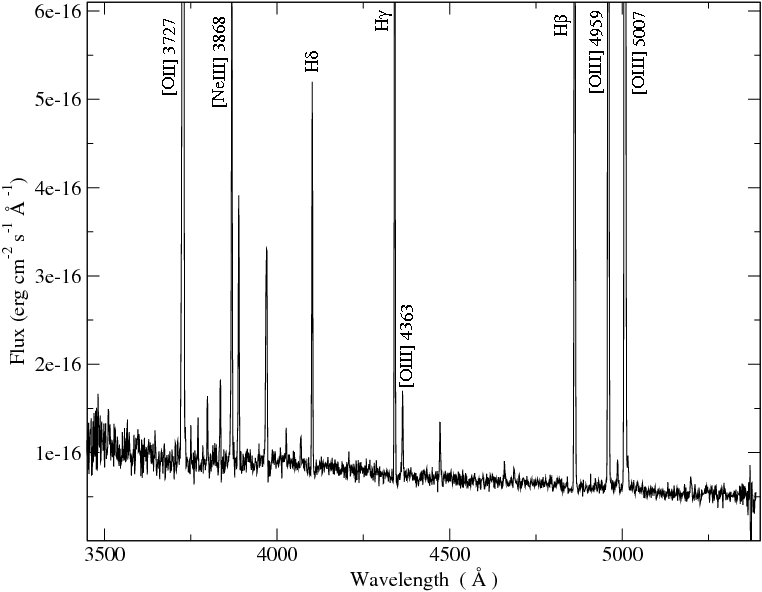}\hspace{0.5cm}
\includegraphics[width=.48\textwidth,height=.31\textwidth,angle=0]{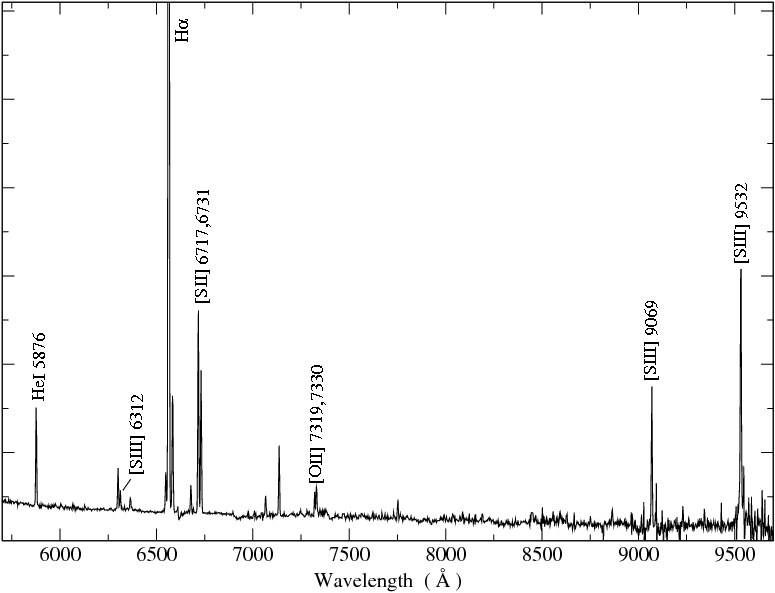}\\
\vspace{0.2cm}
\includegraphics[width=.48\textwidth,height=.31\textwidth,angle=0]{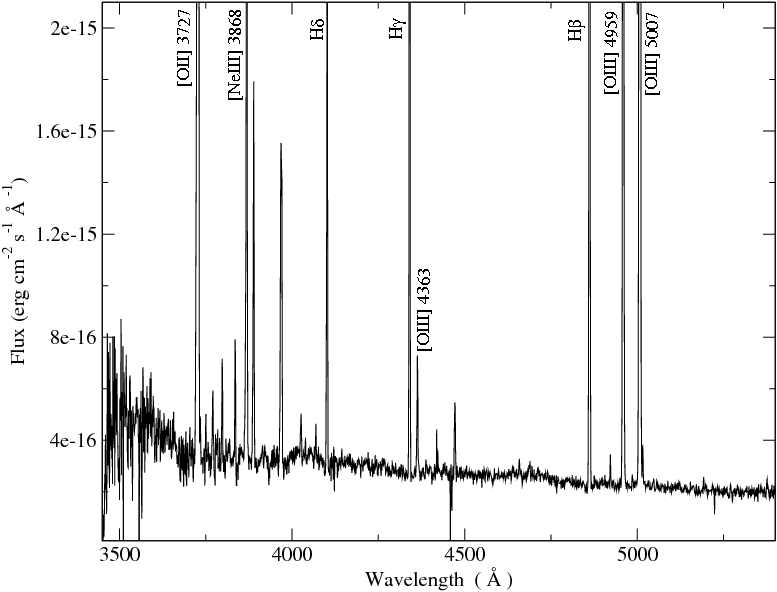}\hspace{0.5cm}
\includegraphics[width=.48\textwidth,height=.31\textwidth,angle=0]{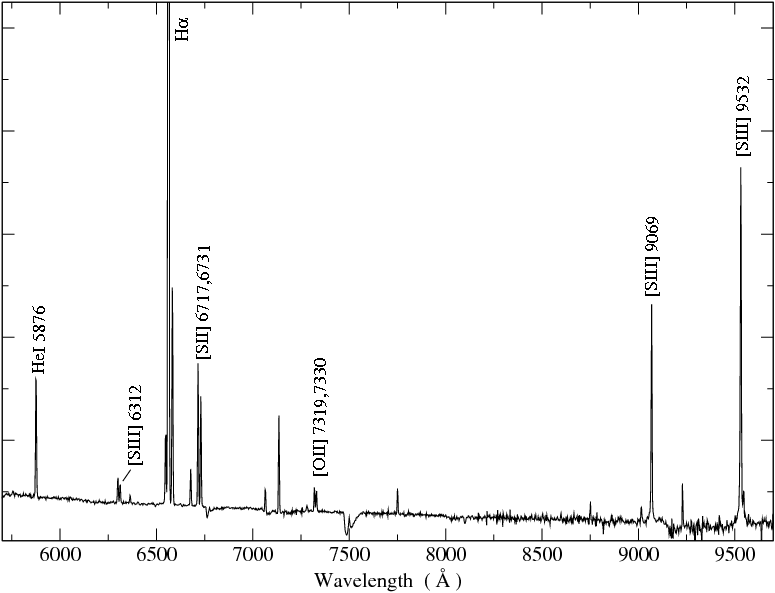}
\caption[$(cont)$ Blue and red CAHA spectra]{({\it cont}) Blue and red CAHA
spectra of \cincoc, \seisc\ and \sietec\ in the rest frame. The flux scales
are the same in both spectral ranges.} 
\end{figure}

The  emission line fluxes were measured using the SPLOT task of IRAF. We have
used two different ways to integrate the flux of a given line: (1) in the case
of an isolated line or two blended and unresolved lines the intensity was
calculated integrating between two points given by the position of the local
continuum placed by eye; (2) if two lines are blended, but they can be
resolved, we have used a multiple Gaussian fit procedure to estimate
individual fluxes.  Following Gonz\'alez-Delgado et al.\
\cite*{1994ApJ...437..239G}, Catellanos, D\'iaz and Terlevich
\cite*{2002MNRAS.329..315C} and  P\'erez-Montero and D\'iaz
\cite*{2003MNRAS.346..105P}, the statistical errors associated with the observed
emission fluxes have been calculated using the expression

\[
\sigma_{l}\,=\,\sigma_{c}N^{1/2}[1 + EW/(N\Delta)]^{1/2} 
\]

\noindent where $\sigma_{l}$ is  
the error in the observed line flux, $\sigma_{c}$ represents the standard
deviation in a box near the measured emission line and stands for the error in
the continuum placement, N is the number of pixels used in the measurement of 
the line flux, EW is the line equivalent width, and $\Delta$ is the wavelength
dispersion in angstroms per pixel. 

There are several lines affected by bad pixels, internal reflections or charge
transfer in the CCD, telluric emission lines or atmospheric absorption
lines. These cause the errors to increase, and, in some cases, they are
impossible to quantify. In these cases we do not include these lines in the
Tables nore in our calculations, and these lines were excluded from any
subsequent analysis.  The  only exception is the 
emission line [S{\sc iii}]\,$\lambda$\,9069 for the WHT object \whttresc\
which is affected by the strong narrow water-vapor lines present in the
$\lambda$\,9300\,-\,9500 wavelength region \cite{1985MNRAS.212..737D}.
We have listed the value of the measurement of this line 
but all the physical parameters depending on its intensity
were calculated using the theoretical ratio between this line and
[S{\sc iii}]\,$\lambda$\,9532, I(9069)\,$\approx$\,2.44\,$\times$\,I(9532)
\cite{1989agna.book.....O}. In the case of the CAHA object \tresc, the [Ar{\sc
iii}]\,$\lambda$\,7136 is affected by a 
sky absorption band. Its observed flux has been scaled to that of the [Ar{\sc
    iii}]\,7751\,\AA\ line according to the theoretical relation, [Ar{\sc
    iii}]\,7136/[Ar{\sc iii}]\,7751\,=\,4.17, derived from the IONIC
task in the STSDAS package of IRAF for a wide temperature range, from 5000 to
50000\,K. For the CAHA data of \cincoc\ the [S{\sc iii}]\,$\lambda$\,9532 line
is affected by strong narrow water-vapour lines and therefore its value has
been set to its theoretical ratio to the weaker [S{\sc
    iii}]\,$\lambda$\,9069\,\AA\ line.

In some cases there is an observable line (e.g., [Cl{\sc
    iii}]\,$\lambda\lambda$\,5517,5537, several carbon recombination lines,
Balmer or Paschen lines) for which it is impossible to give a precise
measurement. This might be due to a low signal to noise between the line and the
surrounding continuum. This is also the case for the Balmer jump for the CAHA
data and the Paschen jump in all the cases, that could not be measured 
even though it was observed, because it was very difficult to locate the
continuum at both sides of the discontinuity with an acceptable accuracy.  

The spectrum of \whtunoc\ presents very wide lines
(FWHM\,$\approx$\,7.5\,\AA\,$\approx$\,340\,km\,s$^{-1}$ 
for $\lambda$\,$\approx$\,6600\,\AA) for the
expected velocity dispersion in a low mass galaxy of this type. This could be
due to an intrinsic velocity dispersion in this object, the interaction with
another unobservable object or a projection effect on the line of sight. There
are \HII\ galaxies that in fact are multiple systems, with two or more 
components, despite their ``a priory" assumption of compactness
\cite{1966ApJ...143..192Z,1970ApJ...162L.155S}. In some cases, 
these systems show some evidence of interaction among their
components \cite{1997MNRAS.288...78T}. For instance, IIZw40, which was
first classified as a compact emission line galaxy by
Sargent \cite*{1970ApJ...160..405S}, when observed with enough spatial resolution
showed to be the merge of two separate subsystems \cite{1982MNRAS.198..535B}.
As a consequence, lines that should be resolved are blended in the
spectrum of \whtunoc. Such is the case of H$\alpha$ and 
[N{\sc ii}]\,$\lambda$\,6548\,\AA\
emission lines. We have resorted to the theoretical ratio, I(6584)
$\approx$ 3\,$\times$\,I(6548), to decontaminate the observed flux of H$\alpha$ by
the emission of [N{\sc ii}]\,$\lambda$\,6548 and to derive the electron
temperature of [N{\sc ii}].

\begin{figure}
\centering
\includegraphics[width=.6\textwidth,angle=270,clip=]{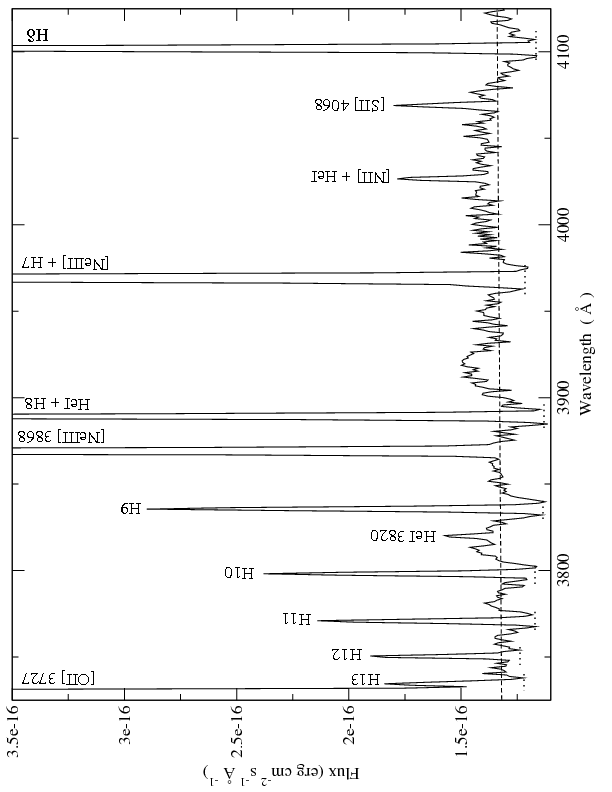}\\
\includegraphics[width=.6\textwidth,angle=270,clip=]{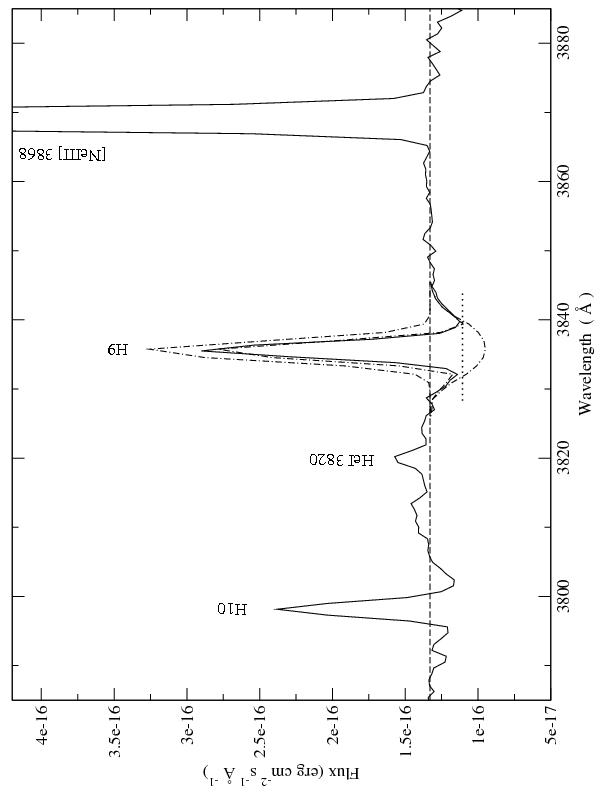}
\caption[Underlying stellar population]{Upper panel: Section of the spectrum of \whtdosc\ taken
  with the WHT. The spectrum is in the rest frame and its spectral range is from
  3725 to 4125\,\AA. We can appreciate the presence of absorption features
  originated in an oldish stellar population which mainly affect the Balmer
  emission lines. Lower panel: Section of the same spectrum with a spectral range
  from 3785 to 3885\,\AA. We have superposed the fit to H9 made using the
  ngaussfit task from IRAF (dashed-dotted lines). For both panels: The
  dashed line traces the continuum and the dotted lines show the pseudo-continuum
  used to measure the Balmer emission lines.}
\label{under}
\end{figure}

A conspicuous underlying stellar population is easily appreciable in the
spectra by the
presence of absorption features that depress the Balmer and Paschen emission
lines. The upper panel of Figure \ref{under} shows an example of this effect for the
Balmer lines (H13 to H$\delta$) on an enlargement of the spectrum of \whtdosc,
the object that presents the most prominent and appreciable 
absorption lines. A pseudo-continuum has been defined at the base of the
hydrogen emission lines to measure the line intensities and minimize the
errors introduced by the underlying population.
The pseudo-continuum used to measure the line fluxes is
also shown in the Figure. We can clearly see the wings of the absorption lines
implying that, 
even though we have used a pseudo-continuum, there is an absorbed fraction of
the emitted flux that we are not able to measure with an acceptable
accuracy (see discussion in D\'iaz 1988). This fraction is not the same
for all lines, nor are the ratios between the absorbed fractions and the
emissions. In order to quantify the effect of the underlying absorption 
on the measured emission line intensities, we have performed a multi-Gaussian
fit to the absorption and emission components seen in this galaxy. The fitting
can be seen in the lower panel of Figure \ref{under}. The difference between the
measurements of the absorption subtracted lines and the ones obtained with the
use of the pseudo-continuum is, for all Balmer lines, within the observational
errors and, in fact, the additional fractional error introduced by the
subtraction of the absorption component is almost inappreciable for the
stronger lines. In the other galaxies, the absorption wings in the Balmer
lines are not prominent enough as to provide sensible results by the
multi-Gaussian component fitting. Therefore, we have doubled the error derived
using the expression for the statistical errors associated with the
observed emission fluxes, ($\sigma_{l}$), as a conservative approach to
include the uncertainties introduced by the presence of the underlying stellar
population.

\begin{figure}
\centering
\includegraphics[width=.75\textwidth,angle=0,clip=]{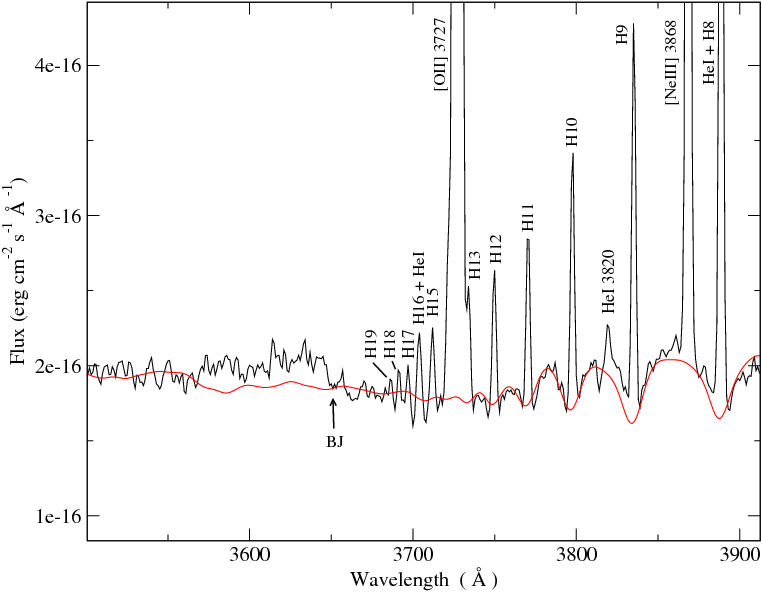}\\\vspace{0.5cm}
\includegraphics[width=.75\textwidth,angle=0,clip=]{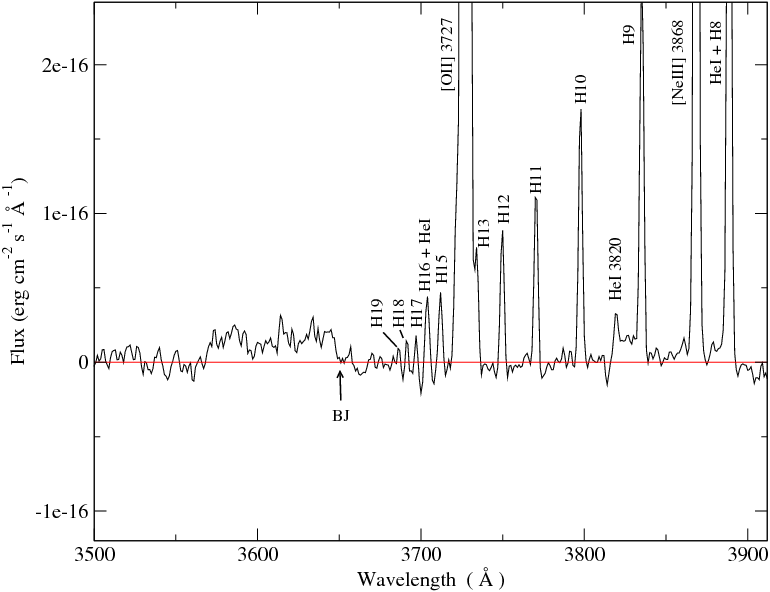}
\caption[Upper panel: spectral fitting made using the STARLIGHT spectral
  synthesis code (between 3500 and 3912\,\AA). Lower panel: subtraction of
  this fitting to the WHT spectrum]{Upper panel: enlargement of the
  spectrum of \whttresc\ between 3500 
  and 3912\,\AA, around the Balmer jump and the high order Balmer
  series, together with the spectral fitting (red lines) made using the
  STARLIGHT spectral synthesis code. Lower panel: subtraction of this fitting
  to the WHT spectrum in the same spectral ranges, and the solid red
  line shows the zero.}
\label{starlight1}
\end{figure}

\begin{figure}
\centering
\includegraphics[width=.75\textwidth,angle=0,clip=]{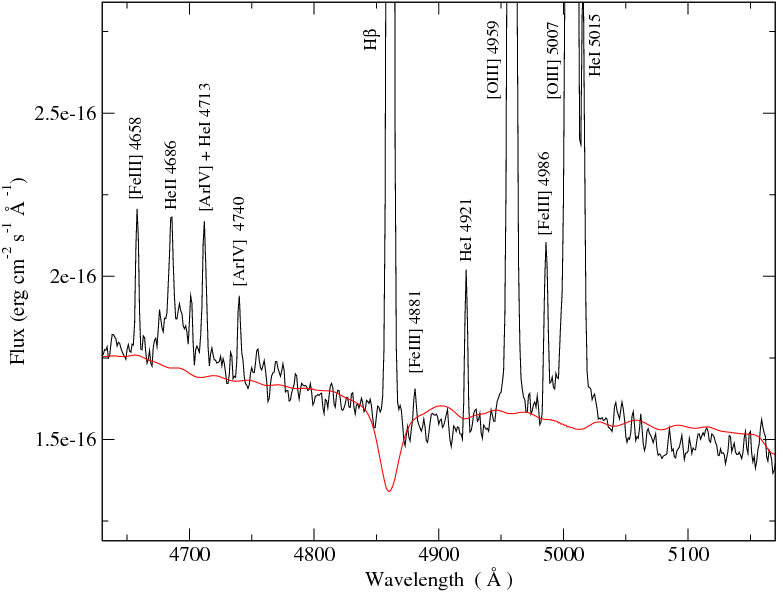}\\\vspace{0.5cm}
\includegraphics[width=.75\textwidth,angle=0,clip=]{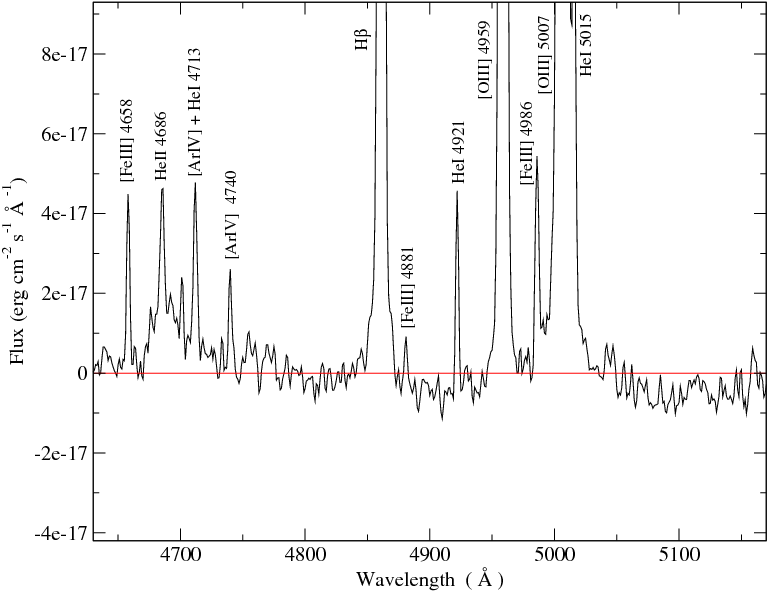}
\caption[Upper panel: spectral fitting made using the STARLIGHT
  spectral synthesis code (between 4630 and 5170\,\AA). Lower panel:
  subtraction of this fitting to the WHT spectrum]{Upper panel:
  enlargement of the spectrum of \whttresc\ between 4630 
  and 5170\,\AA, around H$\beta$, together with the spectral fitting (red
  lines) made using the STARLIGHT spectral synthesis code. Lower panel:
  subtraction 
  of this fitting to the WHT spectrum in the same spectral ranges,
  and the solid red line shows the zero.}
\label{starlight2}
\end{figure}

Another approach used 
to separate the emission spectra from the underlying stellar absorptions was 
to use  spectral synthesis codes for the stellar population like STARLIGHT
\cite{2005MNRAS.358..363C}. 
In the upper panels of Figures \ref{starlight1} and \ref{starlight2} we show
an enlargement of the spectrum of \whttresc\ between 3500 and 3912\,\AA\
(around the Balmer jump and the high order Balmer series) and between 4630 and
5170\,\AA\, (around H$\beta$), respectively, together with the spectral fitting
(red lines)
made using the STARLIGHT spectral synthesis code\footnote{available from
  www.starlight.ufsc.br}, version 04 \cite{2005MNRAS.358..363C}. The
synthetic stellar population used in the fitting was obtained using Starburst99
\cite{1999ApJS..123....3L,2005ApJ...621..695V} with the Geneva stellar
evolutionary tracks for continuous star formation with 
high mass loss \cite{1994A&AS..103...97M}, the Kroupa Initial Mass Function
(IMF; \citeplain{2002Sci...295...82K}) in two intervals (0.1-0.5 and
0.5-100\,M$_\odot$) with different exponents (1.3 and 2.3, respectively), the
theoretical wind model \cite{1992ApJ...401..596L}, with the model
atmospheres from \citetex{2002MNRAS.337.1309S}, and the stellar cluster
metallicity being the closest to the nebular one.
In the lower panels of those Figures we plot the subtraction of this fitting
to the WHT spectrum in the same spectral ranges, and the solid red
lines show the zero. In the case of the high Balmer series we can appreciate
that small errors in the fit of the underlying stellar population could become
a great unquantifiable error in the emission line fluxes. The wings of the
H$\beta$ absorption line are over estimated, yielding an unreal excess in the
emission line flux. Nevertheless, for the strongest emission lines the
differences between the measurements made after the subtraction of the
STARLIGHT fit and the ones made using our pseudo-continuum are well below the
observational errors, giving almost the same result. It must be highlighted
that in this type of objects the absorption lines, used to fit the underlying
stellar populations, are mostly affected by the presence of emission
lines. There are a few that are not affected, such as some calcium and
magnesium absorption lines. To make the STARLIGHT fit it is necessary to mask
the contaminated lines, thus the fit is based on the continuum shape and take
into account only a few absorption lines. Hence, it is not surprising that
the STARLIGHT results are not so good for this kind of objects.

It must be noted that the STARLIGHT methodology is very advantageous for
statistical studies. Those studies only consider the results derived from
the strongest emission lines
\cite{2005MNRAS.358..363C,2006MNRAS.366..480G,2006MNRAS.370..721M,2007MNRAS.374.1457M,2007MNRAS.375L..16C,2007MNRAS.381..263A}
which have small proportional errors. However, we must be careful when the aim
of our work is the detailed and precise study of a reduced sample of objects
because, as we saw, we can not quantify the error introduced by the
fitting. Moreover, these errors are higher for the weakest lines, including
the auroral ones.

The absorption features of the underlying stellar population may also affect
the helium emission lines to some extent. However, these absorption lines are
narrower than those of hydrogen (see, for example, Gonz\'alez-Delgado et al.,
2005). \nocite{2005MNRAS.357..945G} Therefore it is difficult to set adequate
pseudo-continua at both sides of the lines to measure their fluxes.

The reddening coefficient [$c$(H$\beta$)] has been calculated assuming the
galactic extinction law of Miller and Mathews \cite*{1972ApJ...172..593M} with
$R_v$=3.2 and 
obtained  by performing a least square fit to the difference between the
theoretical and observed Balmer and Paschen decrements vs.\ the reddening law
whose slope is the logarithmic reddening at the H$\beta$ wavelength: 
\[
log\Big(\frac{I(\lambda)}{I(H\beta)}\Big)\,=\,log\Big(\frac{F(\lambda)}{F(H\beta)}\Big)+c(H\beta)\,f(\lambda)
\]
The theoretical Balmer line intensities have been computed using
Storey and Hummer \cite*{1995MNRAS.272...41S} with an iterative method to
estimate T$_e$ (electronic temperature) and N$_e$ (electronic density) in each
case. As N$_e$ introduces only a second order effect, for 
simplicity we assume N$_e$ equal to N([S{\sc ii}]). Due to the large error
introduced by the presence of the underlying stellar population, only the four
strongest Balmer emission lines (H$\alpha$, H$\beta$, H$\gamma$ and H$\delta$)
have been used. 

In Tables \ref{ratiostot wht1} to \ref{ratiostot 7}, at the end of the present
Chapter, we have listed the equivalent widths and 
the reddening corrected emission lines for each observed galaxy together with
the reddening 
constant and its error taken as the uncertainties of the least square fit and
the reddening corrected H$\beta$ intensity. Column 1 of each Table shows the
wavelength and the name of the measured lines, as referred to in Garc\'ia-Rojas
et al. \cite*{2004ApJS..153..501G}. The adopted reddening curve,
$f(\lambda)$, normalized to H$\beta$, is given in column 2 of the Tables. The
errors in the emission lines were obtained by propagating in quadrature the
observational errors in the emission line fluxes and the reddening constant
uncertainties.  We have not taken into account errors
in the theoretical intensities since they are much lower than the observational
ones. Finally, the values listed in Column 5 of the Tables indicate the
fractional error in the line intensities calculated as explained above. 

A detailed comparison has been made between the results of measurements and
data analysis as obtained from the SDSS spectra and the WHT ones, which have
comparable S/N ratios. The equivalent widths and the reddening corrected
emission lines for each SDSS spectrum of the observed WHT galaxy together with
the reddening constant and its error are listed in Tables \ref{ratiostot wht1}
to \ref{ratiostot wht3}. 

The relative errors in the emission lines vary from a few percent for the more
intense nebular emission lines (e.g. [O{\sc
    iii}]\,$\lambda\lambda$\,4959,5007, [S{\sc
    ii}]\,$\lambda\lambda$\,6717,6731 or the strongest Balmer emission lines)
to 10-35\,\% for the weakest lines that have less contrast with the continuum
noise (e.g. He{\sc i}\,$\lambda\lambda$\,3820,7281, [Ar{\sc
    iv}]\,$\lambda$\,4740 or O{\sc i}\,$\lambda$\,8446). For the auroral
lines, the fractional errors are between $\sim$3 and $\sim$10\,\%.

\section{Physical conditions of the gas}
\label{physiHIIgal}

\subsection{Electron densities and temperatures from forbidden lines}
\label{electden-temp}

The physical conditions of the ionized gas, including electron temperatures and
electron density, have been derived from the emission line data using the same
procedures as in P\'erez-Montero and D\'\i az (2003), based on the five-level
statistical equilibrium atom approximation in the task TEMDEN, of the software
package IRAF \cite{1987JRASC..81..195D,1995PASP..107..896S}. The atomic
coefficients used here are the  same as in P\'erez-Montero and D\'\i az (2003),
except in the case of O$^+$ for which we have used 
the transition probabilities from Zeippen \cite*{1982MNRAS.198..111Z} and the
collision strengths from Pradham (1976), which offer more reliable nebular
diagnostics results for this species \cite{2004A&A...427..873W}.  The
references for the different ion atomic coefficients are given in Table
\ref{atomic}. We have taken as sources of error the
uncertainties associated with the measurement of the emission-line fluxes and
the reddening correction, and we have propagated them through our calculations. 

The electron density, N$_e$, has been derived from the [S{\sc
    ii}]\,$\lambda\lambda$\,6717\,/\,6731\,\AA\ line ratio, which is
representative of the low-excitation zone of the ionized gas. In all the
observed galaxies  the electron densities have been found to be lower than 200
cm$^{-3}$, well below the critical density for collisional de-excitation. 
We have tried to derive the electron densities from the [Ar{\sc
    iv}]\,$\lambda\lambda$\,4713\,/\,4740\,\AA\ line ratio, by decontaminating
the first one from the He{\sc i} contribution, but the derived density values
had unacceptable errors due to their large and sensitive dependencies on the
emission line intensities and the errors of the observed fluxes. 
Then, we were not able to estimate the density from line ratios representative
of the higher ionization zones, hence we are not able to determine any
existing distribution in density.


%
%

\begin{table}
\vspace{-0.3cm}
\normalsize
\caption{Sources of the effective collision strengths of each ion.}
\label{atomic}
\begin{center}
\begin{tabular}{@{}ll}
\hline
\hline
  Ion & references  \\ 
\hline
 O{\sc ii}   &  Pradhan \cite*{1976MNRAS.177...31P} \\ 

 O{\sc iii}, N{\sc ii}   &  Lennon and Burke \cite*{1994AAS..103..273L}  \\ 

 S{\sc ii}   &  Ramsbottom et al.\ \cite*{1996ADNDT..63...57R}  \\ 

 S{\sc iii}  &  Tayal and Gupta \cite*{1999ApJ...526..544T}  \\ 

 Ne{\sc iii} & Butler and Zeippen \cite*{1994AAS..108....1B}  \\ 

 Ar{\sc iii} & Galavis et al.\ \cite*{1995A+AS..111..347G}  \\ 

 Ar{\sc iv}  & Zeippen et al.\ \cite*{1987A+A...188..251Z}  \\ 
\hline
\end{tabular}
\end{center}
\end{table}

%
%

\begin{table}
\normalsize
\caption{Emission-line ratios used to derive electron densities and temperatures.}
\label{ratiost}
\begin{center}
\begin{tabular}{@{}ll}
\hline
\hline
  & ratios  \\
\hline
n$_e$([S{\sc ii}])     &    R$_{S2}$\,=\,I(6717)\,/\,I(6731) \\
t$_e$([O{\sc iii}])    &    R$_{O3}$\,=\,(I(4959)+I(5007))\,/\,I(4363) \\
t$_e$([O{\sc ii}])     &    R$_{O2}$\,=\,I(3727)\,/\,(I(7319)+I(7330)) \\
t$_e$([S{\sc iii}])    &    R$_{S3}$\,=\,(I(9069)+I(9532))\,/\,I(6312) \\
t$_e$([S{\sc ii}])     &    R$_{S2}'$\,=\,(I(6717)+I(6731))\,/\,(I(4068)+I(4074)) \\
t$_e$([N{\sc ii}])     &    R$_{N2}$\,=\,(I(6548)+I(6584))\,/\,I(5755) \\
\hline
\end{tabular}
\end{center}
\end{table}

\begin{figure}
\centering
\includegraphics[width=.6\textwidth]{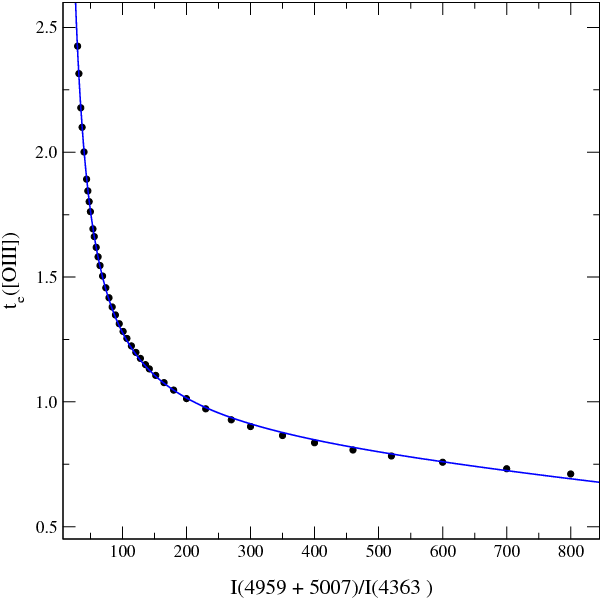}\\
\caption{Relation between  R$_{O3}$ and t$_e$([O{\sc iii}]).}
\label{temdenfit}
\end{figure}

For all the objects we have derived the electron temperatures of [O{\sc ii}],
[O{\sc iii}], [S{\sc ii}] and [S{\sc iii}]. Only for three objects, \whtunoc,
\whtdosc\ and \sietec, it was possible to derive T$_e$([N{\sc ii}]) from the
[N{\sc ii}] line at 5755\,\AA.  The emission-line ratios used to 
calculate the electron density and each temperature are summarized in Table
\ref{ratiost}. Adequate fitting functions have been derived from the TEMDEM
task and are given below:

\[
n_e([S{\textrm{\sc ii}}])\,=\,10^3\,\frac{R_{S2}\,a_0(t)+a_1(t)}{R_{S2}\,a_2(t)+a_3(t)}
\]
where
\begin{eqnarray*}
&&a_0(t) = 2.21-1.3/t-1.25t+0.23t^2\\
&&a_1(t) = -3.35+1.94/t+1.93t-0.36t^2\\
&&a_2(t) = -4.33+2.33/t+2.72t-0.57t^2\\
&&a_3(t) = 1.84-1/t-1.14t+0.24t^2
\end{eqnarray*}
here $t$ is generally t$_e$([O{\sc iii}]), where t$_e$\,=\,10$^{-4}$\,T$_e$,
although an iterative process could be 
used to calculate it with t$_e$([S{\sc ii}]) given that this temperature, like
t$_e$([O{\sc ii}]), a type np$^3$ ion, is density dependent.

\[
t_e([O{\textrm{\sc iii}}])\, =\, 0.8254\, - \,0.0002415\,R_{O3} \,+\,
\frac{47.77}{R_{O3}}
\]
\[
t_e([S{\textrm{\sc iii}}])\,=\,\frac{R_{S3}\,+\,36.4}{1.8\,R_{S3}\,-3.01}\\
\]
\[
t_e([O{\textrm{\sc ii}}]) \,=\,0.23 +0.0017\,R_{O2}\,+\,\frac{38.3}{R_{O2}}\,+\,f_1(n_e)\\
\]
\[
t_e([S{\textrm{\sc ii}}] \,=\,1.92\,-\,0.0375\,R_{S2}'\,-\,\frac{14.5}{R_{S2}'}\,+\,\frac{105.64}{R_{S2}'^2}\,+\,f_2(n_e)\\
\]
\[
t_e([N{\textrm{\sc ii}}])\,=\,0.537\,+0.000253\,R_{N2}\,+\,\frac{42.13}{R_{N2}}
\]

\noindent where n$_e$\,=\,10$^{-4}$\,N$_e$ and $f_1(n_e)$,$f_2(n_e)$\,$<<$\,1
for N$_e$\,$<$\,1000\,cm$^{-\rm 3}$. The above expressions are valid in the
temperature range between 7000 and 23000\,K and the errors involved in the
fittings are always lower than observational errors by factors between 5 and
10. In Figure \ref{temdenfit} we show an example of the fits.

Both the [O{\sc ii}]\,$\lambda\lambda$\,7319,7330\,\AA\ and the [N{\sc
    ii}]\,$\lambda$\,5755\,\AA\  lines have a contribution by direct
recombination which increases with temperature. Such emission, however, can be
quantified and corrected for as: 
\begin{eqnarray*}
\frac{I_R(7319+7330)}{I(H\beta)}\,=\,9.36\,t^{0.44}\,\frac{O^{2+}}{H^+} \\
\frac{I_R(5755)}{I(H\beta)}\,=\,3.19\,t^{0.30}\,\frac{N^{2+}}{H^+} 
\end{eqnarray*}
\noindent  where $t$ denotes the electron temperature in units of 10$^4$\,K
\cite{2000MNRAS.312..585L}. Using the calculated [O{\sc iii}] electron
temperatures, we have estimated these contributions to be less than 4\,\% in all
cases and therefore we have not corrected for this effect, but we have
included it as an additional source of error. In the worst cases this amounts
to about 10\,\% of the total error. The expressions above, however, are valid
only in the range of temperatures between 5000 and 10000\,K in the case of
[O{\sc ii}] and between 5000 and 20000 K in the case of [N{\sc ii}] according
to the authors. While the
[O{\sc iii}] temperatures found in our objects are inside the range of
validity for [N{\sc ii}], they are slightly over that range for [O{\sc
    ii}]. At any rate, the relative contribution of recombination to
collisional intensities decreases rapidly with increasing temperature,
therefore for the high T$_e$  values found in our objects this contribution is
expected to be small.

%
%


 \begin{table*}
 {\small
 \caption{Electron densities and temperatures for the observed galaxies.}
 \label{temden}
 \begin{center}
 \begin{tabular}{@{}lcccccc@{}}
 \hline
 \hline
  & n([S{\sc ii}]) & t$_e$([O{\sc iii}]) &  t$_e$([O{\sc ii}]) & t$_e$([S{\sc iii}]) & t$_e$([S{\sc ii}]) &  t$_e$([N{\sc ii}])
  \\
 \hline

  \multicolumn{7}{c}{WHT - ISIS} \\[2pt]
\whtunoc   & 120\,$\pm$\,68 & 1.25\,$\pm$\,0.02 & 1.03\,$\pm$\,0.02 & 1.31\,$\pm$\,0.05 & 0.86\,$\pm$\,0.06 & 1.19\,$\pm$\,0.05 \\
\whtdosc   &  54:         & 1.28\,$\pm$\,0.02 & 1.35\,$\pm$\,0.04 & 1.36\,$\pm$\,0.07 & 1.03\,$\pm$\,0.05 &       ---       \\
\whttresc  &  58:         & 1.24\,$\pm$\,0.01 & 1.31\,$\pm$\,0.03 & 1.26\,$\pm$\,0.04 & 1.04\,$\pm$\,0.07 & 1.42\,$\pm$\,0.08 \\
  \multicolumn{7}{c}{CAHA - TWIN} \\[2pt]
SDSS J1455 &  94\,$\pm$\,40 & 1.40\,$\pm$\,0.02 & 1.33\,$\pm$\,0.07 & 1.37\,$\pm$\,0.05 & 1.31\,$\pm$\,0.11 &       ---       \\
SDSS J1509 &  85\,$\pm$\,45 & 1.09\,$\pm$\,0.01 & 1.18\,$\pm$\,0.05 & 1.02\,$\pm$\,0.04 & 0.89\,$\pm$\,0.07 &       ---       \\
SDSS J1528 &  60:         & 1.16\,$\pm$\,0.01 & 1.17\,$\pm$\,0.05 & 1.21\,$\pm$\,0.06 & 0.99\,$\pm$\,0.07 &       ---       \\
SDSS J1540 &  47\,$\pm$\,38 & 1.13\,$\pm$\,0.02 & 1.15\,$\pm$\,0.06 & 0.97\,$\pm$\,0.04 & 0.85\,$\pm$\,0.05 &       ---       \\
SDSS J1616 &  54:         & 1.30\,$\pm$\,0.01 & 1.29\,$\pm$\,0.09 & 1.29\,$\pm$\,0.06 & 1.21\,$\pm$\,0.12 &       ---       \\
SDSS J1657 &  30:         & 1.23\,$\pm$\,0.02 & 1.33\,$\pm$\,0.07 & 1.45\,$\pm$\,0.08 & 0.88\,$\pm$\,0.05 &       ---       \\
SDSS J1729 & 109\,$\pm$\,47 & 1.26\,$\pm$\,0.02 & 1.16\,$\pm$\,0.04 & 1.13\,$\pm$\,0.05 & 0.82\,$\pm$\,0.06 & 1.40\,$\pm$\,0.09 \\

 \hline
 \multicolumn{7}{l}{densities in $cm^{-3}$                 and temperatures in 10$^4$\,K}
 \end{tabular}
 \end{center}}
 \end{table*}

\begin{table*}
{\small
\caption[Comparison between the WHT and SDSS derived electron densities and
  temperatures.]{Electron densities and temperatures for the observed WHT  
galaxies using WHT and SDSS spectroscopy. Temperatures marked with asterisks
have been deduced using equations from grids of photoionization models.} 
\label{temdenwht}
\begin{center}
\begin{tabular}{@{}l cc cc cc@{} }
\hline
\hline
               & \multicolumn{2}{c}{\whtunoc} & \multicolumn{2}{c}{\whtdosc} &   \multicolumn{2}{c}{\whttresc}\\
               &   WHT          &      SDSS        &   WHT            &  SDSS            &   WHT         & SDSS              \\
\hline                                    
n([S{\sc ii}]) &   120$\pm$68   &   77$\pm$58      &   54:            &   56:            &   58:         &   66:       \\
t$_e$([O{\sc iii}])&  1.25$\pm$0.02 & 1.13$\pm$0.02    & 1.28$\pm$0.02    & 1.28$\pm$0.03    & 1.24$\pm$0.01 & 1.16$\pm$0.01     \\ 
t$_e$([O{\sc ii}]) &  1.03$\pm$0.02 & 1.05$\pm$0.02    & 1.35$\pm$0.04    &   ---            & 1.31$\pm$0.03 & 1.23$\pm$0.03     \\ 
$\langle$t$_e$([O{\sc ii}])$\rangle^*$ &  --- & --- & --- & 1.36 & --- & ---   \\ 
t$_e$([S{\sc iii}])&  1.31$\pm$0.05 & ---              & 1.36$\pm$0.07    &   ---            & 1.26$\pm$0.04 &     ---           \\ 
$\langle$t$_e$([S{\sc iii}])$\rangle^*$ & --- & 1.02 & ---   & 1.21    & --- & 1.06    \\  
t$_e$([S{\sc ii}]) &  0.86$\pm$0.06 & 0.76$\pm$0.06    & 1.03$\pm$0.05    & 0.92$\pm$0.06    & 1.04$\pm$0.07 & 1.02$\pm$0.09     \\ 
t$_e$([N{\sc ii}]) &  1.19$\pm$0.05 & 1.36$\pm$0.06    &   ---            &   ---            & 1.42$\pm$0.08 &     ---           \\ 
$\langle$t$_e$([N{\sc ii}])$\rangle^*$ & --- & --- & 1.35   & 1.36   & --- & 1.23    \\ 
T(Bac) & 1.24$\pm$0.27 & --- & 0.96$\pm$0.16    & ---    & 1.23$\pm$0.22 & ---    \\ 
T(H$\beta$)   & 1.24$\pm$0.31 & --- & 1.02$\pm$0.20    & ---    & 1.24$\pm$0.26 & ---    \\ 
T(He{\sc ii}) & 1.24$\pm$0.29 & --- & 0.98$\pm$0.19    & ---    & 1.24$\pm$0.23 & ---    \\ 
\hline
\multicolumn{7}{l}{densities in $cm^{-3}$ and temperatures in 10$^4$\,K}
\end{tabular}
\end{center}}
\end{table*}




The derived electron densities and temperatures for the observed 
objects are given in Table \ref{temden} along with their corresponding
errors. In Table \ref{temdenwht} we have compared the values derived using the
WHT and the SDSS data for the WHT objects. The derived electron densities and
temperatures using the WHT data are listed in columns 2, 4 and 6 of that Table
along with their corresponding errors, while the values derived
using the SDSS data are listed in columns 3, 5, and 7.

\subsection{Balmer temperature}
\label{balmer}

\begin{figure}
\vspace{3.5cm}
\centering
\includegraphics[width=.75\textwidth,angle=270]{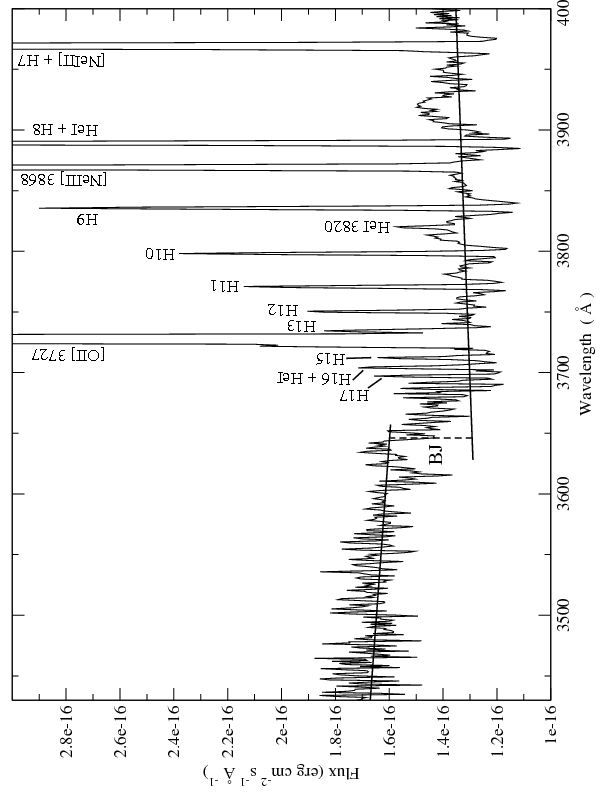}
\caption[Balmer jump of \whtdosc]{Enlargement of the spectrum of \whtdosc\ taken with the
  WHT. Its spectral range is from 3430 to 4000\,\AA, and it is in the rest frame.
  The solid lines trace the continuum to both sides of the Balmer jump and the
  dashed line depict the value of its measurement.} 
\label{Bjump}
\end{figure}

The Balmer temperature depends on the value of the Balmer jump (BJ) in
emission. To measure this value we have adjusted the continuum at both sides of
the discontinuity ($\lambda_B$\,=\,3646\,\AA). Figure \ref{Bjump} shows an example
of the procedure for \whtdosc. The contribution of the
underlying population (see Section 3) affect, among  others, the hydrogen
emission lines near the Balmer jump. The increase of the number of lines toward
shorter wavelengths produces blends which tend to depress the continuum level to
the right of the discontinuity and precludes the application of
multi-Gaussian component fittings. We have taken special care in the definition
of this continuum by using a long baseline and spectral windows free of
absorption lines.  The uncertainties due to the 
presence of the underlying stellar population (different possible continuum
placements) have been included in the errors of the measurements of the
discontinuities. They are actually smaller than the error introduced by the
fitting of stellar templates. Once the Balmer jump is measured, the Balmer
continuum 
temperature [T(Bac)] is determined from the ratio of the Balmer jump flux to the
flux of the H11 Balmer emission line using equation (3) in Liu et al.\
\cite*{2001MNRAS.327..141L}: 
\[
T(Bac)\,=\,368\times(1\,+\,0.259y^+\,+\,3.409y^{2+})\Big(\frac{BJ}{H11}\Big)^{-3/2}\,K
\]

\noindent
where $y^+$ and $y^{2+}$ are the ionic abundances of singly and doubly ionized
helium, He$^+$/H$^+$ and He$^{2+}$/H$^+$ (see \S \ref{chem-abund-der}), respectively,
and BJ is in ergs\,cm$^{-2}$\,s$^{-1}$\,\AA$^{-1}$. We can only measure this
discontinuity for the WHT objects due to the low signal-to-noise ratio of the
continuum at both sides of the jump in the CAHA data. The derived values for
T(Bac) are also listed in Table \ref{temdenwht}.

\section{Chemical abundance derivation}
\label{chem-abund-der}

We have derived the ionic chemical abundances of the different species using
the strongest available emission lines detected in the analyzed spectra and
the task IONIC of the STSDAS package in IRAF. This package is also based on
the five-level statistical equilibrium atom approximation (de Robertis, Dufour
and Hunt, 1987; Shaw and Dufour, 1995).  

The total abundances have been derived by taking into account, when required,
the unseen ionization stages of each element, using the appropriate ionization
correction factor (ICF) for each species: 
\[\frac{X}{H} = ICF(X^{+i}) \frac{X^{+i}}{H^+}\]

\subsection{Ionic abundances}

\subsubsection{Helium}

\begin{figure}
\centering
\includegraphics[width=.6\textwidth,angle=270]{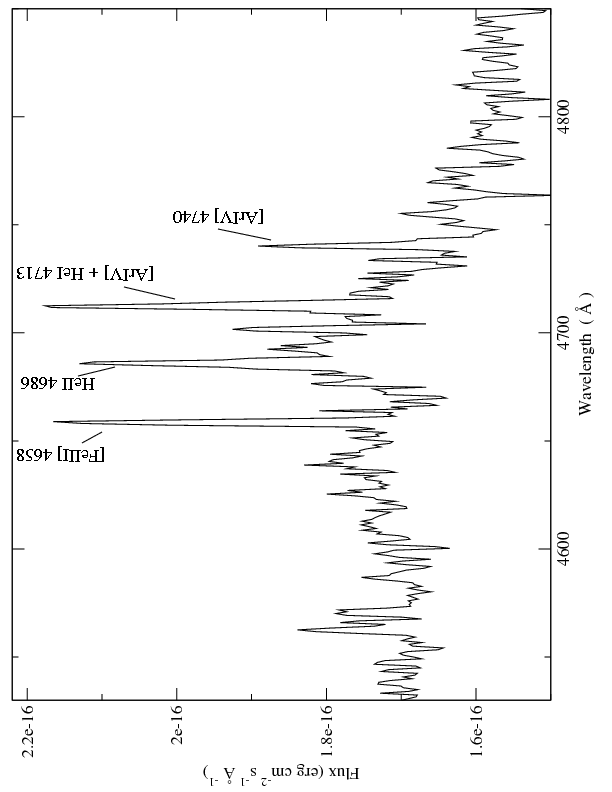}\\
\hspace{0.2cm}\includegraphics[width=.76\textwidth,angle=0]{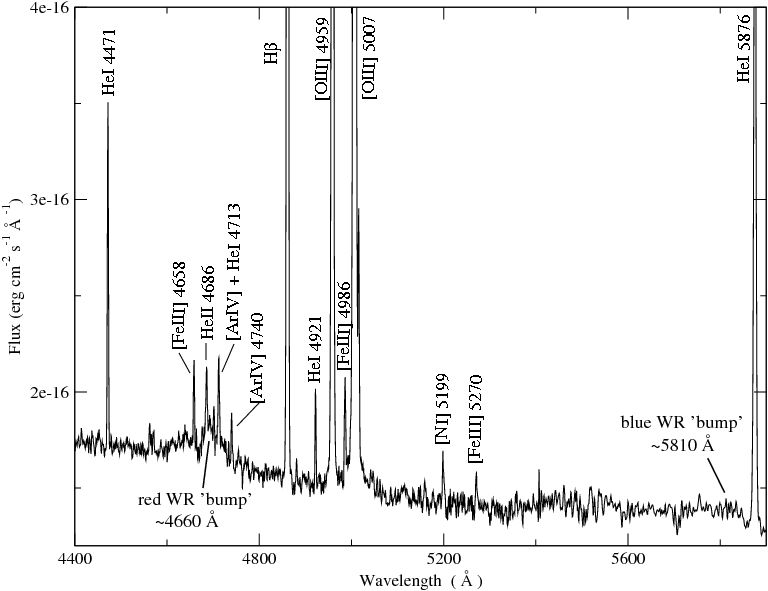}\\
\caption{Wolf-Rayet features in \whttresc.}
\label{WRbumps}
\end{figure}

We have used the well detected and measured He{\sc i}\,$\lambda\lambda$\,4471,
5876, 6678 and 7065\,\AA\ lines,  to calculate the abundances of once ionized
helium. For six of the objects (all the WHT and three of CAHA) also the He{\sc
  ii}\,$\lambda$\,4686\,\AA\ 
line was measured allowing the calculation of twice ionized He. The He lines
arise mainly from pure recombination, although they could have some
contribution from collisional excitation and be affected by self-absorption
(see Olive and Skillman, 2001, 2004, for a complete treatment of
these effects) \nocite{2001NewA....6..119O,2004ApJ...617...29O}. We have taken
the electron temperature of [O{\sc iii}] 
as representative of the zone where the He emission arises since at any rate
ratios of recombination lines are weakly sensitive to electron temperature. We
have used the equations given by Olive and Skillman  to derive the
He$^{+}$/H$^{+}$ value, using the theoretical emissivities scaled to H$\beta$
from Benjamin et al.\ \cite*{1999ApJ...514..307B} and the expressions for the
collisional correction factors from Kingdon and Ferland
\cite*{1995ApJ...442..714K}. We have not made, however, 
any corrections for fluorescence (three of the used helium lines have a small
dependence with optical depth effects but the observed  objects have low
densities) nor for the presence of an underlying stellar population. 
The three WHT galaxies and four of the CAHA objects (\dosc, \cuatroc, \cincoc\
and \sietec) show in their spectra the signature of the presence of Wolf-Rayet
(WR) stars by the blue `bump' around $\lambda$\,4600\,\AA, and three of them
(\whtdosc, \whttresc\ and \dosc) by the red 'bump' around
$\lambda$\,5808\,\AA (see Figure \ref{WRbumps}). Therefore we have to be 
careful when measuring the emission line flux of He{\sc
  ii}\,$\lambda$\,4686\,\AA. 
To calculate the abundance of twice ionized helium we have used equation (9)
from Kunth and Sargent \cite*{1983ApJ...273...81K}. The results obtained for
each line and their corresponding errors are presented in Tables
\ref{absHeWHT} and \ref{absHeCAHA} for the WHT and CAHA objects, respectively,
along with the adopted value for 
He$^{+}$/H$^{+}$ that is the average, weighted by the errors, of the different
ionic abundances derived from each He{\sc i} emission 
line. This value is dubbed ``adopted" in the Table.

For the CAHA objects, observed in 2006,
we have also calculated the average values of He$^+$/H$^+$ from the He{\sc i}
lines 
$\lambda\lambda$4471, 5876, 6678, 7065\,\AA\footnote{Although measured, the
He{\sc i} line at $\lambda$3889\,\AA\ is a blend with H8 so we decided not to
include it for this work.} using Olive
and Skillman (2004) minimization technique combined with the new He
emissivities by Porter et al.\ (2005; hearafter P05) 
\nocite{2005ApJ...622L..73P}, in order to evaluate their effect on the derived
values. We solved simultaneously for underlying stellar absorptions and
optical depth. The values are shown in Table \ref{absHeCAHA} under P05. We
found no 
significant differences for these objects in the Helium abundances
obtained using the two different sets of He emissivities. On the other hand
following this method is crucial when trying to determine He abundances to
better than 2 percent (like e.g.~for determining a value of the primordial
He). We chose to wait for a complete error budget determination from the
atomic physics parameters (Porter in preparation, private communication)
before adopting the latter values for the He abundances.

\begin{figure}
\centering
\vspace{0.3cm}
\includegraphics[width=.6\textwidth]{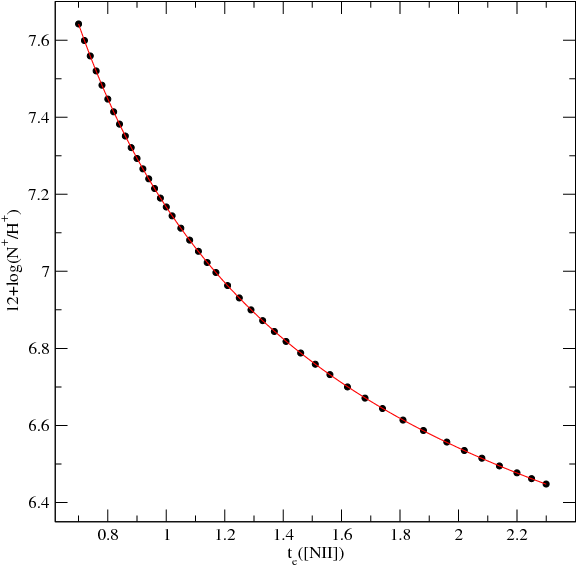}\\
\caption[N$^+$ abundance as a function of electron temperature]{N$^+$
  abundance as a function of electron temperature, for a fixed line value.} 
\label{abundfit}
\end{figure}

\landscape


\begin{table*}
{\footnotesize
\caption{Ionic and total chemical abundances for helium for the WHT objects.} 
\label{absHeWHT}

\begin{center}
\begin{tabular}{@{}cccccccc@{}}
\hline
\hline
     & & \multicolumn{2}{c}{\whtunoc} & \multicolumn{2}{c}{\whtdosc}&   \multicolumn{2}{c}{\whttresc}  \\
                  & $\lambda$\,(\AA)   &      WHT      &        SDSS   &     WHT       &        SDSS   &   WHT         &      SDSS       \\
\hline
$He^+/H^+$        & 4471 & 0.086$\pm$0.006 & 0.077$\pm$0.006 & 0.076$\pm$0.004 & 0.074$\pm$0.006 & 0.086$\pm$0.003 & 0.079$\pm$0.002   \\
                  & 5876 & 0.096$\pm$0.005 & 0.094$\pm$0.006 & 0.096$\pm$0.004 & 0.088$\pm$0.003 & 0.104$\pm$0.002 & 0.086$\pm$0.002   \\
                  & 6678 & 0.087$\pm$0.004 & 0.089$\pm$0.006 & 0.089$\pm$0.004 & 0.079$\pm$0.001 & 0.094$\pm$0.005 & 0.091$\pm$0.005   \\
                  & 7065 & 0.094$\pm$0.005 & 0.102$\pm$0.006 & 0.086$\pm$0.009 & 0.076$\pm$0.006 & 0.104$\pm$0.008 & 0.099$\pm$0.007   \\
               & adopted & 0.090$\pm$0.005 & 0.090$\pm$0.009 & 0.087$\pm$0.007 & 0.080$\pm$0.005 & 0.098$\pm$0.007 & 0.085$\pm$0.009   \\
       
$He^{2+}/H^+$     & 4686 & 0.0006$\pm$0.0001 & 0.0004$\pm$0.0001 & 0.0011$\pm$0.0001 & 0.0013$\pm$0.0001 & 0.0008$\pm$0.0001 & 0.0007$\pm$0.0001   \\
       
\bf{(He/H)}       &    & 0.091$\pm$0.005 &  0.090$\pm$0.009 & 0.088$\pm$0.007 & 0.081$\pm$0.005 & 0.098$\pm$0.007 & 0.085$\pm$0.009   \\

\hline
\end{tabular}
\end{center}}
\end{table*}

\begin{table*}
{\footnotesize
\caption{Ionic and total chemical abundances for helium for the CAHA objects.} 
\label{absHeCAHA}
\begin{center}
\begin{tabular}{@{}cc ccccccc@{}}
\hline
\hline
  & $\lambda$\,(\AA)   & \unoc & \dosc & \tresc & \cuatroc & \cincoc & \seisc & \sietec  \\
\hline
$He^+/H^+$        & 4471    & 0.079$\pm$0.004 & 0.092$\pm$0.006 & 0.095$\pm$0.005 & 0.094$\pm$0.006 & 0.084$\pm$0.006 & 0.091$\pm$0.007 & 0.106$\pm$0.004
  \\
                  & 5876    & 0.089$\pm$0.002 & 0.094$\pm$0.004 & 0.093$\pm$0.002 & 0.087$\pm$0.002 & 0.082$\pm$0.006 & 0.086$\pm$0.003 & 0.096$\pm$0.003
  \\
                  & 6678    & 0.098$\pm$0.010 & 0.103$\pm$0.004 & 0.094$\pm$0.002 & 0.084$\pm$0.004 & 0.081$\pm$0.005 & 0.086$\pm$0.005 & 0.097$\pm$0.003
  \\
                  & 7065    & 0.104$\pm$0.005 & 0.122$\pm$0.008 & 0.120$\pm$0.007 & 0.082$\pm$0.004 & 0.087$\pm$0.006 & 0.093$\pm$0.005 & 0.099$\pm$0.008
  \\
               & adopted    & 0.090$\pm$0.010 & 0.100$\pm$0.012 & 0.095$\pm$0.013 & 0.086$\pm$0.005 & 0.083$\pm$0.002 & 0.088$\pm$0.003 & 0.098$\pm$0.004
  \\
               & P05        & 0.096$\pm$0.014 &0.102 $\pm$0.011 & 0.096$\pm$0.010& 0.089$\pm$0.010 & 0.084$\pm$0.012 & 0.089$\pm$0.009& 0.100$\pm$0.010
  \\
$He^{2+}/H^+$     & 4686    & 0.0007$\pm$0.0001 &         ---       &         ---       &         ---       & 0.0029$\pm$0.0004 & 0.0011$\pm$0.0001 &         ---      
  \\
\bf{(He/H)}       &         & 0.090$\pm$0.010 &        ---      &        ---      &        ---      & 0.086$\pm$0.002 & 0.089$\pm$0.003 &        ---     
  \\

\hline
\end{tabular}
\end{center}}
\end{table*}


\endlandscape

\subsubsection{Forbidden lines}

We have derived appropriate fittings to the IONIC task results following the
functional form given by Pagel et al.\ \cite*{1992MNRAS.255..325P}. In Figure
\ref{abundfit} we show an example of these fittings, the abundance of the
nitrogen one time ionized as function of its corresponding temperatures. The
expressions for these functions are listed below:

\begin{eqnarray*}
12\,+\,log(O^+/H^+)\,=\,log\frac{I(3727+3729)}{I(H\beta)}\,+\,5.992\,+\\
+\,\frac{1.583}{t_e}\,
-\,0.681\,log\,t_e\,+\,log(1+2.3\,n_e)\\
12\,+\,log(O^{2+}/H^+)\,=\,log\frac{I(4959+5007)}{I(H\beta)}\,+\,6.144\,+\\
+\,\frac{1.251}{t_e}\,
-\,0.550\,log\,t_e\\
12\,+\,log(S^+/H^+)\,=\,log\frac{I(6717+6731)}{I(H\beta)}\,+\,5.423\,+\\
+\,\frac{0.929}{t_e}\,
-\,0.280\,log\,t_e\,+\,log(1+1.0\,n_e)\\
12\,+\,log(S^{2+}/H^+)\,=\,log\frac{I(9069+9532)}{I(H\beta)}\,+\,5.80\,+\\
+\,\frac{0.77}{t_e}\,
-\,0.22\,log\,t_e\\
12\,+\,log(N^+/H^+)\,=\,log\frac{I(6548+6584)}{I(H\beta)}\,+\,6.273\,+\\
+\,\frac{0.894}{t_e}\,
-\,0.592\,log\,t_e\\
12\,+\,log(Ne^{2+}/H^+)\,=\,log\frac{I(3868)}{I(H\beta)}\,+\,6.486\,+\\
+\,\frac{1.558}{t_e}\,
-\,0.504\,log\,t_e\\
12\,+\,log(Ar^{2+}/H^+)\,=\,log\frac{I(7137)}{I(H\beta)}\,+\,6.157\,+\\
+\,\frac{0.808}{t_e}\,
-\,0.508\,log\,t_e\\
12\,+\,log(Ar^{3+}/H^+)\,=\,log\frac{I(4740)}{I(H\beta)}\,+\,5.705\,+\\
+\,\frac{1.246}{t_e}\,
-\,0.156\,log\,t_e\\
12\,+\,log(Fe^{2+}/H^+)\,=\,log\frac{I(4658)}{I(H\beta)}\,+\,3.504\,+\\
+\,\frac{1.298}{t_e}\,
-\,0.483\,log\,t_e)
\end{eqnarray*}
\noindent where $t_e$ denotes the appropriate line electron temperature, in
units of 10$^{\rm 4}$ K, corresponding to the assumed ionization structure as
explained below.

\noindent {\bf -Oxygen} The oxygen ionic abundance ratios, O$^{+}$/H$^{+}$ and
O$^{2+}$/H$^{+}$, have been derived from the [O{\sc
    ii}]\,$\lambda\lambda$\,3727,29\,\AA\ and  [O{\sc
    iii}]\,$\lambda\lambda$\,4959, 5007\,\AA\ lines respectively using for
each ion its corresponding temperature. 

\noindent {\bf -Sulphur.} In the same way,  we have derived S$^+$/H$^{+}$ and
S$^{2+}$/H$^{+}$, abundances using T$_e$([S{\sc ii}]) and T$_e$([S{\sc iii}])
values and the fluxes of the [S{\sc ii}] emission lines at
$\lambda\lambda$\,6717,6731\,{\AA} and  the near-IR [S{\sc
    iii}]\,$\lambda\lambda$\,9069, 9532\,\AA\ lines respectively.

\noindent {\bf -Nitrogen.} The ionic abundance of nitrogen, N$^{+}$/H$^{+}$
has been derived from the intensities of the $\lambda\lambda$\,6548,
6584\,\AA\ lines and  the derived electron temperature of [N{\sc ii}] in the
case of \whtunoc\ and \whttresc\ for the WHT objects, and \sietec\ for the CAHA
objects. For the rest of the objects the assumption
T$_e$([N{\sc ii}])\,$\approx$\,T$_e$([O{\sc ii}]) has been made.  

\noindent {\bf -Neon.} Neon is only visible in all the spectra via the [Ne{\sc iii}] emission line at
$\lambda$3868\,{\AA}, then Ne$^{2+}$ has been derived using this line. For
this ion we have taken the electron temperature of [O{\sc iii}], as
representative of the high excitation zone (T$_e$([Ne{\sc
    iii}])\,$\approx$\,T$_e$([O{\sc iii}]); Peimbert and Costero
1969). \nocite{1969BOTT....5....3P}

\noindent {\bf -Argon.} The main ionization states of Ar in ionized regions
are Ar$^{2+}$ and Ar$^{3+}$. The abundance of Ar$^{2+}$ has been calculated
from the measured [Ar{\sc iii}]\,$\lambda$\,7136\,\AA\ line emission assuming
that T$_e$([Ar{\sc iii}])\,$\approx$\,T$_e$([S{\sc iii}])
\cite{1992AJ....103.1330G}, while the  ionic abundance of Ar$^{3+}$ has been
calculated from the emission line of [Ar{\sc iv}]\,$\lambda$\,4740\,\AA\ under
the assumption that T$_e$([Ar{\sc iv}])\,$\approx$\,T$_e$([O{\sc iii}]). 
We could have used the blended emission line [Ar{\sc iv}]+He{\sc
  I}\, at $\lambda$\,4713\,\AA\  subtracting the helium contribution. However,
due to the relative abundance of these species and the signal-to-noise ratio
of this blended line, we prefer not to estimate this ionic abundance with such
a  large error.

\noindent {\bf -Iron.} Finally, for iron we have used the emission line of
[Fe{\sc iii}]\,$\lambda$\,4658\,\AA\ to calculate Fe$^{2+}$ assuming
T$_e$([Fe{\sc iii}])\,$\approx$\,T$_e$([O{\sc iii}]).  

The ionic abundances of the different elements with respect to ionized
hydrogen along with their corresponding errors are given in Tables
\ref{ion-absWHT} and \ref{ion-absCAHA} for the WHT and CAHA objects,
respectively.  

\landscape


\begin{table*}
{\footnotesize
\caption{Ionic chemical abundances derived from forbidden emission lines for
  the WHT objects.}
\label{ion-absWHT}

\begin{center}
\begin{tabular}{ccccccc}
\hline
\hline
     & \multicolumn{2}{c}{\whtunoc} & \multicolumn{2}{c}{\whtdosc}&   \multicolumn{2}{c}{\whttresc}  \\
                      &      WHT      &        SDSS   &     WHT       &        SDSS   &   WHT         &      SDSS       \\
\hline
12+$\log(O^+/H^+)$    & 7.73$\pm$0.06 & 7.73$\pm$0.05 & 7.15$\pm$0.06 & 7.10 & 7.18$\pm$0.06 &  7.30$\pm$0.05  \\
12+$\log(O^{2+}/H^+)$ & 7.86$\pm$0.02 & 8.01$\pm$0.02 & 7.86$\pm$0.02 & 7.87$\pm$0.05 & 7.98$\pm$0.02 &  8.08$\pm$0.01  \\
12+$\log(S^+/H^+)$    & 5.93$\pm$0.11 & 6.13$\pm$0.12 & 5.80$\pm$0.06 & 5.88$\pm$0.08 & 5.69$\pm$0.08 &  5.67$\pm$0.10  \\
12+$\log(S^{2+}/H^+)$ & 5.91$\pm$0.05 & 6.55 & 6.16$\pm$0.06 & 6.34 & 6.14$\pm$0.04 &  6.49  \\
12+$\log(N^+/H^+)$    & 6.68$\pm$0.04 & 6.62$\pm$0.03 & 6.03 & 5.94 & 5.92$\pm$0.06 &  5.98  \\
12+$\log(Ne^{2+}/H^+)$& 7.27$\pm$0.03 & 7.42$\pm$0.03 & 7.30$\pm$0.06 & 7.27$\pm$0.05 & 7.33$\pm$0.03 &  7.44$\pm$0.02  \\
12+$\log(Ar^{2+}/H^+)$& 5.50$\pm$0.05 & 5.67 & 5.65$\pm$0.06 & 5.62 & 5.66$\pm$0.04 &  5.70  \\
12+$\log(Ar^{3+}/H^+)$& 4.37$\pm$0.05 &   ---         & 4.32$\pm$0.07 & 4.37$\pm$0.05 & 4.36$\pm$0.05 &  4.43$\pm$0.05  \\
12+$\log(Fe^{2+}/H^+)$& 5.50$\pm$0.05 & 5.66$\pm$0.07 & 5.42$\pm$0.06 & 5.46$\pm$0.06 & 5.48$\pm$0.05 &  5.59$\pm$0.05  \\

\hline
\end{tabular}
\end{center}}
\end{table*}


\begin{table*}
{\footnotesize
\caption{Ionic chemical abundances derived from forbidden emission lines for
  the CAHA objects.}
\label{ion-absCAHA}

\begin{center}
\begin{tabular}{@{}cccccccc}
\hline
\hline
  & \unoc & \dosc & \tresc & \cuatroc & \cincoc & \seisc & \sietec  \\
\hline
 12+$\log(O^+/H^+)$        &  7.16$\pm$0.09 &  7.48$\pm$0.08 &  7.67$\pm$0.09 &  7.67$\pm$0.09 &  7.07$\pm$0.12 &  7.37$\pm$0.09 &  7.57$\pm$0.07
  \\
 12+$\log(O^{2+}/H^+)$     &  7.87$\pm$0.02 &  8.10$\pm$0.02 &  8.00$\pm$0.02 &  7.84$\pm$0.02 &  7.96$\pm$0.02 &  7.87$\pm$0.02 &  7.92$\pm$0.02
  \\
 12+$\log(S^+/H^+)$        &  5.36$\pm$0.08 &  6.02$\pm$0.10 &  5.89$\pm$0.10 &  6.19$\pm$0.08 &  5.30$\pm$0.10 &  6.07$\pm$0.08 &  5.95$\pm$0.10
  \\
 12+$\log(S^{2+}/H^+)$     &  5.98$\pm$0.05 &  6.44$\pm$0.05 &  6.18$\pm$0.07 &  6.47$\pm$0.06 &  6.13$\pm$0.05 &  6.00$\pm$0.06 &  6.31$\pm$0.06
  \\
 12+$\log(N^+/H^+)$        &  5.90$\pm$0.06 &  6.28$\pm$0.06 &  6.43$\pm$0.06 &  6.47$\pm$0.07 &  5.67$\pm$0.09 &  6.15$\pm$0.06 &  6.30$\pm$0.07
  \\
 12+$\log(Ne^{2+}/H^+)$    &  7.20$\pm$0.03 &  7.44$\pm$0.03 &  7.47$\pm$0.04 &  7.17$\pm$0.04 &  7.24$\pm$0.03 &  7.22$\pm$0.04 &  7.35$\pm$0.04
  \\
 12+$\log(Ar^{2+}/H^+)$    &  5.50$\pm$0.05 &  5.94$\pm$0.05 &  5.73$\pm$0.09 &  5.95$\pm$0.06 &  5.59$\pm$0.06 &  5.49$\pm$0.06 &  5.78$\pm$0.06
  \\
 12+$\log(Ar^{3+}/H^+)$    &  4.58$\pm$0.07 &       ---      &       ---      &       ---      &  4.49$\pm$0.13 &       ---      &  4.28$\pm$0.12
  \\
 12+$\log(Fe^{2+}/H^+)$    &  4.81$\pm$0.08 &  5.71$\pm$0.06 &  5.69$\pm$0.07 &  5.52$\pm$0.09 &       ---      &  5.54$\pm$0.08 &  5.44$\pm$0.08
  \\

\hline
\end{tabular}
\end{center}}
\end{table*}

\endlandscape

\subsection{Ionization correction factors and total abundances}

For the three WHT objects and for three of the CAHA objects for which the He{\sc
  ii} line has been measured, the total abundance of He has been found by adding
directly the two ionic abundances:  
\[
\frac{He}{H}\,=\,\frac{He^++He^{2+}}{H^+}
\]

As was pointed out by Skillman et al.\ \cite*{1994ApJ...431..172S}, the
potential fraction of 
unobservable neutral helium is a long lasting problem and represents a source
of uncertainty in the derivation of the helium total abundance. The correction
factor for He$^0$ can be approximated by 1.0 for \HII\ regions ionized by very
hot stars (T$_{eff}$\,$\geq$\,40000\,K). An estimate of the ionizing stellar
temperature for our objects can be obtained from the $\eta$ parameter
\cite{1988MNRAS.231..257V}\footnote{The $\eta$ parameter is defined as the
ratio of the O$^+$/O$^{2+}$ to the S$^+$/S$^{2+}$ ionic ratios.}. The values of
log\,($\eta$) are -0.35 for two of the observed WHT objects and -0.15 for the
other one, and for these three CAHA objects are -0.10, -0.07 and -0.57. In all
the cases much smaller than the upper limit of log\,($\eta$), 0.9, for which
Pagel et al.\ \cite*{1992MNRAS.255..325P} 
claim that the correction factor for neutral helium is equal to 1.0.  

In Tables \ref{absHeWHT} and \ref{absHeCAHA} we present the total helium
abundance values for the WHT objects and for these three CAHA objects,
respectively, together with their corresponding errors. 

\label{oxygen}
At the temperatures derived for our observed galaxies,
most of the oxygen is in the form of O$^+$ and O$^{2+}$, therefore the
approximation: 
\[ 
\frac{O}{H}\,=\,\frac{O^++O^{2+}}{H^+}
\]
\noindent has been used. 

\label{sulphur}
This is not however the case for sulphur for which a relatively important
contribution from S$^{3+}$ may be expected depending on the nebular
excitation. The total sulphur abundance has been calculated using an ICF for
S$^+$+S$^{2+}$ 
according to Barker's (1980) formula, which is based on the photo-ionization
models by Stasi\'nska \cite*{1978A&A....66..257S}: 
\nocite{1980ApJ...240...99B}
 
\[
ICF(S^++S^{2+}) = \left[ 1-\left( 1-\frac{O^+}{O^{+}+O^{2+}}
  \right)^\alpha\right]^{-1/\alpha} 
\]
\noindent where $\alpha$\,=\,2.5 gives the best fit to the scarce observational
data on S$^{3+}$ abundances (P\'erez-Montero et al., 2006). 
\nocite{2006A&A...449..193P}

Although it is customary to write Barker's expression as a function of the 
O$^{+}$/(O$^{+}$+O$^{2+}$) ionic fraction, it can be reformulated in terms of 
O$^{2+}$/(O$^{+}$+O$^{2+}$)  since the errors associated to O$^{2+}$ are
considerably smaller than for O$^{+}$. Then, the ICF for
S$^+$+S$^{2+}$ is:  
\[
ICF(S^++S^{2+}) = \left[ 1-\left(\frac{O^{2+}}{O^{+}+O^{2+}}
  \right)^{2.5}\right]^{-0.4} 
\]

\label{N/O}
We have derived the N/O abundance ratio under the assumption
that   
\[
\frac{N}{O}\,=\,\frac{N^+}{O^+}
\]
\noindent and the N/H ratio as:
\[ log \frac{N}{H} = log\frac{N}{O} + log\frac{O}{H} \]

\label{Neon}

Classically, the total abundance of neon has been calculated assuming that:
\[
\frac{Ne}{O}\,=\,\frac{Ne^{2+}}{O^{2+}}
\]
Izotov et al.\ (2004) point out that this assumption can
lead to an overestimate of 
Ne/H in objects with low excitation, where the charge transfer between O$^{2+}$
and H$^0$ becomes important. Nevertheless, in our case, this assumption is
probably justified given the high excitation of the observed  objects.

\nocite{2004A&A...415...87I}

For the CAHA objects, the ionization correction factor for neon has been
calculated according to the expression given in Chapter \S \ref{neon}: 
\[
ICF(Ne^{2+})\,=\,0.142\,x+0.753+\frac{0.171}{x}
\]
where $x$\,=\,O$^{2+}$/(O$^{+}$+O$^{2+}$). This expression has been derived
from photo-ionization models \cite{1998PASP..110..761F},
taking as ionizing sources the spectral energy distribution of O and B stars
\cite{2001A&A...375..161P}.

Given the high excitation of the observed objects there are no significant
differences between the total neon abundance derived using this ICF and those
estimated using the classical approximation:
Ne/O\,$\approx$\,Ne$^{2+}$/O$^{2+}$. 

\label{Argon}
The total abundance of argon for the WHT objects, following the published
values, has been calculated  using the
ICF(Ar$^{2+}$+Ar$^{3+}$) given by Izotov et al.\ (1994) which, in turn, has
been derived from the photo-ionization models by Stasi\'nska
\cite*{1990A&AS...83..501S} as:  
\[
ICF(Ar^{2+}+Ar^{3+})\,=\,\Big[0.99+0.091\Big(\frac{O^+}{O}\Big)-1.14\Big(\frac{O^+}{O}\Big)^2+0.077\Big(\frac{O^+}{O}\Big)^3\Big]^{-1}
\]

\landscape


\begin{table*}
{\footnotesize
\caption{ICFs and total chemical abundances for elements heavier than
  Helium for WHT objects.}
 
\label{total-abs-WHT}

\begin{center}
\begin{tabular}{@{}ccccccc@{}}
\hline
\hline
     & \multicolumn{2}{c}{\whtunoc} & \multicolumn{2}{c}{\whtdosc}&   \multicolumn{2}{c}{\whttresc}  \\
     &      WHT      &        SDSS   &     WHT       &        SDSS   &   WHT         &      SDSS       \\
\hline
\bf{ 12+log(O/H)}     & 8.10$\pm$0.04 & 8.19$\pm$0.03 & 7.93$\pm$0.03 & 7.93 & 8.05$\pm$0.02 &  8.14$\pm$0.02  \\
ICF($S^++S^{2+}$)     & 1.12$\pm$0.10 & 1.18 & 1.50$\pm$0.15 & 1.57 & 1.61$\pm$0.13 &  1.58  \\
\bf{ 12+log(S/H)}     & 6.27$\pm$0.12 & 6.77 & 6.49$\pm$0.11 & 6.67 & 6.48$\pm$0.09 &  6.75  \\
 \bf{log(S/O)}        & -1.83$\pm$0.15& -1.42& -1.44$\pm$0.14&-1.26 &-1.57$\pm$0.11 &  -1.39 \\
\bf{log(N/O)}         & -1.06 $\pm$0.10 & -1.12 $\pm$0.08 & -1.11 & -1.16 & -1.26$\pm$0.12 &  -1.32  \\
\bf{log(Ne/O)}        & -0.59$\pm$0.06 & -0.59$\pm$0.06 & -0.63$\pm$0.06 & -0.60$\pm$0.07 & -0.65$\pm$0.05 &  -0.64$\pm$0.03  \\
ICF($Ar^{2+}$+$Ar^{3+}$)& 1.21$\pm$0.06 & ---         & 1.01$\pm$0.01 & 1.02 & 1.01$\pm$0.01 &  1.02$\pm$0.01  \\
\bf{ 12+log(Ar/H)}    & 5.59$\pm$0.08 & --- & 5.66$\pm$0.06 & 5.63 & 5.66$\pm$0.05 &  5.70  \\
ICF($Fe^{2+}$)        & 3.09$\pm$0.33 & 3.52$\pm$0.32 & 6.04$\pm$0.84 & 6.63 & 7.00$\pm$0.83&  6.72$\pm$0.63  \\
\bf{ 12+log(Fe/H)}    & 5.99$\pm$0.10 & 6.20$\pm$0.11 & 6.20$\pm$0.12 & 6.28 & 6.32$\pm$0.10 &  6.42$\pm$0.09  \\
\bf{log(S$_{23}$)}    & -0.22$\pm$0.02 & ---  & -0.02$\pm$0.03  & ---   & -0.10$\pm$0.02  & ---  \\

\hline
\end{tabular}
\end{center}}
\end{table*}


\begin{table*}
{\footnotesize
\caption{ICFs and total chemical abundances for elements heavier than
  Helium for CAHA objects.}
\label{total-abs-CAHA}

\begin{center}
\begin{tabular}{@{}cccccccc@{}}
\hline
\hline
  & \unoc & \dosc & \tresc & \cuatroc & \cincoc & \seisc & \sietec  \\
\hline
 \bf{12+log(O/H)}          &  7.94$\pm$0.03 &  8.19$\pm$0.03 &  8.17$\pm$0.04 &  8.07$\pm$0.05 &  8.01$\pm$0.03 &  7.99$\pm$0.04 &  8.08$\pm$0.04
  \\
 ICF($S^++S^{2+}$)         &  1.51$\pm$0.08 &  1.42$\pm$0.06 &  1.22$\pm$0.04 &  1.14$\pm$0.03 &  1.71$\pm$0.16 &  1.32$\pm$0.05 &  1.23$\pm$0.03
  \\
 \bf{12+log(S/H)}          &  6.25$\pm$0.06 &  6.73$\pm$0.07 &  6.45$\pm$0.08 &  6.71$\pm$0.07 &  6.42$\pm$0.07 &  6.46$\pm$0.07 &  6.55$\pm$0.07
  \\
 \bf{log(S/O)}             & -1.69$\pm$0.06 & -1.46$\pm$0.07 & -1.72$\pm$0.09 & -1.36$\pm$0.08 & -1.59$\pm$0.07 & -1.53$\pm$0.08 & -1.52$\pm$0.08
  \\
 \bf{log(N/O)}             & -1.26$\pm$0.10 & -1.21$\pm$0.10 & -1.24$\pm$0.11 & -1.20$\pm$0.11 & -1.40$\pm$0.15 & -1.23$\pm$0.11 & -1.27$\pm$0.10
  \\
 ICF($Ne^{2+}$)            &  1.08$\pm$0.01 &  1.08$\pm$0.01 &  1.10$\pm$0.01 &  1.12$\pm$0.01 &  1.07$\pm$0.01 &  1.09$\pm$0.01 &  1.10$\pm$0.01
  \\
 \bf{12+log(Ne/H)}         &  7.23$\pm$0.03 &  7.48$\pm$0.03 &  7.51$\pm$0.04 &  7.22$\pm$0.04 &  7.27$\pm$0.03 &  7.25$\pm$0.04 &  7.39$\pm$0.04
  \\
 \bf{log(Ne/O)}            & -0.71$\pm$0.04 & -0.72$\pm$0.05 & -0.66$\pm$0.06 & -0.84$\pm$0.07 & -0.74$\pm$0.04 & -0.74$\pm$0.06 & -0.69$\pm$0.05
  \\
 ICF($Ar^{2+}$)            &       ---      &  1.18$\pm$0.03 &  1.11$\pm$0.01 &  1.11$\pm$0.01 &       ---      &  1.13$\pm$0.02 &       ---     
  \\
 ICF($Ar^{2+}$+$Ar^{3+}$)  &  1.03$\pm$0.01 &       ---      &       ---      &       ---      &  1.02$\pm$0.01 &       ---      &  1.06$\pm$0.01
  \\
 \bf{12+log(Ar/H)}         &  5.56$\pm$0.06 &  6.01$\pm$0.05 &  5.78$\pm$0.07 &  6.00$\pm$0.06 &  5.64$\pm$0.08 &  5.54$\pm$0.06 &  5.82$\pm$0.06
  \\
 \bf{log(Ar/O)}            & -2.39$\pm$0.06 & -2.19$\pm$0.06 & -2.39$\pm$0.08 & -2.07$\pm$0.08 & -2.38$\pm$0.08 & -2.45$\pm$0.07 & -2.25$\pm$0.07
  \\
 ICF($Fe^{2+}$)            &  6.08$\pm$1.10 &  5.28$\pm$0.88 &  3.77$\pm$0.66 &  3.21$\pm$0.57 &       ---      &  4.49$\pm$0.79 &  3.82$\pm$0.56
  \\
 \bf{12+log(Fe/H)}         &  5.59$\pm$0.10 &  6.43$\pm$0.09 &  6.27$\pm$0.10 &  6.03$\pm$0.12 &       ---      &  6.19$\pm$0.11 &  6.03$\pm$0.10
  \\
 \bf{log(S$_{23}$)}        & -0.21$\pm$0.02 &  0.05$\pm$0.02 & -0.04$\pm$0.03 &  0.08$\pm$0.02 & -0.15$\pm$0.02 & -0.05$\pm$0.02 & -0.04$\pm$0.02
  \\

\hline
\end{tabular}
\end{center}}
\end{table*}


\endlandscape

For the CAHA objects, as in the case of neon, the total abundance of argon has
been calculated using the ionization correction factors (ICF(Ar$^{2+}$) and the
ICF(Ar$^{2+}$+Ar$^{3+}$)) given in Chapter \S \ref{neon}. We have used
the first one only when we cannot derive a value for Ar$^{3+}$. The
expressions for these ICFs are:   
\[
ICF(Ar^{2+})\,=\,0.507\,(1-x)+0.749+\frac{0.064}{(1-x)}
\]
\[
ICF(Ar^{2+}+Ar^{3+})\,=\,0.364\,(1-x)+0.928+\frac{0.006}{(1-x)}
\]
where $x$\,=\,O$^{2+}$/(O$^{+}$+O$^{2+}$).

\label{Iron}

The ICF for iron twice ionized has been taken from Rodr\'iguez and Rubin
\cite*{2004IAUS..217..188R}:  
\[
ICF(Fe^{2+})\,=\,\Big(\frac{O^+}{O^{2+}}\Big)^{0.09} \cdot
\Big[1+\frac{O^{2+}}{O^{+}}\Big]
\]

In Tables \ref{total-abs-WHT} and \ref{total-abs-CAHA} we list all the total
chemical abundances and the ICFs derived for elements heavier than helium for
the WHT and CAHA objects, respectively.

\section{Discussion}
\label{discu}

\subsection{Comparison between the WHT and SDSS estimated parameters}

In order to compare our results for the WHT objects with those provided by
SDSS spectra, we have measured the emission line intensities and equivalent
widths on the SLOAN spectra of
the three observed objects in the same way as described in Section \S
\ref{results}. In order 
to allow an easy comparison between the results on both sets of spectra, we
have listed these values in Columns 6-8 of Tables \ref{ratiostot
  wht1}-\ref{ratiostot wht3}. 

Strong emission line fluxes relative to H$\beta$ measured on WHT and SDSS
spectra differ by less than 10\% for \whtdosc\ and \whttresc\ and about 30\%
for \whtunoc. This is 
partially compensated by differences in the derived reddening constant so that
reddening corrected emission line intensities relative to H$\beta$ differ by
less that 10\% for the [O{\sc ii}]\,$\lambda\lambda$\,3727,29\,\AA\ line, about
15\% for the weak  [O{\sc iii}] $\lambda$\,4363\,\AA\ and  [S{\sc
    iii}]\,$\lambda$\,6312\,\AA\ lines and only a few percent for the
strong  [O{\sc iii}]\,$\lambda$\,5007\,\AA\ line.  In fact, given the
difference in aperture between both sets of 
observations, 0.5 and 3 arcsec for WHT and SDSS respectively, some differences
are to be expected. While we have probably observed the bright cores of the
galaxies where most of the light and present star formation is concentrated,
the SDSS observations map a more extensive area which could include external
diffuse zones. This is evidenced by the more conspicuous underlying stellar
population detected in the WHT spectra which leads to lower values of the
emission line equivalent widths. 
The best agreement between the two sets of measurements is found for
\whtdosc\, which is probably the most compact object of the three. This is 
consistent with the fact that the difference in the measured H$\beta$ fluxes
is only a factor of 2 and there is close agreement between the measured
equivalent widths on the two spectra.

\begin{figure}
\centering
\includegraphics[width=.82\textwidth,angle=0,clip=]{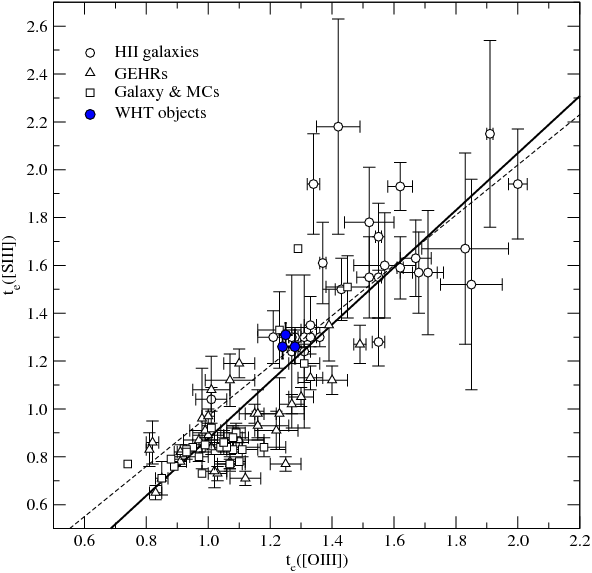}\\
\caption[Relation between t$_e$([S{\sc iii}{\textrm]}) and t$_e$([O{\sc
  iii}{\textrm]}) for \HII\ like objects]{Relation between t$_e$([S{\sc
  iii}]) and t$_e$([O{\sc iii}]) for the observed WHT
  objects (solid blue circles), and \HII\ galaxies (open circles), Giant Extragalactic
  \HII\ regions (upward triangles) and diffuse \HII\ regions in the Galaxy and
  the Magellanic Clouds (squares), for which data on the auroral and nebular lines of
  [O{\sc iii}] and [S{\sc iii}]  exist (see P\'erez-Montero et al.\ 2006). The
  temperatures are in units of 10$^4$\,K.} 
\label{T([SIII]-T([OIII])-wht}
\end{figure}

The SDSS spectra have been analyzed following the same methodology as explained
in Sections \S \ref{physiHIIgal} and \S \ref{chem-abund-der} for the
derivation of temperatures and abundances although, 
due to the different nature of the observations: different spectral coverage,
signal-to-noise ratio and spectral resolution, some further assumptions had to
be made. These refer mainly to the temperatures of the different ions.  
Regarding sulphur, the SDSS data do not reach the 9000-9600\,\AA\ range covered
by the WHT spectra and therefore it was not possible to determine directly
T$_e$([S{\sc iii}]). The relation between t$_e$([S{\sc iii}]) and t$_e$([O{\sc
    iii}]) is reproduced in Figure \ref{T([SIII]-T([OIII])-wht}. 
The sample used for comparison is a compilation of published data for which
measurements of the nebular and auroral lines of [O{\sc iii}] and [S{\sc iii}]
exist, thus allowing the simultaneous determination of T$_e$([O{\sc iii}]) and 
T$_e$([S{\sc iii}]) (P\'erez-Montero et al., 2006). 
The dashed line in the plot corresponds to the theoretical
relation based on the grids of photo-ionization models described in
P\'erez-Montero and D\'\i az (2005),  
\[
t_e([S\textrm{\sc iii}])\,=\,1.05\,t_e([O\textrm{\sc iii}])\,-\,0.08
\]
\noindent which  differs slightly from the semi-empirical relation by Garnett
(1992) mostly due to the introduction of the new atomic coefficients for
S$^{2+}$ from Tayal and Gupta (1999). 
The solid line in Figure \ref{T([SIII]-T([OIII])-wht} corresponds to the
actual least-square fit to the data: 
\[t_e([S\textrm{\sc iii}]) = (1.19 \pm 0.08)\,t_e([O\textrm{\sc iii}]) - (0.32 \pm 0.10)\]
The individual errors have not been taken into account in performing the fit. 
This fit is different from that found by Garnett (1992) that seems to
reproduce well the M101 \HII\ region data analyzed by Kennicutt, Bresolin and
Garnett (2003; KBG03). This is mostly due to the larger temperature baseline
that we use. The object with the highest T$_e$([S{\sc iii}]) in KBG03 is NGC5471A
(12800\,K); our sample includes high excitation \HII\ galaxies with
T$_e$([S{\sc iii}]) up to 22000\,K, while including at the same time KBG03
sample. The introduction of 
the high excitation objects make the relation steeper and increases the error
of the calibration. This illustrates the danger of extrapolating relations
found for a restricted range of values. 
We have used our empirical calibration in order to obtain T$_e$([S{\sc
iii}]) for the SDSS spectra. The estimated errors  introduced by the
calibration are of the order of 12\% for T$_e$([S{\sc iii}]),  i.e. between
1400 and 1500\,K for the observed WHT objects.  

\nocite{2003ApJ...591..801K}


\begin{figure}
\centering
\includegraphics[width=.82\textwidth,angle=0,clip=]{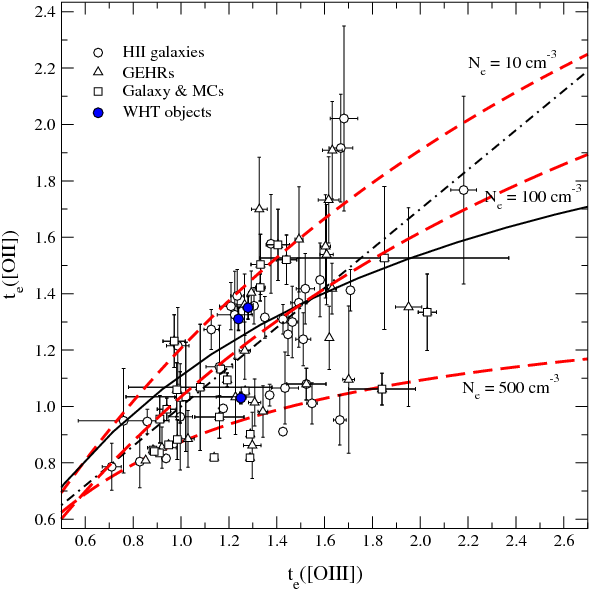}\\
\caption[Relation between t$_e$([O{\sc ii}{\textrm]}) and t$_e$([O{\sc
  iii}{\textrm]}) for \HII\ like objects]{Relation between t$_e$([O{\sc ii}])
  and t$_e$([O{\sc iii}]) for the observed   objects (solid circles) and \HII\
  galaxies (open circles), Giant Extragalactic 
  \HII\ regions (upward triangles) and diffuse \HII\ regions in the Galaxy and
  the Magellanic Clouds (squares) from P\'erez-Montero and D\'\i az (2005). The 
  dashed lines correspond to photo-ionization models from P\'erez-Montero and
  D\'\i az (2003) for electron densities N$_{e}$\,=\,10, 100 and
  500\,cm$^{-3}$. The model sequences from Stasi\'nska (1980; solid line) and 
  Stasi\'nska (1990; dashed-dotted line) are also shown. The temperatures are in units of
  10$^4$\,K.}
\label{T([OII]-T([OIII])-wht}
\end{figure}


\nocite{1980A&A....84..320S}

Likewise, the SDSS spectrum of \whtdosc\ does not include the lines of
[O{\sc ii}] at $\lambda\lambda$\,3727,29\,\AA\ and therefore it was not
possible to derive T$_e$([O{\sc ii}]). Again, we have resorted to the model
predicted relationship between  T$_e$([O{\sc ii}]) and T$_e$([O{\sc iii}])
found by P\'erez-Montero and D\'iaz (2003) that takes explicitly into account
the dependence of T$_e$([O{\sc ii}]) on electron density. Three model
sequences are represented in Figure \ref{T([OII]-T([OIII])-wht} corresponding
to three different values of the density: 10, 100 and 500\,cm$^{-3}$ (dashed
lines in the plot). The model sequence for N$_e$\,=\,100\,cm$^{-3}$ is very
similar to the one derived from the models by Stasi\'nska (1980). The sample
used for comparison comprises the objects from P\'erez-Montero and D\'\i az
(2005) for which the derivation of T$_e$([O{\sc ii}]) and T$_e$([O{\sc iii}])
has been possible.  In this case, due to the dependence of T$_e$([O{\sc ii}])
on electron density, there is no a single empirical calibration and it is not
possible to give an estimate of the error introduced by the application of
this procedure. 

In the case of nitrogen, the [N{\sc ii}]\,$\lambda$\,5755\,\AA\ line could be
measured only in the SDSS spectrum of \whtunoc, due to poor signal to
noise in the other two cases where the assumption T([N{\sc
    ii}])\,$\approx$\,T([O{\sc ii}]) 
has been made. This assumption is usually made in standard analysis techniques;
however, there are not enough data for \HII\ galaxies to test it empirically
(see discussion below and Figure \ref{T([NII]-T([OII])}). 

Finally, the Balmer continuum temperature could not be calculated from the SDSS
spectra due to lack of spectral coverage. 

Concerning  abundances, the S$^{2+}$/H$^{+}$ abundance ratios had to be
calculated using the intensity of the weak auroral [S{\sc iii}] line at
$\lambda$\,6312\,\AA\ and, in the case of \whtdosc, the
O$^{+}$/H$^{+}$ abundance ratio was derived using the [O{\sc
    ii}]\,$\lambda\lambda$\,7319,7330\, \AA\ lines following the procedure
described by Kniazev et al. (2004).  

The values of electron density and temperatures, ionic and total abundances
derived from the SDSS spectra are listed in Columns 3, 5 and 7 of Tables
\ref{temdenwht}, \ref{ion-absWHT} and \ref{total-abs-WHT}, respectively, along
with their 
corresponding errors in the cases where they have 
been derived from measured emission lines intensities. Otherwise, since the
uncertainties introduced by the different assumptions made and the theoretical
models used are impossible to quantify, no formal errors are given. These
quantities should be considered as estimates and be used with caution. 

For \whtdosc\ the values we have obtained for densities, temperatures and
abundances from the WHT and SDSS spectra are in excellent agreement within the
observational errors, as expected from the 
close agreement between the measured emission line intensities. For the other
two objects, the agreement can be considered as satisfactory, taking into
account the difference in aperture between both sets of observations.

\subsection{Comparison with previous published data for the CAHA objects}
\label{previous-data}

Five of the seven CAHA \HII\ galaxies presented here (\unoc, \dosc,\cuatroc,
\cincoc\ and \sietec) have been previously studied by
Izotov et al.\ \cite*{2006A&A...448..955I} from SDSS/DR3 spectra. \dosc\ was
also analyzed by Peimbert and Torres-Peimbert \cite*{1992A&A...253..349P}
together with \tresc\ using spectra in the $\lambda\lambda$ 3400-7000\,\AA\
range obtained with the 2.1\,m telescope at KPNO through a 3.2\,arcsec
slit. \cincoc\ and \sietec\ have also been studied 
by Kniazev at al. \cite*{2004ApJS..153..429K} from SDSS/DR1 spectra. For each
observed object the reddening corrected emission line intensities reported in
these studies are given for comparison in Tables \ref{ratiostot
  1}-\ref{ratiostot 7}. Only the line intensities with errors less than 40 per
cent are listed. 

A good general agreement between our measurements and the ones in the
literature for the strong emission lines and most of the weak ones is
found. There are however some noticeable differences. 

For \dosc , published values for the intensities of the lines bluer than
H$\beta$ are larger than measured in this work. For the [O{\sc
    ii}]\,$\lambda$\,3727\,\AA\ line the discrepancy amounts to 37 \% for the
data by Izotov et al.\ \cite*{2006A&A...448..955I} and 22 \%  for those by
Peimbert and Torres-Peimbert \cite*{1992A&A...253..349P}. In all cases the
derived values of the reddening constant are similar. 
It is worth noting that in the SDSS image the object seems to have a 3\,arcsec
size (see Figures \ref{images-wht} and \ref{images-caha}), thus the SDSS fiber
aperture and the KPNO observations might contain a 
substantial part of the galaxy, while our data comes from the brightest
central knot. This could explain part of the found differences.  

In the case of \tresc\ the values given by Peimbert and Torres-Peimbert
\cite*{1992A&A...253..349P} for the [O{\sc i}]\,$\lambda$\,6300\,\AA\ and
[S{\sc ii}]\,$\lambda\lambda$\,6717,6731\,\AA\ lines are larger and smaller
respectively than measured in this work. This is most noticeable for the
[O{\sc i}] line with their value being larger than ours by almost a factor of
three. 

For \cincoc, the [Ne{\sc iii}]\,$\lambda$\,3868\,\AA\ line intensity given in
Izotov et al.\ \cite*{2006A&A...448..955I} is larger than measured here by
33\%.  On the contrary, a smaller intensity than given here by about 30\% is
measured for this line in \sietec\ by Izotov et al.\
\cite*{2006A&A...448..955I}.

\subsection{Properties of the ionized gas: Gaseous physical conditions and element abundances}

\subsubsection{Densities and temperatures}

Four electron temperatures -- T$_e$([O{\sc iii}]), T$_e$([O{\sc ii}]),
T$_e$([S{\sc iii}]) and  T$_e$([S{\sc ii}])-- have been estimated in all the
observed objects.  In addition, T$_e$([N{\sc ii}]) has been estimated in two
of the WHT objects, \whtunoc\ and \whttresc, and one of the CAHA objects,
\sietec; [N{\sc ii}]\,$\lambda$\,5755 is detected, but has poor signal, in
\whtdosc, \unoc, \dosc, \tresc\ and \seisc\  and falls in the gap between the
blue and 
red spectra for the other two CAHA objects (\cuatroc\ and \cincoc). The good
quality of the data allows us to reach accuracies of the order of 1\% for
T$_e$([O{\sc iii}]), 3\% for T$_e$([O{\sc ii}]) and 5\% in the case of
T$_e$([N{\sc ii}]), 
T$_e$([S{\sc ii}]) and T$_e$([S{\sc iii}]) for the WHT data, while in the case
of the CAHA spectra the accuracies reach the order of 2\%, 4\%,
5\% and 7\%  for T$_e$([O{\sc iii}]), T$_e$([O{\sc ii}]), T$_e$([S{\sc iii}]),
and  both T$_e$([S{\sc ii}]) and T$_e$([N{\sc ii}]), respectively. The worse
measurement of a line temperature is T$_e$([S{\sc ii}]) for \cincoc, with a
$\sim$\,10\% error.

\begin{figure}
\centering
\includegraphics[width=.48\textwidth,angle=0,clip=]{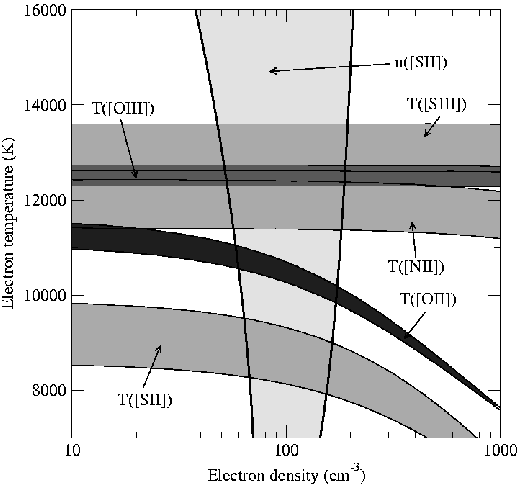}\hspace{0.2cm}
\includegraphics[width=.48\textwidth,angle=0,clip=]{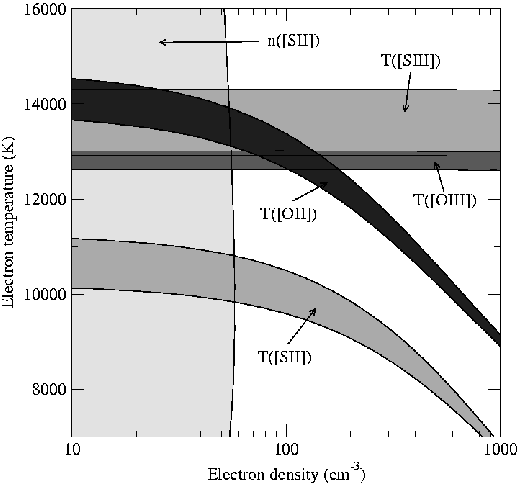}\\
\vspace{0.5cm}
\includegraphics[width=.48\textwidth,angle=0,clip=]{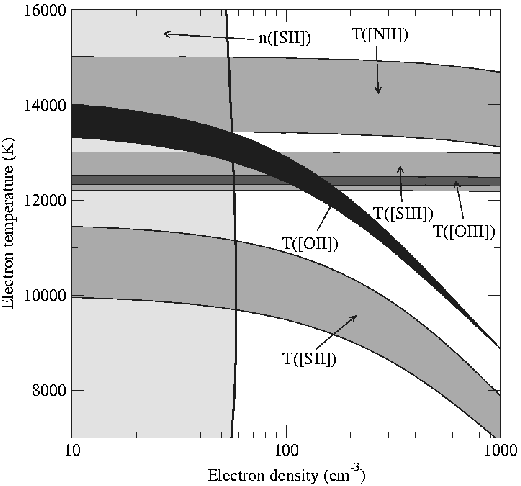}\\
\caption[Measured line temperatures and electron density for the WHT
  objects]{Measured line temperatures and electron density for the WHT objects 
  \whtunoc\ (upper left panel), \whtdosc\ (upper right panel) and \whttresc\
  (lower panel). The width of the bands corresponds to one $\sigma$
  error. Note: n([S{\sc ii}])\,$\equiv$\,N$_e$ and T\,$\equiv$\,T$_e$.}
\label{temperaturas}
\end{figure}

Figure \ref{temperaturas} shows the range of the measured
line temperatures and electron density for the WHT objects, \whtunoc,
\whtdosc\ and \whttresc. The data
for the latter two are very similar, although T$_e$([N{\sc ii}]) could not be
measured for \whtdosc. The width of the bands 
corresponds to one $\sigma$ error. In the two cases T$_e$([O{\sc iii}]) and
T$_e$([S{\sc iii}]) overlap to some extent, while T$_e$([O{\sc ii}]) is lower
than T$_e$([O{\sc iii}]) in the first case and slightly higher than
T$_e$([O{\sc iii}]) in the second.

\begin{figure}
\centering
\includegraphics[width=.82\textwidth,angle=0,clip=]{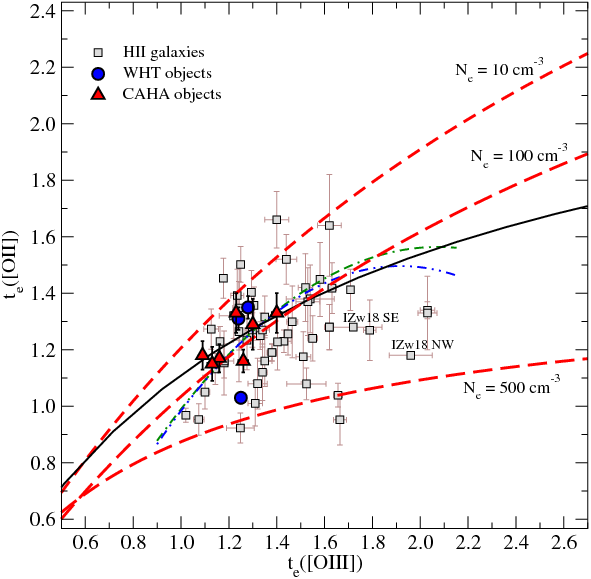}\\
\caption[Relation between t$_e$([O{\sc ii}{\textrm ]}) and t$_e$([O{\sc
  iii}{\textrm ]}) for all the observed objects and \HII\ galaxies from the
  literature]{The same as Figure \ref{T([OII]-T([OIII])-wht}, relation between 
  t$_e$([O{\sc ii}]) and t$_e$([O{\sc iii}]), for the WHT and CAHA objects and
  the \HII\ galaxies from the literature presented in Table \ref{temp}. The
  red dashed lines correspond to 
  photo-ionization models from P\'erez-Montero and D\'iaz (2003) for electron 
  densities N$ _{e} $\,=\,10, 100 and 500\,cm$^{-3}$. The model sequences from
  Stasi\'nska (1990, solid line) \nocite{1990A&AS...83..501S} and  Izotov et
  al.\ (2006) for low and intermediate metallicity \HII\ regions (green double
  dashed-dotted line and blue dashed-double dotted line) are also
  shown. Temperatures are in units of 10$^4$\,K.}
\label{T([OII]-T([OIII])}
\end{figure}

All the observed objects show temperatures within a relatively narrow range,
between 10900 and 14000\,K for T$_e$([O{\sc iii}]). This could be due to the
adopted selection 
criteria, high H$\beta$ flux and large equivalent width of H$\alpha$, which
tends to select objects with abundances and electron temperatures close to the
median values shown by \HII\ galaxies. In Table \ref{comp} we have listed the
previously reported t$_e$([O{\sc iii}]) values for the CAHA objects. We
find a general good agreement between those values and our measurements. Only
for two objects, \dosc\ and \sietec, we find differences of 900\,K --in
average-- and 1100\,K, respectively. Figure \ref{T([OII]-T([OIII])} shows the
relation between the [O{\sc ii}] and [O{\sc iii}] temperatures measured for
all these objects. Also shown in the Figure are the corresponding values for
\HII\ 
galaxies as derived from the emission line intensities compiled from the
literature. This derivation has been done following the same prescriptions
given in the present work. These values are given in Table \ref{temp} together
with the references that have been used. We have restricted our compilation to
objects for which the temperatures could be derived with an accuracy better
than 10\%. An exception has been made for IZw18. In this case, the large
errors in the derived [O{\sc ii}] temperatures (30\% and 40\% for T$_e$([O{\sc
    ii}]) in the NW and SE knots respectively), are probably related to the
low oxygen abundance  in this galaxy and hence the weakness of the involved
emission lines. These points are labelled in the Figure and, due to their
large values, no error bars in t$_e$([O{\sc ii}]) are shown.

\begin{table*}
\centering
{\scriptsize
\caption{Previous to our work published [O{\sc iii}] temperature and
  abundances for the observed CAHA objects.}
\label{comp}
\begin{tabular} {@{}l c c c c c c c}
\hline
          &   t$_e$([O{\sc iii}]) &  12+log(O/H)      &  log(S/O) & log(N/O)& log(Ne/O) & log(Ar/O)   ref.\\
\hline	   							        	     
 \unoc    & 1.36\,$\pm$\,0.04 & 8.03\,$\pm$\,0.03 &     -1.65 &  -1.37  &  -0.73    &  -2.45    & $^{1}$ \\
 \dosc    & 1.16\,$\pm$\,0.05 & 8.22\,$\pm$\,0.04 &     -1.78 &  -1.40  &  -0.69    &  -2.50    & $^{1}$ \\
          & 1.20\,$\pm$\,0.08 & 8.17\,$\pm$\,0.10 &     ---   &  -1.24  &  -0.69    &    ---    & $^{2}$ \\
          &      1.10$^a$     & 8.26$^a$          &     ---   &  -1.24$^a$ &  ---   &    ---    & $^{2}$ \\
 \tresc   &      1.10$^a$     & 8.26$^a$          &     ---   &  -1.46$^a$ &  ---   &    ---    & $^{2}$ \\
 \cuatroc & 1.04\,$\pm$\,0.07 & 8.26\,$\pm$\,0.07 &     -1.78 &  -1.34  &  -0.80    &  -2.36    & $^{1}$ \\
          & 1.06\,$\pm$\,0.08 & 8.13\,$\pm$\,0.08 &     ---   &  ---    &    ---    &   ---     & $^{3}$ \\
 \cincoc  & 1.39\,$\pm$\,0.05 & 7.98\,$\pm$\,0.04 &     -1.52 &  -1.56  &  -0.69    &  -2.33    & $^{1}$ \\
 \sietec  & 1.15\,$\pm$\,0.04 & 8.21\,$\pm$\,0.03 &     -1.76 &  -1.18  &  -0.74    &  -2.40    & $^{1}$ \\
          & 1.15\,$\pm$\,0.03 & 8.17\,$\pm$\,0.03 &     ---   &  ---    &    ---    &   ---     & $^{3}$ \\
\hline
\multicolumn{6}{l}{$^{1}$Izotov et al.\ (2006); $^{2}$Peimbert and Torres-Peimbert (1992); $^{3}$Kniazev et al.\ (2004).}\\
\multicolumn{6}{l}{$^a$based on an empirical method from Pagel et al.\ \cite*{1979MNRAS.189...95P}.}
\end{tabular}}
\end{table*}

%
%

\begin{table*}
\centering
{\scriptsize
\caption[Electron temperatures for \HII\ Galaxies from the
  literature]{Electron temperatures for \HII\ Galaxies from the literature in
  units of 10$^{-4}$ K.} 
\label{temp}
\begin{tabular}{l c c c c l}
\hline
Object  & t$_e$([O{\sc ii}])  & t$_e$([O{\sc iii}]) & t$_e$([S{\sc ii}]) & t$_e$([S{\sc iii}]) & Ref.  \\
\hline
SBS0335-052    & 1.34$\pm$0.03 & 2.03$\pm$0.03 &          --           &          --            & ICS01 \\
SBS0832+699    & 0.95$\pm$0.09 & 1.66$\pm$0.03 &          --           &          --            & ITL94 \\
SBS1135+581   & 1.36$\pm$0.07 & 1.31$\pm$0.02 & 1.01$\pm$0.03 &          --            & ITL94 \\
SBS1152+579   & 1.42$\pm$0.09 & 1.63$\pm$0.02 & 1.70$\pm$0.10 &          --            & ITL94 \\
SBS1415+437   & 1.41$\pm$0.07 & 1.71$\pm$0.01 & 1.39$\pm$0.06 &          --            & IT98 \\
SBS0723+692A & 1.45$\pm$0.13 & 1.58$\pm$0.01 & 1.74$\pm$0.14 &          --            & ITL97 \\
SBS0749+568   & 1.08$\pm$0.06 & 1.52$\pm$0.08 &          --            & 1.78$\pm$0.23 & ITL97, PMD03 \\ 
SBS0907+543   & 1.52$\pm$0.09 & 1.44$\pm$0.04 &          --            &          --            & ITL97 \\
SBS0917+527   & 1.18$\pm$0.09 & 1.51$\pm$0.03 & 1.19$\pm$0.08 &          --            & ITL97 \\
SBS0926+606   & 1.31$\pm$0.05 & 1.43$\pm$0.02 & 1.07$\pm$0.06 & 1.49$\pm$0.13 & ITL97, PMD03 \\
SBS0940+544N & 1.33$\pm$0.13 & 2.03$\pm$0.04 &          --           &          --            & ITL97 \\
SBS1222+614   & 1.26$\pm$0.08 & 1.45$\pm$0.01 & 1.79$\pm$0.16 &          --            & ITL97 \\
SBS1256+351   & 1.32$\pm$0.07 & 1.35$\pm$0.01 & 0.97$\pm$0.04 &          --            & ITL97 \\
SBS1319+579A & 1.40$\pm$0.08 & 1.29$\pm$0.01 & 0.65$\pm$0.04 &          --            & ITL97 \\
SBS1319+579C & 1.27$\pm$0.07 & 1.23$\pm$0.03 & 0.77$\pm$0.06 &          --            & ITL97 \\
SBS1358+576   & 1.30$\pm$0.13 & 1.47$\pm$0.02 & 1.18$\pm$0.08 &          --            & ITL97 \\
SBS1533+574B & 1.39$\pm$0.09 & 1.24$\pm$0.03 & 0.78$\pm$0.07 &          --            & ITL97 \\
Pox36               & 0.92$\pm$0.05 & 1.25$\pm$0.06 & 0.79$\pm$0.05 &          --            & IT04 \\
CGC007-025      & 1.04$\pm$0.04 & 1.66$\pm$0.02 & 1.72$\pm$0.10 &          --            & IT04 \\ 
Mrk 450-1          & 1.23$\pm$0.05 & 1.16$\pm$0.01 & 1.04$\pm$0.04 &          --            & IT04 \\
Mrk 450-2          & 1.35$\pm$0.13 & 1.24$\pm$0.03 & 1.07$\pm$0.10 &          --            & IT04 \\
HS0029+1748    & 1.23$\pm$0.11 & 1.28$\pm$0.06 &          --            &          --            & IT04 \\
HS0122+0743    & 1.27$\pm$0.11 & 1.79$\pm$0.03 & 1.30$\pm$0.13 &          --            & IT04 \\
HS0128+2832    & 1.50$\pm$0.06 & 1.25$\pm$0.01 & 1.22$\pm$0.07 &          --            & IT04 \\
HS1203+3636A  & 0.95$\pm$0.06 & 1.07$\pm$0.02 & 1.51$\pm$0.12 &          --            & IT04 \\
HS1214+3801    & 1.25$\pm$0.05 & 1.33$\pm$0.01 & 1.18$\pm$0.06 &          --            & IT04 \\
HS1312+3508    & 1.01$\pm$0.08 & 1.31$\pm$0.03 &          --            &          --            & IT04 \\
HS2359+1659    & 1.15$\pm$0.11 & 1.18$\pm$0.02 &          --            &          --            & IT04 \\
Mrk 35               & 0.97$\pm$0.03 & 1.02$\pm$0.01 & 1.06$\pm$0.02 &          --            & IT04 \\
UM 238              & 0.83$\pm$0.08 & 1.24$\pm$0.02 & 0.83$\pm$0.08 &          --            & IT04 \\
UM 439              & 1.23$\pm$0.10 & 1.28$\pm$0.06 &       --              &          --             & IT04 \\
IIZw 40              & 1.27$\pm$0.06 & 1.34$\pm$0.03 &       --               & 1.30$\pm$0.04 & GIT00,PMD03 \\
Mrk 22               & 1.16$\pm$0.09 & 1.35$\pm$0.03 & 0.96$\pm$0.08 &  1.94$\pm$0.21& ITL94, PMD03 \\
Mrk 36               & 1.37$\pm$0.12 & 1.53$\pm$0.05 &        --              & 1.55$\pm$0.17 & IT98, PMD03 \\
UM 461              & 1.64$\pm$0.16 & 1.62$\pm$0.05 &        --              & 1.93$\pm$0.10 & IT98, PMD03 \\
UM 462              & 1.19$\pm$0.03 & 1.38$\pm$0.02 & 1.00$\pm$0.07 & 1.61$\pm$0.17 & IT98, PMD03 \\
Mrk5                  & 1.32$\pm$0.08 & 1.22$\pm$0.06 &        --              & 1.30$\pm$0.11 & IT98, PMD03 \\
VIIZw 403          &  1.42$\pm$0.12 & 1.52$\pm$0.03 &        --              & 1.28$\pm$0.10 & ITL97, PMD03 \\
Mrk 209             &  1.28$\pm$0.08 & 1.62$\pm$0.01 & 1.25$\pm$0.13 & 1.59$\pm$0.13 & ITL97, PMD03 \\ 
Mrk 1434           &  1.24$\pm$0.08 & 1.55$\pm$0.02 &        --              & 1.72$\pm$0.14 & ITL97, PMD03 \\
Mrk 709             &  1.50$\pm$0.15 & 1.67$\pm$0.06 &        --              & 1.62$\pm$0.16 & T91, PMD03 \\
UGC 4483          &         --             & 1.68$\pm$0.06 &        --              & 1.57$\pm$0.17 & S94 \\  
IZw18NW           & 1.28$\pm$0.40  & 1.96$\pm$0.09 &        --              & 2.49$\pm$0.51 & SK93 \\
IZw18SE            & 1.18$\pm$0.50  & 1.72$\pm$0.12 &        --              & 1.97$\pm$0.34 & SK93 \\
KISSR 1845       & 1.08$\pm$0.09 & 1.32$\pm$0.04 &       --               &          --            & M04 \\  
KISSR 396         & 1.66$\pm$0.10 & 1.40$\pm$0.05 &       --               &          --            & M04 \\
KISSB 171         & 1.16$\pm$0.06 & 1.18$\pm$0.04 &       --               &          --            & L04 \\
KISSB 175         & 1.12$\pm$0.10 & 1.34$\pm$0.02 &       --               &          --            & L04 \\
KISSR 286         & 1.05$\pm$0.06 & 1.10$\pm$0.02 &       --               &          --            & L04 \\
\hline
\multicolumn{6}{l}{{\small T91: Terlevich et al.\ \cite*{1991A&AS...91..285T};
SK93: Skillman and Kennicutt \cite*{1993ApJ...411..655S}; }}\\
\multicolumn{6}{l}{{\small S94: Skillman et al.\ \cite*{1994ApJ...431..172S};
ITL94: Izotov et al.\ \cite*{1994ApJ...435..647I}; }}\\
\multicolumn{6}{l}{{\small ITL97: Izotov et al.\ \cite*{1997ApJS..108....1I};
IT98: Izotov and Thuan \cite*{1998ApJ...500..188I};}}\\ 
\multicolumn{6}{l}{{\small ICS01: Izotov et al.\ \cite*{2001A&A...378L..45I};
GIT00: Guseva et al.\ \cite*{2000ApJ...531..776G}; }}\\
\multicolumn{6}{l}{{\small IT04: Izotov et al.\ \cite*{2004A&A...415...87I};
    PMD03: P\'erez-Montero and D\'iaz \cite*{2003MNRAS.346..105P};}}\\
\multicolumn{6}{l}{{\small M04: Melbourne et al.\ \cite*{2004AJ....127..686M};
    L04: Lee et al.\ \cite*{2004ApJ...616..752L}}}\\ 
\end{tabular}}
\end{table*}



The relation between the [O{\sc ii}] and [O{\sc iii}] temperatures does not
show a clear trend, showing a scatter which is larger than observational
errors. Given that the [O{\sc ii}] temperature is somewhat dependent on
density one could be tempted to adscribe this scatter to a density effect. The
density effect can be seen by looking at the model sequences which are
overplotted and which correspond to photo-ionization models from
P\'erez-Montero and D\'\i az (2003) for electron densities N$_{e}$\,=\,10, 100
and 500\,cm$^{-3}$. Higher density models show lower values of t$_e$([O{\sc
    ii}]) for a given  t$_e$([O{\sc iii}]). The effect is more noticeable at
high electron temperatures. In fact, the data points populate the region of
the diagram spanned  by model sequences with most objects located between the
model sequences corresponding to N$_e$ 100 and 500 cm$^{\rm -3}$. Our observed
objects, however,  lie between the model sequences for N$_{e}$ 10 and 100
cm$^{\rm -3}$. This is actually consistent with the derived values of
N$_e$([S{\sc ii}]).


All in all, the data show that there is not a unique relation between the
[O{\sc ii}] and [O{\sc iii}] temperatures which allows a reliable derivation
of one of this temperatures when the other one can not be secured. This is
actually a standard procedure in principle adopted for the analysis of low
resolution and poor signal-to-noise data and now extended to data of much
higher quality. The solid line in Figure \ref{T([OII]-T([OIII])} shows the
relation based on the photo-ionization models by Stasi\'nska (1990), adopted in
many abundance studies of ionized nebulae. A substantial part of the sample
objects show [O{\sc ii}] temperatures which are lower or higher than predicted
by this relation for as much as 3000 K. At a value of T$_e$([O{\sc iii}] of
10000 K, this differences translate into higher and lower O$^+$/H$^+$ ionic
ratios, respectively, by a factor of 2.5. However, when using model sequences
to predict [O{\sc ii}] temperatures no uncertainties are attached to the
t$_e$([O{\sc ii}]) vs.\ t$_e$([O{\sc iii}]) relation and the outcome is a
reported T$_e$([O{\sc ii}]) which carries only the usually small observational
error of T$_e$([O{\sc iii}]) which translates into very small errors in the
oxygen ionic and total abundances. Thus it is possible to find in the
literature values of T$_e$([O{\sc ii}]) with quoted fractional errors lower
than 1\% and absolute errors actually less than that quoted for T$_e$([O{\sc
    iii}]) \cite{1998ApJ...500..188I}, which translate into ionic
O$^+$/H$^+$ ratios with errors of only 0.02\,dex. 

Recently, this procedure has been justified by Izotov et al.\
\cite*{2006A&A...448..955I} based on the comparison of a selected SDSS  data
sample of "\HII-region like" objects with photo-ionization models computed by
the authors. Different expressions of T$_e$([O{\sc ii}]) as a function of
T$_e$([O{\sc iii}]) are given for different metallicity regimes and it is
argued that, despite a large scatter, the relation between T$_e$([O{\sc ii}])
and T$_e$([O{\sc iii}]) derived from observations follows generally the one
obtained by models. However, no clear trend is shown by the data and the large
errors attached to the electron temperature determinations, in many cases
around $\pm$\,2000\,K for T$_e$([O{\sc ii}]), actually preclude the test of
such a statement. In fact, most of the data with the smallest error bars lie
below and above the theoretical relation. While it may well be that these
objects belong to a different family from typical \HII\ galaxies, this has not
been actually shown to be the case. The model sequences of Izotov et al.\
\cite*{2006A&A...448..955I} for the cases of low and intermediate
metallicities are shown in Figure \ref{T([SIII]-T([OIII])} as green double
dashed-dotted and the blue dashed-double dotted lines respectively. These
models diverge from previous sequences  
at temperatures below 10000 K and above 18000 K. Of great concern is the model
degenerate behaviour at high temperatures. Unfortunately, only one object in
our sample has an [O{\sc iii}] temperature larger than 20000 K (SBS0940+544N,
T$_e$([O{\sc iii}])\,=\,20300\,$\pm$\,400\,K) and its [S{\sc iii}] temperature
has a 10\% error. Therefore it is not possible with the present data to
address this important issue.

\begin{figure}
\vspace{2.5cm}
\centering
\includegraphics[width=.82\textwidth,angle=0,clip=]{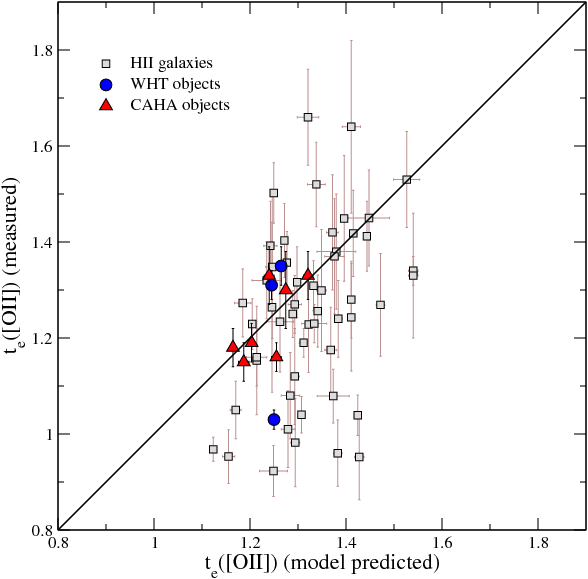}\\
\caption[Comparison between the estimated and model derived t$_e$([O{\sc
  ii}{\textrm ]})]{Comparison between the t$_e$([O{\sc ii}]) values derived
  from direct measurements with those derived from t$_e$([O{\sc iii}]) using
  Stasi\'nska (1990) photo-ionization models.}  
\label{tmodel}
\end{figure}

Figure \ref{tmodel} shows the comparison between the t$_e$([O{\sc ii}]) values
derived from direct measurements with those derived from t$_e$([O{\sc iii}])
using \cite{1990A&AS...83..501S} photo-ionization models. These values agree
for seven of our observed objects, the value is higher than the measured one by
2200\,K for the WHT object \whtunoc, and lower and higher by 900\,K
for the CAHA objects \seisc\ and \sietec, respectively. 
This difference for the WHT object translate into a lower O$^+$/H$^+$ ionic
ratio by a factor of 4 and a lower total oxygen abundance by 0.17\,dex in the
case of \whtunoc. In the case of the CAHA objects \seisc\ and \sietec,
these differences translates into higher 
and lower O$^+$/H$^+$ ionic ratios by factors of $\sim$1.32 and $\sim$1.26,
respectively, and higher and lower total oxygen abundances by $\sim$0.05 and
$\sim$0.06\,dex, respectively.

In general, model predictions overestimate t$_e$([O{\sc ii}]) and hence
underestimate the O$^+$/H$^+$ ratio. This is of relatively little concern in
objects of high excitation for which O$^+$/O is less than $\sim$ 10\%, but
caution should be taken when dealing with lower excitation objects where total
oxygen abundances could be underestimated by up to 0.2 dex.

%
%

\begin{figure}
\vspace{3.5cm}
\centering
\includegraphics[width=.82\textwidth,angle=0,clip=]{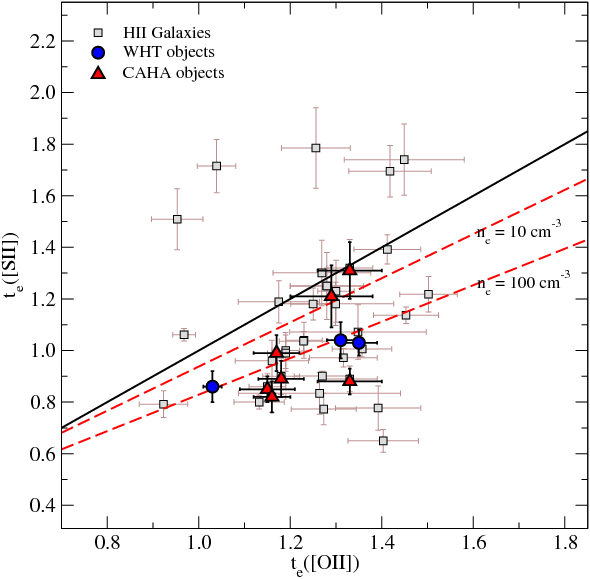}\\
\caption[Relation between t$_e$([S{\sc ii}{\textrm ]}) and t$_e$([O{\sc
  ii}{\textrm ]}) for all 
  the objects and \HII\ galaxies from the literature]{Relation between
  t$_e$([S{\sc ii}]) and t$_e$([O{\sc ii}]) for the WHT and CAHA
  objects and those from Table \ref{temp}. The solid line
  represents the   one to one relation. The red dashed lines correspond to the
  photo-ionization models from 
  P\'erez-Montero and D\'\i az (2003) for electron densities N$ _{e}$\,=\,10
  and 100\,cm$^{-3}$. Temperatures are in units of  10$^4$\,K.} 
\label{T([SII]-T([OII])}
\end{figure}

In the usually assumed structure of ionized nebulae, low ionization lines
arise from the same region and therefore the temperatures of [O{\sc ii}],
[S{\sc ii}] and [N{\sc ii}] are expected to show similar values, once
allowance is made for a possible density effect for the first two. In
Figure \ref{T([SII]-T([OII])} we show the relation between t$_e$([S{\sc ii}])
and t$_e$([O{\sc ii}])  for the WHT and CAHA objects and those from the
literature shown in Table
\ref{temp}. The relations derived by P\'erez-Montero and D\'\i az (2003) using
photo-ionization models for electron densities N$ _{e} $\,=\,10 and
100\,cm$^{-3}$ are also shown. Our measurement for the [S{\sc ii}] and [O{\sc
    ii}] temperatures are located in the region predicted by the models,
although any dependence on density is difficult to appreciate. Some objects
however are seen to lie well above the one-to-one relation. These objects are:
SBS1222+614,  HS1203+3636A and CGC007-025. The latter one shows a value of the
[O{\sc iii}] temperature close to the [S{\sc ii}] one, with the [O{\sc ii}]
temperature being lower by about 6000 K. On the other hand, the other two
objects show [O{\sc iii}] and [O{\sc ii}] temperatures fitting nicely on the
model sequence for N$_e$\,=\,100\,cm$^{-3}$. 


\begin{figure}
\centering
\includegraphics[width=.82\textwidth,angle=0,clip=]{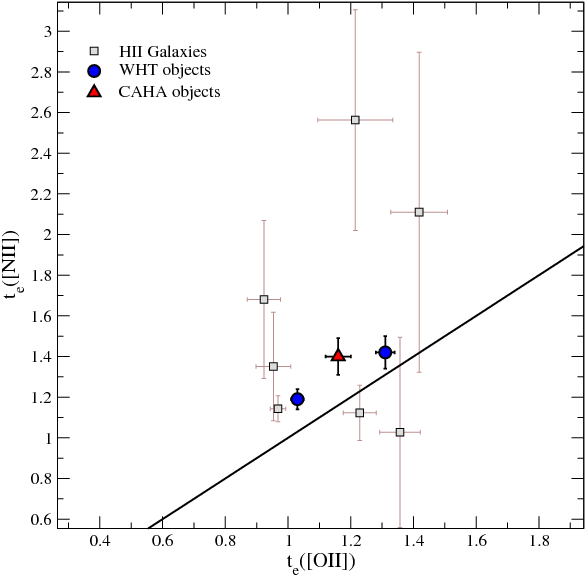}\\
\caption[Relation between t$_e$([N{\sc ii}{\textrm ]}) and t$_e$([O{\sc
  ii}{\textrm ]}) for all 
  the objects and \HII\ galaxies from the literature]{Relation between
  t$_e$([N{\sc ii}]) and t$_e$([O{\sc ii}]) for the WHT and CAHA
  objects and from \citetex{1994ApJ...435..647I} and
  \citetex{2004ApJ...602..200I} (squares).
  The solid line represents the one to one relation. Temperatures are in units
  of 10$^4$\,K.}
\label{T([NII]-T([OII])}
\end{figure}

\begin{figure}
\centering
\includegraphics[width=.82\textwidth,angle=0,clip=]{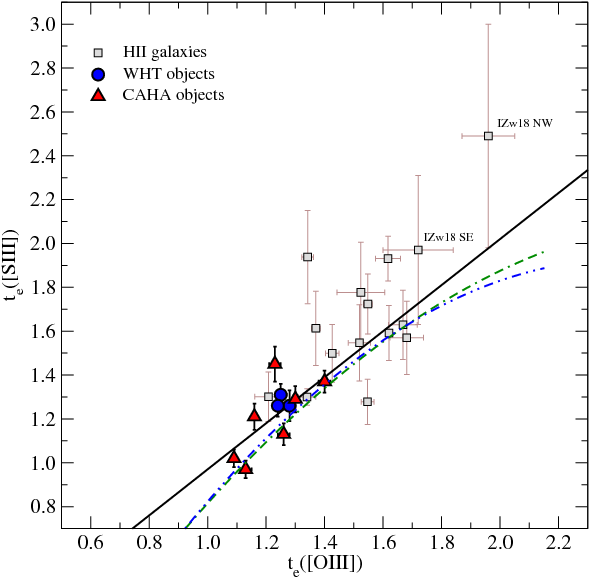}\\
\caption[Relation between t$_e$([S{\sc iii}{\textrm ]}) and t$_e$([O{\sc
      iii}{\textrm ]}) for all the observed objects and \HII\ galaxies from
  the literature]{The same as Figure \ref{T([SIII]-T([OIII])-wht}, relation
  between t$_e$([S{\sc iii}]) and t$_e$([O{\sc iii}]) for the WHT and CAHA
  objects and those from the literature presented in Table \ref{temp}. The
  solid line corresponds to the photo-ionization model sequence of
  P\'erez-Montero and D\'\i az (2005). The green double dashed-dotted line and
  the blue dashed-double dotted  line represent the models presented by Izotov
  et al.\ (2006) for low and intermediate metallicity \HII\ regions. The
  temperatures are in units of 10$^4$\,K.}
\label{T([SIII]-T([OIII])}
\end{figure}

Regarding T$_e$([N{\sc ii}]), the $\lambda$ 5575 \AA\ line is usually very
weak and difficult to measure in \HII\ galaxies due to their low
metallicities, and therefore only a few measurements exist and with very large
errors, larger than 50 \% in some cases (see Figure
\ref{T([NII]-T([OII])}). If we restrict ourselves to the data 
with errors of less than 10 \%, the [O{\sc ii}] and [N{\sc ii}] temperatures
differ by at most 1500 K. More high quality data would be needed in order to
confirm this usually assumed relation. 
 

The situation seems to be better for the [S{\sc iii}] temperature. Figure
\ref{T([SIII]-T([OIII])} shows the relation between t$_e$([S{\sc iii}]) and
t$_e$([O{\sc iii}]) for our objects and the
compilation of published data on \HII\ galaxies for which measurements of the
nebular and auroral lines of [O{\sc iii}] and [S{\sc iii}] exist, thus
allowing the simultaneous determination of T$_e$([O{\sc iii}]) and
T$_e$([S{\sc iii}]) (see Table \ref{temp}). The solid line in the
figure shows the model sequence from P\'erez-Montero and D\'\i az (2005), which
differs slightly from the semi-empirical relation by
Garnett \cite*{1992AJ....103.1330G}, while the other two lines correspond to
the relations given by Izotov et al.\ \cite*{2006A&A...448..955I} for low and
intermediate metallicity \HII\ regions. 
The three model sequences are coincident for temperatures in the range from
12000 and 17000 K, and very little, if any, metallicity dependence is
predicted. Although the number of objects is small (only 25, including the
galaxies in the present work) and the errors for the data found in the
literature are large, most objects seem to follow the trend shown by model
sequences. The most discrepant object is Mrk~22 that shows a [S{\sc iii}]
temperature larger than predicted by about 6000 K. Obviously, more high
quality data are needed in order to confirm the relation between the [O{\sc
    iii}] and [S{\sc iii}] temperatures and obtain an empirical fit with well
determined {\em rms} errors.


\subsubsection*{Abundances}

The abundances derived for the observed objects show the characteristic low
values found in strong line \HII\ galaxies (Terlevich et al., 1991; Hoyos and
D\'iaz, 2006; 12+log(O/H)\,=\,8.0), within the errors.
\nocite{1991A&AS...91..285T,2006MNRAS.365..454H} These values of 12+log(O/H)
are between 7.93 and 8.19. The mean error values for the oxygen and neon
abundances are 0.04 dex and slightly larger, 0.07, for sulphur and argon. 

The three WHT objects have previous oxygen abundance determinations. They are part
of the first edition of the SDSS \HII\ galaxies with oxygen abundance catalog,
presented by Kniazev et al.\ \cite*{2004ApJS..153..429K}. These authors
derived total oxygen 
abundances of 12+log(O/H)\,=\,8.18$\pm$0.04 for \whtunoc,
12+log(O/H)\,=\,8.07$\pm$0.02 for \whtdosc\ and
12+log(O/H)\,=\,8.17$\pm$0.01 for \whttresc, higher than ours by
0.17\,dex, but consistent with the values we obtain from the analysis of the SDSS
spectra for two of the objects: \whtunoc\
(12+log(O/H)\,=\,8.19$\pm$0.03) and \whttresc\
(12+log(O/H)\,=\,8.14$\pm$0.02). For \whtdosc, our analysis of
its SDSS spectrum yields a total oxygen abundance 12+log(O/H)\,=\,7.93$\pm$0.03,
lower than theirs by  0.14\,dex and closer to the value derived by
Ugryumov et al.\ \cite*{2003A&A...397..463U} in their Hamburg/SAO Survey. The
spectrum analyzed by Kniazev et al.\ \cite*{2004ApJS..153..429K} was extracted
from the first SDSS data release. This 
might point to a difference in the calibration routines of SDSS spectra from one
release to the other.

In the case of the CAHA objects, six of them have published abundance
determinations which are listed in 
Table \ref{comp}. Our results are in general good agreement with those in the
literature. In the case of \cuatroc\ the difference between the value derived
by Izotov et al.\ \cite*{2006A&A...448..955I} and ours is 0.19\,dex, but our
derived value is in agreement --within the errors-- with that obtained by
Kniazev et al.\ \cite*{2004ApJS..153..429K}. We have also found a small
difference of 0.13\,dex 
--in average-- in the oxygen abundance of \sietec. These differences are
similar to those found for the WHT objects between our
derived abundances using WHT spectra and the values estimated by
Kniazev et al.\ \cite*{2004ApJS..153..429K}.

\begin{figure*}
\centering
\includegraphics[width=.6\textwidth,angle=0,clip=]{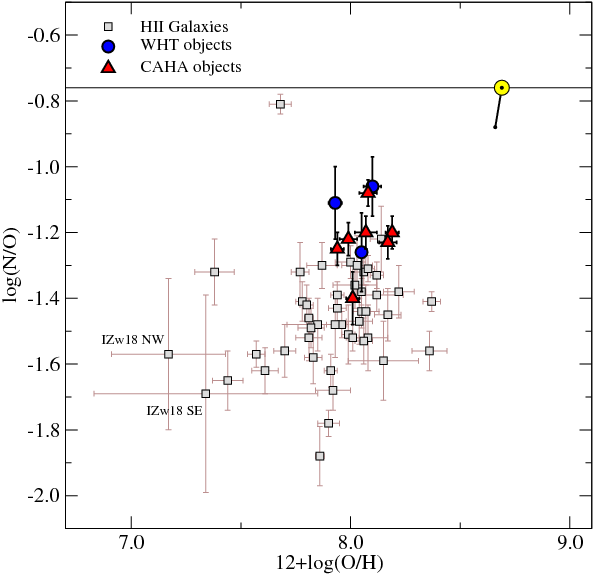}\\
\vspace{0.3cm}
\includegraphics[width=.6\textwidth,angle=0,clip=]{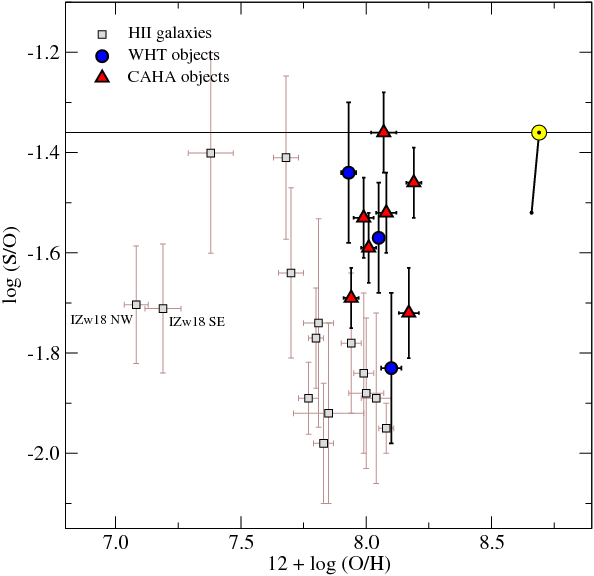}
\caption[Upper panel: N/O ratio as a function of 12+log(O/H). Lower panel: S/O
  ratio as a function of 12+log(O/H)]{Upper panel: N/O ratio as a function of
  12+log(O/H) for the observed WHT and CAHA objects, filled blue circles and
  red triangles, respectively, and the \HII\ 
  galaxies (open squares) from Table \ref{temp}. Lower panel: same as in the
  upper panel, but for the S/O ratio. The solar values are shown with the usual
  sun symbol  oxygen from Allende-Prieto et al.\ (2001), nitrogen from
  Holweger (2001) and sulphur from Grevesse and Sauval (1998). These values
  are linked by a solid line with  the solar ratios from Asplund et al.\
  (2005).}
\label{abundratios}
\end{figure*}

The logarithmic N/O ratios found for the galaxies for which there are
T$_e$([O{\sc ii}]) and T$_e$([N{\sc ii}]) determinations are -1.06$\pm$0.10
and -1.26$\pm$0.12 for the WHT data and is -1.27$\pm$0.05 for the CAHA one. 
It is worth noting that an analysis of the data along the lines discussed
above, \textit{i. e. } T$_e$([O{\sc ii}]) derivation from T$_e$([O{\sc iii}])
according to Stasi\'nska (1990) models and the assumption t$_e$([O{\sc
ii}])\,=\,t$_e$([N{\sc ii}]), would provide for  
\whtunoc, an N/O ratio larger by a factor of 3 (log
N/O\,=\,-0.64) and by a factor of 1.5 for \sietec. 
For the rest of the galaxies, for
which this assumption has been made, log(N/O) ratios are between -1.40 and
-1.11 with an average error of 0.05. For all the objects, the derived values
are on the high log(N/O) side of the distribution for this kind of objects
(see upper panel of Figure \ref{abundratios}). In general, the common
procedure of obtaining 
t$_e$([O{\sc ii}]) from t$_e$([O{\sc iii}]) using Stasi\'nska's (1990)
relation and assuming t$_e$([O{\sc ii}])\,=\,t$_e$([N{\sc ii}]), yields N/O
ratios larger than using the measured t$_e$([O{\sc ii}]) values, since
according to Figure \ref{tmodel}, in most cases, the model sequence
over-predicts t$_e$([O{\sc ii}]). An over-prediction of this temperature by a
30\,\% at T$_e$([O{\sc ii}])\,=\,13000\,K would increase the N/O ratio by a
factor of 
2. Therefore, the effect of our observed objects showing relatively high N/O
ratios seems to be real. 


\nocite{2001ApJ...556L..63A}
\nocite{2001AIPC..598...23H}
\nocite{1998SSRv...85..161G}
\nocite{2005ASPC..336...25A}


The log(S/O) ratios found for the objects are also listed in Tables
\ref{total-abs-WHT} and \ref{total-abs-CAHA}. These values vary between  -1.83
and -1.36  with an average error of 0.09, consistent with solar
(log(S/O)$_{\odot}$\,=\,-1.36\footnote{Oxygen from Allende-Prieto et al.\
  (2001) and sulphur from \citetex{1998SSRv...85..161G}.}) within the
observational errors, except for \whtunoc, \unoc\ and \tresc\  for  which S/O
is lower by a factor of about 2.7 for the WHT object, the first one, and is
lower by a factor of about 1.8 for the CAHA objects, the other two (see
lower panel of Figure \ref{abundratios}).  
Comparing with the Izotov et al.\ \cite*{2006A&A...448..955I} derived S/O
logarithmic ratios for the CAHA objects (see Table \ref{comp}) we find that
for three of the observed objects their
S/O ratios are lower than ours by us much as 0.4 dex or a factor of about
2.5.


\begin{figure*}
\centering
\includegraphics[width=.6\textwidth,angle=0,clip=]{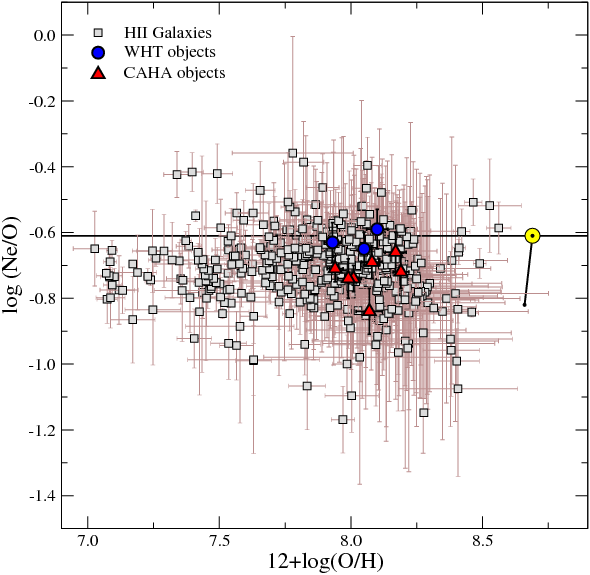}\\
\vspace{0.3cm}
\includegraphics[width=.6\textwidth,angle=0,clip=]{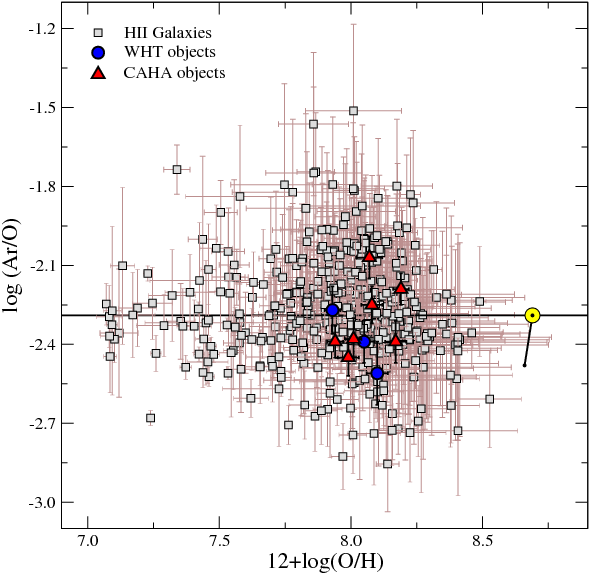}
\caption[Upper panel: Ne/O ratio as a function of 12+log(O/H). Lower panel: Ar/O
  ratio as a function of 12+log(O/H)]{Upper panel: Ne/O ratio as a function of
  12+log(O/H) for the WHT and 
  CAHA objects, filled blue circles and red respectively, and \HII\ galaxies
  (open squares) from P\'erez-Montero et al.\ (2007; see chapter \S
  \ref{neon}). Lower panel: same as in 
  the upper panel, for the Ar/O ratio. The solar values are shown by the usual
  sun symbol, with oxygen as before and neon and argon from Gravesse and
  Sauval (1998).} 
\label{abundratios2}
\end{figure*}


\nocite{2007MNRAS.381..125P}
\nocite{1998SSRv...85..161G}


The logarithmic Ne/O  ratio varies between -0.65 and -0.59 and between -0.84
and -0.66 for the WHT and CAHA objects, respectively. They have a constant
value (see upper panel Figure \ref{abundratios2}) within the errors (Tables
\ref{total-abs-WHT} and \ref{total-abs-CAHA})
consistent with solar one (log(Ne/O)\,=\,-0.61\,dex\footnote{Oxygen from
  Allende-Prieto et al.\ (2001) and neon from
  \citetex{1998SSRv...85..161G}.}), if  the object with the 
lowest ratio is excluded. An excellent agreement with the literature is
found.

The values calculated using the classical approximation for the ICF
(Ne/O\,=\,Ne$^{2+}$/O$^{2+}$), although systematically larger, are within
errors very close to those derived using the ICF for neon (see chapter \S
\ref{neon}). This is to be expected, given the high degree of ionization of
the objects in the sample.


Finally, the Ar/O ratios found for the observed objects show a larger
dispersion than in the case of Ne/O (see lower panel of Figure
\ref{abundratios2}), with a mean value consistent with solar\footnote{Oxygen
from Allende-Prieto et al.\ (2001) and argon from
  \citetex{1998SSRv...85..161G}.}. Comparing our estimations for the
logarithmic Ar/O ratios with those derived by Izotov et al.\
\cite*{2006A&A...448..955I}, we find a good agreement for three objects, and
larger values (0.31 and 0.29) for \dosc\ and \cuatroc, respectively. We must
note that for these two objects we have not been able to measure the ionic
abundances of Ar$^{3+}$.

\subsubsection{S$_{23}$ parameter}

On the other hand, the measurement of the [S{\sc iii}] IR lines allows the
calculation of the sulphur abundance parameter S$ _{23} $
\cite{1996MNRAS.280..720V}:
\[
S_{23}\,=\,\frac{[S\textrm{\sc ii}]\,\lambda\lambda\,6717,31+[S\textrm{\sc
      iii}]\,\lambda\lambda\,9069,9532}{H\beta}
\]

\begin{figure}
\centering
\includegraphics[width=.82\textwidth,angle=0]{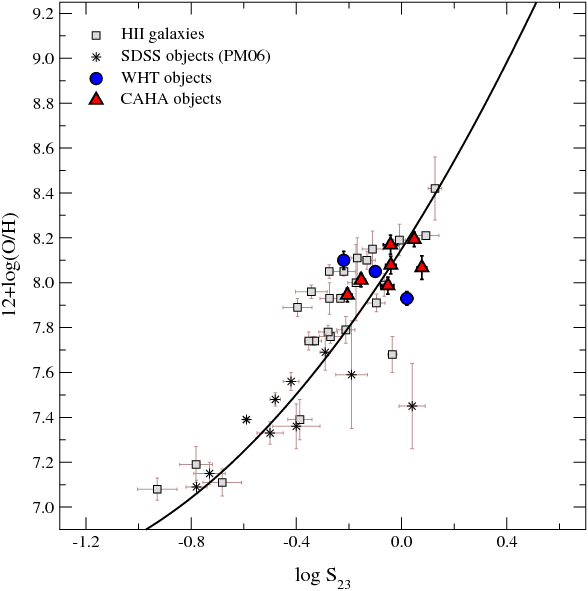}
\caption[Relationship between log\,S$_{23}$ and 12+log(O/H)]{Relationship
  between log\,S$_{23}$ and metallicity, represented by 12+log(O/H), for the
  WHT and CAHA objects, filled blue circles and red triangles respectively,
  and the \HII\ galaxies from P\'erez-Montero and D\'iaz (2005) with data on
  the [S{\sc iii}]\,$\lambda\lambda$\,9069,9532\,\AA\ emission lines
  (squares), and SDSS BCDs galaxies (asterisks) from P\'erez-Montero et al.\
  (2006). The solid line shows the calibration made by P\'erez-Montero and
  D\'iaz (2005).}  
\label{s23-o}
\end{figure}

This parameter constitutes probably the best
empirical abundance indicator for \HII\ 
galaxies, since contrary to what happens for the widely used O$_{23}$
(R$_{23}$) 
parameter  \cite{1979MNRAS.189...95P,1979A&A....78..200A}, the calibration is
linear up to solar abundances, thus solving the degeneration problem usually
presented by this kind of objects. This is particularly dramatic for objects
with log\,O$ _{23}$\,$\geq$\,0.8 and 12+log(O/H)\,$\geq$\,8.0 which can show
the same value of the abundance parameter while having oxygen abundances that
differ 
by up to an order of magnitude. About 40\,\%\ of the observed \HII\ galaxies
belong to this category (see D\'iaz and P\'erez-Montero, 1999).
\nocite{1999cezh.conf..134D} The logarithm of the S$_{23}$ parameter derived
for our objects are given in the last row of Tables \ref{total-abs-WHT} and
\ref{total-abs-CAHA}. Figure \ref{s23-o} shows the points corresponding to the
observed objects in the log\,S$_{23}$ 
vs.\ 12+log(O/H) diagram, together with their observational error. The rest of
the data correspond to \HII\ galaxies  
from the compilation made by P\'erez-Montero and D\'iaz (2005) together with
the data of ten SDSS BCD 
galaxies presented by Kniazev et al.\ \cite*{2004ApJS..153..429K} and analyzed
by P\'erez-Montero et al.\ (2006; these objects are labeled as SDSS
objects). The  
solid line shows the calibration by P\'erez-Montero and D\'iaz (2005). Three
objects are seen to clearly deviate from this  
line. Two of them correspond to galaxies from Kniazev et al.\
\cite*{2004ApJS..153..429K}  
for which no data on the [O{\sc ii}]\,3727 line exist and therefore the
O$^+$/H$^+$ ratio is derived from the red [O{\sc ii}]\,7325 lines. Both objects
show low ionization parameters as estimated from the [S{\sc ii}]/[S{\sc iii}]
ratio and relatively high values 
of O/H as derived from the [N{\sc ii}]/H$\alpha$ calibration
\cite{2002MNRAS.330.69D}. The third object is
Mrk709 that shows similar characteristics (see P\'erez-Montero and
D\'iaz 2003). These objects might be affected by shocks and deserve further
study. Actually the accuracy of the S$_{23}$ calibration for \HII\ galaxies, as
a family, is only 0.10\,dex (see P\'erez-Montero and D\'iaz, 2005). Clearly, 
more observations are needed in order to improve the S$_{23}$ calibration and
truly understand the origin of the observed dispersion.

\subsection{Ionization structure}
\label{ionization}

The ionization structure of a nebula depends essentially on the shape of the
ionizing continuum and the nebular geometry and can be traced  by the ratio of
successive stages of ionization of the different elements. With our data it is
possible to use the O$^+$/O$ ^{2+} $ and the S$ ^{+} $/S$ ^{2+} $ to probe the
nebular ionization structure. In fact, V\'ilchez and Pagel
\cite*{1988MNRAS.231..257V} showed that 
the quotient of these two quantities that they called ``softness parameter" and
denoted by $ \eta $ is intrinsically related to the shape of the ionizing
continuum and depends on geometry only slightly. If the simplifying assumption
of spherical geometry and constant filling factor is made, the geometrical
effect can be represented by the ionization parameter which, in turn can be
estimated from the [O{\sc ii}]/[O{\sc iii}] ratio. Actually, the
[O{\sc ii}]/[O{\sc iii}] ratio depends on stellar effective temperature which,
in turn, 
depends on metallicity, in the sense that for a given stellar mass, stars of
higher metallicity have a lower effective temperature. Since the observed
objects have similar values of this quotient, and
show oxygen abundances in a very narrow range, we can assume they also 
share a common value for their ionization parameter. Under these circumstances,
the value of $\eta$ points to the temperature of the ionizing radiation. 

\begin{figure}
\centering
\includegraphics[width=.6\textwidth,angle=0]{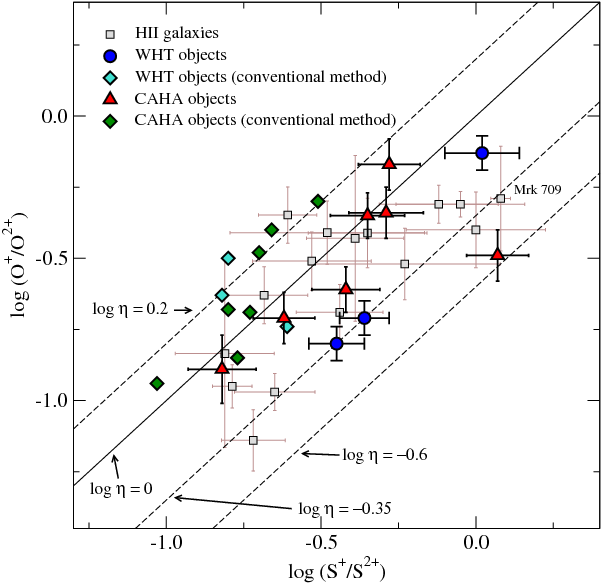}\\
\vspace{0.3cm}
\includegraphics[width=.6\textwidth,angle=0]{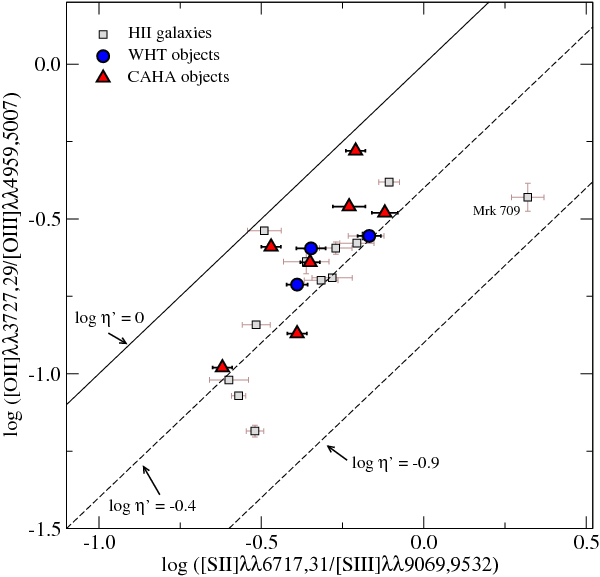}\\
\caption[Upper panel: $\eta$ diagram. Lower panel: $\eta$' diagram]{Upper
  panel: log(O$^+$/O$^{2+}$) 
  vs. log(S$^{+}$/S$^{2+}$) for the WHT and CAHA objects, solid blue circle
  and red triangles respectively, for values calculated using our
  methodology. Solid diamonds 
  represent the values derived using the conventional method for the same
  objects, turquoise and dark green respectively. Open squares are the \HII\
  galaxies in Table \ref{temp} with derived  oxygen and
  sulphur ionic ratios. Diagonals in 
  this diagram correspond to constant values of $\eta$. Lower panel: log([O{\sc
      ii}]/[O{\sc iii}]) vs. log([S{\sc ii}]/[S{\sc iii}]), symbols as in
  upper panel, and for the \HII\ galaxies in Table \ref{temp} that have
  data on the [S{\sc iii}]\,$\lambda\lambda$\,9069,9532\,\AA\ emission
  lines. Diagonals in this diagram correspond to constant values of $\eta$'.} 
\label{eta}
\end{figure}

In Figure \ref{eta}, upper panel, we show the relation between
log(O$^+$/O$^{2+}$) and log(S$^{+}$/S$^{2+}$) for the
WHT and CAHA objects (filled blue circle and red triangles,
respectively, for values calculated using our methodology) and the \HII\
galaxies in Table \ref{temp}. Besides, solid turquoise and dark green diamonds
represent the values for the WHT and CAHA galaxies, respectively, derived
using the conventional method. In this 
diagram diagonal lines correspond to constant values of the $\eta$ parameter
which can be taken as an indicator of the ionizing temperature (V\'ilchez and
Pagel, 1988). In this diagram \HII\ galaxies occupy the region with log\,$\eta$
between -0.35 and 0.2,  which corresponds to high values of the ionizing
temperature according to these authors. One of the observed objects, \seisc,
shows a very low value of $\eta$=-0.6. This object however, had the [O{\sc
    ii}]\,$\lambda\lambda$\,7319,25 \AA\ affected by atmospheric absorption
lines. Unfortunately no previous data of this object exist apart form the SDSS
spectrum. We have retrieved this spectrum and measured the [O{\sc ii}] lines
deriving a t$_e$([O{\sc ii}])\,=\,1.23$\pm$0.21, the large error being due to
the poor signal to noise in the [O{\sc
    ii}]\,$\lambda\lambda$\,7319,25\,\AA. This lower 
temperature would increase the value of O$^+$/O$^{2+}$ moving the data point
corresponding to this object upwards in the upper panel of Figure
\ref{eta}. 
This would be consistent with the position of the object in the
lower panel of the figure which shows log([O{\sc ii}]/[O{\sc iii}]) vs.\
log([S{\sc ii}]/[S{\sc iii}]), which does not require explicit knowledge of
the line temperatures involved in the derivation of the ionic ratios, and
therefore does not depend on the method to derive or estimate these
temperatures. The lower panel in Figure \ref{eta} shows the purely
observational counterpart of the upper panel. In this diagram diagonal  
lines represent constant values of log\,$\eta$' defined by V\'ilchez and Pagel 
(1988) as:

\[
log\,\eta{\textrm'}\,=\,log\Big[ \frac{[O{\textrm {\sc ii}}]\lambda\lambda\,3727,29\,/\,[O{\textrm {\sc iii}}]\lambda\lambda\,4959,5007}{[S{\textrm {\sc ii}}]\lambda\lambda\,6717,31\,/\,[S{\textrm {\sc iii}}]\lambda\lambda\,9069,9532}\Big]\,=\,log\,\eta - \frac{0.14}{t_e} -0.16
\]

\noindent where t$_e$ is the electron temperature in units of 10$^4$. $\eta$ and
$\eta$' are related through the electron temperature but very weakly, so that a
change in temperature from 7000 to 14000\,K implies a change in
log\,$\eta$ by 0.1\,dex, inside observational errors. Also, log\,$\eta$' is
always less than log\,$\eta$. The value of log\,$\eta$' for \seisc\ is -0.36
corresponding to $\eta$\,=\,-0.09 for t$_e$\,=\,1.23. 

Inconsistencies between the values of $\eta$ and $\eta$' are also found if the
ionic ratios are derived using values of electron temperatures obtained
following the prescriptions given by Izotov et al.\ (2006). These values are
represented by solid diamonds in the upper panel of Figure \ref{eta} for the
observed objects. In all cases, higher values of
$\eta$ are obtained which, in conjunction with the measured values of $\eta$', 
would indicate values of electron temperatures much lower than directly
obtained. These higher values of $\eta$ would also imply ionizing temperatures
lower than those shown by the measured $\eta$' values. 

An important conclusion is that metallicity calibrations based on abundances
derived following the conventional method are probably bound to provide
metallicities which are systematically too high and should therefore be
revised.

\subsection{Characteristics of the observed \HII\ galaxies}

We have calculated the H$\alpha$ luminosities [L(H$\alpha$)] for the observed
\HII\ galaxies from our spectra, correcting the measured H$\alpha$ fluxes
[F(H$\alpha$)] for extinction according to the values found from 
the spectroscopic analysis and using their distances, D, taken from the
literature, but we do not correct for aperture effects. The 
resulting values are listed in Table \ref{hiigal_prop}. These values are in
the typical luminosity range found in \HII\ galaxies
\cite{2006MNRAS.365..454H}.


\begin{table}
\centering
\caption{General properties of the observed CNSFR.}
{\scriptsize
\begin{tabular}{@{}l c c c c c c c c@{}}
\hline
\hline
Galaxy &  F(H$\alpha$)            & D$^a$ & L(H$\alpha$) & Q(H$_0$)	& log u & Diameter & M$_{ion}$ &   M(\HII) \\
       & (erg cm$^{-2}$ s$^{-1}$) & (Mpc)  & (erg s$^{-1}$) &  (ph s$^{-1}$) &  & (pc)  &   &  \\
\hline

\whtunoc   & 7.13E-14  & 399.5  &  1.36E+42 &  1.00E+54 & -2.33    &  353  &  648.04 &	480.78\\
\whtdosc   & 3.50E-14  & 69.1   &  2.00E+40 &  1.47E+52 & -2.43    &  681  &  10.20  &	7.82  \\
\whttresc  & 1.41E-13  & 133.4  &  2.99E+41 &  2.20E+53 & -2.57    & 1470  &  138.11 &	165.64\\
\unoc      & 4.14E-14  & 117.9  &  6.88E+40 &  5.06E+52 & -2.61    &  176  &  24.99  &	48.59 \\
\dosc      & 3.78E-14  & 201.3  &  1.83E+41 &  1.35E+53 & -1.99    &   35  &  70.85  &	112.55\\
\tresc     & 5.00E-14  & 268.6  &  4.31E+41 &  3.17E+53 & -2.77    &  923  &  125.87 &	477.42\\
\cuatroc   & 1.52E-14  & 49.4   &  4.43E+39 &  3.26E+51 & -2.19    &  205  &  1.73   &	1.35  \\
\cincoc    & 3.79E-14  & 13.7   &  8.52E+38 &  6.26E+50 & -2.73    & 2207  &  0.46   &	0.24  \\
\seisc     & 1.75E-14  & 161.2  &  5.44E+40 &  4.00E+52 & -2.42    &  281  &  21.91  &	33.45 \\
\sietec    & 6.89E-14  & 66.9   &  3.69E+40 &  2.71E+52 & -2.48    & 1122  &  14.06  &	21.11 \\
\hline
\multicolumn{9}{@{}l}{$^a$\,from \citetex{2000ApJ...529..786Mtot}. Note: masses in 10$^5$\,M$_\odot$.}
\end{tabular}}
\label{hiigal_prop}
\end{table}  


We have derived the total number of ionizing photons from the H$\alpha$
luminosities (see for example \citeplain{1989agna.book.....O}):
\[
Q(H_0)\,=\,7.35\,\times\,10^{11}\,L(H\alpha)
\]

The ionization parameter, $u$\,=\,$Q$(H$_0$)/4$\pi$cR$^2$N$_e$ (where R is the 
radius of the ionized region and c is the speed of light), can be estimated
from the [S{\sc ii}]/[S{\sc iii}] ratio \cite{1991MNRAS.253..245D} as: 
\[ 
log\,u\,=\,-1.68 log([S{\textrm{\sc ii}}]/[S{\textrm{\sc iii}}]) -2.99 
\]
and ranges between -2.77 and -1.99 for our observed \HII\ galaxies.

From the estimated number of hydrogen ionizing photons, ionization parameter
and electron density, it is possible to derive the angular size --in arcsec--
of the emitting regions ($\phi$; see \citeplain{2002MNRAS.329..315C}).
\[
\phi\,=\,0.64\,\left[\frac{F(H\alpha)}{10^{-14}}\right]^{1/2}\bigg(\frac{u}{10^{-3}}\bigg)^{-1/2}\bigg(\frac{N_e}{10^{2}}\bigg)^{-1/2}
\]
Taking into account the object distances, in Table \ref{hiigal_prop} we have
listed these values in parsecs. 

We have also derived the mass of ionizing stars, M$_*$, from the calculated
number of hydrogen ionizing photons with the use of evolutionary models of
ionizing clusters \cite{1994ApJS...91..553G,1996ApJS..107..661S} assuming that
the regions are ionization bound and that no photons are absorbed by dust. A
Salpeter IMF with upper and lower mass limits of 100 and 0.8\,M$_{\odot}$
respectively, has been assumed. According to these models, a relation exists
between the degree of evolution of the cluster, as represented by its H$\beta$
emission line equivalent width and the number of hydrogen ionizing photons per
unit solar mass \cite{1998Ap&SS.263..143D,2000MNRAS.318..462D}. 
\[
M_{ion}\,=\,\frac{7.35\,\times\,10^{11}\,L(H\alpha)}{10^{44.48\,+\,0.86\,log\left[EW(H\beta)\right]}}
\]
The ionizing cluster masses thus derived are given in Table \ref{hiigal_prop}.

The amount of ionized gas (M$_{{\rm HII}}$) associated to each \HII\ galaxy
has been obtained from our derived H$\alpha$ luminosities using the relation
given by \citetex{1990ApJ...356..389M} for an electron temperature of
10$^4$\,K 
\[
M_{{\rm HII}}\,=\,3.32\,\times\,10^{-33}\,L(H\alpha)\,N_e^{-1}
\]

\subsection{The temperature fluctuation scheme}
\label{variations}

At the end of the 60s and beginning of the 70s, Peimbert
\cite*{1967ApJ...150..825P}, Peimbert and Costero \cite*{1969BOTT....5....3P},
and Peimbert \cite*{1971BOTT....6...29P} established a
complete analytical formulation to study the discrepancies between the
abundances relative to hydrogen derived from recombination lines (RLs) and
from collisionally excited lines (CELs) when a constant electron temperature 
is assumed. In the first of these 
works, Peimbert proposed that this discrepancy is due to spatial temperature
variations which can be characterized by two parameters: the average
temperature weighted by the square of the density over the volume considered,
T$_0$, and the root mean square temperature fluctuation, t$^2$. They are given by  
\begin{eqnarray}
\label{t0}
T_0(X^{i+})\,=\,\frac{\int T_e\,N_e\,N(X^{i+})\,dV}{\int N_e\,N(X^{i+})\,dV}
\end{eqnarray}
\noindent
and
\begin{eqnarray}
\label{t2}
t^2(X^{i+})\,=\,\frac{\int(T_e-T_0(X^{i+}))^2\,N_e\,N(X^{i+})\,dV}{T_0(X^{i+})^2\int N_e\,N(X^{i+})\,dV}
\end{eqnarray}
\noindent
where N$_e$ and N$(X^{+i}$) are the local electron and ion densities
of the observed emission lines, respectively;
T$_e$ is the local electron temperature; and $V$ is the observed volume
\cite{1967ApJ...150..825P}.  

It is possible to obtain the values of T$_0$ and t$^2$ using different
methods. One possibility is to compare the electron temperatures obtained using
two independent ways. Generally, temperatures originated in different zones of
the nebula are used, one  that represents the warmest regions and another
representative of the coldest ones. In the absence of  temperatures derived from
recombination lines, the  temperatures estimated from the hydrogen
discontinuities, either from Balmer or Paschen series, representative of the
temperature of the neutral gas, can be used.  

We have followed Peimbert et al.\ (2000, 2002, 2004) and Ru\'iz et al.\ (2003)
to derive the values of T$_0$ and t$^2$ for our WHT spectra by combining the
Balmer temperature, T(Bac), and the temperature derived from the collisional
[O{\sc iii}] lines, T([O{\sc iii}]), therefore assuming a simple one-zone
ionization scheme. Then, the relation for these two temperatures is given by:

\nocite{2000ApJ...541..688P,2002ApJ...565..668P,2004ApJS..150..431P,2003ApJ...595..247R} 

\begin{eqnarray}
\label{toiii}
T([O\textrm{\sc iii}])\,=\,T_0\,\Big[1+\frac{1}{2}\Big(\frac{91300}{T_0}-3\Big)\,t^2\Big]
\end{eqnarray}
\noindent
and
\begin{eqnarray}
\label{tbac}
T(Bac)\,=\,T_0\big(1-1.67t^2\big)
\end{eqnarray}

\noindent
The solution of this system of equations for each WHT galaxy along with their
corresponding errors are listed in table \ref{fluc}. In the case of the CAHA
objects the spectra do not have enough signal-to-noise ratio to measure the
Balmer jump with an acceptable accuracy. 
The temperature fluctuations are almost negligible for two of the WHT
objects. They have t$^2$-values very similar to the ones derived by
Luridiana et al.\ \cite*{2003ApJ...592..846L} by combining observations from
the literature and 
photo-ionization  models for some BCDs and extragalactic \HII\ regions. 
Guseva et al.\ \cite*{2006ApJ...644..890G} used the Balmer and Paschen jumps
to determine the 
temperatures of the H$^+$ zones of 22 low-metallicity \HII\ regions in 18 BCD
galaxies, one extragalactic \HII\ region in M101 and 24 \HII\ emission-line
galaxies selected from the DR3 of the SDSS. They found that these temperatures
do not differ, in a statistical sense, from the temperatures of the [O{\sc iii}]
zones, given t$^2$-values are close to zero.
The greater t$^2$-value obtained for \whtdosc\ is in the range of
the values derived for giant extragalactic \HII\ regions 
(see Gonz\'alez-Delgado et al., 1994; Jamet et al., 2005; Peimbert et al., 2005)
\nocite{1994ApJ...437..239G, 2005ApJ...634.1056P,2005A&A...444..723J} or
\HII\ regions in the Magellanic Clouds, such as 30 Doradus, LMC\,N11B and
SMC\,N66 \cite{2003MNRAS.338..687T}.

\setlength{\minrowclearance}{2pt}

\begin{table}
{\small
\caption{T$_0$ and t$^2$ parameters for the CAHA objects.}
\label{fluc}
\begin{center}
\begin{tabular}{l c c }
\hline
\hline
       name               &     T$_0$         &   t$^2$         \\
\hline
SDSS J002101.03+005248.1  &    1.24$\pm$0.35  &   0.004$^{+0.044}_{-0.004}$  \\[3pt]
SDSS J003218.60+150014.2  &    1.08$\pm$0.21  &   0.066$\pm$0.026  \\[2pt]
SDSS J162410.11-002202.5  &    1.24$\pm$0.30  &   0.001$^{+0.037}_{-0.001}$  \\[2pt]
\hline

\multicolumn{3}{l}{T$_0$ in 10$^4$\,K. Note that t$^2$ is always greater than zero.}

\end{tabular}
\end{center}}
\end{table}

\setlength{\minrowclearance}{0pt}

In the limit of low densities and small optical depths, and for t$^2$ much lower
than one, the electronic temperature for
helium, T(He{\sc ii}), is proportional to $\langle\alpha\rangle$ and $\beta$,
the average value of the power of the temperature for the helium lines used to
calculate the ionic abundances of He$^+$,  and the corresponding one  for H$\beta$,
respectively. Then, for $\langle\alpha\rangle$ different from $\beta$, 
this temperature is given by equation (14) of Peimbert (1967):
\begin{eqnarray}
\label{the}
T(He\textrm{\sc ii})\,=\,T(He\textrm{\sc ii},H\textrm{\sc
  ii})\,=\,T_0\Big[1+\big(\langle\alpha\rangle+\beta-1\big)\frac{t^2}{2}\Big] 
\end{eqnarray}

The value of the power of the temperature for each helium line in the
low density limit has been obtained from \citetex{1999ApJ...514..307B}. We have 
calculated $\langle\alpha\rangle$  as the average value of $\alpha$ weighted
according to the observational errors \cite{2000ApJ...541..688P}. We have
obtained values of $\langle\alpha\rangle$ equal to -1.37, -1.42, and -1.43 for
\whtunoc, \whtdosc\ and \whttresc, respectively. The value of $\beta$ has been
obtained from Storey and Hummer \cite*{1995MNRAS.272...41S} and is equal to
-0.89. The results for T(He{\sc ii}) and their corresponding errors are listed
in the last row of Table \ref{temdenwht}. 

The temperature for H$\beta$, T(H$\beta$), can be calculated from equation (20)
in Peimbert and Costero \cite*{1969BOTT....5....3P} as:  

\begin{eqnarray}
\label{thbeta}
T(H\beta)=T_0\Big[1-0.95\,t^2\Big]
\end{eqnarray}

\noindent where we have taken $\beta$ = -0.89 as above. The derived values for
T(H$\beta$) and their errors are also given in Table \ref{temdenwht}.

The  line temperature for a collisionally excited line, CEL, for t$^2\ll1$,
($\Delta E_{CEL}$/kT$_0$-1/2)\,$\neq$\,0, and $\alpha$\,$\neq$\,0 is given by
equation (20) of Peimbert and Costero \cite*{1969BOTT....5....3P}: 

\begin{eqnarray}
\label{tcel}
T_{CEL}\,=\,T_0\,\Big\{1+\Big[\frac{(\Delta E_{CEL}/kT_0)^2-3\Delta E_{CEL}/kT_0+3/4}{\Delta E_{CEL}/kT_0-1/2}\Big]\frac{t^2}{2}\Big\}
\end{eqnarray}

\noindent where  $\Delta E_{CEL}$\,=\,$\Delta E_{mn}$ is the energy
difference, in eV, between the upper (m) and lower (n)
levels respectively of the atomic transition that produces the line.

For the case of t$^{2}$\,$>$\,0, and assuming a one-zone ionization scheme,
we can derive the ionic abundances using the values calculated for
t$^2$ equal to zero, and equation (15) of Peimbert et al.\
\cite*{2004ApJS..150..431P}: 

\begin{eqnarray}
\label{abunt2}
\lefteqn{\Big[\frac{N(X^{+i})}{N(H^{+})}\Big]_{t^2>0}\,=\,
  \frac{T(H\beta)^{-0.89}T(\lambda_{mn})^{0.5}}{T([O\textrm{\sc iii}])^{-0.37}}
   } \nonumber\\ & & 
  {} \times\, exp\Big[-\frac{\Delta E}{k\,T([O\textrm{\sc iii}])}+\frac{\Delta
  E}{k\,T(\lambda_{mn})}\Big]\,\times\,\Big[\frac{N(X^{+i})}{N(H^{+})}\Big]_{t^2=0} {} 
\end{eqnarray}

\noindent where T($\lambda_{mn}$) is given by equation (\ref{tcel}).
Using this expression, we have calculated the effect of the temperature
fluctuations on the O$ ^{2+} $/H$ ^{+}$ abundance. In fact, given the high
excitation of the object, this ionic abundance carries the highest weight in the
total abundance of oxygen.  The recalculated value  is 12+log(O$ ^{2+}
$/H$^{+}$)\,=\,8.09$\pm$0.12 which yields a  total oxygen abundance
12+log(O/H)\,=\,8.12$\pm$0.11. These values are  higher than those given in
Tables \ref{ion-absWHT} and \ref{total-abs-WHT} by 0.23 and 0.22\,dex,
respectively.

\section{Summary and conclusions}
\label{summ2}

We have performed a detailed analysis of newly obtained spectra of three and
seven \HII\ galaxies observed with the 4.2\,m WHT and the 3.5\, CAHA
telescopes, 
respectively. These galaxies were selected from the Sloan Digital Sky Survey
Data Release 2 and 3, respectively. For the first set of galaxies the spectra
cover from 3200 to 10550\,\AA\ in wavelength, while for the second group the
data cover from 3400 to 10400 \AA\ with a gap of approximately 100\,\AA\
between 5700 and 5800\,\AA. The WHT spectra have a FWHM resolution of 
about 1800 in the blue and 1700 in the red spectral regions, and the CAHA ones
of about 1400 and 1200, respectively. 
 
The high signal-to-noise ratio of the obtained spectra allows the measurement
of four line electron temperatures: T$_e$([O{\sc iii}]), T$_e$([S{\sc iii}]),
T$_e$([O{\sc ii}]) and T$_e$([S{\sc ii}]), for all the objects of the sample
with the addition of T$_e$([N{\sc ii}]) for three of them, and the Balmer
continuum temperature T(Bac) for the WHT objects. These
measurements and a careful and realistic treatment of the observational errors
yield total oxygen abundances with accuracies between 5 and 12\%. The
fractional error is as low as 1\% for the ionic O$^{2+}$/H$^{+}$ ratio due
to the small errors associated with the measurement of the strong nebular
lines of [O{\sc iii}] and the derived T$_e$([O{\sc iii}]), but increases to up
to 30\% for the O$^{+}$/H$^{+}$ ratio. The accuracies are lower in the case of
the abundances of sulphur (of the order of 25\% for S$^+$ and 15\% for
S$^{2+}$) due to the presence of larger observational errors both in the
measured line fluxes and the derived electron temperatures. The error for the
total abundance of sulphur  is also larger than in the case of oxygen (between
15\% and 30\%) due to the uncertainties in the ionization correction factors.

This is in contrast with the unrealistically small errors quoted for line
temperatures other than T$_e$([O{\sc iii}]) in the literature, in part due to
the commonly assumed methodology of deriving them from the measured
T$_e$([O{\sc iii}]) through a theoretical relation and calculating the errors
simply by propagating statistically the T$_e$([O{\sc iii}]) ones. These
relations are found  
from photo-ionization model sequences and no uncertainty is attached to them
although large scatter is found when observed values are plotted; usually the
line temperatures obtained in this way  carry only the observational error
found for the T$_e$([O{\sc iii}]) measurement and does not include the
observed scatter, thus heavily underestimating the errors in the derived
temperature.

In fact, no clear relation is found between T$_e$([O{\sc iii}]) and
T$_e$([O{\sc ii}]) for the existing sample of objects. A comparison between
measured and model derived T$_e$([O{\sc ii}]) 
shows than, in general, model predictions overestimate this temperature and
hence underestimate the O$^+$/H$^+$ ratio. This, though not very important for
high excitation objects, could be of some concern for lower excitation ones
for which total O/H abundances could be underestimated by up to 0.2 dex. It is 
worth noting that the objects observed with double-arm spectrographs,
therefore implying simultaneous and spatially coincident observations over the
whole spectral range, show less scatter in the T$_e$([O{\sc
iii}])\,-\,T$_e$([O{\sc ii}] plane clustering around the N$_e$\,=\,100
cm$^{-3}$ photo-ionization model sequence. On the other hand, this small
scatter could partially be due to the small range of temperatures shown by
these objects due to possible selection effects. This small temperature range
does not allow either to investigate the metallicity effects found in the
relations between the various line temperatures in recent photo-ionization
models by Izotov et al.\ (2006).  

Also, the observed objects, though showing  Ne/O
and Ar/O relative abundances typical of those found for a large \HII\ galaxy
sample (P\'erez-Montero et al., 2007; see Chapter \S \ref{neon}), show higher
than typical N/O abundance 
ratios that would be even higher if the [O{\sc ii}] temperatures would be
found from photo-ionization models. We therefore conclude that the approach of
deriving the O$^+$ temperature from the O$^{2+}$ one should be discouraged if
an accurate abundance derivation is sought. 

These issues could be addressed by re-observing the objects in Table
\ref{temp}, which cover an ample range in temperatures and metal content,
with double arm spectrographs. This sample should be further extended  to
obtain a self consistent sample of about 50 objects with high signal-to-noise
ratio and 
excellent spectrophotometry covering simultaneously from 3600 to 9900\,\AA.
This simple and easily feasible project would provide important scientific
return in the form of critical tests of photo-ionization models. 

For the WHT objects, we have compared our obtained spectra with those
downloaded from the SDSS DR3 finding a satisfactory agreement. The analysis of
these spectra yields values of line temperatures and elemental ionic and total
abundances which are in general agreement with those derived from the WHT
spectra, although for most quantities, they can only be taken as estimates
since, due to the lack of direct measurements of the required lines,
theoretical models had to be used whose uncertainties are impossible to
quantify. Unfortunately, the spectral coverage of SDSS precludes the
simultaneous observation of the [O{\sc ii}]\,$\lambda\lambda$\,3727,29\,\AA\
and [S{\sc iii}]\,$\lambda\lambda$\,9069, 9532\,\AA\ lines, and therefore the
analysis can never be complete.

The ionization structure found for the observed objects from the 
O$^{+} $/O$^{2+} $ and S$^{+} $/S$ ^{2+} $ ratios for all the observed
galaxies, except one,  cluster around a value of the ``softness parameter"
$\eta$ of 1 implying high values of the stellar ionizing temperature. For the
discrepant object, showing a much lower value of $\eta$,  the intensity of the
[O{\sc ii}]\,$\lambda\lambda$\,7319,25\,\AA\ lines are affected by atmospheric
absorption lines. When the observational counterpart of the ionic ratios is
used, this object shows a ionization structure similar to the rest of the
observed ones. This simple exercise shows the potential of checking for
consistency in both the $\eta$ and $\eta$' plots in order to test if a given
assumed ionization structure is adequate. In fact, these consistency checks
show that the stellar ionizing temperatures found for the observed \HII\
galaxies using the ionization structure predicted by state of the art
ionization models result too low when compared to those implied by the
corresponding observed emission line ratios.  Therefore, metallicity 
calibrations based on abundances derived according to this conventional method 
are probably bound to provide metallicities which are systematically too high
and should be revised.

Finally, we have measured the Balmer continuum temperature for the three WHT
objects and derived the temperature fluctuations as defined by
Peimbert \cite*{1967ApJ...150..825P}. Only for one of the objects, the
temperature fluctuation is significant and could lead to higher oxygen
abundances by about 0.20\,dex.

\newpage

\section{Relative reddening-corrected line intensities: Tables}


{\footnotesize
\begin{longtable}{lc|ccc|ccc}
\caption[Relative reddening-corrected lines for
\whtunoc]{Relative reddening-corrected line intensities
[$F(H\beta)$=$I(H\beta)$=10000] for \whtunoc.
\label{ratiostot wht1}} \\
 \hline
   \multicolumn{2}{c}{} & \multicolumn{6}{c}{SDSS J0021                    } \\
  \multicolumn{2}{c}{}  & \multicolumn{3}{c}{WHT} & \multicolumn{3}{c}{SDSS}\\
 \multicolumn{1}{c}{$\lambda$ ({\AA})} & f($\lambda$)
 & -EW  & $I(\lambda)$ & Error  & -EW  & $I(\lambda)$ & Error  \\
  & & (\AA) & & (\%) & (\AA) & & (\%)  \\
 \hline
 \endfirsthead
 
 \multicolumn{ 8}{l}{\small Table~\ref{ratiostot wht1} continued}
  \\
 \hline
   \multicolumn{2}{c}{}   & \multicolumn{6}{c}{SDSS J0021                    } \\
   \multicolumn{2}{c}{}   & \multicolumn{3}{c}{WHT} & \multicolumn{3}{c}{SDSS}\\
 \multicolumn{1}{c}{$\lambda$ ({\AA})} & f($\lambda$)
 & -EW  & $I(\lambda)$ & Error  & -EW  & $I(\lambda)$ & Error  \\
  & & (\AA) & & (\%) & (\AA) & & (\%)  \\
 \hline
 \endhead
 
 \hline\multicolumn{ 8}{r}{\small\sl continued on next page}
 \endfoot
 
 \hline
 \noalign {\centering $^a$\,possibly blend with an unknown line; $^b$\,[O{\sc
 ii}]\,$\lambda\lambda$\,3726\,+\,3729; $^c$\,[Fe{\sc
 iii}]\,$\lambda\lambda$\,4986\,+\,4987; $^d$\,[O{\sc
 ii}]\,$\lambda\lambda$\,7318\,+\,7320; $^e$\,[O{\sc
 ii}]\,$\lambda\lambda$\,7330\,+\,7331.} 
 \endlastfoot

 3697 H17             &   0.279   &    0.3 &    61\,$\pm$\,7     & 11.7  &   ---  &        ---        & --- \\
 3704 H16+He{\sc i}         &   0.277   &    0.7 &   146\,$\pm$\,14    &  9.9  &   ---  &        ---        & --- \\
 3712 H15             &   0.275   &    0.5 &   117\,$\pm$\,18    & 15.5  &   ---  &        ---        & --- \\
 3727 [O{\sc ii}]$^b$       &   0.271   &   71.1 & 16340\,$\pm$\,192   &  1.2  &   95.0 & 17886\,$\pm$\,198 &  1.1\\
 3750 H12             &   0.266   &    1.3 &   288\,$\pm$\,27    &  9.4  &    1.8 &   299\,$\pm$\,37  & 12.4\\
 3770 H11             &   0.261   &    1.7 &   377\,$\pm$\,30    &  7.8  &    2.0 &   329\,$\pm$\,37  & 11.3\\
 3798 H10             &   0.254   &    2.0 &   422\,$\pm$\,28    &  6.7  &    2.2 &   385\,$\pm$\,37  &  9.5\\
 3835 H9              &   0.246   &    3.2 &   654\,$\pm$\,63    &  9.7  &    4.0 &   635\,$\pm$\,71  & 11.1\\
\end{longtable}

 \begin{longtable}{lc|ccc|ccc}
\caption[Relative reddening-corrected lines for
\whtdosc]{Relative reddening-corrected line intensities
[$F(H\beta)$=$I(H\beta)$=10000] for \whtdosc.
\label{ratiostot wht2}
} \\
 \hline
   \multicolumn{2}{c}{}   & \multicolumn{6}{c}{\whtdosc                    } \\
   \multicolumn{2}{c}{}   & \multicolumn{3}{c}{WHT} & \multicolumn{3}{c}{SDSS}\\
 \multicolumn{1}{c}{$\lambda$ ({\AA})} & f($\lambda$)
 & -EW  & $I(\lambda)$ & Error   & -EW  & $I(\lambda)$ & Error  \\
  & & (\AA) & & (\%) & (\AA) & & (\%)  \\
 \hline
 \endfirsthead
 
 \multicolumn{ 8}{l}{\small Table~\ref{ratiostot wht2} continued}
  \\
 \hline
   \multicolumn{2}{c}{}   & \multicolumn{3}{c}{\whtdosc                    } \\
   \multicolumn{2}{c}{}   & \multicolumn{3}{c}{WHT} & \multicolumn{3}{c}{SDSS}\\
 \multicolumn{1}{c}{$\lambda$ ({\AA})} & f($\lambda$)
 & -EW  & $I(\lambda)$ & Error  & -EW  & $I(\lambda)$ & Error  \\
  & & (\AA) & & (\%) & (\AA) & & (\%)  \\
 \hline
 \endhead
 
 \hline\multicolumn{ 8}{r}{\small\sl continued on next page}
 \endfoot
 
 \hline
 \noalign {\centering $^a$\,possibly blend with an unknown line; $^b$\,[O{\sc
 ii}]\,$\lambda\lambda$\,3726\,+\,3729; $^c$\,[Fe{\sc
 iii}]\,$\lambda\lambda$\,4986\,+\,4987; $^d$\,[O{\sc
 ii}]\,$\lambda\lambda$\,7318\,+\,7320; $^e$\,[O{\sc
 ii}]\,$\lambda\lambda$\,7330\,+\,7331.} 
 \endlastfoot

 3687 H19             &   0.281   &    0.6 &   103\,$\pm$\,9   &  8.5 &   ---  &        ---        & --- \\
 3697 H17             &   0.279   &    0.9 &   144\,$\pm$\,12  &  8.6 &   ---  &        ---        & --- \\
 3704 H16+He{\sc i}         &   0.277   &    1.2 &   203\,$\pm$\,17  &  8.4 &   ---  &        ---        & --- \\
 3712 H15             &   0.275   &    0.9 &   159\,$\pm$\,12  &  7.5 &   ---  &        ---        & --- \\
 3727 [O{\sc ii}]$^b$       &   0.271   &   80.0 & 15728\,$\pm$\,138 &  0.9 &   ---  &        ---        & --- \\
 3734 H13             &   0.270   &    1.2 &   197\,$\pm$\,15  &  7.5 &   ---  &        ---        & --- \\
 3750 H12             &   0.266   &    1.3 &   212\,$\pm$\,16  &  7.7 &   ---  &        ---        & --- \\
 3770 H11             &   0.261   &    2.3 &   366\,$\pm$\,17  &  4.6 &    3.0 &   510\,$\pm$\,37  &  7.3\\
\end{longtable}

 \begin{longtable}{lc|ccc|ccc}
\caption[Relative reddening-corrected lines for
\whttresc]{Relative reddening-corrected line intensities
[$F(H\beta)$=$I(H\beta)$=10000] for \whttresc.
\label{ratiostot wht3}
} \\
 \hline
   \multicolumn{2}{c}{}   & \multicolumn{6}{c}{SDSS J1624                    } \\
   \multicolumn{2}{c}{}   & \multicolumn{3}{c}{WHT} & \multicolumn{3}{c}{SDSS}\\
 \multicolumn{1}{c}{$\lambda$ ({\AA})} & f($\lambda$)
 & -EW  & $I(\lambda)$ & Error  & -EW  & $I(\lambda)$ & Error  \\
  & & (\AA) & & (\%)  \\
 \hline
 \endfirsthead
 
 \multicolumn{ 5}{l}{\small Table~\ref{ratiostot wht3} continued}
  \\
 \hline
   \multicolumn{2}{c}{}   & \multicolumn{6}{c}{SDSS J1624                    } \\
   \multicolumn{2}{c}{}   & \multicolumn{3}{c}{WHT} & \multicolumn{3}{c}{SDSS}\\
 \multicolumn{1}{c}{$\lambda$ ({\AA})} & f($\lambda$)
 & -EW  & $I(\lambda)$ & Error   & -EW  & $I(\lambda)$ & Error \\
  & & (\AA) & & (\%) & (\AA) & & (\%)  \\
 \hline
 \endhead
 
 \hline\multicolumn{ 8}{r}{\small\sl continued on next page}
 \endfoot
 
 \hline
 \noalign {\centering $^a$\,possibly blend with an unknown line; $^b$\,[O{\sc
 ii}]\,$\lambda\lambda$\,3726\,+\,3729; $^c$\,[Fe{\sc
 iii}]\,$\lambda\lambda$\,4986\,+\,4987; $^d$\,[O{\sc
 ii}]\,$\lambda\lambda$\,7318\,+\,7320; $^e$\,[O{\sc
 ii}]\,$\lambda\lambda$\,7330\,+\,7331.} 
 \endlastfoot

 3687 H19             &   0.281   &    0.3 &    52\,$\pm$\,6   & 10.9 &   ---  &       ---         & ---   \\
 3692 H18             &   0.280   &    0.4 &    64\,$\pm$\,6   &  9.7 &   ---  &       ---         & ---   \\
 3697 H17             &   0.279   &    0.8 &   114\,$\pm$\,11  &  9.2 &   ---  &       ---         & ---   \\
 3704 H16+He{\sc i}         &   0.277   &    1.5 &   217\,$\pm$\,17  &  7.6 &   ---  &       ---         & ---   \\
 3712 H15             &   0.275   &    1.5 &   222\,$\pm$\,17  &  7.6 &   ---  &       ---         & ---   \\
 3727 [O{\sc ii}]$^b$       &   0.271   &   91.0 & 14707\,$\pm$\,221 &  1.5 &   96.8 & 14804\,$\pm$\,194 &  1.3  \\
 3750 H12             &   0.266   &    1.9 &   278\,$\pm$\,19  &  6.7 &    1.6 &   227\,$\pm$\,21  &  9.3  \\
 3770 H11             &   0.261   &    2.2 &   332\,$\pm$\,22  &  6.7 &    2.0 &   303\,$\pm$\,25  &  8.3  \\
\end{longtable}


 \begin{longtable}{lccccc}
 \caption[Relative reddening-corrected lines for \unoc]{Relative reddening
 corrected line intensities [$F(H\beta)$=$I(H\beta)$=10000] for \unoc.
 \label{ratiostot 1}} \\
 \hline
  &  & \multicolumn{4}{c}{SDSS J1455} \\
 \multicolumn{1}{c}{$\lambda$ ({\AA})} & f($\lambda$) & -EW  & $I(\lambda)$ &
  Error & $I(\lambda)^\dag$  \\
  & & (\AA) & & (\%)  \\
 \hline
 \endfirsthead
 
 \multicolumn{6}{l}{\small Table~\ref{ratiostot 1} continued}
  \\
 \hline
  &  & \multicolumn{4}{c}{SDSS J1455} \\
 \multicolumn{1}{c}{$\lambda$ ({\AA})} & f($\lambda$) & -EW  & $I(\lambda)$ &
  Error & $I(\lambda)^\dag$ \\
  & & (\AA) & & (\%)  \\
 \hline
 \endhead
 
 \hline\multicolumn{6}{r}{\small\sl continued on next page}
 \endfoot
 
 \hline
 \noalign {\centering $^a$\,possibly blend with an unknown line; $^b$\,[O{\sc
 ii}]\,$\lambda\lambda$\,3726\,+\,3729; $^c$\,[Fe{\sc
 iii}]\,$\lambda\lambda$\,4986\,+\,4987; $^d$\,[O{\sc
 ii}]\,$\lambda\lambda$\,7318\,+\,7320; $^e$\,[O{\sc
 ii}]\,$\lambda\lambda$\,7330\,+\,7331.}
 \noalign {\centering  $^\dag$ from Izotov et al.\
 \cite*{2006A&A...448..955I}; they gave $^\gimel$\,[S{\sc
 ii}]\,$\lambda\lambda$\,6717\,+\,6731 and $^\beth$\,[O{\sc
 ii}]\,$\lambda\lambda$\,7319\,+\,7330.}  
 \endlastfoot

 3727 [O{\sc ii}]$^b$        &   0.271   &   91.6 & 11154$\pm$164 &  1.5 & 10920$\pm$400 \\
 3734 H13                    &   0.270   &    1.1 &   130$\pm$31  & 23.5 &   ---         \\
 3750 H12                    &   0.266   &    1.8 &   204$\pm$48  & 23.8 &   270$\pm$70  \\
 3770 H11                    &   0.261   &    1.9 &   222$\pm$27  & 12.2 &   470$\pm$70  \\
 3798 H10                    &   0.254   &    3.7 &   407$\pm$39  &  9.6 &   520$\pm$80  \\
 3820 He{\sc i}              &   0.249   &    0.6 &    81$\pm$16  & 20.2 &   ---         \\
 3835 H9                     &   0.246   &    5.0 &   574$\pm$68  & 11.9 &   750$\pm$70  \\
 3868 [Ne{\sc iii}]          &   0.238   &   37.1 &  4792$\pm$105 &  2.2 &  5230$\pm$200 \\
\end{longtable}

 \begin{longtable}{lcccccc}
 \caption[Relative reddening-corrected lines for \dosc]{Relative reddening
 corrected line intensities [$F(H\beta)$=$I(H\beta)$=10000] for \dosc.
 \label{ratiostot 2}}\\
 \hline
  &  & \multicolumn{5}{c}{SDSS J1509} \\
 \multicolumn{1}{c}{$\lambda$ ({\AA})} & f($\lambda$) & -EW  & $I(\lambda)$ &  Error &  $I(\lambda)^\dag$ &  $I(\lambda)^\ddag$   \\
  & & (\AA) & & (\%) &    \\
 \hline
 \endfirsthead
 
 \multicolumn{ 7}{l}{\small Table~\ref{ratiostot 2} continued}
  \\
 \hline
  &  & \multicolumn{5}{c}{SDSS J1509} \\
 \multicolumn{1}{c}{$\lambda$ ({\AA})} & f($\lambda$) & -EW  & $I(\lambda)$ &  Error &  $I(\lambda)^\dag$ &  $I(\lambda)^\ddag$   \\
  & & (\AA) & & (\%) &    \\
 \hline
 \endhead
 
 \hline\multicolumn{ 7}{r}{\small\sl continued on next page}
 \endfoot
 
 \hline
 \noalign {\centering $^a$\,possibly blend with an unknown line; $^b$\,[O{\sc
 ii}]\,$\lambda\lambda$\,3726\,+\,3729; $^c$\,[Fe{\sc
 iii}]\,$\lambda\lambda$\,4986\,+\,4987; $^d$\,[O{\sc
 ii}]\,$\lambda\lambda$\,7318\,+\,7320; $^e$\,[O{\sc
 ii}]\,$\lambda\lambda$\,7330\,+\,7331.}
 \noalign {\centering $^\dag$ from Izotov et al.\ \cite*{2006A&A...448..955I};
 they gave $^\gimel$\,[S{\sc  ii}]\,$\lambda\lambda$\,6717\,+\,6731 and
 $^\beth$\,[O{\sc  ii}]\,$\lambda\lambda$\,7319\,+\,7330.} 
 \noalign {\centering $^\ddag$ from Peimbert and Torres-Peimbert
 \cite*{1992A&A...253..349P}.}
 \endlastfoot

 3727 [O{\sc ii}]$^b$        &   0.271   & 131.7 & 15318$\pm$182 &  1.2 &  20970$\pm$760 & 18620$\pm$1800   \\
 3734 H13                    &   0.270   &   3.5 &   278$\pm$56  & 20.3 &    ---         &   ---            \\
 3750 H12                    &   0.266   &   2.1 &   203$\pm$48  & 23.8 &    300$\pm$80  &   ---            \\
 3770 H11                    &   0.261   &   2.4 &   236$\pm$38  & 16.1 &    460$\pm$80  &   ---            \\
 3798 H10                    &   0.254   &   4.7 &   410$\pm$60  & 14.7 &    480$\pm$90  &   ---            \\
 3820 He{\sc i}              &   0.249   &   0.6 &    66$\pm$15  & 23.0 &    ---         &   ---            \\
 3835 H9                     &   0.246   &   6.3 &   537$\pm$59  & 11.0 &    710$\pm$90  &   ---            \\
 3868 [Ne{\sc iii}]          &   0.238   &  32.5 &  3501$\pm$109 &  3.1 &   4120$\pm$180 &  3980$\pm$380    \\
\end{longtable}

 \begin{longtable}{lccccc}
 \caption[Relative reddening-corrected lines for \tresc]{Relative reddening
 corrected line intensities [$F(H\beta)$=$I(H\beta)$=10000] for \tresc.
\label{ratiostot 3}} \\
 \hline
  &  & \multicolumn{4}{c}{SDSS J1528} \\
 \multicolumn{1}{c}{$\lambda$ ({\AA})} & f($\lambda$) & -EW  & $I(\lambda)$ &  Error &  $I(\lambda)^\ddag$  \\
  & & (\AA) & & (\%) & \\
 \hline
 \endfirsthead
 
 \multicolumn{ 6}{l}{\small Table~\ref{ratiostot 3} continued}
  \\
 \hline
  &  & \multicolumn{4}{c}{SDSS J1528} \\
 \multicolumn{1}{c}{$\lambda$ ({\AA})} & f($\lambda$) & -EW  & $I(\lambda)$ &  Error &  $I(\lambda)^\ddag$  \\
  & & (\AA) & & (\%) & \\
 \hline
 \endhead
 
 \hline\multicolumn{ 6}{r}{\small\sl continued on next page}
 \endfoot
 
 \hline
 \noalign {\centering$^a$\,possibly blend with an unknown line; $^b$\,[O{\sc
 ii}]\,$\lambda\lambda$\,3726\,+\,3729; $^c$\,[Fe{\sc
 iii}]\,$\lambda\lambda$\,4986\,+\,4987; $^d$\,[O{\sc
 ii}]\,$\lambda\lambda$\,7318\,+\,7320; $^e$\,[O{\sc
 ii}]\,$\lambda\lambda$\,7330\,+\,7331.} 
 \noalign {\centering$^\ddag$ from Peimbert and
 Torres-Peimbert \cite*{1992A&A...253..349P}.}
 \endlastfoot

 3727 [O{\sc ii}]$^b$        &   0.271   &  177.4 & 22882$\pm$294 &  1.3  & 22390$\pm$2160  \\
 3734 H13                    &   0.270   &    2.8 &   295$\pm$56  & 19.1  &   ---  \\
 3750 H12                    &   0.266   &    2.9 &   315$\pm$64  & 20.3  &   ---  \\
 3770 H11                    &   0.261   &    5.0 &   510$\pm$53  & 10.4  &   ---  \\
 3798 H10                    &   0.254   &    7.6 &   709$\pm$70  &  9.9  &   ---  \\
 3820 He{\sc i}              &   0.249   &   ---  &     ---       &  ---  &   ---  \\
 3835 H9                     &   0.246   &    7.6 &   820$\pm$58  &  7.1  &   ---  \\
 3868 [Ne{\sc iii}]          &   0.238   &   39.0 &  4717$\pm$180 &  3.8  & 5370$\pm$520  \\
\end{longtable}

 \begin{longtable}{lcccccc}
 \caption[Relative reddening-corrected lines for \cuatroc]{Relative reddening
 corrected line intensities [$F(H\beta)$=$I(H\beta)$=10000] for \cuatroc.
\label{ratiostot 4}} \\
 \hline
  &  & \multicolumn{5}{c}{SDSS J1540} \\
 \multicolumn{1}{c}{$\lambda$ ({\AA})} & f($\lambda$) & -EW  & $I(\lambda)$ & Error & $I(\lambda)^\dag$ &  $I(\lambda)^\S$  \\
  & & (\AA) & & (\%) &  \\
 \hline
 \endfirsthead
 
 \multicolumn{ 7}{l}{\small Table~\ref{ratiostot 4} continued}
  \\
 \hline
  &  & \multicolumn{5}{c}{SDSS J1540} \\
 \multicolumn{1}{c}{$\lambda$ ({\AA})} & f($\lambda$) & -EW  & $I(\lambda)$ & Error & $I(\lambda)^\dag$ &  $I(\lambda)^\S$  \\
  & & (\AA) & & (\%) &  \\
 \hline
 \endhead
 
 \hline\multicolumn{ 7}{r}{\small\sl continued on next page}
 \endfoot
 
 \hline
 \noalign {\centering$^a$\,possibly blend with an unknown line; $^b$\,[O{\sc
 ii}]\,$\lambda\lambda$\,3726\,+\,3729; $^c$\,[Fe{\sc
 iii}]\,$\lambda\lambda$\,4986\,+\,4987; $^d$\,[O{\sc
 ii}]\,$\lambda\lambda$\,7318\,+\,7320; $^e$\,[O{\sc
 ii}]\,$\lambda\lambda$\,7330\,+\,7331. }
 \noalign {\centering $^\dag$\,from Izotov et al.\ \cite*{2006A&A...448..955I}; they
 gave $^\gimel$\,[S{\sc ii}]\,$\lambda\lambda$\,6717\,+\,6731 and
 $^\beth$\,[O{\sc ii}]\,$\lambda\lambda$\,7319\,+\,7330.} 
 \noalign {\centering $^\S$\,from Kniazev et
 al.\ \cite*{2004ApJS..153..429K}; they gave $^\varkappa$\,[O{\sc
 ii}]\,$\lambda\lambda$\,7319\,+\,7330.} 
 \endlastfoot

 3727 [O{\sc ii}]$^b$       &   0.271   &  232.8 &  21793$\pm$256 &  1.2 &   ---         &   ---           \\
 3734 H13                   &   0.270   &    2.6 &    272$\pm$48  & 17.7 &   ---         &   ---           \\
 3750 H12                   &   0.266   &   ---  &       ---      &  --- &   ---         &   ---           \\
 3770 H11                   &   0.261   &    2.7 &    314$\pm$61  & 19.4 &   ---         &   ---           \\
 3798 H10                   &   0.254   &    4.8 &    581$\pm$93  & 16.0 &   760$\pm$90  &   810$\pm$280   \\
 3820 He{\sc i}             &   0.249   &    1.5 &    166$\pm$23  & 14.1 &   ---         &   ---           \\
 3835 H9                    &   0.246   &    6.2 &    678$\pm$85  & 12.5 &   920$\pm$90  &   940$\pm$260   \\
 3868 [Ne{\sc iii}]         &   0.238   &   15.2 &   2142$\pm$82  &  3.8 &  1860$\pm$90  &   ---           \\
\end{longtable}

 \begin{longtable}{lccccc}
 \caption[Relative reddening-corrected lines for \cincoc]{Relative reddening
 corrected line intensities [$F(H\beta)$=$I(H\beta)$=10000] for \cincoc.
\label{ratiostot 5}} \\
 \hline
  &  & \multicolumn{4}{c}{SDSS J1616} \\
 \multicolumn{1}{c}{$\lambda$ ({\AA})} & f($\lambda$) & -EW  & $I(\lambda)$ & Error  & $I(\lambda)^\dag$  \\
  & & (\AA) & & (\%) &  \\
 \hline
 \endfirsthead
 
 \multicolumn{ 6}{l}{\small Table~\ref{ratiostot 5} continued}
  \\
 \hline
  &  & \multicolumn{4}{c}{SDSS J1616} \\
 \multicolumn{1}{c}{$\lambda$ ({\AA})} & f($\lambda$) & -EW  & $I(\lambda)$ & Error  & $I(\lambda)^\dag$  \\
  & & (\AA) & & (\%) &  \\
 \hline
 \endhead
 
 \hline\multicolumn{ 6}{r}{\small\sl continued on next page}
 \endfoot
 
 \hline
 \noalign {\centering$^a$\,possibly blend with an unknown line; $^b$\,[O{\sc
 ii}]\,$\lambda\lambda$\,3726\,+\,3729; $^c$\,[Fe{\sc
 iii}]\,$\lambda\lambda$\,4986\,+\,4987; $^d$\,[O{\sc
 ii}]\,$\lambda\lambda$\,7318\,+\,7320; $^e$\,[O{\sc
 ii}]\,$\lambda\lambda$\,7330\,+\,7331.}  
 \noalign {\centering$^\dag$ from Izotov et al.\ \cite*{2006A&A...448..955I}; they gave
 $^\gimel$\,[S{\sc ii}]\,$\lambda\lambda$\,6717\,+\,6731 and $^\beth$\,[O{\sc
 ii}]\,$\lambda\lambda$\,7319\,+\,7330.}
 \endlastfoot

 3727 [O{\sc ii}]$^b$       &   0.271   &   33.5 &  8491$\pm$196 &  2.3 &   ---          \\
 3734 H13                   &   0.270   &   ---  &       ---     &  --- &   ---          \\
 3750 H12                   &   0.266   &    1.8 &   377$\pm$101 & 26.7 &   ---          \\
 3770 H11                   &   0.261   &    1.5 &   357$\pm$113 & 31.7 &   ---          \\
 3798 H10                   &   0.254   &    2.1 &   447$\pm$92  & 20.6 &   ---          \\
 3820 He{\sc i}             &   0.249   &   ---  &       ---     &  --- &   ---          \\
 3835 H9                    &   0.246   &    3.1 &   638$\pm$90  & 14.1 &   790$\pm$110  \\
 3868 [Ne{\sc iii}]         &   0.238   &   17.2 &  4105$\pm$166 &  4.0 &  5450$\pm$230  \\
\end{longtable}

 \begin{longtable}{lcccc}
 \caption[Relative reddening-corrected lines for \seisc]{Relative reddening
 corrected line intensities [$F(H\beta)$=$I(H\beta)$=10000] for \seisc.
\label{ratiostot 6}} \\
 \hline
  &  & \multicolumn{1}{c}{SDSS J1657} \\
 \multicolumn{1}{c}{$\lambda$ ({\AA})} & f($\lambda$) & -EW  & $I(\lambda)$ & Error \\
  & & (\AA) & & (\%) \\
 \hline
 \endfirsthead
 
 \multicolumn{ 5}{l}{\small Table~\ref{ratiostot 6} continued}
  \\
 \hline
  &  & \multicolumn{2}{c}{SDSS J1657} \\
 \multicolumn{1}{c}{$\lambda$ ({\AA})} & f($\lambda$) & -EW  & $I(\lambda)$ & Error \\
  & & (\AA) & & (\%) \\
 \hline
 \endhead
 
 \hline\multicolumn{ 5}{r}{\small\sl continued on next page}
 \endfoot
 
 \hline
 \noalign {\centering$^a$\,possibly blend with an unknown line; $^b$\,[O{\sc
 ii}]\,$\lambda\lambda$\,3726\,+\,3729; $^c$\,[Fe{\sc
 iii}]\,$\lambda\lambda$\,4986\,+\,4987; $^d$\,[O{\sc
 ii}]\,$\lambda\lambda$\,7318\,+\,7320; $^e$\,[O{\sc
 ii}]\,$\lambda\lambda$\,7330\,+\,7331.} 
 \endlastfoot

 3727 [O{\sc ii}]$^b$       &   0.271   &  120.1 & 18832$\pm$230  &  1.2 \\
 3734 H13                   &   0.270   &   ---  &       ---      &  --- \\
 3750 H12                   &   0.266   &    1.9 &   232$\pm$47   & 20.2 \\
 3770 H11                   &   0.261   &    2.3 &   293$\pm$40   & 13.8 \\
 3798 H10                   &   0.254   &    4.1 &   500$\pm$68   & 13.5 \\
 3820 He{\sc i}             &   0.249   &   ---  &       ---      &  --- \\
 3835 H9                    &   0.246   &    7.1 &   780$\pm$93   & 11.9 \\
 3868 [Ne{\sc iii}]         &   0.238   &   23.1 &  3262$\pm$132  &  4.0 \\
\end{longtable}

 \begin{longtable}{lcccccc}
 \caption[Relative reddening-corrected lines for \sietec]{Relative reddening
 corrected line intensities [$F(H\beta)$=$I(H\beta)$=10000] for \sietec.
\label{ratiostot 7}} \\
 \hline
  &  & \multicolumn{5}{c}{SDSS J1729} \\
 \multicolumn{1}{c}{$\lambda$ ({\AA})} & f($\lambda$) & -EW  & $I(\lambda)$ & Error & $I(\lambda)^\dag$ &  $I(\lambda)^\S$ \\
  & & (\AA) &  & (\%) \\
 \hline
 \endfirsthead
 
 \multicolumn{7}{l}{\small Table~\ref{ratiostot 7} continued}
  \\
 \hline
  &  & \multicolumn{5}{c}{SDSS J1729} \\
 \multicolumn{1}{c}{$\lambda$ ({\AA})} & f($\lambda$) & -EW  & $I(\lambda)$ & Error & $I(\lambda)^\dag$ &  $I(\lambda)^\S$ \\
  & & (\AA) &  & (\%) \\
 \hline
 \endhead
 
 \hline\multicolumn{7}{r}{\small\sl continued on next page}
 \endfoot
 
 \hline
 \noalign {\centering$^a$\,possibly blend with an unknown line; $^b$\,[O{\sc
 ii}]\,$\lambda\lambda$\,3726\,+\,3729; $^c$\,[Fe{\sc
 iii}]\,$\lambda\lambda$\,4986\,+\,4987; $^d$\,[O{\sc
 ii}]\,$\lambda\lambda$\,7318\,+\,7320; $^e$\,[O{\sc
 ii}]\,$\lambda\lambda$\,7330\,+\,7331.} 
 \noalign {\centering $^\dag$ from Izotov et al. \cite*{2006A&A...448..955I};
 they gave $^\gimel$\,[S{\sc ii}]\,$\lambda\lambda$\,6717\,+\,6731 and
 $^\beth$\,[O{\sc ii}]\,$\lambda\lambda$\,7319\,+\,7330.}
 \noalign {\centering $^\S$ from Kniazev et al.\
 \cite*{2004ApJS..153..429K}; they gave $^\varkappa$\,[O{\sc
 ii}]\,$\lambda\lambda$\,7319\,+\,7330.} 
 \endlastfoot

 3727 [O{\sc ii}]$^b$       &   0.271   &  135.8 & 17622$\pm$243 &  1.4 &   ---         & --- \\
 3734 H13                   &   0.270   &    1.4 &   175$\pm$54  & 31.0 &   ---         & --- \\
 3750 H12                   &   0.266   &    4.2 &   397$\pm$89  & 22.4 &   370$\pm$70  & --- \\
 3770 H11                   &   0.261   &    7.6 &   713$\pm$88  & 12.4 &   480$\pm$70  & --- \\
 3798 H10                   &   0.254   &    5.2 &   598$\pm$71  & 11.8 &   620$\pm$60  &   370$\pm$200 \\
 3820 He{\sc i}             &   0.249   &   ---  &       ---     &  --- &   ---         & --- \\
 3835 H9                    &   0.246   &    7.6 &   721$\pm$64  &  8.8 &   900$\pm$60  &   640$\pm$140 \\
 3868 [Ne{\sc iii}]         &   0.238   &   34.9 &  4787$\pm$172 &  3.6 &  3730$\pm$140 & --- \\
\end{longtable}

}

\addcontentsline{toc}{section}{\numberline{}Bibliography}

\bibliographystyle{astron}
\bibliography{tesis}

\chapter{Neon and Argon optical emission lines in ionized gaseous nebulae}
\label{neon}

\section{Introduction}

The chemical history of the Universe can be investigated by studying the
behaviour of abundance ratios of different chemical species as a function of
metallicity, which is the main indicator of the chemical evolution of a
galaxy.  If two elements are produced by stars of the same mass range, they
will appear simultaneously in the interstellar medium (ISM) and hence their
relative abundance will be constant. But if they are produced by stars of
different mass ranges, they will be ejected into the ISM in different time
scales. The chemical abundances of the elements heavier than hydrogen can be
studied by measuring the fluxes of the absorption and emission lines in the
spectra of stars and galaxies. Unluckily, only the collisional emission lines
emitted by the ionized gas surrounding massive star clusters are detectable in
most of galaxies. Since the brightest  emission lines in the optical part of
the spectrum are emitted by oxygen, this element has been taken as the main
tracer of metallicity for these objects. Nevertheless, the depletion of some
of the most important elements, including oxygen, onto dust grains, whose
composition is difficult to ascertain, makes the determination of these
abundances more uncertain. This is not the case for the elements that occupy
the last group in the periodic table, that have an electronic configuration
with the outer shell completely filled and they seldom associate to other
elements and do not constitute part of the dust grains in the ISM. Therefore,
although the presence of these grains can affect the determination of
metallicity in other ways, affecting the photo-ionization equilibrium of the
gas, the uncertainty due to depletion factors has not to be considered in the
determination of the chemical abundances of the noble gases.

Neon and argon are products of the late stages in the evolution of massive
stars. Neon is produced by carbon burning and is expected to track oxygen
abundances very closely. The measurement of neon abundances in extragalactic
\HII\ regions \cite{2002ApJ...581.1019G} and planetary nebulae
\cite{1989MNRAS.241..453H} confirms this trend, despite the uncertainties in
the derivation of the ionization correction factor of Ne. On the other hand,
argon, like sulphur, is produced by oxygen burning and, again, it is expected
to track O abundances. Nevertheless, as it is also the case for sulphur, there
is some evidence of decreasing values of Ar/O for higher metallicities
\cite{2002ApJ...581.1019G}. The same trend has been observed for sulphur in
halo metal poor stars \cite{2001ApJ...557L..43I} and extragalactic \HII\
regions
\cite{1991MNRAS.253..245D,2002ApJ...581.1019G,2006A&A...449..193P}. This
problem, perhaps, is related to the proximity of the production site of these
elements to the stellar core and the yields would be sensitive to the
conditions during the supernova explosion \cite{1993PhR...227...65W}. 

The emission lines of Ne and Ar are not as intense as some of the other strong
lines in the optical spectrum but there is a growing number of \HII\ regions
for which there are measurements with good signal-to-noise ratio. This is the
case of [Ne{\sc iii}] at 3869\,\AA, whose blue wavelength makes it
observable in the optical spectrum of bright objects even at high redshifts,
and of the emission line of [Ar{\sc iii}] at 7136\,\AA. The emission lines of
these elements in the IR will supply a great deal of worthy additional
information but, at the moment, the objects with observations of the
fine-structure lines of [Ne {\sc ii}], [Ne {\sc iii}], [Ar {\sc ii}], [Ar {\sc
    iii}] are still very scarce. The derivation of the chemical abundances of
these elements in \HII\ regions can be calculated by means of the previous
determination of some electron collisional temperature and the previous
knowledge of the ionization structure of the element whose abundance  is
required. The results of photo-ionization models point to a similar ionization
structure of Ne and O on the one hand and S and Ar on the other ({\em e.g.}
P\'erez-Montero and D\'\i az, 2007). These facts make these lines suitable to
be used as substitutes of the bright emission lines of oxygen and sulphur in
empirical indicators of metallicity ({\em e.g.} Nagao et al., 2006) or
diagnostic ratios to distinguish starbursts galaxies from active galactic
nuclei ({\em e.g.} Rola et al., 1997). 

\nocite{2007MNRAS.377.1195P,2006A&A...459...85N,1997MNRAS.289..419R}

In this Chapter, we investigate the element abundances of neon, argon and
oxygen 
for a sample of objects that is described in Section  \S \ref{sampleE}. In
Section  \S \ref{physi}, we derive the physical conditions of these objects,
including the calculation of the electron temperatures involved in the
derivation of Ne and Ar abundances. In Section \S \ref{photomod}, we describe
the grid of photo-ionization models that we have used to derive a new set of
ionization correction factors for these two elements. These new ICFs are
described in the next Section, along with the discussion of the behaviour of
Ne/O and Ar/O ratios and the use of the brightest emission lines of these two
elements as empirical calibrators of metallicity and diagnostic
ratios. Finally, we summarize our results and we present our conclusions.

\section{Description of the sample}
\label{sampleE}

The sample includes emission-line objects observed in the optical part of the
spectrum with the detection of, at least, one auroral line with high S/N ratio
in order to derive electron temperatures directly and, hence, ionic chemical
abundances with less uncertainty. We have compiled for this sample the strong
lines of [O{\sc ii}] at 3727\,\AA\ and [O{\sc iii}] at 4959\,\AA\ and
5007\,\AA\ in order to derive oxygen abundances.  We have compiled as well the
[Ne{\sc iii}] emission line at 3869\,\AA, the [Ar{\sc iii}] emission line at
7136\,\AA\ and the [Ar{\sc iv}] at 4740\,\AA\ to measure the respective ionic
abundances of neon and argon. The compilation includes the objects used in
\cite{2005MNRAS.361.1063P}, including \HII\ regions in our Galaxy and
the Magellanic Clouds (MCs), Giant Extragalactic \HII\ regions (GEHRs) and
\HII\ galaxies, 
with the addition of new objects which are listed in Table \ref{refs}.

This list includes 12 \HII\ galaxies from the Sloan Digital Sky Survey
(SDSS\footnote{The SDSS site is at {\tt www.sdss.org}}) with very low
metallicities  identified by Kniazev et al.\ \cite*{2003ApJ...593L..73K} and
whose spectra have been re-analyzed to obtain the required line emission
data. Among these galaxies, 11 do not have observations of the [O{\sc ii}]
emission line at 3727\,\AA, which is the case as well for 183 SDSS galaxies
of the sample compiled by Izotov et al.\ \cite*{2006A&A...448..955I}. The
compilation has therefore 633 \HII\ galaxies, 176 GEHRs and 44 \HII\ regions
of the Galaxy and the Magellanic Clouds with a good determination of oxygen,
neon and argon ionic abundances for those ions that have been observed.


\begin{table}
\caption{Bibliographic references for the emission line fluxes of the compiled sample}
\begin{center}
\begin{tabular}{lcc}
\hline
\hline
Reference & Object type$^a$  & Objects \\
\hline
Bresolin et al., 2004 & M51 GEHRs & 10 \\
Bresolin et al., 2005 & GEHRs & 32 \\
Bresolin, 2007 & M101 GEHRs & 3 \\
Crockett et al., 2006 & M33 GEHRs & 6 \\
Garnett et al., 2004 & M51 GEHRs & 2 \\
Guseva et al., 2003a & SBS 1129+576 & 2 \\
Guseva et al., 2003b & HS 1442+650 & 2 \\
Guseva et al., 2003c & SBS 1415+437 & 2 \\
H\"agele et al., 2006 (Chapter \S \ref{HIIgal-obs}) & \HII G & 3 \\
H\"agele et al., 2008 (Chapter \S \ref{HIIgal-obs}) & \HII G & 7 \\
Izotov et al., 1997 & IZw18  & 2 \\
Izotov and Thuan, 1998 & SBS 0335-018 & 1 \\
Izotov et al., 1999 & \HII G & 3 \\
Izotov et al., 2001 & \HII G & 2 \\
Izotov et al., 2004a & \HII G & 3 \\
Izotov and Thuan, 2004 & \HII G & 33 \\
Izotov et al., 2006 & SDSS galaxies & 309 \\
Kniazev et al., 2004 & SDSS galaxies & 11 \\
Lee et al., 2004 & KISS galaxies & 13 \\
Melbourne et al., 2004 & KISS galaxies & 12 \\
P\'erez-Montero and D\'\i az, 2005 & All & 361 \\
Van Zee, 2000 & UGCA92 & 2 \\
Vermeij et al., 2002 & DRH & 5 \\

\hline
\multicolumn{3}{l}{$^a$GEHR denotes Giant Extragalactic \HII\ Regions, 
\HII G, \HII\ Galaxies}\\
\multicolumn{3}{l}{and DRH, Diffuse \HII\ Regions}
\end{tabular}
\end{center}
\label{refs}
\end{table}


\nocite{2007ApJ...656..186B,2005A&A...441..981B,2004ApJ...615..228B,2006ApJ...637..741C,2004ApJ...607L..21G,2003A&A...407..105G,2003A&A...407...91G,2003A&A...407...75G,2006MNRAS.372..293H,2008MNRAS.383..209H,1997ApJ...476..698I,1998ApJ...497..227I,1999ApJ...527..757I,2001ApJ...562..727I,2004A&A...421..539I,2004ApJ...602..200I,2006A&A...448..955I,2004ApJS..153..429K,2004ApJ...616..752L,2004AJ....127..686M,2005MNRAS.361.1063P,2000ApJ...543L..31V,2002A&A...391.1081V,2004A&A...415...87I}

The emission line intensities have been taken directly from the literature
once reddening corrected. The aperture effects of all the compiled
observations are negligible due to the compact nature of the observed objects
and the high excitation state of the more important involved emission lines.

\section{Physical conditions}
\label{physi}

\subsection{Electron density and temperatures}

In order to derive oxygen, neon and argon abundances for the ions presenting
the corresponding strong emission line we have selected objects allowing a
direct determination of the electron temperature. All physical conditions have
been derived from the appropriate ratios of emission lines and using the
fittings to the values obtained from the task TEMDEN, in the case of electron
densities and temperatures, and IONIC, in the case of ionic abundances, as
was described in detail in Chapter \S \ref{HIIgal-obs}.

Next, following the same procedure described in P\'erez-Montero and D\'\i az
\cite*{2003MNRAS.346..105P}, we have used for the calculation of the chemical
abundance the electron temperature associated to the zone where each ionic
species stays. To do this we have applied different expressions for the
relations between the electron temperature of each zone (see Chapter \S
\ref{HIIgal-obs}).  

Electron densities have been calculated for a subsample of 361 objects from
the ratio of [S{\sc ii}] emission lines I(6717\,\AA)/I(6731\,\AA). It is
representative of the low excitation zone of the ionized gas, and it is
therefore used for the calculation of electron temperatures in this zone which
depend on density. For the remaining objects, for which no data about
[S{\sc ii}] lines exist or for which non valid values are obtained, we have
adopted a mean density of 50 particles per cm$^3$, according to the low
density values found in the rest of \HII\ regions.

The electron temperature of O$^{2+}$, representative of the high excitation
zone of the ionized gas, has been derived from the ratio of [O{\sc iii}]
emission lines (I(4959\,\AA)+I(5007\,\AA))/I(4363\,\AA) for a subsample of 771
objects, while for the rest we have used the relation with other measured
electron temperatures in an inverse way as described below, 33 objects from
T$_e$([S{\sc iii}]), 20 from T$_e$([O{\sc ii}]) and 4 from
T$_e$([N{\sc ii}]). In the latter case, the objects are M51 GEHRs of high
metallicity from Bresolin et al.\ \cite*{2004ApJ...615..228B}. We have
assumed as well the T$_e$([O{\sc iii}]) to be equal to T$_e$([Ne{\sc iii}])
and T$_e$([Ar{\sc iv}]) in order to calculate Ne$^{2+}$ and Ar$^{3+}$
abundances, respectively.

The electron temperature of O$^+$, representative of the low excitation zone
of the ionized gas, has been derived from the ratio of the [O{\sc ii}]
emission lines  (I(3727\,\AA)/I(7319\,\AA)+I(7330\,\AA)) for a subsample of
311 objects. The auroral lines of [O{\sc ii}] present higher uncertainties
because of their lower signal-to-noise ratio and a higher dependence on
reddening correction, due to their larger wavelength distance to the closest
hydrogen recombination emission line. Besides, they present a small 
contribution due to recombination, although in quantities not larger than the
usual reported errors. Additionally, this ratio has a strong dependence on
electron density (see Chapter \S \ref{HIIgal-obs}), which makes the
determination of this temperature and, 
hence, the determination of O$^+$ abundances very uncertain. This can increase
the uncertainty of total oxygen abundances, mostly in high metallicity - low
excitation \HII\ regions \cite{2005MNRAS.361.1063P}. For the rest of the
objects, we have used the grid of relations between T$_e$([O{\sc ii}]) and
T$_e$([O{\sc iii}]) presented in P\'erez-Montero and D\'\i az
\cite*{2003MNRAS.346..105P}, which takes into account the dependence of
T$_e$([O{\sc ii}]) on density and its corresponding uncertainty. For the same
4 objects of M51 for which neither T$_e$([O{\sc iii}]) nor T$_e$([S{\sc iii}]) have
been measured we have taken T$_e$([N{\sc ii}]) as representative of the
electron temperature of the low excitation zone using the ratio of [N{\sc ii}]
emission lines (I(6548\,\AA)+I(6584\,\AA))/I(5755\,\AA).  

In the case of the electron temperature of Ar$^{2+}$, we have assumed that the
ionization structure of this ion is  rather similar to that of S$^{2+}$ (see
Chapter \S \ref{HIIgal-obs}), whose 
electron temperature has intermediate values between the high and low
excitation zones \cite{1992AJ....103.1330G}. We have calculated directly the
temperature of S$^{2+}$ from the ratio of [S{\sc iii}] emission lines
(I(9069\,\AA)+I(9532\,\AA))/I(6312\,\AA) for a subsample of 299 objects. For
the rest of the objects we have taken relations of this temperature with
t$_e$([O{\sc iii}]). In the case of low excitation objects
(T$_e$\,$<$\,10000\,K), based on photo-ionization models
\cite{2005MNRAS.361.1063P}:
\[
t_e([S {\textrm {\sc iii}})]\,=\,1.05\,t_e[(O {\textrm {\sc iii}})]\,-\,0.08
\]
\noindent while for the high excitation objects we have taken the empirical
relation derived in Chapter \S \ref{HIIgal-obs} that accounts
better for the high dispersion found in this relation for \HII\ galaxies: 
\[
t_e([S{\textrm {\sc iii}}]\,=\,(1.19\pm0.08)\,t_e([O{\textrm {\sc iii}}])\,-\,(0.32\pm0.10)
\]

\subsection{Ionic abundances}

Each ionic abundance has been calculated using the most prominent emission
lines for each ion and the appropriate electron temperature. In the case of
O$^+$, we have used the expression presented in Chapter \S \ref{HIIgal-obs}
using the intensity of the [O{\sc ii}] line at 3727\,\AA, the electron
temperature t$_e$([O{\sc ii}]), or any other value associated to the low
excitation zone, and the electron density, N$_e$.

In the case of 11 of the 12 very low metallicity \HII\ galaxies identified by
Kniazev et al.\ \cite*{2003ApJ...593L..73K} and also for a subsample of 183
\HII\ galaxies compiled from Izotov et al.\ \cite*{2006A&A...448..955I}, which
have no data on the [O{\sc ii}]\,$\lambda$\,3727\,\AA\ line, the O$^+$
abundances have been measured from the emission lines of [O{\sc ii}] at 7319
and 7330\,\AA\ using the following expression also derived from a fitting to
the results of the IONIC task:
\begin{eqnarray*}
12\,+\,log(O^+/H^+)\,=\,log\frac{I(7319+7330)}{I(H\beta)}\,+\,6.895\,+\\
+\,\frac{2.44}{t_e}\,
-\,0.58\,log\,t_e\,+\,log(1+4.7\,n_e)\\
\end{eqnarray*}
where n$_e$\,=\,10$^{-4}$\,N$_e$.

For O$^{2+}$ abundances we use the expression for the [O{\sc iii}] emission
lines at 4959 and 5007\,\AA, along with the electron temperature of [O{\sc
    iii}] (see Chapter \S \ref{HIIgal-obs}). Then, we calculate the total
abundance of oxygen in relation to hydrogen directly from these two ionic
abundances. 

Abundances of Ne$^{2+}$ have been calculated from the [Ne{\sc iii}] emission
line at 3869\,\AA\ and the electron temperature of [O{\sc iii}] for a
subsample of 773 objects using the same relation given in Chapter \S
\ref{HIIgal-obs}. 

Finally, regarding Ar, we can measure Ar$^{2+}$, from the [Ar{\sc iii}]
emission line at 7136\,\AA\ and the electron temperature of [S{\sc iii}] for a
subsample of 572 objects using the expression previously given. It is possible
to measure as well the lines of [Ar{\sc iv}] at 4713 and
4740\,\AA. Nevertheless, the first one usually appears blended with a line of
He{\sc i} at 4711\,\AA\ that is difficult to correct, so it is better to use
the second one to calculate the ionic abundance of Ar$^{3+}$ through the
electron temperature of [O{\sc iii}]. This abundance has been calculated for a
subsample of 253 objects with a simultaneous measurement of the [Ar{\sc iii}]
and [Ar{\sc iv}] emission lines.

\section{Photo-ionization models}
\label{photomod}

A large grid of photo-ionization models has been calculated in order to check
the validity of the ionization correction factors for Ne and Ar.

We have used the photo-ionization code Cloudy 06.02 \cite{1998PASP..110..761F},
taking as ionizing  source the spectral energy distributions of O and B stars
calculated with the code WM-Basic (version 2.11\footnote{Available at
  http://www.usm.uni-muenchen.de/people/adi/Programs/Programs.html}, Pauldrach
et al., 2001). We have assumed a spherical geometry, a constant density of
100 particles per cm$^{3}$ and a standard fraction of dust grains in the
interstellar medium. We have assumed as well that the gas has the same
metallicity as the ionizing source, covering the values 0.05Z$_\odot$,
0.2Z$_\odot$, 0.4Z$_\odot$, Z$_\odot$ and 2Z$_\odot$, taking as the solar
metallicity the oxygen abundance measured by Allende-Prieto et al.\ (2001;
12+log(O/H) = 8.69). The rest of ionic abundances have been set to their solar
proportions. A certain amount of depletion has been taken into account for the
elements C, O, Mg, Si and Fe, considered for the formation of dust grains in
the code. Although it is expected that the dust-to-gas ratio scales with
metallicity \cite{1995ApJ...454..807S}, we have checked that different values
of this ratio do not lead to variations in our results about the computed ICFs
in quantities larger than the reported errors. Regarding other functional
parameters we have considered different values of the ionization parameter
(log U\,=\,-3.5, -3.0, -2.5 and -2.0) and the stellar effective temperature
(T$_*$\,=\,35000\,K, 40000\,K, 45000\,K and 50000\,K). This gives a total of
80 photo-ionization models to cover the conditions of different ionized gas
nebulae. Atomic data, including collision strengths in the models are
consistent with those used in the calculation of chemical abundances. We have
checked as well the influence of varying dielectronic recombination rate
coefficients in the code according to the most recent values (Badnell, 2006
and references therein), but we have not found any relevant variation in our
results concerning the calculation of ICFs. 
 
\nocite{2006A&A...447..389B}
\nocite{2001A&A...375..161P,2001ApJ...556L..63A}

\section{Discussion}

\subsection{Ionization correction factors (ICF)}

ICFs stand for the unseen ionization stages of each element.

\[\frac{X}{H} = ICF(X^{+i}) \frac{X^{+i}}{H^+}\]
where $X$ is the element whose ICF is required and $X^{+i}$ is the ionic
species whose abundance is calculated by means of the detected emission
lines. 

In the case of neon, only the [Ne{\sc iii}] line at 3869\,\AA\ is detected in
the optical spectrum in a large sample of objects. Since its ionization
structure is quite similar to that of oxygen, the following approximation can
be used: 
\[\frac{Ne^{2+}}{Ne} \approx \frac{O^{2+}}{O}\]
that leads to:
\[ICF(Ne^{2+}) = O/O^{2+} \approx (O^++O^{2+})/O^{2+}\]

\begin{figure}
\centering
\includegraphics[width=.75\textwidth,angle=0,clip=]{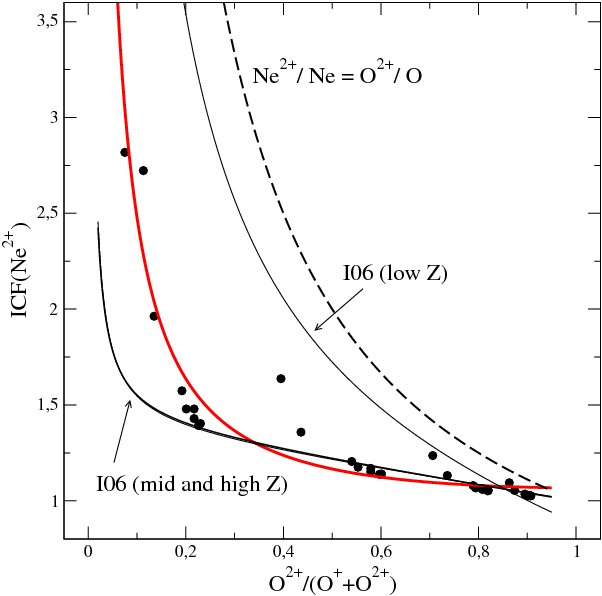}
\caption[Different relations between the ICF(Ne$^{2+}$) and the ratio of
  O$^{2+}$/O]{Different relations
  between the ICF(Ne$^{2+}$) and the ratio of O$^{2+}$/O, approximated by
  O$^{2+}$/(O$^{+}$+O$^{2+}$). The points correspond to the models described
  in the text for stellar atmospheres of 45\,kK and 50\,kK. The thick solid
  red line represents the best fit to these points. The thin and thick black
  solid lines 
  correspond to different fits of photo-ionization models as a function of
  metallicity, {\it low} and {\it mid and high} metallicity respectively, in
  Izotov et 
  al.\ (2006). Finally the thick dashed line represents the classical
  approximation.} 
\label{icf_ne}
\end{figure}

This relation is shown as a thick dashed line in Figure
\ref{icf_ne}. Nevertheless, Izotov et al.\ \cite*{2004A&A...415...87I} point
out from a set of photo-ionization models that this approximation is less
accurate in regions with a lower ionization parameter where the charge
transfer reaction between  O$^{2+}$ and H$^0$ becomes more efficient. In this
type of objects it is expected to find larger abundances of Ne$^{2+}$ in
relation to O$^{2+}$ and, hence, the O$^{2+}$/O ratio provides larger ICFs.
This trend is confirmed with the set of photo-ionization models used by Izotov
et al.\ (2006) that, besides, find a slight dependence of the ICF(Ne$^{2+}$) on
metallicity, with larger values of this ICF for high metallicity regions. This
dependence is due to the addition of X-rays sources in their models with lower
metallicity and not to a real dependence on metallicity or ionization
parameter. 
The ICFs from Izotov et al.\ (2006) are shown as thin and thick black solid
lines in Figure \ref{icf_ne}, {\it low} and {\it mid and high} metallicity,
respectively. The fit of our models for the hottest stars (45000\,K and 
50000\,K), whose effective temperatures reproduce better the radiation field
in the most massive ionizing cluster is shown as a thick solid red line in
Figure 
\ref{icf_ne}. It reveals the same overestimate of the classical approximation
in low excitation objects. However, we do not find any noticeable dependence
on the metallicity. This fit gives:
\[ICF(Ne^{2+}) = 0.753 + 0.142\cdot x + 0.171/x \]
where x\,=\,O$^{2+}$\,/\,(O$^{+}$+O$^{2+}$), with a rms of 0.074 only.

\begin{figure}
\centering
\includegraphics[width=.62\textwidth,angle=0,clip=]{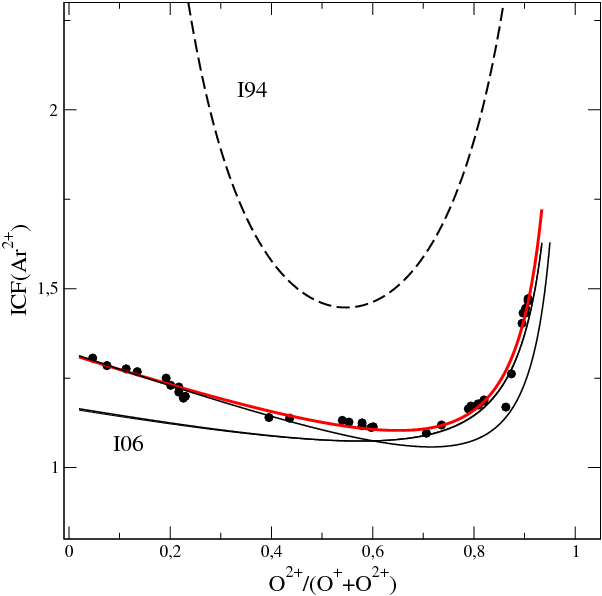}\\\vspace{0.5cm}
\includegraphics[width=.62\textwidth,angle=0,clip=]{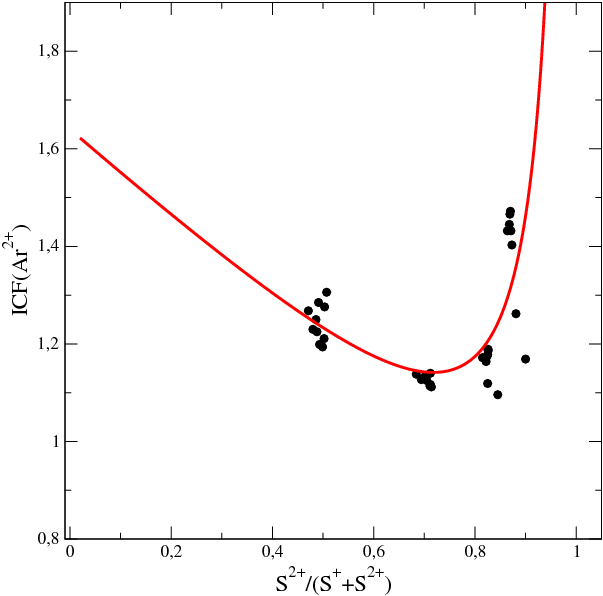}
\caption[Representations of the ICF for Ar$^{2+}$  as functions of 
  O$^{2+}$/(O$^+$+O$^{2+}$) and S$^{2+}$/(S$^+$+S$^{2+}$)]{Representation of
  the ICF for Ar$^{2+}$ as a function of the O$^{2+}$/(O$^+$+O$^{2+}$) ratio in
  the upper panel, and as a function of  S$^{2+}$/(S$^+$+S$^{2+}$) in the
  lower one. Dashed line represent the expression given by Izotov et
  al.\ (1994), the thin solid lines represent those in Izotov et
  al.\ (2006). Finally, the points represent the models described here for
  cluster effective temperatures of 45\,kK and 50\,kK. The thick red line is the
  best fit to these points.}
\label{icf_ar1}
\end{figure}


\nocite{1994ApJ...435..647I}

\begin{figure}
\centering
\includegraphics[width=.62\textwidth,angle=0,clip=]{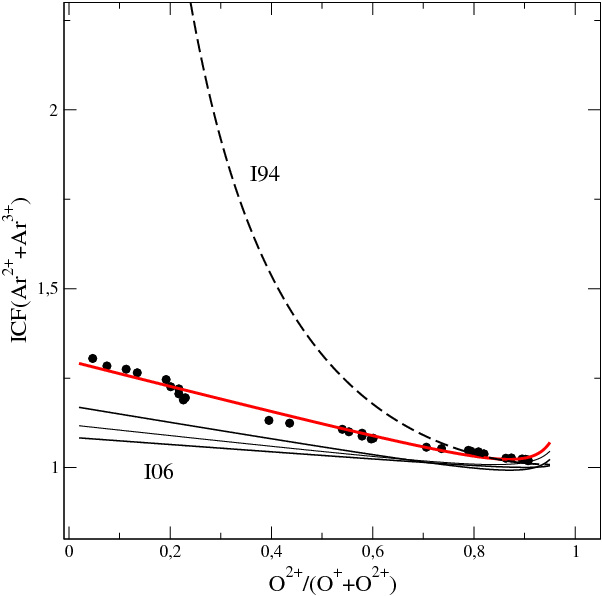}\\\vspace{0.5cm}
\includegraphics[width=.62\textwidth,angle=0,clip=]{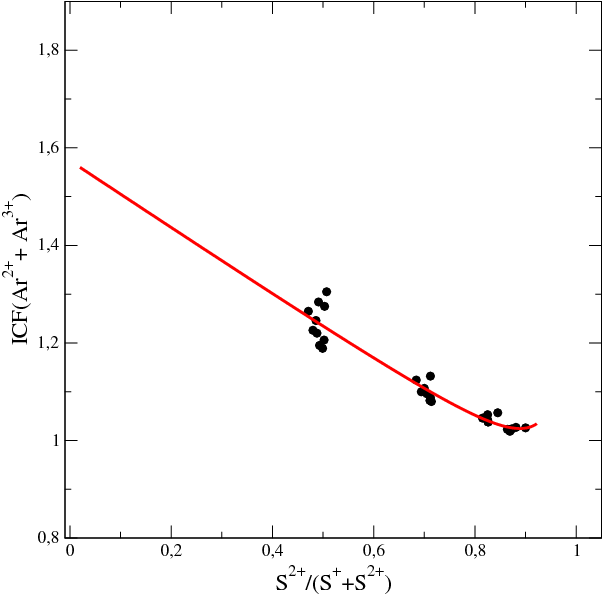}
\caption[Representations of the ICF for Ar$^{2+}$+ Ar$^{3+}$ as functions of 
  O$^{2+}$/(O$^+$+O$^{2+}$) and S$^{2+}$/(S$^+$+S$^{2+}$)]{Idem as Figure
  \ref{icf_ar1} for the ICF of Ar$^{2+}$+ Ar$^{3+}$.}
\label{icf_ar2}
\end{figure}


Regarding argon, the ICF can vary depending on the availability of Ar$^{3+}$
abundances. Izotov et al.\ (1994) propose the following formula to calculate
the total abundance of Ar using the ionic abundances of these two ions: 
\[
ICF(Ar^{2+}+Ar^{3+})\,=\,\Big[0.99+0.091\Big(\frac{O^+}{O}\Big)-1.14\Big(\frac{O^+}{O}\Big)^2+0.077\Big(\frac{O^+}{O}\Big)^3\Big]^{-1}
\]

On the other hand, if only the emission line of [Ar{\sc iii}] is available,
they propose the following expression:

\[
ICF(Ar^{2+})\,=\,\Big[0.15+2.39\Big(\frac{O^+}{O}\Big)-2.64\Big(\frac{O^+}{O}\Big)^2\Big]^{-1}
\]

Both formulae are shown as dashed lines in the upper panels of Figure
\ref{icf_ar1} and \ref{icf_ar2}, respectively. Nevertheless, these fits
clearly overestimate the amount of total Ar compared with the expressions
proposed by Izotov et al.\ (2006) who, as in the case of Ne, find a dependence
with metallicity. These fits are shown in the upper panels of Figure
\ref{icf_ar1} and \ref{icf_ar2} as thin solid lines for low, intermediate and
high metallicities. Finally, the fits to our own models (thick red line in the
Figures) are closer to these
latter, but we do not find any relevant dependence of our models on
metallicity. The expressions for the fits of these models are, as a function
of x\,=\,O$^{2+}$\,/\,(O$^{+}$+O$^{2+}$): 
\[
ICF(Ar^{2+}+Ar^{3+})\,=\,0.928 + 0.364\,(1-x) + 0.006/(1-x) 
\]
\noindent with a rms of 0.011, shown as a thick solid red line in the upper
panel of Figure \ref{icf_ar1}, and
\[
ICF(Ar^{2+})\,=\,0.596 + 0.967\,(1-x) + 0.077/(1-x) 
\]
\noindent with a rms of 0.067, shown as a thick solid red line in the upper
panel of Figure \ref{icf_ar2}. It is possible, as well, to express these ICFs 
as a function of sulphur abundances. This allows to derive total argon
abundances using red or near-IR observations only, in the case of ICF of
Ar$^{2+}$. Besides, these ICFs can be used in the subsample of SDSS objects
whose [O{\sc ii}] line at 3727\,\AA\ is not detected but which, on the
contrary, have good detection of the [S{\sc iii}] at 9069\,\AA. The fits of
our models, as a function of x\,=\,S$^{2+}$\,/\,(S$^{+}$+S$^{2+}$) are: 
\[
ICF(Ar^{2+}+Ar^{3+})\,=\,0.870 + 0.695\,(1-x) + 0.0086/(1-x) 
\]
\noindent with a rms of 0.019, which is shown as a thick solid red line in
the lower panel of Figure \ref{icf_ar1}, and 
\[
ICF(Ar^{2+})\,=\,0.596 + 0.967\,(1-x) + 0.077/(1-x) 
\]
\noindent with a rms of 0.068, shown as a thick solid red line in the lower
panel of Figure \ref{icf_ar2}.

\subsection{Behaviour of Ne/O and Ar/O with metallicity}

\begin{figure}
\centering
\vspace{3.5cm}
\includegraphics[width=.87\textwidth,angle=0,clip=]{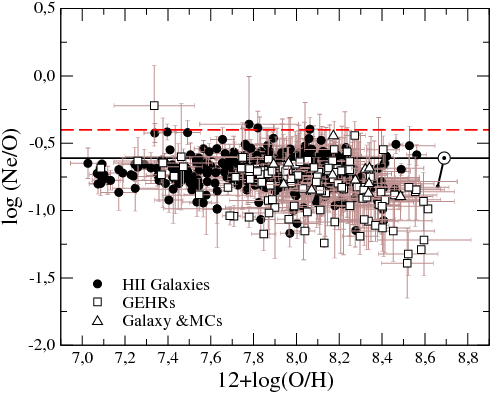}
\caption[Relation between the abundance ratio Ne/O and the metallicity for the
  sample of objects]{Relation between the abundance ratio Ne/O and the
  metallicity for the sample of objects described in the text. Solid circles
  represent \HII\ galaxies, open squares Giant Extragalactic \HII\ regions and
  open triangles \HII\ regions in the Galaxy and the Magellanic Clouds. We show
  as well the solar chemical abundances for O (Allende-Prieto et al.\ 2001) and
  for Ne (Grevesse and Sauval, 1998), linked by a solid line with the solar
  Ne/O ratio from Asplund et al.\ (2005). The dashed red line represents the
  ratio calculated by Drake and Testa (2005) for a set of stellar coronae.}
\label{o_neo}
\end{figure}
 
We have applied the ICF(Ne$^{2+}$) derived from our set of photo-ionization
models to a total sample of 578 \HII\ galaxies, 117 GEHRs and 12 \HII\ regions
in our Galaxy and the Magellanic Clouds. In Figure \ref{o_neo}, we represent
the ratio of Ne/O as a function of the total oxygen abundance, and we compare
the obtained values with the solar one, taking the solar oxygen abundance to
be 12+log(O/H) = 8.69 \cite{2001ApJ...556L..63A} and the solar neon abundance
to be 12+log(Ne/H) = 8.08 \cite{1998SSRv...85..161G}.  Although there is a
high dispersion, probably due to chemical inhomogeneities and different
observational conditions, the values agree quite well with the assumption of a
constant value of Ne/O, at least for low and intermediate metallicities. At
higher metallicities, there is a slight decrease of Ne/O with O/H, perhaps due
to the underestimate of the corresponding ICF in this regime, as pointed out
by Izotov et al.\ (2006). Nevertheless, since we have not used additional X-ray
sources in our models, this problem has to be further investigated. At this
point,  it is necessary to stress that all the results from photo-ionization
models are in contradiction with Vermeij and Van der Hulst
\cite*{2002A&A...391.1081V} who find Ne$^+$ abundances even larger that those
predicted by the classical approximation in a set of \HII\ regions in the
Galaxy and the Magellanic Clouds from direct ISO observations of the emission
lines of [Ne{\sc ii}] and [Ne{\sc iii}] in the mid-infrared. Regarding the
constant value of Ne/O, the average value of the sample, which is -0.72 $\pm$
0.13,  is lower, although within the error, than our assumed solar one, but
higher than the value reported by \citetex{2005ASPC..336...25A}, connected to
the value assumed here with a solid line in Figure \ref{o_neo} (log
(Ne/O)\,=\,-0.82 in the photosphere). Nevertheless, all the derived values are
far from the value found by \citetex{2005Natur.436..525D} using Chandra X-ray
spectra in the coronae of 21 stars, which is log(Ne/O)\,$\approx$\,-0.4. If
this value is correct, then the neon abundances as derived from optical
collisional lines are clearly underestimated.

\begin{figure}
\centering
\vspace{3.5cm}
\includegraphics[width=.87\textwidth,angle=0,clip=]{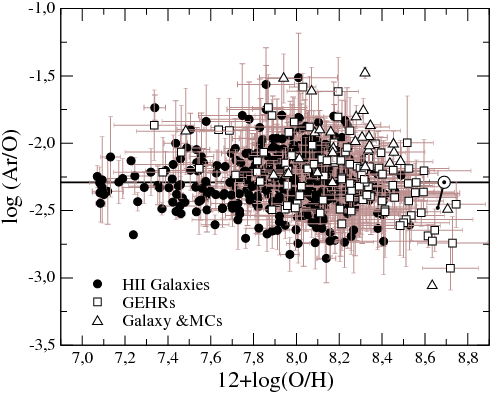}
\caption[Relation between the abundance ratio Ar/O and the metallicity for the
  sample of objects]{Relation between the abundance ratio  Ar/O and the
  metallicity for the sample of objects described in the text. We show as well
  the solar chemical abundances for O (Allende-Prieto et al., 2001) and for Ar
  (Grevesse and Sauval, 1998) linked by a solid line with the solar Ar/O ratio
  from Asplund et al.\ (2005).} 
\label{o_aro}
\end{figure}

\begin{figure}
\centering
\includegraphics[width=.48\textwidth,angle=0,clip=]{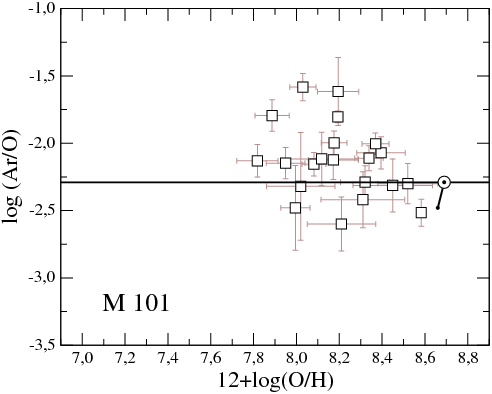}
\includegraphics[width=.48\textwidth,angle=0,clip=]{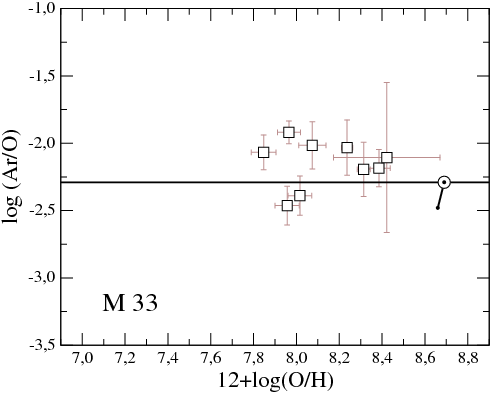}
\includegraphics[width=.48\textwidth,angle=0,clip=]{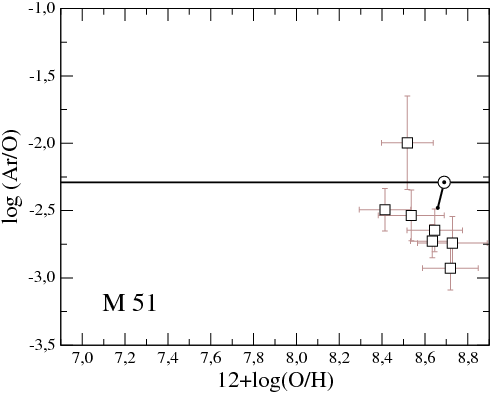}
\includegraphics[width=.48\textwidth,angle=0,clip=]{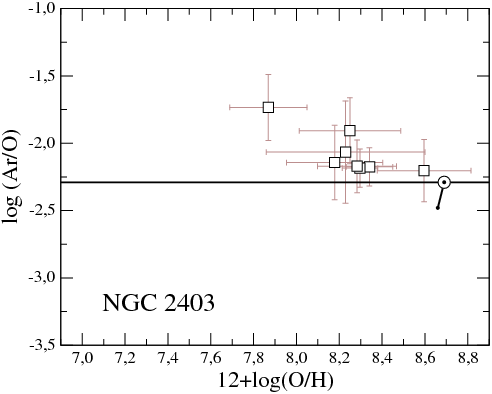}
\caption[Relation between the abundance ratio Ar/O and the metallicity for
  the discs of some spiral galaxies]{Relation between the abundance ratio Ar/O
  and the metallicity for the discs of some spiral galaxies. From left to
  right and up to bottom: M101, M51, M33 and NGC 2403.} 
\label{gradients_aro}
\end{figure}

Regarding the relation between the Ar/O ratio and oxygen abundance, our
results are shown in Figure \ref{o_aro}. In order to calculate total argon
chemical abundances we have used ICFs based on the O$^{2+}$/O ratio for
(Ar$^{2+}$\,+\,Ar$^{3+}$) (104 objects) and Ar$^{2+}$ (344 objects), and ICFs
based on the S$^{2+}$/(S$^{+}$+S$^{2+}$) ratio for the SDSS galaxies from
Izotov et al.(2006) which do not have observation of the [O{\sc
    ii}]\,3727\,\AA\ line but for which measurements of the [S{\sc iii}]
9069\,\AA\ line exist: 44 galaxies in the case of (Ar$^{2+}$\,+\,Ar$^{3+}$)
and 101 in the case of Ar$^{2+}$. The value of Ar/O presents a larger
dispersion than in the case of Ne/O, with an average value similar to the
solar value from Allende-Prieto et al. (2001) and Grevesse and Sauval (1998),
represented by the solar symbol in Figure \ref{o_aro} and slightly higher than
the value from Asplund et al. (2005), connected to it with a solid line. The
relation appears to be slightly different for \HII\ galaxies, whose behaviour
is quite similar to that in the Ne/O diagram, and for Giant Extragalactic
\HII\ regions, for which there exists a slight trend of decreasing Ar/O for
increasing metallicities. A more careful analysis for individual disc galaxies,
as can be seen in Figure \ref{gradients_aro}, shows this trend more clearly,
contrary to what is expected for the production of an alpha element like
Ar. This is compatible with the existence of larger Ar/O at lower effective
radius, since there is evidence of negative gradients of metallicity in all
these galaxies (M101: Kennicutt and Garnett, 1996; M51: D\'\i az et al., 1991;
M33 and NGC2403: Garnett et al., 1997). This result agrees quite well with
that obtained for sulphur, both for Extragalactic \HII\ regions (D\'\i az et
al., 1991; P\'erez-Montero et al., 2006) and halo metal-poor stars (Israelian
and Rebolo, 2001). 

\nocite{1997ApJ...489...63G,1996ApJ...456..504K}

\subsection{Empirical parameters based on Ne and Ar lines}

Although the [Ne{\sc iii}] emission line at 3869\,\AA\ and the [Ar{\sc iii}]
emission line at 7136\,\AA\ are fainter than the oxygen emission lines, which
are commonly used in different empirical calibrations of chemical abundances,
they can be useful to ascertain metallicities when these lines are not
available, either because they shift out of the optical region due to the
object redshift or because the instrumental configuration does not cover the
blue-green region of the spectrum. 

The empirical calibrators are commonly used in objects whose low
signal-to-noise and/or high metallicities do not allow the accurate
measurement of any of the auroral lines and, therefore, it is not possible to
derive the electron temperature and the ionic chemical abundances from the
strong collisional lines. 

\begin{figure}
\centering
\includegraphics[width=.87\textwidth,angle=0,clip=]{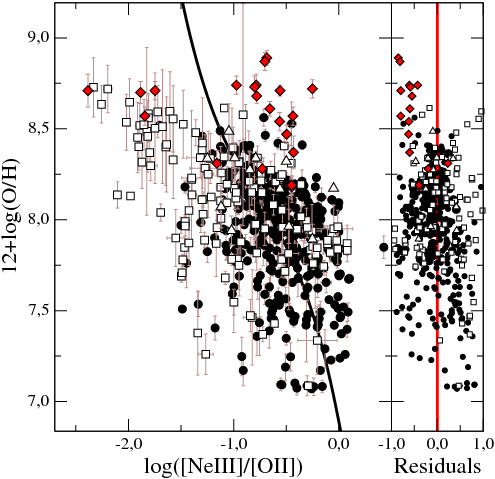}
\caption[Relation between the logarithm of the ratio of the [Ne{\sc
  iii}{\textrm ]}\,3869\,\AA\ and [O{\sc ii}{\textrm ]}\,3727\,\AA\ lines and
  the oxygen abundance]{Relation between the logarithm of the ratio of the
  [Ne{\sc iii}]\,3869\,\AA\ and [O{\sc ii}]\,3727\,\AA\ lines and the oxygen
  abundance for the sample described in Section \S \ref{sampleE}, along with a
  subsample of objects in our Galaxy and the Magellanic Clouds with oxygen
  abundance determinations from oxygen recombination lines (solid red
  diamonds). In solid line, it is shown the relation proposed by Nagao et
  al. (2006) and in the right panel the residuals between the directly derived
  abundances and those obtained from this calibration.} 
\label{o2_ne3}
\end{figure}

This is the case for the emission line ratio I([Ne{\sc iii}]
3869\,\AA)/I([O{\sc ii}]\,3727\AA), proposed by Nagao et
al.\ \cite*{2006A&A...459...85N} as an empirical calibrator useful for
high-redshift galaxies (up to z=1.6 in the optical part of the spectrum) and
relatively independent of reddening due to the proximity of the two lines. We
show in Figure \ref{o2_ne3} the relation between this ratio and the oxygen
abundance for the sample of objects described in Section \S \ref{sampleE}. This
Figure shows a very high dispersion in all the metallicity range, 
especially large in the high metallicity regime. This is confirmed by the
objects in the Galaxy and the Magellanic Clouds compiled by Peimbert et al.\
\cite*{2007RMxAC..29...72P} for which a direct determination of the oxygen
abundance by means of recombination lines (RLs) exists. The residuals between
the  
direct determination of the oxygen abundance and those obtained from the
relation based on the [Ne{\sc iii}]/[O{\sc ii}] ratio are shown in the right
panel of the same Figure. The standard deviation of these residuals in the
whole metallicity regime reaches 0.83\,dex.

\begin{figure}
\centering
\includegraphics[width=.87\textwidth,angle=0,clip=]{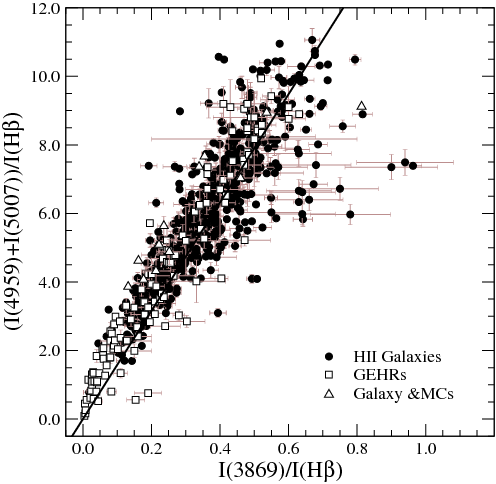}
\caption[Relation between the intensity of the emission line of [Ne{\sc
      iii}{\textrm ]} at 3869\,\AA\ and the sum of the lines of [O{\sc
      iii}{\textrm ]} at 4959 and 5007\,\AA]{Relation between the intensity of
  the emission line of [Ne{\sc iii}] at 3869\,\AA\ and the sum of the lines of
  [O{\sc iii}] at 4959 and 5007\,\AA\ for the sample of objects described in
  Section \S \ref{sampleE}. The solid line represents the best linear fit to
  the sample.} 
\label{o3_ne3}
\end{figure}

The reason for this huge dispersion can be found in the high dependence of the
[Ne{\sc iii}]/[O{\sc ii}] ratio on ionization parameter, as it is stressed by
\citetex{2006A&A...459...85N}. This dependence however is due to the tight
relation 
between the flux of [Ne{\sc iii}] and [O{\sc iii}] emission lines (see Figure
\ref{o3_ne3}), as a consequence of the quite similar ionization structure of
O$^{2+}$ and Ne$^{2+}$ and the constant value of the Ne/O ratio.  The best
linear fit between the fluxes of the lines for this sample yields: 

\[
I([O{\textrm{\sc iii}}] 4959+5007\,{\textrm\AA})=(15.37 \pm 0.25) I([Ne{\textrm{\sc iii}}] 3869\,{\textrm\AA})
\]

Although, there is a deviation from this trend for some \HII\ galaxies showing
abnormally high values of the line of [Ne{\sc iii}] ({\em e.g.} UM382;
Terlevich et al., 1991, HS1440+4302, HS1347+3811; Popescu and Hopp, 2000),
most of the objects are quite close to this relation. Even for the objects
with low intensities, which correspond to high metallicity/low excitation
regions and for which the charge transfer reaction between O$^{2+}$ and H$^0$
becomes more important, there exists a very good agreement with this linear
fit. A possible cause to the deviation from this relation in some \HII\
galaxies comes from the fact that Ne$^+$ has an ionization potential 5.85 eV
larger than O$^+$ and therefore is more sensitive to the stellar effective
temperature, which is higher in these \HII\ galaxies. One of the consequences
of this relation is that the [Ne{\sc iii}]/[O{\sc ii}] ratio is, in fact,
almost equivalent to [O{\sc iii}]/[O{\sc ii}] which is highly dependent on
ionization parameter and effective temperature (P\'erez-Montero and D\'\i az,
2005) and presents a lower dependence on metallicity.

\nocite{1991A&AS...91..285T,2000A&AS..142..247P}

The relation between [O{\sc iii}] and [Ne{\sc iii}] can be used as well in all
the other diagnostic ratios involving [O{\sc iii}] lines in a similar
way. This is the case of the O$_{23}$ parameter, also known as R$_{23}$,
defined by \citetex{1979MNRAS.189...95P} as the relative sum of the strong
lines of [O{\sc ii}] at 3727\,\AA\ and [O{\sc iii}] at 4959 and 5007\,\AA\ in
relation to H$\beta$. The relation between this parameter and the oxygen
abundance is widely used for objects at large redshifts due to the relatively
blue wavelength and high intensities of the involved lines. There exist many
different calibrations of this parameter, whose main drawbacks are well known:
(i) its relation with metallicity is double-valued with increasing values of
O$_{23}$ for increasing values of metallicity in the low metallicity regime
and decreasing values of O$_{23}$ for increasing metallicity in the high
metallicity regime, requiring external methods to distinguish the upper from
the lower branch and with a high dispersion in the middle-range, that makes
metallicities quite uncertain in the range 8.0 $<$ 12+log(O/H) $<$ 8.4; (ii)
the dependence of O$_{23}$ on other functional parameters like ionization
parameter or effective temperature. This is solved with the calibration of the
parameter as a function of other quantities that reduce this dependence, as in
the case of the [O{\sc ii}]/[O{\sc iii}] ratio (Kobulnicky et al., 1999 from
the models of McGaugh, 1991) or the P parameter \cite{2000A&A...362..325P};
(iii) the lack of objects with directly derived abundances in the high
metallicity regime makes the upper branch calibration to rely heavily on
photo-ionization model results, which differ appreciably depending on the model
atmospheres used \cite{2004A&A...415..577M} and the chosen input
conditions. Besides, these models predict higher metallicities than those
derived from electron temperatures of [N{\sc ii}] or [S{\sc iii}] in inner
disc regions ({\em e.g.} Bresolin, 2007), but not than those derived from
oxygen recombination lines. The difficulties found to derive accurate
metallicities using this calibration and, in general, all the other empirical
relations, make them appropriate to study distributions of metallicity in a
statistical way in large surveys of emission lines objects, but they do not
offer reliable results for individual determinations. 

\nocite{1999ApJ...514..544K,1991ApJ...380..140M}

\begin{figure}
\centering
\includegraphics[width=.87\textwidth,angle=0,clip=]{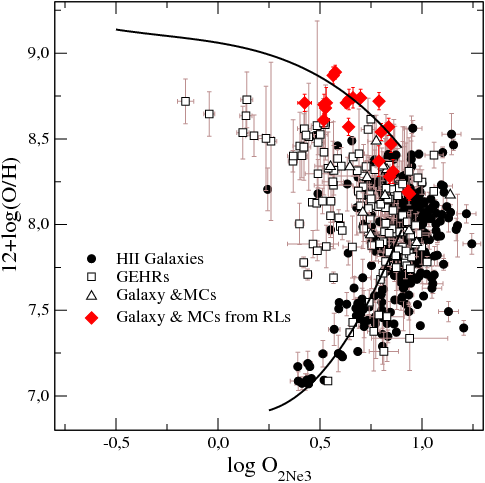}
\caption[Relation between oxygen abundance and the O$_{2Ne3}$ parameter for
  the sample of objects]{Relation between oxygen abundance and the O$_{2Ne3}$
  parameter for the sample of objects described in the text (symbols) and the
  calibration of McGaugh of the O$_{23}$ parameter for an average value of the
  ionization parameter (log([O{\sc iii}]/[O{\sc ii}])\,=\,1, solid line). The
  red diamonds represent the sample of objects with a direct determination of
  the oxygen abundances based on recombination lines.}  
\label{o2ne3}
\end{figure}

We can then define a proxy for the O$_{23}$ parameter, using the relation
between [O{\sc iii}] and [Ne{\sc iii}] lines, previously discussed:
\[
O_{2Ne3}\,=\,\frac{I([O{\textrm{\sc ii}}]\,3727\,{\textrm\AA})+15.37\,I([Ne{\textrm{\sc iii}}]\,3869\,{\textrm\AA)}}{I(H\beta)}\,\approx\,O_{23}
\]
\noindent that is shown in Figure \ref{o2ne3} for the sample described in
Section \S \ref{sampleE}. We also show as a solid line the equivalent to the
calibration of the O$_{23}$ parameter based on the models of McGaugh (1991),
for an average value of the ionization parameter (log([O{\sc iii}]/[O{\sc
    ii}])\,=\,1). This calibration gives the lowest dispersion in relation
with the sample of objects in both the lower and the upper branch
(P\'erez-Montero and D\'\i az, 2005). We also show the sample of objects with
a direct determination of the oxygen abundances based on recombination lines
(red diamonds)
in order to illustrate how this parameter has the same problems described
above for O$_{23}$. We can redefine the O$_{2Ne3}$ parameter relative to the 
closest and brightest hydrogen recombination line, H$\delta$. These lines are
closer in wavelength, which makes its relation to be little reddening and flux
calibration dependent, and useful to larger redshifts (up to z\,$\approx$\,1.3).
\[
O_{2Ne3'} = \frac{I([O{\textrm{\sc ii}}] 3727{\textrm\AA})+
  15.37\,I([Ne{\textrm{\sc iii}}] 3869 {\textrm\AA})}{I(H\delta)}\approx
O_{23} \frac{I(H\beta)}{I(H\delta)}
\]

This parameter being equivalent to O$_{23}$ suffers from its same
problems, with the additional difficulty of being based on weaker lines and
therefore, more difficult to measure with good signal-to-noise ratio. 
In addition, the hydrogen recombination line H$\delta$ in emission is more
strongly affected by underlying stellar absorption and its measurement has
to be carried out carefully ({\em e.g.} H\"agele et al., 2006; see Chapter \S
\ref{HIIgal-obs}).

\begin{figure}
\centering
\includegraphics[width=.87\textwidth,angle=0,clip=]{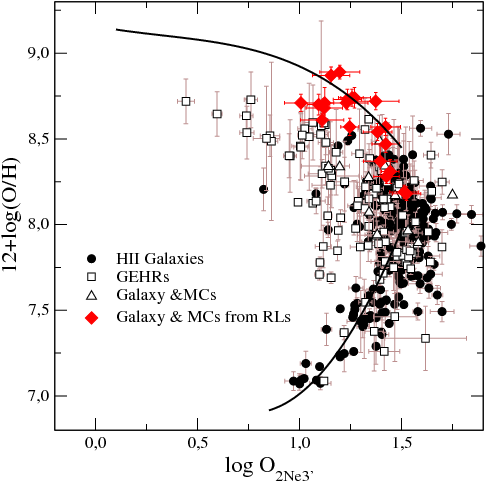}
\caption[Relation between oxygen abundance and the O$_{2Ne3'}$ parameter for
  the sample of objects]{Relation between oxygen abundance and the O$_{2Ne3'}$
  parameter for the sample of objects described in the text (symbols) and the
  empirical calibration of O$_{23}$ of McGaugh for an average value of the
  ionization parameters (log([O{\sc iii}]/[O{\sc ii}])\,=\,1, solid
  line). This calibration has been displaced in the $x$ axis 0.6 dex to take
  into the account the mean value of the H$\beta$/H$\delta$ ratio.} 
\label{o2ne3p}
\end{figure}

\begin{figure}
\centering
\includegraphics[width=.82\textwidth,angle=0,clip=]{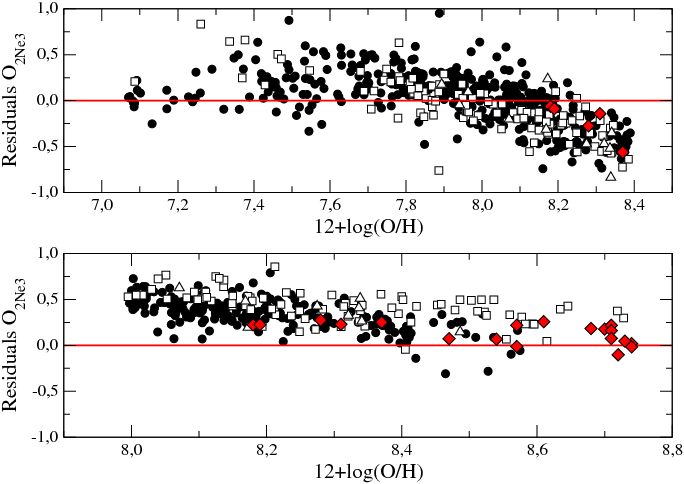}\\\vspace{0.5cm}
\includegraphics[width=.82\textwidth,angle=0,clip=]{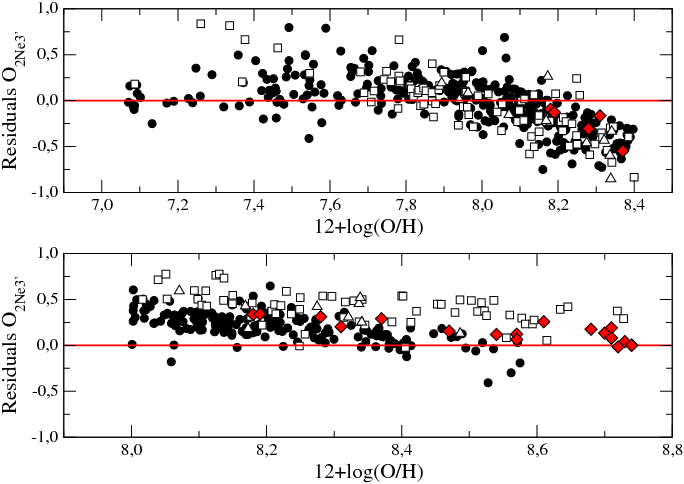}
\caption[Residuals between the oxygen abundances derived from the McGaugh
  (1991) calibration of O$_{23}$ applied to O$_{2Ne3}$ and to
  O${_2Ne3'}$]{Residuals between the oxygen abundances derived from the
  McGaugh 
  (1991) calibration of O$_{23}$ applied to O$_{2Ne3}$ (the two upper plots)
  and to O${_2Ne3'}$ (the two lower) and the oxygen abundances derived from
  the direct method. The first and the third panels from the top show the
  calibration for low metallicities and the second and the fourth ones for high
  metallicities.}
\label{o2ne3_res}
\end{figure}

We show this parameter in Figure \ref{o2ne3p} for the sample of objects
described in Section \S \ref{sampleE} and the sample of \HII\ regions with a
determination of the oxygen abundance based on recombination lines. We show as
well as a solid line the empirical calibration from McGaugh but in this case
we take into account the factor H$\beta$/H$\delta$ in the $x$ axis.  As in the
case of O$_{23}$, it is necessary to distinguish between the upper and lower
branch to derive the oxygen abundance from it. In the case of O$_{23}$, this
is usually done by means of the [N{\sc ii}] lines, but they are not available
in the blue part of the spectrum. Other possibility is using the [Ne{\sc
    iii}]/[O{\sc ii}] ratio, although it is very uncertain. From Figure
\ref{o2_ne3} we can see that while objects with log([Ne{\sc iii}]/[O{\sc
    ii}])\,$<$\,-1.0 probably have high metallicities, the contrary is not
true. The residuals of the oxygen abundances derived with these new parameters
as a function of the abundances measured using the direct method for both the
lower and the upper branches are shown in Figure \ref{o2ne3_res}.
We can see that, as in the case of O$_{23}$, both the upper and lower branch
calibrations show large differences with the derived metallicities in the
turnover region (8.0\,$<$\,12+log(O/H)\,$<$\,8.4). The standard deviation of
the residuals in the other ranges are quite similar to those obtained for the
O$_{23}$ parameter in P\'erez-Montero and D\'\i az (2005) for the McGaugh
calibration of the upper branch (0.20 dex for O$_{2Ne3}$ and 0.18\,dex for
O$_{2Ne3'}$ in the  12+log(O/H) $>$ 8.4 range). The standard deviation of the
residuals between the metallicities derived from the O$_{23}$ parameter are
0.06\,dex for both  O$_{2Ne3}$ and O$_{2Ne3'}$ parameters. However, in the low
metallicity regime, 12+log(O/H) $<$ 8.0, the dispersion of the residuals 
relative to the direct method is 0.24\,dex for the O$_{2Ne3}$ parameter and
0.22\,dex for the O$_{2Ne3'}$ parameter, much higher than the
0.13\,dex calculated for the O$_{23}$ parameter in P\'erez-Montero and D\'\i
az (2005). In this case the residuals between this calibration and those based
on the neon emission line are 0.14\,dex and 0.16\,dex, respectively.

\begin{figure}[p]
\centering
\includegraphics[width=.87\textwidth,angle=0,clip=]{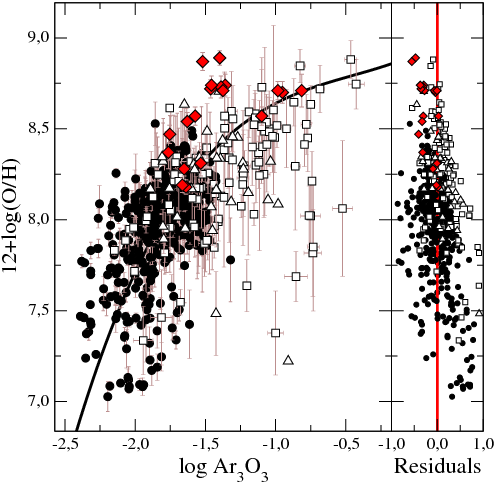}
\caption[Relation between the Ar$_3$O$_3$ parameter and the oxygen abundances
  derived from the direct method]{Relation between the Ar$_3$O$_3$ parameter
  and the oxygen abundances derived from the direct method, along with the
  calibration proposed by Stasi{\'n}ska (2006) for this parameter. In the
  right panel, we represent the residuals between the directly derived
  abundances and those derived from the calibration.}
\label{ar3o3}
\end{figure}

Among the other empirical calibrators based on Ne or Ar emission lines, we can
analyze also the Ar$_3$O$_3$ parameter, defined by Stasi{\'n}ska
\cite*{2006A&A...454L.127S} as the ratio of the fluxes of the emission lines
of [Ar{\sc iii}] at 7136\,\AA\ and [O{\sc iii}] at 5007\,\AA. In Figure
\ref{ar3o3}, we show the relation between the oxygen abundances and the
logarithm of this parameter for the sample of objects described in Section \S
\ref{sampleE} having a measurement of the [Ar{\sc iii}] line. The standard
deviation of the residuals, which are shown in the right panel of the same
Figure as a function of metallicity, gives a value of 0.35 dex. Besides, the
behaviour of this parameter in the high metallicity regime is quite unclear
due to the position of some of the objects with a direct determination of the
oxygen abundance based on recombination lines, that show lower values of
Ar$_3$O$_3$ than those expected for their metallicity. 

\nocite{2006A&A...454L.127S}

\subsection{Diagnostic diagrams for emission-line objects}

Strong lines are used also in emission-line-like objects to find out the
nature of the main ionization mechanism: photo-ionization due to the absorption
by the surrounding gas of UV photons emitted by massive stars, power law
spectral energy distribution associated with active galactic nuclei or shock
excitation. Since some of these diagnostic diagrams for the blue part of the
spectrum use the emission lines of [O{\sc iii}] at 5007 and 4959\,\AA, we
could use as well the [Ne{\sc iii}]\,3869\,\AA\ emission line and the closest
and brightest hydrogen recombination line to redefine these diagrams that
would be useful up to larger redshifts and would also be less reddening
dependent. This is the case of the relations proposed by
\citetex{2004MNRAS.350..396L}, as for instance the relation between ([O{\sc
    iii}] 5007\,\AA/H$\beta$) and ([O{\sc ii}] 3727\,\AA/H$\beta$). Using the
appropriate relations, the analytical expression proposed by these authors
take the following expression: 
\[
\log\left(\frac{[Ne{\textrm{\sc iii}}]\,3869\,{\textrm\AA}}{H\delta}\right)\,=\,\frac{0.14}{\log
  ([O{\textrm{\sc ii}}]\,3727\,{\textrm\AA}/H\delta)-2.05}+0.37
\]

\begin{figure}
\centering
\includegraphics[width=.62\textwidth,angle=0,clip=]{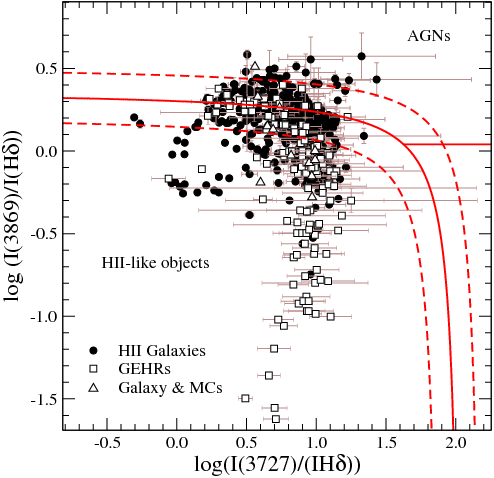}\\\vspace{0.5cm}
\includegraphics[width=.62\textwidth,angle=0,clip=]{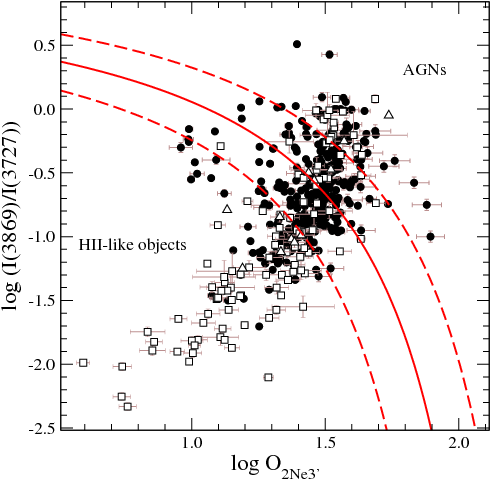}
\caption[Diagnostic diagrams based on neon emission lines for the sample of
  objects]{Diagnostic diagrams based on neon emission lines for the sample of
  objects described in Section \S \ref{sampleE}. Upper panel: the relation between
  the ratios [Ne{\sc iii}]/H$\delta$ and [O{\sc ii}]/H$\delta$. Lower panel:
  the relation between the ratio [Ne{\sc iii}]/[O{\sc ii}] and the O$_{2Ne3'}$
  parameter. Solid red lines represent the analytical division between star
  forming objects and AGNs. Dashed red lines represent the limit of the bands of
  uncertainty 0.15 dex both sides of each relation.}
\label{diag}
\end{figure}

In the upper panel of Figure \ref{diag} we show this diagram for the objects
described in Section  \S \ref{sampleE}. Starbursts galaxies and star forming
regions lie below the solid red line. This diagnostic ratio is, in fact,
equivalent to the relation proposed by Rola et al.\
\cite*{1997MNRAS.289..419R} between ([Ne{\sc iii}]\,3869\,\AA/H$\beta$) and
([O{\sc ii}]\,3727\,\AA/H$\beta$).  The rate of coincidences between this
diagnostic diagram and its predecessor based on [O{\sc iii}] lines reaches 83
\%. This diagram can be used also to separate Seyfert 2 galaxies from LINERS
using the theoretical expression proposed by Lamareille et al.\
\cite*{2004MNRAS.350..396L} adapted for [Ne{\sc iii}] and H$\delta$: 
\[
\log\left(\frac{[Ne{\textrm{\sc iii}}]}{H\delta}\right)\,=\,0.04
\]

All the \HII\ galaxies which present abnormally high values of the emission
line of [Ne{\sc iii}] shown in Figure \ref{o3_ne3} appear in this diagram in
the Seyfert 2 zone, what is consistent with a possibly higher dependence of
the  [Ne{\sc iii}] line on stellar effective temperature, as it has been
previously suggested. This is supported by the fact that none of these
galaxies are classified as Seyfert 2 when other diagnostics diagrams, like
[S{\sc ii}]/H$\alpha$ vs.\ [O{\sc iii}]/H$\beta$ or [N{\sc ii}]/H$\alpha$
vs.\ [O{\sc iii}]/H$\beta$, are used.

The other blue diagnostic ratio proposed by Lamareille et al.\
\cite*{2004MNRAS.350..396L} is the relation ([O{\sc
    iii}]\,4959+5007\,\AA)/([O{\sc ii}]\,3727\,\AA) versus O$_{23}$. The
analytical expressions proposed by these authors for this diagram can be
expressed in terms of [Ne{\sc iii}] and H$\delta$ adopting the form: 
\[
log\left(\frac{[Ne{\textrm{\sc iii}}]\,3869\,{\textrm\AA}}{[O{\textrm{\sc
	ii}}]\,3727\,{\textrm\AA}}\right)\,=\,\frac{1.5}{\log
  O_{2Ne3'}-2.3}+1.21
\]

The diagram is shown in the lower panel of Figure \ref{diag} and, again,
starbursts galaxies and star forming regions lie below this relation, which is
shown as a solid line. Although the rate of coincidences between this relation
and the one based on [O{\sc ii}] and [O{\sc iii}] is very high (92\,\%), a
larger number of objects is found  lying on the AGN region as compared to
other diagnostics which somewhat questions its application.

\section{Summary and conclusions}

We have performed an analysis of a large sample of emission line objects with
a direct determination of the oxygen abundance, via the calculation of the
electron temperature, and an accurate measurement of the emission lines of
[Ne{\sc iii}] at 3869\,\AA\ and [Ar{\sc iii}] at 7136\,\AA. We have
recalculated oxygen, neon and argon abundances taking into account the
electron temperature most representative for each ionic species. The total
chemical abundances have been calculated with the aid of new ionization
correction factors for neon and argon based on a new grid of photo-ionization
models computed using Cloudy v06.02 and WM-Basic stellar model atmospheres. 

The new ICF for Ne yields lower abundances for low excitation objects in
relation with the classical approximation O$^{2+}$/O $\approx$ Ne$^{2+}$/Ne,
which does not take into account the charge transfer reaction between O$^{2+}$
and H$^0$. This new ICF agrees quite well with the fits proposed by Izotov et
al. (2006) for the high and intermediate metallicity regime. Nevertheless,
although we do not find any relevant dependence of this ICF on metallicity as
we have not considered different X-ray sources in our models, there is some
evidence of the underestimate of the ICF in the high metallicity
regime. Firstly, the values found by Vermeij and Van der Hulst (2002) for a
sample of Galactic \HII\ regions from the Ne emission lines in the mid-IR
point to larger abundances of Ne$^+$ in these objects. Secondly, the study of
the Ne/O ratio as a function of metallicity shows lower values of Ne/O in the
high metallicity regime. The average value of this ratio agrees better with
the solar abundances than the values recently obtained from X-ray observations
in stellar coronae (Drake and Testa, 2005). 

Regarding Ar, we have obtained new ICFs for both Ar$^{2+}$ and
Ar$^{2+}$+Ar$^{3+}$ quite similar to the values found by Izotov et al.\ (2006)
but, again, we do not find, any relevant dependence on metallicity. According
to these new ICFs, the values proposed by Izotov et al.\ (1994) clearly
overestimate the total abundances of Ar. We propose as well new ICFs based on
the ratio S$^{2+}$/(S$^{+}$+S$^{2+}$), in order to calculate total Ar
abundances using the red and far-IR wavelength range only. The study of the
Ar/O ratio as a function of metallicity gives contradictory results for \HII\
galaxies and Giant \HII\ regions in spirals discs. For the first ones we find
a constant value of the Ar/O ratio, in agreement with the expected results for
the stellar production of this element. However, for GEHRs we find evidence
for decreasing Ar/O with increasing metallicity. This result is found as well
for the S/O ratio both in GEHRs (D\'\i az et al., 1991; P\'erez-Montero et
al., 2006) and halo massive stars of our Galaxy (Israelian and Rebolo,
2001). Taking into account that the ionization structure of Ar and S are quite
similar and that they are produced in the same stellar cores it is not
surprising that their ratios behave in similar ways. 

We have studied some empirical parameters of metallicity based on Ne and Ar
emission lines. This is the case of the [Ne{\sc iii}]/[O{\sc ii}] ratio
proposed by Nagao et al. (2006). We have shown that this parameter is indeed
much more sensitive to ionization parameter and effective temperature than to
metallicity due to the tight relation existing between [Ne{\sc iii}] and
[O{\sc iii}] emission lines. This is due to the constant value of the Ne/O
ratio and the identical ionization structures of Ne$^{2+}$ and O$^{2+}$. In
fact [O{\sc iii}] can be substituted by [Ne{\sc iii}]  in empirical
metallicity calibrations and diagnostic diagrams. This offers the possibility
of extending the range of applicability of these relations to objects with
higher redshift minimizing at the same time the effects of reddening
corrections and flux calibration. We have then defined an abundance parameter
O$_{2Ne3'}$ equivalent to the commonly used O$_{23}$ but using the sum of the
[O{\sc ii}] and [Ne{\sc iii}] lines relative to H$\delta$. Although this
parameter has the same problems as O$_{23}$ (double-valued, dependence on $U$,
calibration of the upper branch) it constitutes a new tool to derive oxygen
abundances in large deep optical surveys of galaxies up to high redshifts
($\approx$\,1.3). Using the same principle, we have defined new diagnostic
methods based only on [O{\sc ii}] and [Ne{\sc iii}], similar to those proposed
by Rola et al. (1997),  that could be used in the same surveys to separate
star forming galaxies from active galactic nuclei. The results obtained from
these diagrams are very similar to those obtained from the other ones in the
blue part of the spectrum.

\addcontentsline{toc}{section}{\numberline{}Bibliography}

\bibliographystyle{astron}
\bibliography{tesis}

\chapter{Star Formation in Circumnuclear Regions:\\ On the derivation of dynamical masses of the stellar clusters}
\label{cnsfr-obs-kine}

\section{Introduction}

The inner ($\sim$1\,Kpc) parts of some spiral galaxies show high star formation
rates and this star formation is frequently arranged in a ring or pseudo-ring
pattern around their nuclei. This fact seems to correlate with the presence of
bars
\cite{1979ApJ...233...67R,1995A&A...301..649F,1999ApJ...525..691S,2005ApJ...632..217S,2005ApJ...630..837J}
and, in fact, computer models which simulate the behaviour of gas in 
galactic potentials have shown that nuclear rings may appear as a consequence
of matter infall owing to resonances present at the bar edges
\cite{1985A&A...150..327C,1992MNRAS.259..328A}. 

These CNSFRs, with sizes going from a few tens to a few hundreds of parsecs
(e.g. \citeplain{2000MNRAS.311..120D}) seem to be made of several \HII\ regions
ionized by luminous compact stellar clusters whose sizes, as measured from
high spatial resolution {\it Hubble Space Telescope} (HST) images, are seen to
be of only a few parsecs. Their 
masses, as derived with the use of population synthesis models in
circumnuclear regions of different galaxies, suggest that these  clusters are
gravitationally bound and that they might evolve into globular cluster
configurations \cite{1996AJ....111.2248M}. The luminosities of CNSFRs are
rather large with absolute visual magnitudes (M$_v$) between -12 and -17 and
H$\alpha$ luminosities 
between 2\,$\times$\,10$^{38}$ and 7\,$\times$\,10$^{40}$ erg s$^{-1}$ . These
values are comparable to those shown by 30 Dor, the largest \HII\ region in the
Large Magellanic Cloud (LMC), and overlap with those shown by \HII\ galaxies  
(Melnick et al., 1988; D\'iaz et al., 2000a; Hoyos and D\'iaz, 2006, and
references therein).

\nocite{1988MNRAS.235..297M,2000MNRAS.311..120D,2006MNRAS.365..454H}
\nocite{2000MNRAS.311..120D} 

Although these \HII\ regions are very luminous not much is known about their 
kinematics or dynamics for both the ionized gas and the stars. It could be
said that the worst known parameter of these ionizing clusters is their mass. 
There are different methods to estimate the mass of a stellar
cluster. Classically one assumes that the system is virialized and determines
the total mass inside a radius by applying the virial theorem to the observed 
velocity dispersion of the stars ($\sigma_{\ast}$). The stellar velocity
dispersion is however hard to measure in young stellar clusters (a few
million-years old) due to the 
shortage of prominent stellar absorption lines. The optical continuum between
3500 and 7000\,\AA\ shows very few lines since the light at  these wavelengths
is dominated by OB stars which have weak absorption lines at the same
wavelengths of the nebular emission lines (Balmer H and He{\sc i} lines). As
pointed out by several authors (e.g.\ Ho and Filippenko 1996a), at longer 
wavelengths (\,8500\,\AA) the contamination due to nebular lines is much
smaller  and since red supergiant stars, if present, dominate the
near-infrared (IR)
light where the Ca{\sc ii} $\lambda\lambda$\,8498, 8542, 8662\,\AA\ triplet
(CaT) lines are found, these should be easily observable allowing the
determination of $\sigma_{\ast}$
\cite{1990MNRAS.242P..48T,1994A&A...288..396P}. We have previously detected
the CaT lines in CNSFRs but at a spectral resolution that was below that
required to measure accurately their velocity dispersions
(e.g.\ Terlevich et al., 1990). 

\nocite{1996ApJ...466L..83H,1990MNRAS.242P..48T}

These CNSFRs are also expected to be amongst the highest metallicity regions as
corresponds to their position near the galactic center. These facts taken
together make the analysis of these 
regions complicated since, in general, their low excitation makes any
temperature sensitive line too weak to be measured, particularly against a
strong underlying stellar continuum.  In fact, in most cases, the [O{\sc
    iii}]\,$\lambda$\,5007\,\AA\ line, which is typically  one hundred times
more intense than the auroral [O{\sc iii}]\,$\lambda$\,4363\,\AA\ one, can be
barely seen.

To analyze and derive the dynamical properties of the stars and
gas in these type of regions we use high resolution spectra with two very
narrow spectral ranges of the central zone of three early type barred spiral
galaxies: NGC\,2903, NGC\,3310 and NGC\,3351.
In Section \S \ref{sample} we present the properties of the observed galaxies,
and in the next one the details of the observations
and the data reduction. \S \ref{resul-kine} presents the results concerning
the kinematics of gas and stars and the emission line ratios in each of the
observed regions. In sections \S \ref{masses} and \S \ref{ionmass} we present
the techniques used in the determination of their dynamical masses and in the
derivation of the ionizing star cluster properties, respectively. Section
\ref{disc-kine} is devoted to the discussion of these results, and  
the summary and conclusions of this part of the work are presented in Section
\S \ref{summa-kine}.

\section{Properties of the observed galaxies}
\label{sample}

In this Chapter we present high-resolution  far-red spectra
($\sim$0.39\,\AA\,px$^{-1}$\,$\sim$\,13.66\,km\,s$^{-1}$\,px$^{-1}$ at
central wavelength, $\lambda_c$=8563\,\AA ) and  stellar velocity dispersion 
measurements ($\sigma_{\ast}$) along the line of sight for several
CNSFRs and the nuclei of the spiral galaxies NGC\,2903, NGC\,3310 and
NGC\,3351. We have also measured 
the ionized gas velocity dispersions ($\sigma_{g}$) from high-resolution blue
spectra
($\sim$0.21\,\AA\,px$^{-1}$\,$\sim$\,12.63\,km\,s$^{-1}$\,px$^{-1}$ 
at $\lambda_c$\,=\,4989\,\AA) using Balmer H$\beta$ and [O{\sc iii}] emission
lines. The comparison between $\sigma_{\ast}$ and $\sigma_{g}$  might throw
some light on the yet unsolved issue about the validity of the gravitational
hypothesis for the origin of the supersonic motions observed in the ionized
gas in Giant \HII\ regions \cite{1999MNRAS.302..677M}. 

\begin{figure}
\centering
\includegraphics[width=.98\textwidth,angle=0]{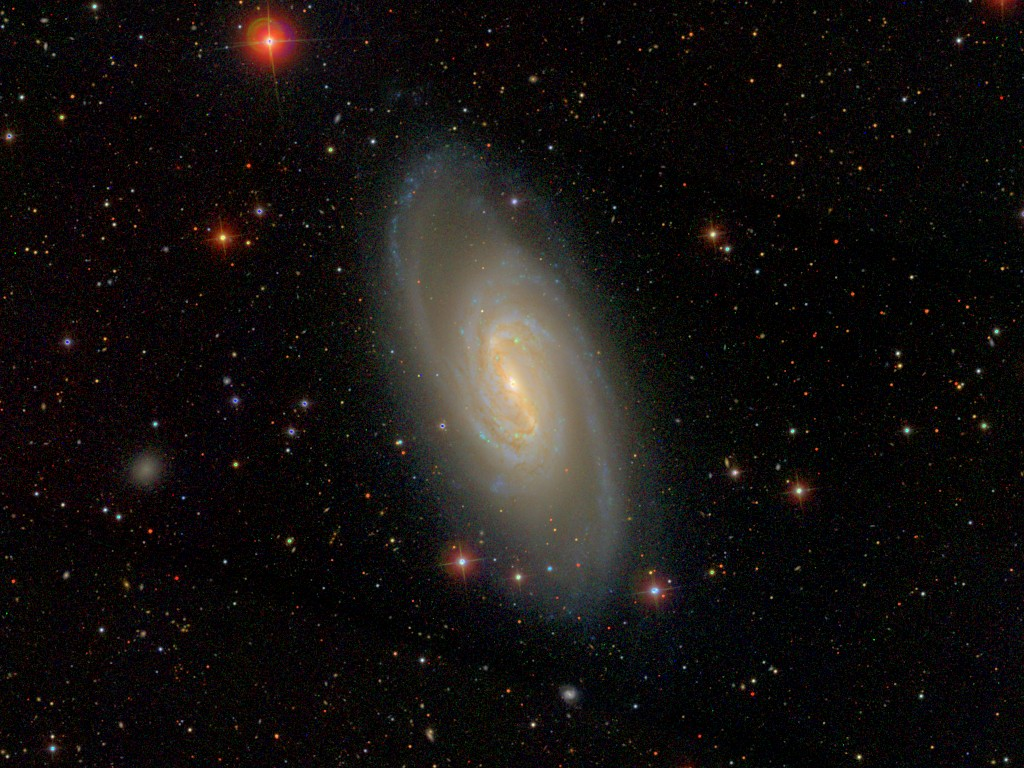}
\includegraphics[width=.55\textwidth,angle=0]{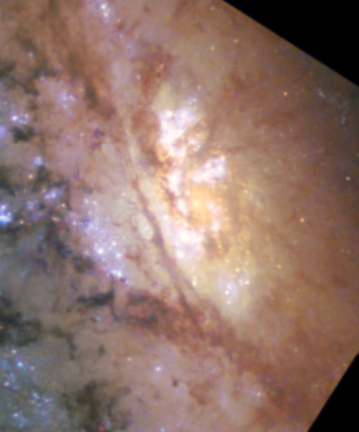}
\caption[False colour images of the barred spiral galaxy NGC\,2903]{False
  colour images of the barred spiral galaxy NGC\,2903. Upper 
  panel: full view from the SDSS; lower panel: enlargement of the central zone
  from the HST (WFPC2-PC1).}
\label{col2903}
\end{figure}

NGC\,2903 (UGC\,5079; see Figure \ref{col2903}) is a well studied galaxy. The
Paschen $\alpha$ image obtained with 
the Hubble Space Telescope (HST) reveals the presence of a nuclear ring-like
morphology with an apparent diameter of approximately 15\,\arcsec\,=\,625\,pc
\cite{2001MNRAS.322..757A}. This structure is also seen, though less
prominent, in the H$\alpha$ observations from Planesas et al.\
\cite*{1997A&A...325...81P}. A large number of stellar clusters are identified
on  high resolution infrared images in the K' and H bands, which do not
coincide spatially with the bright \HII\ regions. A possible interpretation of
this is that the stellar clusters are the result of the evolution of giant
\HII\ regions (e.g. \citeplain{2001MNRAS.322..757A}). The global star
formation rates in the nuclear ring, as derived from  its H$\alpha$ luminosity
is found to be 0.1\,M$_\odot$\,yr$^{-1}$ by \citetex{1997A&A...325...81P} and
0.7\,M$_\odot$\,yr$^{-1}$ by \citetex{2001MNRAS.322..757A}. From CO emission 
observations, Planesas et al.\ derive a mass of molecular gas (H$_2$) of
1.8\,$\times$\,10$^{8}$\,M$_\odot$ inside a circle 1\,Kpc in diameter.

\begin{figure}
\centering
\includegraphics[width=.98\textwidth,angle=0]{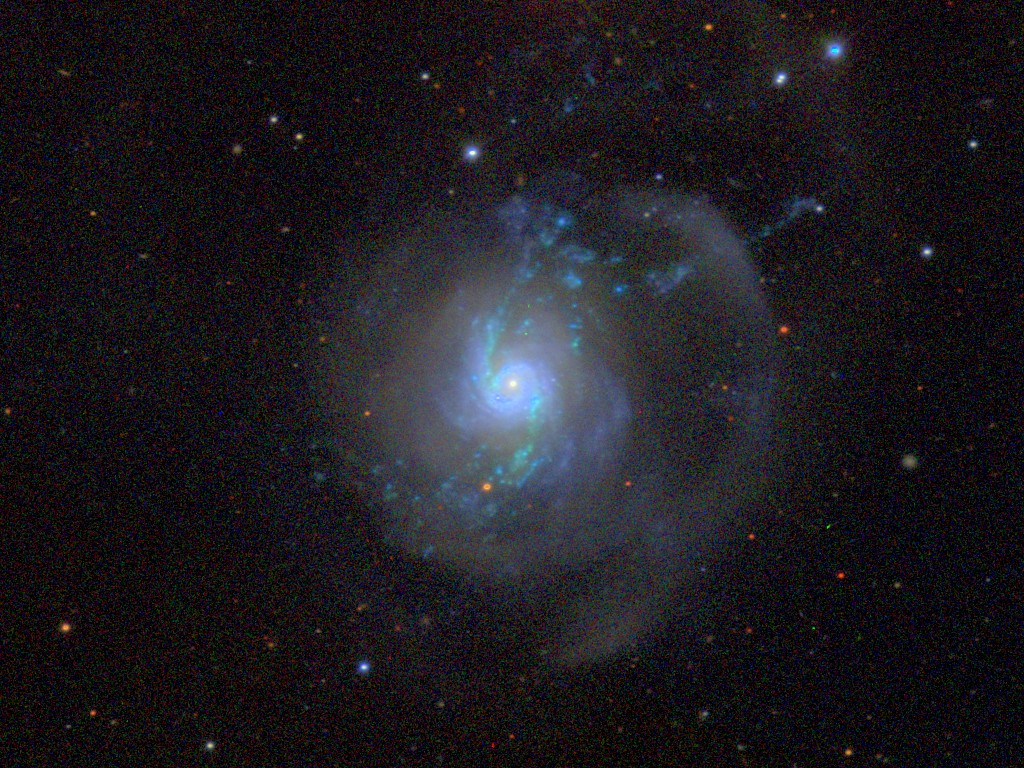}\vspace{-0.45cm}
\includegraphics[width=.6\textwidth,angle=180]{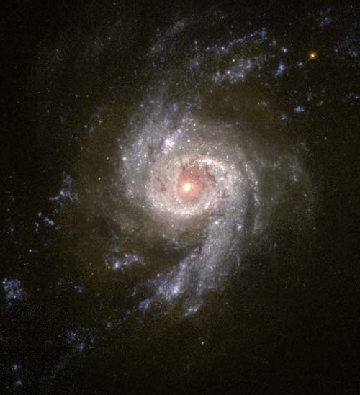}
\caption[False colour images of the barred spiral galaxy NGC\,3310]{False
  colour images of the barred spiral galaxy NGC\,3310. Upper 
  panel: full view from the SDSS; lower panel: enlargement of the central zone
  from the HST (WFPC2-PC1).} 
\label{col3310}
\end{figure}

NGC\,3310 (UGC\,5786, Arp217; see Figure \ref{col3310}) is a starburst galaxy
classified as a SAB(r)bc by 
\citetex{1991trcb.book.....D}, with an inclination of the galactic disc of
about i\,$\sim$\,40 \cite{2000MNRAS.312....2S}. This is a good example of an
overall low 
metallicity galaxy, with a high rate of star formation and very blue
colours. This galaxy has a ring of star forming regions whose 
diameter ranges from 8\,\arcsec to 12\,\arcsec and shows two tightly
wound spiral arms \cite{2002AJ....123.1381E,1976A&A....48..373V} filled with
giant \HII\ regions. The ages indicated by the colours and magnitudes of the
star formation regions are lower than 10\,Myr \cite{2002AJ....123.1381E}. From
near-IR J and K photometry these authors derived an average age of
$\sim$10$^7$\,yr for the large scale ``hot-spots'' (star forming
complexes). From the observed CaT line 
in the Jumbo \HII\ region \citetex{1990MNRAS.242P..48T} derived an age around
5 to 6\,Myr. \citetex{2002AJ....123.1381E}, comparing their data with
Starburst99 models \cite{1999ApJS..123....3L}, estimated masses of the large
``hot-spots'' ranging from 10$^4$ to several times 10$^5$\,M$_\odot$. They
found 
17 candidate super star clusters (SSCs) with absolute magnitudes between
M$_B$\,=\,-11 to -15\,mag, and with colours similar to those measured for SSCs
in other galaxies (see for example
\citeplain{1995AJ....110.2665M,1995AJ....110.1009B,1996AJ....111.2248M,1997AJ....114.2381M,1998AJ....115.1778C,2001AJ....121.3048M,2001AJ....121..182E,2001AJ....122..815O,2001ApJ...556..801L,2002ApJ...579..545C,2003ApJ...596..240M,2007ApJ...663..844M,2008MNRAS.385..468P}).

\begin{figure}
\centering
\includegraphics[width=.98\textwidth,angle=0]{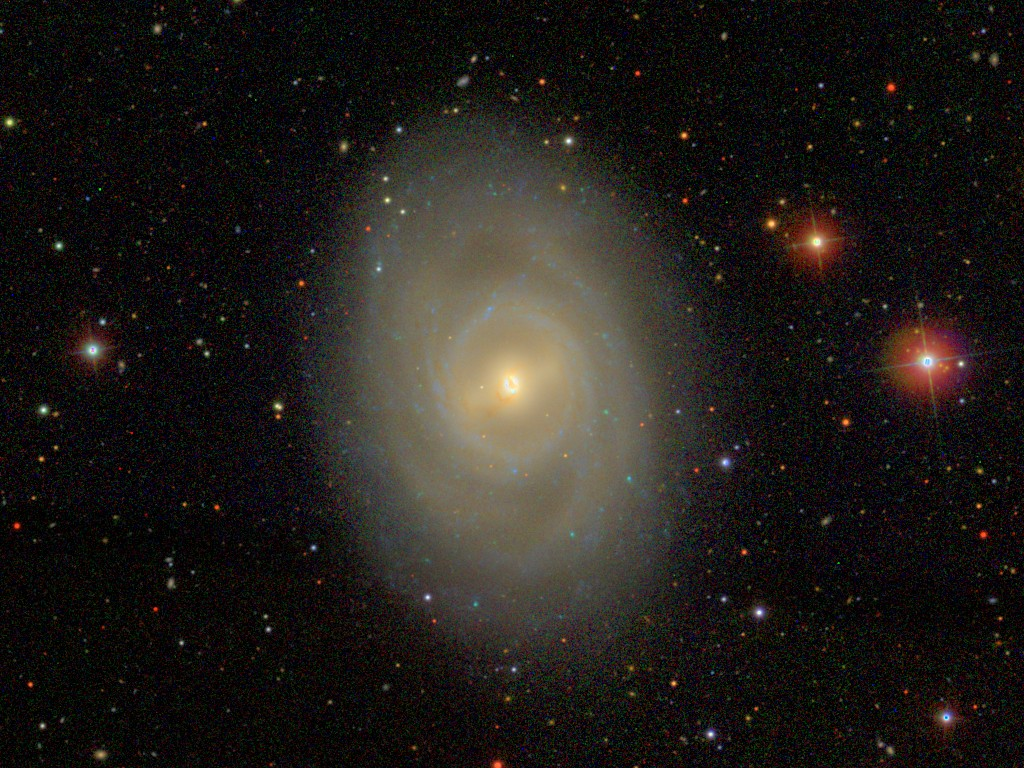}\vspace{-0.45cm}
\includegraphics[width=.75\textwidth,angle=180]{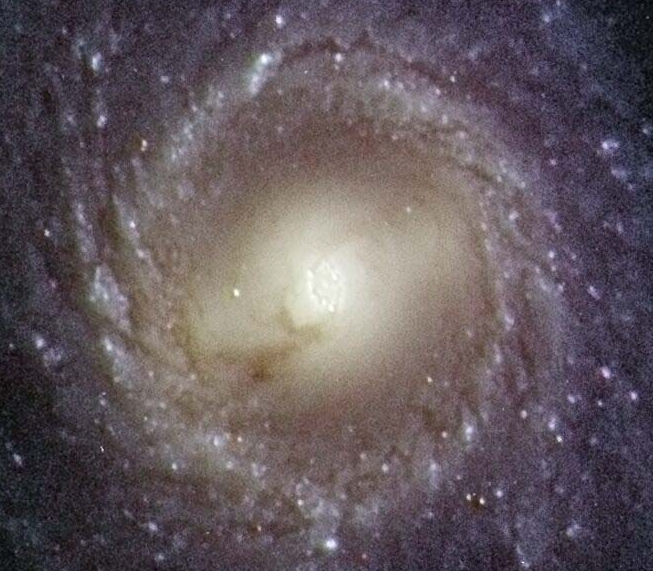}
\caption[False colour images of the barred spiral galaxy NGC\,3351]{False
  colour images of the barred spiral galaxy NGC\,3351. Upper  
  panel: full view from the SDSS; lower panel: enlargement of the central zone
  from the Issac Newton Group of Telescopes.} 
\label{col3351}
\end{figure}

NGC\,3351 (M95, UGC\,5850; see Figure \ref{col3351}) is another well known
``hot-spot'' galaxy 
\cite{1967PASP...79..152S}. Early detailed studies of its nuclear regions
\cite{1982A&AS...50..491A} concluded that NGC\,3351  harbours high-mass
circumnuclear star formation. In fact, the star formation rate per unit area
in the nuclear region is significantly increased over that observed in the
disc \cite{1992AJ....103..784D}. Elmegreen at al.\ \cite*{1997AJ....114.1850E}
from near infrared photometry in  the J and K bands derive a circumnuclear
star formation rate of 0.38\,M$_\odot$\,yr$^{-1}$.  Planesas et al.\
\cite*{1997A&A...325...81P}, from the H$\alpha$ emission,  derive a total star
formation rate for the circumnuclear region of 0.24\,M$_\odot$\,yr$^{-1}$ and
a mass of molecular gas of 3.5\,$\times$\,10$^{8}$\,M$_\odot$ inside a circle
of 1.4 Kpc in diameter, from CO emission observations. 

The main properties of these galaxies are given in Table \ref{propgal1}.

%

\begin{table}
\begin{center}
\caption{The galaxy sample for the kinematical study.}
 \begin{tabular}{cccc}
\hline
Property                 & NGC\,2903     & NGC\,3310   & NGC\,3351   \\
\hline				             
R. A. (2000)$^a$         &  09 32 10.1   & 10 38 45.9  & 10 43 57.7  \\
Dec (2000)$^a$           & +21 30 03     & +53 30 12   & +11 42 14   \\ 
Morph. Type              & SBbc          &  SABbc      & SBb         \\
Distance (Mpc)           &  8.6$^b$      &  15$^a$     &    10$^c$   \\
pc/ \arcsec\             &  42           &  73         &    49       \\
B$_{T}$ (mag)$^a$        &  9.7          &  11.2       &    10.1     \\
E(B-V)$_{gal}$(mag)$^a$  & 0.031         &  0.030      & 0.028       \\
\hline
\multicolumn{3}{l}{$^a$~\citetex{1991trcb.book.....D}}\\
\multicolumn{3}{l}{$^b$~\citetex{1984AAS...56..381B}}\\
\multicolumn{3}{l}{$^c$~Graham et al.\ (1997)}\\
\end{tabular}
\label{propgal1}
\end{center}
\end{table}


\nocite{1997ApJ...477..535G}

\section{Observations and data reduction}
\label{obs-kine}

\subsection{Observations}

High resolution blue and far-red spectra were acquired as part of an observing
run in 2000. They were obtained simultaneously using the blue and red arms of
the Intermediate dispersion Spectrograph and Imaging System (ISIS) on the
4.2-m William Herschel Telescope (WHT) of the Isaac Newton Group (ING) at the
Roque de los Muchachos Observatory on the Spanish island of La Palma. The
H2400B and R1200R gratings were used to cover 
the wavelength ranges from 4779 to 5199\,\AA\ ($\lambda_c$\,=\,4989\,\AA) in
the blue and  from 8363 to 8763\,\AA\ ($\lambda_c$\,=\,8563\,\AA) in the red
with spectral dispersions  of 0.21 and 0.39 \AA\ per pixel, equivalent to a
spectral resolution (R\,=\,$\lambda$\,/\,$\Delta\lambda$) of $\sim$\,23800 and
$\sim$\,22000, respectively, and 
providing a comparable velocity resolution of about 13\,km\,s$ ^{-1}$. The CCD
detectors EEV12 and TEK4 were used for the blue and red arms with a factor of
2 binning in both the ``x" and ``y" directions in the blue with spatial
resolutions of 0.38 and 0.36 \,arcsec\,px$^{-1}$ for the blue and red
configurations respectively. A slit width of 1\,arcsec was used which,
combined with the spectral dispersions, yielded spectral resolutions of about
0.4 and 0.7\,\AA\ FWHM in the blue and the red, respectively, measured on the
sky lines. The spectral ranges and grating resolutions [spectral dispersion in
\AA\,px$^{-1}$ and full width at half-maximum (FWHM) in \AA] attained are
given in Table \ref{journal1} containing the journal of observations.


\begin{table}
\centering
\caption{Journal of Observations}
\begin{tabular} {l c c c c c c}
\hline
\hline
 Date & Spectral range &       Disp.          & R$_{\mathrm{FWHM}}^a$ & Spatial res.            & PA   & Exposure Time \\
         &     (\AA)          & (\AA\,px$^{-1}$) &     & (\arcsec\,px$^{-1}$) &  ($ ^{o} $) & (sec)   \\[2pt]
\hline
\multicolumn{7}{c}{NGC\,2903} \\
2000 February 4 & 4779-5199  &  0.21  &  12500  &  0.38  &    50    &  3\,$\times$\,1200 \\
2000 February 4 & 8363-8763  &  0.39  &  12200  &  0.36  &    50    &  3\,$\times$\,1200  \\
2000 February 5 & 4779-5199  &  0.21  &  12500  &  0.38  &   345    &  3\,$\times$\,1200 \\
2000 February 5 & 8363-8763  &  0.39  &  12200  &  0.36  &   345    &  3\,$\times$\,1200 \\[2pt]
\multicolumn{7}{c}{NGC\,3310} \\
2000 February 4 & 4779-5199  &  0.21  &  12500  &  0.38  &    52    &  3\,$\times$\,1200 \\
2000 February 4 & 8363-8763  &  0.39  &  12200  &  0.36  &    52    &  3\,$\times$\,1200  \\
2000 February 5 & 4779-5199  &  0.21  &  12500  &  0.38  &   100    &  4\,$\times$\,1200 \\
2000 February 5 & 8363-8763  &  0.39  &  12200  &  0.36  &   100    &  4\,$\times$\,1200 \\[2pt]
\multicolumn{7}{c}{NGC\,3351} \\
2000 February 4 & 4779-5199  &  0.21  &  12500  &  0.38  &   355    &  4\,$\times$\,1200 \\
2000 February 4 & 8363-8763  &  0.39  &  12200  &  0.36  &   355    &  4\,$\times$\,1200  \\
2000 February 5 & 4779-5199  &  0.21  &  12500  &  0.38  &    45    &  3\,$\times$\,1200 \\
2000 February 5 & 8363-8763  &  0.39  &  12200  &  0.36  &    45    &  3\,$\times$\,1200 \\
2000 February 5 & 4779-5199  &  0.21  &  12500  &  0.38  &   310    &  3\,$\times$\,1200 \\
2000 February 5 & 8363-8763  &  0.39  &  12200  &  0.36  &   310    &  3\,$\times$\,1200  \\
\hline
\multicolumn{7}{l}{$^a$R$_{\rm{FWHM}}$\,=\,$\lambda$/$\Delta\lambda_{\rm{FWHM}}$}
\end{tabular}
\label{journal1}
\end{table}


In the cases of NGC\,2903 and NGC\,3310 two different slit positions were
chosen in each case to observe 4 and 8 CNSFRs, respectively, which we have  
labelled S1 and S2 for each galaxy. Besides, for NGC\,3310 we observed the
conspicuous Jumbo region, labelled J, which is the same as R19 of
\citetex{2000MNRAS.311..120D} and region A of
\citetex{1993MNRAS.260..177P}. The name of 
this region, Jumbo, comes from the fact that it is 10 times more luminous than
30 Dor \cite{1984ApJ...284..557T}. For the third galaxy, NGC\,3351, three
different slit positions (S1, S2 and S3) were chosen to observe 5 CNSFRs. In
all the cases one of the slits passes across the nucleus.

Several bias and sky flat field frames were taken at the beginning and the end
of each night in both arms. In addition, two lamp flat field and one
calibration lamp exposure per each telescope position were performed. The
calibration lamp used was CuNe+CuAr.


\begin{figure}
\centering
\includegraphics[width=0.98\textwidth]{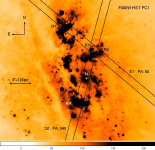}
\caption[F606W (wide V) image centred on NGC\,2903 obtained with the WFPC2
  camera]{F606W (wide V) image centred on NGC\,2903 obtained with the WFPC2
  camera (PC1) of the HST. The orientation is north up, east to the left. The
  location and P.A. of the WHT-ISIS slit positions, together with
  identifications of the CNSFRs extracted, are marked.} 
\label{hst-slits-2903-1}
\end{figure}

\begin{figure}
\centering
\includegraphics[width=0.98\textwidth]{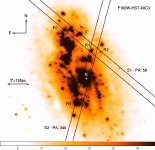}
\caption[HST-NICMOS image centred on NGC\,2903 obtained through the F160W
 filter]{HST-NICMOS image centred on NGC\,2903 obtained through the F160W
 filter. The orientation is north up, east to the left. The location and
 P.A. of the WHT-ISIS slit positions, together with identifications of the
 CNSFRs extracted, are marked.}
\label{hst-slits-2903-2}
\end{figure}

\begin{figure}
\vspace{3.5cm}
\centering
\includegraphics[width=0.98\textwidth]{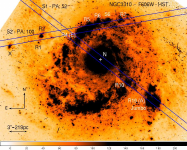}
\caption[Idem as Figure \ref{hst-slits-2903-1} for NGC\,3310]{Idem as Figure
  \ref{hst-slits-2903-1} for NGC\,3310.} 
\label{hst-slits-3310-1}
\end{figure}

\begin{figure}
\vspace{3.5cm}
\centering
\includegraphics[width=0.98\textwidth]{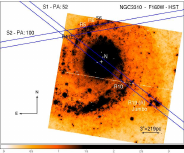}
\caption[Idem as Figure \ref{hst-slits-2903-2} for NGC\,3310]{Idem as Figure
  \ref{hst-slits-2903-2} for NGC\,3310.} 
\label{hst-slits-3310-2}
\end{figure}

\begin{figure}
\centering
\includegraphics[width=0.98\textwidth]{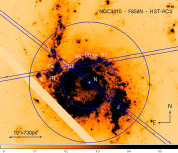}
\caption[F658N image centred on NGC\,3310 obtained with the ACS-HST
  camera]{F658N (narrow-band [N{\sc ii}] filter, at the redshift of NGC\,3310, 
  z\,=\,0.003312, equivalent to the H$\alpha$ narrow band filter) image
  centred on NGC\,3310 obtained with the ACS camera of the HST. The
  orientation is north up, east to the left. The location and P.A. of the
  WHT-ISIS slit positions, together with identifications of the CNSFRs
  extracted, are marked. The radii of the circles, centred at the position of
  the nucleus, are 8 and 18\,\arcsec.}
\label{hst-slits-3310-ACS}
\end{figure}

\begin{figure}
\centering
\includegraphics[width=0.98\textwidth]{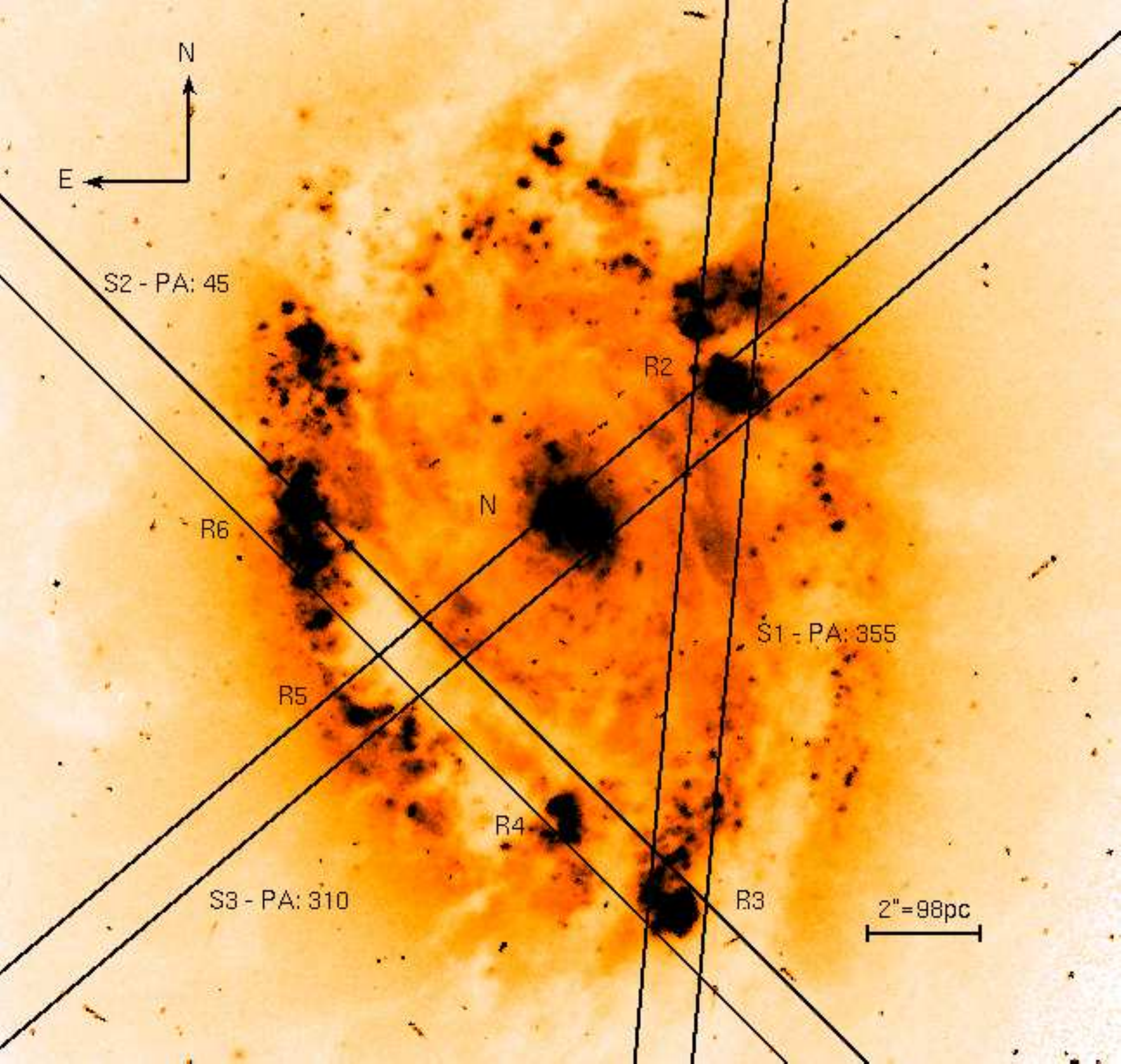}
\caption[Idem as Figure \ref{hst-slits-2903-1} for NGC\,3351]{Idem as Figure
  \ref{hst-slits-2903-1} for NGC\,3351.} 
\label{hst-slits-3351-1}
\end{figure}

\begin{figure}
\centering
\includegraphics[width=0.98\textwidth]{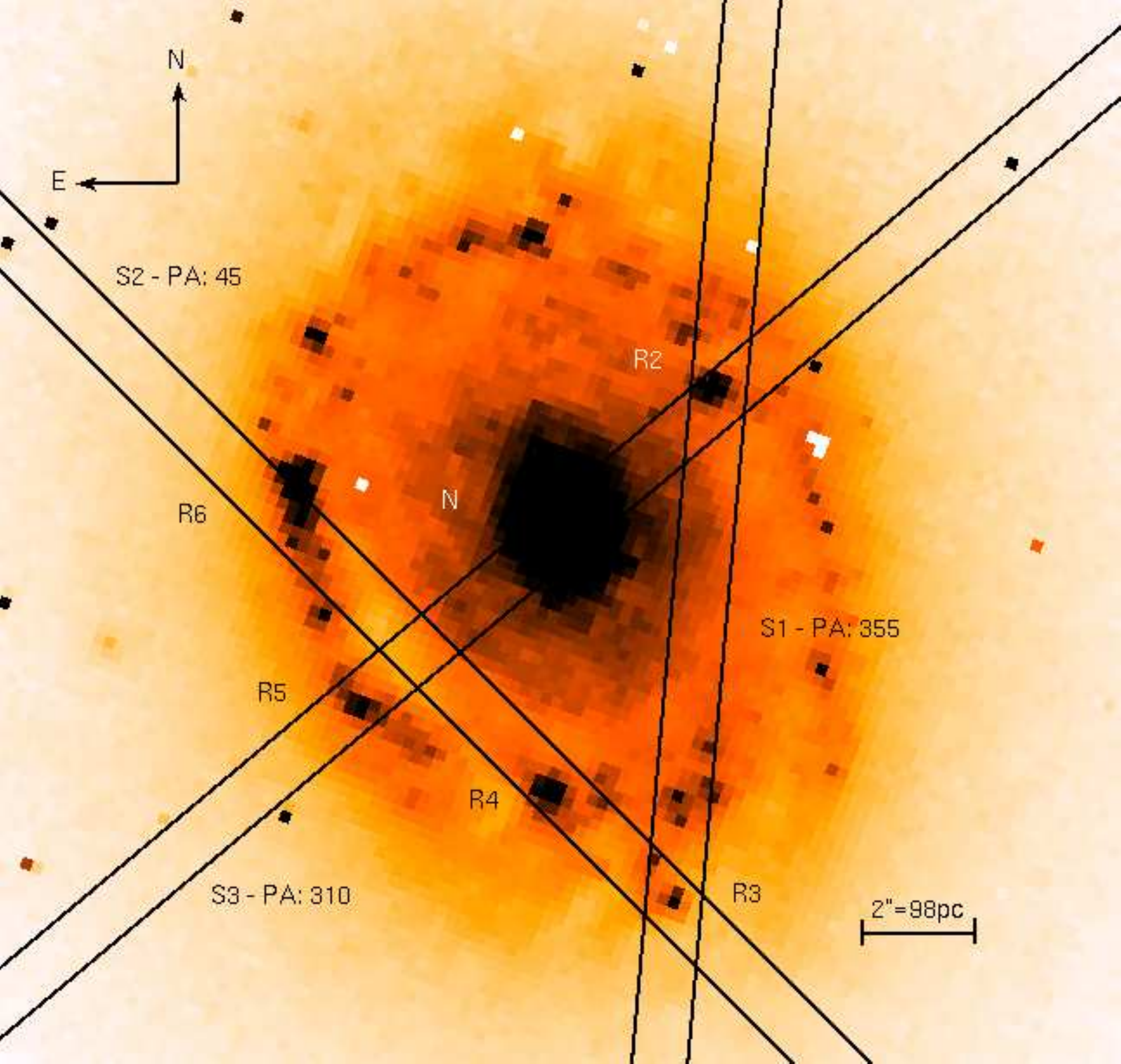}
\caption[Idem as Figure \ref{hst-slits-2903-2} for NGC\,3351]{Idem as Figure
  \ref{hst-slits-2903-2} for NGC\,3351.} 
\label{hst-slits-3351-2}
\end{figure}


We have also downloaded two astrometrically and photometrically calibrated
broad-band images of the central part of each observed galaxy from the
Multimission Archive at Space  
Telescope\footnote{http://archive.stsci.edu/hst/wfpc2}. The images were taken
through the F606W (wide V)  and the F160W (H) filters with the
Wide Field and Planetary Camera 2 (WFPC2; PC1) and the Near-Infrared Camera
and Multi-Object Spectrometer (NICMOS) 2 (NIC2; for NGC\,3310) and 3
(NIC3; for NGC\,2903 and NGC\,3351), both cameras on-board the HST. They
are displayed in Figures \ref{hst-slits-2903-1}-\ref{hst-slits-3310-2} and
\ref{hst-slits-3351-1}-\ref{hst-slits-3351-2}, with
the slit positions overlaid and the CNSFRs and the nucleus labelled. In the
case of NGC\,3310 we also download the F658N narrow band image (equivalent to
H$\alpha$ filter at the NGC\,3310 redshift) taken with the Advanced Camera for
Surveys (ACS) of the HST (see Figure \ref{hst-slits-3310-ACS}).

\subsection{Data reduction}

The data was processed and analyzed using IRAF\footnote{IRAF: the Image
  Reduction and Analysis Facility is distributed by the National Optical
  Astronomy Observatories, which is operated by the Association of
  Universities for Research in Astronomy, Inc. (AURA) under cooperative
  agreement with the National Science Foundation (NSF).} routines in the usual
manner. The procedure includes the removal of cosmic rays, bias subtraction,
division by a normalized flat field and wavelength calibration. Wavelength
fits were performed using 20-25 arc lines in the blue and 10-15 lines in the
far-red by a polynomial of second to third order. These fits have been done at
50 and 60 locations along the slit in the blue and far-red, respectively, and
they have yielded rms residuals between $\sim$0.1 and $\sim$0.2\,px. 

In the red, we also performed the wavelength calibration using sky lines
following the work by \citetex{1996PASP..108..277O}. However, although  more
lines (20-25) were available for the calibration, the fits gave higher rms
residuals, between $\sim$0.3 and $\sim$0.45\,px since the low intensities
of some of the lines did not allow a good Gaussian fit. We have therefore
adopted the calibration made using only the arc lines.

Background subtraction was performed using the spectra at both ends of the
slit. This background includes  light from the disc and bulge of the galaxy. 
It was almost impossible to neatly subtract the background bright emission
lines in the far-red spectra due to the non-uniform and extended nebular
emission surrounding each cluster and the variation over time 
of the sky emission lines. It is worth noting that these spurious features do
not affect the CaT absorption lines.

We have not corrected the spectra for atmospheric extinction or performed any
flux calibration, since our purpose was to measure radial velocities and
velocity dispersions.

In addition to the galaxy frames, observations of 11 template velocity stars
were made (four during the first night and seven during the second) to provide
good stellar reference frames in the same system as the galaxy spectra for the
kinematic analysis in the far-red. They are late-type giant and supergiant
stars which have strong CaT features (see \citeplain{1989MNRAS.239..325D}). In
Table \ref{templates} we list the spectral types and luminosity classes of the
stars and the dates of observation.

\begin{table}
\centering
\caption{Stellar reference frames.}
\begin{tabular} {l l l}
\hline
Star      & ST\,-\,LC & date   \\
\hline
HD71952  &   K0\,I     &    04-02-2000   \\
HD129972 &   G6\,I     &       \\
HD134047 &   G6\,III   &       \\
HD144063 &   G4\,III   &       \\
\hline
HD16400 & G5\,III     &    05-02-2000  \\
HD22007  & G5\,I       &   \\
HD22156  & G6\,III     & \\
HD92588  &  K1\,I       & \\
HD102165 & F7\,I       & \\
HD115004 &  G8\,III     & \\
HD116365 &  K3\,III     & \\
\hline
\end{tabular}
\label{templates}
\end{table}


\section{Results}
\label{resul-kine}

Figures \ref{hst-slits-2903-1}-\ref{hst-slits-3310-2} and
\ref{hst-slits-3351-1}-\ref{hst-slits-3351-2} show the selected 
slits for each galaxy, superimposed on photometrically calibrated optical and
infrared  images of the circumnuclear region of these galaxies acquired with
the Wide Field and Planetary Camera 2 (WFPC2; PC1) and the Near-Infrared
Camera and Multi-Object Spectrometer (NICMOS) Camera 2 and 3 (NIC2 and NIC3)
on board the HST. These images have been downloaded  from the Multimission
Archive at STScI (MAST)\footnote{http://archive.stsci.edu/hst/wfpc2}. The
optical image was obtained through the F606W (wide V) filter, and the near-IR
one, through the F160W (H). For NGC\,2903 and NGC\,3351, the CNSFRs have been
labelled following the same nomenclature as in \citetex{1997A&A...325...81P},
and for NGC\,3310 following \citetex{2000MNRAS.311..120D}, with the
nomenclature given by \citetex{1993MNRAS.260..177P} for the regions in common 
[E for R4 and A for R19 (the main knot of the Jumbo
region)]. In both  studies the regions 
observed are identified on the H$\alpha$ maps. In Figures
\ref{hst-slits-2903-1} and \ref{hst-slits-2903-2} the plus symbol represents
the position of the nucleus as given by the Two Micron All Sky Survey Team,
2MASS Extended Objects - Final Release \cite{2003AJ....125..525J}. In the case
of NGC\,3310, the position of the nucleus mark in Figures
\ref{hst-slits-3310-1} and \ref{hst-slits-3310-2} is given by
\citetex{1999PASP..111..438F}.

\begin{figure}
\centering
\vspace{0.3cm}
\includegraphics[width=.41\textwidth,angle=0]{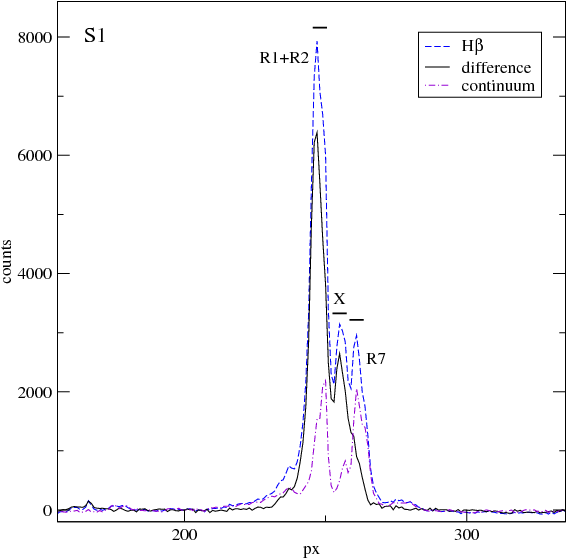}
\hspace{0.2cm}
\includegraphics[width=.41\textwidth,angle=0]{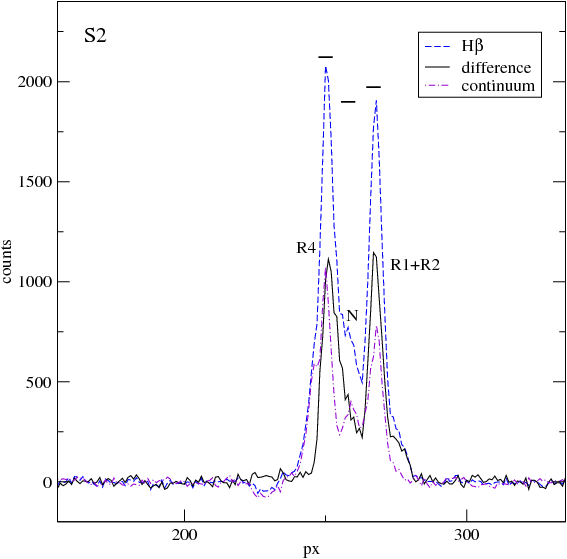}\\
\vspace{0.3cm}
\includegraphics[width=.41\textwidth,angle=0]{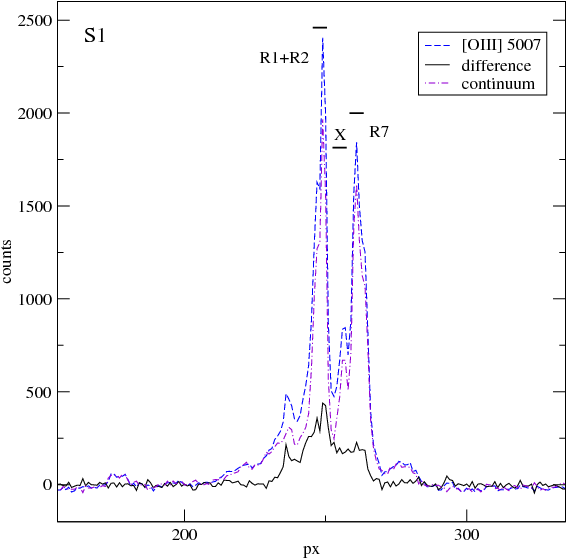}
\hspace{0.2cm}
\includegraphics[width=.41\textwidth,angle=0]{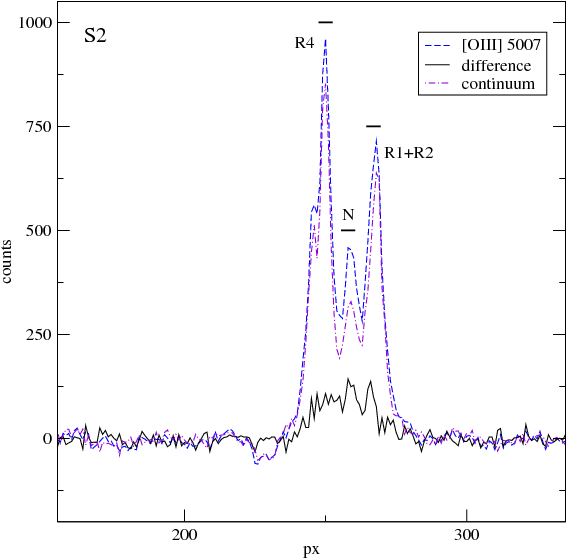}\\
\vspace{0.3cm}
\includegraphics[width=.41\textwidth,angle=0]{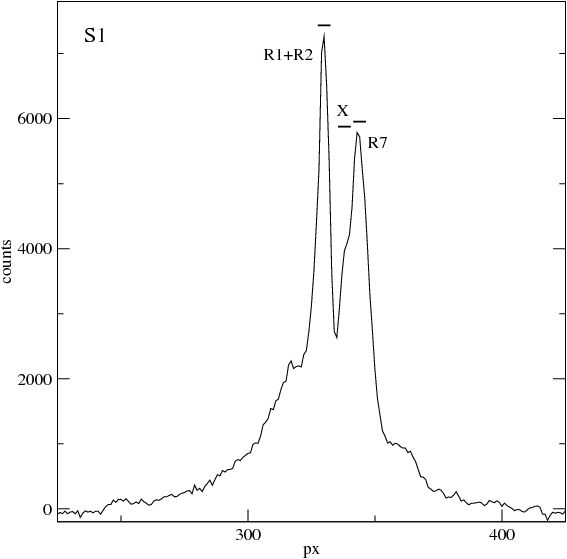}
\hspace{0.2cm}
\includegraphics[width=.41\textwidth,angle=0]{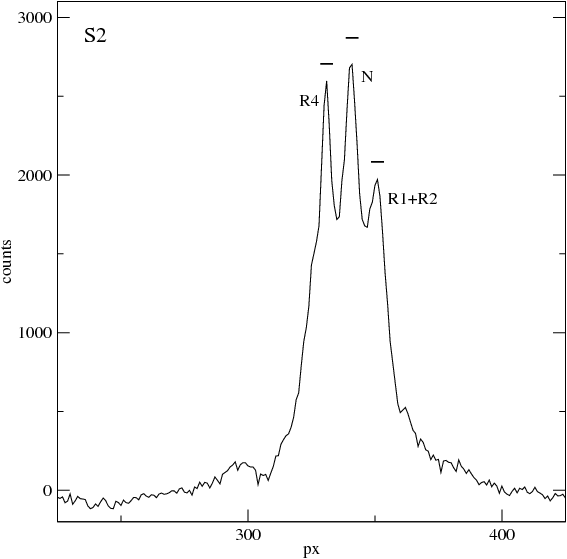}\\
\caption[Spatial profiles of H$\beta$, [O{\sc iii}{\textrm
  ]}\,$\lambda$\,5007\AA\ and the far red for each slit of NGC\,2903.]{Spatial
  profiles of H$\beta$, [O{\sc iii}]\,$\lambda$\,5007\AA\ and the far red
  (upper, middle and lower panels respectively) for each slit of
  NGC\,2903. For the emission lines, the profiles correspond to  
  line+continuum (dashed line), continuum (dashed-dotted line) and the 
  difference between them (solid line), representing the pure emission from
  H$\beta$ and [O{\sc iii}] respectively. For the far red profiles, the solid
  lines represent the continuum. Pixel number increases to the
  North. Horizontal small lines show the location of the CNSFRs and nuclear
  apertures.} 
\label{profiles2903}
\end{figure}

\begin{figure}
\centering
\includegraphics[width=.41\textwidth,angle=0]{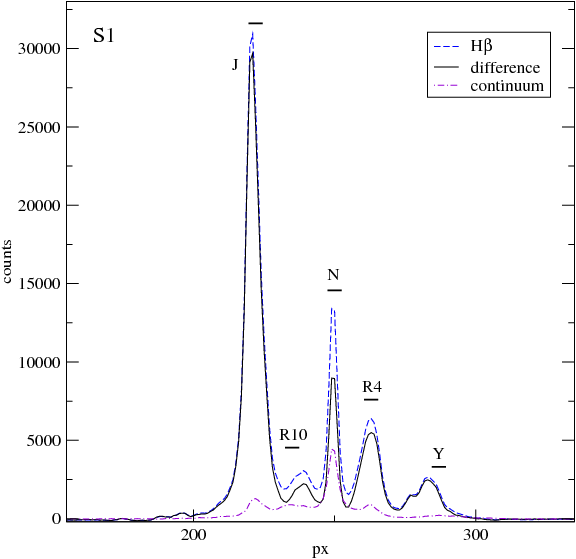}
\hspace{0.2cm}
\includegraphics[width=.41\textwidth,angle=0]{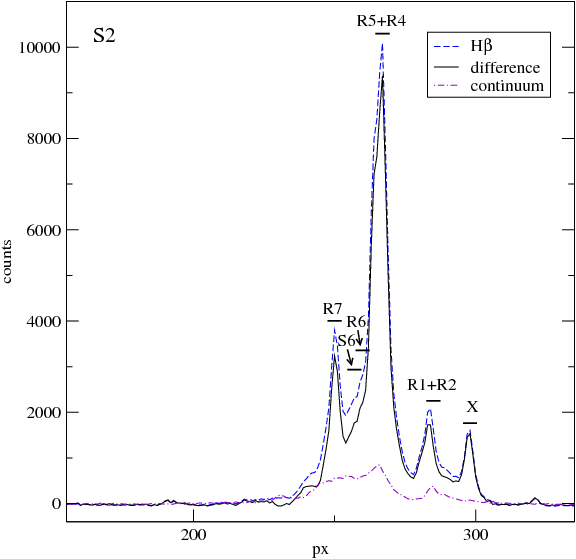}\\
\vspace{0.3cm}
\includegraphics[width=.41\textwidth,angle=0]{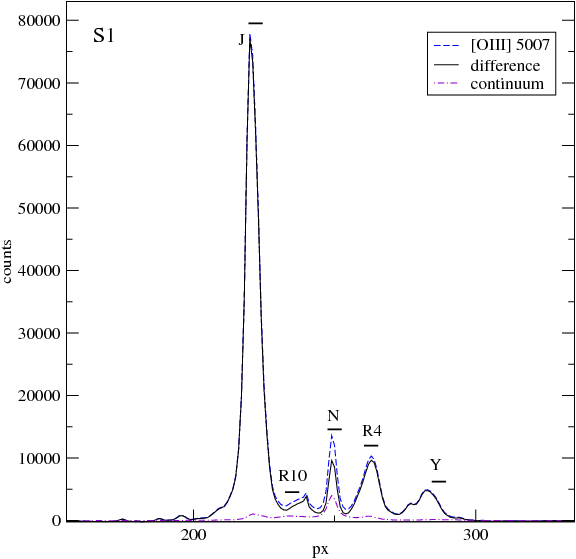}
\hspace{0.2cm}
\includegraphics[width=.41\textwidth,angle=0]{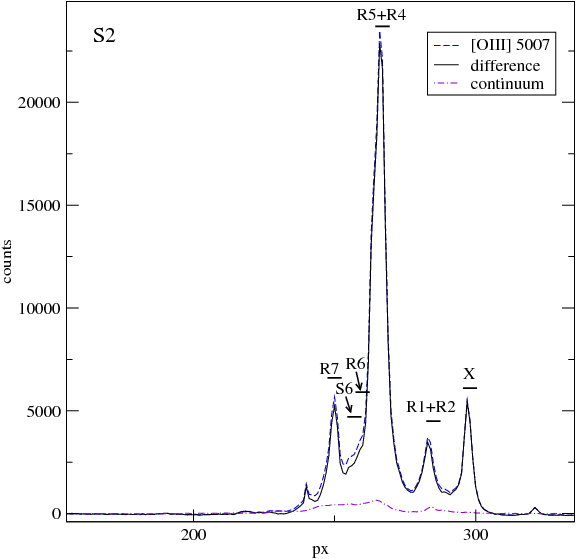}\\
\vspace{0.3cm}
\includegraphics[width=.41\textwidth,angle=0]{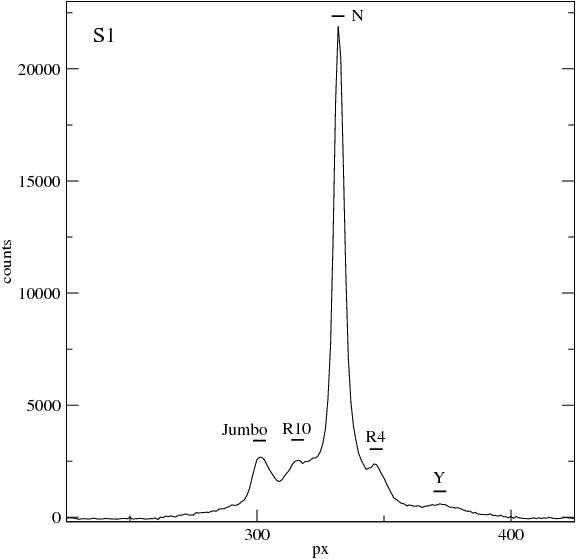}
\hspace{0.2cm}
\includegraphics[width=.41\textwidth,angle=0]{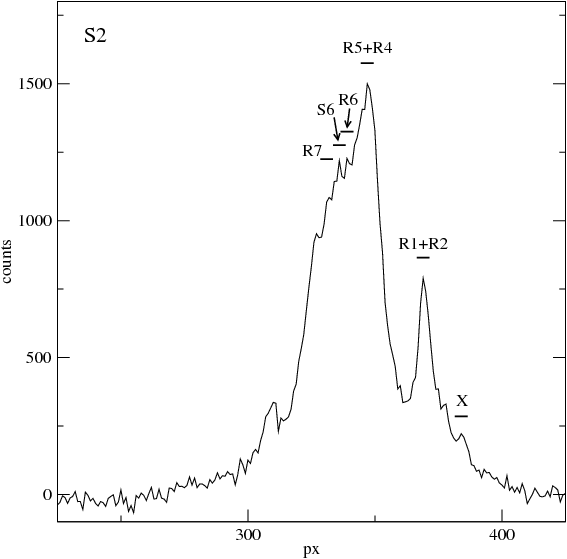}\\
\caption{Idem as Figure \ref{profiles2903} for NGC\,3310.}
\label{profiles3310}
\end{figure}

\begin{figure}
\centering
\includegraphics[width=.41\textwidth,angle=0]{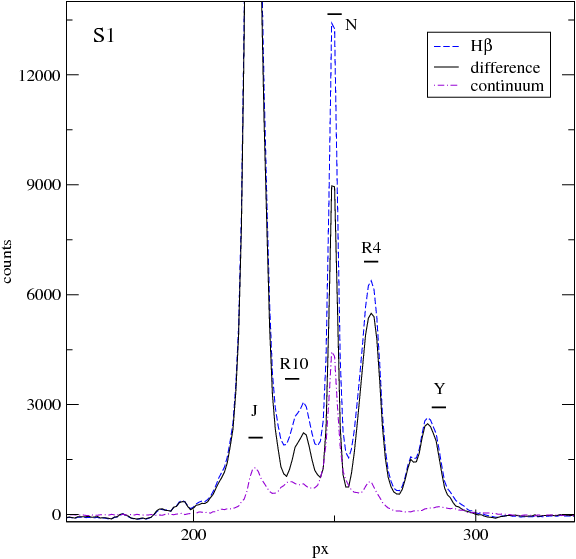}
\hspace{0.2cm}
\includegraphics[width=.41\textwidth,angle=0]{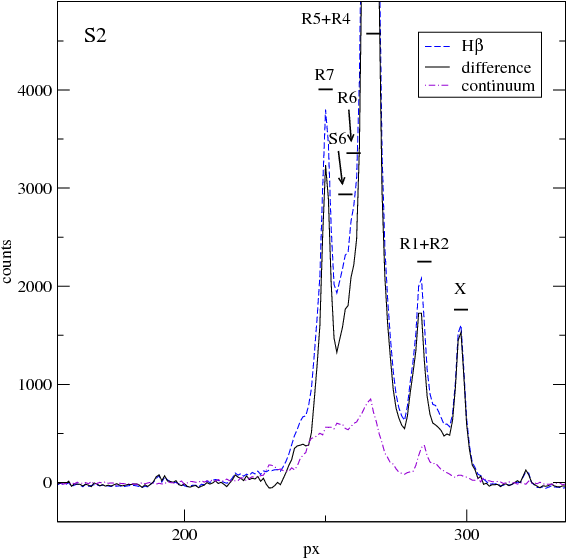}\\
\vspace{0.3cm}
\includegraphics[width=.41\textwidth,angle=0]{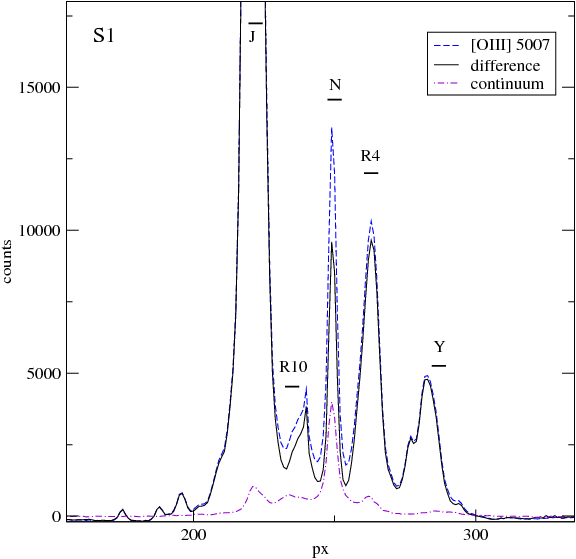}
\hspace{0.2cm}
\includegraphics[width=.41\textwidth,angle=0]{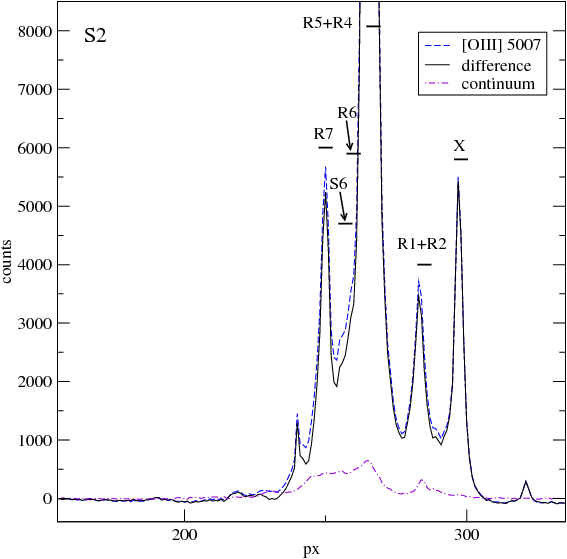}\\
\vspace{0.3cm}
\includegraphics[width=.41\textwidth,angle=0]{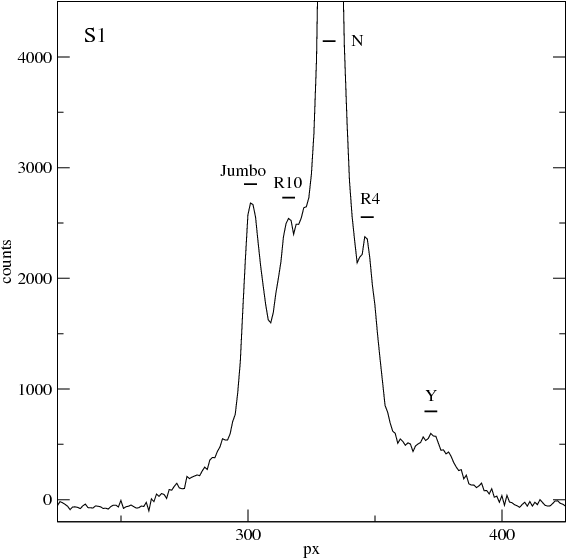}
\hspace{6.5cm}
\caption[An enlargement of the spatial profiles presented in Figure
  \ref{profiles3310}]{An enlargement of the spatial profiles presented in
  Figure \ref{profiles3310}, except in the far red range of S2.} 
\label{profiles3310-2}
\end{figure}

Figures \ref{profiles2903}, \ref{profiles3310} and \ref{profiles3351} show the
spatial profiles in the H$\beta$ and  [O{\sc iii}]\,5007\,\AA\ emission lines
(upper and middle panels) and the far-red continuum (lower panel) along each
slit position for each observed galaxy. 
For NGC\,3310 (\ref{profiles3310}) due to
the presence of intense regions (J in the blue and the galaxy nucleus
in the red range in the case of S1, and R5+R4 in the blue range for S2) the
profile details are very 
difficult to appreciate, therefore we show some enlargements of these profiles
in Figure \ref{profiles3310-2}. In all cases, the emission line
profiles have been generated by collapsing 11 pixels of the spectra in the
direction of the resolution at the central position of the lines in the rest
frame, $\lambda$\,4861 and $\lambda$\,5007\AA\ respectively, and are plotted
as dashed lines. Continuum profiles were generated by collapsing 11 resolution
pixels centred at 11\,\AA\ to the blue of each emission line and are
plotted  
as dash-dotted lines. The difference between the two, shown by a solid line, 
corresponds to the pure emission. The far-red
continuum has been generated by collapsing 11 pixels centered at
$\lambda$\,8620\AA.


\begin{figure}
\centering
\includegraphics[width=.41\textwidth,angle=0]{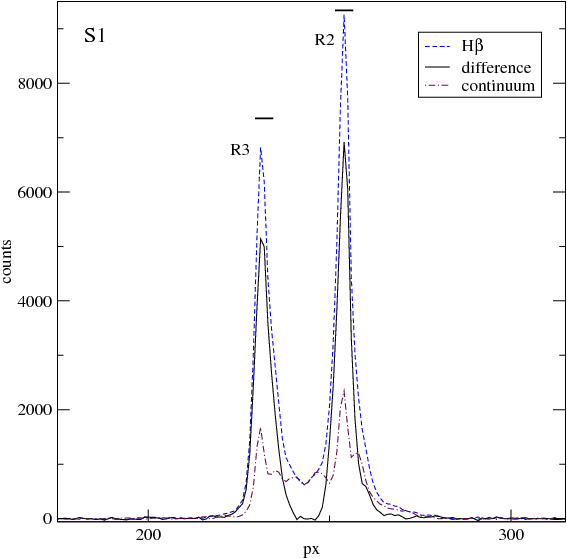}
\hspace{0.2cm}
\includegraphics[width=.41\textwidth,angle=0]{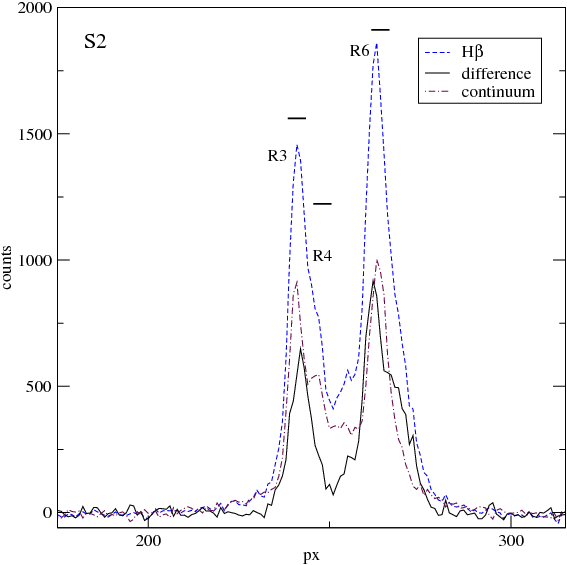}\\
\vspace{0.3cm}
\includegraphics[width=.41\textwidth,angle=0]{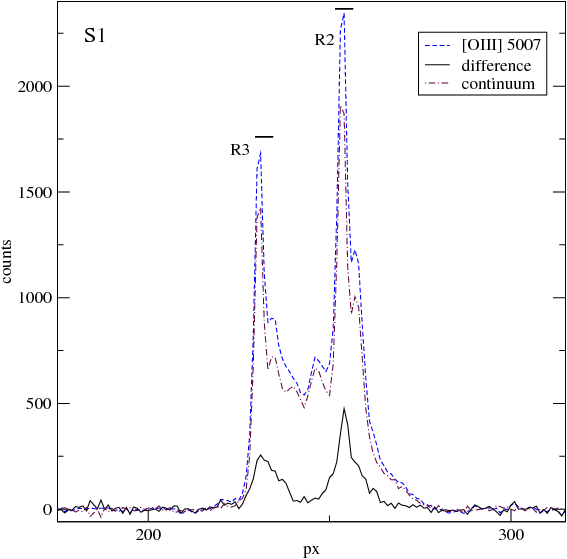}
\hspace{0.2cm}
\includegraphics[width=.41\textwidth,angle=0]{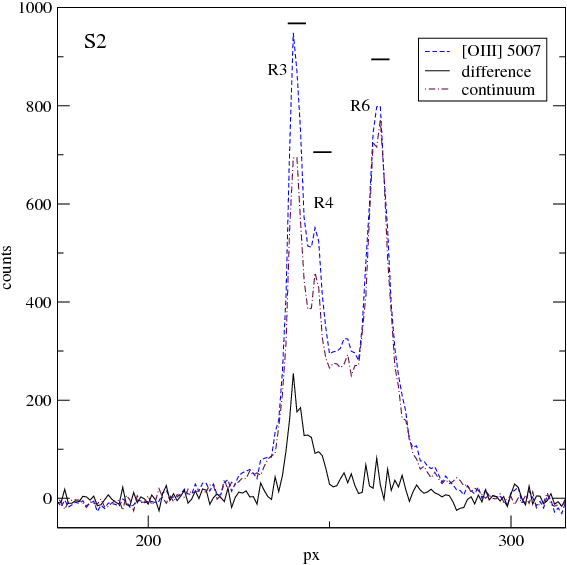}\\
\vspace{0.3cm}
\includegraphics[width=.41\textwidth,angle=0]{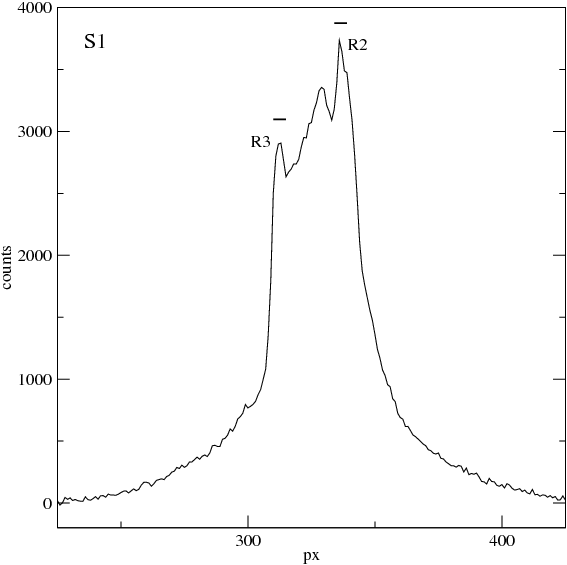}
\hspace{0.2cm}
\includegraphics[width=.41\textwidth,angle=0]{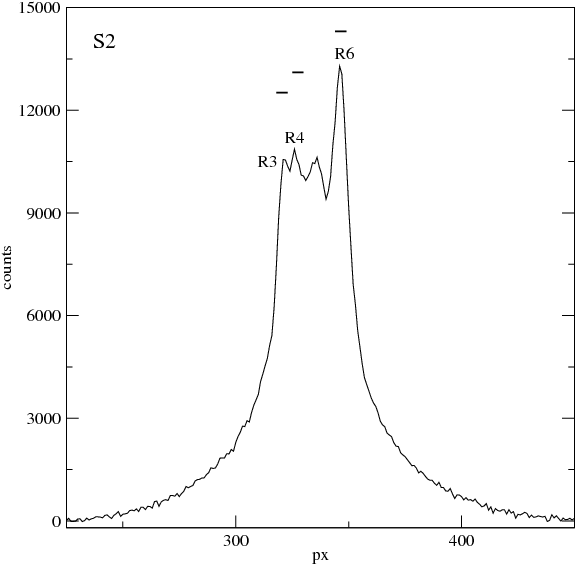}\\
\caption{Idem as Figure \ref{profiles2903} for S1 and S2 of NGC\,3351.}
\label{profiles3351}
\end{figure}

\setcounter{figure}{13}

\begin{figure}
\centering
\includegraphics[width=.41\textwidth,angle=0]{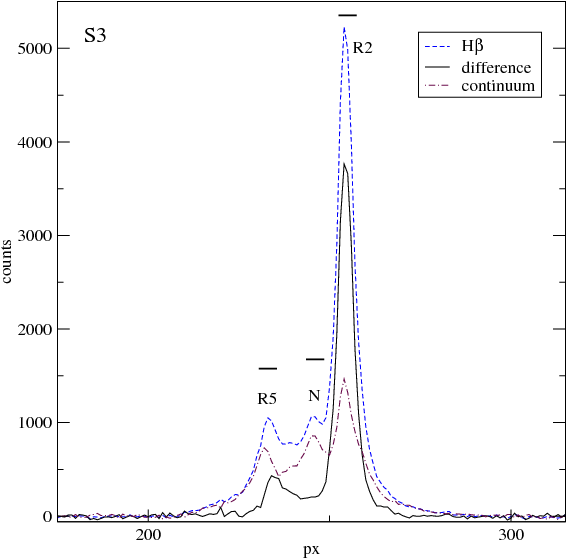}\\
\vspace{0.3cm}
\includegraphics[width=.41\textwidth,angle=0]{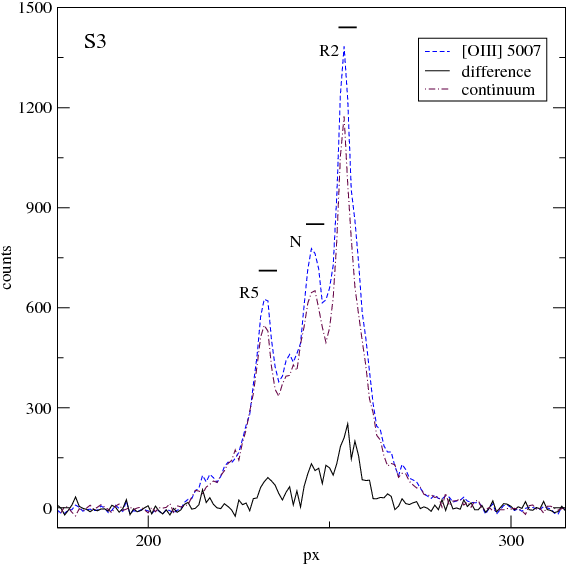}\\
\vspace{0.3cm}
\includegraphics[width=.41\textwidth,angle=0]{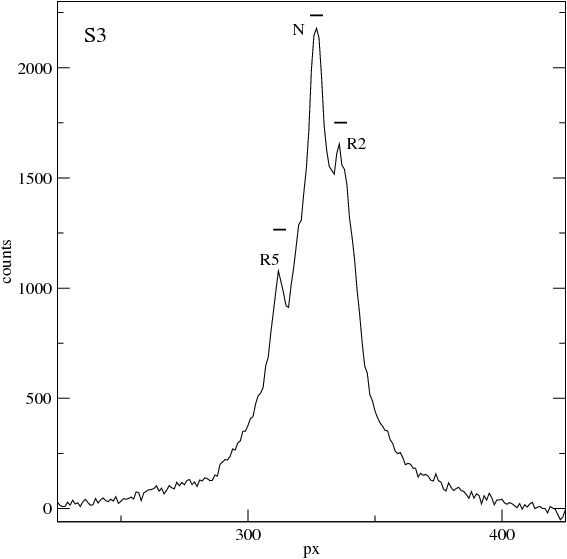}\\
\caption{({\it cont.}) Idem as Figure \ref{profiles2903} for S3 of NGC\,3351.}
\label{profiles3351}
\end{figure}


The regions of the frames to be extracted into one-dimensional spectra 
corresponding to each of the identified CNSFRs, were selected on the continuum
emission profiles both in the blue and in the 
red. These regions are marked by horizontal lines and labelled in the
corresponding 
Figures. In the H$\beta$ profile of slit position S1 of NGC\,2903, an almost
pure emission knot between regions R1+R2 and R7 can be seen with a very weak
continuum. We labelled this region as X. In the case 
of the [O{\sc iii}]\,5007\,\AA\ emission line profile, the main contributor
comes from  the continuum since the actual emission line is very weak as
expected for high metallicity \HII\ regions (\citeplain{2007MNRAS.382..251D},
see Chapter \S \ref{abundan}). For NGC\,3310, we find another two pure emission
knots, one for each slit position, labelled Y and X by us (see Figure
\ref{hst-slits-3310-ACS}), respectively. The former of these regions seems
to be located at the tip of one vertical arm formed by pure emission regions,
since it can be easily appreciable in the H$\alpha$ image from the ACS but it
is almost invisible in the WFPC2-PC1 V-band image (see Figures
\ref{hst-slits-3310-1} and \ref{hst-slits-3310-ACS}).

Spectra in slit position S1 of NGC\,2903 and S2 of NGC\,3310 are placed along 
the circumnuclear regions located to the North and North-East of the nuclei, 
and therefore any contribution from the underlying galaxy bulge is difficult
to assess. This is also the case for slit positions S1 and S2 of
NGC\,3351. Slit position S2 of NGC\,2903, S1 of NGC\,3310 and
S3 of NGC\,3351 cross the galactic nuclei. This can be used to
estimate the underlying bulge contribution. For the blue spectra the light
from the underlying 
bulge is almost negligible amounting to, at most, 5 per cent at the H$\beta$
line for NGC\,2903 and NGC\,3351, and 10 per cent for NGC\,3310. For the red
spectra, the bulge contribution is more important. From Gaussian fits to the
$\lambda$\,8620\,\AA\ continuum profile of S2 in NGC\,2903 we find that it
amounts to about 15 per cent for the lowest surface brightness region,
R1+R2. In the case of S1 in NGC\,3310, the contribution of the underlying
bulge is even more important, reaching 25 per cent for R4, the
weakest region, and the one closest to the nucleus. From the red profile of
slit position S3 of NGC\,3351 we find that this contribution amounts to about
20 per cent for the lowest surface brightness region, R5.

On the other hand, the analysis of the broad near-IR HST-NICMOS images
shown in Figures \ref{hst-slits-2903-2}, \ref{hst-slits-3310-2} and
\ref{hst-slits-3351-2} shows less contrast between the emission from the
regions and the underlying bulge which is very close to the image background
emission. In the case of NGC\,2903, its contribution is about 25 per cent for
the weak regions, R2 and R7 along position slit S1. We find a very similar
behaviour for the central zone of NGC\,3310, that is in very good
agreement with the cluster identification made by
\citetex{2002AJ....123.1381E} using the equivalent J and K-band HST-NICMOS
images and the ground-based data from KPNO (J and K-bands). For NGC\,3351,
similar values to those derived from the profiles are found from inspection
of the HST-NICMOS image.


\begin{figure}
\hspace{0.0cm}
\vspace{0.2cm}
\includegraphics[width=.48\textwidth,height=.30\textwidth]{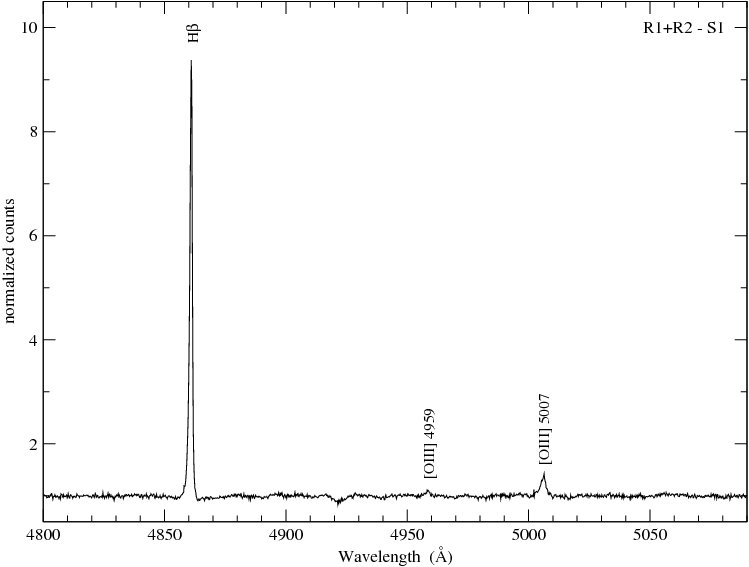}
\includegraphics[width=.48\textwidth,height=.30\textwidth]{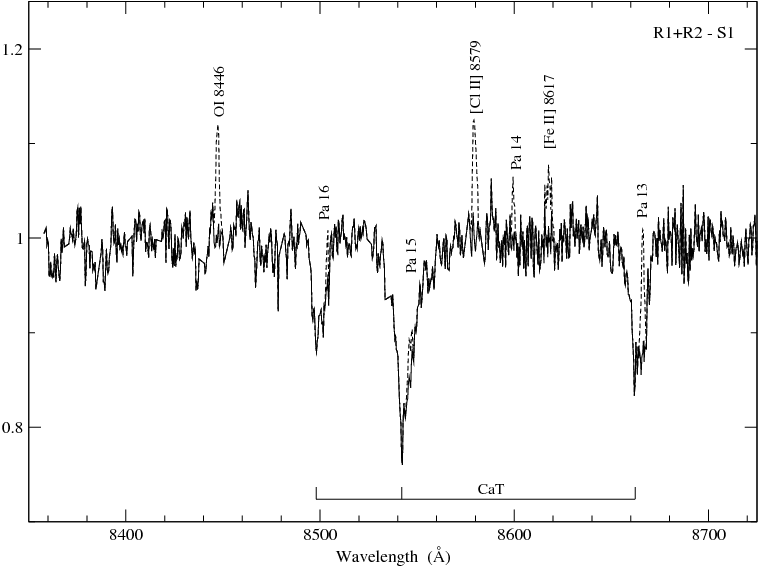}\\
\vspace{0.2cm}
\includegraphics[width=.48\textwidth,height=.30\textwidth]{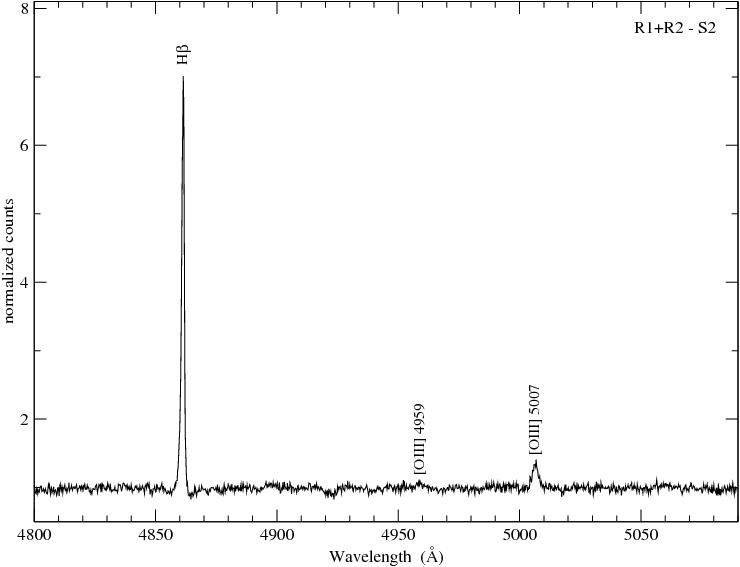}
\includegraphics[width=.48\textwidth,height=.30\textwidth]{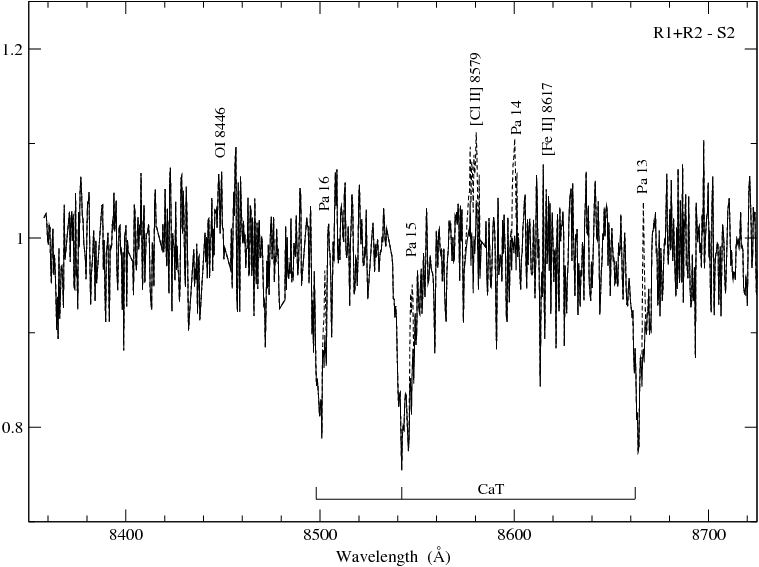}\\
\vspace{0.2cm}
\includegraphics[width=.48\textwidth,height=.30\textwidth]{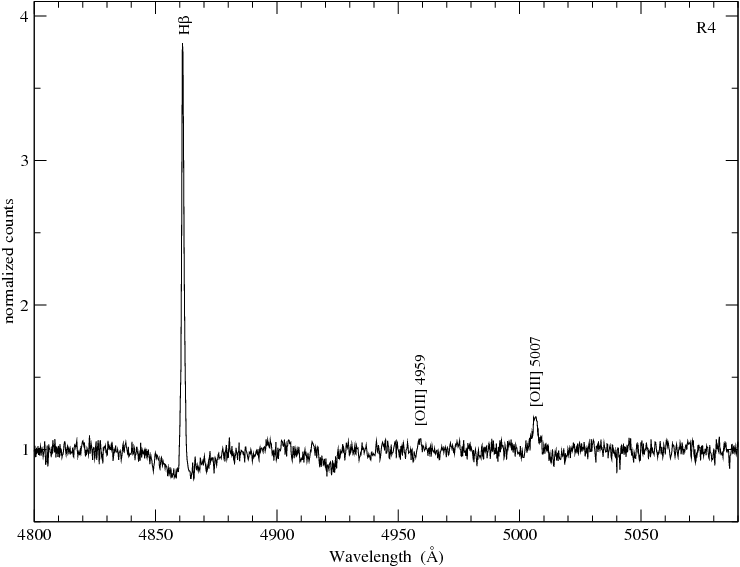}
\includegraphics[width=.48\textwidth,height=.30\textwidth]{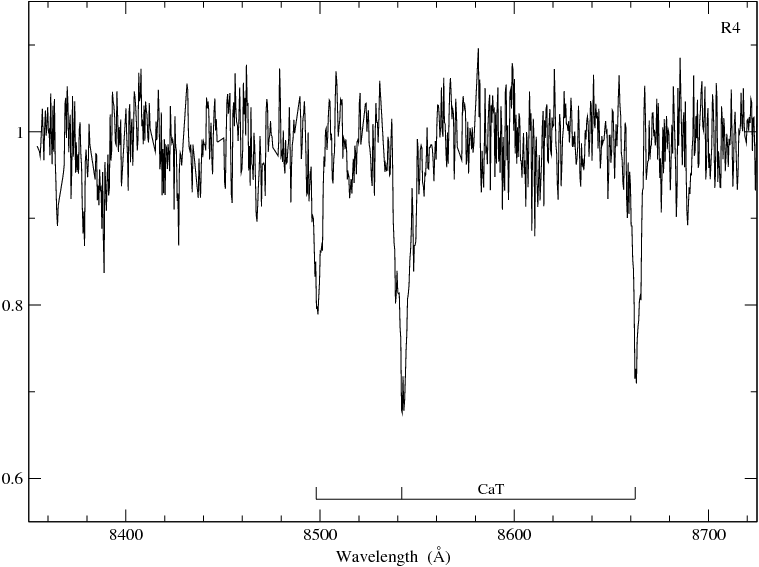}\\
\vspace{0.2cm}
\includegraphics[width=.48\textwidth,height=.30\textwidth]{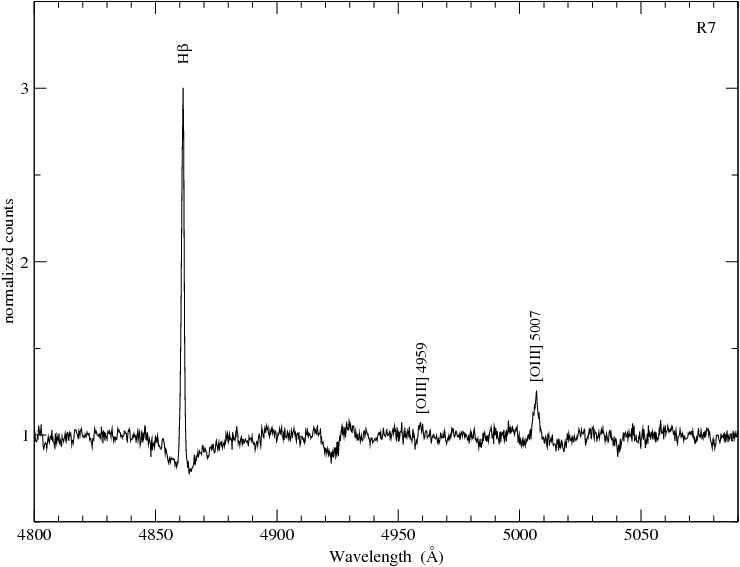}
\includegraphics[width=.48\textwidth,height=.30\textwidth]{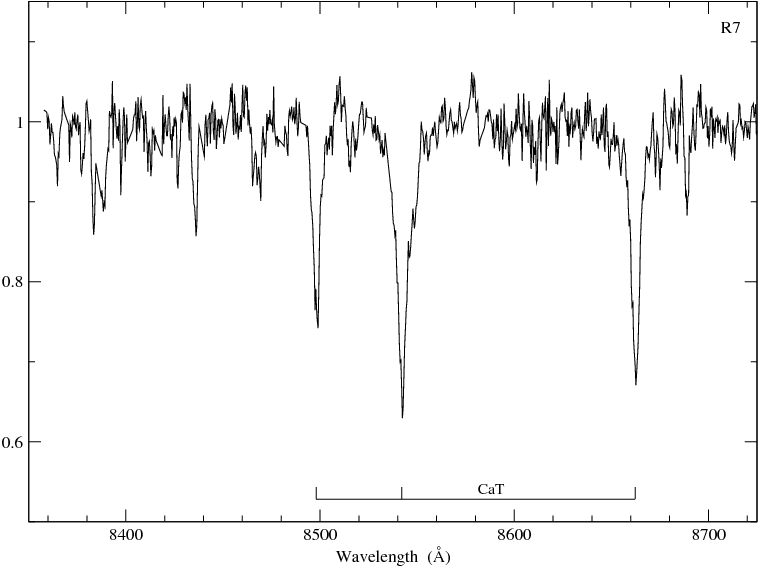}
\caption[Blue and red rest frame normalized spectra of the observed
  CNSFRs of NGC\,2903.]{Blue (left) and red (right) rest frame normalized
  spectra of the observed CNSFRs of NGC\,2903. For R1+R2 in the red range, the
  dashed line shows the obtained spectrum; the solid line represents the
  spectrum after subtracting the emission lines (see text).} 
\label{spectra2903}
\end{figure}


\begin{figure}
\hspace{0.0cm}
\vspace{0.2cm}
\includegraphics[width=.48\textwidth,height=.30\textwidth]{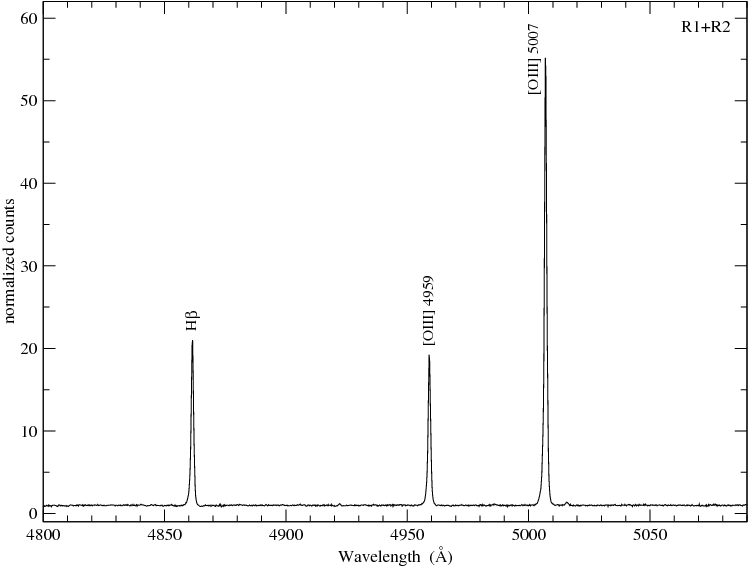}
\includegraphics[width=.48\textwidth,height=.30\textwidth]{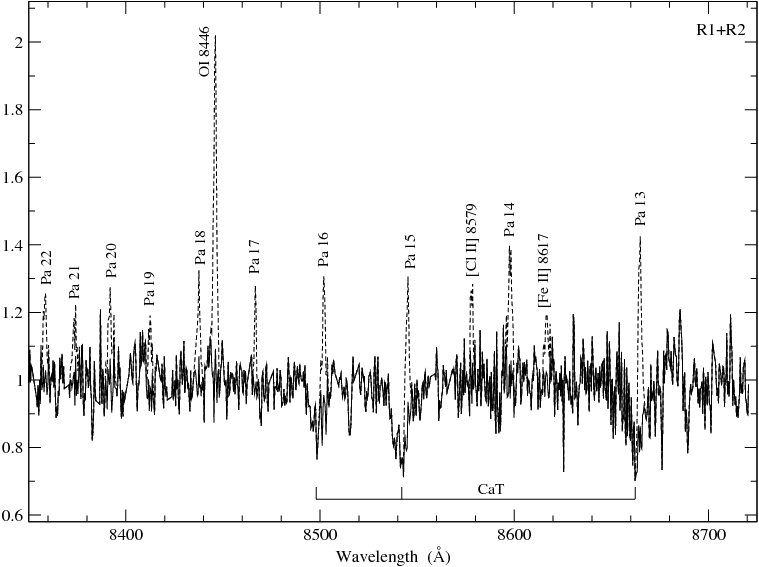}\\
\vspace{0.2cm}
\includegraphics[width=.48\textwidth,height=.30\textwidth]{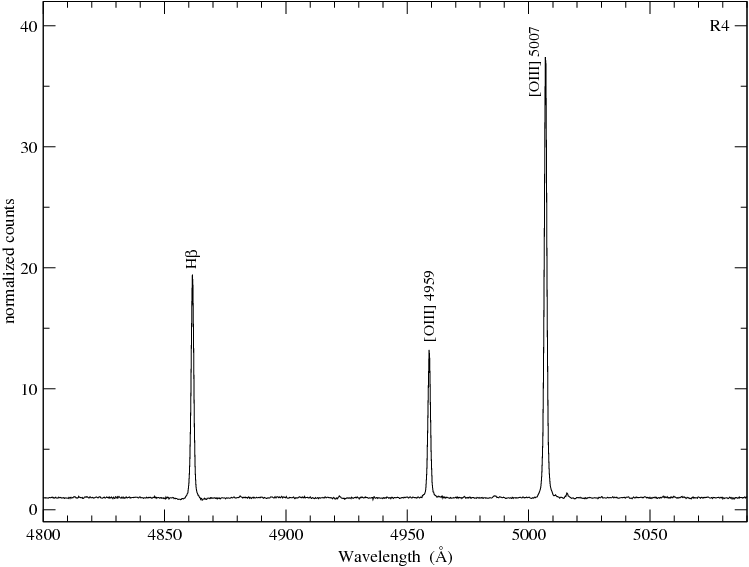}
\includegraphics[width=.48\textwidth,height=.30\textwidth]{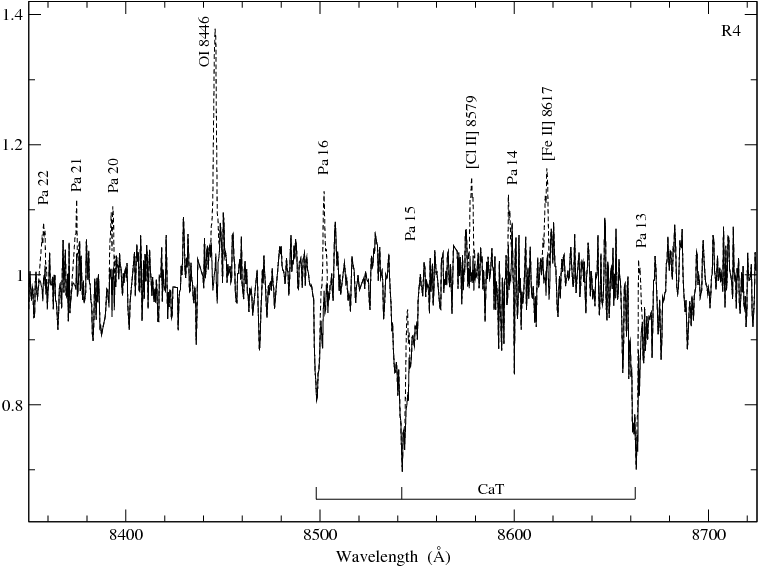}\\
\vspace{0.2cm}
\includegraphics[width=.48\textwidth,height=.30\textwidth]{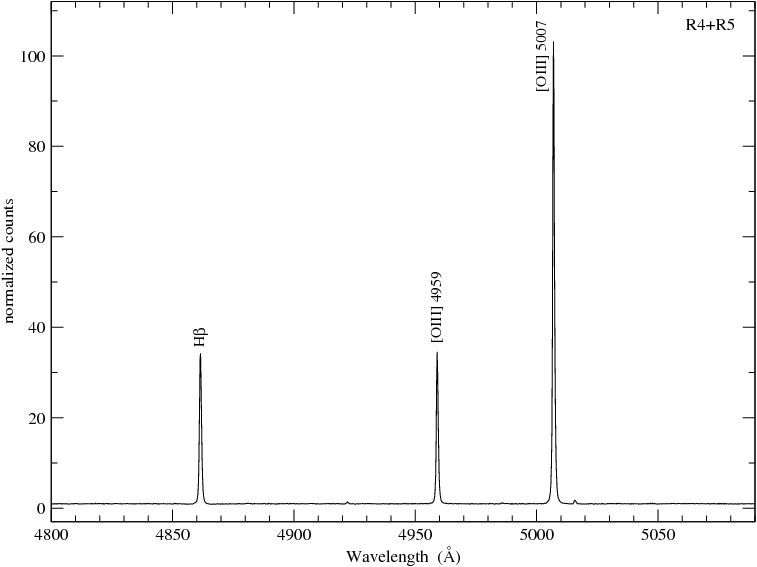}
\includegraphics[width=.48\textwidth,height=.30\textwidth]{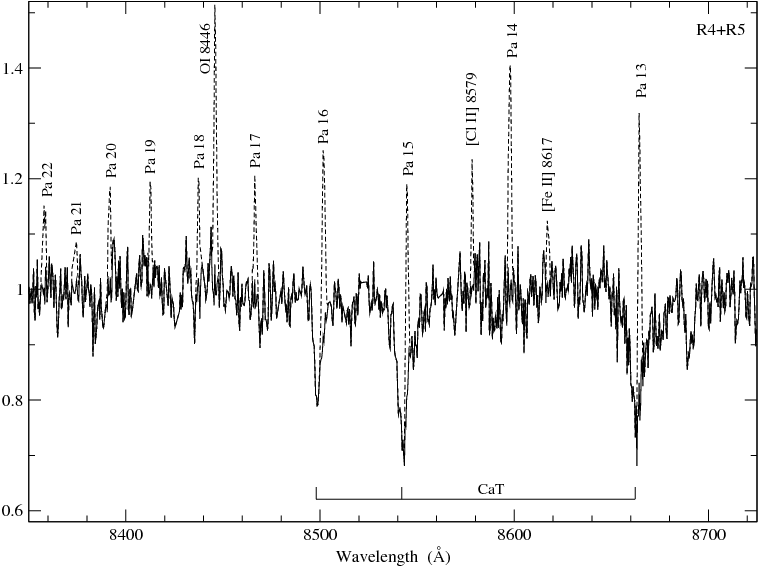}\\
\vspace{0.2cm}
\includegraphics[width=.48\textwidth,height=.30\textwidth]{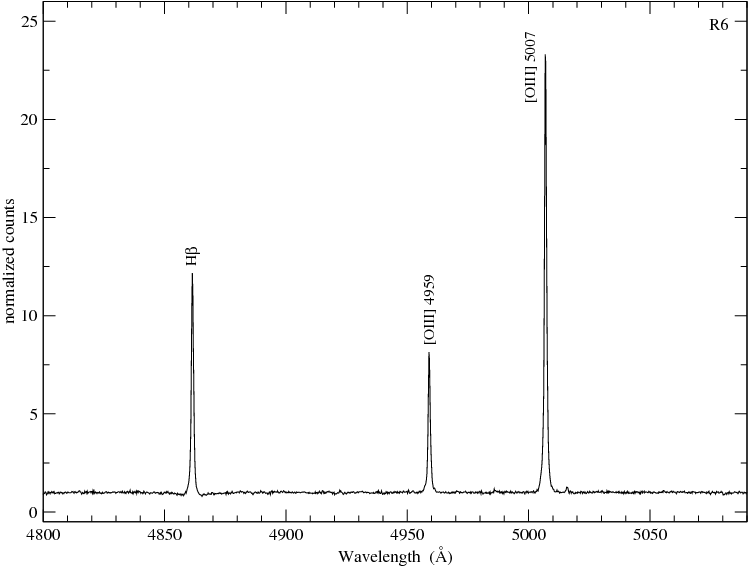}
\includegraphics[width=.48\textwidth,height=.30\textwidth]{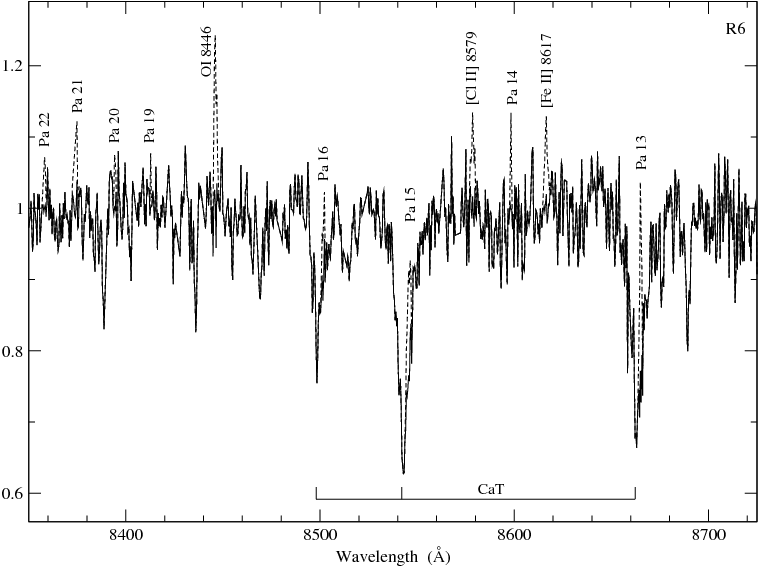}
\caption[Idem as Figure \ref{spectra2903} for the observed CNSFRs of
  NGC\,3310.]{Idem as Figure \ref{spectra2903} for the
  observed CNSFRs of NGC\,3310. For the red range, the dashed line shows the
  obtained spectrum; the solid line represents the spectrum after subtracting
  the emission lines (see text).}
\label{spectra3310}
\end{figure}

\setcounter{figure}{15}

\begin{figure}
\hspace{0.0cm}
\vspace{0.2cm}
\includegraphics[width=.48\textwidth,height=.30\textwidth]{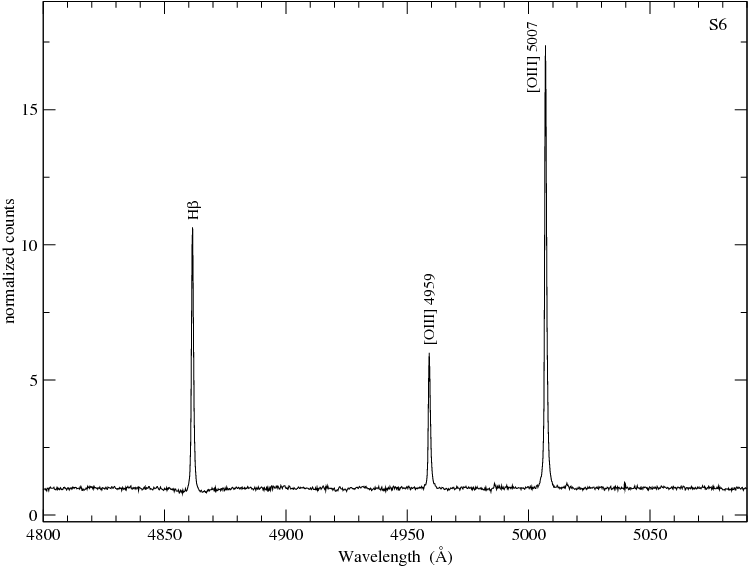}
\includegraphics[width=.48\textwidth,height=.30\textwidth]{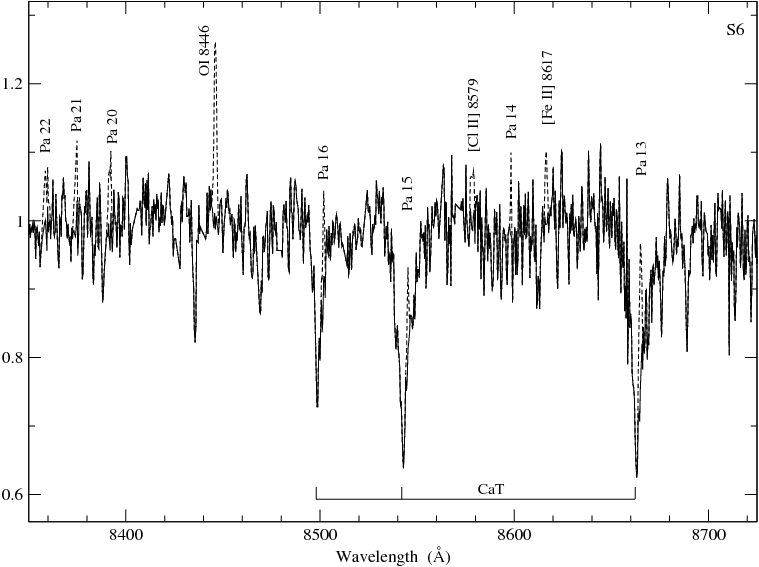}\\
\vspace{0.2cm}
\includegraphics[width=.48\textwidth,height=.30\textwidth]{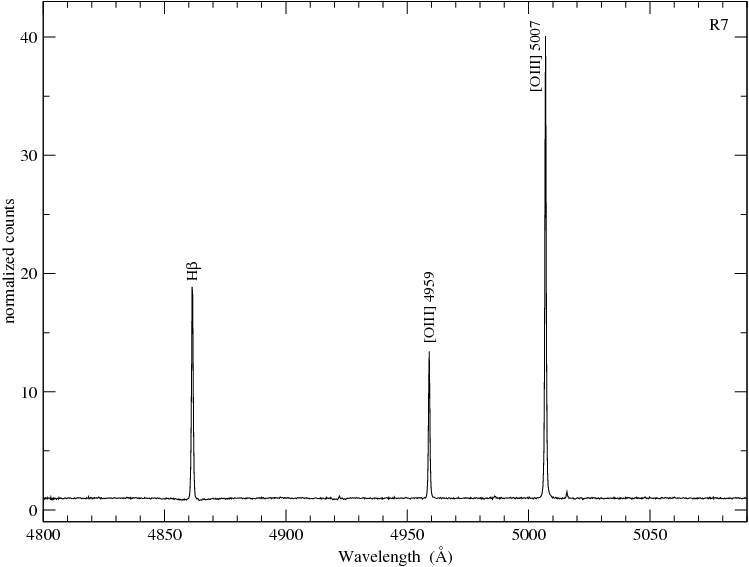}
\includegraphics[width=.48\textwidth,height=.30\textwidth]{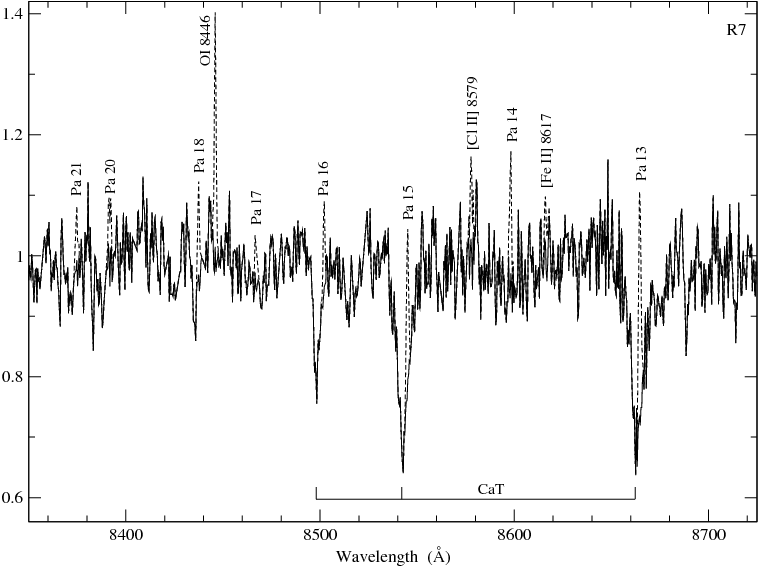}\\
\vspace{0.2cm}
\includegraphics[width=.48\textwidth,height=.30\textwidth]{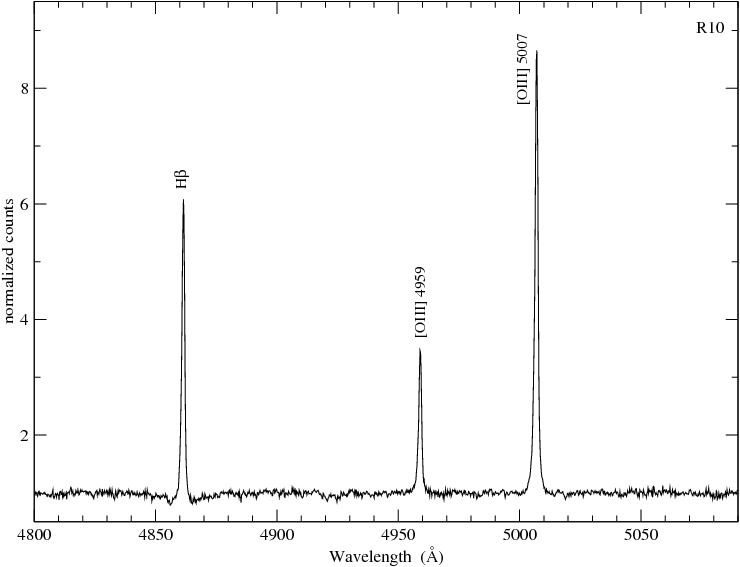}
\includegraphics[width=.48\textwidth,height=.30\textwidth]{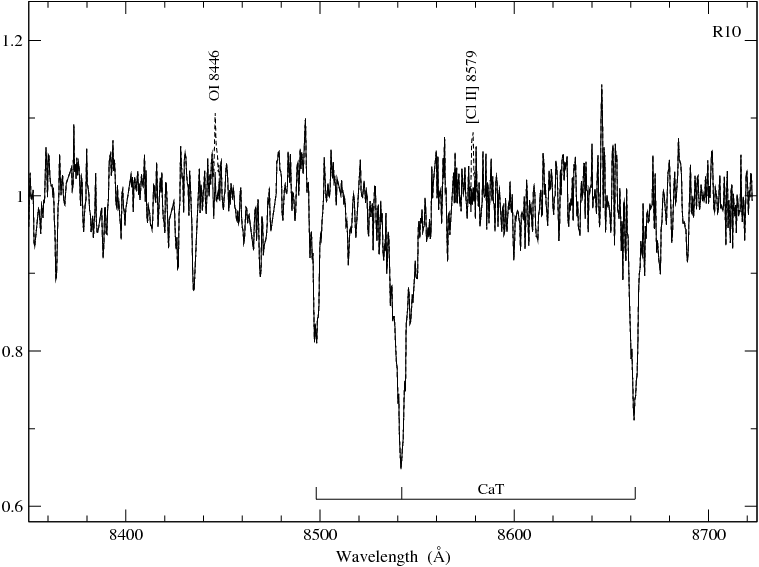}\\
\caption[({\it cont}) Idem as Figure \ref{spectra2903} for the
  observed CNSFRs of NGC\,3310.]{({\it cont}) Idem as Figure \ref{spectra2903}
  for the observed CNSFRs of NGC\,3310. For the red 
  range, the dashed line shows the obtained spectrum; the solid line
  represents the spectrum after subtracting the emission lines (see text).} 
\label{spectra3310}
\end{figure}


\begin{figure}
\hspace{0.0cm}
\vspace{0.2cm}
\includegraphics[width=.48\textwidth,height=.30\textwidth]{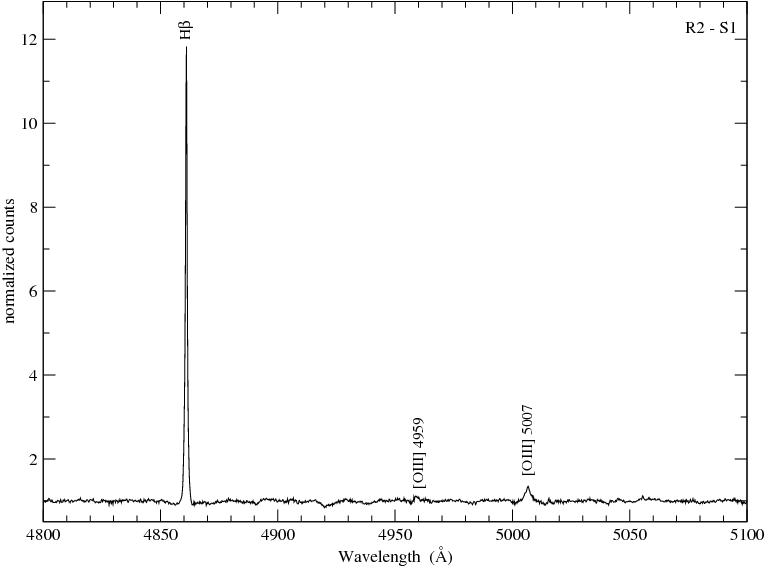}
\includegraphics[width=.48\textwidth,height=.30\textwidth]{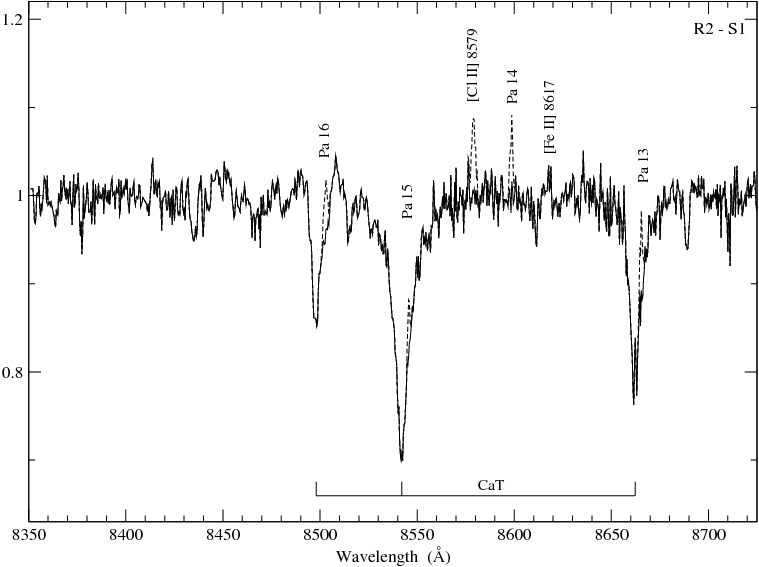}\\
\vspace{0.2cm}
\includegraphics[width=.48\textwidth,height=.30\textwidth]{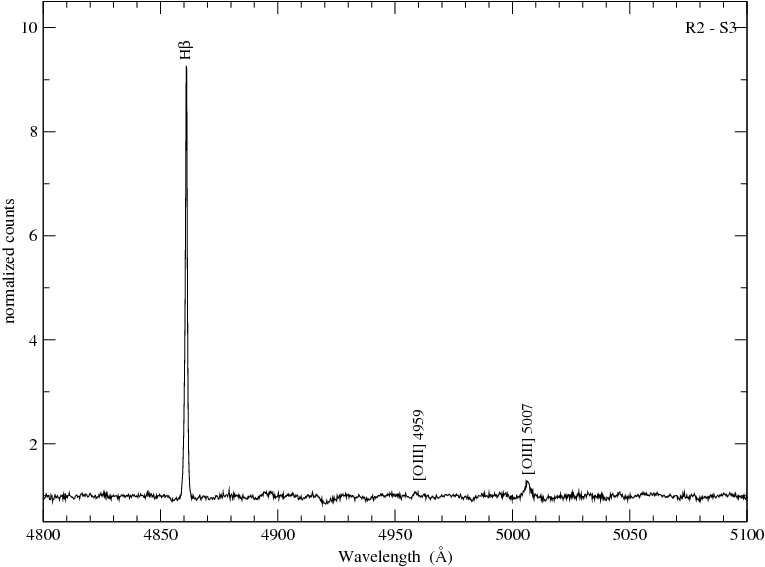}
\includegraphics[width=.48\textwidth,height=.30\textwidth]{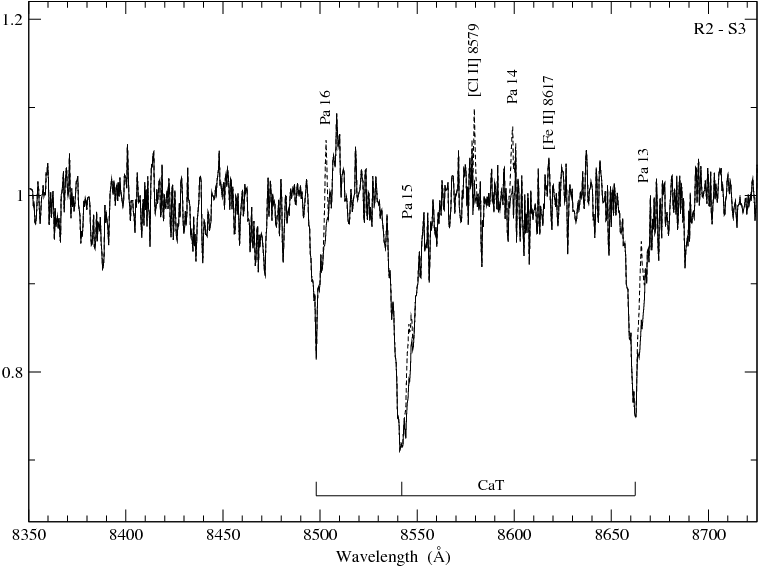}\\
\vspace{0.2cm}
\includegraphics[width=.48\textwidth,height=.30\textwidth]{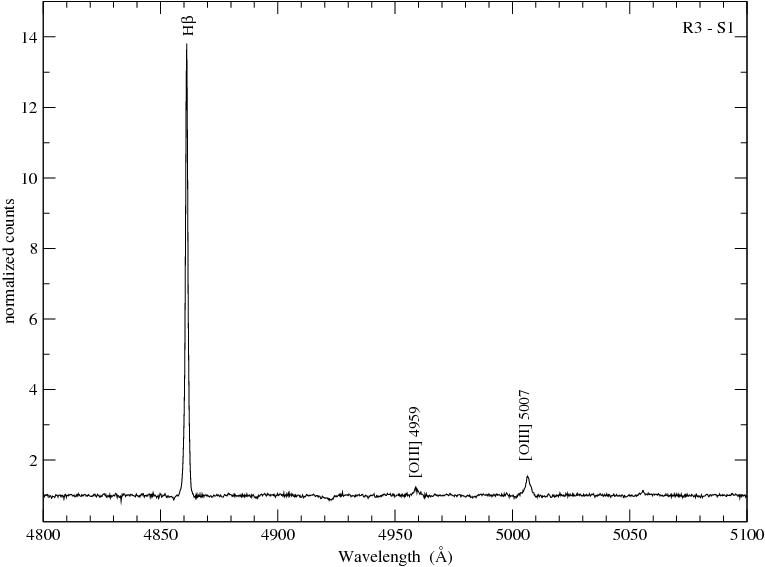}
\includegraphics[width=.48\textwidth,height=.30\textwidth]{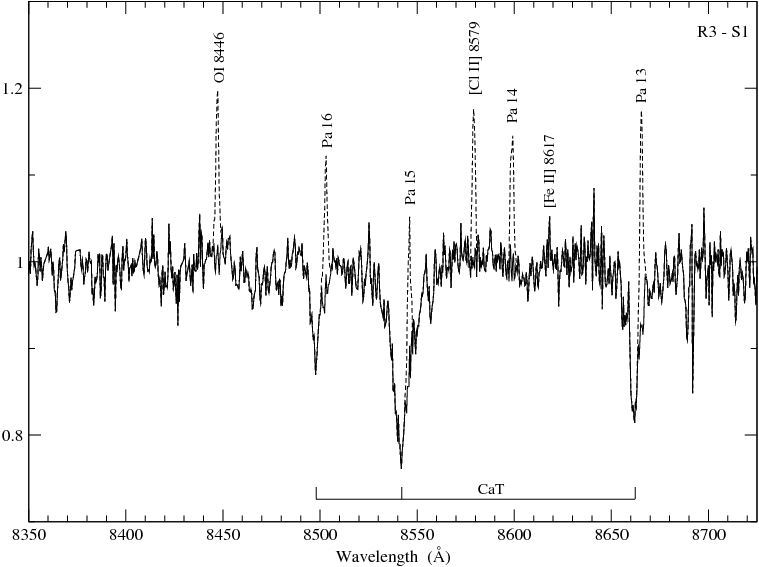}\\
\vspace{0.2cm}
\includegraphics[width=.48\textwidth,height=.30\textwidth]{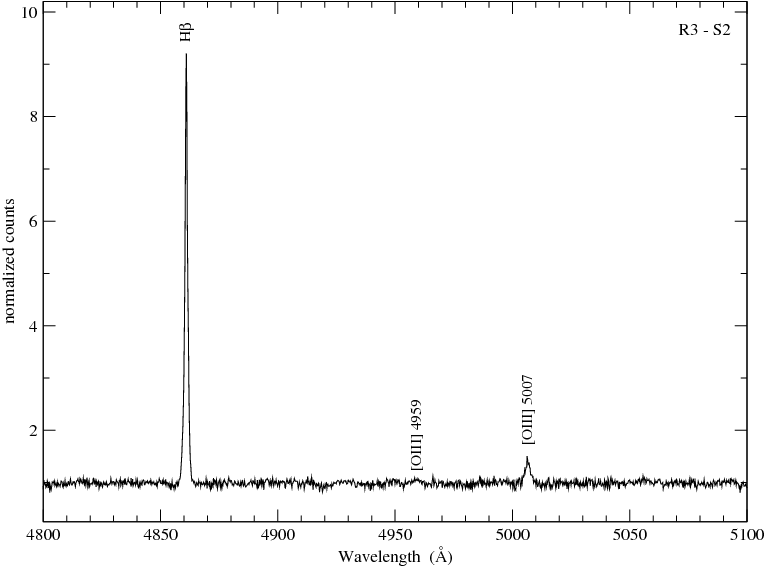}
\includegraphics[width=.48\textwidth,height=.30\textwidth]{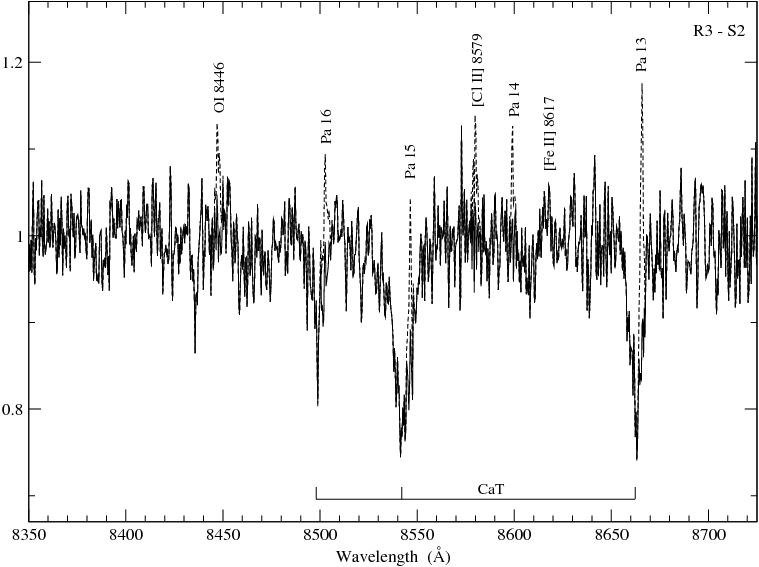}
\caption[Idem as Figure \ref{spectra2903} for the observed
  CNSFRs of NGC\,3351.]{Idem as Figure \ref{spectra2903} for the observed
  CNSFRs of NGC\,3351. For R2 and R3 in the red range, 
  the dashed line shows the obtained spectrum; the solid line represents the 
  spectrum after subtracting the emission lines (see text).} 
\label{spectra3351}
\end{figure}

\setcounter{figure}{16}

\begin{figure}
\hspace{0.0cm}
\vspace{0.2cm}
\includegraphics[width=.48\textwidth,height=.30\textwidth]{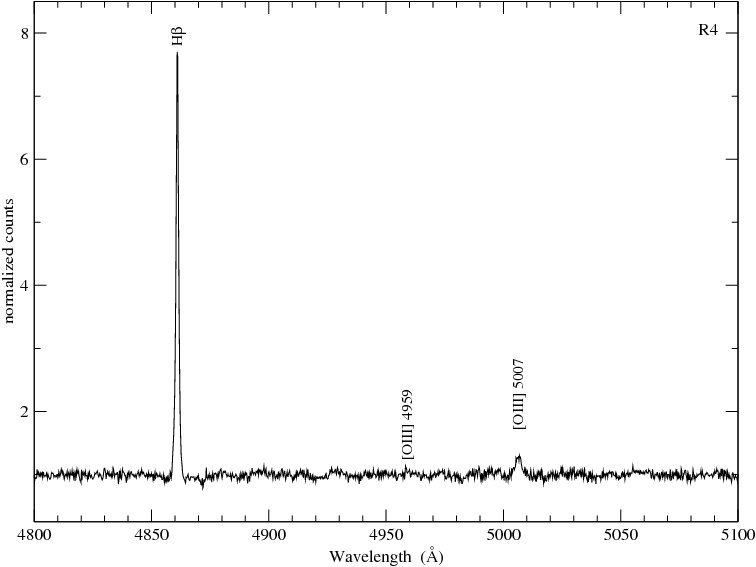}
\includegraphics[width=.48\textwidth,height=.30\textwidth]{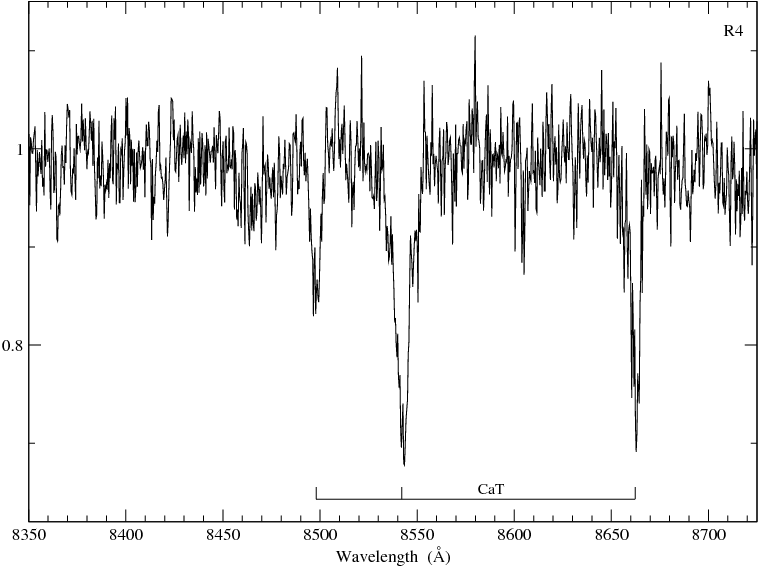}\\
\vspace{0.2cm}
\includegraphics[width=.48\textwidth,height=.30\textwidth]{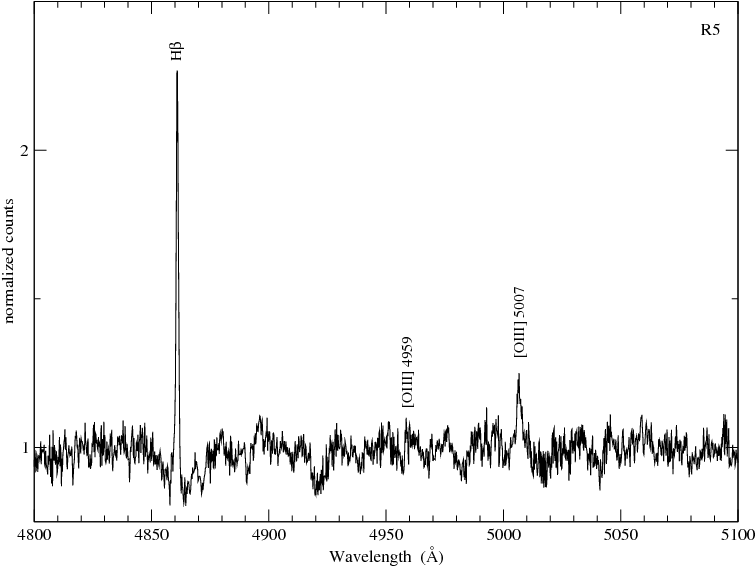}
\includegraphics[width=.48\textwidth,height=.30\textwidth]{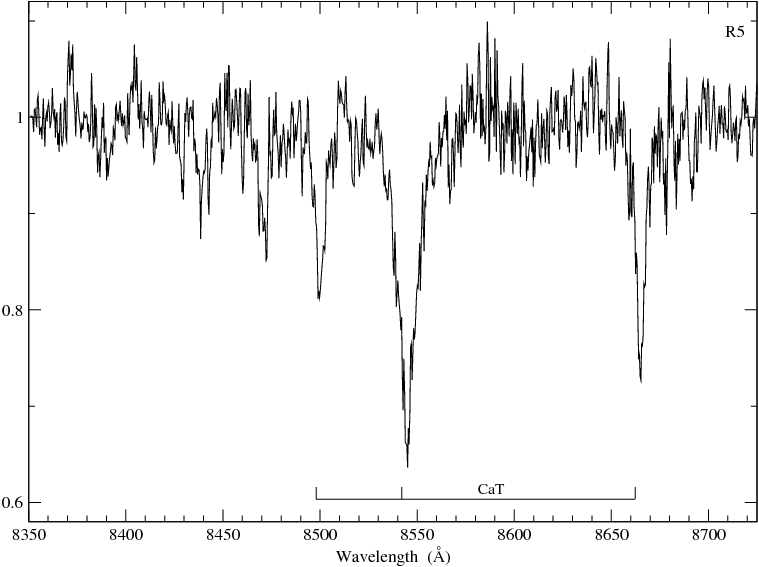}\\
\vspace{0.2cm}
\includegraphics[width=.48\textwidth,height=.30\textwidth]{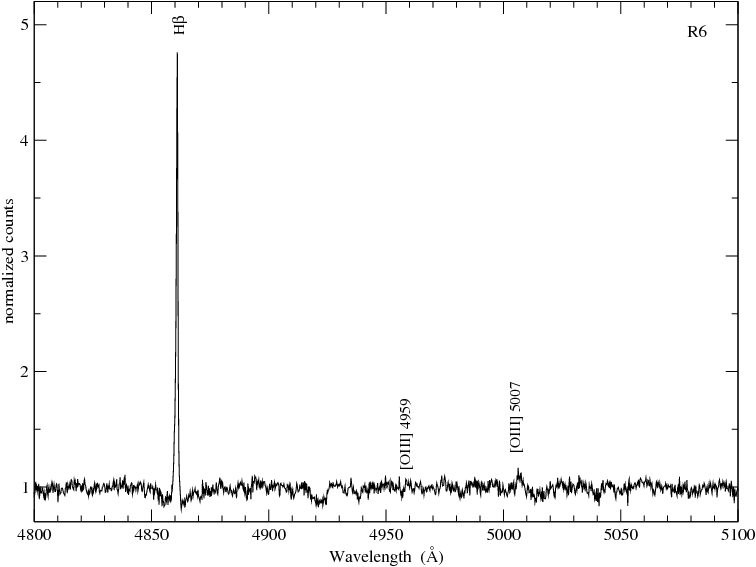}
\includegraphics[width=.48\textwidth,height=.30\textwidth]{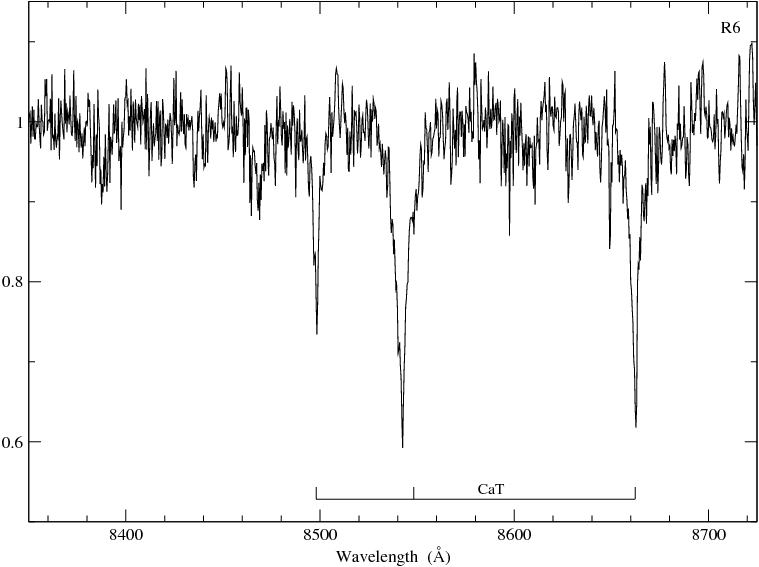}
\caption[({\it cont}) Idem as Figure \ref{spectra2903} for the observed CNSFRs
  of NGC\,3351.]{({\it cont}) Idem as Figure \ref{spectra2903} for the
  observed CNSFRs of NGC\,3351.}
\label{spectra3351}
\end{figure}


\begin{figure}
\hspace{0.0cm}
\vspace{0.2cm}
\includegraphics[width=.48\textwidth,height=.30\textwidth]{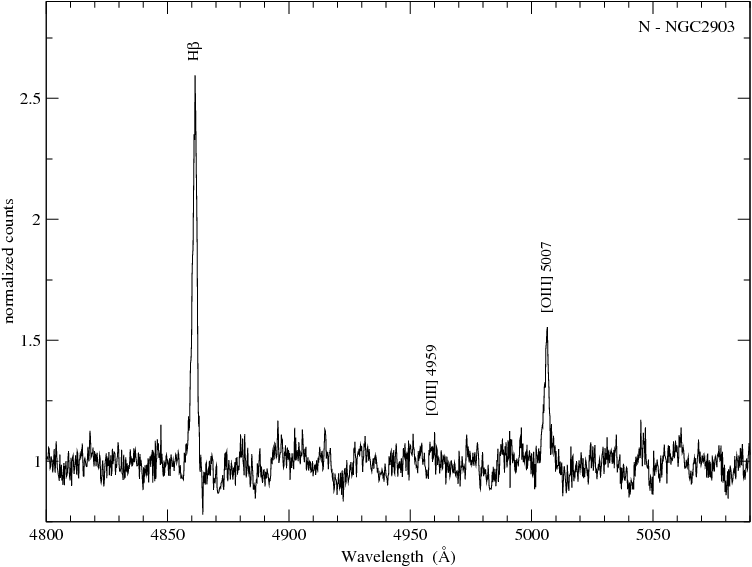}
\includegraphics[width=.48\textwidth,height=.30\textwidth]{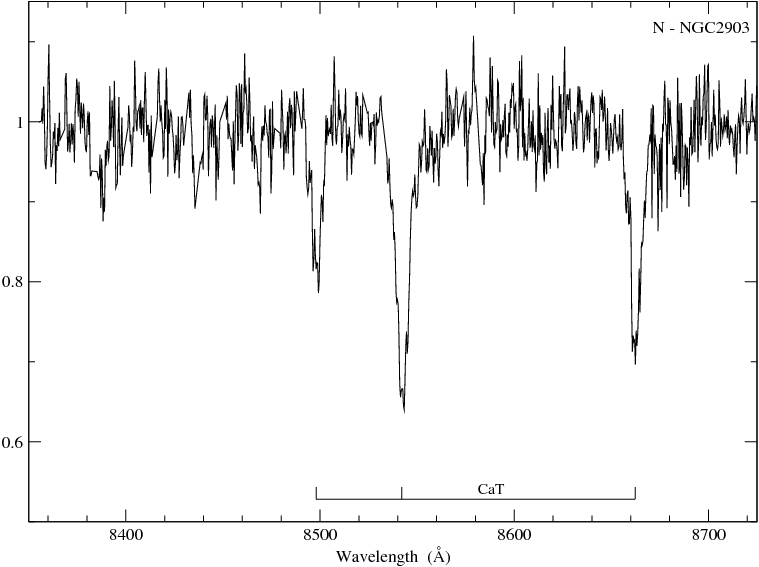}\\
\vspace{0.2cm}
\includegraphics[width=.48\textwidth,height=.30\textwidth]{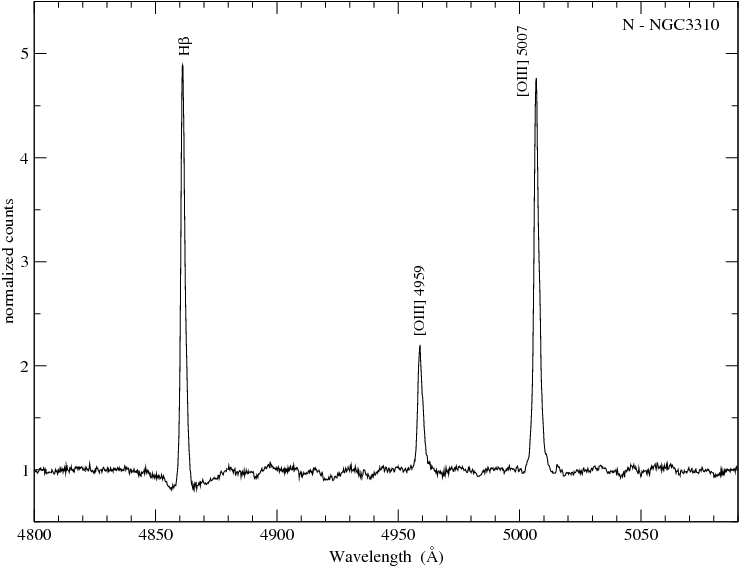}
\includegraphics[width=.48\textwidth,height=.30\textwidth]{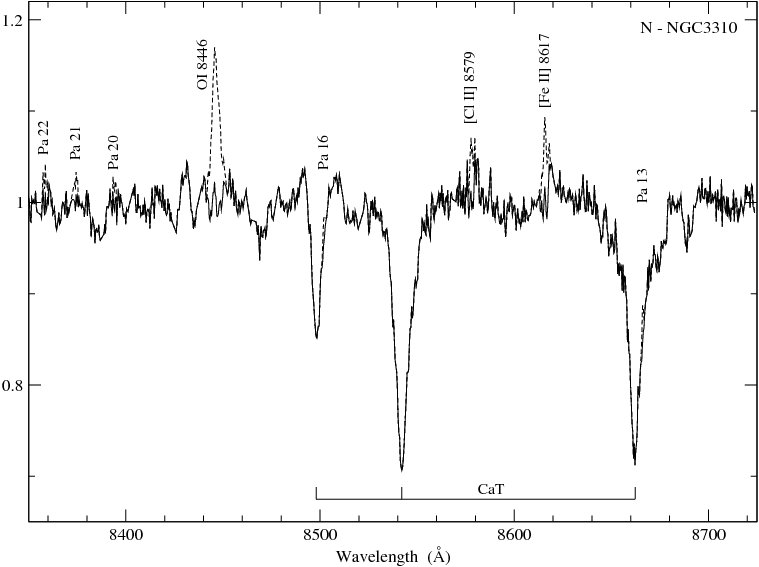}\\
\vspace{0.2cm}
\includegraphics[width=.48\textwidth,height=.30\textwidth]{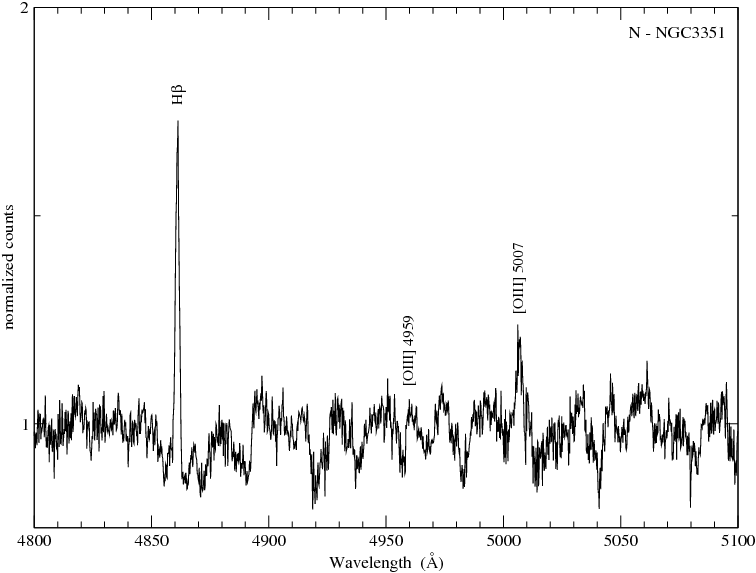}
\includegraphics[width=.48\textwidth,height=.30\textwidth]{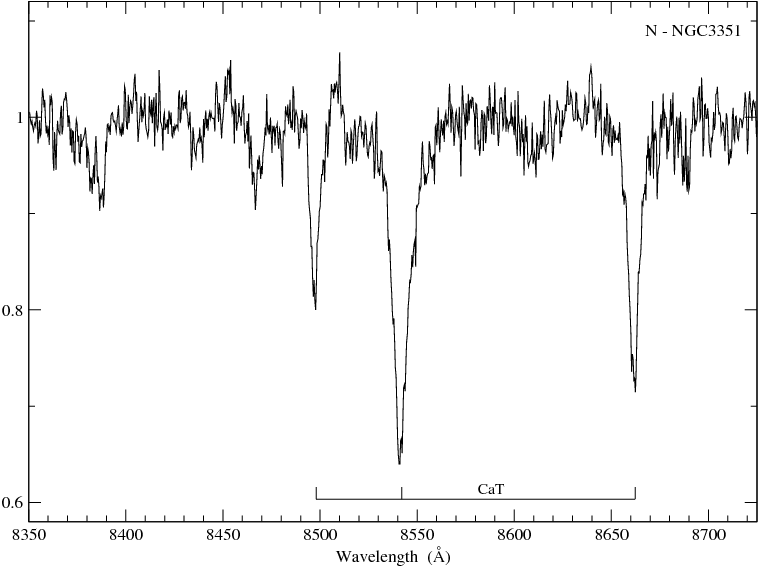}
\caption[Idem as Figure \ref{spectra2903} for the nuclei of
  NGC\,2903, NGC\,3310 and NGC\,3351]{Idem as Figure \ref{spectra2903} for the
  nuclei of NGC\,2903, NGC\,3310 and NGC\,3351 
  (upper, middle and lower panel, respectively). For the nucleus of NGC\,3310
  in the red range, the dashed line shows the obtained spectrum; the solid
  line represents the spectrum after subtracting the emission lines (see
  text).}
\label{spectraN}
\end{figure}


\begin{figure}
\vspace{3.5cm}
\centering
\includegraphics[width=.48\textwidth,height=.30\textwidth,angle=0]{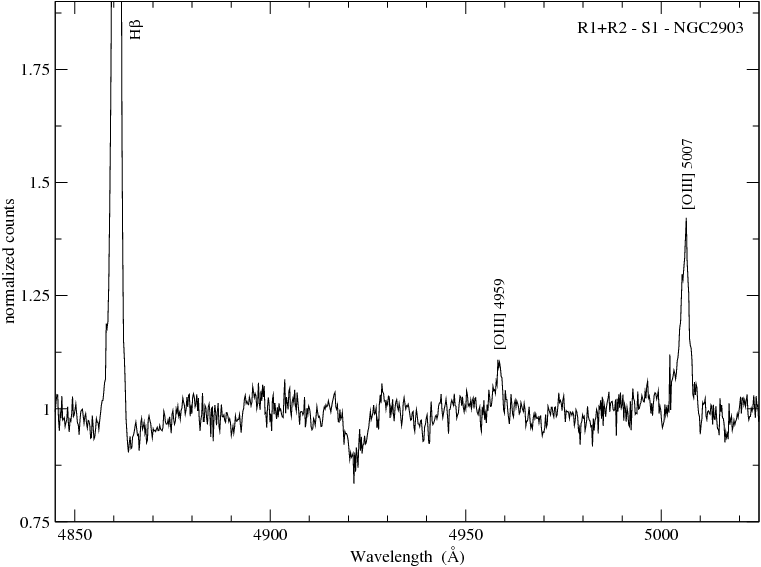}
\includegraphics[width=.48\textwidth,height=.30\textwidth,angle=0]{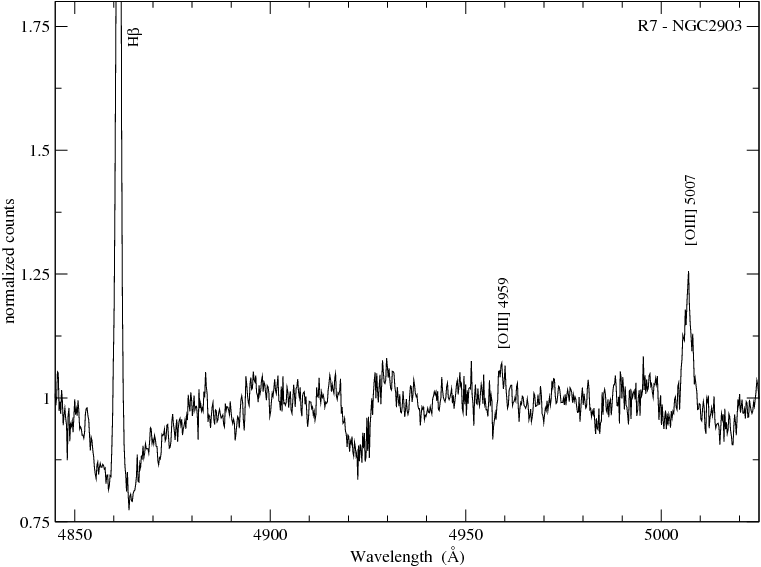}\\
\vspace{0.3cm}
\includegraphics[width=.48\textwidth,height=.30\textwidth,angle=0]{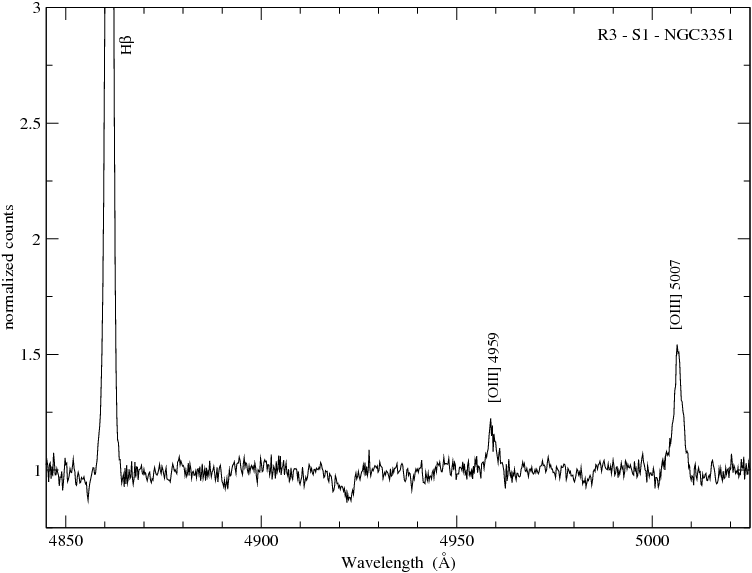}
\includegraphics[width=.48\textwidth,height=.30\textwidth,angle=0]{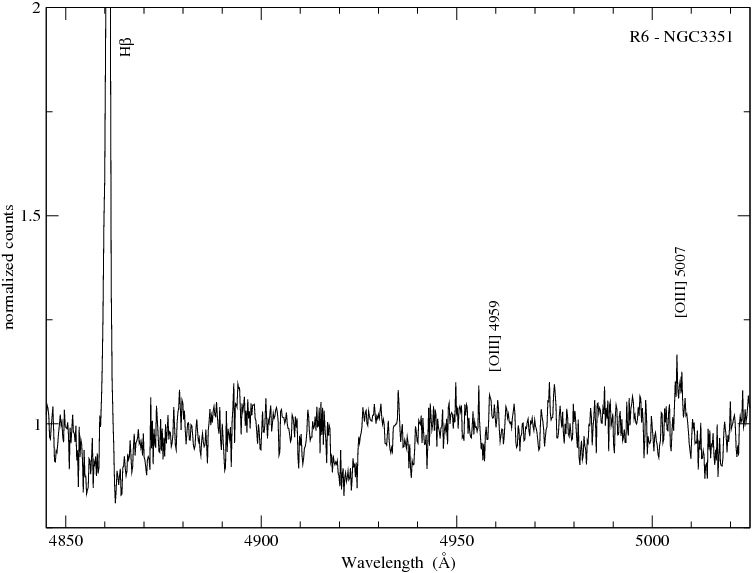}
\caption[Enlargements of the blue rest frame normalized spectra]{Enlargement
  of the blue rest frame normalized spectra. Upper panels: R1+R2 (left) 
  and R7 (right) of NGC\,2903. Lower panels: R6 (left) and R3 (right) of NGC\,3351.}
\label{enlarg}
\end{figure}


Figures \ref{spectra2903}, \ref{spectra3310} and \ref{spectra3351} show the
spectra of the observed circumnuclear regions of NGC\,2903, NGC\,3310 and
NGC\,3351, respectively, split into two panels corresponding to the blue and
the red spectral ranges. The spectra of the nuclei of these galaxies are shown
in Figure \ref{spectraN}.
The blue spectra show the Balmer H$\beta$
recombination line and the collisionally excited [O{\sc iii}] lines at
$\lambda\lambda$\,4959,\,5007\,\AA. For NGC\,2903 and NGC\,3351 we can
appreciate that these forbidden lines are very weak (see 
Figure \ref{enlarg}), and, in some cases, only the strongest $\lambda$ 5007
\AA\ is detected (right-hand panel of this figure). However, in the case of
NGC\,3310 they are very strong, due to the low abundance of the
CNSFRs located in this galaxy, with values between 0.2-0.4\,Z$_\odot$
(\citeplain{1993MNRAS.260..177P}; see Chapter \S \ref{abundan}). These low
values of the abundances of these CNSFRs can be explained by the probably
unusual interaction history of the galaxy
\cite{2002AJ....123.1381E,2001A&A...376...59K,1996A&A...309..403M,1996ApJ...473L..21S,1981A&A....96..271B},
fuelling the ring with accreted neutral gas, as modeled by
\citetex{1992MNRAS.259..345A} and \citetex{1995ApJ...449..508P}. The red
spectra show the 
stellar Ca{\sc ii} triplet lines in absorption (CaT) at $\lambda\lambda$ 8498,
8542, 8662\,\AA. In some cases, for example R1+R2 of NGC\,2903 or R3 in
NGC\,3351,  these lines are contaminated by Paschen emission which occurs at
wavelengths very close to those of the CaT lines. Other emission features,
such as O{\sc i}\,$\lambda$\,8446, [Cl{\sc ii}]\,$\lambda$\,8579, Pa\,14 and
[Fe{\sc ii}]\,$\lambda$\,8617 are also present. This is especially important
in  the case of NGC\,3310. In Figure
\ref{spectra3310} we can easily appreciate, for example in R4+R5, the Paschen
series from Pa13 to Pa22,  as well as the previously mentioned lines of O, Cl
and Fe. In all cases, a single Gaussian fit to the emission lines was
performed and the lines were subsequently subtracted (see also
\citeplain{2004A&A...419L..43O}) after checking that the 
theoretically expected ratio between the Paschen lines was satisfied. The
observed red spectra are plotted in the Figures with a dashed line.   
The solid line shows the subtracted spectra.

\begin{figure}
\hspace{0.0cm}
\vspace{0.2cm}
\includegraphics[width=.48\textwidth,height=.30\textwidth]{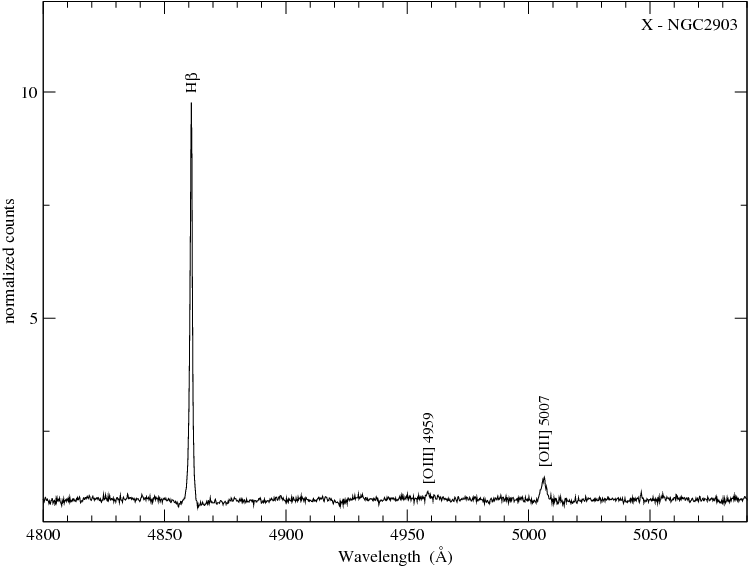}
\includegraphics[width=.48\textwidth,height=.30\textwidth]{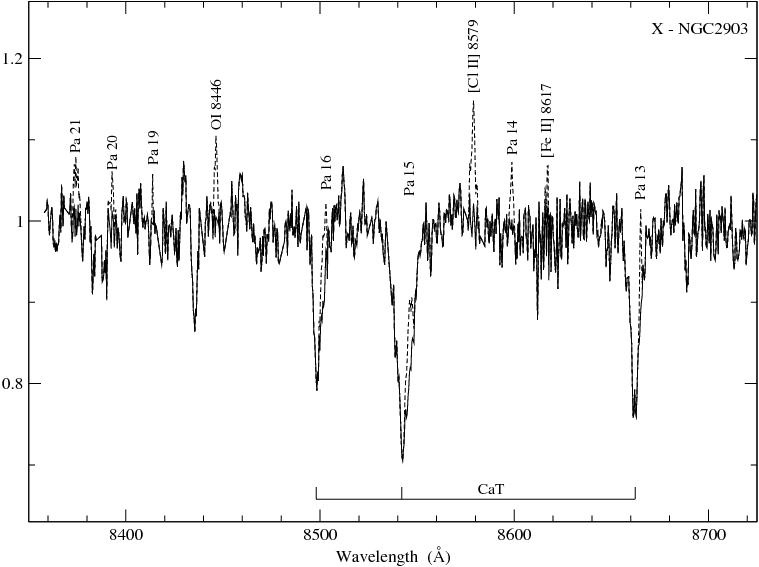}\\
\vspace{0.2cm}
\includegraphics[width=.48\textwidth,height=.30\textwidth]{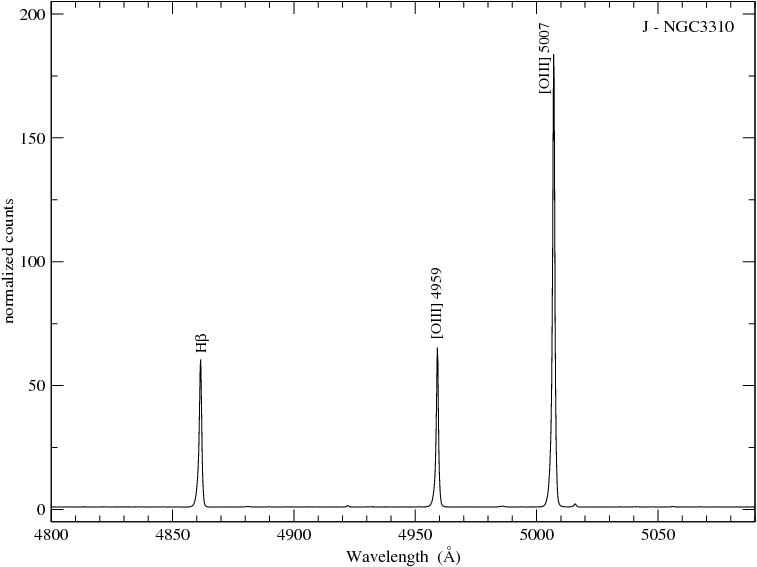}
\includegraphics[width=.48\textwidth,height=.30\textwidth]{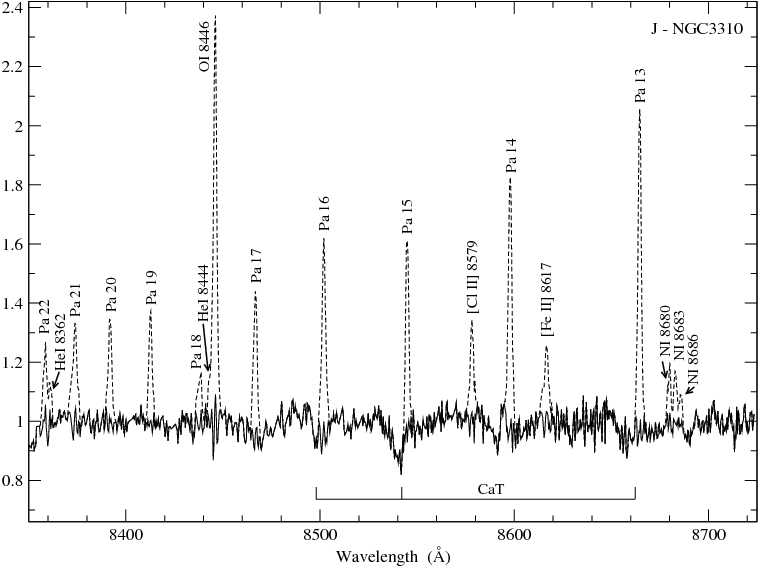}\\
\vspace{0.2cm}
\includegraphics[width=.48\textwidth,height=.30\textwidth]{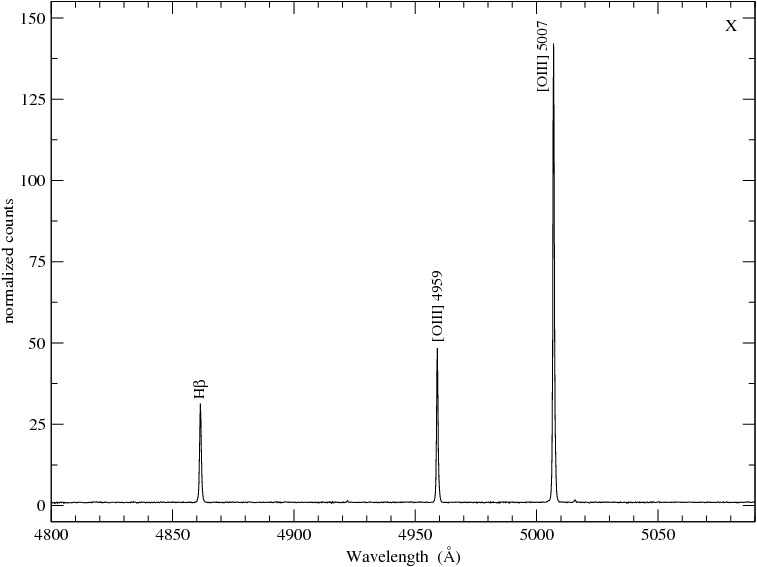}\\
\vspace{0.2cm}
\includegraphics[width=.48\textwidth,height=.30\textwidth]{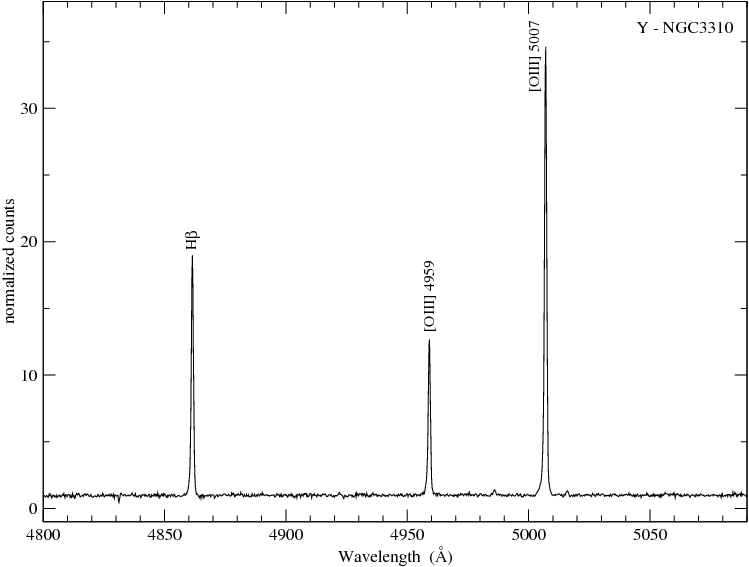}
\includegraphics[width=.48\textwidth,height=.30\textwidth]{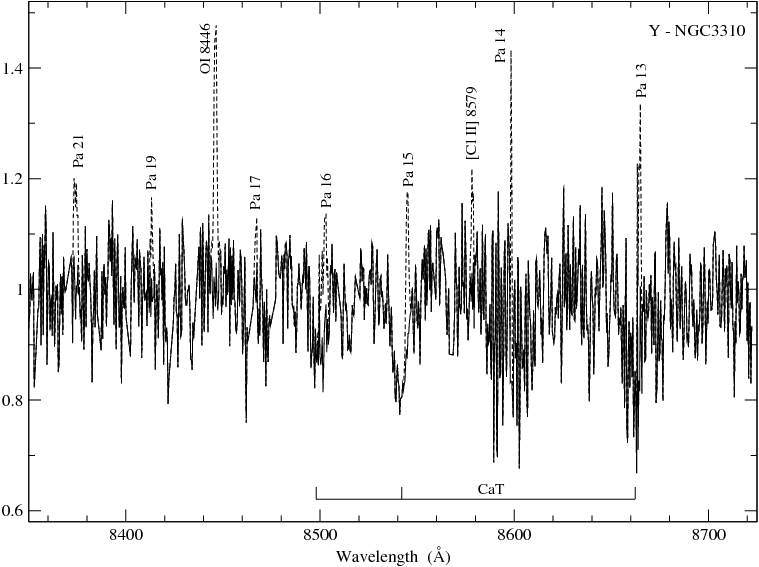}
\caption[Idem as Figure \ref{spectra2903} for region X of NGC\,2903
  and regions J, X and Y of NGC\,3310.]{Idem as Figure
  \ref{spectra2903} for region X of NGC\,2903 and regions
  J, X (only blue) and Y of NGC\,3310. The dashed line shows the obtained
  spectrum; the solid line represents the spectrum after subtracting the
  emission lines (see text).} 
\label{spectraemiknot}
\end{figure}

Figure \ref{spectraemiknot} shows the spectra of the almost pure emission
knot marked in the profiles of NGC\,2903 and NGC\,3310, and those of the Jumbo
region. In all of these cases the blue range of the spectra presents very
intense emission lines. In the case of region X in NGC\,2903, in the red range
we can appreciate a spectrum similar to those shown by the CNSFRs studied by
Planesas et al.\ (1997). It could then correspond to another region with low
continuum surface 
brightness, but with similar characteristics. For the regions of NGC\,3310,
the red spectral range presents a very weak and noisy continuum, and in
the case of region X only noise is detected therefore no spectrum is shown in 
Figure \ref{spectraemiknot}. In the other two regions of this galaxy we can
see a set of emission lines. The 
Jumbo region presents many strong lines, even He and N{\sc i} in
emission. Due to the low signal-to-noise ratio of the continuum and the
presence of the strong emission lines in these regions of NGC\,3310, we could
not obtain stellar spectra with enough signal in the CaT absorption
feature to allow an accurate measurement of velocity dispersions. Regions X
and Y seem to be pure emission knots.


\subsection{Kinematics of stars and ionized gas} 
\label{Method}

\subsubsection*{Stellar analysis}

\begin{figure}
\centering
\includegraphics[width=.83\textwidth,height=.41\textwidth,angle=0]{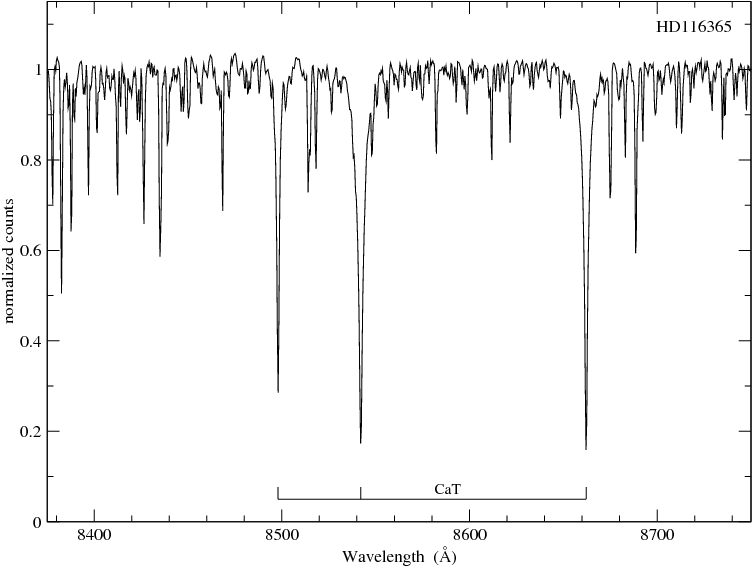}
\caption{Red rest frame normalized spectrum of HD116365.}
\label{temspec}
\end{figure}

Stellar radial velocities and velocity dispersions were obtained from the
absorption CaT lines using the cross-correlation technique  described in
detail by \citetex{1979AJ.....84.1511T}. This method requires the comparison
with a stellar template (which can be synthetic or observed)  that represents
the stellar population that best reproduces the conspicuous feature used to
perform these measurements. An example of the red spectrum of a template star
(HD116365) is shown in Figure \ref{temspec} with the prominent features of CaT
indicated. The line-of-sight velocity dispersions are calculated from the
width of the primary peak of the cross-correlation function (CCF) after
deconvolution of the instrumental profile. A filtering of high frequencies of
the Fourier transform spectrum is usually included in this procedure to avoid
noise contamination and a low frequency filtering is usually made to eliminate
the residual continuum.


\begin{figure}
\vspace{3.5cm}
\centering
\includegraphics[width=.75\textwidth]{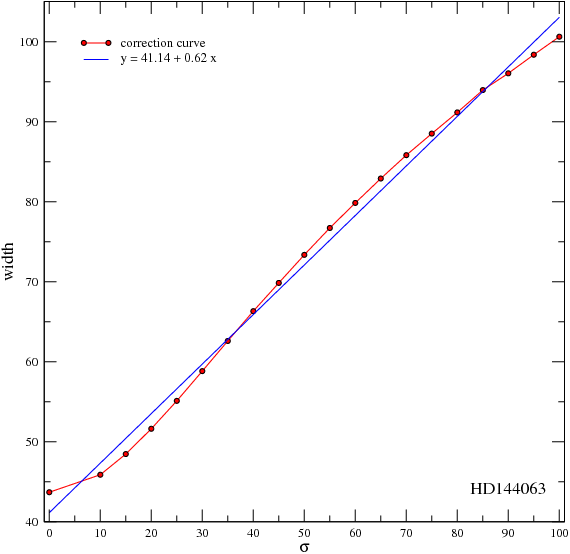}
\caption[Velocity dispersions correction curve for HD144063]{Velocity
  dispersions correction curve for HD144063 (circles). The solid line is a
  linear fit to the curve.} 
\label{correction}
\end{figure}

Minor changes/improvements with respect to the cross-correlation technique
originally proposed by \citetex{1979AJ.....84.1511T} were introduced as
described below. In order to apply this method, we have used XCSAO, an
external package of IRAF within the RVSAO, which implements the
cross-correlation method of Tonry and Davies and is a direct descendant of the
system built by them \cite{1998PASP..110..934K}. We used late-type giant and
supergiant stars that have strong CaT absorption lines (Figure \ref{temspec})
as stellar velocity templates.
We normalized the stellar spectra dividing by a fitted continuum and convolved
each stellar spectrum template with a set of Gaussian functions of different
$\sigma$ simulating a wide range in velocity dispersions from 10 to
100\,km\,s$^{-1}$ with a bin size of 5\,km\,s$^{-1}$. The obtained spectra
are cross-correlated with the 
original template obtaining a relation between the width of the main peak of
the cross-correlation and $\sigma$ of the input Gaussian. This relation
constitutes a correction curve for each template that is applied as described
below to obtain the stellar velocity dispersion for each CNSFR as described in
\citetex{1995ApJS...99...67N}. This procedure will allow us to correct for the
known possible mismatches between template stars and the region composite
spectrum. In Figure \ref{correction}  we show an example of these correction
curves, in particular for HD 144063, together with a linear fit to it. 


To determine the line-of-sight stellar velocity and velocity dispersion along
each slit, extractions were made every two pixels for slit position S1 of each
galaxy and every three pixels for slit position S2 of each galaxy and S3 of
NGC\,3351, with one pixel overlap between consecutive extractions in this
latter case. In this way the S/N ratio and the spatial resolution were
optimized. 

\begin{figure}
\vspace{0.3cm}
\centering
\includegraphics[width=.75\textwidth,height=.50\textwidth]{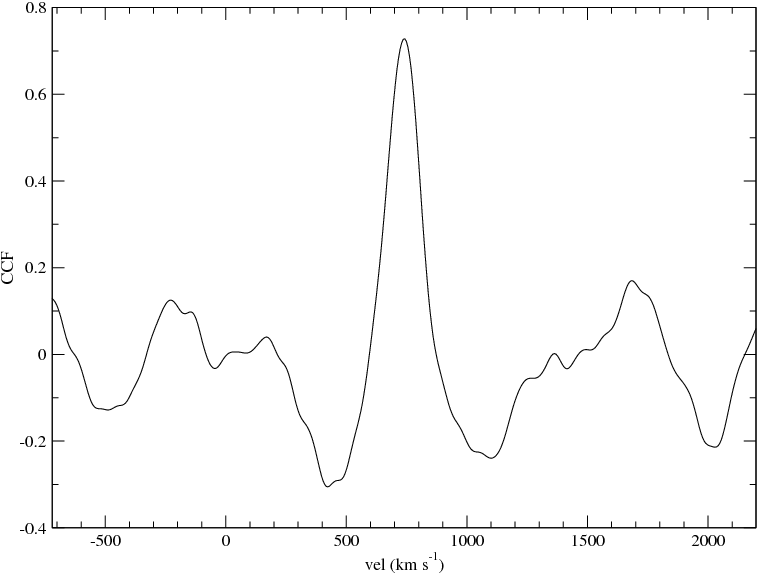}
\caption{Example of the cross correlation function.}
\label{cross}
\end{figure}

The stellar velocity dispersion was estimated at the position of each CNSFR
and the nuclei using apertures of five pixels in all cases, which correspond
to 1.0\,$\times$\,1.8\,arcsec$^2$. A set of 11 templates of different spectral
types and luminosity classes were used following \citetex{1995ApJS...99...67N}
with the variation introduced by \citetex{1997A&A...323..749P} of using the
individual stellar templates instead of an average. To measure the velocity
dispersion in a galactic spectrum we have convolved it with each stellar
template, correcting the width of the main peak of the CCF (see Figure
\ref{cross}) with the corresponding correction curve. Although the linear fits
to the curves are very good approximations (see Figure\ \ref{correction}), we
used a linear interpolation between the two nearest values to estimate the
corrected width. The $\sigma$ of the stars ($\sigma_{\ast}$) is the average of
the $\sigma$ values  found for each stellar template and its error is given by
the dispersion of the individual values of $\sigma$ and the rms of the
residuals of the wavelength fit. This procedure allows a more accurate
estimate of $\sigma_{\ast}$ \cite{1997A&A...323..749P}. The radial velocities
are the average of the radial velocities determined directly from the position
of the main peak of the cross-correlation of each galaxy spectrum with each
template in the rest frame.


The stellar velocity dispersions are listed in column 3 of Table \ref{disp}
along with their corresponding errors.
For region X of NGC\,2903 we estimated a stellar velocity dispersion, but in
the cases of X, Y and the Jumbo region of NGC\,3310 it could not be done due
to the low signal-to-noise ratio of the continua and the CaT absorption
features. For region R1+R2 of NGC\,3310, we have a similar situation, the
red continuum and the CaT features after the emission line subtractions has a
low signal-to-noise ratio, although enough to try to give an estimate of the
stellar velocity dispersion.


\begin{table}
\centering
{\footnotesize
\caption{Velocity dispersions.}
\begin{center}
\begin{tabular} {@{}l c c c c c c c c r@{}}
\hline
\hline
        &      &                  &  \multicolumn{2}{c}{{\it 1 component}} &
        \multicolumn{4}{c}{{\it 2 components}}    \\
        &      &           &       &      &  \multicolumn{2}{c}{{\it narrow}}
        & \multicolumn{2}{c}{{\it broad}} &   \\
 Region & Slit & $\sigma_{\ast}$  & $\sigma_{gas}$(H$\beta$) &
 $\sigma_{gas}$([O{\sc iii}]) &  $\sigma_{gas}$(H$\beta$)   &
        $\sigma_{gas}$([O{\sc iii}]) & $\sigma_{gas}$(H$\beta$)  &
        $\sigma_{gas}$([O{\sc iii}]) &  $\Delta$v$_{nb}$  \\
\hline

\multicolumn{9}{c}{NGC\,2903}\\
R1+R2 & S1 & 60$\pm$3 & 34$\pm$2 & 73$\pm$8  & 23$\pm$2 & 26$\pm$8 & 51$\pm$3  & 93$\pm$9   & 30   \\
R1+R2 & S2 & 64$\pm$3 & 35$\pm$2 & 71$\pm$9  & 27$\pm$2 & 27$\pm$7 & 53$\pm$4  & 95$\pm$10  & 35   \\
R4    & S2 & 44$\pm$3 & 32$\pm$2 & 76$\pm$10 & 20$\pm$2 & 35$\pm$9 & 47$\pm$4  & 89$\pm$8   & -10  \\
R7    & S1 & 37$\pm$3 & 32$\pm$4 & 59$\pm$10 & 29$\pm$5 & 17$\pm$5 & 34$\pm$8  & 67$\pm$8   & 35   \\[2pt]
N     & S2 & 65$\pm$3 & 59$\pm$3 & 59$\pm$7  & 47$\pm$4 & 27$\pm$7 & 99$\pm$13 & 83$\pm$12  & 20   \\[2pt]
X     & S1 & 38$\pm$3 & 32$\pm$2 & 66$\pm$7  & 22$\pm$2 & 29$\pm$5 & 54$\pm$4  & 75$\pm$8   & 15  \\[4pt]

\multicolumn{9}{c}{NGC\,3310}\\
R1+R2 & S2 &   80:    & 33$\pm$4 & 31$\pm$4 & 24$\pm$5 & 22$\pm$5 & 54$\pm$7 & 50$\pm$4  & -10  \\
R4    & S1 & 36$\pm$3 & 34$\pm$3 & 32$\pm$3 & 28$\pm$4 & 26$\pm$3 & 55$\pm$5 & 52$\pm$4  & 10 \\
R4+R5 & S2 & 38$\pm$3 & 27$\pm$4 & 22$\pm$3 & 22$\pm$4 & 18$\pm$3 & 46$\pm$5 & 40$\pm$2  & 15 \\
R6    & S2 & 35$\pm$5 & 30$\pm$3 & 28$\pm$3 & 23$\pm$3 & 21$\pm$3 & 54$\pm$7 & 56$\pm$3  & 0   \\
S6    & S2 & 31$\pm$4 & 27$\pm$3 & 26$\pm$3 & 20$\pm$3 & 19$\pm$3 & 47$\pm$4 & 47$\pm$3  & 10 \\
R7    & S2 & 44$\pm$5 & 21$\pm$5 & 17$\pm$4 & 18$\pm$4 & 14$\pm$3 & 41$\pm$4 & 36$\pm$3  & 0   \\
R10   & S1 & 39$\pm$3 & 38$\pm$3 & 40$\pm$3 & 26$\pm$2 & 26$\pm$3 & 54$\pm$2 & 59$\pm$4  & -20  \\[2pt]
N     & S1 & 73$\pm$3 & 55$\pm$3 & 66$\pm$3 & 35$\pm$3 & 35$\pm$3 & 73$\pm$3 & 83$\pm$4  & 20 \\[2pt]
J     & S1 &    ---   & 34$\pm$2 & 30$\pm$2 & 25$\pm$2 & 22$\pm$2 & 61$\pm$3 & 57$\pm$3  & -25  \\
X     & S2 &    ---   & 22$\pm$4 & 18$\pm$3 & 18$\pm$4 & 14$\pm$3 & 40$\pm$5 & 30$\pm$3  & -5   \\
Y     & S1 &    ---   & 28$\pm$3 & 28$\pm$3 & 22$\pm$4 & 25$\pm$3 & 44$\pm$4 & 61$\pm$5  & 10 \\[4pt]

\multicolumn{9}{c}{NGC\,3351}\\
R2  &  S1   &  50$\pm$1   & 26$\pm$1  & 72$\pm$7 & 17$\pm$3  & 21$\pm$4  &  45$\pm$3 & 74$\pm$5  & 0  \\
R2  &  S3   &  51$\pm$6   & 29$\pm$3  & 69$\pm$9 & 16$\pm$2  & 23$\pm$5  &  43$\pm$2 & 76$\pm$8  & 0  \\
R3  &  S1   &  55$\pm$5   & 35$\pm$1  & 67$\pm$7 & 25$\pm$3  &  28$\pm$4 &  59$\pm$4 & 71$\pm$4  & 0  \\
R3  &  S2   &  59$\pm$7   & 39$\pm$5  & 70$\pm$7 & 24$\pm$3  &  24$\pm$6 &  59$\pm$3 & 74$\pm$9  & 0  \\
R4  &  S2   &  66$\pm$4   & 37$\pm$4  & 76$\pm$8 & 29$\pm$3  &  ---      &  65$\pm$4 &   ---     & 0  \\
R5  &  S3   &  47$\pm$4   & 34$\pm$2  & 56$\pm$7 & 30        &  ---      &  76       & ---       & 0  \\
R6  &  S2   &  39$\pm$6   & 29$\pm$6  & 46$\pm$7 & 16$\pm$3  &  ---      &  46$\pm$4 & ---       & 0  \\[2pt]
N   &  S3   &  67$\pm$1   & 53$\pm$3  & 73$\pm$6 & 41$\pm$5  &  ---      &  67$\pm$7 & ---       & 0  \\

\hline
\multicolumn{9}{@{}l}{velocity dispersions in km\,s$^{-1}$} \\

\end{tabular}
\end{center}
\label{disp}
}
\end{table}


\subsubsection*{Ionized gas analysis}

The wavelength and the width of the H$\beta$ and [O{\sc
iii}]\,$\lambda$\,5007\,\AA\ emission lines were measured to determine both
the radial velocities and the velocity dispersions of the ionized gas. 

The velocity dispersion of the gas was estimated at the position of each CNSFR
and the nuclei using five-pixel apertures, corresponding to
1.0\,$\times$\,1.8\,arcsec$^2$. Following \citetex{2000MNRAS.317..907J} we
adjusted three different suitable continua chosen by visual inspection and
fitted a single Gaussian to the whole line. Positions and widths of the
emission lines are the average of the corresponding measurements and their
errors are calculated as the dispersion of these measurements taking into
account the rms of the residuals of the wavelength calibration. Thus, the
error is associated with the continuum placement. 

The velocity dispersions of the gas are calculated as
\[
\sigma_{gas}\,=\,\sqrt{\sigma_m^2\,-\,\sigma_i^2}
\]
\noindent where $\sigma_m$ and $\sigma_i$ are the measured and 
instrumental dispersions respectively. $\sigma_i$ was  measured directly from
the sky emission lines and is about 10.5\,km\,s$^{-1}$ at $\lambda$\,4861\,\AA.

\begin{figure}
\centering
\includegraphics[width=.75\textwidth,height=.45\textwidth,angle=0]{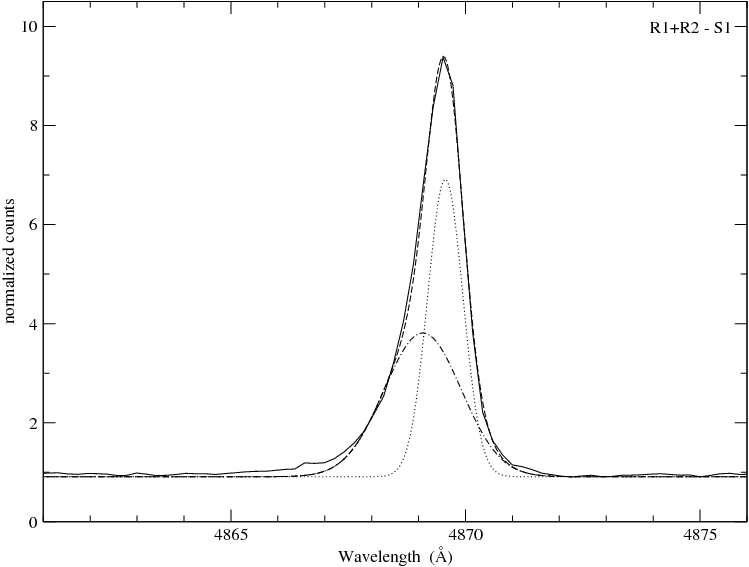}\\
\vspace{0.3cm}
\includegraphics[width=.75\textwidth,height=.45\textwidth,angle=0]{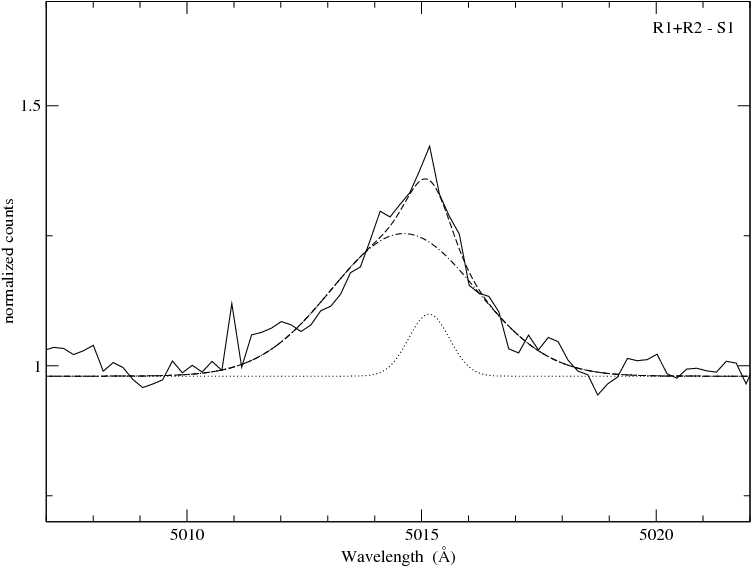}
\caption[Sections of the normalized spectrum of R1+R2 of NGC\,2903 with the fits
 from the ngaussfit superimpose]{Sections of the normalized spectrum of R1+R2 of
 NGC\,2903 (solid line). The upper panel shows from 4861 to 4876\,\AA\
 containing  H$\beta$ and the lower panel shows from 5007 to 5022\,\AA\
 containing the [O{\sc iii}]\,$\lambda$\,5007\,\AA\ emission line. For both
 we have superposed the fits from the ngaussfit task in IRAF; the
 dashed-dotted line is the broad component, the dotted line is the narrow
 component and the dashed line is the sum of both.}
\label{ngauss2903}
\end{figure}

\begin{figure}
\centering
\includegraphics[width=.75\textwidth,height=.45\textwidth,angle=0]{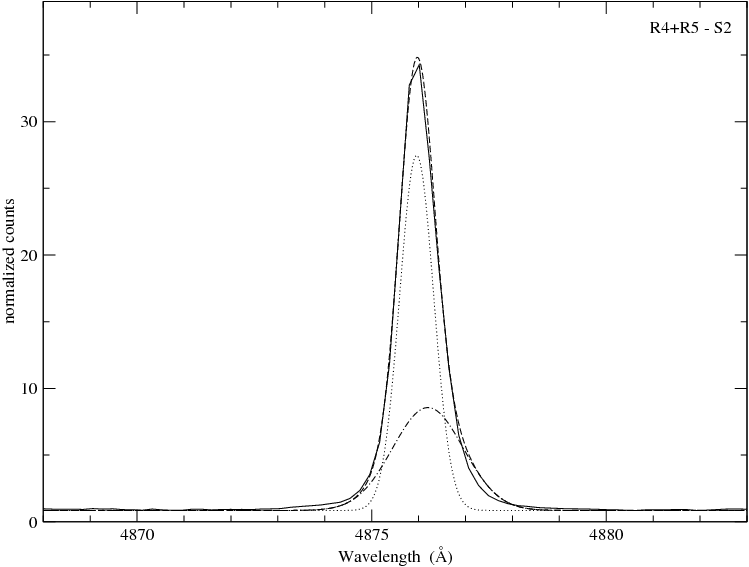}\\
\vspace{0.3cm}
\includegraphics[width=.75\textwidth,height=.45\textwidth,angle=0]{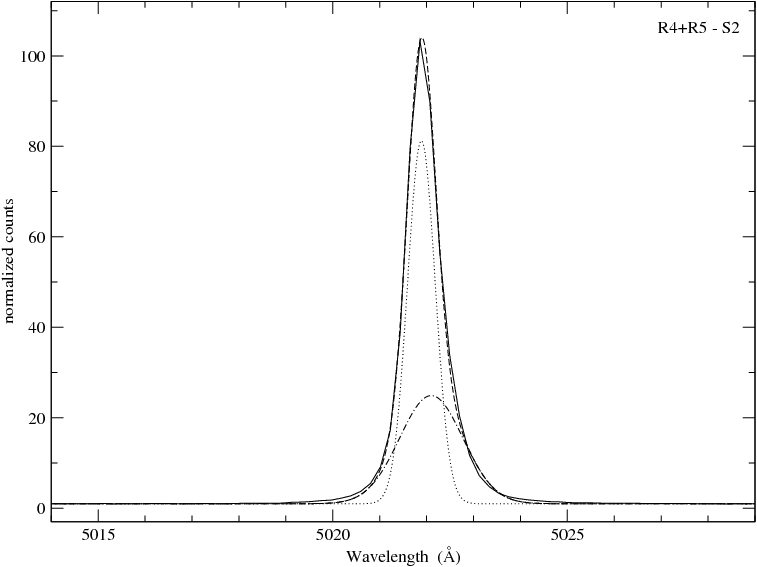}
\caption[Sections of the normalized spectrum of R4+R5 of NGC\,3310 with the
 fits from the ngaussfit superimpose]{Sections of the normalized spectrum of
 R4+R5 of 
 NGC\,3310 (solid line). The upper panel shows from 4868 to 4883\,\AA\
 containing  H$\beta$ and the lower panel shows from 5014 to 5029\,\AA\
 containing the [O{\sc iii}]\,$\lambda$\,5007\,\AA\ emission line. For both
 we have superposed the fits from the ngaussfit task in IRAF; the
 dashed-dotted line is the broad component, the dotted line is the narrow
 component and the dashed line is the sum of both.}
\label{ngauss3310}
\end{figure}

\begin{figure}
\centering
\includegraphics[width=.75\textwidth,height=.45\textwidth,angle=0]{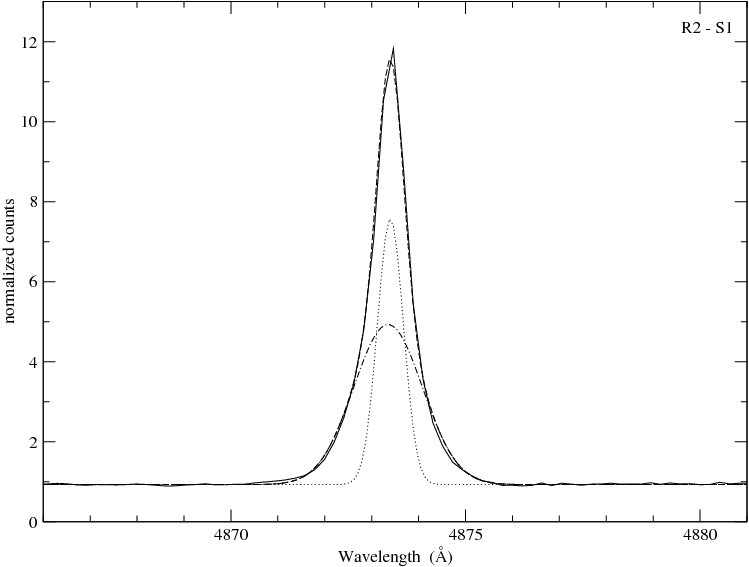}\\
\vspace{0.3cm}
\includegraphics[width=.75\textwidth,height=.45\textwidth,angle=0]{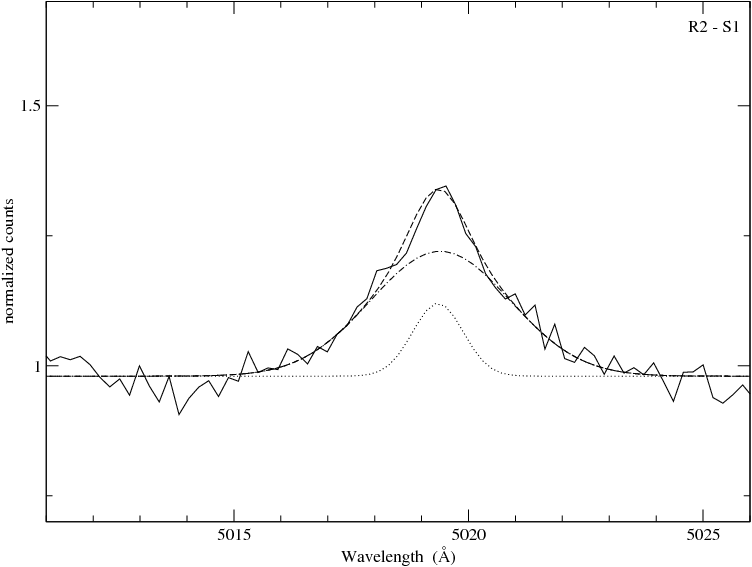}
\caption[Sections of the normalized spectrum of R2 of NGC\,3351 with the fits
  from the ngaussfit superimposed]{Sections of the normalized spectrum of R2 of
  NGC\,3351 (solid line). The upper panel shows from 4866 to 4881\,\AA\
  containing  H$\beta$ 
  and the lower panel shows from 5011 to 5026\,\AA\ containing the [O{\sc
  iii}]\,$\lambda$\,5007\,\AA\ emission line. For both  we have superposed the
  fits from the ngaussfit task in IRAF; the dashed-dotted line is the broad
  component, the dotted line is the narrow component and the dashed line is
  the sum of both.} 
\label{ngauss3351}
\end{figure}

Unexpectedly, the Gaussian fit just described revealed the presence of more
than one component in the H$\beta$ lines. The optimal fit was found for two
different components for all the regions. We  used the widths of those
components as an initial approximation to fit the [O{\sc iii}] lines which,
due to their intrinsic weakness, show  lower S/N ratio, and found them to
provide also an optimal fit. In the case of the regions of NGC\,3351, the
radial velocities found for the narrow and broad components of both H$\beta$
and [O{\sc iii}], are the same within the errors. For this galaxy, we can make
the multi Gaussian fit for the [O{\sc iii}] 
for regions R2 and R3 in the four analyzed spectra corresponding to two
different slit positions. For the rest of the regions of NGC\,3351 the
two-component fit did not show any improvement over the single component one,
probably due to the low S/N ratio. For the other two galaxies, NGC\,2903 and
NGC\,3310, we find a different behaviour. The two components show different
radial velocities (see Figures \ref{ngauss2903} and \ref{ngauss3310}). 
The velocity shifts found reach 35\,km\,s$^{-1}$ for NGC\,2903 and
25\,km\,s$^{-1}$ for NGC\,3310. 
In the case of NGC\,3351, the radial velocities found using this method are the
same, within the errors, as those found by fitting a single
component. Examples of the fits for H$\beta$ and [O{\sc
    iii}]\,$\lambda$\,5007\,\AA\ can be seen in Figures \ref{ngauss2903},
\ref{ngauss3310} and \ref{ngauss3351} for NGC\,2903, NGC\,3310 and NGC\,3351,
respectively. 


For each CNSFR, the gas velocity dispersions for the H$\beta$ and [O{\sc
iii}]\,$\lambda$\,5007\,\AA\ lines derived using single and  double line
Gaussian fits, and their corresponding errors are listed in Table
\ref{disp}. Columns 4 and 5, labelled `One component', give the results for
the single Gaussian fit. Columns 6 and 7, and 8 and 9,  labelled `Two
components - Narrow' and `Two components - Broad' respectively,  list the
results for the two component fits. The last column of the table, labelled
$\Delta$v$_{nb}$, gives the velocity difference between the narrow and broad
components. This is calculated as the average of shifts found in the H$\beta$
and [O{\sc iii}] fits. Taking into account the errors in the two component
fits, the 
errors in these velocity differences vary from  10 to 15\,km\,s$^{-1}$.

As in the case of the stars, we determined the radial velocity of the gas in
the line-of-sight along each slit, every two pixels for position S1 of each
galaxy and every three pixels, with one pixel overlap for consecutive
extractions, for position S2 of each galaxy and position S3 of
NGC\,3351. These spectra, however, do not have the required S/N ratio to allow
an acceptable two-component fit, therefore a single-Gaussian component was
used. The goodness of this procedure is discussed below.

\subsection{Emission line ratios}
\label{lineratios}

We have used two different ways to integrate the intensity of a given line:
(1) in the cases of a single Gaussian fit the  emission line intensities
were  measured using the SPLOT task of IRAF. The 
positions of the local continua are placed by eye. For the H$\beta$
emission lines a conspicuous underlying stellar population is easily
appreciable by the presence of absorption features that depress the lines
(see discussion in \citeplain{1988MNRAS.231...57D} and
\citeplain{2007MNRAS.382..251D}). Examples of this effect can 
be appreciated in Figure \ref{enlarg}. We have then defined a pseudo-continuum
at the base  of the line to measure the line intensities and minimize
the errors introduced by the underlying population
(\citeplain{2006MNRAS.372..293H} and \citeplain{2008MNRAS.383..209H}; details
in Chapter \S \ref{HIIgal-obs}). (2) In the cases of a fit by two 
Gaussians the individual intensities  of the narrow and broad components
are estimated from the fitting parameters using the expression
I\,=\,1.0645\,A\,$\times$\,FWHM 
(=\,$\sqrt{2\pi}$\,A\,$\sigma$), where I is the Gaussian intensity,
A is the amplitude of the Gaussian and FWHM is the full width at half-maximum
($\sigma$ is the dispersion of the Gaussian). A pseudo-continuum for 
the H$\beta$ emission line was also defined in these cases.

Following \citetex{1994ApJ...437..239G}, \citetex{2002MNRAS.329..315C} and 
\citetex{2003MNRAS.346..105P}, the statistical errors associated with the observed
emission fluxes have been calculated using the expression
$\sigma_{l}$\,=\,$\sigma_{c}$N$^{1/2}$[1 + EW/(N$\Delta$)]$^{1/2}$, where
$\sigma_{l}$ is  the error in the observed line flux, $\sigma_{c}$ represents
the standard 
deviation in a box near the measured emission line and stands for the error in
the continuum placement, N is the number of pixels used in the measurement of 
the line intensity, EW is the line equivalent width and $\Delta$ is 
the wavelength dispersion in \AA\,px$^{-1}$. For the H$\beta$ emission 
line we have
doubled the derived error, $\sigma_{l}$, in order to take into account the
uncertainties introduced by the presence of the underlying stellar population 
\cite{2006MNRAS.372..293H}.

The logarithmic ratio between the emission line intensities of [O{\sc
iii}]\,$\lambda$\,5007\,\AA\ and H$\beta$ and their corresponding errors
are presented in Table \ref{ratios}. We have also listed the
logarithmic ratio between the emission line fluxes of [N{\sc
ii}]\,$\lambda$\,6584\,\AA\ and H$\alpha$ together with their corresponding
errors from \citetex{2007MNRAS.382..251D} (see Chapter \S \ref{abundan}) for
R1+R2 and R4 of NGC\,2903 and all 
the CNSFRs and the nucleus of NGC\,3351, and from
\citetex{1997A&A...325...81P} for R7 and the nucleus of NGC\,2903.
For these last two objects, the [N{\sc ii}] and H$\alpha$ emission are
derived from the H$\alpha$+[N{\sc ii}] narrow band images. Planesas and
collaborators estimated the relative contribution of the [N{\sc
    ii}]\,$\lambda\lambda$\,6548,6584\,\AA\ lines from the H$\alpha$ and
[N{\sc ii}]\,$\lambda$\,6584\,\AA\ equivalent width measurements made by
\citetex{1982ApJS...50..517S}. The logarithmic ratios for R1+R2
and R4 of NGC\,2903 derived using this procedure are similar
(-0.64\,$\pm$\,0.04 and -0.65\,$\pm$\,0.04). For the X region we assumed a
value of -0.38 (without error) for this ratio as given
by D\'iaz et al.\ (2007; see Chapter \S \ref{abundan}) for the other CNSFRs of
this galaxy. For NGC\,3310, the values listed for R4, R4+R5, S6, J and the
nucleus are from Pastoriza et al.\ (1993; S6 seems to be their region L). For
the rest of the regions we  
used an extrapolation of the results given by them from the spatial profiles of
these emission lines. They reported a constant value of 0.2 for the [N{\sc
ii}]\,/\,H$\alpha$ ratio for the \HII\ regions in their slit positions 2 and
3, while this value changes from 0.2 to 0.5 along position 1 
over the nucleus of the galaxy, being 0.23 and 0.27 for regions B and L,
respectively. We adopt for R1+R2, R6, R7, R10, X and Y, a value of 0.2
without assigning any error to it.


\begin{table}
\centering
{\footnotesize
\caption{Line ratios.}
\label{ratios}
\begin{tabular} {@{}l c c c c c@{}}
\hline
        &      &         One component &
        \multicolumn{2}{c}{Two components}   & \\
        &      &    &  Narrow  & Broad &  \\
 Region & Slit & log([O{\sc iii}]5007/H$\beta$) & log([O{\sc
        iii}]5007/H$\beta$) & log([O{\sc iii}]5007/H$\beta$) & log([N{\sc
        ii}]6584/H$\alpha$) \\
\hline
\multicolumn{6}{c}{NGC\,2903}\\[2pt]

R1+R2 &  S1   &  -1.03$\pm$0.05   & -1.64$\pm$0.09  & -0.76$\pm$0.10 & -0.37$\pm$0.01$^a$\\
R1+R2 &  S2   &  -0.92$\pm$0.07   & -1.51$\pm$0.11  & -0.57$\pm$0.14 &                   \\
R4    &  S2   &  -0.78$\pm$0.11   & -1.01$\pm$0.10  & -0.71$\pm$0.15 & -0.38$\pm$0.01$^a$\\
R7    &  S1   &  -0.78$\pm$0.10   & -1.79$\pm$0.12  & -0.40$\pm$0.18 & -0.66$\pm$0.04$^b$\\[2pt]
N     &  S2   &  -0.50$\pm$0.07   & -0.92$\pm$0.10  & -0.23$\pm$0.18 & -0.68$\pm$0.04$^b$\\[2pt]
X     &  S1   &  -0.93$\pm$0.06   & -1.45$\pm$0.09  & -0.77$\pm$0.11 & -0.38:$^c$ \\[4pt]

\multicolumn{6}{c}{NGC\,3310}\\

R1+R2 &  S2   &   0.42$\pm$0.01   &  0.39$\pm$0.01  &  0.45$\pm$0.03 &  -0.70:$^d$ \\
R4    &  S1   &   0.28$\pm$0.01   &  0.28$\pm$0.01  &  0.28$\pm$0.04 &  -0.69$\pm$0.01$^d$ \\
R4+R5 &  S2   &   0.42$\pm$0.01   &  0.39$\pm$0.01  &  0.44$\pm$0.03 &  -0.69$\pm$0.01$^d$ \\
R6    &  S2   &   0.28$\pm$0.01   &  0.30$\pm$0.01  &  0.28$\pm$0.06 &  -0.70:$^d$ \\
S6    &  S2   &   0.21$\pm$0.01   &  0.21$\pm$0.01  &  0.22$\pm$0.08 &  -0.56$\pm$0.01$^d$ \\
R7    &  S2   &   0.23$\pm$0.01   &  0.21$\pm$0.01  &  0.32$\pm$0.09 &  -0.70:$^d$ \\
R10   &  S1   &   0.19$\pm$0.01   &  0.20$\pm$0.04  &  0.20$\pm$0.08 &  -0.70:$^d$ \\[2pt] 
N     &  S2   &   0.04$\pm$0.02   & -0.13$\pm$0.06  &  0.11$\pm$0.07 &  -0.30$\pm$0.01$^d$ \\[2pt]
J     &  S1   &   0.45$\pm$0.01   &  0.45$\pm$0.01  &  0.45$\pm$0.01 &  -0.80$\pm$0.01$^d$ \\
X     &  S2   &   0.60$\pm$0.01   &  0.55$\pm$0.01  &  0.67$\pm$0.04 &  -0.70:$^d$ \\
Y     &  S1   &   0.28$\pm$0.01   &  0.43$\pm$0.01  &  0.00$\pm$0.04 &  -0.70:$^d$ \\[4pt]

\multicolumn{6}{c}{NGC\,3351}\\
R2  &  S1   &  -1.07$\pm$0.06   & -1.66$\pm$0.08  & -0.93$\pm$0.12 & -0.43$\pm$0.01$^a$\\
R2  &  S3   &  -1.01$\pm$0.06   & -1.55$\pm$0.08  & -0.96$\pm$0.13 & \\
R3  &  S1   &  -1.10$\pm$0.06   & -1.57$\pm$0.07  & -0.93$\pm$0.10 & -0.42$\pm$0.01$^a$\\
R3  &  S2   &  -1.00$\pm$0.06   & -1.52$\pm$0.09  & -0.89$\pm$0.10 & \\
R4  &  S2   &  -1.03$\pm$0.07   &    ---          &        ---     & -0.49$\pm$0.01$^a$\\
R5  &  S3   &  -0.85$\pm$0.12   &    ---          &        ---     & -0.37$\pm$0.03$^a$\\
R6  &  S2   &  -1.09$\pm$0.11   &    ---          &        ---     & -0.52$\pm$0.02$^a$\\[2pt]
N   &  S3   &  -0.28$\pm$0.05   &    ---          &        ---     & \\

\hline

\multicolumn{6}{@{}l}{$^a$From D\'iaz et al.\ (2007; see Chapter \S \ref{abundan}).}\\
\multicolumn{6}{@{}l}{$^b$From Planesas et al.\ (1997).}\\
\multicolumn{6}{@{}l}{$^c$Assumed from the values given by
  D\'iaz et al.\ (2007; see Chapter \S \ref{abundan}) for the other}\\
\multicolumn{6}{@{}l}{CNSFRs of this galaxy.}\\
\multicolumn{6}{@{}l}{$^d$From Pastoriza et al.\ (1993).}

\end{tabular}}
\end{table}


\nocite{2003MNRAS.346.1055K}
\nocite{2001ApJ...556..121K}
\nocite{1997ApJS..112..315H}

\section{Dynamical mass derivation}
\label{masses}

Two parameters are needed in order to determine the mass of a virialized
stellar system, namely its velocity dispersion and its size.

Previously estimated radii (R) for the regions, such as those given by
Planesas et al.\ (1997) for NGC\,2903 and NGC\,3351, by
\citetex{2000MNRAS.311..120D} for NGC\,3310 from H$\alpha$ images, or by
\citetex{1997ApJ...484L..41C} from UV Space Telescope Imaging Spectrograph
(STIS)-HST images for those of NGC\,3351 were defined  to include the total
integrated emission flux of the regions, and therefore they are not
appropriate for calculating their dynamical masses, as what is
needed for this is a measurement of the size of the mass distribution, i.e. the
star-cluster size. This is traditionally done measuring the effective radius
using images obtained in bands where the light is dominated by the stellar
contribution and where the contamination by gaseous emission is either small or
can be estimated and corrected.

\begin{figure}
\centering
\begin{minipage}[c]{5.6in}
\includegraphics[width=1.42\textwidth,angle=90]{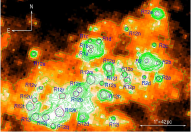}
\end{minipage}
\begin{minipage}[c]{0.4in}
\rotcaption[Enlargements of the F606W image around regions R1 and R2 of
  NGC\,2903 with the contours overlapped]{Enlargements of the F606W image
  around regions R1 and R2 of NGC\,2903 with the contours overlapped. The blue
  circles correspond to the adopted radius for each region and the red
  contours corresponding to the half light brightness are in red.}
\label{sizes2903R1}
\end{minipage}
\end{figure}


\begin{figure}
\centering
\begin{minipage}[c]{5.6in}
\includegraphics[width=1.42\textwidth,angle=90]{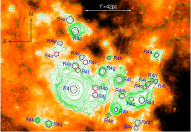}
\end{minipage}
\begin{minipage}[c]{0.4in}
\rotcaption{Idem as Figure \ref{sizes2903R1} for region R4 of NGC\,2903.} 
\label{sizes2903R4}
\end{minipage}
\end{figure}


\begin{figure}
\centering
\begin{minipage}[c]{5.6in}
\includegraphics[width=1.42\textwidth,angle=90]{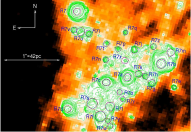}
\end{minipage}
\begin{minipage}[c]{0.4in}
\rotcaption{Idem as Figure \ref{sizes2903R1} for region R7 of NGC\,2903.} 
\label{sizes2903R7}
\end{minipage}
\end{figure}


\begin{figure}
\centering
\begin{minipage}[c]{5.6in}
\includegraphics[width=1.42\textwidth,angle=90]{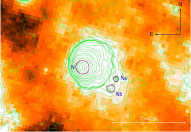}
\end{minipage}
\begin{minipage}[c]{0.4in}
\rotcaption{Idem as Figure \ref{sizes2903R1} for the nucleus of NGC\,2903.} 
\label{sizes2903N}
\end{minipage}
\end{figure}


\begin{figure*}
\centering
\begin{minipage}[c]{5.8in}
\includegraphics[width=1.22\textwidth,angle=90]{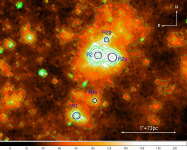}
\end{minipage}
\begin{minipage}[c]{0.2in}
\rotcaption{Idem as Figure \ref{sizes2903R1} for regions R1 and R2 of NGC\,3310.} 
\label{sizes3310R1}
\end{minipage}
\end{figure*}


\begin{figure*}
\centering
\begin{minipage}[c]{5.8in}
\includegraphics[width=1.22\textwidth,angle=90]{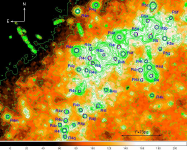}
\end{minipage}
\begin{minipage}[c]{0.2in}
\rotcaption{Idem as Figure \ref{sizes2903R1} for regions R4 and R5 of NGC\,3310.} 
\label{sizes3310R4}
\end{minipage}
\end{figure*}


\begin{figure*}
\centering
\begin{minipage}[c]{5.8in}
\includegraphics[width=1.22\textwidth,angle=90]{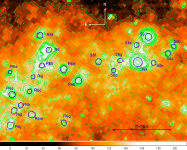}
\end{minipage}
\begin{minipage}[c]{0.2in}
\rotcaption{Idem as Figure \ref{sizes2903R1} for regions R6 and S6 of NGC\,3310.} 
\label{sizes3310R6}
\end{minipage}
\end{figure*}


\begin{figure*}
\centering
\begin{minipage}[c]{5.8in}
\includegraphics[width=1.22\textwidth,angle=90]{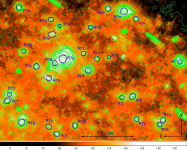}
\end{minipage}
\begin{minipage}[c]{0.2in}
\rotcaption{Idem as Figure \ref{sizes2903R1} for region R7 of NGC\,3310.} 
\label{sizes3310R7}
\end{minipage}
\end{figure*}


\begin{figure*}
\centering
\begin{minipage}[c]{5.8in}
\includegraphics[width=1.22\textwidth,angle=90]{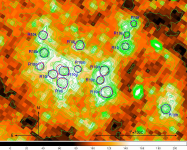}
\end{minipage}
\begin{minipage}[c]{0.2in}
\rotcaption{Idem as Figure \ref{sizes2903R1} for region R10 of NGC\,3310.} 
\label{sizes3310R10}
\end{minipage}
\end{figure*}


\begin{figure*}
\centering
\begin{minipage}[c]{5.8in}
\includegraphics[width=1.22\textwidth,angle=90]{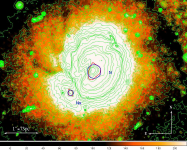}
\end{minipage}
\begin{minipage}[c]{0.2in}
\rotcaption{Idem as Figure \ref{sizes2903R1} for the nucleus of NGC\,3310.} 
\label{sizes3310N}
\end{minipage}
\end{figure*}


\begin{figure}
\centering
\begin{minipage}[c]{5.6in}
\includegraphics[width=1.22\textwidth,angle=90]{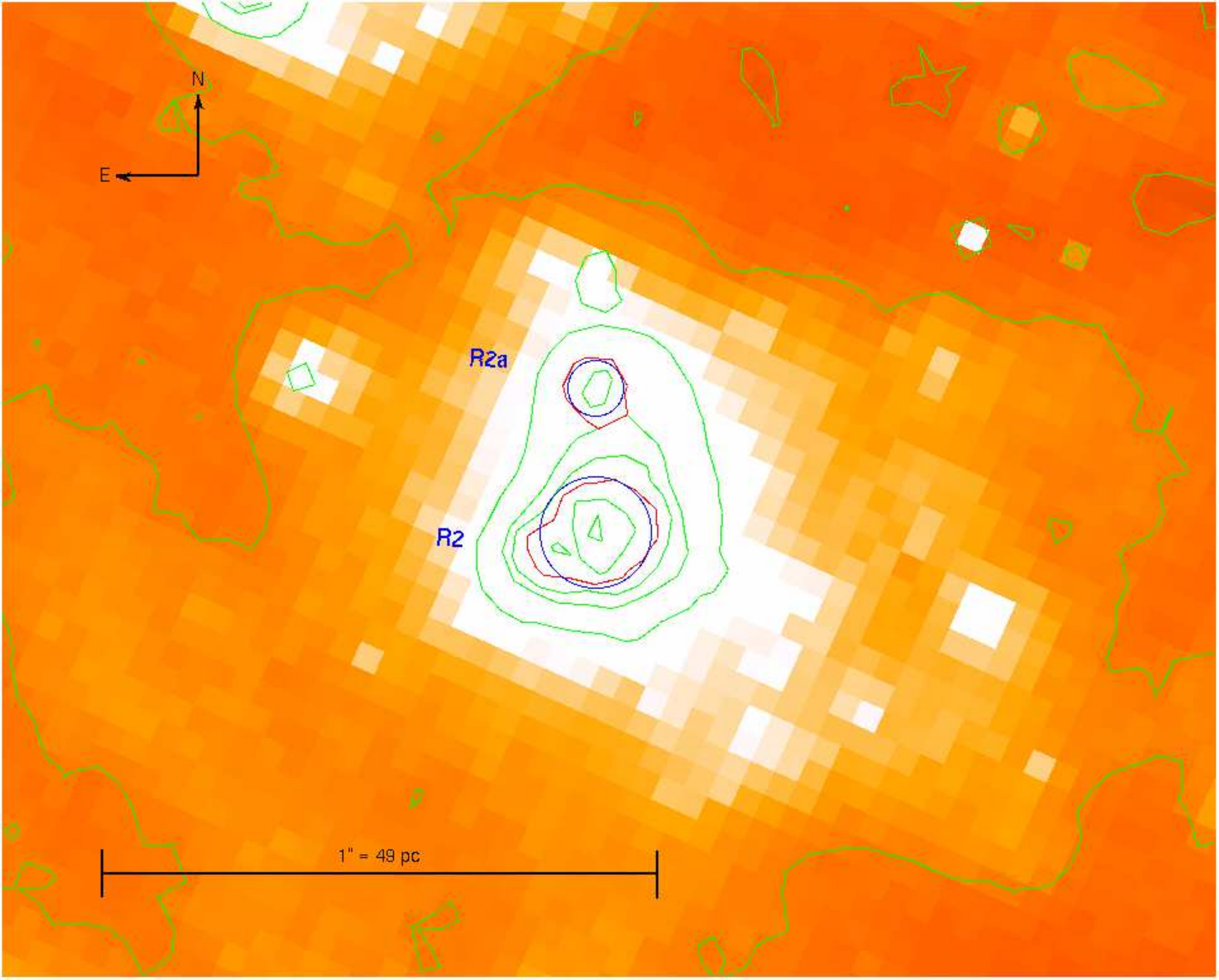}
\end{minipage}
\begin{minipage}[c]{0.4in}
\rotcaption{Idem as Figure \ref{sizes2903R1} for region R2 of NGC\,3351.} 
\label{sizes3351R2}
\end{minipage}
\end{figure}


\begin{figure}
\centering
\begin{minipage}[c]{5.6in}
\includegraphics[width=1.22\textwidth,angle=90]{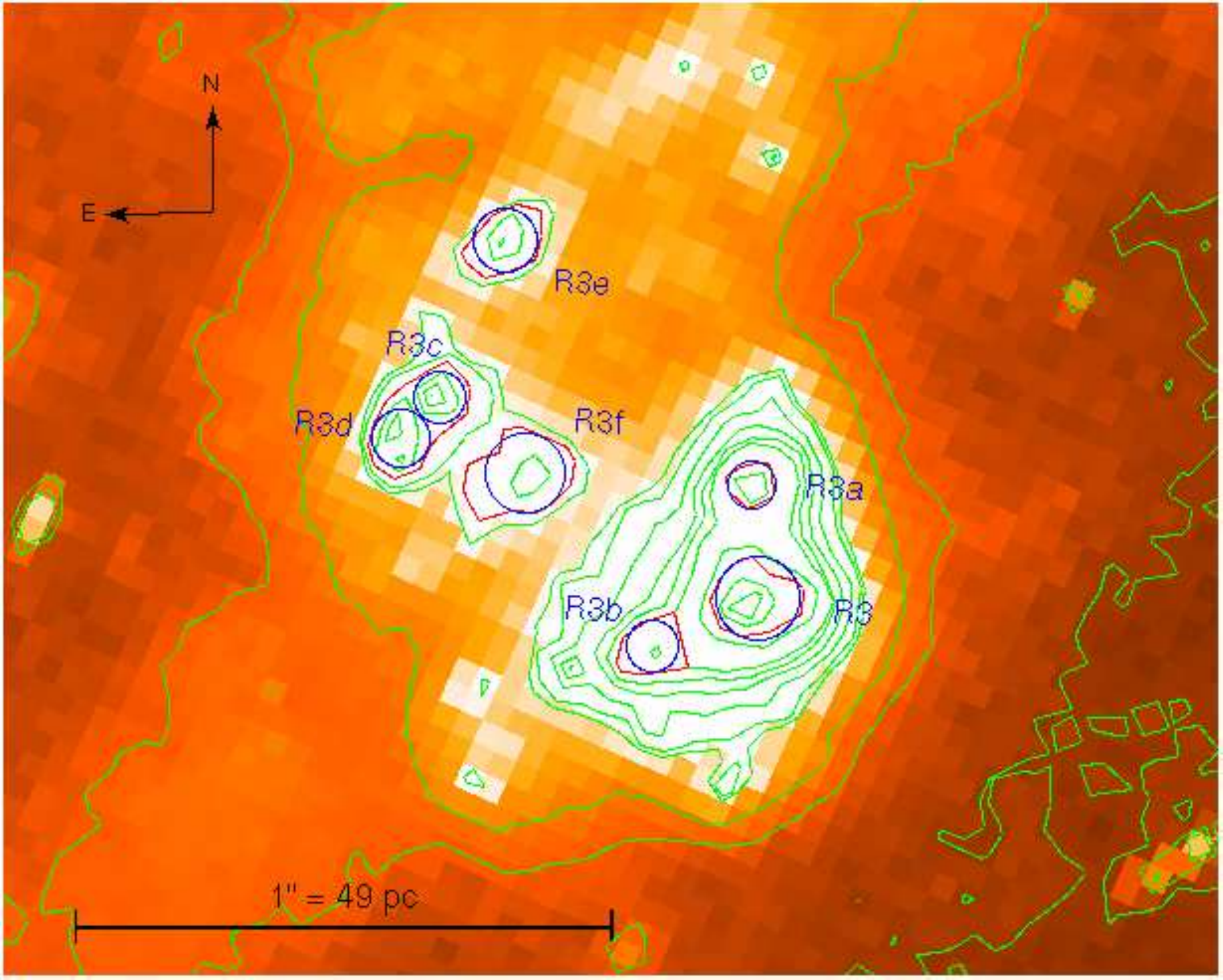}
\end{minipage}
\begin{minipage}[c]{0.4in}
\rotcaption{Idem as Figure \ref{sizes2903R1} for region R3 of NGC\,3351.} 
\label{sizes3351R3}
\end{minipage}
\end{figure}


\begin{figure}
\centering
\begin{minipage}[c]{5.6in}
\includegraphics[width=1.22\textwidth,angle=90]{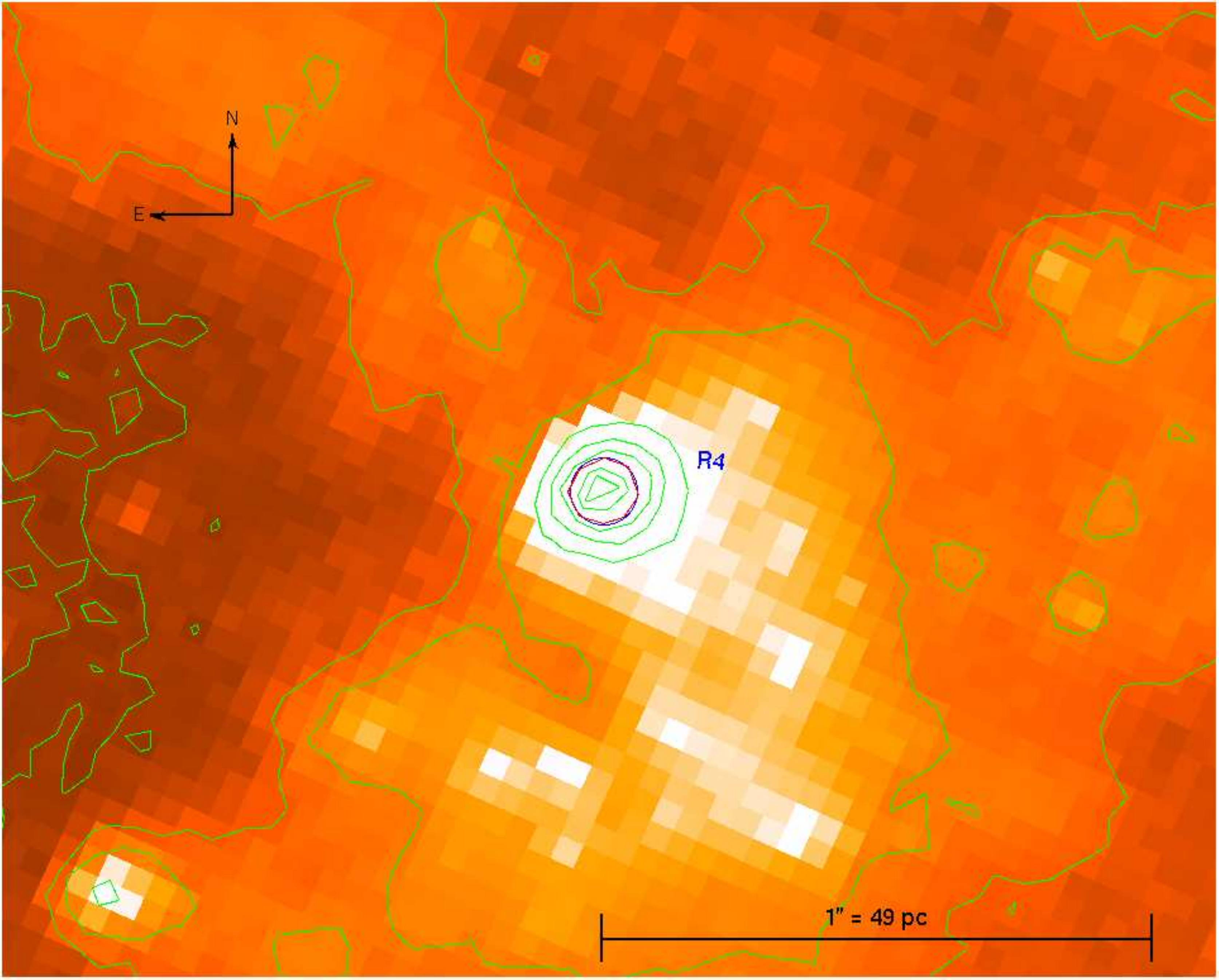}
\end{minipage}
\begin{minipage}[c]{0.4in}
\rotcaption{Idem as Figure \ref{sizes2903R1} for region R4 of NGC\,3351.} 
\label{sizes3351R4}
\end{minipage}
\end{figure}


\begin{figure}
\centering
\begin{minipage}[c]{5.6in}
\includegraphics[width=1.22\textwidth,angle=90]{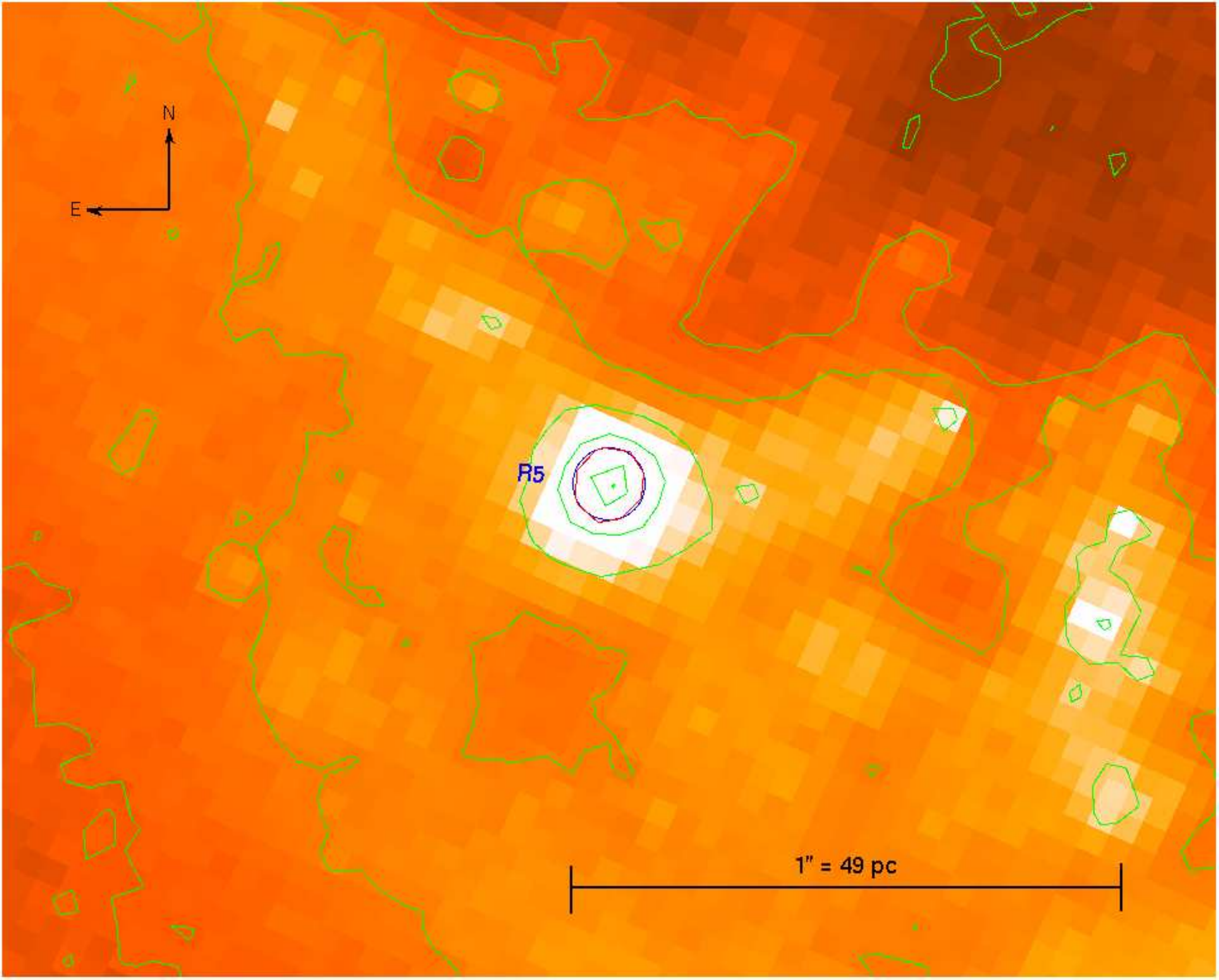}
\end{minipage}
\begin{minipage}[c]{0.4in}
\rotcaption{Idem as Figure \ref{sizes2903R1} for region R5 of NGC\,3351.} 
\label{sizes3351R5}
\end{minipage}
\end{figure}


\begin{figure}
\centering
\begin{minipage}[c]{5.6in}
\includegraphics[width=1.22\textwidth,angle=90]{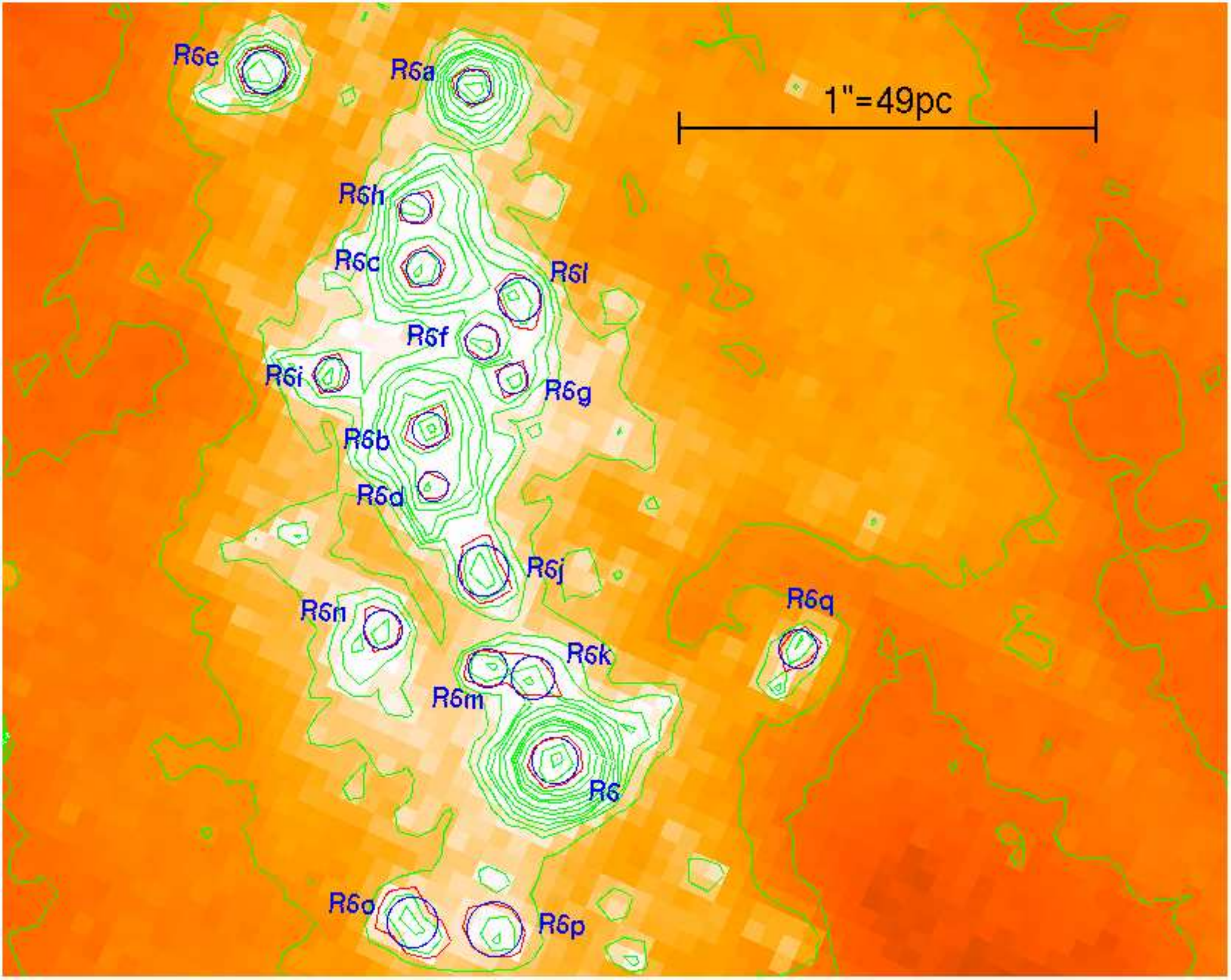}
\end{minipage}
\begin{minipage}[c]{0.4in}
\rotcaption{Idem as Figure \ref{sizes2903R1} for region R6 of NGC\,3351.} 
\label{sizes3351R6}
\end{minipage}
\end{figure}


\begin{figure}
\centering
\begin{minipage}[c]{5.6in}
\includegraphics[width=1.22\textwidth,angle=90]{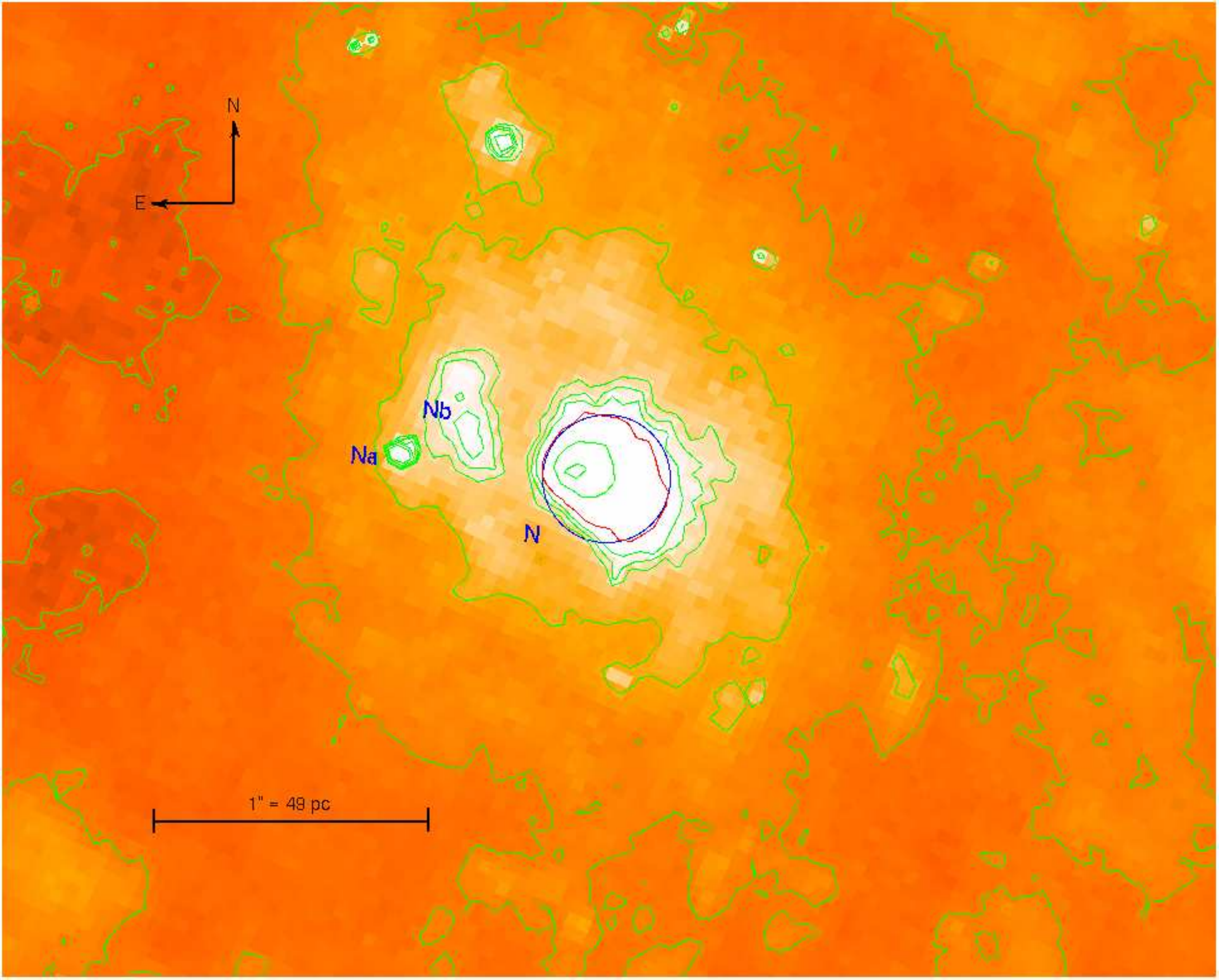}
\end{minipage}
\begin{minipage}[c]{0.4in}
\rotcaption{Idem as Figure \ref{sizes2903R1} for the nucleus of NGC\,3351.} 
\label{sizes3351N}
\end{minipage}
\end{figure}


In order to determine the sizes of the stellar clusters within our observed
CNSFRs, we have used the retrieved wide V HST images. Following
\citetex{1996ApJ...466L..83H} we have defined the radius (R) of a given
structure as in \citetex{1995AJ....110.2665M} assuming that they have an
intrinsically circularly symmetric Gaussian profile. As we can see in Figures
\ref{sizes2903R1}-\ref{sizes3351N} this is a rather good approximation. We
have determined the FWHM, $W_{50}$, by superposing the contours in the F606W
WFPC2-HST image and determining the half light strength between the peak of
the intensity for each region and the galaxy background. We have fitted
circular regions (blue circles in the Figures) to the intensity contours
corresponding to this half light brightness distribution (red contours) of
each single structure. Then,
for this model, we take R as the half light radius,
R\,=\,0.5$\times$W$_{50}$. While \citetex{1995AJ....110.2665M} measured the
sizes from an H$\alpha$ image, we use the F606W (wide V) image dominated by
the stellar continuum (although it contains the H$\alpha$ emission line) and
with the appropriate spatial resolution.

We find, as expected, that almost all of these regions are composed by
more than one knot. Only R4 and R5 in NGC\,3351 seem to have a unique
principal knot, at least at the PC1 spatial resolution. In NGC\,2903, regions
R1 and R2 form an interrelated complex with two main structures: R1 and R2
surrounded by, at least, 31 secondary structures. R1 is resolved into two
separate clusters, labelled R1a and R1b and shows other two associated knots,
labelled R1c and R1d. R2 is made up of a main cluster, labelled R2, and four
secondary ones labelled ``a'' to ``d''. The rest of the knots can not be
directly 
related to any of the two main structures, and have been labelled as R12
followed by a letter from ``a'' to ``v''. The same criterion has been applied
for 
regions R4 and R7 for which we find 28 and 27 individual clusters,
respectively. On the other hand, the galaxy nucleus seems to be formed by one
main structure with two small secondary ones to the South West, and some very
weak knots which are difficult to appreciate in the picture. The secondary
structures have been designated by Na and Nb. When we analyze the F606W-PC1
image of NGC\,3310, we find another region very close to R6, not classified 
by \citetex{2000MNRAS.311..120D}, probably due to the lower spatial resolution
of the data used in that work, and labelled S6 by us, which seems to be
coincident with 
knot L of \citetex{1993MNRAS.260..177P}. For all the regions of this
galaxy we find a principal knot and several secondary knots with lower peak
intensities, except for R1 and R2, which present only one and two secondary
knots, respectively. The same criterion as for NGC\,2903 has been
applied to name the regions. For the other regions of NGC\,3310, R4, R5, R6,
S6, R7 and R10, we find 31, 8, 14, 9, 29 and 15 individual clusters,
respectively. For the third galaxy, NGC\,3351, apart from the single knots R4
and R5, and R2 formed by a principal and a secondary structure, we also find
that the regions are 
composed by several individual small clusters, 7 and 18 for R3 and R6,
respectively. In all cases the knots have been found with a detection
level of 10\,$\sigma$ above the background. All these knots are within the
radius of the regions defined by Planesas et al.\ (1997) for NGC\,2903 and
NGC\,3351, and by \citetex{2000MNRAS.311..120D} for NGC\,3310, except for S6.
We have to remark that our search for knots has not been exhaustive since that
is not the aim of this work. The radii of the single knots vary between 1.5
and 4.0\,pc for NGC\,2903, between 2.2 and 6.2\,pc for NGC\,3310, and between
1.7 and 4.9\,pc for NGC\,3351. Obviously, the spatial resolution and, at
least, the estimates of the smaller radii depend on the distance to each
galaxy.

Table \ref{knots} gives, for each identified knot, the position, as given by
the astrometric calibration of the HST image; the radius of the circular
region defined as described above together with its error;  and the peak
intensity in counts, as measured from the WFPC2 image. The nucleus of
NGC\,2903 is rather compact, but resolved, with a radius of 3.8\,pc. For
the nucleus of NGC\,3310 and NGC\,3351 we measure radii of 14.2 and 11.3\,pc,
respectively.

{\footnotesize
\begin{longtable}{@{}lcccc@{}}
\caption[Positions, radii and peak intensities of the CNSFRs and the
 nuclei]{Positions, radii and peak intensities of the CNSFRs and the nuclei
 derived from the HST-F606W image. \label{knots}} \\ 
 \hline
 \hline
Region &  \multicolumn{2}{c}{position} &  R & I \\
  &  $\alpha_{J2000.0}$ & $\delta_{J2000.0}$ & (pc) & (counts)  \\
 \hline
 \endfirsthead
 
 \multicolumn{5}{l}{\small Table~\ref{knots} continued}
  \\
 \hline
 \hline
Region &  \multicolumn{2}{c}{position} &  R & I \\
  &  $\alpha_{J2000.0}$ & $\delta_{J2000.0}$ & (pc) & (counts)  \\
 \hline
 \endhead
 
 \hline\multicolumn{ 5}{r}{\small\sl continued on next page}
 \endfoot
 
 \hline
 \noalign {\centering } 
 \endlastfoot
 \multicolumn{5}{c}{NGC\,2903}\\[4pt]
R1a & 9$^{h}$32$^{m}$10\fsec06 & +21$^\circ$30\arcmin06\farcs86 & 1.6$\pm$0.1 & 4083  \\
R1b & 9$^{h}$32$^{m}$10\fsec06 & +21$^\circ$30\arcmin06\farcs95 & 1.6$\pm$0.2 & 4015  \\
R1c & 9$^{h}$32$^{m}$10\fsec07 & +21$^\circ$30\arcmin07\farcs10 & 1.5$\pm$0.1 &  363  \\
R1d & 9$^{h}$32$^{m}$10\fsec05 & +21$^\circ$30\arcmin07\farcs08 & 2.1$\pm$0.1 &  254  \\[2pt]
R2  & 9$^{h}$32$^{m}$09\fsec98 & +21$^\circ$30\arcmin06\farcs65 & 1.9$\pm$0.1 & 2892  \\
R2a & 9$^{h}$32$^{m}$09\fsec96 & +21$^\circ$30\arcmin06\farcs28 & 1.9$\pm$0.1 &  390  \\
R2b & 9$^{h}$32$^{m}$09\fsec96 & +21$^\circ$30\arcmin06\farcs61 & 1.9$\pm$0.2 &  307  \\
\end{longtable}
}

We have used the virial theorem to estimate upper limits to the dynamical
masses (M$_{\ast}$) inside the half light radius (R) for each observed knot in
the F606W WFPC2-HST image. In order to do this, we have assumed that the
systems are spherically symmetric, gravitationally  bound and have an
isotropic 
velocity distribution [$\sigma^2$(total)\,=\,3 $\sigma_{\ast}^2$]. Then, the
dynamical mass is given by M$_{\ast}$\,=\,3\,$\sigma_{\ast}^2$\,R/G
(Ho and Filippenko 1996a,b).
\nocite{1996ApJ...466L..83H,1996ApJ...472..600H}

It must be noted that we have measurements for the size of each knot
(typically between 3 and 5\,pc), but we do not have direct access 
to the stellar velocity dispersion of each individual clusters, since
our spectroscopic measurements encompass a wider area
(1.0\,$\times$\,1.9\,arcsec$^2$, which corresponds approximately to
42\,$\times$\,76\,pc$^2$ for NGC\,2903, 73\,$\times$\,131\,pc$^2$ for
NGC\,3310, and 49\,$\times$\,88\,pc$^2$ for NGC\,3351) that includes the whole
CNSFRs to which each group of knots belong. The use of these wider size scale
velocity dispersion measurements to estimate the mass of each knot, leads us to
overestimate the mass of the individual clusters, and hence of each CNSFR.

As we can not be sure that we are actually measuring their velocity dispersion,
we prefer to say that our measurements of $\sigma_{\ast}$, and hence the
dynamical masses, constitute upper limits. Although we are well aware of the
difficulties, still we are confident that these upper limits are valid and
important for comparison with the gas kinematical measurements. 

The estimated dynamical masses and their corresponding errors for each knot
are listed in Table \ref{mass}. For the regions that have been observed in
more than one slit position, we list the derived values using the two
different stellar velocity dispersions. The dynamical masses in the rows
labelled  ``sum''  have been found by adding the individual masses for each
knot of the corresponding CNSFR, when there are more than one. We have taken
the average of the R12sum estimated from slit positions S1 of NGC\,2903 as the
``adopted'' dynamical mass for R12. The ``adopted'' dynamical 
mass for R2 in NGC\,3351 is directly the sum of the dynamical masses of the
two knots derived using the velocity dispersion estimated from S1, since the
mass calculated from S3 is in agreement with the first one (see Table
\ref{mass}) but it has a much greater error. This greater error is due to the
lower S/N ratio of the spectrum of R2 extracted from S3. 
In contrast, the derived dynamical masses for R3 in this galaxy are not in
complete agreement, although they coincide within the errors,
which are comparable. In this case we have taken  the
average  as the ``adopted'' value. The fractional errors of the dynamical
masses of the individual knots and of the CNSFRs are listed in column 4. 
As explained above, since the stellar velocity dispersion for
the R1+R2 region of NGC\,3310 has a large error,
we do not list the errors of the dynamical masses of the individual knots or
of the whole region R12sum.


{\footnotesize
\begin{longtable}{@{}lcccrlccc@{}}
\caption{Dynamical masses. \label{mass}} \\ 
 \hline
 \hline
 Region & Slit & M$_{\ast}$ & error(\%) & & Region & Slit & M$_{\ast}$ & error(\%) \\ 
 \hline
 \endfirsthead
 
 \multicolumn{5}{l}{\small Table~\ref{mass} continued}
  \\
 \hline
 \hline
 Region & Slit & M$_{\ast}$ & error(\%) & & Region & Slit & M$_{\ast}$ & error(\%) \\ 
 \hline
 \endhead
 
 \hline\multicolumn{8}{r}{\small\sl continued on next page}
 \endfoot
 
 \hline
 \multicolumn{8}{l}{masses in 10$^5$\,M$_\odot$.}\\
 \endlastfoot

 \multicolumn{9}{c}{NGC\,2903}\\[2pt]
R1a	& S1 & 40$\pm$5   & 12 &  & R4         & S2 & 46$\pm$6 & 13   \\
R1b 	& S1 & 39$\pm$6   & 17 &  & R4a        & S2 & 26$\pm$4 & 16   \\
R1c	& S1 & 37$\pm$5   & 13 &  & R4b        & S2 & 23$\pm$4 & 17   \\
R1d	& S1 & 52$\pm$6   & 12 &  & R4c        & S2 & 26$\pm$4 & 14   \\
R2	& S1 & 46$\pm$6   & 12 &  & R4d        & S2 & 37$\pm$5 & 15   \\
R2a	& S1 & 47$\pm$6   & 12 &  & R4e        & S2 & 46$\pm$8 & 17   \\
R2b	& S1 & 47$\pm$7   & 15 &  & R4f        & S2 & 40$\pm$5 & 13   \\
R2c	& S1 & 46$\pm$6   & 12 &  & R4g        & S2 & 40$\pm$7 & 19   \\
R2d	& S1 & 44$\pm$7   & 16 &  & R4h        & S2 & 20$\pm$4 & 19   \\
R12a	& S1 & 47$\pm$6   & 12 &  & R4i        & S2 & 27$\pm$5 & 20   \\
R12b	& S1 & 63$\pm$7   & 11 &  & R4j        & S2 & 32$\pm$5 & 15   \\
R12c	& S1 & 63$\pm$7   & 11 &  & R4k        & S2 & 27$\pm$4 & 16   \\
R12d	& S1 & 52$\pm$6   & 12 &  & R4l        & S2 & 29$\pm$5 & 19   \\
R12e	& S1 & 40$\pm$5   & 12 &  & R4m        & S2 & 30$\pm$6 & 18   \\
R12f	& S1 & 45$\pm$5   & 12 &  & R4n        & S2 & 20$\pm$5 & 24   \\
R12g	& S1 & 79$\pm$10  & 12 &  & R4o        & S2 & 31$\pm$6 & 18   \\
R12h	& S1 & 89$\pm$14  & 15 &  & R4p        & S2 & 22$\pm$4 & 18   \\
R12i	& S1 & 79$\pm$11  & 14 &  & R4q        & S2 & 36$\pm$6 & 17   \\
R12j	& S1 & 81$\pm$13  & 16 &  & R4r        & S2 & 23$\pm$4 & 17   \\
R12k	& S1 & 100$\pm$16 & 16 &  & R4s        & S2 & 36$\pm$6 & 17   \\
R12l	& S1 & 47$\pm$6   & 12 &  & R4t        & S2 & 24$\pm$4 & 17   \\
R12m	& S1 & 94$\pm$13  & 13 &  & R4u        & S2 & 33$\pm$6 & 18   \\
R12n	& S1 & 44$\pm$5   & 12 &  & R4v        & S2 & 29$\pm$5 & 19   \\
R12o	& S1 & 64$\pm$8   & 13 &  & R4w        & S2 & 27$\pm$4 & 16   \\
R12p	& S1 & 70$\pm$9   & 13 &  & R4x        & S2 & 43$\pm$8 & 18   \\
R12q	& S1 & 58$\pm$8   & 14 &  & R4y        & S2 & 26$\pm$5 & 20   \\
R12r	& S1 & 68$\pm$12  & 18 &  & R4z        & S2 & 27$\pm$5 & 20   \\
R12s	& S1 & 52$\pm$6   & 12 &  & R4$\alpha$ & S2 & 27$\pm$5 & 20   \\
R12t	& S1 & 88$\pm$14  & 16 &  & R4sum      & S2 & 853$\pm$28 & 3  \\
R12u	& S1 & 52$\pm$7   & 14 &  & R7         & S1 & 26$\pm$5 & 18   \\
R12v	& S1 & 42$\pm$7   & 16 &  & R7a        & S1 & 26$\pm$5 & 19   \\
R12sum	& S1 & 1816$\pm$48& 3  &  & R7b        & S1 & 32$\pm$6 & 17   \\
R1a     & S2 & 45$\pm$5   & 10 &  & R7c        & S1 & 26$\pm$5 & 20   \\
R1b     & S2 & 44$\pm$6   & 14 &  & R7d        & S1 & 25$\pm$5 & 19   \\
R1c     & S2 & 41$\pm$4   & 10 &  & R7e        & S1 & 14$\pm$3 & 22   \\
R1d     & S2 & 59$\pm$6   & 9  &  & R7f        & S1 & 38$\pm$8 & 21   \\
R2      & S2 & 52$\pm$5   & 10 &  & R7g        & S1 & 24$\pm$5 & 19   \\
R2a     & S2 & 53$\pm$5   & 10 &  & R7h        & S1 & 21$\pm$4 & 19   \\
R2b     & S2 & 53$\pm$7   & 13 &  & R7i        & S1 & 22$\pm$4 & 18   \\
R2c     & S2 & 52$\pm$5   & 10 &  & R7j        & S1 & 25$\pm$5 & 19   \\
R2d     & S2 & 50$\pm$6   & 13 &  & R7k        & S1 & 25$\pm$5 & 21   \\
R12a    & S2 & 53$\pm$5   & 10 &  & R7l        & S1 & 31$\pm$7 & 23   \\
R12b    & S2 & 71$\pm$6   & 9  &  & R7m        & S1 & 27$\pm$5 & 19   \\
R12c    & S2 & 71$\pm$6   & 9  &  & R7n        & S1 & 20$\pm$5 & 22   \\
R12d    & S2 & 59$\pm$6   & 9  &  & R7o        & S1 & 20$\pm$4 & 20   \\
R12e    & S2 & 45$\pm$5   & 10 &  & R7p        & S1 & 19$\pm$3 & 18   \\
R12f    & S2 & 51$\pm$5   & 10 &  & R7q        & S1 & 17$\pm$3 & 18   \\
R12g    & S2 & 89$\pm$9   & 10 &  & R7r        & S1 & 18$\pm$4 & 20   \\
R12h    & S2 & 101$\pm$13 & 13 &  & R7s        & S1 & 36$\pm$8 & 22   \\
R12i    & S2 & 89$\pm$11  & 12 &  & R7t        & S1 & 21$\pm$5 & 25   \\
R12j    & S2 & 91$\pm$13  & 14 &  & R7u        & S1 & 28$\pm$6 & 22   \\
R12k    & S2 & 113$\pm$16 & 14 &  & R7v        & S1 & 21$\pm$4 & 19   \\
R12l    & S2 & 53$\pm$5   & 10 &  & R7w        & S1 & 19$\pm$4 & 20   \\
R12m    & S2 & 107$\pm$12 & 11 &  & R7x        & S1 & 23$\pm$4 & 19   \\
R12n    & S2 & 50$\pm$5   & 10 &  & R7y        & S1 & 17$\pm$3 & 18   \\
R12o    & S2 & 72$\pm$8   & 11 &  & R7z        & S1 & 21$\pm$4 & 18   \\
R12p    & S2 & 79$\pm$8   & 10 &  & R7sum      & S1 & 642$\pm$26& 4   \\
R12q    & S2 & 65$\pm$7   & 11 &  & N          & S2 & 112$\pm$10 & 9  \\
R12r    & S2 & 77$\pm$12  & 15 &  & Na         & S2 & 43$\pm$7   & 16 \\
R12s    & S2 & 59$\pm$6   & 9  &  & Nb         & S2 & 62$\pm$8   & 13 \\
R12t    & S2 & 100$\pm$13 & 13 &  & Nsum       & S2 & 217$\pm$15 & 7  \\
R12u    & S2 & 59$\pm$7   & 12 &  &\\
R12v    & S2 & 47$\pm$6   & 13  \\
R12sum  & S2 & 2054$\pm$45& 2   \\
R12 (adopted) & & 1935$\pm$66 & 3  \\[4pt]

 \multicolumn{9}{c}{NGC\,3310}\\[2pt]
R1	    & S2 & 202:          & ---& & R6	    & S2 & 28$\pm$8	 & 30 \\
R1a	    & S2 & 115:          & ---& & R6a	    & S2 & 18$\pm$6	 & 32 \\
R2	    & S2 & 208:          & ---& & R6b	    & S2 & 32$\pm$9	 & 29 \\
R2a	    & S2 & 241:          & ---& & R6c	    & S2 & 22$\pm$7	 & 30 \\
R2b	    & S2 & 144:          & ---& & R6d	    & S2 & 35$\pm$10	 & 30 \\  
R12sum	    & S2 & 911:          & ---& & R6e	    & S2 & 32$\pm$10	 & 31 \\  
R4	    & S1 & 26$\pm$4	 & 17 & & R6f	    & S2 & 30$\pm$9	 & 29 \\  
R4a	    & S1 & 27$\pm$5	 & 17 & & R6g	    & S2 & 28$\pm$9	 & 31 \\  
R4b	    & S1 & 22$\pm$4	 & 18 & & R6h	    & S2 & 26$\pm$8	 & 30 \\	  
R4c	    & S1 & 21$\pm$4	 & 18 & & R6i	    & S2 & 21$\pm$6	 & 30 \\  
R4d	    & S1 & 36$\pm$6	 & 16 & & R6j	    & S2 & 33$\pm$10	 & 31 \\	  
R4e	    & S1 & 38$\pm$6	 & 17 & & R6k	    & S2 & 25$\pm$8	 & 30 \\	  
R4f	    & S1 & 38$\pm$6	 & 16 & & R6l	    & S2 & 31$\pm$10	 & 31 \\	  
R4g	    & S1 & 26$\pm$4	 & 17 & & R6m	    & S2 & 35$\pm$11	 & 31 \\	  
R4h	    & S1 & 25$\pm$5	 & 21 & & R6sum	    & S2 & 397$\pm$33	 & 8  \\	  
R4i	    & S1 & 37$\pm$6	 & 16 & & S6	    & S2 & 28$\pm$6	 & 23 \\	  
R4j	    & S1 & 27$\pm$5	 & 19 & & S6a	    & S2 & 37$\pm$8	 & 23 \\	  
R4k	    & S1 & 24$\pm$4	 & 17 & & S6b	    & S2 & 18$\pm$5	 & 25 \\	  
R4l	    & S1 & 30$\pm$7	 & 21 & & S6c	    & S2 & 23$\pm$6	 & 24 \\	  
R4m	    & S1 & 25$\pm$5	 & 19 & & S6d	    & S2 & 18$\pm$5	 & 25 \\  
R4n	    & S1 & 30$\pm$5	 & 18 & & S6e	    & S2 & 20$\pm$5	 & 23 \\	  
R4o	    & S1 & 28$\pm$5	 & 17 & & S6f	    & S2 & 24$\pm$6	 & 23 \\	  
R4p	    & S1 & 21$\pm$4	 & 18 & & S6g	    & S2 & 20$\pm$5	 & 25 \\	  
R4q	    & S1 & 28$\pm$6	 & 20 & & S6h	    & S2 & 19$\pm$5	 & 25 \\	  
R4r	    & S1 & 27$\pm$5	 & 18 & & S6sum	    & S2 & 210$\pm$17	 & 8  \\	  
R4s	    & S1 & 24$\pm$5	 & 19 & & R7	    & S2 & 42$\pm$9	 & 22 \\	  
R4t	    & S1 & 26$\pm$5	 & 21 & & R7a	    & S2 & 71$\pm$15	 & 21 \\	  
R4u	    & S1 & 34$\pm$6	 & 19 & & R7b	    & S2 & 46$\pm$11	 & 23 \\	  
R4v	    & S1 & 36$\pm$7	 & 19 & & R7c	    & S2 & 64$\pm$14	 & 21 \\	  
R4w	    & S1 & 26$\pm$5	 & 21 & & R7d	    & S2 & 40$\pm$9	 & 22 \\	  
R4x	    & S1 & 28$\pm$5	 & 18 & & R7e	    & S2 & 32$\pm$8	 & 24 \\	  
R4y	    & S1 & 26$\pm$5	 & 21 & & R7f	    & S2 & 52$\pm$11	 & 22 \\	  
R4z	    & S1 & 31$\pm$6	 & 18 & & R7g	    & S2 & 60$\pm$13	 & 22 \\	  
R4$\alpha$  & S1 & 29$\pm$6	 & 20 & & R7h	    & S2 & 51$\pm$11	 & 22 \\	  
R4$\beta$   & S1 & 33$\pm$6	 & 18 & & R7i	    & S2 & 40$\pm$9	 & 22 \\	  
R4$\gamma$  & S1 & 33$\pm$6	 & 18 & & R7j	    & S2 & 53$\pm$13	 & 24 \\	  
R4$\delta$  & S1 & 23$\pm$5	 & 19 & & R7k	    & S2 & 44$\pm$10	 & 23 \\	  
R4sum	    & S1 & 886$\pm$30	 & 3  & & R7l	    & S2 & 54$\pm$12	 & 22 \\	  
R4	    & S2 & 30$\pm$5	 & 18 & & R7m	    & S2 & 40$\pm$9	 & 23 \\	  
R4a	    & S2 & 31$\pm$6	 & 18 & & R7n	    & S2 & 44$\pm$10	 & 23 \\	  
R4b	    & S2 & 25$\pm$5	 & 19 & & R7o	    & S2 & 52$\pm$13	 & 25 \\	  
R4c	    & S2 & 25$\pm$5	 & 19 & & R7p	    & S2 & 43$\pm$9	 & 22 \\	  
R4d	    & S2 & 42$\pm$7	 & 17 & & R7q	    & S2 & 50$\pm$11	 & 22 \\	  
R4e	    & S2 & 43$\pm$8	 & 18 & & R7r	    & S2 & 52$\pm$12	 & 22 \\	  
R4f	    & S2 & 43$\pm$7	 & 17 & & R7s	    & S2 & 53$\pm$12	 & 22 \\	  
R4g	    & S2 & 30$\pm$5	 & 18 & & R7t	    & S2 & 32$\pm$8	 & 24 \\	  
R4h	    & S2 & 28$\pm$6	 & 22 & & R7u	    & S2 & 48$\pm$11	 & 24 \\	  
R4i	    & S2 & 42$\pm$7	 & 17 & & R7v	    & S2 & 48$\pm$11	 & 23 \\	  
R4j	    & S2 & 30$\pm$6	 & 19 & & R7w	    & S2 & 52$\pm$11	 & 22 \\	  
R4k	    & S2 & 27$\pm$5	 & 18 & & R7x	    & S2 & 46$\pm$10	 & 22 \\	  
R4l	    & S2 & 35$\pm$8	 & 22 & & R7y	    & S2 & 52$\pm$12	 & 22 \\	  
R4m	    & S2 & 29$\pm$6	 & 20 & & R7z	    & S2 & 44$\pm$11	 & 24 \\	  
R4n	    & S2 & 35$\pm$7	 & 19 & & R7$\alpha$& S2 & 63$\pm$15	 & 25 \\	  
R4o	    & S2 & 32$\pm$6	 & 18 & & R7$\beta$ & S2 & 44$\pm$11	 & 24 \\	  
R4p	    & S2 & 25$\pm$5	 & 19 & & R7sum	    & S2 & 1413$\pm$60	 & 4  \\	  
R4q	    & S2 & 32$\pm$7	 & 21 & & R10	    & S1 & 33$\pm$5	 & 17 \\	  
R4r	    & S2 & 31$\pm$6	 & 19 & & R10a	    & S1 & 53$\pm$10	 & 18 \\	  
R4s	    & S2 & 27$\pm$6	 & 20 & & R10b	    & S1 & 40$\pm$6	 & 16 \\	  
R4t	    & S2 & 30$\pm$6	 & 22 & & R10c	    & S1 & 37$\pm$6	 & 18 \\	  
R4u	    & S2 & 39$\pm$8	 & 20 & & R10d	    & S1 & 43$\pm$8	 & 18 \\	  
R4v	    & S2 & 41$\pm$8	 & 19 & & R10e	    & S1 & 29$\pm$5	 & 19 \\	  
R4w	    & S2 & 30$\pm$6	 & 22 & & R10f	    & S1 & 29$\pm$6	 & 21 \\
R4x	    & S2 & 32$\pm$6	 & 19 & & R10g	    & S1 & 46$\pm$9	 & 21 \\
R4y	    & S2 & 30$\pm$6	 & 22 & & R10h	    & S1 & 43$\pm$7	 & 17 \\
R4z	    & S2 & 36$\pm$7	 & 19 & & R10i	    & S1 & 50$\pm$9	 & 19 \\
R4$\alpha$  & S2 & 33$\pm$7	 & 21 & & R10j	    & S1 & 42$\pm$8	 & 19 \\
R4$\beta$   & S2 & 38$\pm$7	 & 19 & & R10k	    & S1 & 40$\pm$7	 & 17 \\
R4$\gamma$  & S2 & 38$\pm$7	 & 19 & & R10l	    & S1 & 40$\pm$7	 & 17 \\
R4$\delta$  & S2 & 27$\pm$5	 & 20 & & R10m	    & S1 & 32$\pm$6	 & 18 \\
R5	    & S2 & 35$\pm$7	 & 19 & & R10n	    & S1 & 33$\pm$7	 & 20 \\ 
R5a	    & S2 & 32$\pm$7	 & 21 & & R10sum    & S1 & 588$\pm$28	 & 5  \\ 
R5b	    & S2 & 36$\pm$7	 & 20 & & N         & S1 & 526$\pm$57	 & 11 \\ 
R5c	    & S2 & 63$\pm$12	 & 19 & & Na  	    & S1 & 216$\pm$28	 & 13 \\ 
R5d	    & S2 & 36$\pm$7	 & 19 & & Nsum	    & S1 & 742$\pm$64	 & 9  \\ 
R5e	    & S2 & 37$\pm$7	 & 18 & & \\ 
R5f	    & S2 & 30$\pm$5	 & 18 & & \\ 
R5g	    & S2 & 33$\pm$6	 & 19 & & \\ 
R45sum	    & S2 & 1317$\pm$41	 & 3  & & \\[4pt]

 \multicolumn{9}{c}{NGC\,3351}\\[2pt]

R2          &  S1   &  85$\pm$5   & 6     & & R6          &  S2   &  29$\pm$9   & 32   \\
R2a         &  S1   &  44$\pm$4   & 9     & & R6a         &  S2   &  21$\pm$8   & 37   \\
R2sum       &  S1   & 129$\pm$6   & 3     & & R6b         &  S2   &  21$\pm$8   & 38   \\
R2          &  S3   &  87$\pm$21  & 24    & & R6c         &  S2   &  21$\pm$8   & 37   \\ 
R2a         &  S3   &  44$\pm$11  & 25    & & R6d         &  S2   &  18$\pm$7   & 37   \\ 
R2sum       &  S3   & 131$\pm$23  & 18    & & R6e         &  S2   &  26$\pm$10  & 38   \\ 
R2 (adopted)&       & 129$\pm$6   & 3     & & R6f         &  S2   &  21$\pm$8   & 37   \\ 
R3          &  S1   &  82$\pm$14  & 17    & & R6g         &  S2   &  18$\pm$7   & 37   \\
R3a         &  S1   &  46$\pm$8   & 17    & & R6h         &  S2   &  18$\pm$7   & 37   \\
R3b         &  S1   &  42$\pm$9   & 22    & & R6i         &  S2   &  21$\pm$8   & 38   \\
R3c         &  S1   &  52$\pm$11  & 20    & & R6j         &  S2   &  31$\pm$12  & 38   \\
R3d         &  S1   &  57$\pm$11  & 18    & & R6k         &  S2   &  26$\pm$10  & 39   \\
R3e         &  S1   &  61$\pm$11  & 20    & & R6l         &  S2   &  26$\pm$10  & 38   \\
R3f         &  S1   &  78$\pm$15  & 20    & & R6m         &  S2   &  24$\pm$9   & 39   \\
R3sum       &  S1   & 417$\pm$31  & 7     & & R6n         &  S2   &  24$\pm$9   & 38   \\
R3          &  S2   &  94$\pm$23  & 24    & & R6o         &  S2   &  31$\pm$12  & 38   \\ 
R3a         &  S2   &  53$\pm$13  & 24    & & R6p         &  S2   &  33$\pm$13  & 38   \\ 
R3b         &  S2   &  48$\pm$14  & 28    & & R6q         &  S2   &  24$\pm$9   & 39   \\ 
R3c         &  S2   &  60$\pm$16  & 27    & & R6sum       &  S2   & 434$\pm$39  & 9    \\ 
R3d         &  S2   &  65$\pm$17  & 26    & & N           &  S3   & 350$\pm$11  & 3    \\ 
R3e         &  S2   &  70$\pm$17  & 25    & & \\ 
R3f         &  S2   &  89$\pm$5   & 26    & & \\ 
R3sum       &  S2   & 477$\pm$47  & 10    & & \\ 
R3 (adopted)&       & 447$\pm$56  & 13    & & \\ 
R4          &  S2   &  87$\pm$12  & 14    & & \\ 
R5          &  S3   &  49$\pm$8   & 16    & & \\ 

 \end{longtable}
}

\section{Ionizing star cluster properties}
\label{ionmass}

We have derived the masses of the ionizing star clusters (M$_{ion}$) from the
total number of ionizing photons using solar metallicity single burst models
by \citetex{1995A&AS..112...35G}, assuming a Salpeter initial mass function
(\citeplain{1955ApJ...121..161S}; IMF) with lower and upper mass limits of 0.8
and 120 M$_\odot$ which provide the number of ionizing photons per unit mass,
[$Q(H_0)/M_{ion}$].  In order to
take into account the evolution of the \HII\ region, we have made use of the
fact that a relation exists between the degree of evolution of the cluster, as
represented by the equivalent width of the H$\beta$ emission line, and the
number of Lyman continuum photons per unit solar mass
(e.g.\ \citeplain{2000MNRAS.318..462D}). This number decreases with the age of
the region. We have 
used the following relation between $Q(H_0)/M_{ion}$ and the equivalent
width (EW) of H$\beta$, EW(H$\beta$), derived from the models, in order to
take into account the evolutionary state of the region
\cite{1998Ap&SS.263..143D}: 

\[
log\big(Q(H_0)/M_{ion}\big)\,=\,44.48\,+\,0.86\,log\big(EW(H\beta)\big)
\]

We have  measured the EW(H$\beta$) from our spectra  
(see Table \ref{parameters}) following the same procedure as in
(\citeplain{2006MNRAS.372..293H,2008MNRAS.383..209H}; see Chapter \S
\ref{HIIgal-obs}), that is defining a pseudo-continuum to 
take into account the absorption  from the underlying stellar
population. This correction would 
increase the values of EW(H$\beta$) thus decreasing the calculated M$_{ion}$.
However, since there is a contribution to the continuum by the
older stellar population (see discussion in D\'\i az et al.\ 2007; see Chapter
\S \ref{abundan}) the final derived values of the equivalent widths are still
underestimated.
The total number of ionizing photons has been derived from the H$\alpha$
luminosities (see for example \citeplain{1989agna.book.....O}):
\[
Q(H_0)\,=\,7.35\,\times\,10^{11}\,L(H\alpha)
\]

For NGC\,2903 and NGC\,3351 we have taken the total observed H$\alpha$
luminosities from Planesas et al.\ (1997) and for NGC\,3310 from
\citetex{2000MNRAS.311..120D} and \citetex{1993MNRAS.260..177P}. Corrections
were applied to take into account the different assumed distances. In the
cases of the regions of NGC\,2903 and NGC\,3351 we have corrected the
H$\alpha$ luminosities given by Planesas and co-authors for internal
extinction using the colour excess [E(B-V)] estimated by \cite{tesisdiego}
from optical spectroscopy and assuming  the  galactic extinction law of 
\cite{1972ApJ...172..593M} with $R_v$\,=\,3.2. \citetex{2000MNRAS.311..120D}
and \citetex{1993MNRAS.260..177P} give the already corrected luminosities for
NGC\,3310. In the case of NGC\,2903, Planesas et al.\ (1997) estimated a
diameter of 2.0\,arcsec for regions R1, R2, R7 and the nucleus and 2.4\,arcsec
for region R4. In the case of R1+R2 we added their H$\alpha$
luminosities. Likewise, for NGC\,3351, these authors estimated a diameter of
2.4\,arcsec for each whole region and the nucleus, except for R4 for which
they used 2.2\,arcsec. \citetex{2000MNRAS.311..120D} estimated a diameter of
2\,arcsec for regions R2, R4, R6, R7 and R10, and 1.4\,arcsec for R1 and R5 of
NGC\,3310, and 3.4 for the Jumbo region. For this last one we give the values
derived using the different  quantities for the luminosities given by
\citetex{2000MNRAS.311..120D} (region R19) and \citetex{1993MNRAS.260..177P}
(region A of that work). Again, for R1+R2 and R4+R5 (for S2) we added their
H$\alpha$ luminosities. No values are found in the literature for the
H$\alpha$ luminosity of region X  of NGC\,2903 and regions X and Y of
NGC\,3310. Our derived values of Q(H$_0$) constitute lower limits since we
have not taken into account the fraction of photons that may have been
absorbed by dust or may have escaped the region.

The final expression for the derivation of M$_{ion}$ is: 

\[
M_{ion}\,=\,\frac{7.35\,\times\,10^{11}\,L(H\alpha)}{10^{44.48\,+\,0.86\,log\left[EW(H\beta)\right]}}
\]

The derived masses for the ionizing populations of the observed CNSFRs are
given in column 8 of Table \ref{parameters} and are between 2 and 8 per cent
of the dynamical mass for those regions of NGC\,2903, between 1 and 7 per cent
for those of NGC\,3310, and between 2 and 16 per cent in the cases of
NGC\,3351 (see column 11 of the table).


\begin{table}
\centering
\caption{Physical parameters.}
{\footnotesize
\begin{tabular} {@{}l c c c c c c c c c c@{}}
\hline
\hline
 Region & L$_{obs}$(H$\alpha$) & E(B-V)$^b$ & c(H$\alpha$) &
 L(H$\alpha$) & Q(H$_0$) & EW(H$\beta$) & M$_{ion}$ & N$_e$ & M$_{{\rm
 HII}}$ & M$_{ion}$/M$_{\ast}$ 
 \\ 
  & & & & & & & & & & (per cent) \\
\hline

\multicolumn{11}{c}{NGC\,2903}\\[2pt]
R1+R2 & 20.3$^{a}$ & 0.53 & 0.75 & 114.0 & 83.7 & 12.1 & 32.5 & 280$^c$& 1.35 & 1.7 \\
R4    &  8.9$^{a}$ & 0.66 & 0.93 &  76.3 & 56.1 &  4.8 & 48.2 & 270$^c$& 0.94 & 5.7 \\
R7    &  6.4$^{a}$ & 0.71 & 1.01 &  64.5 & 47.4 &  3.9 & 48.8 & 350$^b$& 0.61 & 7.6 \\[2pt]
N$^d$ &  2.8$^{a}$ &  --- & ---  &   2.8 &  2.0 &  3.8 &  2.1 & ---    & 0.03 & 1.0 \\[4pt]

\multicolumn{11}{c}{NGC\,3310}\\[2pt]
R1+R2 &  --- &  --- & 0.33$^e$ & 102.0$^e$ & 74.9 & 28.6 & 13.9 & 100$^f$ & 3.38 & 1.5 \\
R4    &  --- &  --- & 0.33$^e$ & 144.0$^e$ &106.0 & 32.4 & 17.6 & 100$^g$ & 4.78 & 1.6 \\ 
R4+R5 &  --- &  --- & 0.29$^e$ & 218.0$^e$ &160.0 & 41.7 & 21.4 & 100$^g$ & 7.24 & 1.1 \\
R6    &  --- &  --- & 0.25$^e$ &  57.3$^e$ & 42.1 & 16.7 & 12.4 & 100$^f$ & 1.90 & 3.5 \\
S6    &  --- &  --- & 0.35$^g$ &  62.5$^g$ & 45.9 & 12.5 & 17.4 & 100$^g$ & 2.07 & 6.6 \\
R7    &  --- &  --- & 0.25$^e$ &  45.5$^e$ & 33.5 & 19.4 &  8.7 & 100$^f$ & 1.51 & 1.0 \\
R10   &  --- &  --- & 0.33$^e$ &  45.5$^e$ & 33.5 &  9.7 & 15.7 & 100$^f$ & 1.51 & 2.4 \\
N     &  --- &  --- & 0.42$^g$ & 113.0$^g$ & 82.9 & 11.0 & 34.9 & 8000$^g$ & 0.05 & 4.7\\[2pt]
J     &  --- &  --- & 0.33$^e$ & 573.0$^e$ &421.0 & 82.5 & 31.4 & 200$^g$ & 9.52 & --- \\
      &  --- &  --- & 0.19$^g$ & 236.0$^g$ &174.0 & 82.5 & 12.9 & 200$^g$ & 3.93 & --- \\[4pt]

\multicolumn{11}{c}{NGC\,3351}\\[2pt]
R2  & 19.3$^{a}$ & 0.17 &  0.24 & 33.6  & 24.7 &  9.5  &  11.8 & 440$^c$ & 0.25 &  9.0 \\
R3  & 25.0$^{a}$ & 0.46 &  0.65 & 112.0 & 82.3 &  16.5 &  24.5 & 430$^c$ & 0.87 &  5.5 \\
R4  & 12.7$^{a}$ & 0.27 &  0.38 & 30.6  & 22.5 &  13.0 &  8.2  & 310$^c$ & 0.33 &  9.5 \\
R5  & 5.9$^{a}$  & 0.25 &  0.35 & 13.3  & 9.8  &  5.1  &  8.0  & 360$^c$ & 0.12 &  16.2\\
R6  & 7.5$^{a}$  & 0.0  &  0.00 & 7.5   & 5.5  &  2.3  &  8.9  & 360$^c$ & 0.07 &  2.0 \\[2pt]
N   & 3.3$^{a}$  & 0.07 &  0.10 & 4.1   & 3.0  &  1.8  &  6.0  & 650$^b$ & 0.02 &  1.7 \\

\hline
\multicolumn{11}{@{}l}{Luminosities in 10$^{38}$\,erg\,s$^{-1}$, masses in 10$^5$
  M$_\odot$, ionizing photons in 10$^{50}$\,photon\,s$^{-1}$ and densities in
cm$^{-3}$}\\
\multicolumn{11}{@{}l}{$^a$From Planesas et al.\ (1997) corrected for
 the different adopted distances.} \\ 
\multicolumn{11}{@{}l}{$^b$From \citetex{tesisdiego}.}\\
\multicolumn{11}{@{}l}{$^c$From D\'iaz et al.\ (2007; see Chapter \S \ref{abundan}).}\\
\multicolumn{11}{@{}l}{$^d$We assume a value of 0.0 and 300 for E(B-V) and N$_e$,
respectively.}\\
\multicolumn{11}{@{}l}{$^e$From \citetex{2000MNRAS.311..120D}.} \\
\multicolumn{11}{@{}l}{$^f$Assumed from the values given by
  \citetex{1993MNRAS.260..177P} for the other CNSFRs of this galaxy studied by
  }\\ 
\multicolumn{11}{@{}l}{them.}\\
\multicolumn{11}{@{}l}{$^g$From \citetex{1993MNRAS.260..177P}.}
\end{tabular}
}
\label{parameters}
\end{table}


The amount of ionized gas (M$_{{\rm HII}}$) associated to each  star-forming
complex has been obtained from our derived H$\alpha$ luminosities
using the relation given by \citetex{1990ApJ...356..389M} for an electron
temperature of 10$^4$\,K 

\[
M_{{\rm HII}}\,=\,3.32\,\times\,10^{-33}\,L(H\alpha)\,N_e^{-1}
\]

\noindent where N$_e$ is the electron density. The electron density for each
region (obtained from  the [S{\sc ii}]\,$\lambda\lambda$\,6717\,/\,6731\,\AA\
line ratio) has been taken from D\'iaz et al.\ (2007; see Chapter \S
\ref{abundan}) for the CNSFRs of NGC\,2903 and NGC\,3351, and from
\citetex{1993MNRAS.260..177P} for the regions in common and the nucleus of
NGC\,3310. For the rest of the regions of NGC\,3310 we assume a value of N$_e$
equal to 100\,cm$^{-3}$, the value given by \citetex{1993MNRAS.260..177P} for
the other CNSFRs of this galaxy studied by them. For the nucleus of NGC\,2903,
N$_e$ equal to 300\,cm$^{-3}$, typical of nuclear \HII\ regions, has been
assumed. The electron density of the nucleus of NGC\,3351 has been taken from
\citetex{tesisdiego} (see Table \ref{parameters}).

\section{Discussion}
\label{disc-kine}

\subsection*{NGC\,2903}

\begin{figure}
\centering
\vspace*{0.3cm}
\includegraphics[width=.6\textwidth,angle=0]{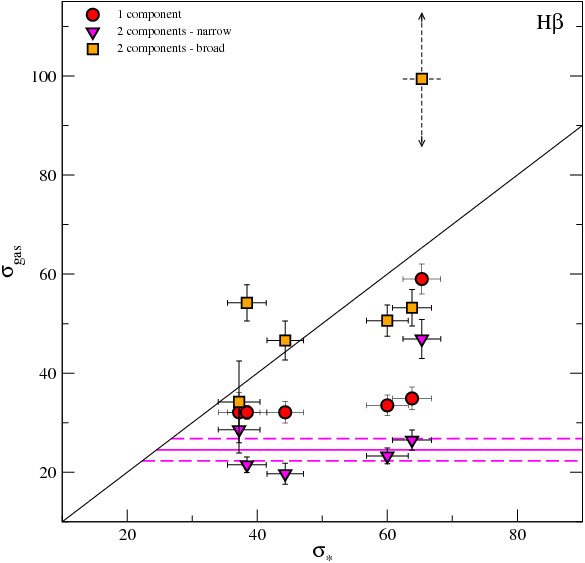}\\\vspace*{0.5cm}
\includegraphics[width=.6\textwidth,angle=0]{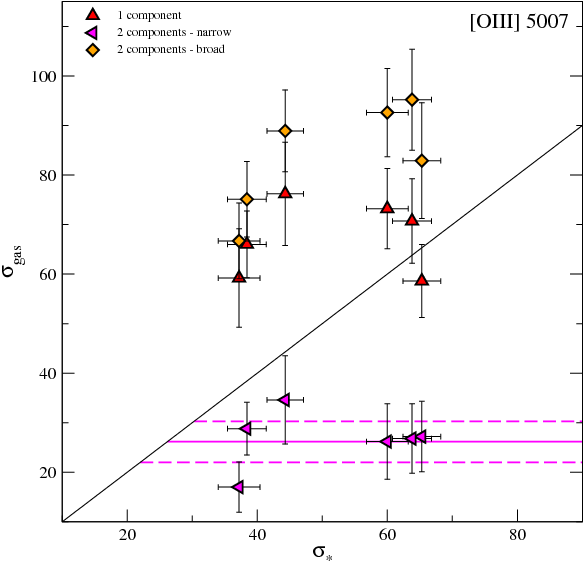}
\caption[Relation between velocity dispersions of the gas, derived
  from H$\beta$ and [O{\sc iii}{\textrm ]}, and stars (CaT) in
  NGC\,2903]{Upper panel: relation between velocity 
  dispersions of the gas (derived from H$\beta$) and stars (CaT) for the
  CNSFRs and the nucleus of NGC\,2903. Symbols are as follows: single Gaussian
  fit, red 
  circles; two Gaussian fit, broad component, orange squares; narrow component,
  magenta downward triangles. Lower panel: as the upper panel for the
  [O{\sc iii}] line. Red upward triangles correspond to the estimates using a
  single Gaussian fit, orange diamonds represent the broad components of the two
  Gaussian fit and magenta left triangles, the narrow components. The 
  magenta line represents the average velocity dispersion of the narrow
  component of the gas (H$\beta$ upper and [O{\sc iii}] lower panel) for the
  CNSFRs, and the magenta dashed lines represent their estimated errors.}
\label{dispersions2903}
\end{figure}

The relations between the velocity dispersions of gas (as measured by the
H$\beta$ and the [O{\sc iii}] emission lines) and stars (as measured by the IR
Ca{\sc ii} triplet in absorption) in the case of NGC\,2903 are shown in Figure
\ref{dispersions2903}. H$\beta$ is shown in the upper panel, [O{\sc iii}] in
the lower one. The straight line shows the one-to-one relation in both
panels. Red circles in the upper panel correspond to gas velocity dispersions
measured from the H$\beta$ emission line using a single Gaussian fit. Orange
squares and magenta downward triangles correspond to the broad and narrow
component respectively, for measurements performed using two-component
Gaussian fits. The deviant point, marked with arrows, corresponds to the
nucleus, which has low signal-to-noise ratio and for which the fits do not
provide 
accurate results for the broad component. In the lower panel, red upward
triangles, orange diamonds and magenta left triangles correspond to the
values obtained by a single Gaussian fit, and to the broad and narrow
components of the two Gaussian fits, respectively. 

In general, the H$\beta$ velocity dispersions of the CNSFRs of NGC\,2903
derived by a single Gaussian fit are lower than the stellar ones by about
25\,km\,s$^{-1}$, except for regions R7 and X, for which these values are 
very similar. These two regions also have the lowest velocity dispersions, and
in the first of them the values derived using the two different
techniques are very similar. On the other hand, a good
agreement is found between the stellar velocity dispersions of the CNSFRs and
those of the broad component of H$\beta$. The narrow component shows velocity
dispersions even lower than those obtained by single Gaussian fits, and 
similar to each other in all cases, with an average value equal to
24.5\,$\pm$\,2.2\,km\,s$^{-1}$, with the error given by the dispersion of the
individual values. This average is represented as a magenta horizontal line
in the upper panel of Figure \ref{dispersions2903}, with its error as 
magenta dashed lines. This narrow component could be identified with ionized
gas in a rotating disc. The broad component should then correspond to the gas
response to the gravitational potential of the stellar cluster and therefore
should, in principle, coincide with the stellar velocity dispersion.

This seems to be the case for the H$\beta$ line (see upper panel of Figure
\ref{dispersions2903}). In the case of the [O{\sc iii}] however, there seems
to be an excess in the broad component velocity dispersion (orange diamonds
in the Figure) which is found to be larger than the stellar one by about
35\,km\,s$^{-1}$ (see Table \ref{disp}). This extra component 
could be identified with peculiar velocities in the high ionization gas
related to massive star winds or even supernova remnants. Since the H$\beta$
line is dominated by the narrow component, this peculiar velocity contributes
negligibly to the line width.

To test the possibility of finding a spurious results due to the low
signal-to-noise ratio of this weak [O{\sc iii}] emission line, we 
generates a synthetic spectrum with the measured characteristics and we added
an artificial noise with a rms twice the observed one. We then measured the
synthetic 
spectra with the same technique as the data, using also a double Gaussian
fitting with the same initial parameters, and we obtain the same results
within the observational errors.

The widths of the narrow [O{\sc iii}] component (magenta left triangles) are
comparable to those of the  H$\beta$ lines, and again show a relatively
constant value, but in this case with a larger
dispersion, giving $\sigma_g$\,=\,26.2\,$\pm$\,4.2\,km\,s$^{-1}$.

The gas velocity dispersions derived for the nucleus from its H$\beta$
emission line is closer to those estimated for the CNSFRs from their [O{\sc
iii}]\,5007\,\AA\ emission lines than to those derived 
from H$\beta$. This broad nuclear component has a velocity dispersion higher
than $\sigma_\ast$ by about 35\,km\,s$^{-1}$, and also the narrow 
one is higher than the average of the narrow components of the CNSFRs this
time by approximately 20\,km\,s$^{-1}$ (see upper panel of Figure
\ref{dispersions2903}).  
However, we have to take into account that this values
are derived with a large error (in particular for the broad component) due to
the low signal-to-noise ratio of our blue spectrum of the nucleus of NGC\,2903.
The single Gaussian fit to this recombination line gives a value similar to
the stellar one (see Table \ref{disp}). 
The nuclear and CNSFRs velocity dispersions estimated from the [O{\sc iii}]
line have a similar behaviour, with the narrow component of the nucleus giving
a gas velocity dispersion (27.2\,$\pm$\,7.1\,km\,s$^{-1}$) almost equal to the
average of the CNSFRs.

The ratios between the fluxes in the narrow and broad
components of the H$\beta$ emission lines of the CNSFRs are
between 0.65 and 0.95, except for the weakest knot R7, for which it is about
1.46.  
For the [O{\sc iii}] emission line the ratio between the narrow and broad
component fluxes vary between 0.06 (for R7) and 0.24.
For the nucleus, the ratio between the narrow and broad component of the
H$\beta$ emission line fluxes are approximately the same as in R7, 1.44, and
the ratio between the narrow and broad [O{\sc iii}] fluxes is 0.29.

\label{Radial velocities 2903}

\begin{figure}
\centering
\includegraphics[width=.70\textwidth,angle=0]{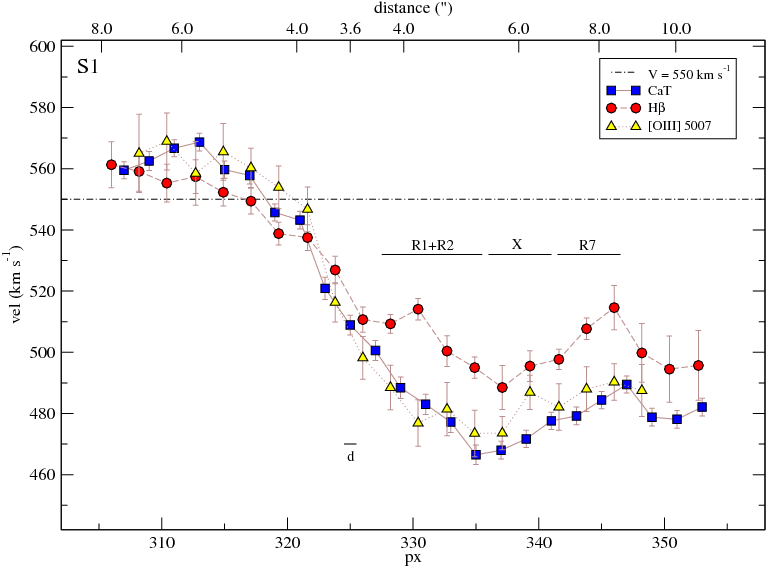}\\
\vspace*{0.3cm}
\includegraphics[width=.70\textwidth,angle=0]{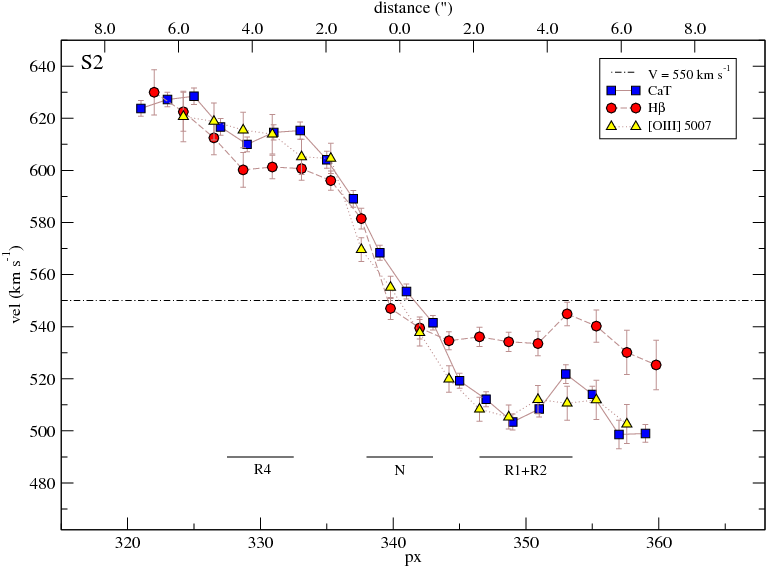}
\caption[Radial velocities along the slit versus pixel number for each slit
  position of NGC\,2903]{Radial velocities along the slit versus pixel number
  for each slit position of NGC\,2903 (upper panel: S1; lower panel: S2) as
  derived from the gas emission lines (red circles: H$\beta$; upward triangles:
  [O{\sc iii}]) and the stellar absorption ones (blue squares). The individual
  CNSFRs and the nucleus, ``N'', or the closest position to it, ``d'', are
  marked in the plots. The dashed-dotted line is the systemic velocity of
  NGC\,2903 derived by Planesas et al.\ (1997). The distance in arcsec from
  the nucleus is displayed in the upper x-axis of each panel.}
\label{velocities-2903}
\end{figure}

The radial velocities along the slit for each angular position of NGC\,2903 as
derived from the ionized gas emission lines, H$\beta$ and [O{\sc
iii}]\,5007\,\AA, and the stellar CaT absorptions are shown in both panels of
Figure \ref{velocities-2903}. The rotation curves seem to have the turnover
points at the same positions as the star-forming ring, specially for the S2
slit position across the nucleus, as found in other galaxies (see
\citeplain{1988ApJ...334..573T,1999ApJ...512..623D} and references
therein). For the systemic velocity of NGC\,2903, the derived values are
consistent with those previously obtained by Planesas et al.\ (1997) and
\citetex{1998AJ....115...62H}, and with the velocity  distribution expected
for this type of galaxies \cite{1987gady.book.....B}.

\begin{figure}
\centering
\includegraphics[width=.70\textwidth,angle=0]{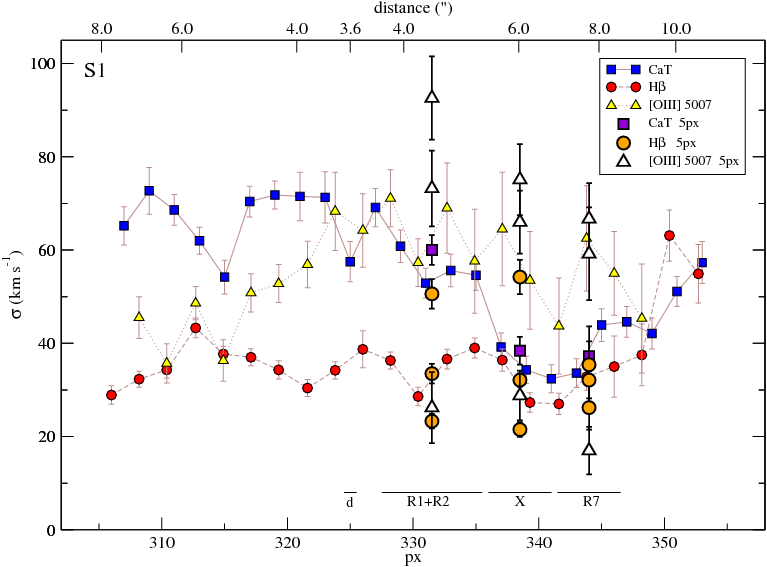}\\
\vspace*{0.3cm}
\includegraphics[width=.70\textwidth,angle=0]{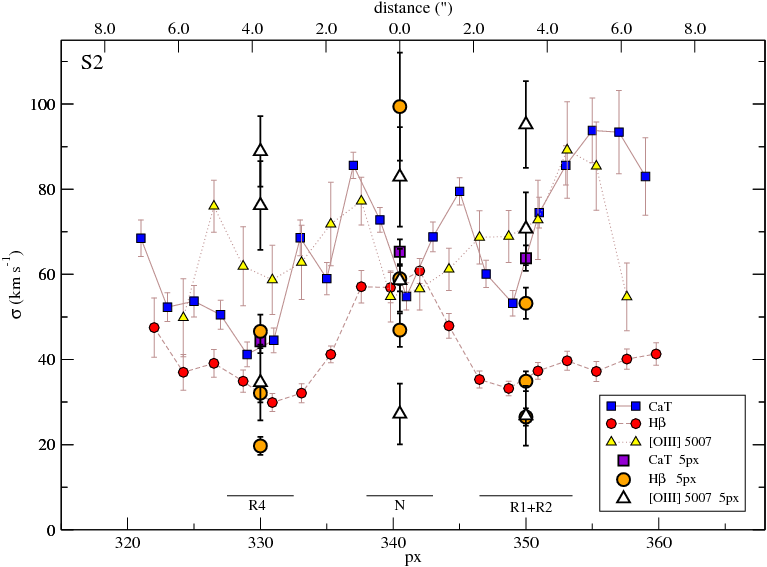}
\caption[Velocity dispersions along the slit versus pixel number for each slit
  position of NGC\,2903]{Velocity dispersions along the slit versus pixel
  number for each slit position of NGC\,2903 (upper panel: S1; lower panel:
  S2) as derived from the gas emission lines (small red circles: H$\beta$;
  small yellow triangles: [O{\sc iii}]) and the stellar absorption ones (small
  blue squares). The velocity dispersions derived for each region (using the
  5\,px aperture) and the
  nucleus are also plotted with large orange circles, white triangles and
  violet squares for H$\beta$, [O{\sc iii}] and CaT, respectively. The
  individual CNSFRs and the nucleus, ``N'', or the closest position to it,
  ``d'', are marked in the plots. The distance in arcsec from the nucleus is
  displayed in the upper x-axis of each panel.}
\label{dispersions-px-2903}
\end{figure}

The rotation velocities derived for both stars and [O{\sc iii}]\,5007\,\AA,
are in good agreement. The H$\beta$ velocities differ from the stellar ones in
quantities similar to those differences shown by the two Gaussian components
($\Delta$v$_{nb}$; see Table \ref{disp}). This could be explained if the broad
component dominates the single fit in  [O{\sc iii}], and hence the measured
velocity corresponds mainly to this component,
while for H$\beta$ the dominant one is the narrow component.

In the upper and lower panels of Figure \ref{dispersions-px-2903} we plot the
velocity dispersions along the slit versus pixel number for slit position S1
and S2 of NGC\,2903, respectively. These velocity dispersions have been
derived from the gas emission lines, H$\beta$ and [O{\sc iii}], and the
stellar absorption ones using the 2 and 3\,px apertures for S1 and S2,
respectively. We have also plotted the gas and stellar velocity 
dispersions derived for each region and the nucleus of NGC\,2903 using the
5\,px apertures. The values derived with wider apertures using a single
Gaussian fit are approximately the average of the velocity dispersions
estimated using the narrower apertures. 

The estimated dynamical masses were derived through the virial theorem using
the 
stellar velocity dispersions, estimated from the CaT absorptions features, and
the cluster sizes, measured in the high spatial resolution WFPC2-PC1 HST image.
For the individual clusters of NGC\,2903 these masses are in the range between
1.4\,$\times$\,10$^6$ and  1.13\,$\times$\, 10$^7$\,M$_\odot$ (see table
\ref{mass}), 
with fractional errors between $\sim$9 and $\sim$25 per cent. The dynamical
masses estimated for the whole CNSFRs (``sum'') are between
6.4\,$\times$\,10$^7$ and  1.9\,$\times$\,10$^8$\,M$_\odot$, with fractional
errors between $\sim$3 and $\sim$4 per cent. The dynamical mass derived for
the nuclear region inside the inner 3.8\,pc is
1.1\,$\times$\,10$^7$\,M$_\odot$, with a fractional error of about 9 per
cent.

On the other hand, the masses of the ionizing stellar clusters of the CNSFRs,
as derived from their H$\alpha$ luminosities, are between 3.3 and
4.9\,$\times$\,10$^6$\,M$_\odot$ for the star-forming regions, and
2.1\,$\times$\,10$^5$\,M$_\odot$ for the nucleus (see Table
\ref{parameters}). In column 11 of Table \ref{parameters} we show a comparison
(in percentage) between ionizing stellar masses of the circumnuclear regions
and their dynamical masses. These values are approximately between 2\,-\,8 per
cent for the CNSFRs, and  1 per cent for the nucleus of NGC\,2903. 

Finally, the masses of the ionized gas, also derived from their H$\alpha$
luminosities, range between 6.1\,$\times$\,10$^4$ and
1.3\,$\times$\,10$^5$\,M$_\odot$ for the CNSFRs, and
3\,$\times$\,10$^3$\,M$_\odot$ for the nucleus (see Table
\ref{parameters}). They make up a small fraction of the total mass of the
regions. 

The masses of these young clusters, alternatively, have been obtained from the
fitting of broad-band colours or spectra using stellar population synthesis
models. \citetex{2001MNRAS.322..757A} from near-IR photometry in the H band
and the model mass-to-light ratios from \citetex{1993ApJ...412...99R} with a
single burst of star formation with Gaussian FWHM values of 1, 5 and 100\,Myr,
and a Salpeter IMF with lower and upper mass cut-offs of 1 and 80\,M$_\odot$
(the parameters of the model are described in detail in
\citeplain{2000ApJ...532..845A,2001ApJ...546..952A}) estimated a total mass in
the young stellar clusters of 2.1-3.6\,$\times$\,10$^8$\,M$_\odot$ within the
central $\sim$625\,pc. This range in masses takes into account the spread of
the age estimates, and thus a variation in the mass-to-light ratio, of these
young stellar clusters. Using the mass-to-light ratio found by
\citetex{1988ApJ...327..671T}, Alonso-Herrero and collaborators estimated a
mass in old stars of 2.7\,$\times$\,10$^{9}$\,M$_\odot$ within the same region
of NGC\,2903. Likewise, they estimated the masses of the ionizing stellar
population (which are even younger than the young stellar clusters defined by
them) using the output from the evolutionary synthesis models. Our
estimates of the masses of the bright \HII\ regions in the star-forming ring
are similar to those found by them for the young clusters.

\subsection*{NGC\,3310}

\begin{figure}
\centering
\vspace*{0.3cm}
\includegraphics[width=.6\textwidth,angle=0]{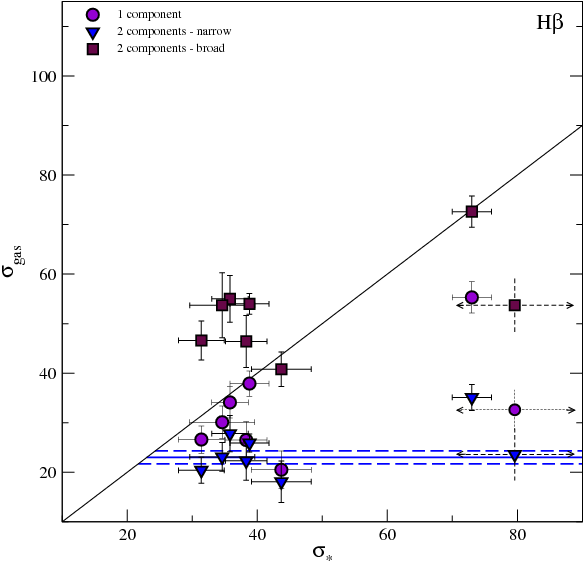}\\\vspace*{0.5cm}
\includegraphics[width=.6\textwidth,angle=0]{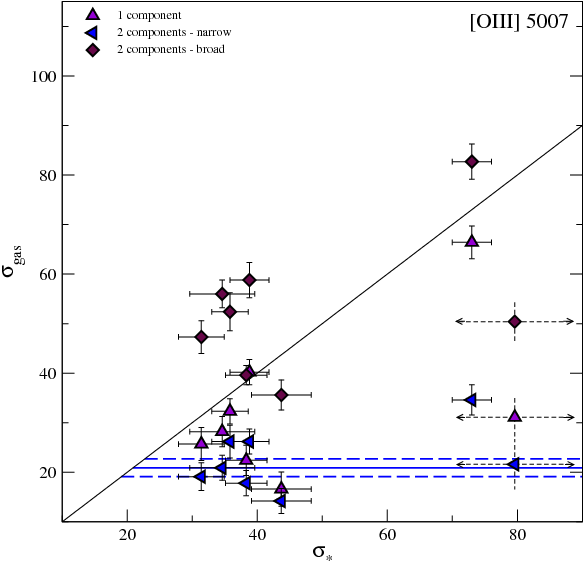}
\caption[Idem as Figure \ref{dispersions2903} for NGC\,3310]{Idem as Figure
  \ref{dispersions2903} for NGC\,3310. Symbols are as follows: Upper panel:
  single Gaussian fit, violet circles; two Gaussian fit, broad component,
  maroon squares; narrow component, blue downward triangles; lower
  panel: violet upward triangles correspond to the estimates using a single
  Gaussian fit, maroon diamonds represent the broad components of the two 
  Gaussian fit and blue left triangles, the narrow components. The
  blue solid line represents the average velocity dispersion of the
  narrow component of the gas (H$\beta$ upper and [O{\sc iii}] lower panel),
  for the CNSFRs, and the blue dashed lines represent their estimated
  errors.} 
\label{dispersions3310}
\end{figure}

Figure \ref{dispersions3310} shows the relations between the velocity
dispersions of gas and stars for NGC\,3310. In the upper panel we plot the
relation derived using the H$\beta$ emission line, and in the lower one, that
from the [O{\sc iii}]. In the first of these panels the violet circles, maroon
squares and blue downward triangles correspond to the single Gaussian fit, and
the broad and 
narrow component of the two-component Gaussian fits, respectively. The deviant
points, marked with arrows, corresponds to the 
R1+R2 region, which has low signal-to-noise ratio and the cross-correlation
does not provide accurate results. On the other hand, in the lower panel,
violet upward triangles, maroon diamonds and blue left triangles correspond to
the values obtained by a single Gaussian fit, and to the broad and narrow
components of the two Gaussian fits, respectively. Again, the deviant points
(R1+R2) are marked with arrows.

The H$\beta$ velocity dispersions of the CNSFRs of NGC\,3310 derived by a
single-component Gaussian fit are the same, within the errors, as the 
stellar ones, except for R1+R2 and R7, for which they are lower by about 50
and 25\,km\,s$^{-1}$, respectively. For R1+R2 this can be it can be due to an
overestimation of $\sigma_\ast$. Given the relatively low metal 
abundance of NGC\,3310 \cite{1993MNRAS.260..177P} generally the emission lines
have a very good signal-to-noise ratio in the spectra of the CNSFRs, while the
red continuum depends on the particular case (see Figures \ref{profiles3310}
and \ref{spectra3310}, the spatial profiles and the spectra,
respectively). On the other hand, the stellar velocity dispersions of the
CNSFRs are lower than those of the broad component of H$\beta$ by about
20\,km\,s$^{-1}$, again except for R1+R2 where $\sigma_\ast$ is greater by
about 35\,km\,s$^{-1}$, and for R7 where $\sigma_\ast$ and H$\beta$ broad are
in good agreement. The narrow component of the CNSFRs shows velocity
dispersions very similar for all the regions, with an average value of
23.0\,$\pm$\,1.3\,km\,s$^{-1}$, and is represented as a blue solid line 
in the upper panel of Figure \ref{dispersions3310}, while the dashed lines of
the same colour represent its error. We find the same behaviour for the two
different components of the ionized gas and the stars both from H$\beta$ and
[O{\sc iii}]. In all cases the gas narrow component has velocity dispersion
lower than the stellar one, and the broad component are slightly over it.
Again the narrow component may be associated with rotating gas 
within a disc-like structure. The broad component of the
[O{\sc iii}]\,5007\,\AA\ emission line (maroon diamonds in the Figure) is
wider than the stellar lines by about 20\,km\,s$^{-1}$, except for R4+R5 (S2
slit) and R7 for which they are approximately similar. The narrow component of
[O{\sc iii}] (blue left triangles in the 
Figure) shows a relatively constant value with an average of
20.9\,$\pm$\,1.8\,km\,s$^{-1}$ (blue lines in the Figure).

In the nuclear region of the galaxy the stars and the broad component of the
ionized gas, both for H$\beta$ and [O{\sc iii}] have the same width, with a
$\sigma$ of about 70\,km\,s$^{-1}$. The gas narrow component however shows a
substantially lower velocity dispersion but still higher than shown by the
CNSFRs by about 15\,km\,s$^{-1}$. 

For regions J, X and Y we
can not derive a stellar velocity dispersion due to the low signal-to-noise
ratio of their continuum and the noise added by subtracting the large amount
of emission lines present in their red spectra
(see right panels of Figure \ref{spectra3310}). However, if we analyze the
velocity dispersions derived from their strong emission lines in the blue
spectral range we find values very similar to those of the
CNSFRs (see Table \ref{disp}).

In general, the broad and narrow component of the H$\beta$ have comparable
fluxes. The ratios between the fluxes in the narrow and broad
components of the H$\beta$ emission lines of the regions (including J, X and
Y) vary from 0.95 to 1.65, except for R10 and R7 for which we find ratios of
0.65 and 2.5, respectively. For the nucleus we estimate a value of 0.55. The
ratio of the narrow to broad [O{\sc iii}] fluxes is between 0.85 and 1.5,
except again for R10 and R7, for which this value is 
0.65 and 1.98, respectively. The nucleus of NGC\,3310 presents a comparatively
low value (0.33) for this ratio with respect to the CNSFRs. 

\label{Radial velocities 3310}

\begin{figure}
\centering
\includegraphics[width=.70\textwidth,angle=0]{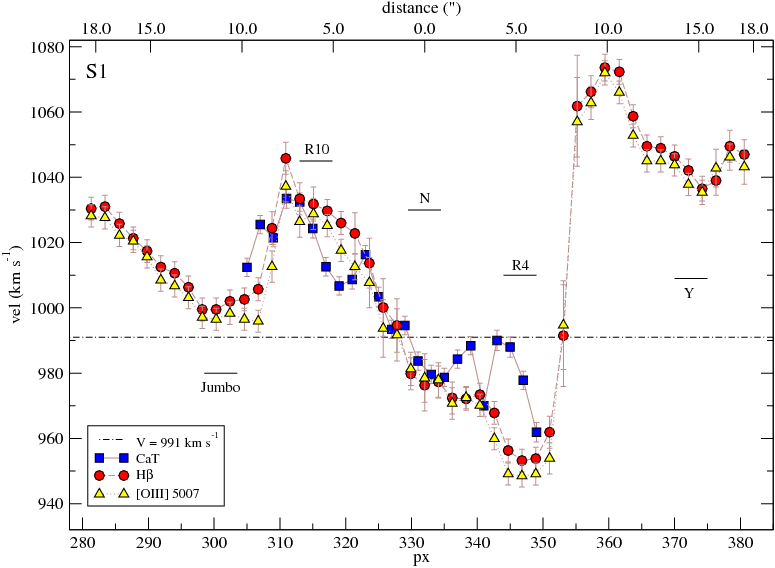}\\
\vspace*{0.3cm}
\includegraphics[width=.70\textwidth,angle=0]{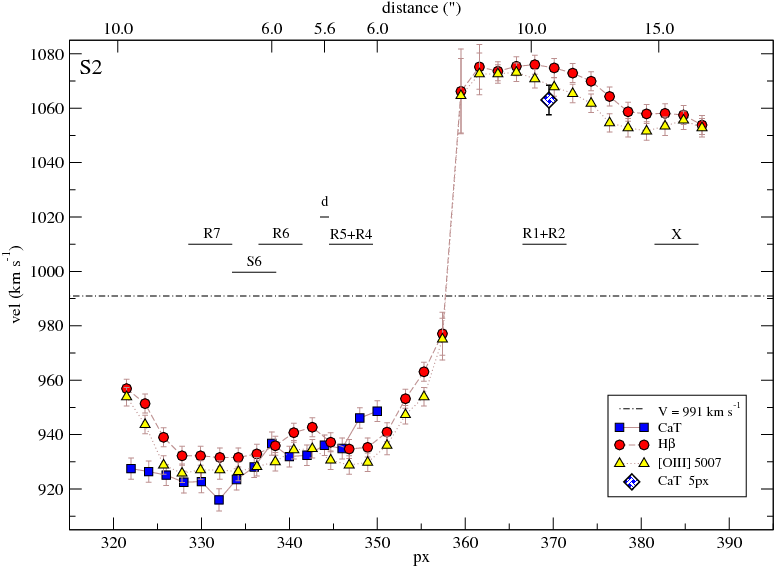}
\caption[Idem as Figure \ref{velocities-2903} for each slit position of
  NGC\,3310]{Idem as Figure \ref{velocities-2903} for each slit position of
  NGC\,3310 (upper panel: S1; lower panel: S2). The criss-crossed white-blue
  diamond in the lower panel correspond to the stellar velocity of R1+R2
  derived using the 5\,px aperture. The dashed-dotted line is the
  velocity of the nucleus of NGC\,3310 derived by
  \citetex{1976A&A....49..161V}.}
\label{velocities-3310}
\end{figure}

Figure \ref{velocities-3310} presents the radial velocities derived from the 
H$\beta$ and [O{\sc iii}] emission lines and the CaT absorptions along the
slit for each angular slit position of NGC\,3310, S1 in the upper panel and S2
in the lower one. The rotation curves seem to have the turnover points located
in or near the star-forming ring. Due to the relatively low
metallicity of NGC\,3310 \cite{1993MNRAS.260..177P}, the gas emission lines
are very strong in the central zone of the galaxy and we can derive the
radial velocity of the gas much further (up to 18\,\arcsec) than for NGC\,2903
and NGC\,3351. However, the measurements of the radial velocities using the
CaT absorption feature are confined approximately to a zone similar to in the
other two cases. 
The gas H$\beta$ and [O{\sc iii}]\,5007\,\AA\ radial velocities and the
stellar ones are in very good agreement, except in the zone located around R4,
where we find a difference between the gas and stars of about 30 to
40\,km\,s$^{-1}$. We must note that these differences can be due to a
low signal-to-noise ratio of the stellar continuum emission in this zone of
the galaxy.

In both slit positions we can appreciate a very steep slope in the radial
velocity curve as measured from the H$\beta$ and [O{\sc iii}], 
in very good agreement with each other. These strong gradients occur 
more or less at the same distance from the nucleus ($\sim$8\,\arcsec)
toward the north-east of the galaxy, almost at the position where the two
slits intersect (see Figures \ref{hst-slits-3310-1} and
\ref{hst-slits-3310-ACS}). These curves show a step in the radial velocity of
about 70 and 90\,km\,s$^{-1}$ for S1 and 
S2, respectively. In Figure \ref{hst-slits-3310-ACS} we indicate with two 
circles with radii of 8 and 18\arcsec, the approximate position where this
step is located and the distance from the nucleus up to where we can
derive the radial velocities for S1. The stellar radial velocity of R1+R2
derived using the 5\,px aperture (criss-crossed white-blue diamond in the
lower panel of the Figure \ref{velocities-3310}) are in very good agreement
with the values derived from the gas emission lines. 

A very similar result was found by
\citetex{2007A&A...469..405P} using high spatial resolution spectra
($\sim$\,0.07\,\arcsec\,px$^{-1}$) and moderate spectral resolution
(R\,=\,$\Delta\lambda$/$\lambda$\,$\sim$\,6000; given a resolution of
$\Delta\lambda$\,$\sim$\,1.108\,\AA\,px$^{-1}$ at H$\alpha$) from the
STIS-HST, and using a 
2\,$\times$\,2 on-chip binning for the non nuclear spectra. The position
angle of their slits is 170$^\circ$ and three parallel positions were used,
one of them passing across the nucleus and the other two with 0.2\,\arcsec\
offset to both sides. However, it must be noted that the 
H$\alpha$ narrow band-filter image in Figure 2 of that work (the same ACS
image presented by us) is rotated by 180$^\circ$ with respect to their quoted
position, since the Jumbo region appears placed at the north-east of the
nucleus (according to the orientation of the image given by these authors)
while this region is located at the south-west (see e.g.\
\citeplain{1984ApJ...284..557T,1990MNRAS.242P..48T,1993MNRAS.260..177P}).
As was pointed out by Pastorini and collaborators, NGC\,3310 shows a typical
rotation curve expected for a rotating disc (a typical $S$ feature,
\citeplain{2006A&A...448..921M}) but the curve is disturbed. Besides, we can
appreciate that in the zone 
near the Jumbo region there is a deviation from the circular rotation
motion. The step in the radial velocity found by us is an effect of the
spatial resolution. In Figure 3 of \citetex{2007A&A...469..405P}, the better
spatial resolution of the STIS-HST instrument resolves the steep gradient and
completes the information missing in our data. The disturbed behaviour of the
rotation curve 
of the gas in the galactic disc, characterized by non-circular motions, was
also found by \citetex{1995A&A...300..687M} using H{\sc i} radio data;
\citetex{1996A&A...309..403M} from H{\sc i} and H{\sc i} and H{\sc ii} radio
data and  and \citetex{2001A&A...376...59K} using optical data. They found a
strong streaming along the spiral arms, supporting the 
hypothesis of a recent collision with a dwarf galaxy that triggered the
circumnuclear star formation during the last 10$^8$\,yr.
Besides, \citetex{1976A&A....49..161V} shows that the rotation centre of the
gas is offset with respect to the stellar continuum (by $\sim$\,1.5\,\arcsec). 
The central velocity of NGC\,3310 derived by us is in very good agreement with
that previously estimated by
\citetex{1976A&A....49..161V,1995A&A...300..687M,1996A&A...309..403M,1998AJ....115...62H,2001A&A...376...59K}.
The velocity distribution is also in very good agreement with that expected
for this type of galaxies \cite{1987gady.book.....B}. 


\begin{figure}
\centering
\includegraphics[width=.70\textwidth,angle=0]{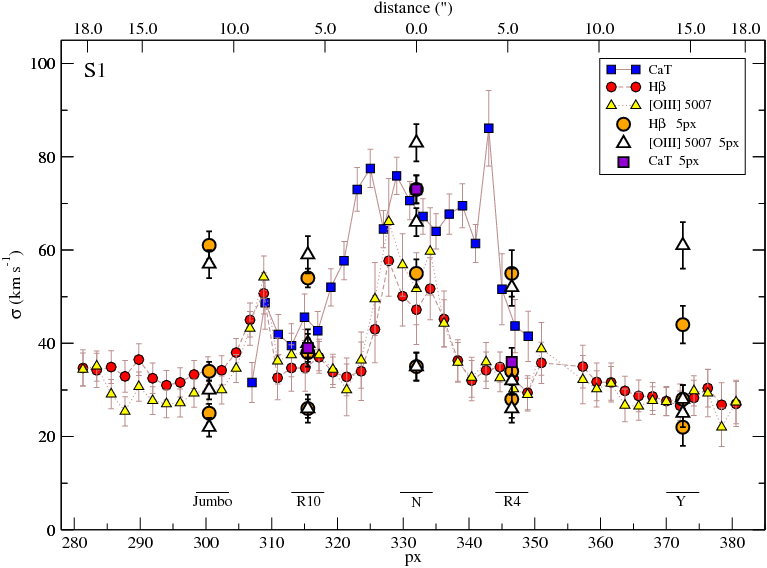}\\
\vspace*{0.3cm}
\includegraphics[width=.70\textwidth,angle=0]{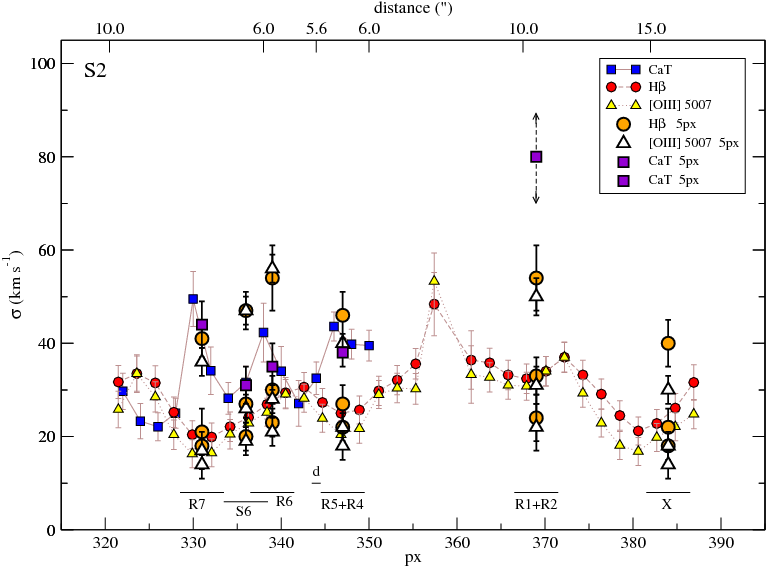}
\caption[Idem as Figure \ref{dispersions-px-2903} for each slit position of
  NGC\,3310]{Idem as Figure \ref{dispersions-px-2903} for each slit position of
  NGC\,3310 (upper panel: S1; lower panel: S2).}
\label{dispersions-px-3310}
\end{figure}

The velocity dispersions along the slit versus pixel number for slit position
S1 and S2 of NGC\,3310 are plotted in the upper and lower panel of Figure
\ref{dispersions-px-3310}, respectively. The gas velocity dispersions are
derived from H$\beta$ and [O{\sc iii}] emission lines, and the
stellar velocity dispersions from the CaT absorption features. For S1 we have
plotted the results derived using the 2\,px apertures; for S2, we used
the 3\,px ones, while for the actual regions and the nucleus we plot in
the Figure the values derived using the 5\,px apertures. It can be seen that
the values estimated from the single Gaussian fit using this last aperture are
approximately the average of 
those derived with the narrower apertures. The velocity dispersions, both for
stars and gas, show a behaviour characteristic of a regular circular motion in
a   rotating disc, a smooth plateau and a peak in the centre. A similar result
was found by \citetex{2007A&A...469..405P}.

The estimated dynamical masses derived using the virial theorem 
for the individual clusters of NGC\,3310, except for R1+R2, are
between 1.8 and 7.1\,$\times$\,10$^6$\,M$_\odot$ (see table 
\ref{mass}), with fractional errors between $\sim$\,16 and $\sim$\,32 per
cent. The dynamical masses estimated for the whole CNSFRs (``sum''), except
for R1+R2, are between  
2.1\,$\times$\,10$^7$ and 1.4\,$\times$\,10$^8$\,M$_\odot$, with fractional
errors between $\sim$3 and $\sim$8 per cent. R1+R2 stellar velocity dispersion
has a big associated error, so we do not give an explicit error for its
derived mass. The individual clusters of this CNSFR derived using this 
value of the velocity dispersion present greater masses, between
1.2 and 2.4\,$\times$\,10$^7$\,M$_\odot$. However, if we assume that the
relation between the gas and the stellar velocity dispersions follows a
behaviour similar to those of the other CNSFRs of NGC\,3310 (see Figure
\ref{dispersions3310}) these masses could be reduced by a factor of 2.3,
giving individual masses between 5.2\,$\times$\,10$^6$ and
1.0\,$\times$\,10$^7$\,M$_\odot$. In the same way, the total dynamical mass of
R1+R2 is 9.1\,$\times$\,10$^8$\,M$_\odot$, which is reduced to
4.0\,$\times$\,10$^8$\,M$_\odot$ assuming the same 2.3 factor.
The dynamical mass derived for the nuclear region inside the inner 14.2\,pc is
5.3\,$\times$\,10$^7$\,M$_\odot$, with a fractional error of about 11 per
cent. This value is a somewhat higher than that derived by
\citetex{2007A&A...469..405P} under the assumption of the presence of a super
massive black hole (BH) in the centre of this galaxy. When they take into
account the allowed disc inclinations, this BH mass varies in the range
5.0\,$\times$\,10$^6$\,-\,4.2\,$\times$\,10$^7$\,M$_\odot$.

On the other hand, the masses of the ionizing stellar clusters of the CNSFRs,
derived from their H$\alpha$ luminosities, are between 8.7\,$\times$\,10$^5$
and 2.1\,$\times$\,10$^6$\,M$_\odot$ for the star-forming regions, and
3.5\,$\times$\,10$^6$\,M$_\odot$ for the nucleus (see Table
\ref{parameters}). These ionizing stellar masses are approximately between
1\,-\,7 
per cent of their dynamical masses, and 5 per cent for the nucleus of
NGC\,3310. In the case of the Jumbo region this mass vary between 1.3 and
3.1\,$\times$\,10$^6$\,M$_\odot$ whether we take the H$\alpha$ luminosities
from \citetex{1993MNRAS.260..177P} or \citetex{2000MNRAS.311..120D}
respectively, probably because the estimated reddening constants in these two
works are different (see Table \ref{parameters}). These different values of
the luminosities can be due to the very complex
structure of the region (see Figure 1 of both works) and differences in the
selection of the zone used to measure the H$\alpha$ fluxes. Besides, Pastoriza
and colleagues used spectroscopic data while D\'iaz and collaborators used
photometric images.

Finally, the masses of the ionized gas, range between 1.5 and
7.2\,$\times$\,10$^5$\,M$_\odot$ for the CNSFRs, and
5\,$\times$\,10$^3$\,M$_\odot$ for the nucleus, also derived from their
H$\alpha$ luminosities (see Table 
\ref{parameters}). They make up a small fraction of the total mass of the
regions. As in the case of the masses of the ionizing stellar
clusters of the Jumbo region and for the same reasons, the mass of the
ionized gas derived using the H$\alpha$ luminosities from
\citetex{1993MNRAS.260..177P} or from \citetex{2000MNRAS.311..120D} are 
different, 3.93 and 9.52\,$\times$\,10$^5$\,M$_\odot$, respectively.

Comparing colours of the small-scale clusters from HST data with models,
\citetex{2002AJ....123.1381E} suggest that most of these clusters contain
masses from 10$^4$\,M$_\odot$ to several times 10$^4$\,M$_\odot$. For the
large-scale ``hot-spots'' they estimate masses ranging from 10$^4$ to several
times 10$^5$\,M$_\odot$. For their 108+109 region (almost equivalent to the
Jumbo region) they derived a mass of 6.3\,$\times$\,$10^5$\,M$_\odot$, similar
to the value derived by \citetex{1993MNRAS.260..177P},
7\,$\times$\,$10^5$\,M$_\odot$. These last
authors estimated from the unusually large H and He{\sc ii} emission line
luminosities, that this region must contain 220 WN and 570 OB
stars. Regarding other regions, Elmegreen et al.\ derived a mass of
$10^6$\,M$_\odot$ for 
their region 78 using the NIR data and 7.7\,$\times$\,$10^6$\,M$_\odot$ from
their optical ones. Besides, they found 17 candidate SSCs, with masses in the
range between 2\,$\times$\,$10^4$ to 4.5\,$\times$\,$10^5$\,M$_\odot$.
Some of them coincide with the CNSFRs studied by us, such as 48, 49 and
50, approximately coincident with R5 and R6. These SSCs are mostly in the
innermost southern spiral arm, with some in the northern one or outside the
southern 
arm of the ring \cite{2002AJ....123.1381E}. Their derived (J-H) colours
suggest two different populations of SSCs, a very young of few million years
and an older one in a range between 10 and 50 million years, with the younger
clusters located in the northern part of the ring.

\subsection*{NGC\,3351}

\begin{figure}
\centering
\vspace*{0.3cm}
\includegraphics[width=.6\textwidth,angle=0]{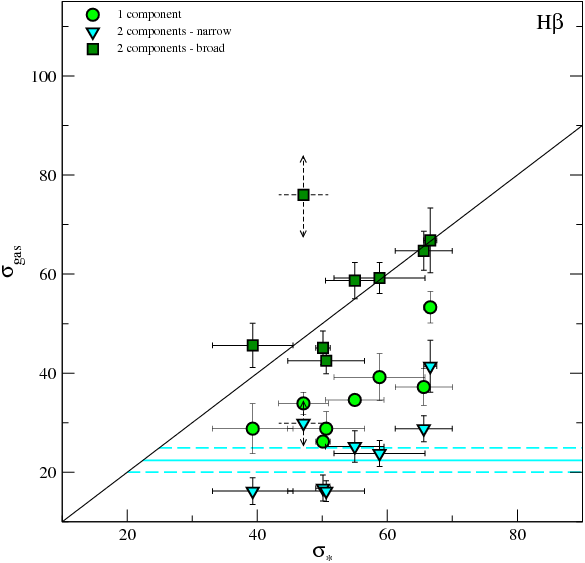}\\\vspace*{0.5cm}
\includegraphics[width=.6\textwidth,angle=0]{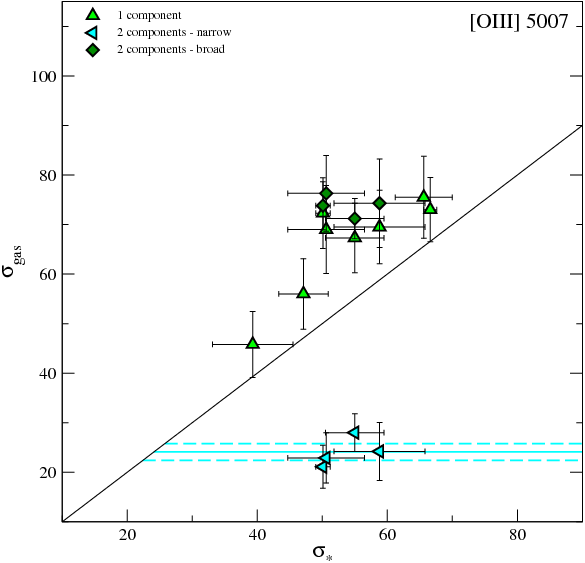}
\caption[Idem as Figure \ref{dispersions2903} for NGC\,3351]{Idem as Figure
  \ref{dispersions2903} for NGC\,3351. Symbols are as follows: Upper panel:
  single Gaussian fit, light green circles; two Gaussian fit, broad component,
  dark green squares; narrow component, cyan downward triangles; lower
  panel: light green upward triangles correspond to the estimates using a single
  Gaussian fit, dark green diamonds represent the broad components of the two 
  Gaussian fit and cyan left triangles, the narrow components. The 
  cyan solid line represents the average velocity dispersion of the narrow
  component of the gas (H$\beta$ upper and [O{\sc iii}] lower panel) for the
  CNSFRs, and the cyan dashed lines represent their estimated errors.}
\label{dispersions3351}
\end{figure}

The upper panel of Figure \ref{dispersions3351} shows the relation between the
velocity dispersions of gas and stars. Light green circles correspond to gas
velocity dispersions measured from the H$\beta$ emission line using a single
Gaussian fit. Dark green squares and cyan downward triangles correspond to
measurements performed using two-component Gaussian fits, squares for the
broad  component and triangles for the narrow one. The straight line shows the
one-to-one relation. As a general result, the H$\beta$ velocity dispersions
derived by a single Gaussian fit are lower than the stellar ones by about 25
km\,s$^{-1}$. This is also the case for the Paschen lines in the two regions
where they could be measured. On the other hand, a good agreement is found
between the stellar velocity dispersions and those of the broad component of
H$\beta$. The deviant point, marked with arrows, corresponds to the region
with the lowest signal-to-noise ratio (R5) for which the fits do not provide
accurate results. The narrow component shows velocity dispersions even lower
than those obtained by single Gaussian fits. The ratio between the fluxes in
the narrow and broad components is between 0.7 and 0.95 (except for the case
of R5 for which no meaningful result is found).

The lower panel of Figure \ref{dispersions3351} shows the relation between the
stellar velocity dispersions and those of  the [O{\sc iii}] emission line
measured by both single- and two-component Gaussian fits, which  due to the
weakness of the line (see Figure \ref{spectra3351}), could be done in only
four cases, 
corresponding to the two slit positions on regions R2 and R3. 
In this case, the broad component seems to dominate the width of the emission
line which agrees with the stellar one. The width of the narrow
component, in the cases in which the two-component fit was possible, is
comparable to that of the narrow component of the H$\beta$ line. 

The narrow components of the H$\beta$ emission line show similar values,
with an average equal to 22.4\,$\pm$\,2.5\,km\,s$^{-1}$. This average is
represented as a cyan horizontal line in the upper panel of Figure
\ref{dispersions3351}, with its error as cyan dashed lines. A very similar
behaviour is found for the [O{\sc iii}] emission line with an average value of
24.1\,$\pm$\,1.7\,km\,s$^{-1}$, and is shown in the lower panel of the Figure.

\begin{figure}
\centering
\includegraphics[width=.60\textwidth,angle=0]{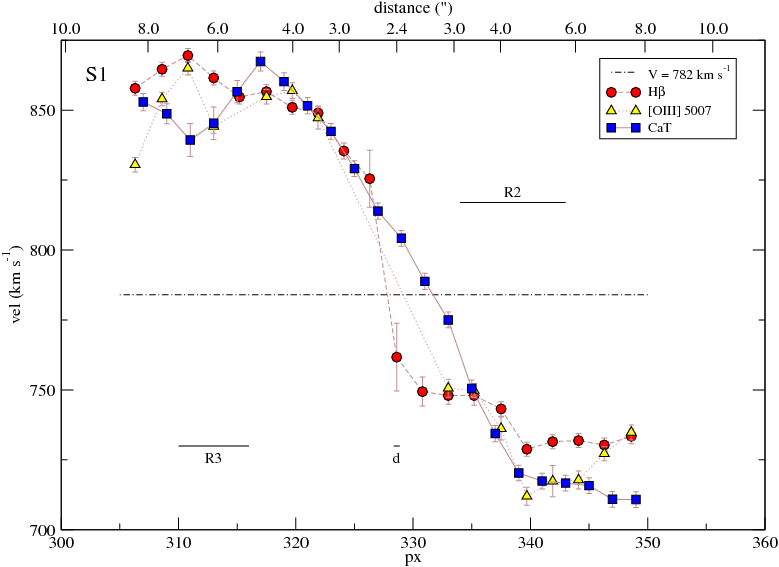}\vspace*{0.3cm}
\vspace*{0.3cm}
\includegraphics[width=.59\textwidth,angle=0]{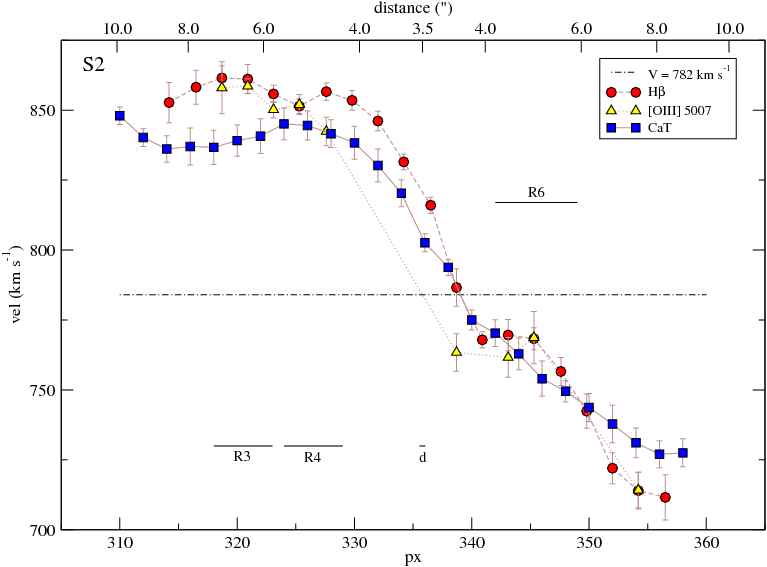}\\
\hspace*{0.1cm}\includegraphics[width=.60\textwidth,angle=0]{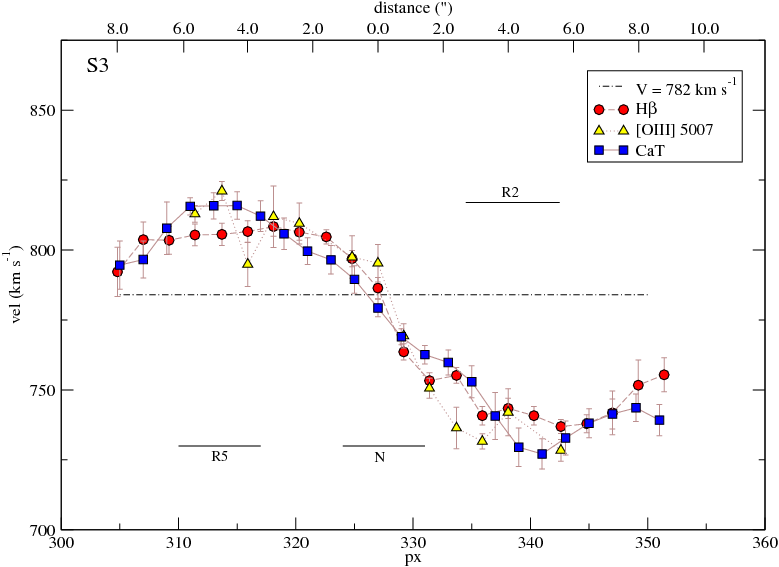}
\caption[Idem as Figure \ref{velocities-2903} for each slit position of
  NGC\,3351]{Idem as Figure \ref{velocities-2903} for each slit position of
  NGC\,3351 (upper panel: S1; middle panel: S2; lower panel: S3). The
  dashed-dotted line is the systemic velocity of NGC\,3351 derived by Planesas
  et al.\ (1997).}
\label{velocities-3351}
\end{figure}

\label{Radial velocities}
Figure\ \ref{velocities-3351} shows the radial velocities along the slit for
each slit position in NGC\,3351 as derived from the ionized gas emission lines
and the stellar absorptions. The turnover points of the rotation curves 
seem to be located at the same position than the star-forming ring, specially
for the S3 slit position that crosses the centre of the galaxy. For the
systemic velocity of NGC\,3351, the derived values are consistent with those
previously obtained by \citetex{1975ApJ...199...39R} and Planesas et al.\
(1997), and with the velocity distribution expected for this type of galaxies
\cite{1987gady.book.....B}.

\begin{figure}
\centering
\includegraphics[width=.60\textwidth,angle=0]{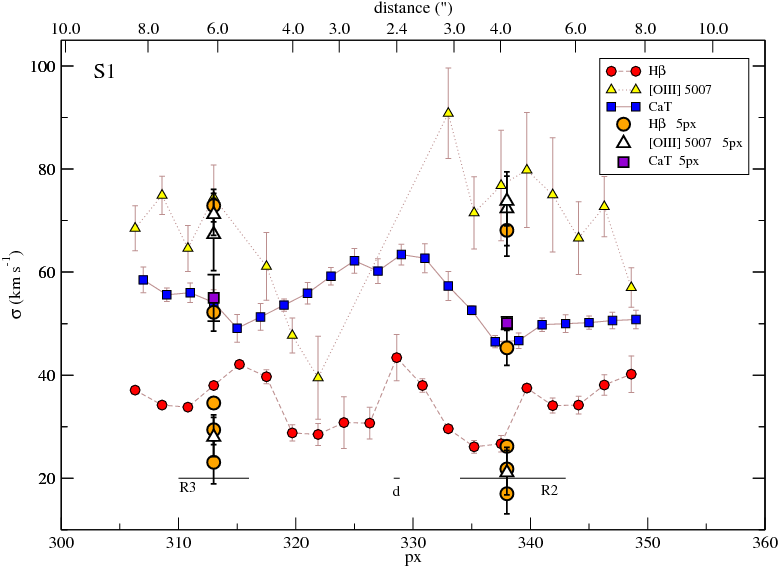}\vspace*{0.3cm}
\vspace*{0.3cm}
\includegraphics[width=.59\textwidth,angle=0]{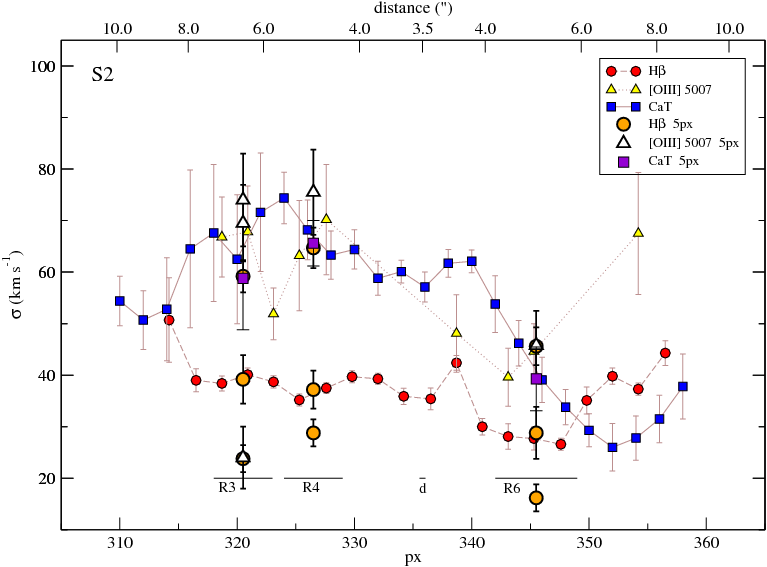}\\
\hspace*{0.1cm}\includegraphics[width=.60\textwidth,angle=0]{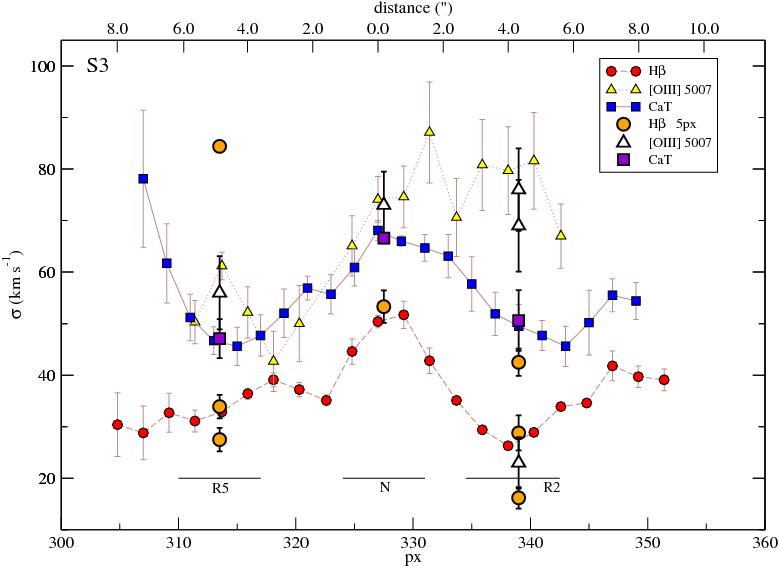}
\caption[Idem as Figure \ref{dispersions-px-2903} for each slit position of
  NGC\,3351]{Idem as Figure \ref{dispersions-px-2903} for each slit position of
  NGC\,3351 (upper panel: S1; middle panel: S2; lower panel: S3).}
\label{dispersions-px-3351}
\end{figure}

The rotation velocities derived for both stars and gas are in reasonable
agreement, although in some cases the gas seems to rotate somewhat faster than
the stars. In fact, in the lower panel of Figure \ref{velocities-3351}, which
corresponds to the slit 
position passing through the nucleus, it is interesting to note that at the
maximum and minimum of the velocity curve, which correspond approximately to
the positions of regions R5 and R2, the H$\beta$ emission line shows
velocities lower and higher than the stars by about 25 and 30\,km\,s$^{-1}$,
respectively. This could be interpreted as motions of the ionized hydrogen
deviating from rotation and consistent with a radial infall to the central
regions of the galaxy. A similar result was found by Rubin et al.\ (1975) from
high dispersion observations of the H$\alpha$ line in the central region of
this galaxy. Their preferred model for the fitting of the kinematical data
consists of gas which is rotating and contracting with V$ _{rot}
$\,=\,126\,$\pm$\,16\,km\,s$^{-1}$ and V$ _{cont} $\,=\,34\,$ \pm
$\,11\,km\,s$^{-1}$. 

In the three panels of Figure \ref{dispersions-px-3351} we plot the
velocity dispersions along the slit versus pixel number for slit position S1
(upper), S2 (middle) and S3 (lower) of NGC\,3351, respectively. Again, these
velocity dispersions have been 
derived from the gas emission lines, H$\beta$ and [O{\sc iii}], and the
stellar absorption ones. We have also plotted the gas and stellar velocity
dispersions derived for each region and the nucleus of NGC\,3351 using the
5\,px aperture. For the
broad H$\beta$ component of R5 in S3 we do not plot the error bars. As in the
cases of NGC\,2903 and NGC\,3310, the values derived with the wider apertures
(five pixels) using a single Gaussian fitting are 
approximately the average of the velocity dispersions estimated using the
extractions made with the narrower apertures.

Our values of the dynamical masses have been derived from stellar velocity
dispersion measurements as mapped by the CaT absorption lines and the sizes
measured on an HST image. They are in the range between 4.9\,$\times$\,10$^6$
and 4.5\,$\times$\,10$^7$\,M$_\odot$ for the CNSFRs and is
3.5\,$\times$\,10$^7$\,M$_\odot$ for the nuclear region inside the inner
11.3\,pc (see Table \ref{mass}). The fractional errors of the dynamical masses
of the CNSFRs are between $\sim$3 and $\sim$\,18 per cent, and  is
about 3 per cent for the nucleus. The masses derived here for the
circumnuclear star-forming individual clusters are between 1.8 and
8.7\,$\times$\,10$^6$\,M$_\odot$. 

Other estimates of the masses of these regions have been obtained from the
fitting of broad-band colours or spectra with the use of stellar population
synthesis models. \citetex{1997AJ....114.1850E} from near-IR photometry in the
J and 
K bands and models by \citetex{1995ApJS...96....9L} for  instantaneous star
formation and solar and twice solar metallicity, derive masses of the
CNSFRs from 1 to
10\,$\times$\,10$^{5}$\,M$_\odot$. \citetex{1997ApJ...484L..41C} from UV [{\it
    International Ultraviolet Explorer} (IUE)] 
spectra and instantaneous burst models by 
\citetex{1991A&AS...88..399M} derived a value of 3\,$\times$\,10$^5$\,M$_\odot$
for the whole SF ring. In both cases, in fact, these observables 
trace the young massive population which constitutes only part of the total 
mass. 

The masses of the ionizing stellar clusters of the CNSFRs of NGC\,3351, as
derived from 
their H$\alpha$ luminosities, are between 8.0\,$\times$\,10$^5$ and
2.5\,$\times$\,10$^6$\,M$_\odot$ for the star-forming regions, and is
6.0\,$\times$\,10$^5$\,M$_\odot$ for the nucleus (see Table
\ref{parameters}). In column 11 of Table \ref{parameters} we show
a comparison (in percentage) between ionizing stellar masses of the
circumnuclear regions and their dynamical masses. These values vary
approximately in the range 2\,-\,16 per cent for the CNSFRs, and is 1.7 per
cent for the nucleus. 
Since the CaT absorption features are dominated by young stars (see
discussion above, Section \ref{masses}), and the M$_{ion}$/M$_{\ast}$ fraction
is still 
remarkably small in the case of the CNSFRs composed of single knots (R4 and
R5, for which the dynamical mass is most robustly estimated) we can assume 
that our upper limits to the dynamical masses, in spite of the limitations of
the method used to derive them, are rather tight. Then, our results concerning
the M$_{ion}$/M$_{\ast}$ fraction are robust.

Finally, the masses of the ionized gas vary between 7\,$\times$\,10$^3$ and
8.7\,$\times$\,10$^4$\,M$_\odot$ for the CNSFRs, and is
2\,$\times$\,10$^3$\,M$_\odot$ for the nucleus (see Table
\ref{parameters}). Again, these masses of the ionized gas make up
a small fraction of the total mass of the regions.

\bigskip

\begin{figure}
\centering
\vspace*{0.3cm}
\includegraphics[width=.6\textwidth,angle=0]{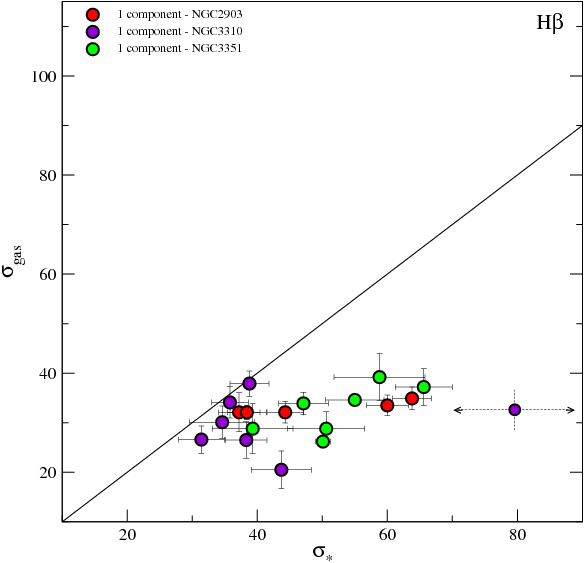}\\\vspace*{0.5cm}
\includegraphics[width=.6\textwidth,angle=0]{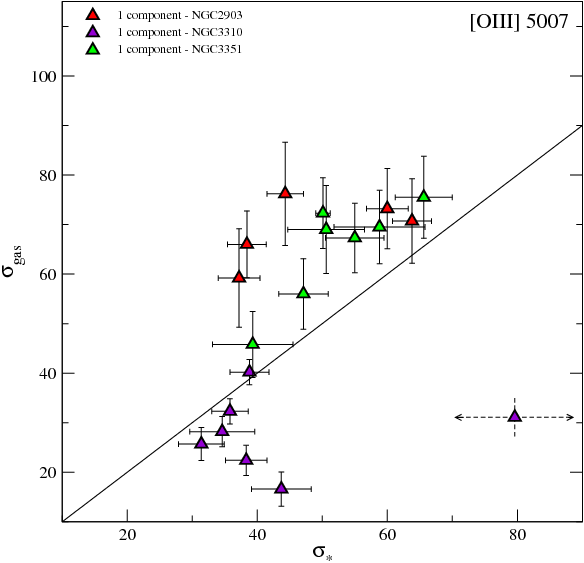}
\caption[Relation between velocity dispersions of the gas derived using a
  single Gaussian fit and the stars for all the CNSFRs studied]{Relation between
  velocity dispersions of the gas (upper panel: H$\beta$; lower panel: [O{\sc
  iii}]\,5007) derived using a single Gaussian fit and the stars (CaT)
  for all the CNSFRs studied.  Symbols are as in Figures \ref{dispersions2903},
  \ref{dispersions3310} and \ref{dispersions3351} for NGC\,2903, NGC\,3310 and
  NGC\,3351, respectively.}
\label{dispersions-1comp}
\end{figure}

\begin{figure}
\centering
\vspace*{0.3cm}
\includegraphics[width=.6\textwidth,angle=0]{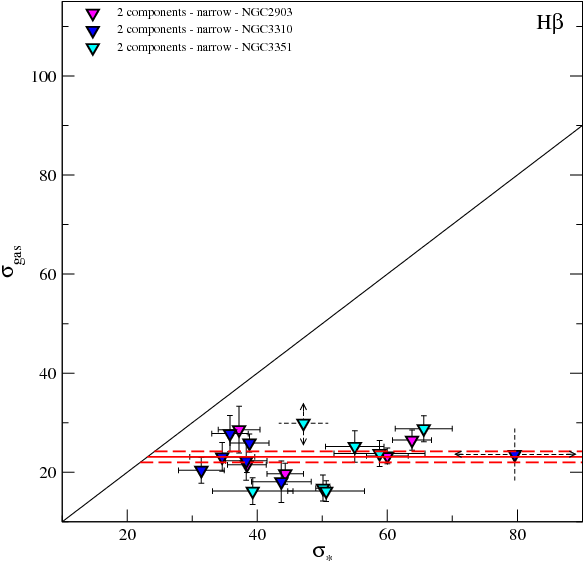}\\\vspace*{0.5cm}
\includegraphics[width=.6\textwidth,angle=0]{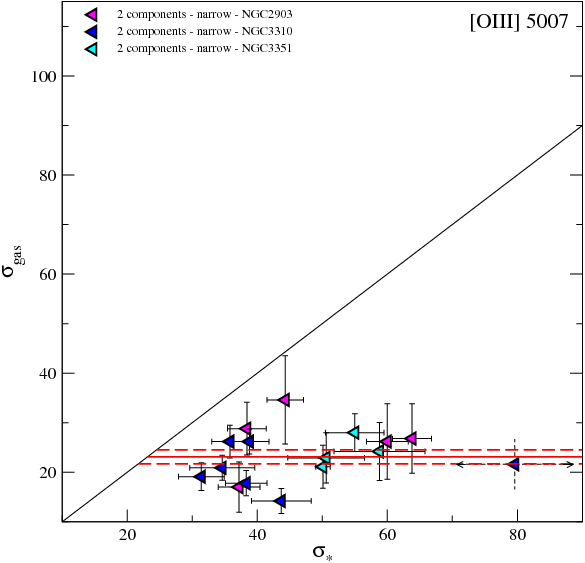}
\caption[Idem as Figure \ref{dispersions-1comp} for the narrow component of
  the gas]{Idem as Figure
  \ref{dispersions-1comp} for the narrow component of the gas derived using a
  two component Gaussian fit. Symbols are as 
  in Figures \ref{dispersions2903}, \ref{dispersions3310} and
  \ref{dispersions3351} for NGC\,2903, NGC\,3310 and NGC\,3351,
  respectively. The red solid line is the average value for all the studied
  CNSFRs and the red dashed ones their corresponding errors.} 
\label{dispersions-narrow}
\end{figure}

\begin{figure}
\centering
\vspace*{0.3cm}
\includegraphics[width=.6\textwidth,angle=0]{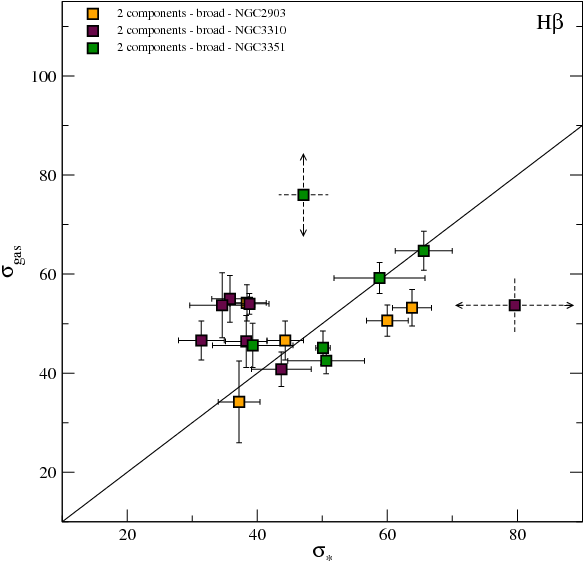}\\\vspace*{0.5cm}
\includegraphics[width=.6\textwidth,angle=0]{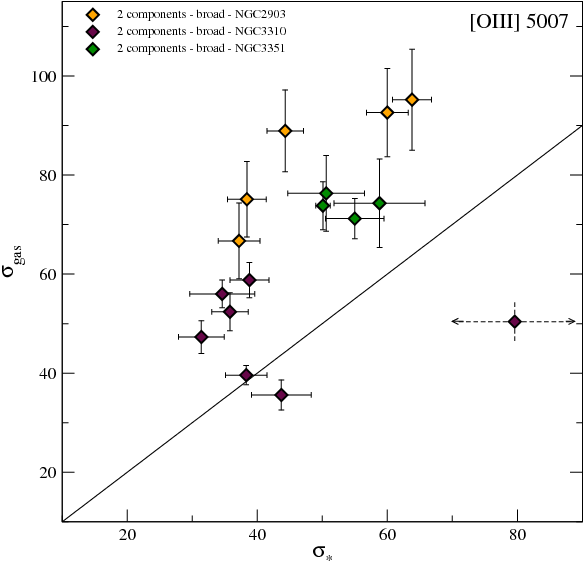}
\caption[Idem as Figure \ref{dispersions-1comp} for the broad component of
  the gas]{Idem as Figure \ref{dispersions-1comp} for the broad component of
  the gas derived using a two component Gaussian fit. Symbols are as in
  Figures \ref{dispersions2903}, \ref{dispersions3310} and
  \ref{dispersions3351} for NGC\,2903, NGC\,3310 and NGC\,3351, respectively.}
\label{dispersions-broad}
\end{figure}

\subsection*{Global analysis}

In Figures \ref{dispersions-1comp}, \ref{dispersions-narrow} and
\ref{dispersions-broad} we present the different relations found between the
velocity dispersions measured for the whole sample of CNSFRs.
Figure \ref{dispersions-1comp} shows the velocity dispersion of the gas as
measured from the H$\beta$ and the [O{\sc iii}] emission lines versus the
stellar one. The gas velocity dispersions have been measured using a single
Gaussian fit. Figure \ref{dispersions-narrow} shows the same graph, but with
the gas velocity dispersion corresponding to the narrow velocity component as
measured from the two-component fit. Finally, Figure \ref{dispersions-broad}
shows this last graph but for the broad velocity component.
For the three Figures the upper panel shows the
values derived from H$\beta$ emission line and the lower one, those
derived from the [O{\sc iii}]\,5007\,\AA\ line. The red solid lines
in Figure \ref{dispersions-narrow} represent the average of the narrow
components of all the CNSFRs and the red
dashed lines their errors. 

The behaviour of the velocity dispersion of the CNSFRs of NGC\,2903 and
NGC\,3351 derived from their H$\beta$ and the [O{\sc iii}] emission lines
measured using a single Gaussian fit are very different (compare upper and
lower panels of Figure \ref{dispersions-1comp}). While the values derived
for the first line are always lower than the stellar velocity dispersion,
approaching the one-to-one relation for the lowest values, in the case of the 
collisional line, the gas velocity dispersions are always
greater than the stellar ones by about 20\,km\,s$^{-1}$. On the other hand,
the values derived using these two emission lines for NGC\,3310 are
in very good agreement and similar to the stellar ones in almost all cases
(lower by about 25\,km\,s$^{-1}$ for R7). R1+R2 in NGC\,3310 seems to be a
special case as discussed above.

The velocity dispersions derived from the narrow components of the
two Gaussian fit show a relatively constant value for the whole sample (see
Figure \ref{dispersions-narrow}).  In both cases, H$\beta$ and 
[O{\sc iii}], the global average value is equal to 23.1\,km\,s$^{-1}$, with
errors of 1.1 and 1.4\,km\,s$^{-1}$, respectively.
In the case of (Figure \ref{dispersions-broad}) the broad components show
velocity 
dispersions derived from the H$\beta$ line are in very good agreement with the
stellar ones, except for some regions (four) of NGC\,3310 whose estimated broad
H$\beta$ components are greater by about 20\,km\,s$^{-1}$, and two deviant
points: one from NGC\,3310 (R1+R2) and another one from NGC\,3351 (R5). The
broad 
components derived from the [O{\sc iii}] emission line show velocity
dispersions also greater than the stellar ones, but in these cases by about
30\,km\,s$^{-1}$, except for a couple of exceptions in NGC\,3310, and the
deviant point of the same galaxy. The difference between  the
narrow component of the ionized gas on one hand, and the broad component and
stars on the other, could be interpreted by assuming that part of the emission
comes from gas that is confined in the disc and supported by rotation while
the stars and that portion of the gas related to the star-forming regions are
mostly supported by dynamical pressure (see \citeplain{2004A&A...424..447P}
and references therein).

\begin{figure}
\centering
\includegraphics[width=.88\textwidth,angle=0]{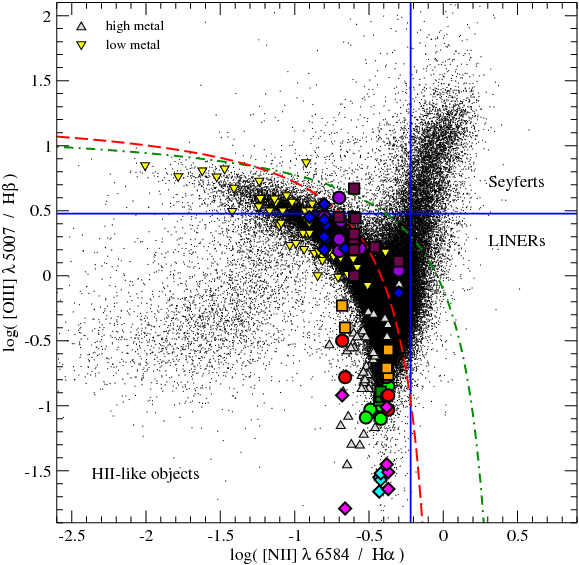}
\caption[BPT diagram [O{\sc iii}\textrm{]}/H$\beta$ vs.\ [N{\sc
  ii}\textrm{]}/H$\alpha$]{BPT diagram [O{\sc iii}]/H$\beta$ vs.\ [N{\sc
  ii}]/H$\alpha$. Circles (NGC\,2903: red; NGC\,3310: violet; NGC\,3351: light
  green) correspond to the ratio of the emission intensities of [O{\sc iii}] 
  and H$\beta$ estimated using a single Gaussian fit, diamonds
  (NGC\,2903: magenta; NGC\,3310: blue; NGC\,3351: cyan) to the narrow and
  squares (NGC\,2903: orange; NGC\,3310: maroon; NGC\,3351: dark green) 
  to the broad components of two Gaussian fits. Green dot-dashed and red
  dashed curves are the boundary between Active Galactic Nuclei (AGNs) and
  \HII\ galaxies defined by Kewley et al.\ (2001) and Kauffmann et al.\ (2003),
  respectively. The blue solid horizontal and vertical lines represent the
  division between Seyfert galaxies and LINERs (low-ionization nuclear
  emission-line region galaxies) according to Ho et al.\ (1997). Dots
  correspond to a subsample of emission line objects, including \HII\ galaxies,
  from SDSS-DR3, from L\'opez (2005). Triangles correspond to \HII\ regions
  from the sample of P\'erez-Montero and D\'iaz (2005). They have been split
  into low-metallicity (gray upwards triangles) and high-metallicity (yellow
  upside down triangles) regions according to the criterion by D\'iaz and 
  P\'erez-Montero (2000) based on oxygen and sulphur abundance parameters.}
\label{bpt}
\end{figure}

\nocite{2000MNRAS.312..130D}

\label{BPT}
The presence of two different gaseous components could have an effect
on the classification of the activity in the central regions of galaxies
through  diagnostic diagrams. We have plotted these two components
individually, together with the value obtained by fitting a single Gaussian,
for the three studied galaxies,
in Figure \ref{bpt}. This figure shows the location of the CNSFRs 
and of the nuclei of the observed galaxies in the [O{\sc iii}]/H$\beta$ versus
[N{\sc ii}]/H$\alpha$ diagram (\citeplain{1981PASP...93....5B}; BPT), together
with a  
sample of emission line galaxies (including \HII-like objects) from the Sloan
Digital Sky Survey - Data Release 3 (SDSS-DR3) (L\'opez, 2005)
\nocite{tesisjesus} 
and \HII\ regions from the sample of \citetex{2005MNRAS.361.1063P}. This BPT
diagram clearly shows that the two systems are segregated in the cases of
NGC\,2903 and NGC\,3351, with the narrow component showing the lowest
excitation and occupying the same position in the diagram as the starburst
systems in the SDSS data set with the lowest excitation found. 
However, in the case of NGC\,3310, the [O{\sc iii}]/H$\beta$ ratios of the
different fittings and components are more or less equal.
The studied regions of NGC\,3310 are located in the BPT diagram among the
low-metallicity  \HII\ regions 
from the sample of P\'erez-Montero and D\'iaz (2005), which is in agreement
with the metal abundances derived by \citetex{1993MNRAS.260..177P}. 
It should be noted that we can discriminate between line ratios of the two
separate velocity components only for [O{\sc iii}]/H$\beta$ since we do not
have the necessary high dispersion data for the [N{\sc ii}] and H$\alpha$
lines. In fact, for the CNSFRs of NGC\,3310, we have applied
an artificial offset to the log([N{\sc ii}]/H$\alpha$) of -0.1 and
0.1\,dex for the narrow and broad components, respectively, for illustrative
purposes only.

It  is clearly of major interest to find out how widespread is the presence of
two distinct components in the emission lines in ionized regions and what is
its influence on the observed line ratios 
for several reasons. Firstly, a change in position in the diagnostic diagrams
would certainly affect the classification of the activity in the central
regions of the concerned galaxies. Secondly,  it will affect the inferences
about the nature of the source of ionization in the two components. Thirdly,
it could have an influence on the gas abundance determinations given 
that the ratio of the narrow to the broad components of H$\beta$ is about
1. Clearly it is not possible to use global line ratios to estimate gaseous
abundances if the permitted and forbidden line fluxes are partially originated
in different kinematical systems. 
To disentangle the origin of these two components it will be necessary to map
these regions with high spectral and spatial resolution and much better S/N
ratio in particular for the O$^{2+}$ lines.  Three-dimensional (3D)
spectroscopy with Integral 
Field Units (IFUs) would be the ideal tool to approach this issue.  

We have found that only two of the observed regions, R4 and R5 of NGC\,3351,
seem to possess just one knot showing up in the continuum image and coincident
with the H$\alpha$ 
emission. The sizes of these knots are 2.9 and 3.2\,pc, respectively. The
rest of the regions are made up of several knots which presumably correspond
to individual star clusters.  Their sizes are between 1.5 and 4.0\,pc for
NGC\,2903, between 2.2 and 6.2\,pc for NGC\,3310, 
and between 1.7 and 4.9\,pc for NGC\,3351. In all the cases the small sizes
are at the limit of the resolution. For comparison, the size of cluster A in
the BCD/amorphous galaxy NGC\,1569 \cite{1985AJ.....90.1163A} is 1.9\,pc, as
given by \citetex{1995AJ....110.2665M}. The sizes and 
absolute visual magnitudes estimated for each individual star cluster (e.g.
M$_v$\,=\,-12.55 for R4 of NGC\,3351 derived from the HST image) are in the
ranges  established by \citetex{1995AJ....110.2665M} in the definition of a
super 
star cluster (SSC).

Using the stellar velocity dispersion measurements as mapped by the CaT
absorption lines, the sizes measured on the HST image and the virial theorem
we have derived values of the dynamical masses, under the assumption that the
systems are spherically symmetric, gravitationally bound and have isotropic
velocity distribution.  Masses derived from the H$\beta$ velocity dispersion
under the assumption of a single component for the gas would have been
underestimated by factors of between approximately 2 and 4.

The values of the dynamical masses of the individual clusters derived by us
for the CNSFRs of these three galaxies are several times
the mass derived for the SSC A in NGC\,1569 by \citetex{1996ApJ...466L..83H}
from stellar velocity dispersion measurements using red ($\sim$6000\,\AA)
spectra. The mass of SSC A is
(3.3$\pm$0.5)\,$\times$\,10$^5$\,M$_\odot$. Our estimated masses are
between 4.2 and 33 times this mass. Besides, they are larger
than masses derived kinematically for SSC in irregular galaxies
\cite{2003ApJ...596..240M,2004AJ....128.2295L} and also larger than
those derived by \citetex{2002AJ....123.1411B} for the individual circumnuclear
clusters of NGC\,4314 from HST imaging data following the procedure of
\citetex{1999AJ....117..764E}, which are in the range
0.2\,$\times$\,10$^4\,\leq$\,M$\leq$\,1.6$\times$\,10$^4$\,M$_\odot$. However,
the H$\beta$ luminosities of the NGC\,4314 clusters are also lower than those
of NGC\,2903, NGC\,3310 and NGC\,3351 by factors of about 40 for the first and
third galaxy, and 50 for the second one.

It must be noted that we have measured the size of each knot (typically
between 3 and 5\,pc), but the stellar velocity dispersion corresponds to the
integrated CNSFR wider area containing several
knots. The use of these wider size scale velocity dispersion measurements to
estimate the mass of each knot, leads us to overestimate the mass of the
individual clusters, and hence of each CNSFR.

However, as can be seen in the HST-NICMOS images (Figures
\ref{hst-slits-2903-2}, \ref{hst-slits-3310-2} and \ref{hst-slits-3351-2}),
the CNSFRs are clearly visible in the IR
and dominate the light inside the apertures observed. All the regions analyzed
show up very prominently in the near-IR and therefore we can assume that the
light at the CaT wavelength region is dominated by the stars in the clusters.
The IR CaT is very strong, in fact the strongest stellar feature, in very
young clusters, i.e. older than 4\,Myr \cite{1990MNRAS.242P..48T}.  Besides,
we detect a minimum in the velocity dispersion at the position of the
clusters, indicating that they are kinematically distinct, and the values
derived with the widest apertures (five pixels) using a single Gaussian
fitting are approximately the average of the velocity dispersions estimated
using the extractions made with the narrower apertures (see
Figures \ref{dispersions-px-2903}, \ref{dispersions-px-3310} and
\ref{dispersions-px-3351}). We can not be sure though that we are actually 
measuring their velocity dispersion and thus prefer to say that our
measurements of $\sigma_{\ast}$ and hence dynamical masses constitute upper
limits. Although we are well aware of the difficulties, still we are confident
that these upper limits are valid and important for comparison with the gas
kinematic measurements.

Another important effect that can affect to the estimated dynamical masses is
the 
presence of binaries among the red supergiant and red giant populations from
which we have estimated the stellar velocity dispersions. On the basis of 
multislit spectroscopy of 180 stars in the ionizing cluster of 30~Doradus in
the LMC, \citetex{2001A&A...380..137B} studied the possible influence of
spectroscopic binaries in the radial velocity dispersion. 
In a recent work, Bosch et al.\ (2008; private communication)
\nocite{2008guille} using GMOS-GEMINI data investigated the presence of binary
stars within the ionizing cluster of 30~Doradus. From a three epoch observing
run they detected a rate of candidate to be binary systems within their OB
stars sample of $\sim$\,48\,\% (25 out of 52 stars). Interestingly enough,
this detection rate is consistent with a spectroscopic population of 100\,\%
binaries, when the observational parameters described in
\citetex{2001RMxAC..11...29B} are set for their observations. Their final
sample of `single' stars (after removing their confirmed or candidate
binaries) decreased to 26 stars, still enough to calculate a representative
value of the stellar radial velocity dispersion, for which they estimated a
value of 8.3\,km\,s$^{-1}$. When they derived $\sigma_{\ast}$ from a single
GMOS observation epoch, they found values as high as 30\,km\,s$^{-1}$,
consistent with the values derived from NTT observations by
\citetex{2001A&A...380..137B}. 

Although the environment of our CNSFRs is very different from that of 30~Dor
and the stellar components of the binary systems studied by Bosch and
collaborators (O and early B type stars) are very different from  those stars
in our regions where the CaT arise (red supergiants), this is a very
illustrative observational example of this problem. The orbital motions of
the stars in binary (multiple) systems produce an overestimate of the
velocity dispersions and hence of the dynamical masses. The single-star
assumption introduces a systematic error that depends on the properties of the
star cluster and the binary population, with an important effect on the
cluster mass if the typical orbital velocity of a binary component is of
the order of, or larger than, the velocity dispersion of the single/binary
stars in the potential of the cluster \cite{2007arXiv0710.1207K}. As was
pointed out by these authors, the
relative weights between the single and binary stars in the velocity dispersion
measurements depend on the binary fraction, which, together with the
semi-major axis or period distribution, are the most important parameters to
determine whether the binary population affects  the estimated dynamical
masses. Their simulations indicate that the dynamical mass is overestimated by
70\,\% for a measured stellar velocity dispersion in the line of sight of
1\,km\,s$^{-1}$, 50\,\% for 2\,km\,s$^{-1}$, and 5\,\% for
10\,km\,s$^{-1}$. They conclude that most of the known dynamical masses
of massive star clusters are only mildly affected by the presence of
binaries. Hence, for our clusters, where the smallest estimated velocity 
dispersion is 31\,km\,s$^{-1}$, we can assume that the contribution of
binaries to 
the stellar velocity dispersions is not that important. Moreover, the binary
fraction of the red supergiants and red giants in this type of circumnuclear
and typically high metal rich environment is not known.

The masses of the ionizing stellar clusters of the circumnuclear regions, as
derived from their H$\alpha$ luminosities under the assumption that the
regions are ionization bound and without taking into account any photon
absorption by dust (see Table \ref{parameters}), are comparable to that
derived by \citetex{1995ApJ...439..604G} for the circumnuclear region A in
NGC\,7714 (5.1$\times$10$^5$ M$_\odot$) and so are the masses of the ionized
gas (3$\times$10$^5$\,M$_\odot$) as can be seen from Table \ref{parameters}. 
In all the cases the masses of the ionized gas
make up a small fraction of the total mass of the regions. We have derived
both the masses of the ionizing stellar clusters and of the ionized gas from
the H$\alpha$ luminosity of the CNSFRs assuming that they consist of one
single component. However, if we consider only the broad component whose
kinematics follows that of the stars in the regions, all derived quantities
would be smaller by a factor of 2.

\section{Summary and conclusions} 
\label{summa-kine}

We have measured gas and stellar velocity dispersions in seventeen CNSFRs and
the nuclei of three barred spirals, NGC\,2903, NGC\,3310 and NGC\,3351 (four,
eight and five CNSFRs, respectively). The stellar velocity dispersions have
been measured from the CaT lines at
$\lambda\lambda$\,8494, 8542, 8662\,\AA, while the gas velocity dispersions
have been measured by Gaussian fits to the H$\beta$\,$\lambda$\,4861\,\AA\ and
the [O{\sc iii}]\,$\lambda$\,5007\AA\ emission lines on high dispersion
spectra.

Stellar velocity dispersions are between 31 and 73\,km\,s$^{-1}$. In the case
of NGC\,2903 and NGC\,3351 ($\sigma_\ast$ between 37 and 65, and between 39
and 67\,km\,s$^{-1}$, respectively) these values are about
25\,km\,s$^{-1}$ larger than those measured for the gas from the H$\beta$
emission line using a single Gaussian fit. For NGC\,3310 the stellar (between
31 and 73\,km\,s$^{-1}$) and gas
velocity dispersion are in relatively good agreement, with the first being
slightly larger. The [O{\sc iii}]\,5007\,\AA\ presents velocity dispersions
almost 
coincident with the stellar ones, or slightly, for NGC\,2903 and
NGC\,3351, while in the case of NGC\,3310 its behaviour is very similar to
that shown by the H$\beta$ line.
However, the best Gaussian fits involved two different components for the gas:
a ``broad component" with a velocity dispersion similar to that measured for
the stars for NGC\,2903 and NGC\,3351, and larger by about 20\,km\,s$^{-1}$
for NGC\,3310, and a ``narrow component" with a velocity dispersion lower than
the stellar 
one by about 30\,km\,s$^{-1}$. This last component seems to have a relatively
constant value for all the studied CNSFRs in these three galaxies, with
estimated values close 
to 25\,km\,s$^{-1}$ for the two gas emission lines. The velocities of the two
components of the multi-Gaussian fits in the CNSFRs of NGC\,3351 are the same
within the observational errors, but in the cases of NGC\,2903 and NGC\,3310
we find a shift between the narrow and the broad component that vary between
-25 and 35\,km\,s$^{-1}$ in radial velocity.

When plotted in a [O{\sc iii}]/H$\beta$ versus [N{\sc ii}]/H$\alpha$
diagram, the two systems are clearly segregated for the high-metallicity
regions of NGC\,2903 and NGC\,3351, with the narrow component
having the lowest excitation and being among the lowest excitation line ratios
detected within the SDSS data set of starburst systems. In the regions of the
low-metallicity galaxy these two components and those values derived using the
single Gaussian fit are very similar but it should be remembered that no
comparable information exists for the [N{\sc ii}]/H$\alpha$ ratio.

The dynamical masses estimated from the stellar velocity dispersion using the
virial theorem for the CNSFRs of NGC\,2903 are in the range between
6.4\,$\times\,10^7$ and 1.9\,$\times\,10^8$\,M$_\odot$ and is
1.1\,$\times\,10^7$ for its nuclear region inside the inner 3.8\,pc. In the
case of NGC\,3310 the masses are in the range between 2.1\,$\times\,10^7$ and
1.4\,$\times\,10^8$\,M$_\odot$ for the CNSFRs and for the nuclear region
inside the inner 14.2\,pc is 5.3\,$\times\,10^7$\,M$_\odot$. For NGC\,3351 the
dynamical masses are in the range between 4.9\,$\times\,10^6$ and
4.5\,$\times\,10^7$\,M$_\odot$, and is 3.5\,$\times\,10^7$ for its nuclear
region inside the inner 11.3\,pc. Masses derived
from the H$\beta$ velocity dispersions under the assumption of a single
component for the gas would have been underestimated by factors between
2 and 4 approximately.

The derived masses for the individual clusters are between
1.4\,$\times$\,10$^6$ and 1.1\,$\times$\,10$^7$\,M$_\odot$, between
1.8 and 7.1\,$\times$\,10$^6$\,M$_\odot$, and between 1.8
and 8.7\,$\times$\,10$^6$\,M$_\odot$ for NGC\,2903, NGC\,3310 and NGC\,3351,
respectively. Then, globally, the masses of these individual clusters vary
between 1.4\,$\times$\,10$^6$ and 1.1\,$\times$\,10$^7$\,M$_\odot$.
These values are between 4.2 and 33 times the mass derived for
the SSC A in NGC\,1569 by \citetex{1996ApJ...466L..83H} and larger than other
kinematically derived SSC masses. 

Masses of the ionizing stellar clusters of the CNSFRs have been derived from
their H$\alpha$ luminosities under the assumption that the regions are
ionization bound and without taking into account any photon absorption by
dust. For the regions of NGC\,2903 these masses are between 3.3 and
4.9\,$\times\,10^6$\,M$_\odot$, and is 2.1\,$\times\,10^5$ for its
nucleus. The values derived in NGC\,3310 are between 8.7\,$\times\,10^5$ and 
2.1\,$\times\,10^6$\,M$_\odot$ for the star-forming regions, and is
3.5\,$\times\,10^6$ for the nucleus. For NGC\,3351 are between
8.0\,$\times$\,10$^5$ and 2.5\,$\times$\,10$^6$\,M$_\odot$ for the
regions, and is 6.0\,$\times$\,10$^5$\,M$_\odot$ for its nuclear region
(see table \ref{parameters}). Thus, the ionizing stellar cluster studied in
these three galaxies vary between 8.0\,$\times$\,10$^5$ and
4.9\,$\times\,10^6$\,M$_\odot$. Therefore, the ratio of the ionizing
stellar population to the total dynamical mass is between 0.01
and 0.16. 
These values of the masses of the ionizing stellar clusters of the CNSFRs are
comparable to that derived by \citetex{1995ApJ...439..604G} for the
circumnuclear region A in NGC\,7714.

Derived masses for the ionized gas, also from their H$\alpha$
luminosities, vary between 6.1\,$\times\,10^4$ and
1.3\,$\times\,10^5$\,M$_\odot$ for the regions and is 
3\,$\times\,10^3$\,M$_\odot$ for the nucleus of NGC\,2903; between 1.5 and
7.2\,$\times\,10^5$\,M$_\odot$ for the CNSFRs and is
5\,$\times\,10^3$\,M$_\odot$ for the nucleus of NGC\,3310; and between
7.0\,$\times$\,10$^3$ 
and 8.7\,$\times$\,10$^4$\,M$_\odot$ for the CNSFRs of NGC\,3351, and is
2\,$\times$\,10$^3$\,M$_\odot$ for its nucleus. These values are also
comparable to that derived by \citetex{1995ApJ...439..604G}.

It is interesting to note that, according to our findings, the SSC in CNSFRs
seem to contain composite stellar populations. Although the youngest one
dominates the UV light and is responsible for the gas ionization, it
constitutes only about 10 per cent of the total. This can explain the low
EWs of emission lines measured in these regions.  This may well apply to
the case of other SSC and therefore conclusions drawn from fits of single
stellar population (SSP) models should be taken with caution 
(e.g.\ \citeplain{2003ApJ...596..240M,2004AJ....128.2295L}).
Furthermore, the composite
nature of the CNSFRs  means that star formation in the rings is a process that
has taken place over time periods much longer than those implied by the
properties of the ionized gas.

The observed stellar and [O{\sc iii}] rotational velocities of NGC\,2903 are
in good agreement, while the H$\beta$ measurements show shifts similar to
those find between the narrow and the broad components. This different 
behaviour can be explained if the positions of the single Gaussian fits are
dominated by the broad component in the case of the [O{\sc iii}] emission line
and by the narrow one in the case of the H$\beta$. The
rotation curve corresponding to the position going through the nucleus shows
maximum and minimum values at the positions of the circumnuclear regions, as
observed in other galaxies with CNSFRs. 

In the case of NGC\,3310 the rotation curve shows a typical $S$ feature, with
the presence of some perturbations, in particular near the location of the
Jumbo region. In this galaxy the values derived from the gas emission lines
and the stellar absorption features are in very good agreement. Again, the
position going through the nucleus shows maximum and minimum values more or
less coincident with the location of the CNSFRs.

For NGC\,3351, the rotation velocities derived for both stars and gas are in
reasonable 
agreement, although in some cases the gas shows a velocity slightly different
from that of the stars. The rotation curve corresponding to the position
going through the centre of the galaxy also shows maximum and minimum values
at the position of the circumnuclear ring. The differences in velocity between
gas and stars can be interpreted 
as motions of the ionized hydrogen deviating from rotation and consistent with
a radial infall to the central regions of the galaxy. Our results are
consistent with those found by Rubin et al.\ (1975) and would yield an infall
velocity of about 25\,km\,s$^{-1}$.\\ 

The existence of more than one velocity component in the ionized
gas corresponding to kinematically distinct systems, deserves further
study. Several results derived from the observations of the different emission
lines could be affected, among others: the classification of the activity in
the central regions of galaxies, the inferences about the nature of the source
of ionization, the gas abundance determinations, the number of ionizing
photons from a given region and any quantity derived from them. To
disentangle the origin of these two components it will be necessary to map
these regions with high spectral and spatial resolution and much better S/N
ratio in particular for the O$^{2+}$ lines. High resolution 3D spectroscopy
with IFUs would be the ideal tool to approach this issue. \\

\addcontentsline{toc}{section}{\numberline{}Bibliography}

\bibliographystyle{astron}
\bibliography{tesis}

\chapter{Star Formation in Circumnuclear Regions:\\ The metal abundance}
\label{abundan}

\section{Introduction}

Despite its difficulty, the importance of an accurate determination of
the abundances of high metallicity \HII\ regions cannot be overestimated since
they constitute most of the \HII\ regions in early spiral galaxies (Sa to Sbc)
and the inner regions of most late  type ones (Sc to Sd)
\cite{1989epg..conf..377D,1992MNRAS.259..121V} without  which our description
of the metallicity distribution in galaxies cannot be  complete. In
particular, the effects of the choice of different calibrations  on the
derivation of abundance gradients can be very important since any abundance
profile fit will be strongly biased towards data points at the ends of the
distribution. It should be kept in mind that abundance gradients are widely
used to constrain chemical evolution models, histories of star formation over
galactic discs or galaxy formation scenarios. The question of how high is the
highest oxygen abundance in the gaseous phase of galaxies is still standing
and extrapolation of known radial abundance gradients would point to CNSFRs as
the most probable sites for these high metallicities.

As was pointed out in Chapter \S \ref{cnsfr-obs-kine}, the [O{\sc iii}]
emission lines in CNSFRs are generally very weak ([O{\sc
    iii}]\,5007/H$\beta$\,$<$\,1; e.g.\ \citeplain{2000A&A...353..893P}),
and in some cases unobservable. This low value of these collisional excited 
lines can be explained by their over-solar metal abundances (e.g.\
\citeplain{1993A&A...277..397B}). The equivalent width of the emission lines
are lower than those shown by the disc \HII\ regions (see for example
\citeplain{1989AJ.....97.1022K,1997AJ....113..975B,1999ApJ...510..104B}).
Kennicutt and collaborators found that the equivalent widths of H$\alpha$ in
CNSFRs are lower by a factor of seven than the ones in disc \HII\
regions. They suggest different mechanisms for this fact, none of them
dominant: (i) strong deficiency of high-mass
stars in the initial mass function; (ii) a long timescale star formation, with
the excess continuum arising from more evolved stars, and the star formation
nearly continuous over the ``hot-spot'' region as a whole; (iii) very high 
dust abundance; (iv) some of the nuclear \HII\ regions are not ionization
bounded; (v) contamination of the continuum by contribution from the bulge or
other underlying stellar populations. 

Combining two-dimensional integral field unit data from GEMINI 
South (GMOS-IFU) and a grid of photo-ionization models
\citetex{2008arXiv0802.2070D} conclude that the contamination of the continua
of CNSFRs by underlying contributions from both old bulge stars and stars
formed in the ring in previous episodes of star formation (10-20\,Myr) yield
the observed low equivalent widths, and correcting for these contributions
they did not find significant differences in ages between the CNSFRs and the
inner disc \HII\ regions.

NGC\,2903, NGC\,3351 and NGC\,3504 have been studied using moderate spectral
resolution data covering a broad spectral range to estimate the chemical
abundances and densities of the CNSFRs. In Section \ref{sampleabund} we
present the sample with emphasis on  NGC\,3504 that has not been studied in
Chapter \$ \ref{cnsfr-obs-kine}. In \S \ref{met-obs}, the observations and the
data reduction procedures. The results are presented in  Section \S
\ref{results-abun}. The method for the derivation of chemical abundances is
described in Section \S \ref{che-abun}. Section \S \ref{dis-abun} is
devoted to the discussion of our results. Finally, Section \S
\ref{summ-abun} summarizes the main conclusions of this part of the work.

\section{Sample selection}
\label{sampleabund}

We have obtained moderate resolution observations of 12 CNSFRs in three
``hot-spot'' galaxies: NGC\,2903, NGC\,3351 and NGC\,3504. The three of them
are early barred spirals and 
show a high star formation rate in their nuclear regions. They are quoted in
the literature as among the spirals with the highest overall oxygen abundances
\cite{1994ApJ...420...87Z,tesisdiego}. High resolution spectroscopy on two of
these objects, NGC\,2903 and NGC\,3351, were presented in Chapter \S
\ref{cnsfr-obs-kine}.

\begin{figure}
\centering
\includegraphics[width=.98\textwidth,angle=0]{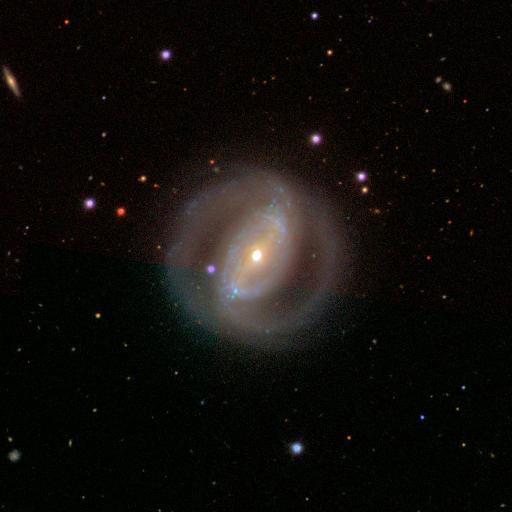}
\caption{False colour image of the barred spiral galaxy NGC\,3504 from the
  SDSS.} 
\label{col3504}
\end{figure}

The third one, NGC\,3504 (see Figure \ref{col3504}), is the
brightest galaxy in the optically selected catalogue of 
starburst galactic nuclei in Balzano \cite*{1983ApJ...268..602B}. It forms a
pair with NGC\,3512. Various studies at different wavelengths confirmed that
it harbours a very intense nuclear starburst
\cite{1989ApJ...346..126D,1990ApJ...364...77P}. Infrared observations in the J
and K bands reveal a ring with five discrete clumps of star formation with
colours indicating ages of about 10$^7$ yr \cite{1997AJ....114.1850E}. The
H$\alpha$ emission from Planesas et al.\ \cite*{1997A&A...325...81P} traces a
compact ring structure with a radius of 2\arcsec\ (200\,pc) around the nucleus
where four separate \HII\ regions can be identified. From the H$\alpha$
emission, these authors derive a global star formation rate for the
circumnuclear region of 0.62\,M$_\odot$\,yr$^{-1}$, while from their CO
observations, a molecular gas mass of 21\,$\times$\,10$^{8}$\,M$_\odot$
inside a circle 2.7\,Kpc in diameter is derived.

The main properties of these galaxies are given in Table \ref{propgal}.

%

\begin{table}
\begin{center}
\caption{The galaxy sample.}
 \begin{tabular}{@{}cccc@{}}
\hline
\hline
Property                 & NGC\,2903     & NGC\,3351  & NGC\,3504   \\
\hline				         
R. A. (2000)$^a$         &  09 32 10.1   & 10 43 57.7 & 11 03 11.2 \\
Dec (2000)$^a$           & +21 30 03     & +11 42 14  & +27 58 21  \\ 
Morph. Type              & SBbc          & SBb        & SABab      \\
Distance (Mpc)$^b$       &  8.6          &    10.1    &  20        \\
pc/ \arcsec\             &  42           &    50      &  100       \\
B$_{T}$ (mag)$^a$        &  9.7          &    10.5    & 11.8       \\
E(B-V)$_{gal}$(mag)$^a$  & 0.031         & 0.028      & 0.027      \\
\hline
\multicolumn{3}{l}{$^a$~\citetex{1991trcb.book.....D}}\\
\multicolumn{3}{l}{$^b$~NGC\,2903: \citetex{1984AAS...56..381B}}\\
\multicolumn{3}{l}{~~NGC\,3351: Graham et al.\ (1997)}\\
\multicolumn{3}{l}{~~NGC\,3504: \citetex{1992ApJ...395L..79K}} 
\end{tabular}
\label{propgal}
\end{center}
\end{table}


\nocite{1997ApJ...477..535G}

\section{Observations and data reduction}
\label{met-obs}

Our spectrophotometric observations were obtained with the 4.2m William
Herschel Telescope at the Roque de los Muchachos Observatory, in 2001 January
26, using the ISIS double spectrograph, with the EEV12 and TEK4 detectors in
the blue and red arm respectively. The incoming light was split by the
dichroic at $\lambda$7500\,\AA.  Gratings  R300B in the blue arm and R600R in
the red arm were used, covering  3400\,\AA\ in  the blue ($\lambda$3650 to
$\lambda$7000) and 800\,\AA\ in the near IR ($\lambda$8850 to $\lambda$9650)
and yielding spectral  dispersions of 1.73\,\AA\ pixel$^{-1}$ in the blue arm
and  0.79\,\AA\ pixel$^{-1}$ in the red arm. With a slit width of 1\farcs05,
spectral resolutions of {$\sim$}2.0\,\AA\ and 1.5\,\AA\ FWHM in the blue and
red arms respectively were attained. This is an optimal configuration which
allows the simultaneous observation of a given region in both frames in a
single exposure.

The nominal spatial sampling is 0\farcs4 pixel$^{-1}$ in each frame and the
average seeing for this night was {$\sim$}1\farcs2. The journal of the
observations is given in Table \ref{journal}. 

%
%
\begin{table}
\centering
\caption{Journal of Observations.}
\begin{tabular} {@{}l c c c c c c@{}}
\hline
\hline
 Galaxy & Spectral range & Grating &      Disp.       &    Spatial resolution & PA   & Exposure Time \\
            &     (\AA)           &            & (\AA\,px$^{-1}$)  & (\arcsec\,px$^{-1}$) &  ($ ^{o} $) & (sec)   \\
\hline
NGC\,2903 & 3650-7000  & R300B &      1.73       &  0.4    &     105    &   2\,$\times$\,1800 \\
NGC\,2903 & 8850-9650  & R600R &      0.79       &  0.4    &     105    &   2\,$\times$\,1800  \\
NGC\,2903 & 3650-7000  & R300B &      1.73       &  0.4    &     162    &   2\,$\times$\,1800 \\
NGC\,2903 & 8850-9650  & R600R &      0.79       &  0.4    &     162    &   2\,$\times$\,1800 \\
NGC\,3351 & 3650-7000  & R300B &      1.73       &  0.4    &      10    &   2\,$\times$\,1800 \\
NGC\,3351 & 8850-9650  & R600R &      0.79       &  0.4    &      10    &   2\,$\times$\,1800  \\
NGC\,3351 & 3650-7000  & R300B &      1.73       &  0.4    &      38    &   2\,$\times$\,1800 \\
NGC\,3351 & 8850-9650  & R600R &      0.79       &  0.4    &      38    &   2\,$\times$\,1800  \\
NGC\,3351 & 3650-7000  & R300B &      1.73       &  0.4    &      61    &   2\,$\times$\,1800 \\
NGC\,3351 & 8850-9650  & R600R &      0.79       &  0.4    &      61    &   2\,$\times$\,1800  \\
NGC\,3504 & 3650-7000  & R300B &      1.73       &  0.4    &     110    &   2\,$\times$\,1800 \\
NGC\,3504 & 8850-9650  & R600R &      0.79       &  0.4    &     110    &   2\,$\times$\,1800  \\
\hline
\end{tabular}
\label{journal}
\end{table}


\begin{figure*}
\centering
\includegraphics[width=.98\textwidth,angle=0]{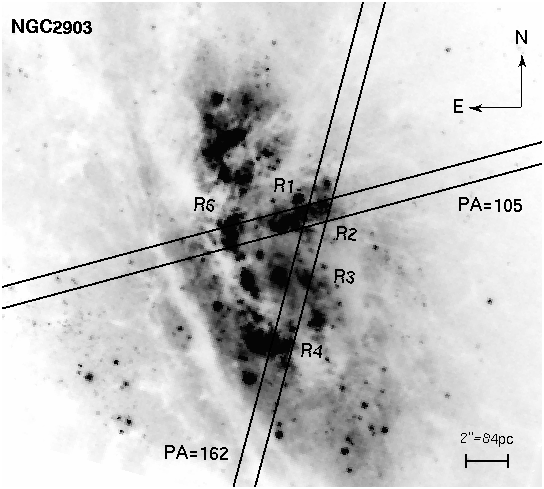}
\caption[Observed CNSFRs in NGC\,2903 for the abundance
  analysis]{Observed CNSFRs in NGC\,2903. The different slit 
  positions are superimposed on an image taken from the HST archive and
  obtained with the WFPC2 camera through the F606W filter. The position
  angles of every 
  slit position are indicated.} 
\label{hst-slits1}
\end{figure*}

\begin{figure*}
\centering
\includegraphics[width=.98\textwidth,angle=0]{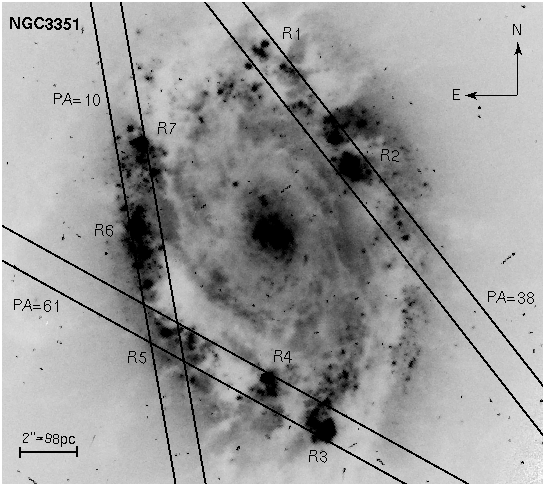}
\caption[Idem as Figure \ref{hst-slits1} for NGC\,3351]{Idem as Figure
  \ref{hst-slits1} for NGC\,3351.}  
\label{hst-slits2}
\end{figure*}

\begin{figure*}
\centering
\includegraphics[width=.98\textwidth,angle=0]{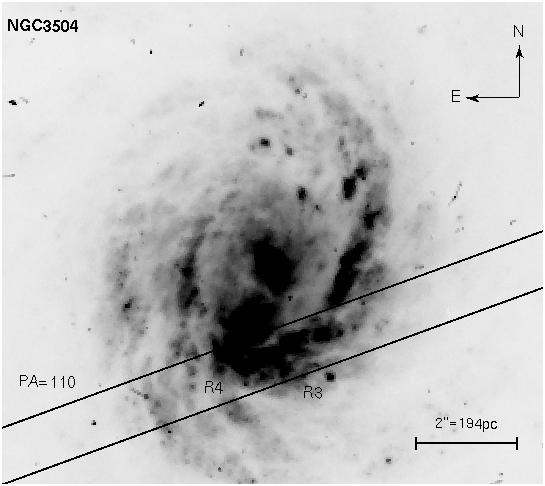}  
\caption[Idem as Figure \ref{hst-slits1} for NGC\,3504]{Idem as Figure
  \ref{hst-slits1} for NGC\,3504.} 
\label{hst-slits3}
\end{figure*}


Two, three and one slit positions in NGC\,2903, NGC\,3351 and NGC\,3504
respectively were chosen to observe a total of 12 CNSFRs. Figures
\ref{hst-slits1}-\ref{hst-slits3} show the different slit positions
superimposed on images 
obtained with the HST-WFPC2 camera and taken from the HST archive.  Some
characteristics of the observed regions, as given by  Planesas et al.\ (1997),
from where the identification numbers have also been taken, are listed in
Table \ref{prop_cn}.

%
%

\begin{table}
\centering
 \caption{Characteristics of the CNSFRs observed}
 \begin{tabular}{@{}lcccc@{}}
\hline
\hline
Galaxy & Region & Offsets from center & Diameter$^a$& F(H$\alpha$)\\
            &            & \arcsec\, \arcsec\       & \arcsec\  &
$\times$10$^{-14}$ erg s$^{-1}$ cm$^{-2}$ \\

\hline
NGC\,2903        & R1+R2    & -1.7,+3.5  & 4.0 & 22.8 \\
                 & R3       & -1.6,-0.3   & 2.0 & 2.6  \\
                 & R4       & -0.3,-3.3   & 2.4 & 10.1 \\
                 & R6       & +2.3,+2.4 & 2.4 & 12.7 \\  
NGC\,3351        & R1       & +0.4,+6.5 & 2.4 & 12.3 \\
                 & R2       & -2.6,+2.6  & 2.4 & 16.0 \\
                 & R3       & -1.5,-6.5   & 2.4 & 20.7 \\
                 & R4       & +0.5,5.4   & 2.2 & 10.5 \\
                 & R5       & +2.9,-3.5  & 2.4 & 4.9 \\
                 & R6       & +4.1,-0.4  & 2.4 & 6.1 \\
                 & R7       & +4.8,+3.6 & 1.8 & 9.5 \\
NGC\,3504        & R3+R4    & -0.5,-1.7   & 1.6 & 21.6 \\

\hline
\multicolumn{5}{@{}l}{$^a$~Size of the circular aperture used to measure fluxes} \\
\end{tabular}
\label{prop_cn}
\end{table}


The data were reduced using the IRAF\footnote{IRAF: the Image Reduction and
Analysis Facility is distributed by the National Optical Astronomy
Observatories, which is operated by the Association of Universities for
Research in Astronomy, Inc. (AURA) under cooperative agreement with the
National Science Foundation (NSF).} package following standard methods. The
two-dimensional wavelength calibration was accurate to 1\,\AA\ in all cases,
obtained 
by means of Cu, Ne and Ar calibration lamps. The two-dimensional frames were
flux calibrated using four spectroscopic standard stars: Feige~34,
BD26+2606, HZ44 and HD84937, observed before and after each program object
with a 3\arcsec\ width slit. For two of the standard stars: Feige~34 and HZ44,
the fluxes have been obtained from the most updated version of the original
Oke's spectra \cite{1990AJ.....99.1621O} and cover the 3200 to 9200\,\AA\
range. Data between 9200 and 9650\,\AA\ have been obtained from stellar
atmosphere models. For the other two stars: BD26+2606 and HD84937, the fluxes
have been taken from Oke and Gunn \cite*{1983ApJ...266..713O} that cover the
whole spectral range. The agreement between the individual calibration curves
was better than 5\% in all cases and a weighted mean calibration curve was 
derived. The spectra were previously corrected for atmospheric extinction
using a mean extinction curve applicable to La Palma observing site. 

Regarding background subtraction, the high spectral dispersion used in the
near infrared allowed the almost complete elimination of the night-sky OH
emission lines and, in fact, the observed [S{\sc
    iii}]\,$\lambda$\,9532/$\lambda$\,9069 
ratio is close to the theoretical value of 2.44 in most cases. Telluric
absorptions have been removed from the spectra of the regions by dividing by
the relatively featureless continuum of a subdwarf star observed in the same
night.

%
%

\begin{figure*}
\centering
\includegraphics[width=.48\textwidth,angle=0]{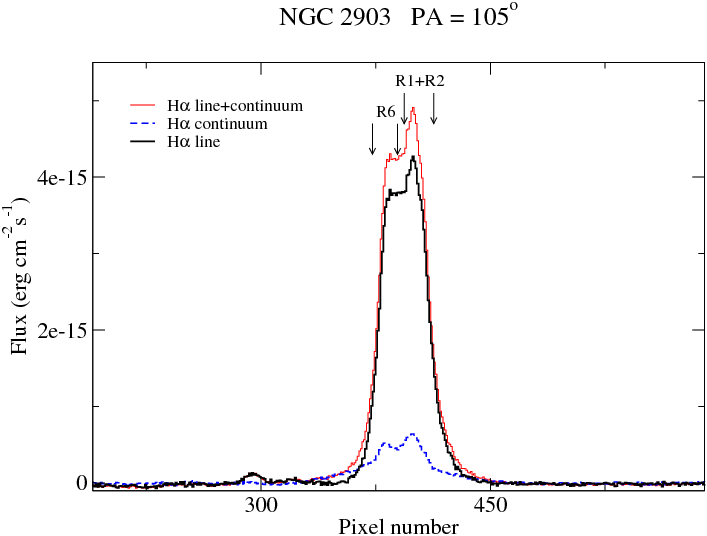}
\hspace{0.2cm}
\includegraphics[width=.48\textwidth,angle=0]{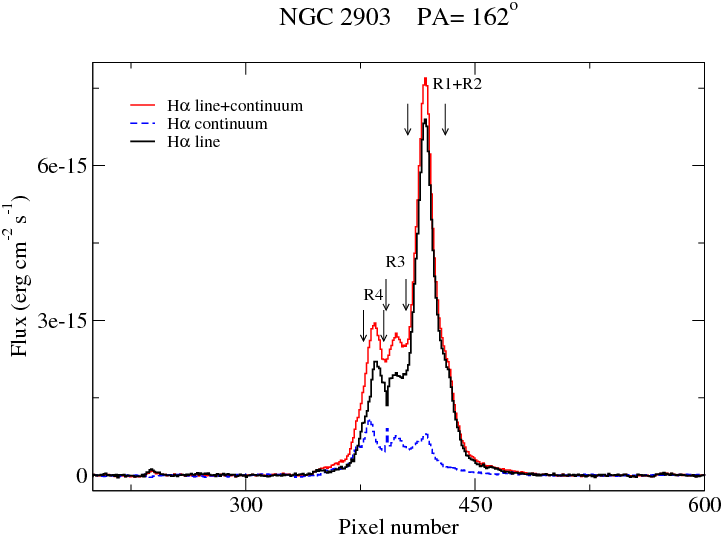}\\
\vspace{0.5cm}
\includegraphics[width=.48\textwidth,angle=0]{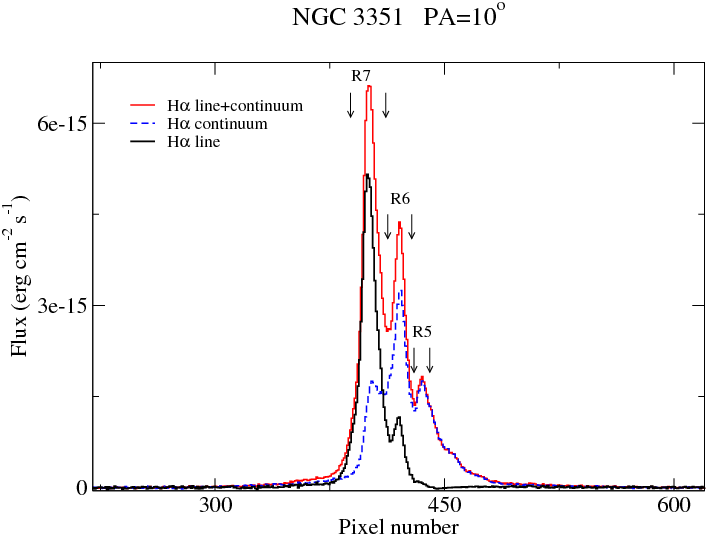}
\hspace{0.2cm}
\includegraphics[width=.48\textwidth,angle=0]{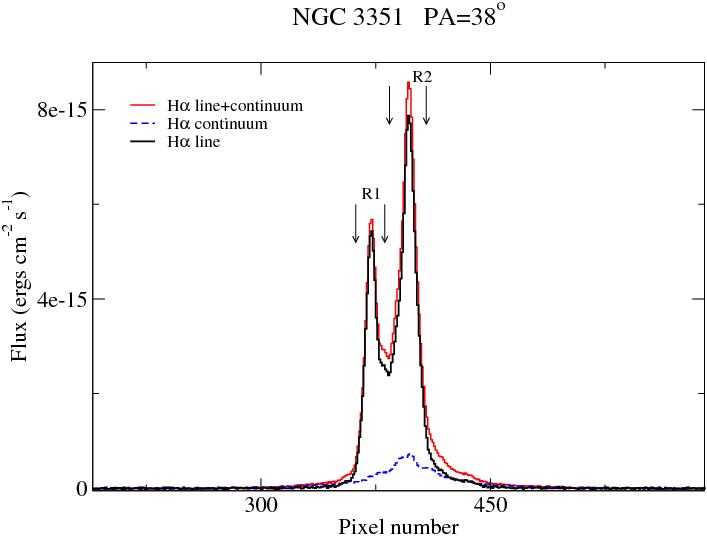}\\
\vspace{0.5cm}
\includegraphics[width=.48\textwidth,angle=0]{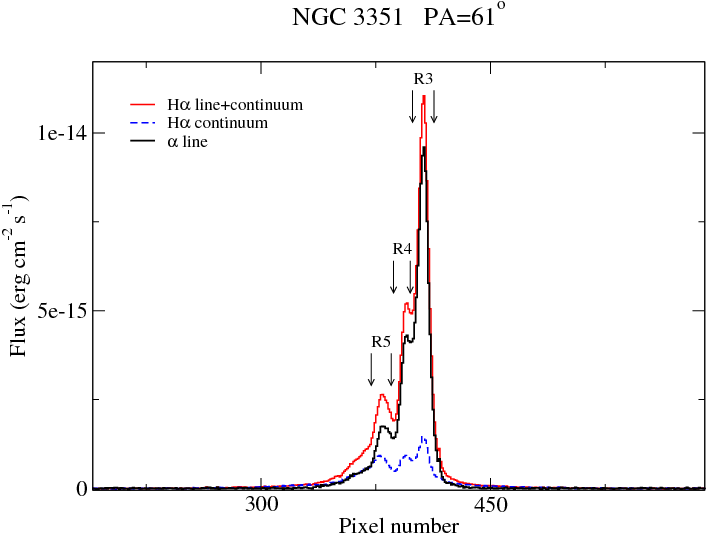}
\hspace{0.2cm}
\includegraphics[width=.48\textwidth,angle=0]{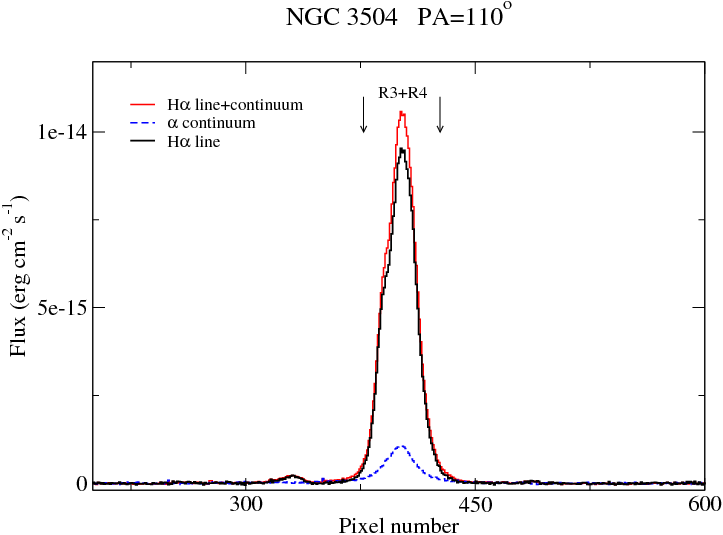}\\
\vspace{0.2cm}
\caption[H$\alpha$ profiles for the observed slit positions]{H$\alpha$
  profiles for the observed slit positions. Each figure includes the name of
  the galaxy, the P.A. of the slit, and the name of the observed regions.}
\label{profiles}
\end{figure*}


\section{Results}
\label{results-abun}

\begin{figure*}
\centering
\vspace{0.5cm}
\includegraphics[width=.45\textwidth,angle=0]{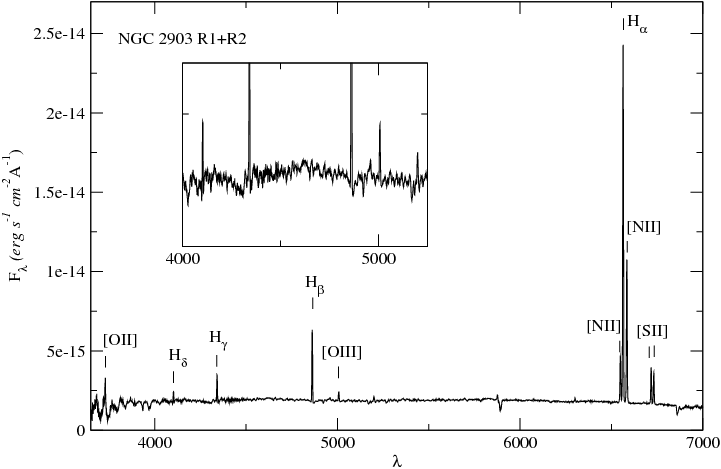}
\hspace{0.2cm}
\includegraphics[width=.45\textwidth,angle=0]{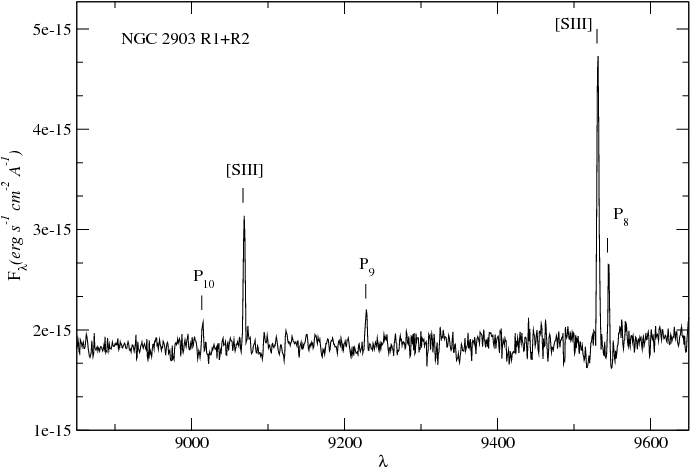}\\
\vspace{0.9cm}
\includegraphics[width=.45\textwidth,angle=0]{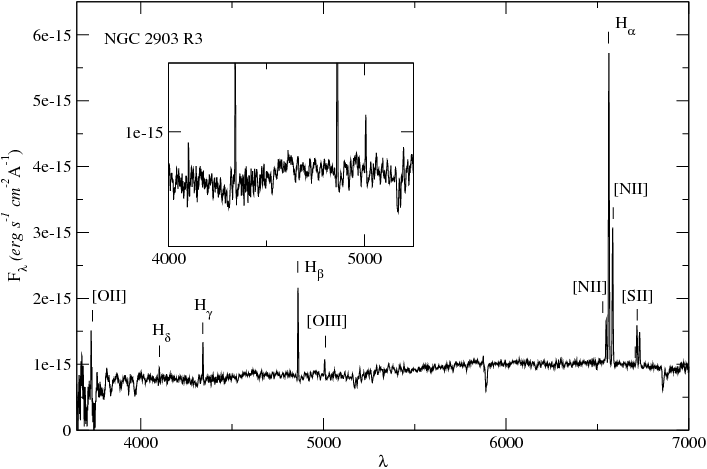}
\hspace{0.2cm}
\includegraphics[width=.45\textwidth,angle=0]{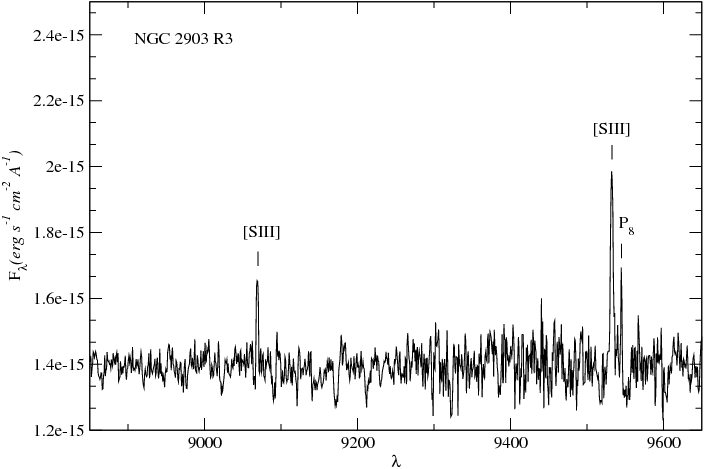}\\
\vspace{0.9cm}
\includegraphics[width=.45\textwidth,angle=0]{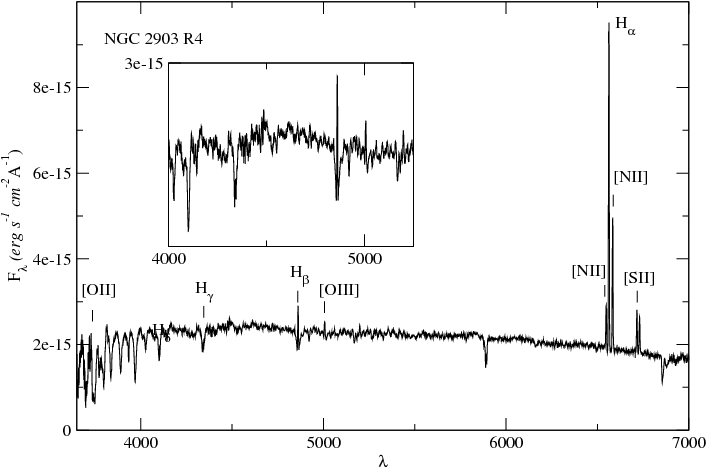}
\hspace{0.2cm}
\includegraphics[width=.45\textwidth,angle=0]{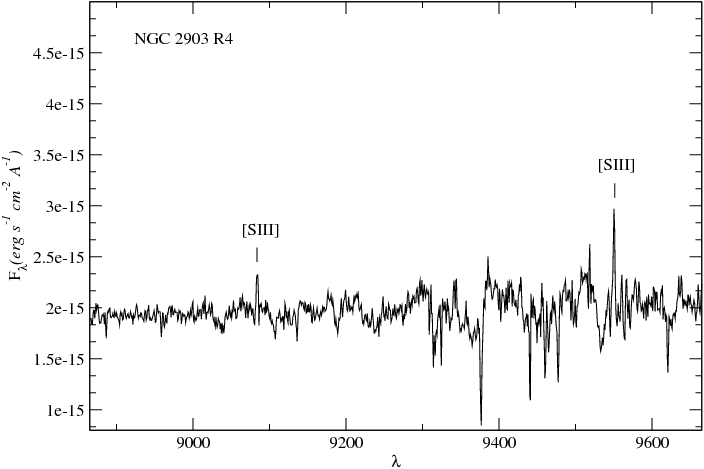}\\
\vspace{0.9cm}
\includegraphics[width=.45\textwidth,angle=0]{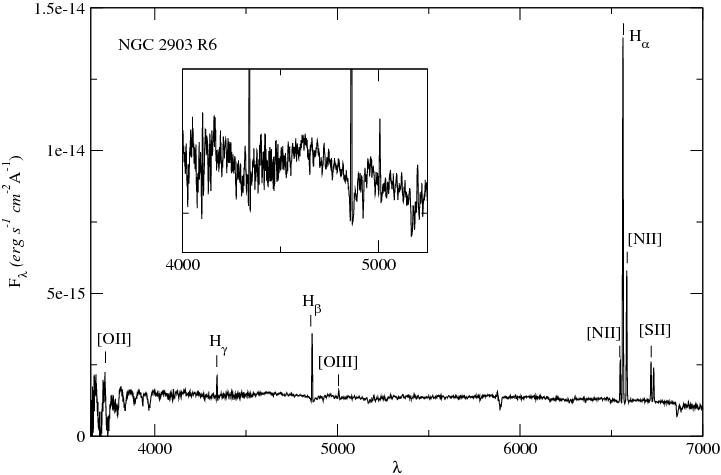}
\hspace{0.2cm}
\includegraphics[width=.45\textwidth,angle=0]{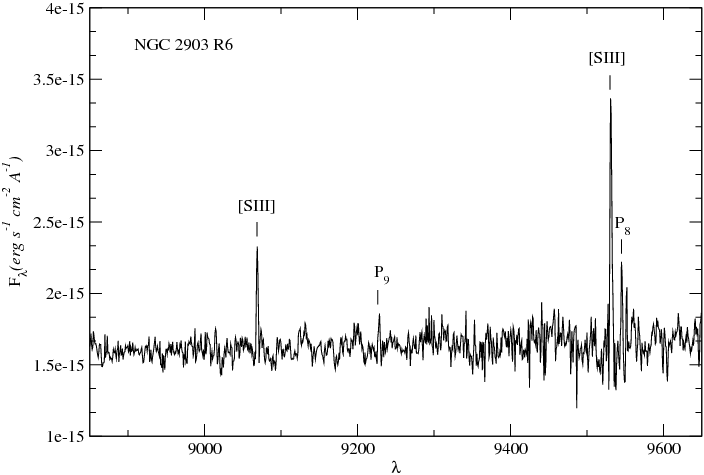}\\
\caption[Extracted spectra for the observed regions of NGC\,2903]{Extracted
  blue (left) and red (right) spectra for the observed regions of
  NGC\,2903. From top to bottom: R1+R2, R3, R4 and R6. Here and in Figures
  \ref{spectra2} and \ref{spectra3}, the insets show an enlarged region to see
  weak spectral features.} 
\label{spectra1}
\end{figure*}

\begin{figure*}
\centering
\vspace{0.7cm}
\includegraphics[width=.45\textwidth,angle=0]{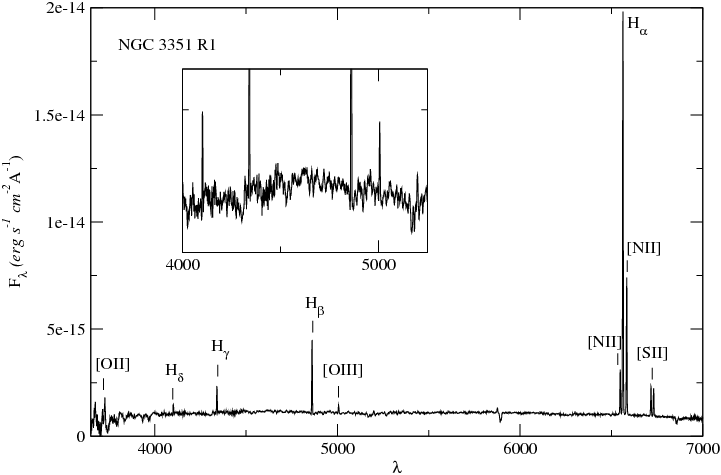}
\hspace{0.2cm}
\includegraphics[width=.45\textwidth,angle=0]{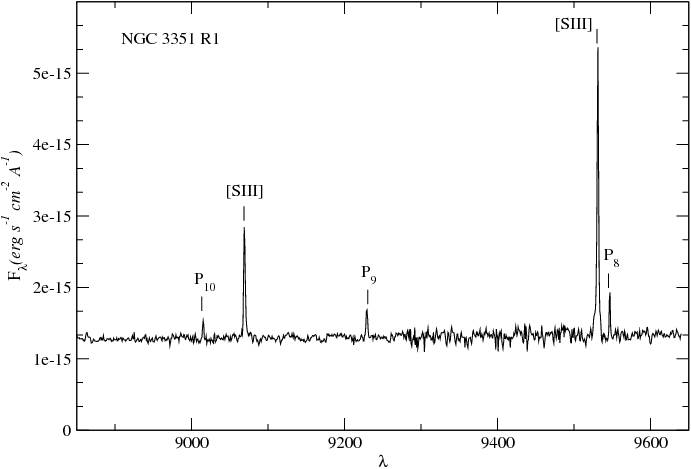}\\
\vspace{0.9cm}
\includegraphics[width=.45\textwidth,angle=0]{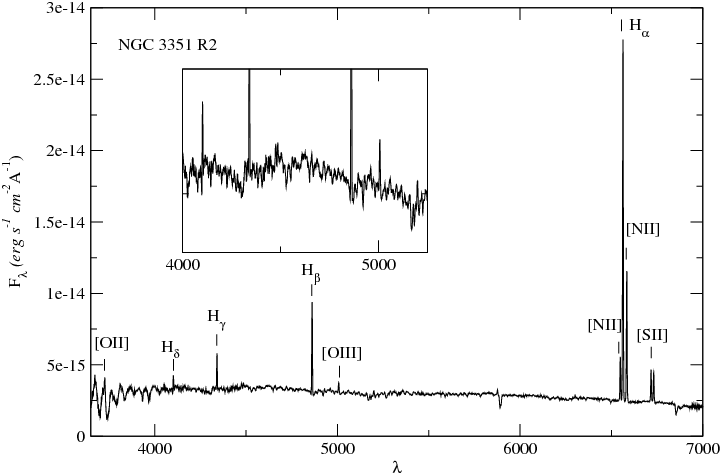}
\hspace{0.2cm}
\includegraphics[width=.45\textwidth,angle=0]{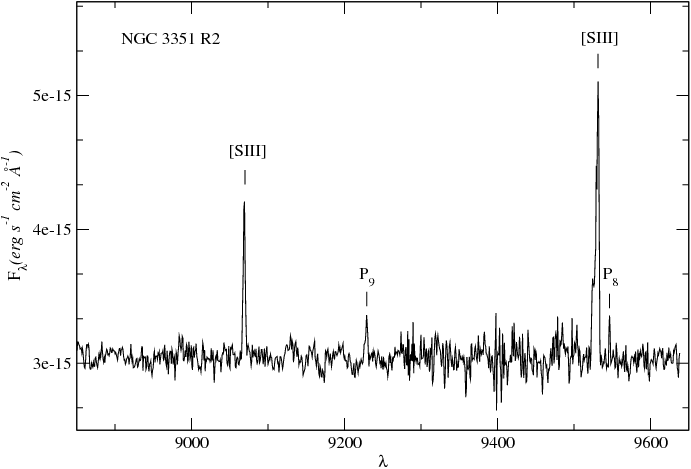}\\
\vspace{0.9cm}
\includegraphics[width=.45\textwidth,angle=0]{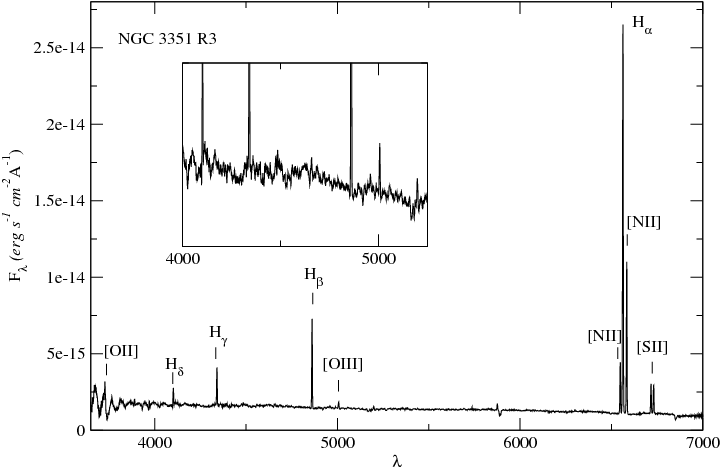}
\hspace{0.2cm}
\includegraphics[width=.45\textwidth,angle=0]{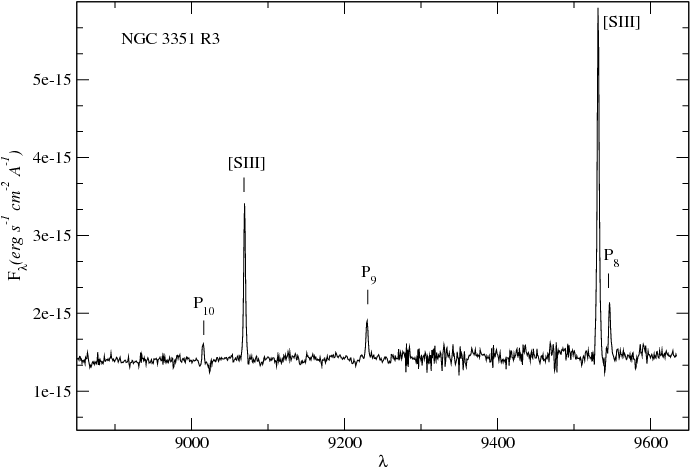}\\
\vspace{0.9cm}
\includegraphics[width=.45\textwidth,angle=0]{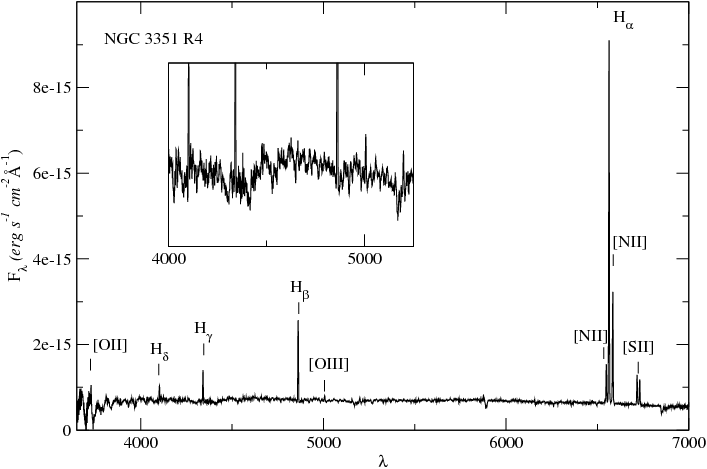}
\hspace{0.2cm}
\includegraphics[width=.45\textwidth,angle=0]{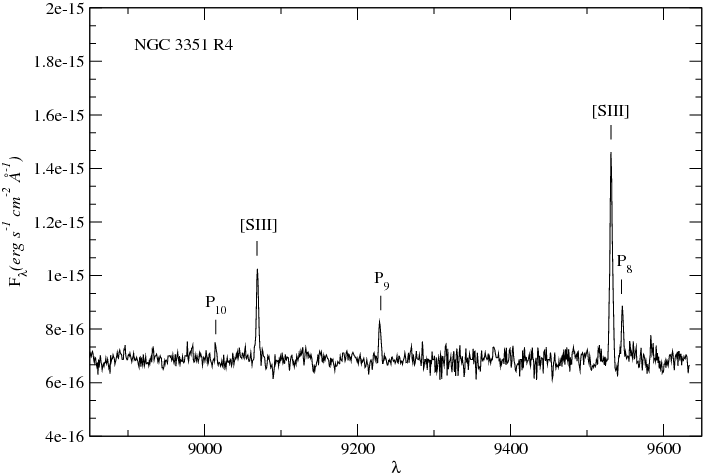}\\
\caption[Extracted spectra for the observed regions of NGC\,3351]{Extracted
  blue (left) and red (right) spectra for the observed regions of
  NGC\,3351. From top to bottom: R1, R2, R3 and R4.} 
\label{spectra2}
\end{figure*}

Figure  \ref{profiles} shows the spatial distribution of the H$\alpha$ flux
along the slit for the six different positions observed, one in 
NGC\,3504, two in NGC\,2903 and three in 
NGC\,3351. The regions that were extracted into 1-D spectra are delimited by
arrows. The spectra corresponding to each of the identified regions are shown
in Figures \ref{spectra1}, \ref{spectra2} and \ref{spectra3} for NGC\,2903, NGC\,3351 and
NGC\,3504, respectively.

%
%

\setcounter{figure}{6}
\begin{figure*}
\centering
\vspace{0.5cm}
\includegraphics[width=.45\textwidth,angle=0]{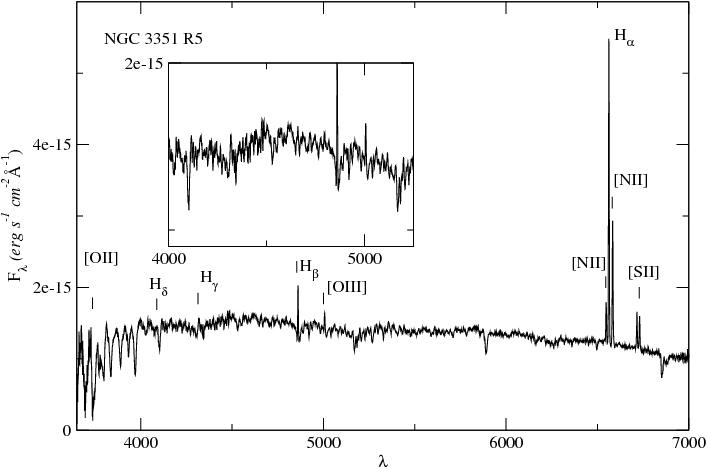}
\hspace{0.2cm}
\includegraphics[width=.45\textwidth,angle=0]{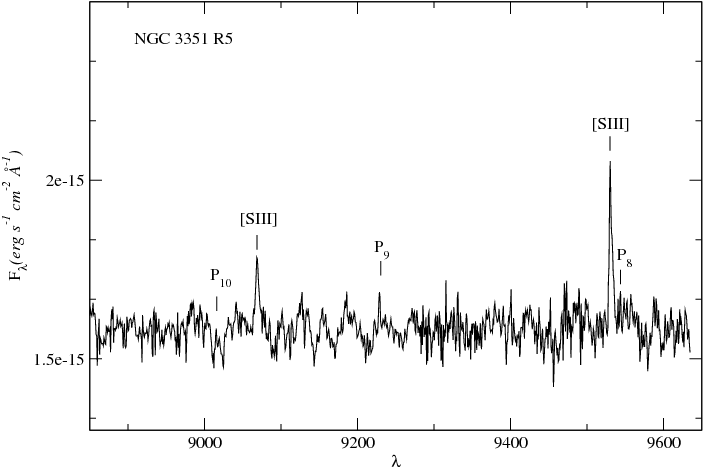}\\
\vspace{0.9cm}
\includegraphics[width=.45\textwidth,angle=0]{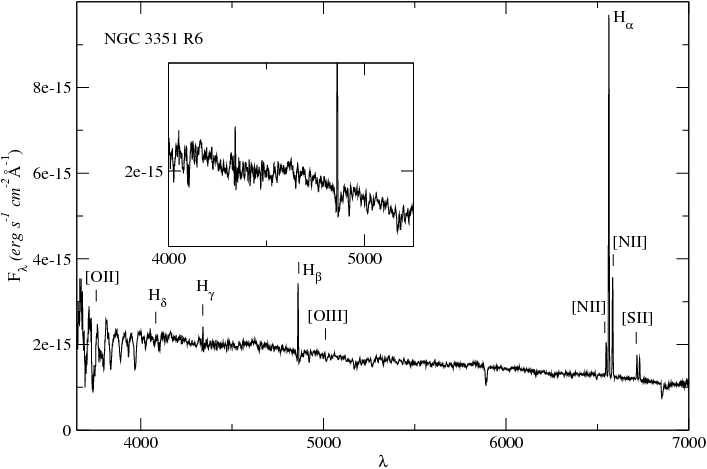}
\hspace{0.2cm}
\includegraphics[width=.45\textwidth,angle=0]{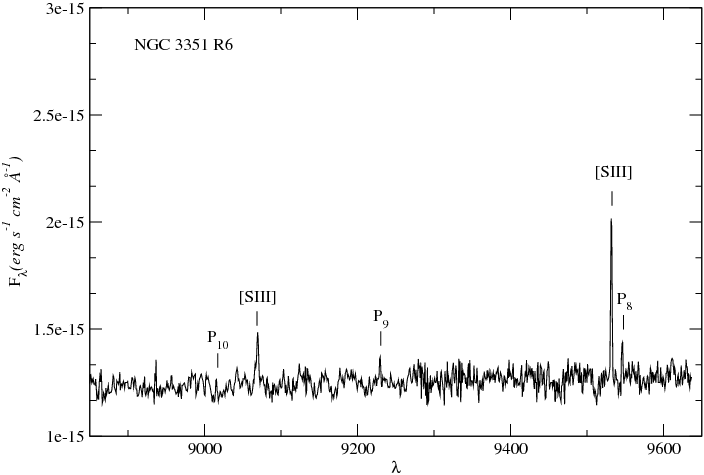}\\
\vspace{0.9cm}
\includegraphics[width=.45\textwidth,angle=0]{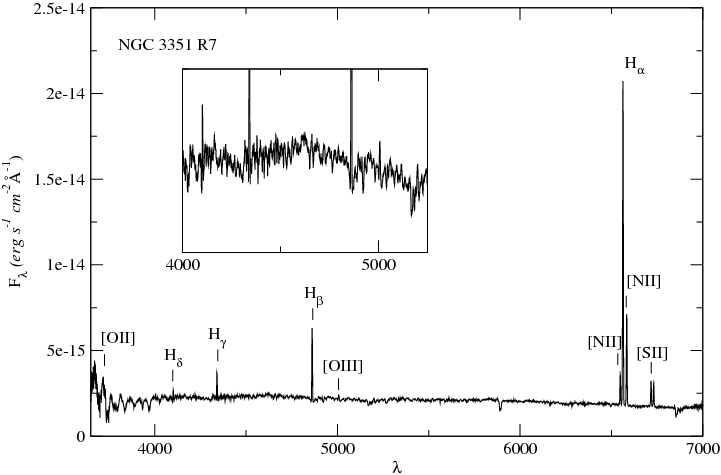}
\hspace{0.2cm}
\includegraphics[width=.45\textwidth,angle=0]{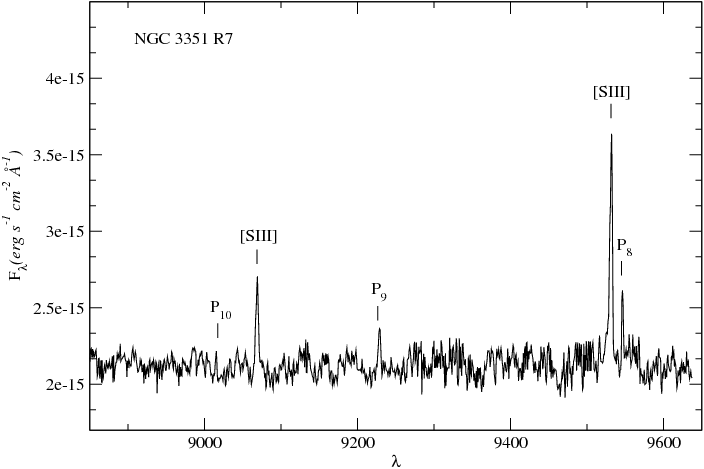}\\
\caption[({\it cont}) Extracted spectra for the observed regions of
  NGC\,3351]{({\it cont}) Extracted blue (left) and red (right) spectra for the
  observed regions of NGC\,3351. From top to bottom: R5, R6 and R7.} 
\end{figure*}

\begin{figure*}
\centering
\vspace{0.7cm}
\includegraphics[width=.45\textwidth,angle=0]{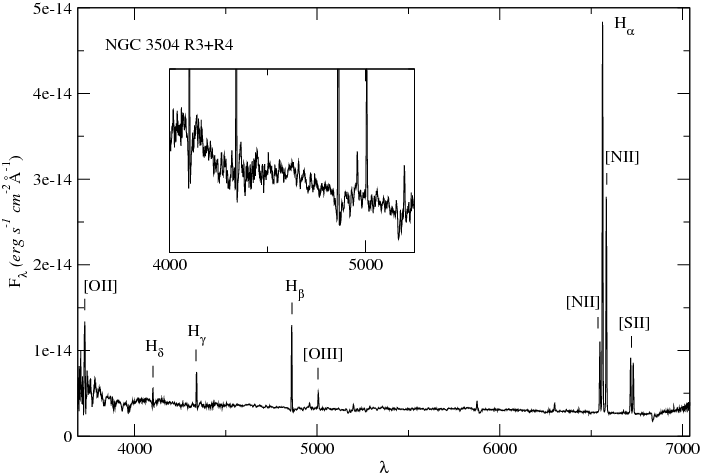}
\hspace{0.2cm}
\includegraphics[width=.45\textwidth,angle=0]{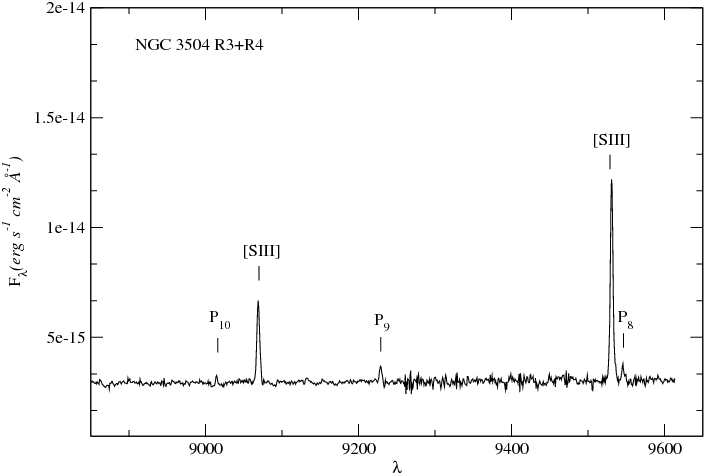}
\caption[Extracted spectra for the observed regions of NGC\,3504]{Extracted
  blue (left) and red (right) spectra for regions R3 + R4 of NGC\,3504.} 
\label{spectra3}
\end{figure*}

\subsection{Underlying population}
\begin{figure*}
\centering
\includegraphics[width=.98\textwidth,height=0.49\textwidth,angle=0]{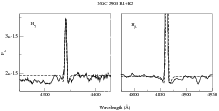}\\
\vspace{0.9cm}
\includegraphics[width=.98\textwidth,height=0.49\textwidth,angle=0]{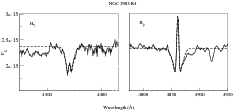}
\caption[Examples of the fitting procedure used to correct the Balmer emission
  line intensities]{Examples of the fitting procedure used to correct the
  Balmer emission line intensities for underlying absorption.}
\label{underlying}
\end{figure*}

The presence of underlying Balmer stellar absorptions is clearly evident in
the blue spectra of the observed regions (see Figures \ref{spectra1},
\ref{spectra2} and \ref{spectra3}) and complicates the measurements. A
two-component -- emission and absorption -- Gaussian fit was performed in
order to correct the Balmer emission lines for this effect. Examples of this
procedure can be seen in Figure \ref{underlying}. The equivalent widths (in
\AA mstrongs) of the Gaussian absorption components resulting from the fits
are given in Table \ref{balmer5} together with the ratio between the line flux
measured after subtraction of the absorption component and the line flux
measured without any correction and using a pseudo-continuum placed at the
bottom of the line. This factor provides a value for the final correction to
the measured fluxes in terms of each line flux. In regions R4 of NGC\,2903 and
R5 of NGC\,3351 the H$\delta$ line is seen only in absorption. In region R5 of
NGC\,3351 also H$\gamma$ is seen only in absorption. No fitting was performed
for these lines, hence no correction is listed for them in Table
\ref{balmer5}.  
No prominent absorption wings are observed in the He{\sc i} and Paschen lines
that would allow the fitting of an absorption component as it was done in
the case of the Balmer lines. These lines were measured with respect to a
local continuum placed at their base, which partially corrects by
underlying absorption.

%
%


%
%

\begin{table}
\begin{center}
 \caption{Equivalent widths of Balmer absorption lines for the observed CNSFRs}
 \begin{tabular}{@{}lcccccccc@{}}
\hline
\hline
Galaxy & Region & H$\delta$ & Corr. factor & H$\gamma$ & Corr. factor & H$\beta$ & Corr. factor\\
 & & (\AA ) & & (\AA ) & & (\AA )& \\
\hline
NGC\,2903        & R1+R2    & 2.3 & 1.15 & 3.1 & 1.17 & 2.9 & 1.09 \\
                 & R3       & 2.8 & 1.17 & 3.6 & 1.21 & 3.5 & 1.09 \\
                 & R4       & 4.5 &   --   & 3.8 & 1.08 & 4.8 & 1.24 \\
                 & R6       & 3.0 & 1.04 & 2.6 & 1.06 & 3.9 & 1.18 \\
NGC\,3351        & R1       & 2.1 & 1.16 & 1.6 & 1.14 & 3.0 & 1.08 \\
                 & R2       & 2.4 & 1.30 & 1.2 & 1.09 & 3.2 & 1.06 \\
                 & R3       & 2.5 & 1.16 & 1.2 & 1.07 & 1.2 & 1.03 \\
                 & R4       & 2.1 & 1.04 & 1.2 & 1.08 & 2.5 & 1.07 \\
                 & R5       & 3.7 & --     & 2.5 &  --    & 3.7 & 1.27 \\
                 & R6       & 2.4 & 1.32 & 2.0 & 1.04 & 3.5 & 1.13 \\
                 & R7       & 2.4 & 1.26 & 2.1 & 1.17 & 3.4 & 1.09 \\
NGC\,3504        & R3+R4    & 4.5 & 1.25 & 3.6 & 1.18 & 4.4 & 1.10 \\
\hline
\end{tabular}
\label{balmer5}
\end{center}
\end{table}


\subsection{Line intensity measurements}
\label{Line intensity}

Emission line fluxes were measured on the extracted spectra using the IRAF
SPLOT software package, by integrating the line intensity over a local fitted
continuum. The errors in the observed line fluxes have been calculated from
the expression $\sigma_{l}$ = $\sigma_{c}$N$^{1/2}$[1 +
  EW/(N$\Delta$)]$^{1/2}$, where $\sigma_{l}$ is the error in the line flux,
$\sigma_{c}$ represents the standard deviation in a box near the measured
emission line and stands for the error in the continuum placement, N is the
number of pixels used in the measurement of the line flux, EW is the line
equivalent width, and $\Delta$ is the wavelength dispersion in \AA ngstroms
per pixel. The first term represents the error in the line flux introduced by
the uncertainty in the placement of the continuum, while the second one scales
the signal-to-noise in the continuum to the line \cite{1994ApJ...437..239G}.

\begin{figure*}
\centering
\vspace{0.5cm}
\includegraphics[width=.85\textwidth,angle=0]{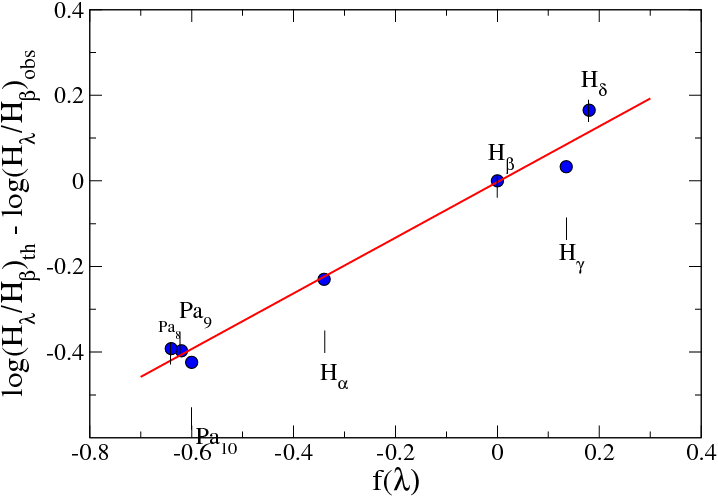}
\caption[Reddening determination for region R1+R2 in NGC\,2903]{Reddening
  determination for region R1+R2 in NGC\,2903. The tight relation found shows
  the goodness of the correction to the Balmer emission lines by the
  underlying absorption continuum.}
\label{reddening}
\end{figure*}

The Balmer emission lines were corrected for the underlying absorption as
explained above. The logarithmic extinction at H$\beta$, c(H$\beta$), was
calculated from the corrected Balmer line decrements assuming the Balmer line
theoretical values for  case B recombination \citetex{1971MNRAS.153..471B}
for a temperature of 6000 K, as expected for high metallicity regions, and an
average extinction law \cite{1972ApJ...172..593M}. 
An example of this procedure is shown in Figure \ref{reddening} for region
R1+R2 of NGC\,2903, where the logarithm of the quotient between observed and
theoretical Balmer decrements is represented against the logarithmic
extinction at the Balmer line wavelengths, f($\lambda$).  The tight relation
found for a baseline from the Paschen lines to H$\delta$ can be taken as
evidence of the reliability of our underlying absorption subtraction procedure
for this region. This 
is also the case for most regions, although in some cases the
H$\gamma$/H$\beta$ ratio lies above or below the reddening line, implying that
in that case the subtraction procedure is not that good. This is not 
surprising since the presence of the G-band and other metal features in the
left wing  of H$\gamma$, makes the fitting less reliable than for the other
Balmer lines. We have used all the available Balmer and Paschen-to-Balmer
ratios, although due to the uncertainties in the emission line intensities of
the higher order Balmer lines, which are difficult to estimate (in fact, in
some regions they are seen only in absorption), these ratios have been given a
lower weight in the fit. This almost amounts to deriving the values of
c(H$\beta$) from the H$\alpha$/H$\beta$ ratio and checking for consistency
against the values measured for the Paschen lines, which are much less
affected by underlying absorption due to the smaller contribution to the
continuum from main sequence AF stars. The errors in c(H$\beta$) have in fact
been derived from the measured errors in the H$\alpha$ and H$\beta$ lines,
since in any weighted average these two lines have by far the largest weight. 

\begin{figure*}
\centering
\includegraphics[width=.85\textwidth,angle=0]{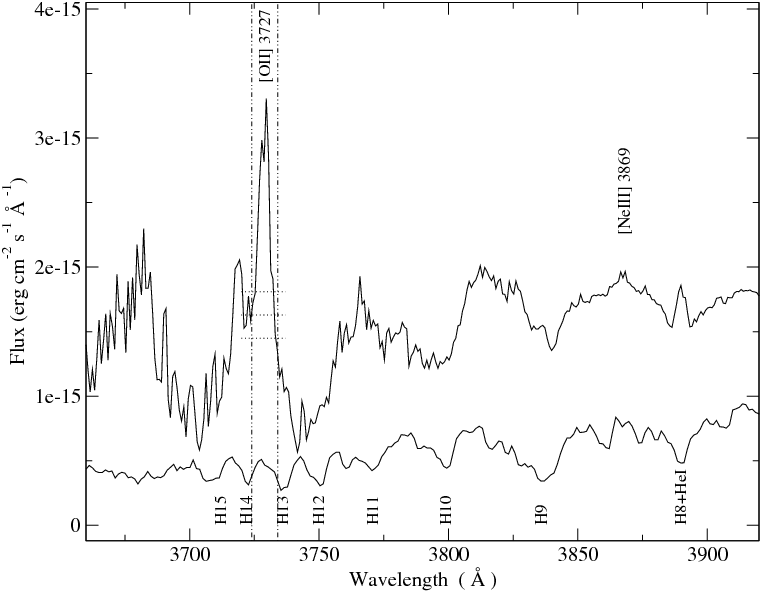}
\caption[Blue spectrum of region R1+R2 in NGC\,2903 showing the location of
  the [O{\sc ii}{\textrm ]}\,$\lambda\lambda$\,3727,29\,\AA\ lines]{Blue
  spectrum of region R1+R2 in NGC\,2903 showing the location of the [O{\sc
  ii}]\,$\lambda\lambda$\,3727,29\,\AA\ lines. The spectrum below corresponds
  to a metal rich globular cluster in M~31 which can be compared to the
  underlying stellar population in the region. The way in which the [O{\sc
  ii}] line has been measured by placing a local pseudo-continuum at its base
  is shown. The different continuum placements used for the computation of
  the errors are shown by horizontal lines.}
\label{medidas1}
\end{figure*}

The blue region of the spectrum near the Balmer discontinuity is dominated by
absorption lines which cause a depression of the continuum and difficult the
measurement of the [O{\sc ii}] lines at $\lambda$ 3727 \AA. None of them
however is 
at the actual wavelength of the [O{\sc ii}] line. This can be seen in Figure
\ref{medidas1} where we show the spectrum of region R1+R2 in NGC\,2903 compared
to that corresponding to a globular cluster of M~31, 337-068, of relatively
high metallicity (\citeplain{2000AJ....119..727B}; Mike Beasley, private
communication). 
We have measured the line using a local continuum at its base as shown in the
figure.  The different continuum placements used for computing  the error are
shown by horizontal dotted lines.

%
%

Once the reddening constant was found, the measured line intensities relative
to H$\beta$ were corrected for interstellar reddening according to
the assumed reddening law. The errors in the reddening corrected line
intensities have been derived by means of error propagation theory. Measured
and reddening corrected emission line fluxes, together with their
corresponding errors,  are given in Tables \ref{intensities1},
\ref{intensities2} and \ref{intensities3} for the observed CNSFRs in NGC\,2903,
NGC\,3351 and NGC\,3504 respectively.  Balmer emission lines are given after
correction for
underlying absorption. Also given in the tables are the assumed reddening law,
the  H$\beta$ intensity underlying  absorption and extinction corrected,  the
H$\beta$ equivalent width, also corrected for absorption, and the reddening
constant.

%
%

%
%

\newpage
\landscape

\begin{table}
\centering

\caption{Reddening corrected emission line intensities for the CNSFRs in NGC\,2903}
\begin{tabular}{@{}l c c c c c c c c c c@{}}
 \hline
\hline
Region & & &\multicolumn{2}{c}{R1+R2} & \multicolumn{2}{c}{R3} & \multicolumn{2}{c}{R4} & \multicolumn{2}{c}{R6}\\
            & &  &                &                  &            &                  &                  &                &              &    \\
 $\lambda$ (\AA ) & Line   & f$_\lambda$& F$_\lambda$& I$_\lambda$ &F$_\lambda$& I$_\lambda$&F$_\lambda$& I$_\lambda$&F$_\lambda$& I$_\lambda$ \\
            & &  &                &                  &            &                  &                  &                &              &     \\
 \hline
 
3727 & [O{\sc ii}]      & 0.271 & 267$\pm$27  & 428$\pm$44 &  333 $\pm$ 33     &  430 $\pm$ 45  &  421 $\pm$ 42 & 841 $\pm$ 88 &   198 $\pm$ 20 &  310  $\pm$ 31 \\ 
4102 & H$_\delta$ & 0.188 &  179$\pm$11 &  248$\pm$16 &  235$\pm$16 &  280$\pm$19 &     --      &     --      &  145$\pm$9  &  197$\pm$13  \\
4340 & H$_\gamma$ & 0.142 &  432$\pm$13 &  552$\pm$17 &  553$\pm$18 &  632$\pm$20 &  164$\pm$17 &  235$\pm$24 &  298$\pm$10 &  377$\pm$13  \\
4686 & He{\sc ii}       & 0.045 &    5$\pm$2  &    6$\pm$3  &     --      &     --      &     --      &     --      &  11$\pm$4 &  12$\pm$4  \\
4861 & H$_\beta$  & 0.000 & 1000$\pm$16 & 1000$\pm$16 & 1000$\pm$21 & 1000$\pm$21 & 1000$\pm$27 & 1000$\pm$28 & 1000$\pm$13 & 1000$\pm$13  \\ 
4959 & [O{\sc iii}]     &-0.024 &   43$\pm$6  &   41$\pm$5  &   78$\pm$5  &   76$\pm$5  &  132$\pm$84 &  125$\pm$79 &   41$\pm$3  &   40$\pm$3   \\ 
5007 & [O{\sc iii}]     &-0.035 &  131$\pm$5  &  123$\pm$5  &  229$\pm$13 &  222$\pm$12 &  391$\pm$25 &  357$\pm$23 &  122$\pm$9  &  115$\pm$8   \\
5199 & [N{\sc i}]       &-0.078 &   56$\pm$8  &   49$\pm$7  &   65$\pm$26 &   60$\pm$24 &  139$\pm$45 &  114$\pm$37 &   60$\pm$15 &   53$\pm$13  \\
5876 & He{\sc i}        &-0.209 &   79$\pm$14 &   55$\pm$10 &   97$\pm$26 &   80$\pm$21 &     --      &    --       &   97$\pm$17 &   69$\pm$12  \\
6300 & [O{\sc i}]       &-0.276 &   39$\pm$6  &   24$\pm$3  &   54$\pm$13 &   41$\pm$10 &   56$\pm$14 &   28$\pm$7  &   40$\pm$13 &   25$\pm$8   \\
6312 & [S{\sc iii}]     &-0.278 &    7$\pm$1  &    4$\pm$1  &     --      &     --      &     --      &    --       &     --      &    --        \\
6548 & [N{\sc ii}]      &-0.311 &  675$\pm$11 &  394$\pm$12 &  536$\pm$16 &  400$\pm$17 &  866$\pm$62 &  391$\pm$31 &  529$\pm$17 &  318$\pm$12  \\
6563 & H$_\alpha$ &-0.313 & 4897$\pm$45 & 2850$\pm$26 & 3826$\pm$43 & 2850$\pm$32 & 6341$\pm$50 & 2850$\pm$22 & 4759$\pm$38 & 2850$\pm$23  \\
6584 & [N{\sc ii}]      &-0.316 & 2082$\pm$19 & 1206$\pm$33 & 1679$\pm$47 & 1247$\pm$54 & 2666$\pm$66 & 1189$\pm$52 & 1707$\pm$26 & 1017$\pm$27  \\
6678 & He{\sc i}        &-0.329 &    8$\pm$3  &    4$\pm$2  &     --      &     --      &     --      &     --      &     --      &     --       \\
6717 & [S{\sc ii}]      &-0.334 &  498$\pm$14 &  280$\pm$11 &  472$\pm$28 &  345$\pm$24 &  811$\pm$34 &  346$\pm$20 &  528$\pm$15 &  306$\pm$11  \\
6731 & [S{\sc ii}]      &-0.336 &  432$\pm$16 &  242$\pm$11 &  391$\pm$24 &  285$\pm$20 &  699$\pm$31 &  297$\pm$18 &  450$\pm$16 &  260$\pm$11  \\
8863 & P11        &-0.546 &   36$\pm$8  &   14$\pm$3  &     --      &     --      &     --      &     --      &     --      &     --       \\
9016 & P10        &-0.557 &   46$\pm$7  &   18$\pm$3  &     --      &     --      &     --      &     --      &   53$\pm$14 &   21$\pm$6   \\
9069 & [S{\sc iii}]     &-0.561 &  157$\pm$10 &   60$\pm$5  &  103$\pm$9  &   61$\pm$6  &  220$\pm$33 &   53$\pm$9  &  137$\pm$11 &   55$\pm$5   \\
9230 & P9         &-0.572 &   61$\pm$8  &   23$\pm$3  &     --      &     --      &     --      &     --      &   87$\pm$16 &   34$\pm$6   \\
9532 & [S{\sc iii}]     &-0.592 &  393$\pm$16 &  141$\pm$9  &  337$\pm$16 &  193$\pm$15 &  508$\pm$53 &   112$\pm$14 &  450$\pm$29 &  171$\pm$13  \\
9547 & P8         &-0.593 &  128$\pm$14 &   46$\pm$5  &  128$\pm$17 &   73$\pm$11 &  305$\pm$58 &   67$\pm$14 &  192$\pm$20 &   73$\pm$8   \\
          
\hline

c(H$\beta$)  & & & 0.75$\pm$0.04 & & 0.41$\pm$0.05 & & 1.11$\pm$0.05 & & 0.71$\pm$0.03 &  \\
I(H$\beta$)$^a$        & & & & 13.7$\pm$0.22 & & 1.80$\pm$0.04 & & 7.76$\pm$0.21 & & 6.84$\pm$0.09   \\
\multicolumn{2}{@{}l}{EW(H$\beta$)(\AA )} & & & 12.7$\pm$0.2 & & 8.2$\pm$0.2 & & 2.6$\pm$0.1& & 9.6$\pm$0.2  \\

\hline
\multicolumn{11}{@{}l}{$^a$in units of 10$^{-14}$ erg s$^{-1}$ cm$^{-2}$}
\end{tabular}
\label{intensities1}
\end{table}
\endlandscape
\newpage

%
%

\landscape

\begin{table}
\centering

\caption{Reddening corrected emission line intensities for the CNSFRs in NGC\,3351}
\begin{tabular}{@{}l c c c c c c c c c c@{}}
 \hline
\hline
Region & & &\multicolumn{2}{c}{R1} & \multicolumn{2}{c}{R2} & \multicolumn{2}{c}{R3} & \multicolumn{2}{c}{R4}\\
            & &  &                &                  &            &                  &                  &                &              &    \\
 $\lambda$ (\AA ) & Line   & f$_\lambda$& F$_\lambda$& I$_\lambda$ &F$_\lambda$& I$_\lambda$&F$_\lambda$& I$_\lambda$&F$_\lambda$& I$_\lambda$ \\
            & &  &                &                  &            &                  &                  &                &              &     \\
 \hline
 
3727 & [O{\sc ii}]      & 0.271   & 282 $\pm$ 28 & 455 $\pm$ 46 & 172 $\pm$ 17 & 230 $\pm$ 23 & 226 $\pm$ 23 & 336 $\pm$ 34 & 188 $\pm$ 19 & 271 $\pm$ 28 \\ 
4102 & H$_\delta$ & 0.188 &  164$\pm$7  &  229$\pm$9  &  270$\pm$7  &  330$\pm$9  &  255$\pm$6  &  335$\pm$8  &  218$\pm$13 &   281$\pm$16  \\
4340 & H$_\gamma$ & 0.142 &  372$\pm$8  &  478$\pm$10 &  388$\pm$8  &  452$\pm$9  &  434$\pm$7  &  534$\pm$9  &  386$\pm$14 &   467$\pm$17  \\
4686 & He{\sc ii}       & 0.045 &     --      &     --      &	    --      &	  --      &     --      &     --      &     --      &      --       \\
4861 & H$_\beta$  & 0.000 & 1000$\pm$10 & 1000$\pm$10 & 1000$\pm$10 & 1000$\pm$10 & 1000$\pm$9  & 1000$\pm$9  & 1000$\pm$18 &  1000$\pm$18  \\ 
4959 & [O{\sc iii}]     &-0.024 &   36$\pm$2  &   34$\pm$2  &   47$\pm$4  &   46$\pm$4  &   27$\pm$2  &   26$\pm$2  &   28$\pm$3  &    27$\pm$3   \\ 
5007 & [O{\sc iii}]     &-0.035 &  106$\pm$6  &  100$\pm$6  &  140$\pm$6  &  135$\pm$6  &   84$\pm$4  &   79$\pm$4  &   83$\pm$8 &    79$\pm$8  \\
5199 & [N{\sc i}]       &-0.078 &   49$\pm$13 &   42$\pm$11 &   61$\pm$16 &   56$\pm$14 &   38$\pm$8  &   34$\pm$7  &   61$\pm$17 &    55$\pm$15  \\
5876 & He{\sc i}        &-0.209 &   65$\pm$12 &   45$\pm$8  &   63$\pm$27 &   51$\pm$21 &   69$\pm$10 &   51$\pm$7  &   55$\pm$10 &    41$\pm$7   \\
6300 & [O{\sc i}]       &-0.276 &   24$\pm$6  &   15$\pm$4  &   31$\pm$10 &   23$\pm$7  &   19$\pm$4  &   13$\pm$3  &   19$\pm$5  &    13$\pm$3   \\
6312 & [S{\sc iii}]     &-0.278 &     --      &     --      &	    --      &	  --      &     --      &     --      &     --      &      --       \\
6548 & [N{\sc ii}]      &-0.311 &  565$\pm$17 &  326$\pm$12 &  476$\pm$16 &  341$\pm$14 &  584$\pm$13 &  372$\pm$11 &  457$\pm$24 &   301$\pm$18  \\
6563 & H$_\alpha$ &-0.313 & 4952$\pm$48 & 2850$\pm$28 & 3981$\pm$42 & 2850$\pm$30 & 4494$\pm$37 & 2850$\pm$24 & 4342$\pm$45 &  2850$\pm$30  \\
6584 & [N{\sc ii}]      &-0.316 & 1739$\pm$30 &  996$\pm$27 & 1489$\pm$21 & 1063$\pm$27 & 1719$\pm$19 & 1086$\pm$23 & 1406$\pm$34 &   919$\pm$35  \\
6678 & He{\sc i}        &-0.329 &   16$\pm$4  &    9$\pm$2  &    5$\pm$2  &    3$\pm$2  &   15$\pm$4  &    9$\pm$3  &     --      &      --       \\
6717 & [S{\sc ii}]      &-0.334 &  385$\pm$9  &  214$\pm$7  &  353$\pm$13 &  247$\pm$11 &  347$\pm$6  &  213$\pm$5  &  366$\pm$18 &   233$\pm$14  \\
6731 & [S{\sc ii}]      &-0.336 &  353$\pm$9  &  195$\pm$7  &   337$\pm$12 &   235$\pm$10  &  330$\pm$8  &  203$\pm$6  &  322$\pm$15 &   205$\pm$11  \\
8863 & P11        &-0.546 &   36$\pm$5  &   14$\pm$2  &	    --      &	  --      &     --      &     --      &     --      &      --       \\
9016 & P10        &-0.557 &   53$\pm$5  &   20$\pm$2  &   18$\pm$1  &   10$\pm$1  &   40$\pm$7  &   18$\pm$3  &   24$\pm$4  &    12$\pm$2   \\
9069 & [S{\sc iii}]     &-0.561 &  264$\pm$15 &   98$\pm$7  &  118$\pm$7  &   65$\pm$5  &  233$\pm$13 &  103$\pm$6  &  139$\pm$13 &    66$\pm$7   \\
9230 & P9         &-0.572 &   92$\pm$6  &   33$\pm$3  &   43$\pm$5  &   23$\pm$3  &   66$\pm$7  &   29$\pm$3  &   66$\pm$51 &    31$\pm$24  \\
9532 & [S{\sc iii}]     &-0.592 &  697$\pm$30 &  245$\pm$14 &  301$\pm$16 &  160$\pm$11 &  581$\pm$30 &  245$\pm$15 &  346$\pm$21 &   156$\pm$13  \\
9547 & P8         &-0.593 &   89$\pm$5  &   31$\pm$2  &  247$\pm$2  &  131$\pm$5  &  103$\pm$1  &   43$\pm$1  &   92$\pm$7  &    42$\pm$4   \\
          
\hline

c(H$\beta$)   & & & 0.77$\pm$0.03 & & 0.46$\pm$0.03 & & 0.63$\pm$0.02 & & 0.58$\pm$0.04 &  \\
I(H$\beta$)$^a$    & & & & 9.63$\pm$0.10 & & 8.87$\pm$0.09 & & 11.4$\pm$0.11 & & 3.41$\pm$0.06   \\
\multicolumn{2}{@{}l}{EW(H$\beta$)(\AA )} & & & 14.6$\pm$0.2 & & 9.5$\pm$0.1 & & 18.1$\pm$0.3 & & 12.3$\pm$0.3  \\

\hline
\multicolumn{11}{@{}l}{$^a$in units of 10$^{-14}$ erg s$^{-1}$ cm$^{-2}$}
\end{tabular}
\label{intensities2}
\end{table}
\endlandscape
\newpage

\setcounter{table}{5}

%
%

\landscape

\begin{table}
\centering

\caption{({\it cont}) Reddening corrected emission line intensities for the CNSFRs in NGC\,3351}
\begin{tabular}{@{}l c c c c c c c c@{}}
\hline
 \hline
Region & & &\multicolumn{2}{c}{R5} & \multicolumn{2}{c}{R6} & \multicolumn{2}{c}{R7} \\
            & &  &                &                  &            &                  &                  &     \\
 $\lambda$ (\AA ) & Line   & f$_\lambda$& F$_\lambda$& I$_\lambda$ &F$_\lambda$& I$_\lambda$&F$_\lambda$& I$_\lambda$ \\
            & &  &                &                  &            &                  &                  &               \\
 \hline
 
3727 & [O{\sc ii}]      & 0.271 & 348 $\pm$ 35 & 521 $\pm$ 61 & 350:  & 495:  & 196:  & 284:   \\ 
4102 & H$_\delta$ & 0.188 &     --       &     --       &   85$\pm$6  &  109$\pm$7  &  193$\pm$9  &  249$\pm$11    \\
4340 & H$_\gamma$ & 0.142 &     --       &     --       &  262$\pm$11 &  315$\pm$13 &  400$\pm$10 &  486$\pm$12    \\
4686 & He{\sc ii}       & 0.045 &     --       &     --       &     --      &     --      &   12$\pm$4  &   12$\pm$5     \\
4861 & H$_\beta$  & 0.000 & 1000$\pm$37  & 1000$\pm$37  & 1000$\pm$20 & 1000$\pm$20 & 1000$\pm$13 & 1000$\pm$13    \\ 
4959 & [O{\sc iii}]     &-0.024 &  138$\pm$13  &  134$\pm$12  &   32:  &   31:  &   34$\pm$2 &   33$\pm$2    \\ 
5007 & [O{\sc iii}]     &-0.035 &  409$\pm$37  &  388$\pm$35  &   96: &   91: &  100$\pm$7  &   96$\pm$7     \\
5199 & [N{\sc i}]       &-0.078 &   78$\pm$33  &   70$\pm$29  &   61$\pm$19 &   55$\pm$17 &   61$\pm$13 &   55$\pm$12    \\
5876 & He{\sc i}        &-0.209 &     --       &     --       &     --      &     --      &   36$\pm$8  &   27$\pm$6     \\
6300 & [O{\sc i}]       &-0.276 &   47$\pm$14  &   31$\pm$10  &     --      &     --      &   14$\pm$5  &    9$\pm$4     \\
6312 & [S{\sc iii}]     &-0.278 &     --       &     --       &     --      &     --      &     --      &     --         \\
6548 & [N{\sc ii}]      &-0.311 &  622$\pm$43  &  391$\pm$39  &  420$\pm$28 &  282$\pm$20 &  412$\pm$17 &  269$\pm$13    \\
6563 & H$_\alpha$ &-0.313 & 4543$\pm$52  & 2850$\pm$95  & 4259$\pm$44 & 2850$\pm$30 & 4382$\pm$34 & 2850$\pm$22    \\
6584 & [N{\sc ii}]      &-0.316 & 1928$\pm$109 & 1204$\pm$109 & 1272$\pm$55 &  848$\pm$45 & 1289$\pm$29 &  835$\pm$26    \\
6678 & He{\sc i}        &-0.329 &     --       &     --       &     --      &     --      &    7$\pm$3  &    4$\pm$2     \\
6717 & [S{\sc ii}]      &-0.334 &  548$\pm$37  &  333$\pm$34  &  275$\pm$18 &  180$\pm$13 &  345$\pm$12 &  218$\pm$9     \\
6731 & [S{\sc ii}]      &-0.336 &  487$\pm$33  &  295$\pm$30  &  248$\pm$17 &  161$\pm$12 &  325$\pm$11 &  205$\pm$8     \\
8863 & P11        &-0.546 &     --       &     --       &     --      &     --      &     --      &     --         \\
9016 & P10        &-0.557 &     --       &     --       &     --      &     --      &   22$\pm$3  &   10$\pm$1     \\
9069 & [S{\sc iii}]     &-0.561 &  191$\pm$20  &   83$\pm$13  &   89$\pm$13 &   43$\pm$7  &  112$\pm$13 &   52$\pm$6     \\
9230 & P9         &-0.572 &   88$\pm$29  &   38$\pm$13  &   36$\pm$11 &   17$\pm$5  &   66$\pm$7  &   30$\pm$4     \\
9532 & [S{\sc iii}]     &-0.592 &  415$\pm$39  &  172$\pm$28  &  202$\pm$12 &   95$\pm$8  &  307$\pm$29 &  136$\pm$14    \\
9547 & P8         &-0.593 &     --       &     --       &   74$\pm$11 &   35$\pm$6  &   87$\pm$1  &   39$\pm$2     \\
          
\hline

c(H$\beta$)   & & & 0.65$\pm$0.10 & & 0.56$\pm$0.043 & & 0.60$\pm$0.03 &   \\
I(H$\beta$)$^a$     & & & & 2.04$\pm$0.07 & & 3.26$\pm$0.07 & & 8.05$\pm$0.10   \\
\multicolumn{2}{@{}l}{EW(H$\beta$)(\AA )} & & &  3.1$\pm$0.1 & & 4.9$\pm$0.1 & &  9.0$\pm$0.2  \\

\hline
\multicolumn{9}{@{}l}{$^a$in units of 10$^{-14}$ erg s$^{-1}$ cm$^{-2}$}
\end{tabular}
\end{table}
\endlandscape

%
%

\begin{table}[!h]
\centering

\caption{Reddening corrected emission line intensities for the CNSFRs in NGC\,3504}
\begin{tabular}{@{}l c c c c@{}}
\hline
\hline
Region & & &\multicolumn{2}{c}{R3+R4} \\
            & & \\
 $\lambda$ (\AA ) & Line   & f$_\lambda$& F$_\lambda$& I$_\lambda$ \\
            & &  \\
 \hline

3727 & [O{\sc ii}]      & 0.271 & 953 $\pm$ 95 & 1395 $\pm$ 141  \\ 
4102 & H$_\delta$ & 0.188 &  250$\pm$5  &  326$\pm$7  \\
4340 & H$_\gamma$ & 0.142 &   460$\pm$6  &   561$\pm$8  \\
4686 & He{\sc ii}       & 0.045 &     --      &     --      \\
4861 & H$_\beta$  & 0.000 & 1000$\pm$8  & 1000$\pm$8  \\ 
4959 & [O{\sc iii}]     &-0.024 &   63$\pm$4  &   61$\pm$4  \\ 
5007 & [O{\sc iii}]     &-0.035 &  192$\pm$8  &  183$\pm$7  \\
5199 & [N{\sc i}]       &-0.078 &   69$\pm$15 &   62$\pm$13 \\
5876 & He{\sc i}        &-0.209 &   87$\pm$11 &   65$\pm$8  \\
6300 & [O{\sc i}]       &-0.276 &   80$\pm$7  &   54$\pm$5  \\
6312 & [S{\sc iii}]     &-0.278 &     --      &     --      \\
6548 & [N{\sc ii}]      &-0.311 &  818$\pm$17 &  529$\pm$13 \\
6563 & H$_\alpha$ &-0.313 & 4423$\pm$29 & 2850$\pm$19 \\
6584 & [N{\sc ii}]      &-0.316 & 2466$\pm$29 & 1583$\pm$30 \\
6678 & He{\sc i}        &-0.329 &   17$\pm$4  &   11$\pm$3  \\
6717 & [S{\sc ii}]      &-0.334 &  652$\pm$13 &  408$\pm$10 \\
6731 & [S{\sc ii}]      &-0.336 &  599$\pm$12 &  374$\pm$9  \\
8863 & P11        &-0.546 &     --      &     --      \\
9016 & P10        &-0.557 &   34$\pm$7  &   15$\pm$3  \\
9069 & [S{\sc iii}]     &-0.561 &  327$\pm$13 &  149$\pm$7  \\
9230 & P9         &-0.572 &   95$\pm$8  &   43$\pm$4  \\
9532 & [S{\sc iii}]     &-0.592 &  802$\pm$25 &  350$\pm$15 \\
9547 & P8         &-0.593 &     --      &     --      \\
          
\hline

c(H$\beta$)   & & & 0.61$\pm$0.02 & \\
I(H$\beta$)$^a$     & & & & 20.8$\pm$0.17 \\
\multicolumn{2}{@{}l}{EW(H$\beta$)(\AA )} & & & 15.2$\pm$0.2 \\

\hline
\multicolumn{5}{@{}l}{$^a$in units of 10$^{-14}$ erg s$^{-1}$ cm$^{-2}$}
\end{tabular}
\label{intensities3}
\end{table}

\newpage

\section{Chemical abundances}
\label{che-abun}

Electron densities for each observed region have been derived from the  [S{\sc
ii}]\,$\lambda\lambda$\,6717, 6731\,\AA\ line ratio, following standard
methods (e.g.\ \citeplain{1989agna.book.....O}). They were found to be, in all
cases, $\le$\,600\,cm$^{-3}$, higher than those usually derived in disc \HII\
regions, but still below the critical value for collisional de-excitation. 

The low excitation of the regions, as evidenced by the weakness of the [O{\sc
    iii}]\,$\lambda$\,5007\,\AA\ line (see left panels of Figures
\ref{spectra1}, \ref{spectra2} and \ref{spectra3}), precludes the detection
and measurement of the auroral [O{\sc iii}]\,$\lambda$\,4363\,\AA\ necessary
for the derivation of the electron temperature. It is therefore impossible to
obtain a direct determination of the oxygen abundances. Empirical calibrations
have to be used instead.  

Different calibrators for strong emission lines have been proposed in the
literature for different kinds of objects involving different chemical
elements, among others, the oxygen abundance parameter R$_{23}$ $ \equiv $
O$_{23}$ \cite{1979MNRAS.189...95P}, the nitrogen N2 parameter
\cite{2002MNRAS.330...69D} and the sulphur abundance parameter  S$_{23}$
\cite{2000MNRAS.312..130D}. Also, calibrators involving a combination of
emission lines of two elements have been proposed, such as [O{\sc iii}]/[N{\sc
    ii}] \cite{1979A&A....78..200A}, [Ar{\sc iii}]$\lambda$ 7135\,\AA/[O{\sc
    iii}] $\lambda$ 5007\,\AA\ and [S{\sc iii}]$\lambda$ 9069\,\AA /[O{\sc iii}]
$\lambda$ 5007\,\AA\ \cite{2006A&A...454L.127S}. Our CNSFRs, show very weak
lines of [O{\sc iii}] which are measured with large errors. This is taken as
evidence for high metallicity in these regions. This, in turn, may imply
values of N/O larger than solar, due to the chemical evolution of the regions
themselves and the increasing production of secondary nitrogen. Hence, we have
considered the use of the N2 parameter unreliable for this kind of objects. On
the other hand, the [S{\sc iii}] lines are seen to be strong as compared to
the [O{\sc iii}] lines (see right panels of Figures \ref{spectra1},
\ref{spectra2} and \ref{spectra3}). Recently, the combination of both the
oxygen abundance, O$_{23}$, and sulphur abundance, S$_{23}$, parameters has
been claimed to be a good metallicity indicator for high metallicity \HII\
regions \cite{2000MNRAS.312..130D,2005MNRAS.361.1063P}. From now onwards we
will call this parameter SO$_{23}$, and is defined as: 
\[
SO_{23}\,=\,\frac{S_{23}}{O_{23}}\,=\,\frac{I([S{\textrm{\sc ii}}]\lambda 6717,31)+I([S{\textrm{\sc iii}}]\lambda 9069,9532)}
{I([O{\textrm{\sc ii}}]\lambda 3727,29)+I([O{\textrm{\sc iii}}]\lambda 4959,5007)} 
\]

\begin{figure*}
\vspace*{4.5cm}
\centering
\includegraphics[width=.90\textwidth,angle=0]{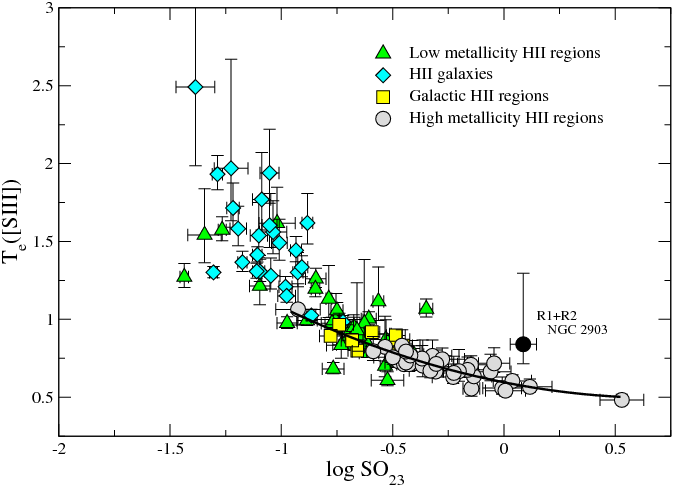}
\caption[Empirical calibration of the [S{\sc iii}{\textrm ]} electron
  temperature as a function of the abundance parameter SO$_{23}$]{Empirical
  calibration of the [S{\sc iii}] electron temperature as a function of the
  abundance parameter SO$_{23}$ defined in the text. The solid line represents
  a quadratic fit to the high metallicity \HII\ region data. References for the
  data are given in the text.} 
\label{te[SIII]cal}
\end{figure*}

This parameter is similar to the S$_3$O$_3$ proposed by
\citetex{2006A&A...454L.127S} but is, at first order, independent of
geometrical (ionization parameter) effects. The amount of available data on
sulphur emission lines is increasingly growing, especially in the high
metallicity regime. This makes possible for the first time to calibrate the
electron temperature of [S{\sc iii}] in terms of the SO$_{23}$ parameter. To perform
this calibration we have compiled all the data so far at hand with sulphur
emission line data both for the auroral and nebular lines at $\lambda$ 6312
\AA\ and $\lambda\lambda$ 9069,9532 \AA\ respectively. The sample comprises
data on galactic \cite{tesisjorge} and extragalactic
\cite{1987MNRAS.226...19D,1988MNRAS.235..633V,1993MNRAS.260..177P,1997ApJ...489...63G,1994ApJ...437..239G,1995ApJ...439..604G,2000MNRAS.318..462D,2002MNRAS.329..315C,2003ApJ...591..801K,2004ApJ...615..228B,2005A&A...441..981B} \HII\ regions and \HII\ galaxies
\cite{1993ApJ...411..655S,1994ApJ...431..172S,2003MNRAS.346..105P,2006MNRAS.372..293H,2008MNRAS.383..209H}.
The data on extragalactic \HII\ regions has been further split into low and
high metallicity \HII\ regions according to the criterion of
\citetex{2000MNRAS.312..130D}, {\it i.e.} log$O_{23}\,\leq$\,0.47 and
-0.5\,$\leq\,log\,S_{23}\,\leq$\,0.28 implies oversolar abundances. For all
the regions the 
[S{\sc iii}] electron temperature has been derived from the ratio between the
auroral and the nebular sulphur lines, using a five-level atom program
\cite{1995PASP..107..896S} and the collisional strengths from
\citetex{1999ApJ...526..544T}, through the task TEMDEN as implemented in the
IRAF package. The calibration is shown in Figure \ref{te[SIII]cal} together
with the quadratic fit to the high metallicity \HII\ region data: 
\[
t_e([S{\textrm{\sc iii}}])\,=\,0.596 - 0.283 log SO_{23} + 0.199 (log SO_{23})^2
\]

\begin{figure*}
\centering
\vspace*{1.5cm}
\includegraphics[width=.80\textwidth,angle=0]{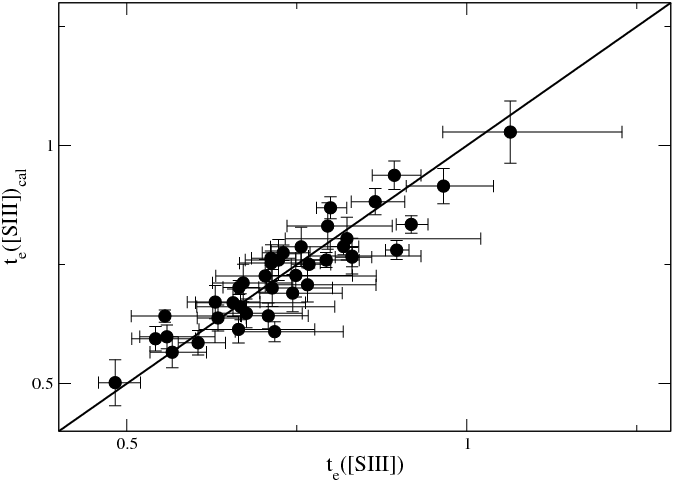}\\
\vspace{1.5cm}
\includegraphics[width=.80\textwidth,angle=0]{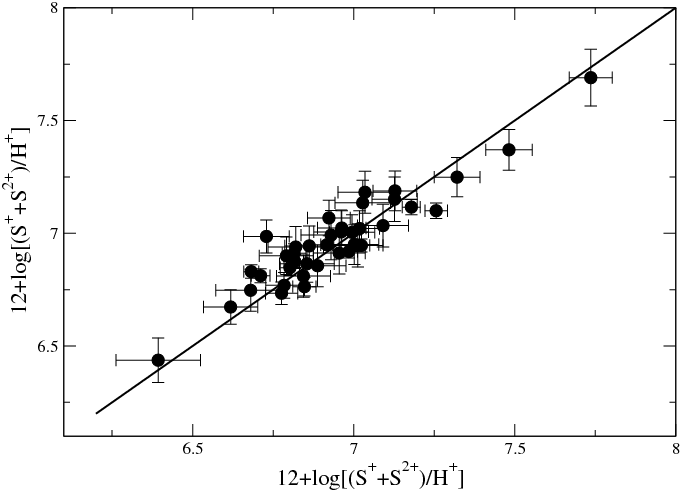}

\caption[Comparison between the electron temperatures and ionic abundances for
  sulphur for the high metallicity \HII\ region sample]{Comparison between the
  electron temperatures (upper) and ionic abundances for sulphur (lower) for
  the high metallicity \HII\ region sample, as derived from measurements of
  the auroral sulphur line $\lambda$\,6312\,\AA\ (abscissa) and from our
  derived calibration (ordinate). The solid line represents the one-to-one
  relation.}
\label{comparison}
\end{figure*}

Figure \ref{comparison} shows the comparison between the electron temperatures
and ionic sulphur abundances derived from measurements of the auroral sulphur
line $\lambda$\,6312\,\AA\ and our derived calibration. The S$^+$/H$^+$ and
S$^{2+}$/H$^+$ ionic ratios (see Chapter \S \ref{HIIgal-obs}) have been derived
from:

\[
12+log\frac{S^+}{H^+}\,=\,log\left( \frac{I(6717)+I(6731)}{I(H\beta} \right) +
5.423 +\frac{0.929}{t}-0.28 logt
\]

\[
12+log\frac{S^{2+}}{H^+}\,=\,log\left( \frac{I(9069)+I(9532)}{I(H\beta}
\right) + 5.8 +\frac{0.771}{t}-0.22 logt 
\]

These expressions have been derived by performing appropriate fittings to the
IONIC task results following the functional form given in
\citetex{1992MNRAS.255..325P}. The assumption has been made that T$_e$([S{\sc
    iii}])\,$\simeq$\,T$_e$([S{\sc ii}]) in the observed regions. This
assumption seems to be justified in view of the results presented by
\citetex{2005A&A...441..981B}.

We have used the calibration above to derive t$_{e}$([S{\sc iii}]) for our
observed CNSFRs. In all cases the values of log$O_{23}$ are inside the range
used in performing the calibration, thus requiring no extrapolation of the
fit. These temperatures, in turn, have been used to derive the S$^+$/H$^+$ and
S$^{2+}$/H$^+$ ionic ratios. The derived T$_e$ and ionic abundances for
sulphur are given in Table \ref{ionic-sulphur}, together with measured values
of the electron density. Temperature and abundance errors are formal errors
calculated from the measured line intensity errors applying error propagation
formulae,  without assigning any error to the temperature calibration itself.

%
%

%
%

%
%
\begin{table}
\centering
\caption[Derived electron densities, electron temperatures and ionic
  abundances from sulphur lines]{Derived electron densities, electron
  temperatures and ionic abundances from sulphur lines in our observed CNSFRs.} 
{\scriptsize
\begin{tabular}{@{}l c c c c c c c c@{}}
\hline
\hline
Galaxy & Region & N$_e$  & T$_e$([S{\sc iii}]) & 12+log (S$^+$/H$^+$) & 12+log (S$^{2+}$/H$^+$) & 12+log[ (S$^+$+S$^{2+}$)/H$^+$] \\
 & & (cm$^{-3}$) &  (K) &  &  &  \\
\hline
NGC\,2903        & R1+R2    & 280 $\pm$  90 & 5731 $\pm$	188  & 6.83 $\pm$ 0.04 &  6.50 $\pm$ 0.05 & 7.01 $\pm$ 0.05\\	
                 & R3       & 230 $\pm$135 & 5737 $\pm$	210  & 6.91 $\pm$ 0.06 &  6.60 $\pm$ 0.06 & 7.09 $\pm$ 0.06 \\  
                 & R4       & 270 $\pm$120 & 6660 $\pm$        331  & 6.68 $\pm$ 0.06 &  6.21 $\pm$ 0.08 & 6.81 $\pm$ 0.06\\   
                 & R6       & 250 $\pm$  80 & 5413 $\pm$	221  & 6.97 $\pm$ 0.05 &  6.64 $\pm$ 0.06 & 7.14 $\pm$ 0.05\\
NGC\,3351        & R1       & 360 $\pm$  80 & 5683 $\pm$	188  & 6.74 $\pm$ 0.04 &  6.75 $\pm$ 0.05 & 7.05 $\pm$ 0.04\\
                 & R2       & 440 $\pm$110 & 5405 $\pm$	210  & 6.90 $\pm$ 0.05 &  6.65 $\pm$ 0.06 & 7.09 $\pm$ 0.05\\   
                 & R3       & 430 $\pm$  70 & 5399 $\pm$	210  & 6.84 $\pm$ 0.04 &  6.83 $\pm$ 0.05 & 7.14 $\pm$ 0.05\\  
                 & R4       & 310 $\pm$120 & 5390 $\pm$	271  & 6.87 $\pm$ 0.06 &  6.64$\pm$ 0.07 & 7.07 $\pm$ 0.07\\  
                 & R5       & 360 $\pm$230 & 6177 $\pm$	286  & 6.78 $\pm$ 0.07 &  6.49 $\pm$ 0.09 & 6.97 $\pm$ 0.08\\  
                 & R6       & 360 $\pm$170 & 6297 $\pm$	295  & 6.49 $\pm$ 0.06 &  6.20 $\pm$ 0.07 & 6.67 $\pm$ 0.07\\  
                 & R7       & 410 $\pm$100 & 5537 $\pm$	220  & 6.80 $\pm$ 0.05 &  6.52 $\pm$ 0.07 & 6.99 $\pm$ 0.06\\  
NGC\,3504        & R3+R4    & 370 $\pm$  60 & 6288 $\pm$	170  & 6.86 $\pm$ 0.03 &  6.77 $\pm$ 0.03 & 7.12 $\pm$ 0.03\\
\hline
\end{tabular}}
\label{ionic-sulphur}
\end{table}


Once the sulphur ionic abundances have been derived, we have estimated the
corresponding oxygen abundances. In order to do that, we have assumed that
sulphur and oxygen electron temperatures follow the relation given by
\citetex{1992AJ....103.1330G} and confirmed by more recent data
(\citeplain{2006MNRAS.372..293H}, see Chapter \S \ref{HIIgal-obs}), and hence
we have derived t$_e$([O{\sc iii}]) according to the expression: 
\[
t_e([O{\textrm{\sc iii}}])\,=\,1.205 t_e([S{\textrm{\sc iii}}])-0.205
\] 
We have also assumed that t$_e$ ([O{\sc ii}]) $\simeq$ t$_e$([S{\sc
    iii}]). The values of T$_e$([O{\sc iii}]) and the ionic abundances for
oxygen are given in Table \ref{ionic-oxygen}. The same comments regarding
errors mentioned above apply. Finally, we have derived the N$^{+}$/O$^{+}$
ratio assuming that t$_e$([O{\sc ii}])\,$\simeq$\,t$_e$([N{\sc
    ii}])\,$\simeq$ t$_e$([S{\sc iii}]). These values are also listed in
Table \ref{ionic-oxygen}.

%
%

\begin{table}
\centering
\caption[Derived T$_e$ and ionic abundances for oxygen and nitrogen]{Derived
  T$_e$ and ionic abundances for oxygen and nitrogen in our observed CNSFRs. } 
{\scriptsize
\begin{tabular}{@{}l c c c c c c c c@{}}
\hline
\hline
Galaxy & Region & T$_e$([O{\sc iii}]) & 12+log(O$^+$/H$^+$) & 12+log(O$^{2+}$/H$^+$) & 12+log[(O$^+$+O$^{2+}$)/H$^+$] & log(N$^+$/O$^+$)\\
 & & (K) & & & & \\
\hline
NGC\,2903        & R1+R2    & 4855 $\pm$   226     &  8.55 $\pm$  0.08   &  8.11 $\pm$  0.08    &  8.69 $\pm$  0.08 & -0.37 $\pm$ 0.07 \\
                 & R3       & 4863 $\pm$   253     &  8.55 $\pm$  0.09   &  8.37 $\pm$  0.09    &  8.77 $\pm$  0.09 & -0.36 $\pm$ 0.08  \\
                 & R4       & 5975 $\pm$   399     &  8.42 $\pm$  0.10	 &  8.04 $\pm$  0.14    &  8.57 $\pm$  0.11 & -0.50 $\pm$ 0.08 \\
                 & R6       & 4473 $\pm$   267     &  8.59 $\pm$  0.10	 &  8.33 $\pm$  0.10    &  8.79 $\pm$  0.10 & -0.38 $\pm$ 0.07 \\
NGC\,3351        & R1       & 4798 $\pm$   227     &  8.60 $\pm$  0.08	 &  8.06 $\pm$  0.08    &  8.72 $\pm$  0.08 & -0.49 $\pm$ 0.07 \\
                 & R2       & 4463 $\pm$   253     &  8.47 $\pm$  0.09	 &  8.41 $\pm$  0.10    &  8.74 $\pm$  0.09 & -0.23 $\pm$ 0.07 \\
                 & R3       & 4456 $\pm$   253     &  8.64 $\pm$  0.09	 &  8.17 $\pm$  0.10    &  8.77 $\pm$  0.09 & -0.38 $\pm$ 0.07 \\
                 & R4       & 4445 $\pm$   327     &  8.55 $\pm$  0.11	 &  8.19 $\pm$  0.13    &  8.71 $\pm$  0.12 & -0.37 $\pm$ 0.08 \\
                 & R5       & 5393 $\pm$   344     &  8.42 $\pm$  0.10	 &  8.34 $\pm$  0.10    &  8.68 $\pm$  0.10 & -0.37 $\pm$ 0.10 \\
                 & R6       & 5538 $\pm$   356     &  8.34 $\pm$  0.10	 &  7.62 $\pm$  0.17    &  8.42 $\pm$  0.11 & -0.47 $\pm$ 0.08 \\
                 & R7       & 4622 $\pm$   266     &  8.48 $\pm$  0.09	 &  8.15 $\pm$  0.10    &  8.65 $\pm$  0.10 & -0.40 $\pm$ 0.07 \\
NGC\,3504        & R3+R4    & 5527 $\pm$   205     &  8.79 $\pm$  0.07	 &  7.94 $\pm$  0.06    &  8.85 $\pm$  0.07 & -0.65 $\pm$ 0.06 \\
\hline
\end{tabular}}
\label{ionic-oxygen}
\end{table}

\begin{figure*}
\vspace*{4.5cm}
\centering
\includegraphics[width=.90\textwidth,angle=0]{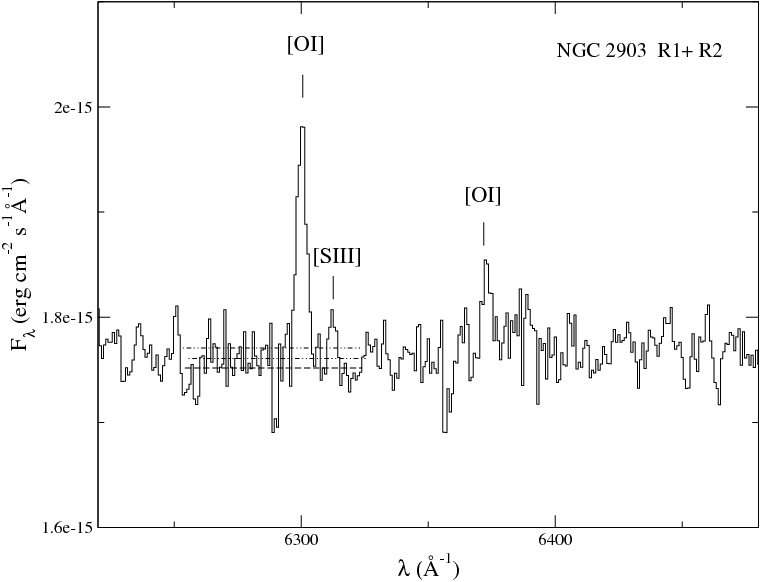}
\caption[Spectrum of region R1+R2 in NGC\,2903 around [S{\sc iii}{\textrm
  ]}\,$\lambda$\,6312\,\AA]{Spectrum of region R1+R2 in NGC\,2903 around
  [S{\sc iii}]\,$\lambda$\,6312\,\AA. The way in which the [S{\sc iii}] line
  has been measured is shown. The different continuum placements used for the
  computation of the errors are shown by horizontal lines. }
\label{medidas2}
\end{figure*}

For one of the observed regions, R1+R2 in NGC\,2903, the [S{\sc
iii}]\,$\lambda$\,6312\,\AA\ has been detected and measured.  Although the
line intensity ratios for this region presented in Table
\ref{intensities1}  have been measured on the combined spectrum of slit
positions 1 and 2, the placement of the continuum for the [S{\sc
iii}]\,$\lambda$\,6312\,\AA\ line was very uncertain. Therefore, we have
performed the temperature measurement on the spectrum extracted from slit
position 1, which shows the best defined continuum. On this spectrum we
have measured the intensity of the [S{\sc iii}]\,$\lambda$\,6312\,\AA\
line with respect to H$\alpha$, and those of  [S{\sc
iii}]\,$\lambda\lambda$\,9069, 9532 \AA\ with respect to
P9\,$\lambda$\,9329, in order to minimize reddening corrections. The
region of this spectrum around the [S{\sc iii}] $\lambda$ 6312 \AA\ line
is shown in Figure \ref{medidas2}  where the horizontal lines show the
different placements of the continuum used to measure the line and
calculate the corresponding errors. The obtained [S{\sc iii}] nebular to
auroral line ratio is: 79.4\,$\pm$\,49.1 which gives a T$_e$([S{\sc
iii}])\,=\,8400$^{+ 4650}_{-1250}$K, slightly higher than predicted by the
proposed fit. This region is represented as a solid black circle in Figure
\ref{te[SIII]cal}. 

%
%

\section{Discussion on metal abundances}
\label{dis-abun}

\subsection{Characteristics of the observed CNSFRs}
\label{Halpha_fluxes}
Planesas et al.\ (1997) provide H$\alpha$ fluxes (from images) for all our
observed regions. These values are larger than those measured inside our slit
by factors between 1.0 and 2.7, depending on the size of the region, something
to be expected given the long-slit nature of our observations. We have
calculated the regions' H$\alpha$ luminosities from our observed
values, correcting for extinction according to the values found from the
spectroscopic analysis. The resulting values are listed in Table
\ref{cnsfr_prop}. These values are larger than the typical ones found for disc
\HII\ regions and overlap with those measured in \HII\ galaxies. The region
with the largest H$\alpha$ luminosity is R3+R4 in NGC\,3504, for which a value
of 2.02\,$\times$\,10$^{40}$\,erg\,s$^{-1}$ is measured.  

We have derived the number of hydrogen ionizing photons [$Q(H_0)$] from the
extinction corrected H$\alpha$ flux as:  
\[ 
log Q(H_0)\,=\,0.802 \times 10^{49}
\left(\frac{F(H\alpha)}{10^{-14}}\right)\left(\frac{D}{10}\right) s^{-1} \] 
\noindent where F(H$\alpha$)  is in erg cm$^{-2}$\,s$^{-1}$ and the distance, D,
is in Mpc. 

\label{ionization_parameter}
The ionization parameter, $u$,  can be estimated from the [S{\sc ii}]/[S{\sc
    iii}] ratio \cite{1991MNRAS.253..245D} as:  
\[ 
log\,u\,=\,-1.68 log([S{\textrm{\sc ii}}]/[S{\textrm{\sc iii}}]) -2.99 
\]
and ranges between  -3.12  and -3.98 for our observed CNSFRs, in the low side
of what is found for disc \HII\ regions, even in the cases of high metallicity
(see \citeplain{1991MNRAS.253..245D}).

\label{sizes, filling factors and M(HII)}
From the calculated values of the number of Lyman $\alpha$ photons, Q(H$_0$),
ionization parameter and electron density, it is possible to derive the size
of the emitting regions as well as the filling factor (see
\citeplain{2002MNRAS.329..315C}). The derived sizes are between 1.5 arcsec for
region R3 in NGC\,3351 and 5.7 arcsec for region R4 in NGC\,2903; these values
correspond to linear dimensions between 74 and 234 pc. The derived filling
factors are low: between 6\,$\times$\,10$^{-4}$ and 1\,$\times$\,10$^{-3}$, lower
than commonly found in giant \HII\ regions ($\sim$\,0.01). Sizes in arcsec,
filling factors and the corresponding masses of ionized hydrogen, M(\HII), are
given in Table \ref{cnsfr_prop}. 

\label{ionizing cluster masses}
We have also derived the mass of ionizing stars, M$_{ion}$, from the calculated
number of hydrogen ionizing photons with the use of evolutionary models of
ionizing clusters \cite{1994ApJS...91..553G,1996ApJS..107..661S} assuming that
the regions are ionization bound and that no photons are absorbed by dust. A
Salpeter IMF with upper and lower mass limits of 100 and 0.8\,M$_{\odot}$
respectively, has been assumed. According to these models, a relation exists
between the degree of evolution of the cluster, as represented by its H$\beta$
emission line equivalent width and the number of hydrogen ionizing photons per
unit solar mass \cite{1998Ap&SS.263..143D,2000MNRAS.318..462D}. The ionizing
cluster masses thus derived are given in Table \ref{cnsfr_prop} and range
between  1.1\,$\times$\,10$^{5}$ and 4.7\,$\times$\,10$^{6}$\,M$_{\odot}$. The
measured H$\beta$ equivalent widths, however, are very low and could be
reflecting the contribution by underlying non-ionizing populations. An
alternative way to take into account the cluster evolution  in the derivation
of the mass is to make use of the existing relation between the ionization
parameter and the H$\beta$ equivalent width for ionized regions
\cite{2006MNRAS.365..454H}. In that case, the derived masses are lower by
factors between 1.5 and 15. At any rate, given the assumptions of no dust
absorption or photon leakage, these masses represent lower limits. 


\begin{table}
\centering
\caption{General properties of the observed CNSFR. }
{\scriptsize
\begin{tabular}{@{}l c c c c c c c c c@{}}
\hline
\hline
Galaxy& Region   &  F(H$\alpha$) & L(H$\alpha$) & Q(H$_0$)	& log u &  Diameter & $\epsilon$ & M$_{ion}$ &   M(\HII) \\
        &  & (erg cm$^{-2}$ s$^{-1}$) & (erg s$^{-1}$) &  (ph s$^{-1}$) &  &  (arcsec) &   &   &  \\
\hline

NGC\,2903        & R1+R2 & 3.06E-13 & 2.71E+39 &	1.98E+51 & -3.69  &     4.66	& 5.93E-04  &  7.39 &	0.144\\
                 & R3       & 3.16E-14 & 2.80E+38 &	2.05E+50 & -3.65  &     1.59	& 2.29E-03  &  1.11 &	0.018\\
                 & R4       & 2.23E-13 & 1.98E+39 &	1.45E+51 & -3.98  &     5.70	& 2.54E-04  &  21.1 &	0.109\\
                 & R6       & 1.48E-13 & 1.31E+39 &	9.60E+50 & -3.66  &     3.33	& 9.89E-04  &  4.55 &	0.078\\
NGC\,3351        & R1       & 2.19E-13 & 2.65E+39 &     1.94E+51 & -3.12  &     1.81	& 3.76E-03  &  6.40 &	0.109\\
                 & R2       & 1.61E-13 & 1.95E+39 &	1.42E+51 & -3.55  &     2.29	& 9.08E-04  &  6.80 &	0.066\\
                 & R3       & 2.35E-13 & 2.84E+39 &	2.08E+51 & -3.12  &     1.72	& 3.31E-03  &  5.70 &	0.098\\
                 & R4       & 6.84E-14 & 8.29E+38 &	6.06E+50 & -3.49  &     1.66	& 2.04E-03  &  2.32 &	0.040\\
                 & R5       & 4.25E-14 & 5.15E+38 &	3.76E+50 & -3.65  &     1.47	& 1.37E-03  &  4.71 &	0.021\\
                 & R6       & 6.43E-14 & 7.79E+38 &	5.70E+50 & -3.65  &     1.81	& 1.11E-03  &  4.81 &	0.032\\
                 & R7       & 1.60E-13 & 1.94E+39 &	1.42E+51 & -3.58  &     2.47	& 8.31E-04  &  7.11 &	0.070\\
NGC\,3504        & R3+R4 & 4.20E-13 & 2.02E+40 &	1.47E+52 & -3.32  &     3.11	& 6.76E-04  &  47.0 &	0.807\\
\hline
\multicolumn{10}{@{}l}{masses in 10$^5$\,M$_\odot$.}
\end{tabular}}
\label{cnsfr_prop}
\end{table}  


\subsection{Metallicity estimates}
The abundances we derive using our T$_e$([S{\sc iii}]) calibration are
comparable to those found by Bresolin et al.\ (2005) for their sample of high
metallicity \HII\ regions. Most of our CNSFRs show total oxygen abundances,
taken to be  O/H= O$^+$/H$^+$ + O$^{2+}$/H$^+$, consistent with solar values
within the errors. The region with the highest oxygen abundance is R3+R4 in
NGC\,3504: 12+log(O/H) = 8.85, about 1.6 solar if the solar oxygen abundance is
set at the value derived by Asplund et al.\ (2005), 12+log(O/H)$_{\odot}$ =
8.66$\pm$0.05. Region R6 in NGC\,3351 has the lowest oxygen abundance of the
sample, about 0.6 times solar. In all the observed CNSFRs the O/H abundance is
dominated by the O$^+$/H$^+$ contribution with 0.18 $\leq$ log(O$^+$/O$^{2+}$)
$\leq$ 0.85. This is also the case for high metallicity disc \HII\ regions
where these values are even higher. For our observed regions, also the
S$^+$/S$^{2+}$ ratios are larger than one, which is at odds with the high
metallicity disc \HII\ regions for which, in general, the sulphur abundances
are dominated by S$^{2+}$/H$^+$.  

\begin{figure*}
\centering
\vspace*{4.5cm}
\includegraphics[width=.9\textwidth,angle=0]{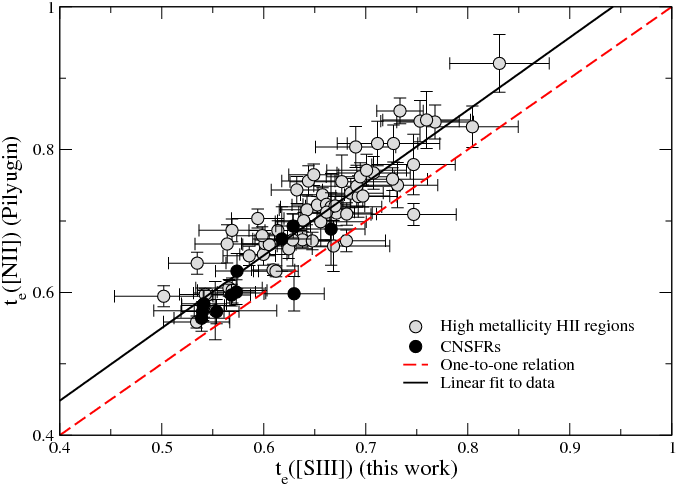}
\caption[t$_e$([N{\sc ii}{\textrm ]}) derived with Pilyugin's method against
  the t$_e$([S{\sc iii}{\textrm ]}) derived from the calibration]{t$_e$([N{\sc
      ii}]) derived with Pilyugin's method against the t$_e$([S{\sc iii}])
  derived using the calibration presented in this work. The dashed line shows
  the one-to-one correspondence.} 
\label{Pilyugin}
\end{figure*}

The fact that both O/H and S/H seem to be dominated by the lower ionization
species, can raise concern about our method of abundance derivation, based on
the calibration of the T$_e$([S{\sc iii}]). In a recent article
\citetex{2007MNRAS.375..685P} addresses this particular problem. He proposes a
calibration of the ratio of the [N{\sc ii}] nebular-to-auroral line
intensities in terms of those of the nebular oxygen lines. Then the [N{\sc
    ii}] electron temperature, thought to properly characterize the low
ionization zone of the nebula, can be obtained. We have applied Pilyugin
(2007) to our observed CNSFRs and to the high metallicity \HII\
sample. Figure \ref{Pilyugin} shows the derived t$_e$([N{\sc ii}]) following
Pilyugin's method against the t$_e$([S{\sc iii}]) derived from our
calibration. The red dashed line shows the one-to-one relation while the
(black) solid line shows the actual fit to all the data. It can be seen that,
in what regards CNSFR, both temperatures are very similar, with the [N{\sc
    ii}] temperature being, in average, 500 K higher than that of [S{\sc
    iii}]. This difference is of the same size of the average errors in the
measured temperatures entering the calibrations, therefore we can assume that
our derived t$_e$([S{\sc iii}]) characterizes the low ionization zone at least
as well as the t$_e$([N{\sc ii}]) derived applying Pilyugin's method.
 
Concerning relative abundances, with our analysis it is possible to derive the
relative N/O value, assumed to be equal to the N$^{+}$/O$^+$ ratio. These
values are in all cases larger than the solar one (log(N/O)$_{\odot}$= -0.88;
Asplund et al., 2005) by factors between 1.7 (R3+R4 in NGC\,3504) and 4.5 (R2 in
NGC\,3351) which are amongst the highest observed N/O ratios (see e.g.\
\citeplain{2006MNRAS.372.1069M}). Regarding S, if -- given the low excitation
of the observed regions -- the fraction of S$^{3+}$ is assumed to be
negligible, the S/O ratio can be obtained as: 
\[ 
\frac{S}{O} = \frac{S^{+}+S^{2+}}{O^{+}+O^{2+}} 
\]
The values of log(S/O) span a very narrow range between -1.76 and -1.63, that
is between 0.6 and 0.8 of the solar value (log(S/O)$_{\odot}$ = -1.52;
\citeplain{2005ASPC..336...25A}).

%
%

\subsection{Comparison with high metallicity \HII\ regions}
\begin{figure*}
\centering
\vspace{0.5cm}
\includegraphics[width=.75\textwidth,angle=0]{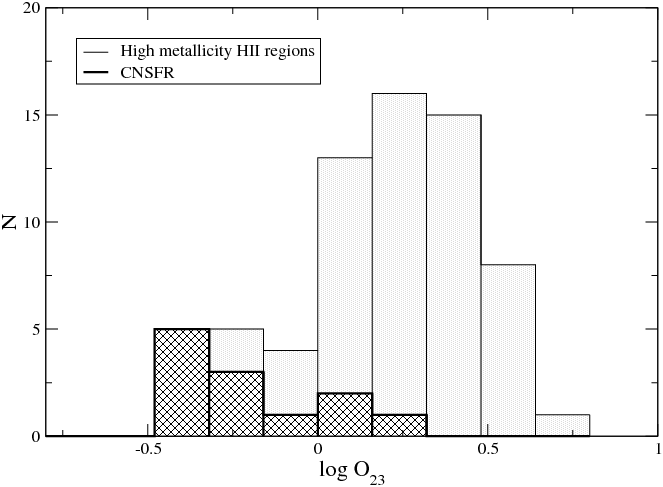}\\
\vspace{1.5cm}
\includegraphics[width=.75\textwidth,angle=0]{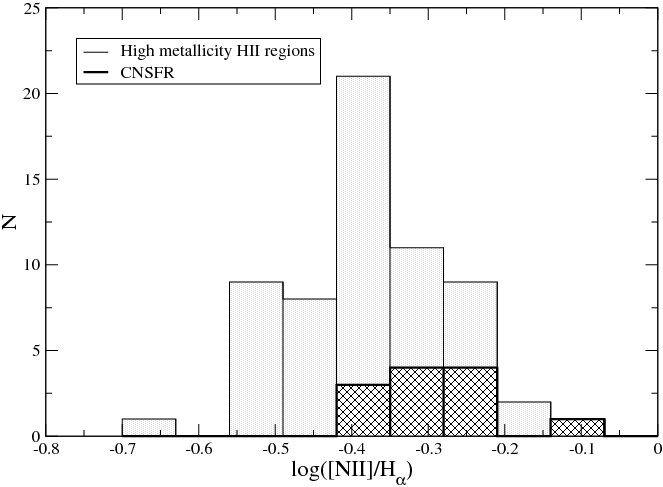}
\caption[Distribution of the empirical abundance parameters O$_{23}$ and
  N2]{Distribution of the empirical abundance parameters O$_{23}$ (upper) and
  N2 (lower) for the observed CNSFRs and the sample of high metallicity disc
  \HII\ regions.} 
\label{histo_O23_N2}
\end{figure*}

\begin{figure*}
\centering
\vspace{3.5cm}
\includegraphics[width=.9\textwidth,angle=0]{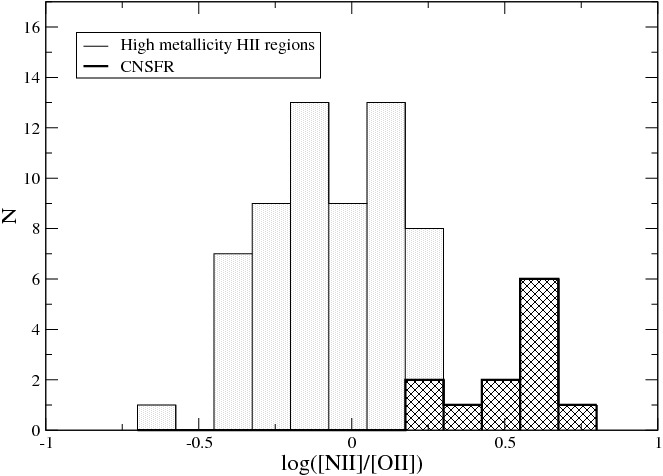}
\caption[Distribution of the [N{\sc ii}{\textrm ]}/[O{\sc ii}{\textrm ]}
  ratio]{Distribution of the [N{\sc ii}]/[O{\sc ii}] ratio for the observed
  CNSFRs (dark) and the sample of high metallicity disc \HII\ regions
  (light).} 
\label{N2-O2}
\end{figure*}

The observed CNSFRs being of high metallicity show however marked differences
with respect to high metallicity disc \HII\ regions. Even though their derived
oxygen and sulphur abundances are similar, they show values of the O$_{23}$
and the N2 parameters whose distributions are shifted to lower and higher
values respectively with respect to the high metallicity disc sample (Figure
\ref{histo_O23_N2}). Hence, if pure empirical methods were used to estimate
the oxygen abundances for these regions, higher values would in principle be
obtained. This would seem to be in agreement with the fact that CNSFR, when
compared to the disc high metallicity regions, show the highest [N{\sc
    ii}]/[O{\sc ii}] ratios.  Figure \ref{N2-O2} shows indeed a bi-modal
distribution of the [N{\sc ii}]/[O{\sc ii}] ratio in disc and circumnuclear
\HII\ regions.

%
%


%
%

%
%

\begin{figure*}
\centering
\vspace{3.5cm}
\includegraphics[width=.9\textwidth,angle=0]{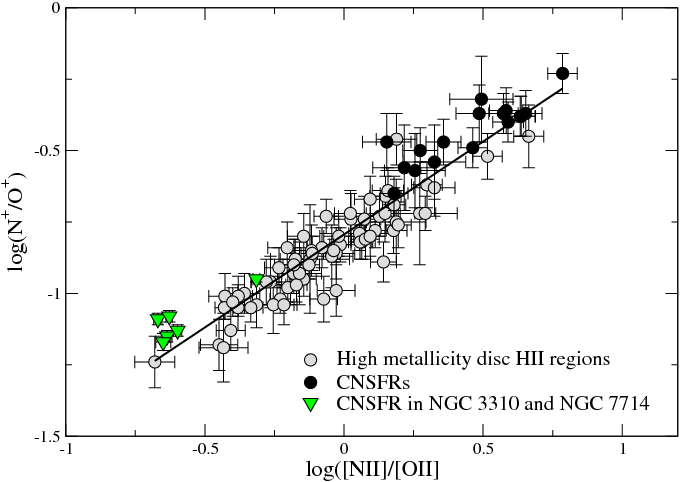}
\caption[The relation between the [N{\sc ii}{\textrm ]}/[O{\sc ii}{\textrm ]}
  emission line intensity ratio and the N$^+$/O$^+$ ratio]{The relation
  between the [N{\sc ii}]/[O{\sc ii}] emission line intensity ratio and the
  N$^+$/O$^+$ ratio for CNSFRs (black  circles) and high metallicity \HII\
  regions (gray circles). Included also (downward triangles) are lower
  metallicity CNSFRs in NGC\,3310 \cite{1993MNRAS.260..177P} and NGC\,7714
  \cite{1995ApJ...439..604G}.} 
\label{NoverO}
\end{figure*}


A good correlation has been found to exist between the [N{\sc ii}]/[O{\sc ii}]
ratio and the N$^+$/O$^+$ ionic abundance ratio, which in turn can be assumed
to trace the N/O ratio  \cite{2005MNRAS.361.1063P}. This relation is shown in
Figure \ref{NoverO} for the observed CNSFRs (black circles) and the high
metallicity \HII\ region sample (gray circles). The CNSFRs sample has been
enlarged with two circumnuclear regions of NGC\,1097 observed by
\citetex{1984MNRAS.210..701P} and other three observed in NGC\,5953 by
\citetex{1996MNRAS.281..781G}, all of them of high metallicity.  Also shown
are data of some CNSFRs in two peculiar galaxies: NGC\,3310
\cite{1993MNRAS.260..177P} and NGC\,7714 \cite{1995ApJ...439..604G} of
reported lower metallicity. In all the cases, ionic and total abundances have
been derived following the same methods as in the CNSFRs in the present study
and described in Section \S \ref{che-abun}. In the case of the latter regions,
abundances 
derived in this way are larger than derived from direct determinations of
t$_e$([O{\sc iii}]) since our estimated values for this temperature are
systematically lower by about 1200 K. This could be due to the fact that our
semi-empirical calibration has been actually derived for high metallicity
regions. However, the values of log(SO$_{23}$) for these regions are between
-0.44 and -0.85 and therefore within the validity range of the calibration and
we have preferred to use the same method for consistency reasons. 

\begin{figure*}
\centering
\vspace{3.5cm}
\includegraphics[width=.9\textwidth,angle=0]{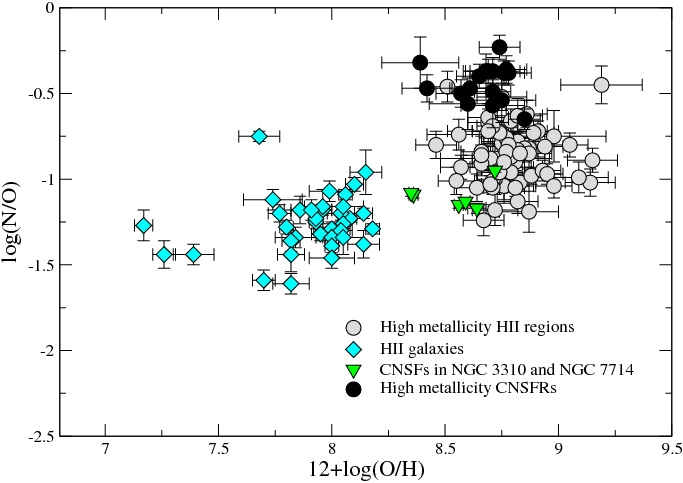}
\caption[Relation between the N/O ratio and the O/H abundance]{Relation
  between the N/O ratio and the O/H abundance for CNSFRs (black circles and
  green upside down triangles for high and low metallicities respectively),
  high metallicity \HII\ regions (gray circles) and \HII\ galaxies (cyan
  diamonds).} 
\label{NoverO-O}
\end{figure*}

We can see that a very tight correlation exists which allows to estimate the
N/O ratio from the measured [N{\sc ii}] and [O{\sc ii}] emission line
intensities. In this relation, high metallicity regions and CNSFRs seem to
follow a sequence of increasing N/O ratio. A linear regression fit to the data
yields the expression: 
\[ 
log(N/O)\,=\,(0.65 \pm 0.02) log ([N{\textrm{\sc ii}}]/[O{\textrm{\sc ii}}]) - (0.79 \pm 0.01)
\] 
This relation is shallower than that found in \citetex{2005MNRAS.361.1063P}
for a sample that did not include high metallicity \HII\ regions. The N/O
ratio for our observed CNSFRs is shown in Figure \ref{NoverO-O} against their
oxygen abundance together with similar data for the high metallicity \HII\
region and \HII\ galaxy samples as described in section \ref{che-abun}. It can
be seen that all the CNSFRs show 
similar oxygen abundances, with the mean value being lower than that shown by
high metallicity disc \HII\ regions, but the observed CNSFRs show larger N/O
ratios and they do not seem to follow the trend of N/O vs.\ O/H which marks the
secondary behaviour of nitrogen.

%
%

\label{What's the true metalicity of CNSFR?}
Equal values, within the errors, are found for the CNSFRs in NGC\,3351
(12+log(O/H)\,= 8.70\,$\pm$\,0.10) except for region R6 which shows an O/H
abundance lower by a factor of about 
2. It is worth noting that this is the region with the highest O$^+$/O ratio
(0.83) and a very low O$^{2+}$/H$^+$ value:
12+log(O$^{2+}$/H$^+$)\,=\,7.62\,$\pm$\,0.17. The average value for the rest
of the regions is in agreement with the central abundance determined by
\citetex{2006MNRAS.367.1139P} by extrapolating the 
galaxy oxygen abundance gradient, 12+log(O/H) = 8.74 $\pm$ 0.02. This value is
somewhat lower than that previously estimated by
\citetex{2004A&A...425..849P}: 8.90 from the same data but following a
different method of analysis. In that same work, the quoted central abundance
for NGC\,2903 is 12+log(O/H) = 8.94. Our results for regions R1+R2, R3 and  R6
in this galaxy yield an average value of 8.75 $\pm$ 0.09 lower than theirs by
0.2 dex. Region R4 shows a lower oxygen abundance but still consistent within
the errors with the average.  Values of N/O ratios for the centres of NGC\,3351
and NGC\,2903 similar to those found here (-0.33 and -0.35 respectively) are
quoted by \citetex{2004A&A...425..849P}.

\begin{figure*}
\centering
\vspace{0.5cm}
\includegraphics[width=.75\textwidth,angle=0]{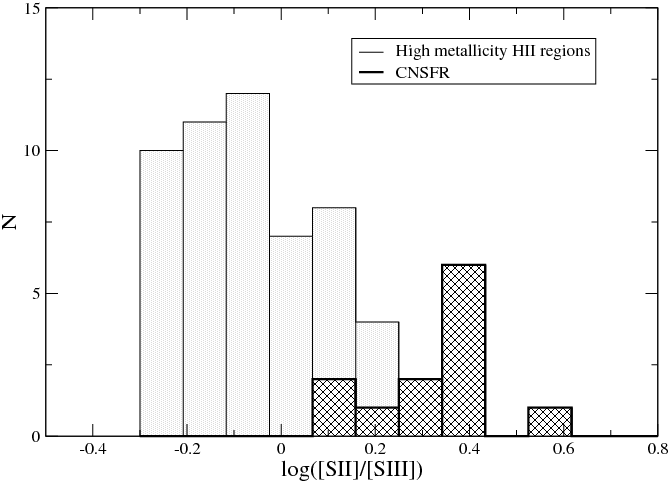}\\
\vspace{1.5cm}
\includegraphics[width=.75\textwidth,angle=0]{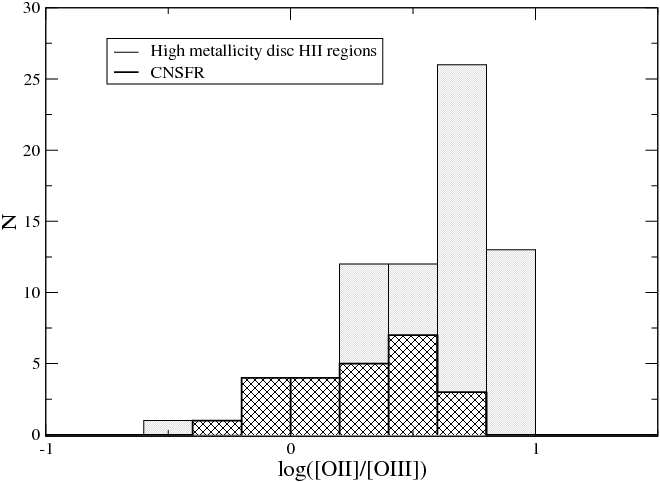}
\caption[Distribution of the [S{\sc ii}{\textrm ]}/[S{\sc iii}{\textrm ]} and
  [O{\sc ii}{\textrm ]}/[O{\sc iii}{\textrm ]}]{Distribution of the [S{\sc
      ii}]/[S{\sc iii}] (upper) and [O{\sc ii}]/[O{\sc iii}] (lower) ratios for
  the observed CNSFRs (dark) and the sample of high metallicity disc \HII\
  regions\ (light).} 
\label{S2-S3}
\end{figure*}

\label{ionization parameter}
Another difference between the high metallicity circumnuclear and disc regions
is related to their average ionization parameter. The upper panel of Figure
\ref{S2-S3} shows the distribution of the [S{\sc ii}]/[S{\sc iii}] ratio for
the two samples. The [S{\sc ii}]/[S{\sc iii}] ratio has been shown to be a
good ionization parameter indicator for moderate to high metallicities
\cite{1991MNRAS.253..245D} with very little dependence on metallicity or
ionization temperature. It can be seen that all the CNSFRs observed show large
[S{\sc ii}]/[S{\sc iii}] ratios which imply extremely low ionization
parameters. On the other hand, a different answer would be found if the [O{\sc
    ii}]/[O{\sc iii}] parameter, also commonly adopted as an ionization
parameter 
indicator, was used. In this case, CNSFRs and high metallicity \HII\ regions
show a much more similar distribution, with CNSFRs showing slightly lower
values of [O{\sc ii}]/[O{\sc iii}] (Figure \ref{S2-S3}, lower panel).  

It should be noted the dependence of the [O{\sc ii}]/[O{\sc iii}] parameter on
metallicity due to the presence of opacity edges of various abundant elements
(O$^+$, Ne$^+$, C$^{2+}$, N$^{2+}$) in the stellar atmospheres  that can
combine to substantially modify the stellar flux of high abundance stars at
energies higher than 35-40 eV and then produce a lower [O{\sc iii}] emission
\cite{1976ApJ...208..336B}. However, if the CNSFRs were ionized by stars of a
higher metallicity than those in disc \HII\ regions this effect would go in
the direction of producing higher [O{\sc ii}]/[O{\sc iii}] ratios for the
CNSFRs, and ionization parameters derived from  [O{\sc ii}]/[O{\sc iii}] ratios
would be found to be lower than those derived from [S{\sc ii}]/[S{\sc iii}]
ratios, contrary to what is actually observed.  

%
%

\begin{figure*}
\vspace{3.5cm}
\centering
\includegraphics[width=.9\textwidth,angle=0]{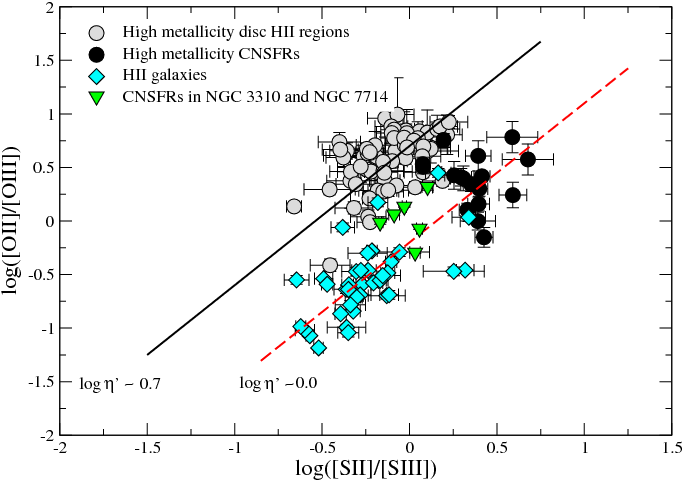}
\caption[The $\eta$' plot: the [O{\sc ii}{\textrm ]}/[O{\sc iii}{\textrm ]}
  ratio vs.\ [S{\sc ii}{\textrm ]}/[S{\sc iii}{\textrm ]} ratio]{The $\eta$'
  plot: the [O{\sc ii}]/[O{\sc iii}] ratio vs.\ [S{\sc ii}]/[S{\sc iii}] ratio
  for different ionized regions: high metallicity CNSFRs (black circles), low
  metallicity CNSFRs (green upside down triangles), high metallicity disc \HII\
  regions (gray) circles and \HII\ galaxies (cyan diamonds). } 
\label{eta-prime-plot}
\end{figure*}

\label{Ionization structure}
The ionization structure can provide important information about the
characteristics of the ionizing source. The relation between the emission line
ratios 
[O{\sc ii}]/[O{\sc iii}] vs.\ [S{\sc ii}]/[S{\sc iii}], in particular, works
as a diagnostic diagram for the nature and temperature of the radiation
field. This 
diagram was used in \citetex{1985MNRAS.212..737D} in order to investigate the
possible contributions by shocks in CNSFRs and LINERs and is the basis of the
definition of the $\eta$' parameter \cite{1988MNRAS.231..257V}. The $\eta$'
parameter, defined as:  
\[ 
\eta '\,=\,\frac{[O{\textrm{\sc ii}}]\,\lambda\lambda\,3727,29 / [O{\textrm{\sc
	iii}}]\,\lambda\lambda\,4959,5007}{[S{\textrm{\sc
	ii}}]\,\lambda\lambda\,6716,6731 /[S{\textrm{\sc
	iii}}]\,\lambda\lambda 9069,9532} 
\] 
is a measure of the ``softness" of the ionizing radiation  and increases with
decreasing ionizing temperature. The ``$\eta$' plot" is shown in Figure
\ref{eta-prime-plot}. In this plot, diagonal lines of slope unity would show
the locus of ionized regions with constant values of $\eta$'. The lines shown
in the plot have slope 1.3 reflecting the second order dependence of $\eta$'
on ionization parameter \cite{1991MNRAS.253..245D}. In this graph CNSFRs are
seen to segregate from disc \HII\ regions. The former cluster around the value
of log$\eta$' = 0.0 (T${ion}\sim$ 40,000 K) while the latter cluster around
log $\eta$' = 0.7 (T${ion}\sim$ 35,000 K). Also shown are the data
corresponding to \HII\ galaxies. Indeed, CNSFRs seem to share more the locus of
the \HII\ galaxies than that of disc \HII\ regions.

%
%


Given the problems involved in the measurement of the [O{\sc ii}] emission
lines mentioned in Section \S \ref{Line intensity} it is interesting to see
the effects that a 
possible underestimate of the intensity of this line would have on our
analysis.  The underestimate of the [O{\sc ii}] line leads to an overestimate
of the SO$_{23}$ parameter and hence to an underestimate of the [S{\sc iii}]
electron temperature. This lower temperature in turn leads to higher O/H
abundances. Therefore, larger values of the intensity of the [O{\sc ii}] line
would lead to lower O/H abundances. However, not by a large amount. Increasing
the [O{\sc ii}] intensity by a factor of 2, would decrease the SO$_{23}$
parameter by a factor between 1.5 and 1.8 depending on the region, which would
lead to increased [S{\sc iii}] electron temperatures by between 500 and
1000\,K. The corresponding O$^+$/H$^+$ would be only slightly smaller, by
about 0.10\,dex, as would also be the total O/H abundance which is dominated
by O$^+$/H$^+$. The O$^{++}$/H$^+$ ionic ratio however would be lower by
between 0.3 and 0.4\,dex, which would produce an ionization structure more
similar to what is found in disc \HII\ regions. The sulphur ionic and total
abundances would be decreased by about 0.17\,dex. Finally, the N$^+$/O$^+$
ratios 
would be decreased by about 0.17\,dex, while the S/O ratios would remain
almost unchanged. In the ``$\eta$' plot" (Figure \ref{eta-prime-plot}) the data
points corresponding to our observed CNSFRs would move upwards by
0.30\,dex. However, we do not find any compelling reason why the [O{\sc ii}]
intensities should be larger than measured by such a big amount (see
Section \S \ref{results-abun} and Figure \ref{medidas1}). 

One possible concern about these CNSFR is that, given their proximity to the
galactic nuclei, they could be affected by  hard radiation coming from a low
luminosity AGN. NGC\,3351 shows a faint UV core . However, the IUE spectrum
that covers the whole central star forming ring, shows broad absorption lines
of Si{\sc iv}\,$\lambda$\,1400\,\AA\ and C{\sc iv}\,$\lambda$\,1549\,\AA\
typical of young stars of high metallicity \cite{1997ApJ...484L..41C}. They
are consistent with a total mass of 3 $\times$ 10$^5$ M$_{\odot}$ of recently
formed stars (4-5 Myr). This is of the order of our derived values for single
CNSFR in this galaxy. Therefore, no signs of activity are found for this
nucleus nor are they reported for the other two galaxy nuclei. On the other
hand, the He{\sc ii}\,$\lambda$\,4686\,\AA\ line is measured in regions R1+R2 and R6
of NGC\,2903  and in region R7 in NGC\,3351. In the first region, there is some
evidence for the presence of WR stars \cite{2002MNRAS.329..315C}. For the other
two, that presence is difficult to assess due to the difficulty in placing the
continuum for which a detailed modeling of the stellar population is needed.  

%
%
\begin{figure*}
\vspace{3.5cm}
\centering
\includegraphics[width=.91\textwidth,angle=0]{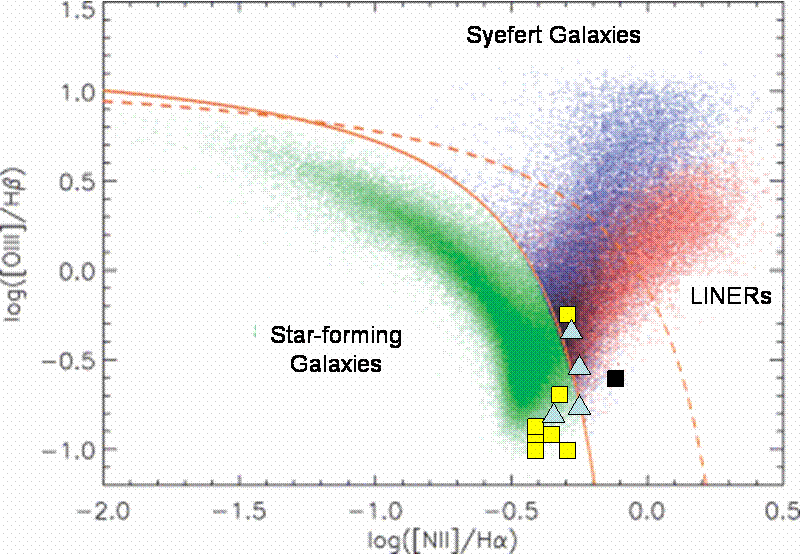}
\caption[The [N{\sc ii}{\textrm ]}/H$\alpha$  vs.\ [O{\sc iii}{\textrm
  ]}/H$\beta$ diagnostics for emission line objects]{The [N{\sc ii}]/H$\alpha$
  vs.\ [O{\sc iii}]/H$\beta$ diagnostics for emission line objects in the SDSS
  (adapted from \citeplain{2006MNRAS.371.1559G}). The location of our observed
  CNSFR is shown overplotted. Light blue triangles correspond to NGC\,2903,
  light yellow squares correspond to NGC\,3351 and the black square
  corresponds to NGC\,3504. Solid \cite{2001ApJ...556..121K} and dashed
  \cite{2003MNRAS.346.1055K} lines separate the star forming from the active
  nucleus galaxy regions.}
\label{diagnostico}
\end{figure*}

Alternatively, the spectra of  these regions harbouring massive clusters of
young stars might be affected by the presence of shocked gas. Diagnostic
diagrams of the kind presented by \citetex{1981PASP...93....5B} can be used to
investigate the possible contribution by either a hidden AGN or by the
presence of shocks to the emission line spectra of the observed CNSFRs. Figure
\ref{diagnostico} shows one of these diagnostic diagrams, log ([N{\sc
    ii}]/H$\alpha$) vs.\ log ([O{\sc iii}])/H$\beta$, for our CNSFRs: light
yellow squares for NGC\,3351, light blue triangles for NGC\,2903 and dark
square for NGC\,3504. The figure has been adapted from
\citetex{2006MNRAS.371.1559G} and 
shows the location of emission line galaxies in the Sloan Digital Sky Survey
(SDSS). Dashed and solid lines correspond to the boundary between Active
Galactic Nuclei (AGN) and \HII\ galaxies defined by
\citetex{2001ApJ...556..121K} and  \citetex{2003MNRAS.346.1055K}
respectively. Some of our observed CNSFRs are found close to the transition
zone between \HII\ region and LINER spectra but only R3+R4 in NGC\,3504 may
show a hint of a slight contamination by shocks. This region, the most
luminous in our sample and also the one with the highest abundance,  should be
studied in more detail.

A final remark concerns the gas kinematics in CNSFRs (see Chapter \S
\ref{cnsfr-obs-kine}; \citeplain{2007MNRAS.378..163H}). As already discussed,
we have studied the 
kinematics of gas and stars in the CNSFRs of NGC\,2903, NGC\,3310 and
NGC\,3351, finding two different components for the
ionized gas in H$\beta$ and [O{\sc iii}] emission: a ``broad component" with a
velocity dispersion similar to that measured for the stars, and a ``narrow
component" with a dispersion lower than the stellar one by about
30\,km\,s$^{-1}$. Obviously the abundance analysis and the location of these
regions on diagnostic diagrams would be affected if more than one velocity
component in the ionized gas corresponding to kinematically distinct systems
are present.

\section{Summary and conclusions on metal abundances}
\label{summ-abun}

We have obtained spectro-photometric observations in the optical
($\lambda\lambda$\,3650-7000\,\AA) and far-red
($\lambda\lambda$\,8850-9650\,\AA) wavelength ranges of a total of 12
circumnuclear \HII\ 
regions in the early type spiral galaxies: NGC\,2903, NGC\,3351 and
NGC\,3504. These regions were expected to be amongst the highest metallicity
regions as corresponds to their position near the galactic centre. At the same
time, this position implies a substantial contribution by the bulge stellar
population to their spectra which represents a major observational problem and
compromises the reliability of the emission line intensities. Its proper
subtraction however requires the detailed modeling of the stellar population
and the disentangling of the contribution by the bulge and by any previous
stellar generations in the CNSFRs themselves. Almost all these regions show
the presence of the CaT lines in the far red (see Chapter \S
\ref{cnsfr-obs-kine}) some of them with equivalent widths that suggest a
certain contribution by red supergiant stars. 

We have derived the characteristics of the observed CNSFRs in terms of size,
H$\alpha$ luminosities and ionizing cluster masses. The derived sizes are
between 1.5 arcsec  and 5.7 arcsec which correspond to linear dimensions
between 74 and 234 pc. The derived filling factors, between
6\,$\times$\,10$^{-4}$ and 1\,$\times$\,10$^{-3}$, are lower than commonly
found in giant \HII\ regions ($\sim$\,0.01). H$\alpha$ luminosities are larger
than the typical ones found for disc \HII\ regions and overlap with those
measured in \HII\ galaxies. The region with the largest H$\alpha$ luminosity
is R3+R4 in NGC~3504, for which a value of 2.02\,$\times$\,10$^{40}$ is
measured.  Ionizing cluster masses range between  1.1\,$\times$\,10$^{5}$ and
4.7\,$\times$\,10$^{6}$\,M$_{\odot}$ but could be lower by factors between 1.5
and 15 if the contribution by the underlying stellar population is taken into
account. 

The low excitation of the regions, as evidenced by the weakness of the [O{\sc
    iii}]\,$\lambda$\,5007\,\AA\ line, precludes the detection and measurement
of the auroral [O{\sc iii}]\,$\lambda$\,4363\,\AA\ necessary for the
derivation of the electron temperature. For one of the regions, the
[S{\sc iii}]\,$\lambda$\,6312\,\AA\ line was detected providing, together with
the nebular [S{\sc iii}] lines at $\lambda\lambda$\,9069, 9532\,\AA, a value
of the electron temperature of T$_e$([S{\sc iii}])\,=\,8400$^{+
  4650}_{-1250}$K. A new method for the derivation of sulphur abundances was
developed based on the calibration of the [S{\sc iii}] electron temperature
    vs.\ 
the empirical parameter SO$_{23}$ defined as the quotient of the oxygen and
sulphur abundance parameters O$_{23}$ and S$_{23}$ and the further assumption
that T([S{\sc iii}])\,$\simeq$\,T([S{\sc ii}]). Using this method, the oxygen
abundances and the N/O and S/O ratios can also be derived.

The derived oxygen abundances are comparable to those found in high
metallicity disc \HII\ regions from direct measurements of electron
temperatures and consistent with solar values within the errors. The region
with the highest oxygen abundance is R3+R4 in NGC~3504, 12+log(O/H) = 8.85,
about 1.6 solar if the solar oxygen abundance is set at the value derived by
Asplund et al.\ (2005), 12+log(O/H)$_{\odot}$ = 8.66$\pm$0.05. Region R7 in
NGC~3351 has the lowest oxygen abundance of the sample, about 0.6 times
solar. In all the observed CNSFRs the O/H abundance is dominated by the
O$^+$/H$^+$ contribution, as is also the case for high metallicity disc \HII\
regions. For our observed regions, however also the  S$^+$/S$^{2+}$ ratio is
larger than one, different from the case of high metallicity disc \HII\
regions for which, in general, the sulphur abundances are dominated by
S$^{2+}$/H$^+$. The derived N/O ratios are in average larger than those found
in high metallicity disc \HII\ regions and they do not seem to follow the
trend of N/O vs.\ O/H which marks the secondary behaviour of nitrogen. On the
other hand, the S/O ratios span a very narrow range between 0.6 and 0.8 of the
solar value. 

When compared to high metallicity disc \HII\ regions, CNSFRs show values of
the O$_{23}$ and the N2 parameters whose distributions are shifted to lower
and higher values respectively, hence, even though their derived oxygen and
sulphur abundances are similar, higher values would in principle be obtained
for the CNSFRs  if pure empirical methods were used to estimate
abundances. CNSFRs also show lower ionization parameters than their disc
counterparts, as derived from the [S{\sc ii}]/[S{\sc iii}] ratio. Their
ionization structure also seems to be different with CNSFRs showing radiation
field properties more similar to \HII\ galaxies than to disc high metallicity
\HII\ regions. The possible contamination of their spectra from hidden low
luminosity AGN and/or shocks, as well as the probable presence  of more than
one velocity component in the ionized gas corresponding to kinematically
distinct systems, should be further investigated.

\addcontentsline{toc}{section}{\numberline{}Bibliography}

\bibliographystyle{astron}
\bibliography{tesis}

\chapter{Conclusions and future work}
\label{genconc}

In this Chapter we summarize the main conclusions of this work and envisage
future ramifications that stem from it.

\medskip

In the present thesis we have studied large scale star formation processes in
galaxies in two very different environments that can be considered extreme in
terms of metal content: relatively compact \HII\ galaxies and circumnuclear
star-forming regions in spiral galaxies. \HII\ galaxies are among the  most
metal-poor objects known while the circumnuclear regions of galaxies are
expected to be very high abundance sites due to their location in the inner
part of galactic bulges. Therefore, even when their emission line spectra are
similar, as they both correspond to ionized gases, they show important
differences. In the first case the spectra show a high excitation, being
dominated by strong forbidden oxygen lines; in the second, the spectra show a
considerably lower excitation, weak oxygen lines and more prominent nitrogen
and sulphur lines when compared to lower abundance gas. In fact, for \HII\
galaxies, [O{\sc iii}]\,5007/H$\beta$\,$>$\,4, and reach values as high as 6
for the objects in this study, while the [O{\sc iii}] emission lines in CNSFRs
are generally very weak ([O{\sc iii}]\,5007/H$\beta$\,$\ll$\,1). 

We have analyzed the physical properties of the gas in 10 \HII\ galaxies and
12 CNSFRs in 3 early type spirals using intermediate resolution spectroscopy
obtained with double arm spectrographs which provide the best
spectrophotometry attainable, since it allows to cover the whole spectral
range from 3600 to 10000 \AA\ simultaneously in the same position in the
object. We have also investigated the kinematical properties of the gas and
stars in a total of 17 CNSFRs using high dispersion spectroscopy. In some
cases, we have used complementary photometric data acquired from data archives
and from the literature. 

The first set of data has been used to infer the properties of the ionizing
stellar populations using photo-ionization and stellar population synthesis
models, including the photometric mass of the ionizing clusters. The second
set of data has been used mainly to determine the dynamical masses of the
CNSFRs, but has yielded some very interesting information about the gas
kinematics in the galactic circumnuclear environments. 

The main findings of our work are summarized below.

\bigskip

\label{intermediate dispersion, HII galaxies}

The star formation processes in \HII\ galaxies are found to occur in low
density environments.  In all the observed galaxies the electron densities
have been found to be lower than 150\,cm$^{-3}$, well below the critical
density for collisional de-excitation. In contrast, in the CNSFRs the
densities are relatively high, with typical values around 300\,cm$^{-3}$, and
in some cases as high as 440\,cm$^{-3}$. These values are higher than those
usually derived for disc \HII\ regions, although still below the critical
value for collisional de-excitation. 

For all our observed \HII\ galaxies we have measured at least four line
temperatures: T$_e$([O{\sc iii}]), T$_e$([S{\sc iii}]), T$_e$([O{\sc ii}]) and
T$_e$([S{\sc ii}]) reaching accuracies of  1\,\% in T$_e$([O{\sc iii}]), 3\,\%
in T$_e$([O{\sc ii}]) and 5\,\% in T$_e$([S{\sc iii}]) in the best
cases. These accuracies are expected to improve as better calibrations based
on more precise measurements, both on electron temperatures and  densities,
are produced.  The temperatures related to the high ionization zone are
between $\sim$\,11000 and $\sim$\,14500\,K, and in the low ionization zone
between $\sim$\,9000 and $\sim$\,13500\,K. This means that the observed
objects are of intermediate excitation, in which the cooling is dominated by
the oxygen forbidden lines.

For three of the \HII\ galaxies of the sample, we measured the Balmer
continuum temperature [T(Bac)] and estimated the temperature fluctuations
according to the scheme introduced by M. Peimbert.  Only one of the objects
shows significant temperature fluctuations which, if taken into account,
would lead to higher oxygen abundances than derived by about 0.20\,dex. 

The temperature measurements for the observed \HII\ galaxies and a careful and
realistic treatment of the observational errors yielded total oxygen
abundances between 7.93 and 8.19 (Z/Z$_\odot$ between $\sim$\,0.19 and
$\sim$\,0.34, using 12+log(O/H)$_{\odot}$\,=\,8.66\,$\pm$\,0.05 from
\citeplain{2005ASPC..336...25A}), with accuracies between 1 and 9\,\%. When
compared to a large sample of \HII\ galaxies, the observed ones show higher
N/O abundance ratios than the average. These ratios would be even higher if the
[O{\sc ii}] temperatures were derived with the use of photo-ionization
models. 

For some elements, like sulphur, neon and argon, the unseen contribution  of
different ionization species can be a source of appreciable error in the
determination of chemical abundances. Using a grid of photo-ionization models
that we have computed, we have calculated new ionization correction factors
for Ne and Ar and we have studied the behaviour of the Ne/O and Ar/O abundance
ratios with metallicity for our observed objects and for a heterogeneous
sample of 853 objects including \HII\ galaxies, giant \HII\ regions in
galactic discs and some \HII\ regions in the Milky Way, but excluding CNSFRs.
A constant value for Ne/O has been found which points to the same
nucleosynthesis sites for both elements  in these objects. The situation
regarding S/O is unclear, since the sample for which the necessary data exist
is still small. For the objects studied in this work, the derived S/O ratios
are consistent with the solar value within the observational errors, except
for \whtunoc, \unoc\ and \tresc\ for  which S/O is lower by a factor of about
2.7 for the first and 1.8 for the other two. On 
the other hand, there seems to be some evidence for the existence of negative
radial gradients of  Ar/O over the discs of some nearby spirals. The origin of
this gradient has to be explored. 

\label{intermediate dispersion, CNSFRs}

The high average metallicity of the CNSFRs require a different methodology for
abundance determinations involving empirical calibrators. We have developed a
scheme for abundance determination using the oxygen and sulphur lines, these
latter ones being much stronger than the first in this kind of objects.  The
scheme involves the calibration of T$_e$([S{\sc iii}]) in terms of the
SO$_{23}$ parameter, and its use in combination with the observed line
intensities of the different elements to derive the ionic abundances. This
scheme has been found "a posteriori" to yield results very similar to those
obtained by L. Pyliugin using oxygen and nitrogen lines. 

In all the observed CNSFRs the estimated T$_e$([S{\sc iii}]) are substantially
lower than for the \HII\ galaxies, ranging between $\sim$\,5400 and
$\sim$\,6700\,K. For one of the observed regions, the [S{\sc
iii}]\,$\lambda$\,6312\,\AA\ auroral emission line was detected providing a
measured electron temperature of T$_e$([S{\sc
    iii}])\,=\,8400\,$^{+4650}_{-1250}$\,K, consistent with our derived
empirical calibration, although slightly higher than predicted. 

The total oxygen abundances derived for our regions are between 8.42 and 8.85
in terms of 12+log(O/H) (Z/Z$_\odot$ between $\sim$\,0.6 and $\sim$\,1.6). In
contrast with the case of \HII\ galaxies for which the O/H abundances are
dominated by O$^{2+}$/H$^+$,  in all the observed CNSFRs,  the O/H abundance
is dominated by the O$^+$/H$^+$ contribution, as it is also the case for high
metallicity disc \HII\ regions. For our observed circumnuclear regions,
however, also the  S$^+$/S$^{2+}$ ratio is larger than one, different from the
case of high metallicity disc \HII\ regions for which, in general, the sulphur
abundances are dominated by S$^{2+}$/H$^+$.

Regarding abundance ratios, for the CNSFRs, the derived N/O ratios are in
average larger than those found in high metallicity disc \HII\ regions and
they do not seem to follow the trend of N/O vs.\ O/H which marks the secondary
behaviour of nitrogen. The S/O ratios span a very narrow range between 0.6 and
0.8 of the solar value.

\label{stellar populations}

In spite of the differences in the environments where the star formation
processes are taking place in the \HII\ galaxies and in the CNSFRs, their
ionization structure and the temperature of their ionizing radiation field, as
measured from their $\eta$ and $\eta$' values, are very similar, yet very
different from those of high metallicity disc \HII\ regions.

Taking into account the distances of our \HII\ galaxies, measured as part of
the HST Key Project on the Extragalactic Distance Scale
\cite{2000ApJ...529..786Mtot}, their H$\alpha$ luminosities are in the range
between 8.52\,$\times$\,10$^{38}$ and
1.36\,$\times$\,10$^{42}$\,erg\,s$^{-1}$, typical of this kind of objects
\cite{2006MNRAS.365..454H}. This luminosity range overlaps with the values
derived for the CNSFRs, ranging between 7.50\,$\times$\,10$^{38}$ and
1.44\,$\times$\,10$^{40}$\,erg\,s$^{-1}$, also typical of this kind of objects
\cite{tesismar}. From these H$\alpha$ luminosities, and using the calibration
by  \citetex{1989ApJ...344..685K}, we find SFRs for our observed \HII\ galaxies
in the range between 0.007 and 10.7\,M$_\odot$\,yr$^{-1}$. In the case of the
studied CNSFRs these values range between 0.006 and
0.11\,M$_\odot$\,yr$^{-1}$, clearly overlapping with those derived for entire
\HII\ galaxies.

There is clear evidence, both in the \HII\ galaxies and in the CNSFRs, of the
presence of  composite stellar populations, a young population responsible for
the ionization of the gaseous medium and an older underlying population
causing the absorption features easily observable in the hydrogen lines and,
in some cases, in the helium ones. The presence of this older population
reduces the equivalent widths of the emission lines increasing the continuum
and absorbing some of the lines. This effect is largest in
CNSFRs. Furthermore, the composite nature of these regions means that star
formation in the rings is a process that has taken place over time periods
much longer than those implied by the properties of the ionized gas.

\label{high dispersion, kinematics}

Our high dispersion observations have allowed the measurement of the line
widths of the gas and stars in CNSFRs in three different spiral galaxies:
NGC\,2903, NGC\,3310 and NGC\,3351. From the stellar absorption lines,
corresponding to the CaT at $\sim$\,8500\,\AA, we have derived the dynamical
masses of the stellar clusters, applying the virial theorem under the
assumption that the systems are spherically symmetric, gravitationally bound
and have isotropic velocity distribution.  This provides upper limits to the
masses inside the half light radius (typically between 3 and 5\,pc) for each
observed knot. Using the measured  stellar velocity dispersions
(31\,-\,66\,km\,s$^{-1}$) we obtained masses for the individual clusters
between 1.4\,$\times$\,10$^6$ and 1.1\,$\times$\,10$^7$\,M$_\odot$. The total
dynamical masses of the CNSFRs are taken as the ``sum'' of these individual
stellar clusters and are between 4.9\,$\times$\,10$^6$ and
1.9\,$\times$\,10$^8$\,M$_\odot$. 

These masses are about 10 times larger than those inferred from the observed
number of ionizing photons emerging from the regions, using simple and robust
stellar population models under the assumption of no photon leakage and
without taking into account any absorption by dust. 

An unexpected result from our study has been the finding of the existence of
more than one velocity component in the ionized gas when Gaussian fits are
performed, both in the hydrogen recombination  and the [O{\sc iii}] lines. A
component with a velocity dispersion substantially lower than that measured
from the stellar lines, is clearly present in the hydrogen recombination lines
and also seems to be present in the [O{\sc iii}] lines. The narrow component
of the two-component Gaussian fits seems to be relatively constant for all the
studied CNSFRs, with an estimated mean value close to 25\,km\,s$^{-1}$. This
narrow component could be identified with ionized gas in a rotating disc,
while the stars and the fraction of the gas (responsible for the broad
component) related to the star-forming regions would be mostly supported by
dynamical pressure \cite{2004A&A...424..447P}. 

The fact that the emission lines show velocity components corresponding to
kinematically distinct systems, could affect somewhat the results derived from
the observations described above, among others, the gas abundance
determinations. Also, masses derived from the H$\beta$ velocity dispersions
under the assumption of a single component for the gas would have been
underestimated by factors of approximately 2 to 4. 

The rotation curve of the central zones of the studied early type spiral
galaxies seem to have the turnover points at the same position as the
star-forming ring, as found in other galaxies
\cite{1988ApJ...334..573T,1999ApJ...512..623D}, and the velocity distribution
is consistent with that expected  for this type of galaxies
\cite{1987gady.book.....B}.

In summary, we have identified two star formation sites which differ widely in
metal content. We conclude  that `massive' star formation that occurs in a high
density, high metallicity environment, like that encountered in CNSFRs, takes
place in systems that comply to the definition of super stellar clusters and
that are arranged in much larger star forming complexes. These complexes 
have H$\alpha$
luminosities, and therefore masses of ionizing stars, that overlap at the
lower end with those found in \HII\ galaxies. The fact that their ionization
structure and the temperature of their ionizing radiation field are very
similar, point to stellar clusters have the same equivalent effective
temperature in these two environments. This is contrary to what is expected from stellar evolution
models which predict lower stellar effective temperatures in high metallicity
regions. This point deserves further study. 

Star forming regions in the two environments show a contribution by underlying
stellar populations. This contribution is larger in CNSFRs and also the
stellar population itself seems to be more evolved as evidenced by the
presence of red supergiants which substantially contribute to the CaT
lines. The fact that the N/O values are much higher in CNSFRs than in \HII\
galaxies and other high metallicity regions also point to the possibility of
chemical evolution and self-enrichment in these regions.

\section*{Future work}

In the course of this study, we have identified several results that inspired
us for further investigation outside the scope of this work. This will entail
the obtention of high quality optical spectroscopy observations   with high
spatial and spectral resolution both  with integral spectroscopy units and
two-arm spectrographs and longslit mode, using large telescopes. These
observations should be complemented with wide electromagnetic spectral
coverage. 

\subsection*{\HII\ galaxies}

One of our immediate aims is to try to {\bf expand the sample of \HII\
  galaxies} for which precise abundances can be obtained. In order to do that,
double arm spectrographs and intermediate spectral resolution are
required. The sample should be designed to encompass objects at the low and
high electron temperature ends. This whole temperature coverage would provide
a much needed constraint of the temperature structure in \HII\ galaxies. 

One important issue is the study of the {\bf spatial variations of
  temperatures and abundances across a given galaxy} since, in many cases,
blue compact galaxies manifest themselves not so compact when observed at high
spatial resolution. There are several nearby galaxies that can be very well
mapped using Integral Field Unit spectrographs attached to 8\,m telescopes
such as GEMINI or VLT. Although these cameras have a small field of view,
these telescopes have great light collector areas and an excellent spatial
resolution. Even the PMAS-IFU instrument at the 3.5\,m telescope of CAHA,
which has a larger field of view but lower spatial resolution and light
collector area, can perform a very complete and excellent work.  

One aspect that deserves deeper study is whether or not {\bf temperature
fluctuations} are present in \HII\ galaxies. Another way to investigate this 
phenomenon is by comparing the abundances obtained for a particular ion both
from collisionally excited and recombination lines. To this end, we plan to
secure deeper observations of nearby and bright \HII\ galaxies in order to
derive their abundances using recombination lines which are observable, but
not measurable, in our spectra. 

Regarding stellar populations, we are at present carrying out a project to
study the properties of the brightest knots of star formation in the  \HII\
galaxies studied in this work using {\bf detailed photo-ionization models}. In
this way we will study the properties of the ionizing stellar population,
including, if present, the Wolf-Rayet stars following the analysis methodology
explained in \citetex{2007MNRAS.377.1195P}.

\subsection*{Circumnuclear star-forming regions}

\begin{figure}
\centering
\includegraphics[width=.98\textwidth,angle=0]{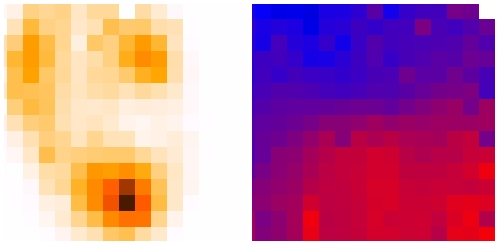}
\caption[H$\alpha$ map of the circumnuclear region of NGC\,3351 and its radial
  velocity map]{Left: H$\alpha$ map of the circumnuclear region of NGC\,3351 
  obtained in March 2007 with PMAS and a single 1200\,s exposure centered at
  6551\,\AA. The main CNSFRs can be identified. Right: radial velocity map
  derived from the data showing the inner galaxy rotation. North is to the top
  and East to the left.}
\label{map}
\end{figure}

Given the importance of establishing the origin of the different kinematical
components of the gas in the circumnuclear regions of galaxies and their
influence in derived quantities that depend on line intensity ratios, we intend
to conduct a campaign to study the velocity field of circumnuclear star forming
rings in early type spirals, by mapping the gas and stellar velocity fields in
these regions with good {\bf spatial sampling}. We have already started this
program using {\bf PMAS} (lens array configuration) at the 3.5\,m telescope of
CAHA. 
We also plan to measure the widths of the [S{\sc iii}] lines at
$\lambda$\,9069, 9532\,\AA, stronger than [O{\sc iii}] in this kind of
objects. Figure \ref{map} presents an H$\alpha$ flux map where the main CNSFRs
can be seen and a radial velocity map showing the inner galaxy rotation, as
part of this project in progress. 

\begin{figure}
\centering
\includegraphics[width=0.4\textwidth]{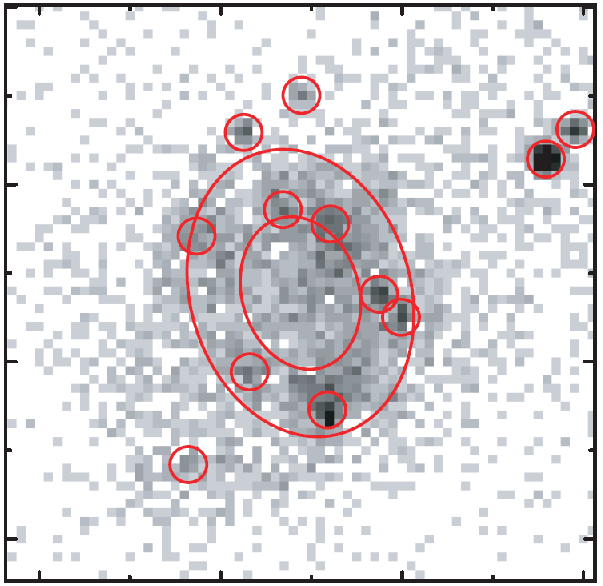}
\includegraphics[width=0.4\textwidth]{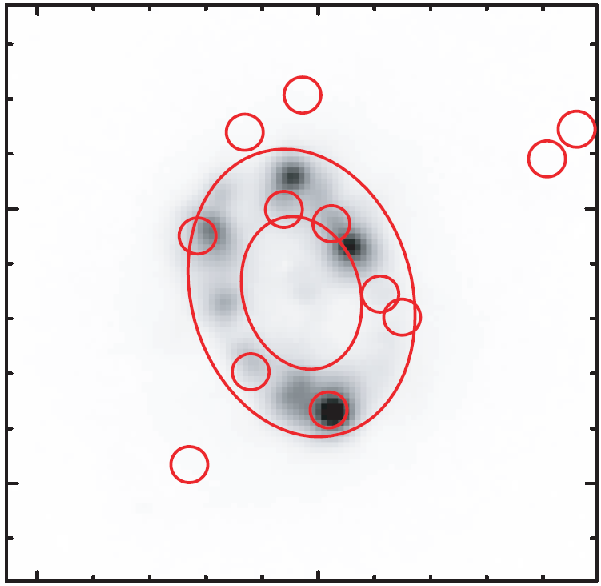}
\caption[Chandra and CTIO-H$\alpha$ images of the central
  32\arcsec$\times$32\arcsec\ region of NGC\,3351]{Images of the central
  32\arcsec$\times$32\arcsec\ region of 
  NGC\,3351 from Swartz et al.\ (2006). Left: {\bf Chandra}
  0.5-8.0\,keV. Right: {\bf CTIO} 4.0\,m Mosaic2 continuum-subtracted
  H$\alpha$. Circles are 1\arcsec\ in radius and denote X-ray source
  positions. Ellipses trace the circumnuclear ring.}
\label{xray}
\end{figure}

The study of the composite stellar populations in CNSFRs constitute a
challenge. Most stellar population models point to a narrow interval of ages
for the ionizing population. However, underlying stellar populations are
clearly evident. In a effort to identify stellar population components not
easily detected in the optical, we have designed observations to reveal the
existence of {\bf supernova remnants (SNRs) and compact \HII\ regions}
associated with evolved and very young stellar populations, respectively, and
closely related to the presence of recent star-formation. Time has been
awarded to conduct these observations  with {\bf MERLIN}, the Multi-Element
Radio Linked Interferometer Network, an array of radio telescopes distributed
around Great Britain, with separations of up to 217\,km.
The requested observations have recently been completed and are being
processed. The sub-arcsecond resolution radio observations of NGC\,3351 will
be sensitive enough not only to detect the already known \HII\ regions and
SNRs, but also to discover new populations of ultra-compact \HII\ regions and
SNRs, which are expected in regions of recent star formation activity. It is
also possible that some of the X-ray sources with no obvious counterparts at
other wavelengths (see Fig.\ \ref{xray}) could be traced in the radio
continuum. With such high resolution observations we will then be able to
constrain the origin of the high energy emission from the point sources
recently detected by \citetex{2006ApJ...647.1030S} using {\bf Chandra}
observations.

We also plan to perform {\bf abundance determinations of the very hot gas} in
the central $\sim$\,1\,kpc of two of the circumnuclear rings studied in the
present thesis combining the simultaneous observations obtained using the {\bf
RGS} and {\bf EPIC-pn} cameras on board the {\bf XMM-Newton} Satellite in a
similar way to that followed by \citetex{2004ApJ...606..862O}. Observations
are in progress for the circumnuclear regions of NGC\,3310. For this galaxy,
the analysis of the ionized gas in the circumnuclear \HII\ regions yields an
oxygen abundance lower than expected from empirical abundance
indicators. These findings point to an effect similar to what is found for
M82, i.e.\ a deficiency of light alpha elements (O, Ne) in the central
regions. If this is common among regions dominated by recent star formation,
the oxygen abundances found for the warm ionized gas might not be
representative of the true metal content of these regions. This could have a
profound effect on the abundance calibrations leading to fundamental relations
like the Mass-Metallicity and Luminosity-Metallicity relations. 

Finally, combining photo-ionization models and synthetic stellar populations
we will study in detail the {\bf stellar population of the CNSFRs}, taking
into account all the physical parameters derived in this thesis, the composite
stellar population, the Wolf-Rayet features observed in their spectra and the
low equivalent widths of the emission lines.

\addcontentsline{toc}{section}{\numberline{}Bibliography}

\bibliographystyle{astron}
\bibliography{tesis}

\setcounter{chapter}{5}

\chapter{Conclusiones}
\label{genconccas}

En este Cap\'itulo resumimos las principales conclusiones de este trabajo.

\medskip

En la presente tesis se presenta un  estudio de procesos de formaci\'on estelar a gran
escala en galaxias en dos entornos muy diferentes que pueden considerarse
extremos en t\'erminos de contenido met\'alico: galaxias \HII\ relativamente
compactas y regiones circunnucleares de formaci\'on estelar (CNSFRs) en galaxias
espirales. Las galaxias \HII\ est\'an entre los objetos m\'as pobres en
metales conocidos, mientras que se espera que las regiones circunnucleares de
galaxias sean sitios de abundancias muy altas debido a su localizaci\'on en la
parte interna de los bulbos gal\'acticos. Por lo tanto, a\'un cuando sus
espectros de l\'ineas de emisi\'on som similares, ya que ambos corresponden a
gas ionizado, muestran importantes diferencias. En el primer caso los
espectros muestran alta excitaci\'on, estando dominados por fuertes l\'ineas
prohibidas de ox\'igeno; en el segundo, los espectros muestran una
excitaci\'on considerablemente m\'as baja, l\'ineas d\'ebiles de ox\'igeno y
l\'ineas de nitr\'ogeno y azufre m\'as prominentes, cuando se comparan con
gas de abundancia m\'as baja. De hecho, para las galaxias \HII, [O{\sc
iii}]\,5007/H$\beta$\,$>$\,4, y alcanza valores tan altos como 6 para los
objetos en este estudio, mientras que las l\'ineas de emisi\'on de [O{\sc 
iii}] en CNSFRs son generalmente d\'ebiles ([O{\sc
iii}]\,5007/H$\beta$\,$\ll$\,1).

Hemos analizado las propiedades f\'isicas del gas en 10 galaxias \HII\ y en 12
CNSFRs de 3 galaxias espirales tempranas utilizando espectroscop\'ia de
resoluci\'on intermedia obtenida con espectr\'ografos de doble brazo que
proporcionan
la mejor espectrofotometr\'ia alcanzable, ya que permite cubrir completamente
el rango espectral desde 3600 a 10000 \AA\ simult\'aneamente en la misma
posici\'on en el objeto. Tambi\'en hemos investigado las propiedades
cinem\'aticas del gas y las estrellas en un total de 17 CNSFRs utilizando
espectroscop\'ia de alta dispersi\'on. En algunos casos, hemos utilizado datos
fotom\'etricos complementarios obtenidos de bases de datos y de la
literatura. 

El primer conjunto de datos se ha utilizado para inferir las propiedades de
las poblaciones estelares ionizantes, incluyendo las masas fotom\'etricas de los
c\'umulos ionizantes, usando modelos de foto-ionizaci\'on y de
s\'intesis de poblaciones estelares. El segundo conjunto de datos ha sido 
utilizado principalmente para determinar las masas din\'amicas de las CNSFRs, pero adem\'as
ha aportado informaci\'on muy interesante sobre la
cinem\'atica del gas en el entorno circunnuclear gal\'actico.

Los principales hallazgos de nuestro trabajo est\'an resumidos a
continuaci\'on. 

\bigskip

\label{intermediate dispersion, HII galaxies}

Se encuentra que los procesos de formaci\'on estelar en galaxias \HII\ ocurren
en ambientes de baja densidad. En todas las galaxias observadas, la densidad
electr\'onica que hemos encontrado es menor que 150\,cm$^{-3}$, menor que la densidad cr\'itica necesaria para la desexcitaci\'on colisional. En contraste,
en las CNSFRs la densidad es relativamente alta, con valores t\'ipicos
alrededor de 300\,cm$^{-3}$, y en algunos casos tan altos como
440\,cm$^{-3}$. Estos valores son mayores que los que se encuentran normalmente
en regiones \HII\ de disco, aunque est\'an todav\'ia por debajo del valor cr\'itico  para la
desexcitaci\'on colisional. 

Para todas las galaxias \HII\ observadas hemos medido al menos cuatro
temperaturas de l\'inea: T$_e$([O{\sc iii}]), T$_e$([S{\sc iii}]), T$_e$([O{\sc
ii}]) y T$_e$([S{\sc ii}]) alcanz\'andose precisiones del 1\,\% en T$_e$([O{\sc
iii}]), 3\,\% en T$_e$([O{\sc ii}]) y 5\,\% en T$_e$([S{\sc iii}]) en los
mejores casos. Se espera que estas precisiones mejoren cuando se realicen
mejores calibraciones basadas en medidas m\'as precisas, tanto de temperaturas
electr\'onicas como de densidades. Las temperaturas relacionadas con la zona
de alta ionizaci\'on est\'an entre $\sim$\,11000 y $\sim$\,14500\,K, y en la
zona de baja ionizaci\'on entre $\sim$\,9000 y $\sim$\,13500\,K. Esto
significa que los objetos observados son de excitaci\'on intermedia, en los
que el enfriamiento esta dominado por las l\'ineas prohibidas del ox\'igeno. 

Para tres de las galaxias \HII\ de la muestra, hemos medido la temperatura del
continuo del Balmer [T(Bac)]  y estimado las fluctuaciones de temperatura de
acuerdo con el esquema introducido por M. Peimbert. S\'olo uno de los objetos
muestra fluctuaciones de temperatura significativas que, si se tienen en
cuenta, podr\'ian conducir a abundancias de ox\'igeno alrededor de 0.20\,dex
mayores que las derivadas.

Las medidas de temperatura para las galaxias \HII\ observadas y un
tratamiento cuidadoso y realista de los errores observacionales, conduce a
abundancias totales de ox\'igeno entre 7.93 y 8.19 (Z/Z$_\odot$ entre
$\sim$\,0.19 y $\sim$\,0.34, utilizando
12+log(O/H)$_{\odot}$\,=\,8.66\,$\pm$\,0.05 de
\citeplain{2005ASPC..336...25A}), con una precisi\'on de entre 1 y 
9\,\%. En comparaci\'on con una muestra mayor de galaxias \HII, las observadas
muestran cocientes de abundancias de N/O mayores que el valor medio. Estos
cocientes podr\'ian ser a\'un mayores si se derivan las temperaturas de [O{\sc
ii}] utilizando los modelos de foto-ionizaci\'on.

Para algunos elementos, como el azufre, ne\'on y arg\'on, la contribuci\'on de
las diferentes especies ionizadas que no se ven puede ser una fuente de error
apreciable en la determinaci\'on de las abundancias qu\'imicas. Utilizando una
red de modelos de foto-ionizaci\'on que hemos calculado, hemos derivado
nuevos factores de correcci\'on por ionizaci\'on para el Ne y el Ar y hemos
estudiado el comportamiento de los cocientes de abundancias Ne/O y Ar/O con la
metalicidad para los objetos observados y para un conjunto heterog\'eneo
de 853 objetos incluyendo galaxias \HII\, regiones \HII\ gigantes en discos
gal\'acticos y algunas regiones \HII\ en la V\'ia L\'actea, pero excluyendo
las CNSFRs. Se encuentra un valor constante para el cociente Ne/O que apunta a que ambos elementos se sinteticen en el mismo tipo de estrellas. Respecto al cociente S/O, la situaci\'on no es clara, ya que la muestra para la
que existen los datos necesarios para su c\'alculo, es a\'un muy peque\~na. Para los objetos
estudiados en este trabajo, las cocientes de S/O derivados son consistentes
con los solares dentro de los errores observacionales, excepto para
\whtunoc, \unoc\ y \tresc\ para los cuales S/O es menor por un factor de
alrededor de 2.7 para el primero y 1.8 para los otros dos. Por otro lado,
parece haber alguna evidencia de la existencia de un 
gradiente radial negativo del Ar/O sobre los discos de algunas espirales
cercanas. El origen de este gradiente est\'a por explorar.

\label{intermediate dispersion, CNSFRs}

La alta metalicidad promedio de las CNSFRs requiere una metodolog\'ia
diferente para la determinaci\'on de abundancias, involucrando calibradores
emp\'iricos. Hemos desarrollado un esquema para la determinaci\'on de
abundancias utlizando las l\'ineas del ox\'igeno y del azufre, siendo estas \'ultimas
mucho m\'as intensas que las primeras en esta clase de objetos. El
esquema propone la calibraci\'on de T$_e$([S{\sc iii}]) en t\'erminos del
par\'ametro SO$_{23}$, y su uso en combinaci\'on con las intensidades de las
l\'ineas observadas de los diferentes elementos para derivar las abundancias
i\'onicas. "A posteriori" se ha encontrado que este esquema produce
resultados muy similares a los obtenidos por L. Pyliugin  utilizando l\'ineas
de ox\'igeno y nitr\'ogeno.

En todas las CNSFRs observadas, las T$_e$([S{\sc iii}]) estimadas son
sustancialmente menores que en las galaxias \HII, estando entre $\sim$\,5400 y
$\sim$\,6700\,K. Para una de las regiones observadas, la l\'inea de emisi\'on
auroral de [S{\sc iii}]\,$\lambda$\,6312\,\AA\ fue detectada proporcionando una medida
de la temperatura electr\'onica de T$_e$([S{\sc
iii}])\,=\,8400\,$^{+4650}_{-1250}$\,K, consistente con nuestra calibraci\'on
emp\'irica, aunque ligeramente mayor que la predicha.

Las abundancias totales de ox\'igeno derivadas para nuestras regiones est\'an
entre 8.42 y 8.85 en t\'erminos de 12+log(O/H) (Z/Z$_\odot$ entre $\sim$\,0.6
y $\sim$\,1.6). En contraste con el caso de las galaxias \HII\ para las cuales
las abundancias de O/H est\'an dominadas por O$^{2+}$/H$^+$, en todas las
CNSFRs, la abundancia de O/H est\'a dominada por la contribuci\'on de
O$^+$/H$^+$, como tambi\'en es el caso para las regiones \HII\ de alta
metalicidad. Para nuestras regiones circunnucleares, sin embargo, tambi\'en
el cociente de S$^+$/S$^{2+}$ es mayor que uno, a diferencia del
caso de las regiones \HII\ de alta metalicidad de disco para las cuales, en
general, las abundancias de azufre est\'an dominadas por  S$^{2+}$/H$^+$.

Con respecto a los cocientes relativos de abundancias, para las CNSFRs, los cocientes de
N/O derivados son, en promedio, mayores que los encontrados en las regiones
\HII\ de alta metalicidad de disco y no parecen seguir la tendencia de N/O
vs.\ O/H que marca el comportamiento secundario del nitr\'ogeno. Los cocientes
de S/O se distribuyen en un intervalo muy estrecho entre 0.6 y 0.8 del valor
solar. 

\label{stellar populations}

A pesar de las diferencias en los entornos donde tienen lugar los procesos de
formaci\'on estelar en las galaxias \HII\ y en las CNSFRs, sus estructuras de
ionizaci\'on y las temperaturas de sus campos de radiaci\'on ionizantes,
medidos a partir de sus valores de $\eta$ y $\eta$', son muy similares, aunque
muy diferentes de los de las regiones \HII\ de alta metalicidad de disco.

Teniendo en cuenta la distancia a nuestras galaxias \HII, medidas como parte
del HST Key Project on the Extragalactic Distance Scale
\cite{2000ApJ...529..786Mtot}, sus luminosidades H$\alpha$ est\'an en el rango
entre 8.52\,$\times$\,10$^{38}$ y 1.36\,$\times$\,10$^{42}$\,erg\,s$^{-1}$,
t\'ipico de esta clase de objetos \cite{2006MNRAS.365..454H}. Este rango en
luminosidades se superpone con los valores derivados para las CNSFRs, que van
de 7.50\,$\times$\,10$^{38}$ a 1.44\,$\times$\,10$^{40}$\,erg\,s$^{-1}$,
tambi\'en t\'ipico de esta clase de objetos \cite{tesismar}. A partir de estas
luminosidades H$\alpha$, y utilizando la calibraci\'on dada por
\citetex{1989ApJ...344..685K}, encontramos tasas de formaci\'on estelar (SFRs)
en el rango entre 0.007 y 10.7\,M$_\odot$\,yr$^{-1}$. En el caso de las CNSFRs
estudiadas estos valores tienen un rango entre 0.006 y
0.11\,M$_\odot$\,yr$^{-1}$, superponi\'endose claramente con los valores
derivados para las galaxias \HII\ en su totalidad.

Hay una clara evidencia, tanto en las galaxias \HII\ como en las CNSFRs, de la
presencia de poblaciones estelares compuestas: una poblaci\'on joven
responsable de la ionizaci\'on del medio gaseoso y una poblaci\'on subyacente, m\'as
vieja, responsable de las caracter\'isticas de absorci\'on f\'acilmente observables en las
l\'ineas del hidr\'ogeno y, en algunos casos, en las del helio. La presencia
de esta poblaci\'on m\'as vieja reduce los anchos equivalentes de las l\'ineas de
emisi\'on incrementando el continuo y absorbiendo algunas de las
l\'ineas. Este efecto es mayor en las CNSFRs. M\'as a\'un, la naturaleza
compuesta de estas regiones significa que la formaci\'on estelar en los
anillos es un proceso que ha tenido lugar durante periodos de tiempo mucho
m\'as largos que los que se deducen de las propiedades del gas ionizado.

\label{high dispersion, kinematics}

Nuestras observaciones de alta disperis\'on nos han permitido medir los anchos
de las l\'ineas del gas y de las estrellas en CNSFRs en tres diferentes
galaxias espirales: NGC\,2903, NGC\,3310 y NGC\,3351. A partir de las l\'ineas
estelares de absorci\'on del CaT en $\sim$\,8500\,\AA, hemos derivado las
masas din\'amicas de los c\'umulos estelares, aplicando el teorema del virial y
suponiendo que los sistemas son esf\'ericamente sim\'etricos, est\'an ligados
gravitatoriamente y tienen una distribuci\'on isotr\'opica de
velocidades. Esto proporciona l\'mites superiores para las masas dentro de los
radios desde donde se emite la mitad de la luz (t\'ipicamente entre 3 y 5\,pc) para cada n\'odulo
observado. Utilizando las medidas de las dispersiones de velocidades estelares
(31\,-\,66\,km\,s$^{-1}$) obtuvimos masas para los c\'umulos individuales
entre 1.4\,$\times$\,10$^6$ y 1.1\,$\times$\,10$^7$\,M$_\odot$. Las masas
din\'amicas totales de las CNSFRs se toman como la ``suma'' de estos
c\'umulos estelares individuales y est\'an entre 4.9\,$\times$\,10$^6$ y
1.9\,$\times$\,10$^8$\,M$_\odot$. 

Estas masas son alrededor de 10 veces m\'as altas que las que se infieren a
partir del n\'umero de fotones ionizantes que emergen de las
regiones, utilizando modelos simples y robustos de poblaciones estelares,
suponiendo que no hay p\'erdida de fotones y sin tener en consideraci\'on la
absorci\'on debida al polvo.

Un resultado inesperado a partir de nuestro estudio ha sido el hallazgo de la
existencia de m\'as de una componente en la velocidad del gas ionizado
cuando se realiz\'o un ajuste de componentes gausianas, tanto en las l\'ineas de
recombinaci\'on del hidr\'ogeno como en las del [O{\sc iii}]. Una componente
con una dispersi\'on de velocidades sustancialmente menor que la medida a
partir de las l\'ineas estelares, est\'a claramente presente en las l\'ineas
de recombinaci\'on de hidr\'ogeno y tambi\'en parece estar presente en las
l\'neas de [O{\sc iii}]. La componente estrecha de los ajustes de
dos componentes gaussianas parece ser relativamente constante para todas las
CNSFRs estudiadas, con un valor promedio estimado pr\'oximo a los
25\,km\,s$^{-1}$. Esta componente estrecha podr\'ia identificarse con el
gas ionizado en un disco rotante, mientras que las estrellas y la fracci\'on
del gas (responsable de la componente ancha) relacionado con las regiones de
formaci\'on estelar estar\'ian principalmente soportadas por presi\'on
din\'amica \cite{2004A&A...424..447P}. 

El hecho de que las l\'ineas de emisi\'on muestren componentes de velocidad
correspondientes a sistemas cinem\'aticamente distintos, podr\'ia afectar algo a
los resultados derivados a partir de las observaciones descritas
anteriormente, entre otras, a las determinaciones de las abundancias del
gas. Adem\'as, las masas derivadas a partir de las dispersiones de velocidades
de H$\beta$ bajo la suposici\'on de una \'unica componente para el gas podr\'ia
haber sido subestimada por un factor entre 2 y 4 aproximadamente.

La curva de rotaci\'on de las zonas centrales de las galaxias espirales de
tipo temprano estudiadas parecen tener m\'aximos y m\'inimos relativos en la misma
posici\'on que los anillos de formaci\'on estelar, como se encuentra en
otras galaxias \cite{1988ApJ...334..573T,1999ApJ...512..623D}, y la
distribuci\'on de velocidades es consistente con la esperada para este tipo de
galaxias \cite{1987gady.book.....B}.

En resumen, se han identificado dos lugares de formaci\'on estelar que difieren
ampliamente en su contenido de metales. Concluimos que la formaci\'on estelar
`masiva' que ocurre en un medio de alta densidad y alta metalicidad, como el
que se encuentra en las CNSFRs, tiene lugar en sistemas que satisfacen la
definici\'on de super c\'umulos estelares y que forman complejos
de formaci\'on estelar mayores. Estos complejos tienen luminosidades H$\alpha$, y por
lo tanto masas de estrellas ionizantes, que se superponen en el extremo
inferior con las encontradas en las galaxias \HII. El hecho de que las
estructuras de ionizaci\'on y las temperaturas de los campos de radiaci\'on
ionizantes de las galaxias \HII\ y de las CNSFRs sean muy similares, apunta a que sus c\'umulos ionizantes tienen la
misma temperatura equivalente efectiva. Esto es contrario a lo que se espera a
partir de modelos de evoluci\'on estelar, que predicen temperaturas estelares
efectivas m\'as bajas en regiones de alta metalicidad. Este punto merece un m\'as detallado
estudio.

Las regiones de formaci\'on estelar en los dos entornos estudiados muestran una
contribuci\'on proveniente de una poblaci\'on estelar subyacente. Esta
contribuci\'on es mayor en las CNSFRs y tambi\'en la poblaci\'on estelar misma
parece ser m\'as evolucionada, como evidencia la presencia de supergigantes
rojas que contribuyen sustancialmente a las l\'ineas del CaT. El hecho de que
los valores de N/O sean mucho mayores en las CNSFRs que en las galaxias \HII\
y en otras regiones de alta metalicidad, tambi\'en apunta a la posibilidad de un cierto grado de  
evoluci\'on qu\'imica y autoenriquecimiento en estas regiones.

\addcontentsline{toc}{section}{\numberline{}Bibliography}

\bibliographystyle{astron}
\bibliography{tesis}

\chapter*{Acknowledgements}

\addcontentsline{toc}{chapter}{\numberline{}Acknowledgements}

This work has been supported by Spanish DGICYT grants AYA-2004-02860-C03 and
AYA- 2007-67965-C03-03. I acknowledge support from the Spanish MEC through FPU
grants AP2003-1821. Furthermore, partial support from the Comunidad
de Madrid under grant S-0505/ESP/ 000237 (ASTROCAM) is acknowledged. I thank
the hospitality of the Institute of Astronomy, Cambridge, where part of this
work was developed, and also thank the hospitality of the INAOE and the
Laboratoire d'Astrophysique de Toulouse-Tarbes. 

The WHT is operated in the island of La Palma by the Isaac Newton Group
in the Spanish Observatorio del Roque de los Muchachos of the Instituto
de Astrof\'isica de Canarias. We thank the Spanish allocation committee
(CAT) for awarding observing time.

This research has made use of the NASA/IPAC Extragalactic Database (NED) which
is operated by the Jet Propulsion Laboratory, California Institute of
Technology, under contract with the National Aeronautics and Space
Administration. 

This research has made use of the SIMBAD database, operated at CDS, Strasbourg,
France.

\medskip
\noindent Chapter \ref{HIIgal-obs}

We are grateful to Jorge Garc\'\i a-Rojas and C\'esar Esteban for calculating 
the effects of the temperature fluctuations over the derived ionic and total
abundances. We wish to express our gratitude to Fabian Rosales for calculating
the ionic He abundances for our objects using Porter's Helium emissivities.

Funding for the creation and distribution of the SDSS Archive has been provided
by the Alfred P. Sloan Foundation, the Participating Institutions, the National
Aeronautics and Space Administration, the National Science Foundation, the US
Department of Energy, the Japanese Monbukagakusho, and the Max Planck
Society. The SDSS Web site is http://www.sdss.org.

The SDSS is managed by the ARC for the Participating Institutions. The
Participating Institutions are the University of Chicago, Fermilab, the
Institute for Advanced Study, the Japan Participation Group, The Johns Hopkins
University, the Korean Scientist Group, Los Alamos National Laboratory, the Max
Planck Institute for Astronomy (MPIA), the Max Planck Institute for
Astrophysics (MPA), New Mexico State University, the University of Pittsburgh,
the University of Portsmouth, Princeton University, the United States Naval
Observatory, and the University of Washington.

We are pleased to thank the staff at Calar Alto, and especially Felipe Hoyo,
for their assistance  during the observations. We also thank the Time
Allocation Committee for awarding observing time. 

When we mentioned to Bernard Pagel the title of the second paper of this
chapter (``Precision abundance analysis of bright \HII\ galaxies'') he said,
with his characteristic cheeky grin: ``precision abundance? but that's an
oxymoron. It will be nice if you can do it". Dear Bernard, you are sadly
missed; we dedicate this work to your memory. 

We are also grateful to the anonymous referees for his/her careful and
constructive revision of the manuscripts of the two papers of this
chapter.

\medskip
\noindent Chapter \ref{neon}

We would like to thank the referee of the paper, G. J. Ferland, for many
valuable suggestions and comments which have helped us to improve this
part of the work.

\medskip
\noindent Chapter \ref{cnsfr-obs-kine}

We are indebted to Jes\'us L\'opez who provided the data on the SDSS sample
prior to publication. We acknowledge fruitful discussions with Horacio
Dottori, Enrique P\'erez, Guillermo Bosch, Nate Bastian, Almudena
Alonso-Herrero, Enrique P\'erez-Montero and Jos\'e V\'ilchez.
We thank very much an anonymous referee for his/her careful examination
of our manuscript of the published paper. We have found the report extremely
thorough and undoubtedly it has helped to improve the contents of this part of
the work.

Some of the data presented in this part of the thesis were obtained from the
Multimission Archive at the Space Telescope Science Institute (MAST). STScI is
operated by the Association of Universities for Research in Astronomy, Inc.,
under NASA contract NAS5-26555. Support for MAST for non-HST data is provided
by the NASA Office of Space Science via grant NAG5-7584 and by other grants
and contracts.

\medskip
\noindent Chapter \ref{abundan}

We would like to thank Mike Beasley for providing the digital spectrum of the
M~31 cluster, Roberto Cid Fernandes for very helpful discussions concerning
subtraction procedures of  the underlying absorptions and Roberto Terlevich
for a careful reading of the manuscript of the paper. We also thank an
anonymous referee 
for a very careful review of this work which lead to the improvement of the
work.  

\bigskip
\bigskip

{\bf This thesis is dedicated to the memory of Bernard Pagel who was always a
source of inspiration and stimuli for me and my collaborators and with whom 
many of
the matters addressed in this work were discussed over the last twenty years
or so.}

\chapter*{Agradecimientos}

\addcontentsline{toc}{chapter}{\numberline{}Agradecimientos}

Hay mucha gente que ha hecho posible que esta tesis doctoral haya llegado no
s\'olo a terminarse, sino que ha sido fundamental a la hora de darme su apoyo
para que tomara la dif\'icil decisi\'on de viajar tan lejos para poder
aprender much\'isimas cosas, o al menos intentarlo, tanto en la parte
relacionada con el trabajo cient\'ifico como en mi crecimiento como
persona. La lista es larga, y lamentablemente es muy posible que me olvide de
alguien, pero a todos les agradezco por su apoyo, consciente o inconsciente,
que me han dado a lo largo de todos estos a\~nos, que no son s\'olo los
\'ultimos cuatro, sino que se remonta a mucho antes de que hubiera terminado
la licenciatura. 

\medskip

En primer lugar quiero agradecerle a Moni, por todo su amor y por haberme
apoyado en los momentos dif\'iciles cuando las cosas se hac\'ian cuesta
arriba. Gracias a sus consejos, su ayuda y su compa\~n\'ia estos a\~nos han
sido 
los mejores de mi vida. Y por si todo esto fuera poco, ahora tenemos a
Vale. Personita incre\'ible que es como si siempre hubiera estado con nosotros
y nos llena de una alegr\'ia inmensa. Ella tambi\'en contribuy\'o estos
\'ultimos meses a que pudiera terminar mi trabajo, esper\'andome en casa con
una sonrisa los primeros meses de su vida, cuando reci\'en hab\'ia nacido y yo
llegaba tarde por la noche despu\'es de un d\'ia largo.

Mis padres, Nelly y Carlos, fueron los primeros en apoyarme para que me
dedicara a lo que m\'as me gustaba, y con sus maneras tan distintas de
preocuparse por m\'i y de interesarse en lo que hac\'ia me incentivaron en
todo momento y me entusiasmaron y me dieron fuerzas para que estudiara
astronom\'ia y llevara adelante esta tesis. Mis hermanos, Jorge y Andr\'es,
compartieron muchos momentos conmigo, incluso en los \'ultimos tiempos que
hemos estado tan lejos, al igual que mis cu\~nadas, Silvina y Adriana, y mis
suegros, Rosa y Ra\'ul. Y quien
podr\'ia olvidarse de mis sobrinitas, Luc\'ia, Camila y Clarita, sobre todo de
Lu, que es la m\'as grande y que siempre pregunta por lo que est\'a haciendo el
t\'io Guiye y por la ``astronom\'ia''. La verdad es que una de las cosas m\'as
duras durante este tiempo lejos de la familia ha sido el perderse un mont\'on
de cosas y el no haberlas visto crecer.

\medskip

Una menci\'on aparte se merecen mis directoras, \'Angeles y Elena. Al
principio no sab\'ia muy bien que iba a pasar, y los miedos de ir casi al otro
lado del mundo a invertir tantos a\~nos de mi vida eran bastante
aterradores, a pesar de las excelentes referencias que ten\'ia de ellas. Sin
embargo, desde un primer momento se comportaron como mucho m\'as que simples
directoras de una tesis doctoral, mostrando que son excelentes personas y
haci\'endome sentir enseguida como parte de un grupo, casi una familia. Sobre
todo \'Angeles, con quien hemos compartido much\'isimas cosas buenas, a la que
consideramos con Moni mucho m\'as que una amiga, ya es parte de nuestra
familia, y es como una t\'ia para Vale.

Tambi\'en tengo que mencionar en esta parte a Roberto y Enrique, quienes han
contribuido a su manera y de distinta forma a la direcci\'on de este
trabajo. Roberto con sus discusiones llenas de ideas y su forma de hacer que
uno deduzca las cosas y lleve su capacidad al l\'imite, siempre generando un
criterio propio y pensando al m\'aximo las cosas tratando de no dejar escapar
nada que se pueda deducir con lo que uno tiene a su disposici\'on para
realizar el trabajo. Enrique, que no s\'olo ha aportado ideas, sino que me ha
ense\~nado 
mucho de su propia experiencia, y me ha dado las pautas para aprovechar las
diferencias entre los lugares donde estudi\'e para sacar el m\'aximo
provecho en mi formaci\'on. Un verdadero amigo de los que duran toda la vida,
aunque sea a la distancia. 

\medskip

En este momento tambi\'en quiero recordar a la gente que primero fueron
profesores cuando estaba en la licenciatura y luego excelentes amigos,
aquellos que fueron los primeros en alentarme y creer que pod\'ia hacer
esto. 

Primero que nadie, Ama, quien fue la primera en sugerirme que la
espectroscop\'ia y el grupo de estrellas masivas era el camino que deb\'ia
seguir, all\'a lejos en el tiempo, cuando ella era Jefe de Trabajos 
pr\'acticos de Estad\'istica Aplicada y el alumno molesto que era yo no dejaba
de hacerle preguntas. Y despu\'es, cuando ya eramos muy buenos amigos, fue la
primera en aconsejarme que hacer el doctorado fuera era lo mejor que
pod\'ia hacer si era lo que realmente quer\'ia y ten\'ia las ganas y la fuerza
necesaria para ello.

Guille y Rudolf, quienes primero fueron mis directores en la licenciatura y
despu\'es en mi comienzos en el doctorado en La Plata, y que siempre me dieron
su apoyo incondicional para venir a Espa\~na, aunque eso significara, de
momento, dejar de trabajar con ellos, y a los que puedo considerar
afortunadamente como excelentes amigos. M\'as a\'un, quiero agradecer de
coraz\'on la confianza de Guille de recomendarme a Elena, y a trav\'es de ella
a \'Angeles, para que me aceptaran como su estudiante de doctorado. Yo s\'e lo
importante que es eso y espero, y creo, no haberlo decepcionado. Sin esa
confianza nada de todo esto podr\'ia haber sido posible.

A Carlos por haber sido mi punto de apoyo en La Plata durante estos a\~nos y
con quien hemos seguido compartiendo muchas cosas a pesar de la distancia. A
Nidia, quien fue mi profesor consejero en los \'ultimos a\~nos de carrera, y
me apoy\'o y anim\'o a trabajar con Guille y Rudolf, sobre todo en los pocos
momentos dif\'iciles que tuvimos al principio, hasta que aprendimos a trabajar
en equipo y acept\'andonos tal cual somos. Tambi\'en quiero hacer un breve
comentario a la memoria de Virpi, quien a su manera me ha dejado un legado
imborrable, como a todos los que en alg\'un momento tuvieron el privilegio de
haber tenido trato con ella.

A los amigos que fueron compa\~neros durante la carrera, Gonzalo, Vero,
Andrea, Javi, Faca, Julia, Richard, Nico, Seba, Diego, Rolo, Face, Claudia,
Erika, Marcelo, Gaby y la gente del grupo de estrellas masivas en general, Ana
Mar\'ia, Patricia y Roberto, y muchos
otros, como tambi\'en Dami\'an, el rosarino cordob\'es. Y a 
los chicos de la Aut\'onoma, muchos de los cuales fueron s\'olo compa\~neros
espor\'adicos debido a situaciones particulares, pero no por eso menos
importantes, Ruben, Jes\'us, Alcione, Chiqui, Latifa, Dani, Mar\'ia, Amelia,
Mike, Ra\'ul, N\'estor, Jairo, Jose, Hector, Manuel, Marta, Yago, Luis,
Marcelo, Mercedes, Marcos y Carlos. 
Y tambi\'en a mucha gente que he conocido durante estos a\~nos y que de alguna
u otra manera han puesto su granito de arena a todo esto, como Pepe V\'ilchez,
Enrique P\'erez, Valentina Luridiana, Miguel Cervi\~no, Mar\'ia
Santos-Lle\'o, Horacio Dottori, Jes\'us L\'opez, Anna Lia Longinotti, Daniel
Rosa-Gonz\'alez. A los amigos de siempre, los de la infancia
y que todav\'ia nos mantenemos en contacto, a pesar de la distancia, Yulian,
Santiago, Pipi. Lamentablemente, me debo estar olvidando de mucha gente,
pero aunque no los mencione en estas pocas l\'ineas, est\'an presentes
t\'acitamente en todo mi trabajo y en el esfuerzo realizado.

Y no me quiero olvidar de los miembros del jurado, aunque est\'e
escribiendo estas l\'ineas despu\'es de haber defendido la tesis, que me han
dado del privilegio de ser los evaluadores de este trabajo a pesar del
esfuerzo y el tiempo que les ha llevado. Ellos han sido
Pepe V\'ilchez, Guillermo Tenorio-Tagle, Evan Skillman, Guille Bosch y Luis
Colina. 

A todos y cada uno, sinceramente, {\sc much\'isimas gracias}.

\end{spacing}

\clearpage

\bibliographystyle{astron}
\bibliography{tesis}

\end{document}